\documentclass[11pt,a4paper,twoside]{report}

\title{Testing the three massive neutrino paradigm}
\author{Jo\~ao Paulo Pinheiro}

\usepackage[T1]{fontenc}
\usepackage[utf8]{inputenc}
\usepackage{lmodern}
\usepackage[main=UKenglish, spanish]{babel}
\usepackage{amsmath, amssymb, amsfonts}
\usepackage{mathtools}
\usepackage{geometry}
\usepackage{graphicx}
\usepackage{subcaption}
\usepackage{booktabs}
\usepackage{fancyhdr}
\usepackage{hyperref}
\usepackage{xcolor}
\usepackage{multirow}
\newcommand{\Author}{Jo\~ao Paulo Pinheiro}
\newcommand{\Title}{Testing the three massive neutrino paradigm}

\usepackage{afterpage}
\usepackage{rotating}

\usepackage{tikz}
\usetikzlibrary{positioning, shapes, arrows}
\usepackage[numbers,sort&compress]{natbib}
\usepackage{ifthen}
\usepackage{enumitem}
\usepackage{slashed}
\setlist[enumerate]{leftmargin=0pt, wide=0.5\parindent, topsep=\parskip, itemsep=\topsep}

\usepackage[normalem]{ulem}

\newcommand{\dd}{\mathrm{d}}
\renewcommand{\Re}{\mathop{\mathrm{Re}}}
\renewcommand{\Im}{\mathop{\mathrm{Im}}}

\DeclareMathOperator{\Tr}{Tr}

\newcommand{\Dmq}{\Delta m^2}
\newcommand{\Eps}{\varepsilon}
\newcommand{\eVq}{\ensuremath{\text{eV}^2}}
\newcommand{\diag}{\mathop{\mathrm{diag}}}
\newcommand{\Nuc}[2][]{{\ensuremath{\ifthenelse{\equal{#1}{}}{}{\mbox{}^{#1}}\text{#2}}}}

\newcommand{\dCP}{\delta_\text{CP}}
\newcommand{\SumNu}{{\textstyle\sum} m_\nu}

\graphicspath{{Figuras/}}

\newcommand{\Epx}{\mathcal{E}}
\newcommand{\Mmed}{M_\text{med}}

\renewcommand{\Re}{\mathop{\mathrm{Re}}}
\renewcommand{\Im}{\mathop{\mathrm{Im}}}

\begin{document}

\pagestyle{fancy}
\fancyhead{} % Clear all headers/footers
\fancyhead[LE]{\nouppercase{\leftmark}}
\fancyhead[RO]{\thepage}
\fancyhead[LO]{\Author. \Title}
\fancyhead[RE]{\thepage}

\newgeometry{bottom=1cm,hmarginratio=1:1}
\begin{titlepage}

	\centering % Centre everything on the title page
	
	{\scshape % Use small caps for all text on the title page
	
	\vspace*{\baselineskip} % White space at the top of the page
	
	%------------------------------------------------
	%	Title
	%------------------------------------------------
{\fontfamily{qtm}\selectfont
\Large
PhD Thesis
\vspace*{1cm}}

	\rule{\textwidth}{1.6pt}\vspace*{-\baselineskip}\vspace*{2pt} % Thick horizontal rule
	\rule{\textwidth}{0.4pt} % Thin horizontal rule    
	\vspace{0.75\baselineskip} % Whitespace above the title
	
	{\Huge Testing the three massive neutrino paradigm\\} % Title
	{ Constraints on Neutrino Properties and Interactions from Recent Experimental Data}
	\vspace{0.75\baselineskip} % Whitespace below the title
	
	\rule{\textwidth}{0.4pt}\vspace*{-\baselineskip}\vspace{3.2pt} % Thin horizontal rule
	\rule{\textwidth}{1.6pt} % Thick horizontal rule
	
	\vspace{2\baselineskip} % Whitespace after the title block
	
	%------------------------------------------------
	%	Subtitle
	%------------------------------------------------
	
	\vspace*{2\baselineskip} % Whitespace under the subtitle
	
	%------------------------------------------------
	%	Editor(s)
	%------------------------------------------------
			
	{\huge Jo\~ao Paulo Pinheiro \\}
	
	\vspace*{4\baselineskip} % Whitespace below the editor list
	
	PhD advisors:\\
	\vspace*{0.2cm}
	{\Large
	  Prof. Dr. Maria Concepcion Gonzalez-Garcia \\
	 Prof. Dr. Michele Maltoni }

	}
	
	\vfill % Whitespace between editor names and publisher logo
	
	%------------------------------------------------
	%	Publisher
	%------------------------------------------------
	
\begin{figure}[b]
\centering
\includegraphics[width=0.5\textwidth]{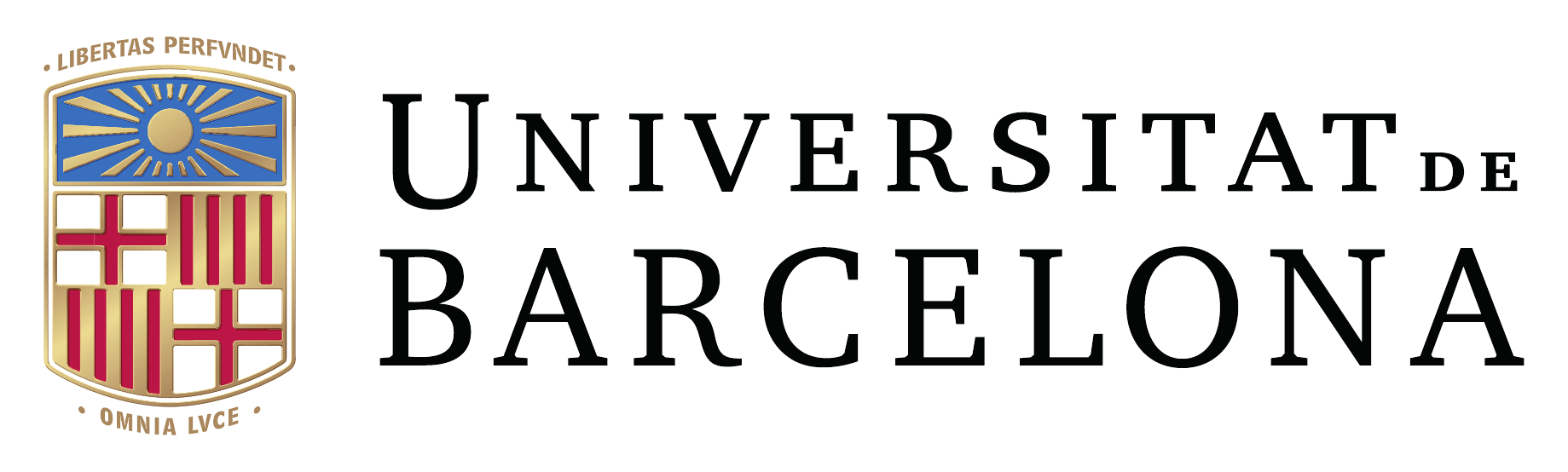}
\end{figure}

\newpage

\pagenumbering{roman}

\thispagestyle{empty}
\begin{center}
 
\vspace*{0.9cm}
{\Huge\textbf{Testing the three massive neutrino paradigm}\\} % Title
	{ \large\textbf{Constraints on Neutrino Properties and Interactions from Recent Experimental Data}\\}\vspace*{1.2cm}
{\large Memoria presentada para optar al grado de doctor por la Universidad de Barcelona \\}
\vspace*{0.2cm}
{\LARGE (Programa de Doctorado en Física)\\}
\vspace*{1.2cm}
{\Large Autor:
Jo\~ao Paulo Pinheiro \\
}
{\Large Directora:
Prof. Dr. Mar\'ia Concepci\'on Gonz\'alez Garc\'ia\\
}
{\Large Co-Director:
Prof. Dr. Michele Maltoni\\
}
{\Large Tutor:
Prof. Dr. Joan Soto Rivera\\
}
\vspace*{1.2cm}

{\large Departament de Física Quàntica i Astrofísica\\
Institut de Ci\`encies del Cosmos\\
Universitat de Barcelona\\
}

%\begin{figure}[hbtp]
%\centering
%\includegraphics[width=0.3\textwidth]{Figuras/Firma.png}
%\end{figure}

\begin{figure}[b]
\centering
\includegraphics[width=0.5\textwidth]{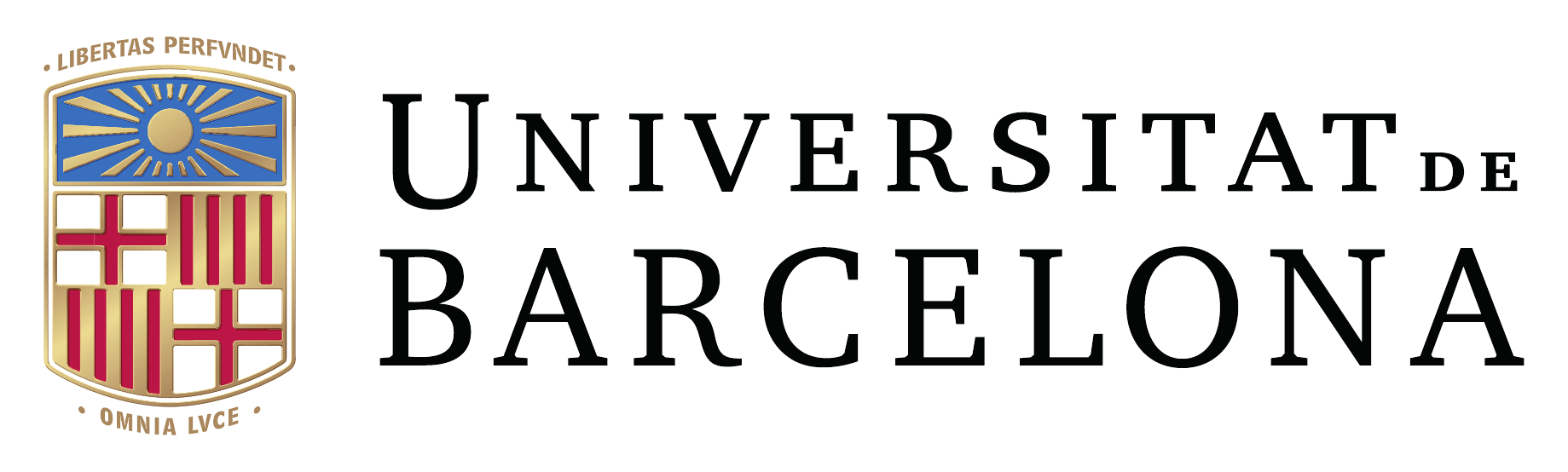}
\end{figure}\end{center}
	
\end{titlepage}
\restoregeometry

\thispagestyle{empty}
\vspace*{21cm}
\begin{flushright}
\textit{Para Eud\'alia, Edn\'olia, Jos\'e, Anna e Arya.}
\end{flushright}
\chapter*{Abstract}
\addcontentsline{toc}{chapter}{Abstract}

Neutrino physics stands at the intersection of particle physics,
astrophysics, and nuclear physics, offering unique insights into
phenomena beyond the Standard Model (BSM). This thesis presents a
synthesis of phenomenological investigations organized around three
interconnected pillars: the consolidation of the three-flavor
oscillation paradigm, the exploration of the viability of new physics
in light of the most up to date experimental data, and the precise
determination of solar neutrino fluxes. By integrating diverse
methodologies—from global oscillation fits to astrophysical flux
calculations and novel BSM constraints—this work demonstrates how
distinct analytical frameworks converge to test the limits of neutrino
physics.

This thesis starts with two introductory chapters to set the state of
the art of the discussion. In Chapter~\ref{chap:theo}, we introduce
the basic elements of the Standard Model and its extension to include
massive neutrinos, leptonic mixing and the implied flavour
transitions. In Chapter~\ref{chap:exp}, we set the
landscape of experimental results which will be included in the
rest of this thesis.  We make special emphasis in
Borexino phases II and III and NOvA experiments, with details of the
statistical analysis of each experiment, which is an original
contribution of the author to the collaboration.

The first pillar establishes the three-flavor oscillation framework
through a comprehensive global analysis of neutrino data from solar,
atmospheric, reactor, and accelerator experiments. In
Chapter~\ref{cap:nufit} we present an updated determination of mixing angles ($\theta_{12}$, $\theta_{13}$,
$\theta_{23}$) and mass-squared differences ($\Delta m^2_{21}$,
$\Delta m^2_{31}$), resolving tensions between datasets and
quantifying persistent ambiguities in the neutrino mass ordering and
$\theta_{23}$ octant. These analyses incorporate results up to
September 2024, solidifying the overall robustness of the three-flavor
paradigm, originally presented in Ref.~\cite{nufit}.

The second pillar conducts a systematic investigation of BSM physics
through recent experimental results, integrating terrestrial and
astrophysical neutrino observations.  Chapters~\ref{chap:theo_bsm}
and~\ref{chap:exp_bsm} are dedicated to laying the ground for the
theoretical framework of the studied BSMs and the experimental
results, based on
Refs.~\cite{Coloma:2022umy, Coloma:2023ixt, Ansarifard:2024zxm}. A
unified analytical framework simultaneously addresses Non-Standard
Interactions (NSI) with electrons and quarks, differentiating between
non-standard matter effects and detection-process
modifications. Precision data analysis from Borexino experiment
constrains electron-coupled NSI parameters, with potential to exclude
Large Mixing Angle-Dark (LMA-D) solutions through matter-enhanced
oscillations. Complementary constraints on quark-coupled NSI emerge
from COHERENT's coherent elastic neutrino-nucleus scattering
(CE$\nu$NS) measurements, sensitive to nuclear recoil kinematics. The
global analysis disentangles parameter degeneracies by combining
oscillation data (solar, atmospheric, reactor, accelerator) with
CE$\nu$NS cross-section measurements, establishing distinct bounds on
propagation and detection couplings. This synthesis demonstrates how
solar neutrino observations primarily constrain flavor-diagonal
electron interactions, while CE$\nu$NS measurements probe universal
quark couplings, with joint analyses excluding previously viable
regions of NSI parameter space. The framework further distinguishes
between light mediator scenarios through energy-dependent signatures
in oscillation versus scattering datasets.

The third pillar advances solar neutrino physics through precision
flux determinations by using updated measurements of pp-chain and
CNO-cycle neutrinos to resolve fundamental questions in stellar
astrophysics. Chapter~\ref{chap:SSM_BX} is dedicated to explain the
state of the art of Standard Solar Models, the determination of solar
fluxes and the incompatibility between the resolution of the gallium
source experiment anomaly and solar neutrino data by means of a light
sterile neutrino. It compiles the results from
Refs.~\cite{Gonzalez-Garcia:2023kva,Gonzalez-Garcia:2024hmf}. By
integrating the latest solar and non-solar neutrino data—including the
first direct detection of CNO-cycle neutrinos—this analysis
disentangles solar model uncertainties from oscillation parameters,
constraining metallicity-dependent energy production mechanisms. The
analysis aims to resolve tensions between high- and low-metallicity
compositional scenarios predicted by different Standard Solar Models
(SSM), while testing sterile neutrino explanations for the Gallium
anomaly. Our analysis resulted in a statistical preference for SSM
that predicts high-metalicity and systematic consistency checks
demonstrate incompatibility between $3+1$ mixing parameters favored by
Gallium experiments and those allowed by solar neutrino observations,
irrespective of astrophysical or reactor flux assumptions. These
results establish solar neutrinos as dual diagnostics of particle
physics and stellar processes, linking precision flux measurements to
both solar composition models and the limits of BSM scenarios.

The convergence of these distinct analytical threads—oscillation
parameter fits, BSM constraints, and astrophysical flux
determinations—forms the core contribution of this work. By rigorously
testing the three-flavor framework while probing its boundaries, this
thesis provides a blueprint for interdisciplinary neutrino
physics. The synthesis of disparate datasets and theoretical
approaches underscores the role of neutrinos as both messengers of
astrophysical processes and laboratories for fundamental physics,
guiding future experiments toward resolving outstanding questions in
mass ordering, CP violation, and dark sector interactions.

\chapter*{Resum}
\addcontentsline{toc}{chapter}{Resum}
\large \textbf{Provant el paradigma dels tres neutrins massius. Restriccions sobre les propietats i interaccions dels neutrins a partir de dades experimentals recents.}
\vspace{1.5cm}

La física de neutrins es troba a la intersecció entre la física de partícules, l'astrofísica i la física nuclear, oferint una visió única de fenòmens més enllà del Model Estàndard (BSM, per les seves sigles en anglès). Aquesta tesi presenta una síntesi d'investigacions fenomenològiques organitzades al voltant de tres pilars interconnectats: la consolidació del paradigma d'oscil·lació de tres sabors, l'exploració de la viabilitat de nova física a la llum de les dades experimentals més recents i la determinació precisa dels fluxos de neutrins solars. Mitjançant la integració de metodologies diverses—des d'ajustos globals d'oscil·lació fins a càlculs de fluxos astrofísics i noves restriccions BSM—aquest treball demostra com diferents marcs analítics convergeixen per provar els límits de la física de neutrins.

Aquesta tesi comença amb dos capítols introductoris per tal d'establir quin és l'estat del debat sobre el tema. Al Capítol~\ref{chap:theo} introduïm els elements bàsics del Model Estàndard i la seva extensió per incloure neutrins massius, la mescla leptònica i les transicions de sabor implicades. Al Capítol~\ref{chap:exp} descrivim l'espectre de resultats experimentals que s'inclouran en la resta de la tesi. Posem especial èmfasi en les fases II i III de Borexino i en l'experiment NOvA, amb detalls de l'anàlisi estadística de cada experiment, que constitueixen una contribució original de l'autor a la col·laboració.

El primer pilar estableix el marc d'oscil·lació de tres sabors mitjançant una anàlisi global exhaustiva de dades de neutrins solars, atmosfèrics, de reactors i d'acceleradors. Al Capítol~\ref{cap:nufit} presentem una determinació actualitzada dels angles de mescla ($\theta_{12}$, $\theta_{13}$, $\theta_{23}$) i les diferències de massa al quadrat ($\Delta m^2_{21}$, $\Delta m^2_{31}$), resolent discrepàncies entre conjunts de dades i quantificant ambigüitats persistents en l'ordenació de masses dels neutrins i l'octant de $\theta_{23}$. Aquestes anàlisis incorporen resultats fins a setembre de 2024, consolidant la robustesa general del paradigma de tres sabors, originalment presentat a la Ref.~\cite{nufit}.

El segon pilar realitza una investigació sistemàtica de física BSM a través de resultats experimentals recents, integrant observacions de neutrins terrestres i astrofísics. Els Capítols~\ref{chap:theo_bsm} i~\ref{chap:exp_bsm} estan dedicats a establir les bases del marc teòric dels BSM estudiats i els resultats experimentals, basats en les Refs.~\cite{Coloma:2022umy, Coloma:2023ixt, Ansarifard:2024zxm}. Un marc analític unificat aborda simultàniament les Interaccions No Estàndard (NSI) amb electrons i quarks, diferenciant entre efectes de matèria no estàndard i modificacions en el procés de detecció. L'anàlisi de dades de precisió de l'experiment Borexino constreny els paràmetres NSI acoblats a electrons, amb el potencial d'excloure solucions LMA-D (Large Mixing Angle-Dark) mitjançant oscil·lacions amplificades per matèria. Restriccions complementàries sobre NSI acoblats a quarks provenen de mesures de dispersió coherent elàstica neutrí-nucli (CE$\nu$NS) de COHERENT, sensibles a la cinemàtica de retrocés nuclear. L'anàlisi global destria degeneracions de paràmetres combinant dades d'oscil·lació (solars, atmosfèriques, de reactors, d'acceleradors) amb mesures de secció eficial CE$\nu$NS, establint límits diferencials per als acoblaments de propagació i detecció. Aquesta síntesi demostra com les observacions de neutrins solars restringeixen principalment les interaccions electòniques diagonals en sabor, mentre que les mesures CE$\nu$NS exploren acoblaments universals amb quarks, amb anàlisis conjuntes que exclouen regions prèviament viables de l'espai de paràmetres NSI. El marc distingeix a més entre escenaris de mediadors lleugers mitjançant signatures dependents de l'energia en conjunts de dades d'oscil·lació versus dispersió.

El tercer pilar anticipa la física de neutrins solars mitjançant determinacions precises de fluxos, utilitzant mesures actualitzades de neutrins de la cadena pp i el cicle CNO per resoldre qüestions fonamentals en astrofísica estel·lar. El Capítol~\ref{chap:SSM_BX} s'encarrega d'explicar quina és la situació més recent dels Models Solars Estàndard (SSM), la determinació de fluxos solars i la incompatibilitat entre la resolució de l'anomalia de l'experiment de font de gal·li i les dades de neutrins solars mitjançant un neutrí estèril lleuger. Recopila els resultats de les Refs.~\cite{Gonzalez-Garcia:2023kva,Gonzalez-Garcia:2024hmf}. Integrant les últimes dades de neutrins solars i no solars—incloent la primera detecció directa de neutrins del cicle CNO—aquesta anàlisi deslliga les incerteses dels models solars dels paràmetres d'oscil·lació, restringint mecanismes de producció d'energia dependents de la metal·licitat. L'anàlisi pretén resoldre tensions entre escenaris de composició d'alta i baixa metal·licitat predits per diferents SSM, mentre prova explicacions basades en neutrins estèrils per a l'anomalia del Gal·li. La nostra anàlisi va resultar en una preferència estadística pels SSM que prediuen alta metal·licitat, i les comprovacions de consistència sistemàtica demostren incompatibilitat entre els paràmetres de mescla $3+1$ afavorits pels experiments de Gal·li i els permesos per les observacions de neutrins solars, independentment de suposicions astrofísiques o de fluxos de reactors. Aquests resultats estableixen els neutrins solars com a diagnòstics duals de física de partícules i processos estel·lars, vinculant mesures precises de fluxos amb models de composició solar i límits d'escenaris BSM.

La convergència d'aquests perspectives analítics diferents—ajustos de paràmetres d'oscil·lació, restriccions BSM i determinacions de fluxos astrofísics—forma la contribució central d'aquest treball. En provar rigorosament el marc de tres sabors mentre s'exploren els seus límits, aquesta tesi proporciona un pla per a la física interdisciplinària de neutrins. La síntesi de conjunts de dades i enfocaments teòrics dispars subratlla el paper dels neutrins tant com a missatgers de processos astrofísics com a laboratoris per a física fonamental, guiant experiments futurs cap a la resolució de qüestions pendents en ordenació de masses, violació de CP i interaccions del sector fosc.
\vspace{1cm}

\hspace{-0.5cm}
\textbf{PARAULES CLAU:} Física de particulas(2203.08), Fisica Teorica(2203.10),Astronomia y Astrofisica(2503.00).

\tableofcontents
\cleardoublepage

\chapter*{List of publications}
\addcontentsline{toc}{chapter}{List of publications}

The original contents of this thesis are based on the following publications:

\begin{itemize}
\item P. Coloma, M. C. Gonzalez-Garcia, M. Maltoni, \textbf{J. P. Pinheiro}, S. Urrea, 
``Constraining new physics with Borexino Phase-II spectral data'',
\textit{J. High Energ. Phys.} \textbf{07} (2022) 138

\item P. Coloma, M. C. Gonzalez-Garcia, M. Maltoni, \textbf{J. P. Pinheiro}, S. Urrea, 
``Global constraints on non-standard neutrino interactions with quarks and electrons'',
\textit{J. High Energ. Phys.} \textbf{08} (2023) 032

\item M. C. Gonzalez-Garcia, M. Maltoni, \textbf{J. P. Pinheiro}, A. M. Serenelli, 
``Status of Direct Determination of Solar Neutrino Fluxes after Borexino'',
\textit{J. High Energ. Phys.} \textbf{02} (2024) 064

\item S. Ansarifard, M. C. Gonzalez-Garcia, M. Maltoni, \textbf{J. P. Pinheiro}, 
``Solar neutrinos and leptonic spin forces'',
\textit{J. High Energ. Phys.} \textbf{07} (2024) 172

\item I. Esteban, M. C. Gonzalez-Garcia, M. Maltoni, I. Martinez-Solar, \textbf{J. P. Pinheiro}, T. Schwetz, 
``NuFit-6.0: Updated global analysis of three-flavor neutrino oscillations'',
\textit{J. High Energ. Phys.} \textbf{12} (2025) 216
\item M. C. Gonzalez-Garcia, M. Maltoni, \textbf{J. P. Pinheiro}, 
``Solar Model Independent Constraints on the Sterile Neutrino Interpretation of the Gallium Anomaly'',
\textit{Phys. Lett. B} \textbf{862} (2025) 139297
\end{itemize}

The following publications have been developed in parallel to the content of this thesis, although they have not been included in this dissertation:

\begin{itemize}
\item \textbf{J. P. Pinheiro}, C. A. S. Pires, 
``Vacuum stability and spontaneous violation of the lepton number at a low-energy scale in a model for light sterile neutrinos'',
\textit{Phys. Rev. D} \textbf{102} (2020) 015015

\item \textbf{J. P. Pinheiro}, C. A. S. Pires, F. Queiroz, Y. Villamizar, 
``Confronting the inverse seesaw mechanism with the recent muon $g-2$ result'',
\textit{Phys. Lett. B} \textbf{823} (2021) 136764

\item \textbf{J. P. Pinheiro}, C. A. S. Pires, 
``On the Higgs spectra of the 3-3-1 model'',
\textit{Phys. Lett. B} \textbf{833} (2022) 137584

\item \textbf{J. P. Pinheiro}, C. A. S. Pires, 
``On the Higgs spectra of the 3-3-1 model with the sextet of scalars engendering the type II seesaw mechanism'',
\textit{Nucl. Phys. B} \textbf{993} (2023) 116293

\item A. Doff, \textbf{J. P. Pinheiro}, C. A. S. Pires, 
``Exploring solutions to the muon $g-2$ anomaly in a THDM-like model under flavor constraints'',
\textit{arXiv:2405.05839} [hep-ph]

\item D. Be\v{c}irevi\'c, F. Jaffredo, \textbf{J. P. Pinheiro}, O. Sumensari, 
``Lepton flavor violation in exclusive $b\to d\ell_i\ell_j$ and $b\to s\ell_i\ell_j$ decay modes'',
\textit{Phys. Rev. D} \textbf{110} (2024) 075004

\item J. Gehrlein, P. A. N. Machado, \textbf{J. P. Pinheiro}, 
``Constraining non-standard neutrino interactions with neutral current events at long-baseline oscillation experiments'',
\textit{JHEP} \textbf{05} (2025) 065

\item A. Doff, \textbf{J. P. Pinheiro}, C. A. S. Pires, 
``Leptoquark-induced radiative masses for active and sterile neutrinos within the framework of the 3-3-1 model'',
\textit{Phys. Lett. B} \textbf{863} (2025) 139375

\item P. Escalona, \textbf{J. P. Pinheiro}, A. Doff, C. A. S. Pires  
``Meson Mixing Bounds on $Z'$ Mass in the Alignment Limit: Establishing the Phenomenological Viability of the 331 Model'',
\textit{arXiv:2503.14653 } [hep-ph]

\end{itemize}

\cleardoublepage
\chapter*{List of abbreviations}
\addcontentsline{toc}{chapter}{List of abbreviations}

\begin{table}[htb!]
\centering
\begin{tabular}{llll}
\textbf{BSM} & Beyond the Standard Model &  & \\
\textbf{CE$\boldsymbol{\nu}$NS} & Coherent elastic neutrino-nucleus scattering &  & \\
\textbf{CP} & Charge Parity &  & \\
\textbf{CC NSI} & Charged Current Non-Standard Interactions &  & \\
\textbf{ESS} & European Spallation Source &  & \\
\textbf{IO} & Inverted ordering &  & \\
\textbf{LBL} & Long baseline &  & \\
\textbf{NC NSI} & Neutral Current Non-Standard Interactions &  & \\
\textbf{NO} & Normal ordering&  & \\
\textbf{NP} & New Physics&  & \\
\textbf{NSI} & Non-Standard Interactions &  & \\
\textbf{QF} & Quenching Factor &  & \\
\textbf{SBL} & Long baseline &  & \\
\textbf{SM} & Standard Model &  & \\
\textbf{SNS} & Spallation Neutron Source &  & \\
\textbf{SSM} & Standard Solar Model &  & \\
\end{tabular}
\end{table}
\cleardoublepage

\pagenumbering{arabic}

\chapter{Introduction}
\label{chap:intro}
%\addcontentsline{toc}{chapter}{Introduction}

There are many ways to start the discussion of the Standard Model and
the role of neutrinos in it. In this thesis, I choose to begin with a
chronological one. Neutrinos are essential to understand all
radioactivity phenomena. As described in detail in
\cite{becquerel_radioactivity_discovery}, in 1896 Henri Becquerel
started to methodologically explore the interaction of Uranium with
different materials. He was intrigued by this ``magical'' behavior of
Uranium crystals that spontaneously emit x-rays from it, discovered
by Wilhelm Röntgen in 1895. He thought that this strange material
would absorb energy emitted by the Sun and after it, releases it in the
form of x-rays. In his diary \cite{becquerel_radioactivity_discovery},
he wrote that one day was very cloudy and he wrapped black paper around
photographic plates together with uranium salt and placed it inside
the drawer of his laboratory, away from sunlight. One day after, to
his surprise, he noticed that the plates had been exposed, indicating
that the Uranium salts emitted radiation spontaneously, without the
need for exposure to light. And how such discovery is associated with
neutrinos? Let us advance a little bit on time.

According to the book ``From X rays to
quarks"\cite{segre_rays_to_quarks} written by Emilio Segr\`e, just
after Becquerel's discovery, experimentalists coupled electric and
magnetic field generator machines in order to separate the content of
the energy released. After applying the fields, three different group
of particles were identified: $\alpha$ particles, a positively charged beam without
penetration properties which could be reflected
by a thin paper; X rays, that are not deflected by electric fields
and have good penetration in some materials, but scatter with bones
and metals; and $\beta$ particles. These could be positively or
negatively charged, with a stronger penetration property than $\alpha$
particles.

In 1914, James Chadwick observed that the primary beta spectrum is
continuum\cite{segre_rays_to_quarks}. This means that, for example, a
Thorium-234 decays into protactinium-234 and a beta particle. From a
simple two-body decay, using conservation of energy, the beta particle
should be monochromatic!

After different interpretations of the strange experimental result, it
appears the genius idea of Wolfgang Pauli in his famous letter for the
``Radioactives"\cite{winter_neutrino_physics}. There, he suggested ``I
have hit upon a desperate remedy to save the “exchange theorem” of
statistics and the law of conservation of energy. Namely, the
possibility that there could exist in the nuclei electrically neutral
particles, that I wish to call neutrons, which have spin $1/2$ and
obey the exclusion principle ... The continuous spectrum would then
become understandable by the assumption that in decay, a neutron is
emitted in addition to the electron such that the sum of the energies
of the neutron and electron is constant. Now the question that has to
be dealt with is which forces act on the neutrons?..." With this the
{\sl neutrino} was born into the theoretical physics.

In fact, some years after Pauli's letter, Chadwick discovered a neutral
spin $1/2$ particle, heavy, that participates in the radioactive
interactions. He named it neutron. But this neutron  was not the light
emitted state postulated by Pauli. In 1934, Enrico Fermi
\cite{winter_neutrino_physics, mohapatra_massive_neutrinos}
renamed the light neutral emitted state as {\sl neutrino} and he 
wrote down the first field theoretical form of an interaction
involving the neutron, the proton, electron and neutrino fields that
would describe weak interactions. Schematically 
\begin{equation}
    {\cal H}_\mathrm{weak}^{CC} =
    \dfrac{4G_F}{\sqrt{2}}\Big[J^\mu(x)J_\mu(x)\Big],
\end{equation}
such that $J^\mu(x)$ represents the charged current (CC) 
\begin{equation}
    J_\mu(x)\sim\bar{p}\gamma_\mu  n + \bar{\ell}\gamma_\mu \nu
\end{equation}
As we will see below  the Lorentz structure
proposed by Fermi was not quite correct. Also we now know that  the
fundamental  interaction can be described not in terms of hadrons
like protons and neutrons, but quarks.

Still from the interaction described above, it was possible to write the
decay of a neutron into a proton, an electron and an electron
antineutrino as:
\begin{equation}
    n(udd)\to p(uud) + e^- +\bar{\nu}_e
\end{equation}  
unraveling the  fundamental understanding of  all radioactive
phenomena.

After all our discussion, we are able to describe all physics behind
what happened to Becquerel on that cloudy day. It is well known that
Uranium is a heavy element whose atomic nucleus is unstable, making it
radioactive. It undergoes radioactive decay processes involving the
emission of subatomic particles.  The uranium nucleus emits an
$\alpha$ particle, which consists of 2 protons and 2 neutrons (a
helium nucleus).

The most common material studied by Becquerel was Uranium-238
(\(^{238}\mathrm{U}\)). It typically decays into thorium-234
(\(^{234}\mathrm{Th}\)):
        \[
        ^{238}\mathrm{U} \rightarrow ^{234}\mathrm{Th} + \alpha
        \].

Thorium-234 (\(^{234}\mathrm{Th}\)), produced via alpha decay,
undergoes beta decay to form protactinium-234 (\(^{234}\mathrm{Pa}\)):
        \[
        ^{234}\mathrm{Th} \rightarrow ^{234}\mathrm{Pa} + \beta^- +
        \bar{\nu}_e
        \]
% A neutron in the nucleus decays into a proton, emitting an electron (\(\beta^-\)) and an electron antineutrino (\(\bar{\nu}_e\))  After alpha or beta decay, the daughter nucleus is often left in an excited state. It releases excess energy as electromagnetic radiation (gamma rays).

%falar do experimento de decaimento beta inverso e deteccao de neutrinos livres feito por Reines e Cowan

%The choice of \(SU(2)_L\) stems from the weak interactions, which are observed to violate parity (\(P\)) and charge conjugation (\(C\)) symmetry. These violations were first detected in the beta decay of cobalt nuclei by Wu et al. (1957), which confirmed that weak interactions only involve left-handed particles. The violation of \(C\) symmetry was later observed in kaon decay. Together, these phenomena suggest an intrinsic asymmetry in the weak interaction, described naturally by \(SU(2)_L\), where only left-handed fermions participate.

In this process, a neutron in the nucleus decays into a proton,
emitting an electron ($\beta^-$) and an electron antineutrino
($\bar{\nu}_e$). After alpha or beta decay, the daughter nucleus is
often left in an excited state, releasing its excess energy as gamma
radiation.

\subsection*{The Inverse Beta Decay Experiment}
The existence of free neutrinos was confirmed through the inverse beta
decay experiment conducted by Frederick Reines and Clyde Cowan in 1956
\cite{reines_cowan_experiment}. They observed the reaction:

\begin{equation}
    \bar{\nu}_e + p \rightarrow e^+ + n,
\end{equation}
where an electron antineutrino ($\bar{\nu}_e$) interacts with a proton
to produce a positron ($e^+$) and a neutron. This experiment was
conducted using a nuclear reactor as a source of antineutrinos and a
detector consisting of a liquid scintillator to capture the positrons
and neutrons. The detection of the positron and neutron provided
direct evidence for the existence of neutrinos and validated Fermi's
theory of weak interactions, which will be described below.

The inverse beta decay experiment confirmed the reality of neutrinos,
which had been postulated by Pauli. The success of this experiment
also demonstrated the power of neutrino detection techniques, which
have since become essential tools in studying weak interactions and
neutrino properties.

\subsection*{Parity Violation and the V-A Structure}

The structure of weak interactions was further clarified by the
discovery of parity violation. In 1956, Chien-Shiung Wu and
collaborators conducted an experiment involving the beta decay of
cobalt-60 nuclei, which demonstrated that weak interactions violate
parity symmetry ($P$) \cite{wu_parity_violation}. This means that weak
interactions are not invariant under spatial inversion. The experiment
measured the angular distribution of electrons emitted from polarized
cobalt-60 nuclei and found an asymmetry in the emission relative to
the nuclear spin direction. This asymmetry indicated that the weak
force distinguishes between left-handed and right-handed systems, a
clear violation of parity symmetry\cite{winter_neutrino_physics}.

The violation of parity was a profound discovery, as it revealed that
the weak interaction is intrinsically chiral. This led to the
formulation of the $V-A$ (vector minus axial-vector) structure of the
weak currents, proposed by Marshak and Sudarshan and independently by
Feynman and Gell-Mann\cite{winter_neutrino_physics}. The $V-A$
structure implies that the weak interaction couples only to
left-handed particles and right-handed antiparticles. Mathematically,
the weak current $J^\mu$ is then expressed as:
\begin{equation}
    J_\mu(x)=\bar{u}\gamma_\mu P_L  d+
    \bar{\ell}\gamma_\mu P_L \nu
\label{eq:ccj}
\end{equation}
where the left and right chirality projection operators are defined as
\begin{eqnarray}
    P_L=\frac{1}{2}(1-\gamma_5),&\,\,\,\,\,&P_R=\frac{1}{2}(1+\gamma_5).
\end{eqnarray}
The $P_L$ in Eq.~\eqref{eq:ccj}   ensures that only
the left-handed components of the fermion fields participate in weak
CC interactions. This chiral nature of the weak force is a fundamental
feature that distinguishes it from other interactions, such as
electromagnetism and the strong force, which are parity-conserving.

Despite its successes, the $V-A$ theory faced several theoretical and
experimental challenges. From a theoretical point of view, the
four-fermion interaction described by the $V-A$ theory is
non-renormalizable. This means that calculations at high energies or
short distances lead to infinities that cannot be systematically
removed, making the theory inconsistent as a fundamental description
of weak interactions. At the same time, at high energies, the
cross-section for weak processes predicted by the $V-A$ theory grows
without bound, violating unitarity.  And since the $V-A$ theory is
effective, it does not explain the mechanism by which weak
interactions are mediated. Unlike electromagnetism, which is mediated
by the photon, the four-fermi theory lacks a corresponding gauge boson.

These issues pointed to the need for a deeper theory that could
address the shortcomings of the $V-A$ framework. This led to the
development of the \textbf{Standard Model (SM)} of particle physics.

\chapter{Electroweak theory and the neutrino mass}
\label{chap:theo}

\section{Electroweak theory of Massless Neutrinos}
The Standard Model (SM) of the elementary particles, proposed by
Weinberg, Glashow and Salam
\cite{weinberg1967model,salam1968elementary,glashow1961partial} is the
most precisely tested fundamental theory with predictions matching
observations to $10^{-11}$ accuracy ~\cite{patrignani2016review}.  In
this thesis, we will not go into all the details of the SM.  We are
going to focus on the Electroweak sector the aspects which are
directly associated with neutrinos.

The Standard Model is a gauge theory constructed under the gauge
symmetry principle which states that the laws of physics arise as
a consequence of invariance under local (aka gauge) transformations
of some symmetry group \cite{cheng1984gauge}. In these constructions 
the interactions are mediated by spin-one particles, called gauge bosons,
in correspondence with  the generators of the gauge group.

How is this machinery associated with neutrinos? To make
a gauge theory comprehending the neutrino interactions one must first
find a gauge symmetry group  that is compatible with what we know
of those interactions, in particular  it  must violate parity
\cite{wu1957experimental} and mediate $V-A$ interactions between the
 left-handed neutrinos and electrons so the mediator must
be electrically charged. This can be achieved with a symmetry group
$SU(2)$ transforming only the left-handed fields, hence labeled
$SU(2)_L$ (\cite{lee1956question}.

More technically, the general way to interpret the 4-fermion
interaction is via the exchange of the gauge boson associated with the
$T^\pm\propto T^1\pm i T^2$ generators of  the group $SU(2)_L$, denoted as
$\hat{W}^\pm$, which transforms $\nu_L$ into $\ell_L$ (and quark $u_L$
into quark $d_L$) and vice versa. This $SU(2)_L$ group is further combined with $U(1)_Y$, a {\sl
hypercharge} symmetry group, with  
$Y$ related to the electric
charge $Q$ through the third generator $T_3$ of $SU(2)_L$
as \cite{gell1964symmetries} 
\begin{equation}
\label{eq:hypercharge_electcharge}
    Q = T_3 + \frac{Y}{2}.
\end{equation}
The resulting {\sl electroweak} unified theory has four generators and 
predicts the existence of three gauge bosons, $W^\pm$ and $Z^0$, as
well as the photon, all of which have been experimentally confirmed
\cite{glashow1970weak}.

The Standard Model (SM) consists then of three generations of fermions,
and four gauge bosons. In addition, the model also contains a fundamental
scalar, the Higgs boson, needed for spontaneous symmetry breaking
and mass generation.  We list their transformation properties under
the gauge group $SU(2)_L \times U(1)_Y$
\cite{halzen1984quarks} in Table
~\ref{tab:smcont}. We notice that the table does not contain a
right-handed neutrino, $\nu_R$,
because such a state would be a total singlet of the gauge
interactions: as seen from Eq.~\eqref{eq:hypercharge_electcharge},
its hypercharge,  charge, colour and weak-isospin would
all be zero.

\begin{table}[!ht]
\centering
\begin{tabular}{|c|c|c|c|c|}
\hline \textbf{Particle} & \textbf{$SU(2)_L$} & \textbf{$Y$} &
\textbf{$Q$} & \textbf{Type} \\ \hline $
Q_{L,i}\equiv \begin{pmatrix} u_{L}
  \\ d_L \end{pmatrix}$,$\begin{pmatrix} c_{ L}
  \\ s_L \end{pmatrix}$,$\begin{pmatrix} t_{L} \\ b_L \end{pmatrix}$ &
Doublet & $+\frac{1}{3}$ & $u: +\frac{2}{3}, \, d: -\frac{1}{3}$ &
Quarks \\ \hline $u_R, c_R, t_R$ & Singlet & $+\frac{4}{3}$ &
$+\frac{2}{3}$ & Quarks \\ \hline $d_R, s_R, b_R$ & Singlet &
$-\frac{2}{3}$ & $-\frac{1}{3}$ & Quarks \\ \hline $L_{L,i}\equiv\begin{pmatrix}
  \nu_{eL} \\ e_L \end{pmatrix}$,$\begin{pmatrix} \nu_{\mu L}
  \\ \mu_L \end{pmatrix}$,$\begin{pmatrix} \nu_{\tau L}
  \\ \tau_L \end{pmatrix}$ & Doublet & $-1$ & $\nu: 0, \, e: -1$ &
Leptons \\ \hline $e_R, \mu_R, \tau_R$ & Singlet & $-2$ & $-1$ &
Leptons \\ \hline $\phi\equiv\begin{pmatrix} G^+ \\ \frac{v + h + i
    G_0}{\sqrt{2}} \end{pmatrix}$ & Doublet & $+1$ & $G^+: +1, \,
h,G_0: 0$ & Higgs \\ \hline
\end{tabular}
\caption{Quantum numbers for the different representations of the
  gauge group, associated to the Higgs boson and the fermions
  belonging to each family in the SM. Fermion families are ordered
  attending to their masses: the third one is the most massive, while
  the first one contains the lightest fermions.}
\label{tab:smcont}
\end{table}

\subsection{The Electroweak Lagrangian}
The electroweak Lagrangian constructed  
with the matter contents of Table~\ref{tab:smcont} and the gauge symmetry group $SU(2)_L \times U(1)_Y$ can be written as:
\begin{eqnarray}
\mathcal{L}_{\text{EW}} = \mathcal{L}_{\text{Gauge}} +
\mathcal{L}_{\text{Fermion}} +
\mathcal{L}_{\text{Higgs}}+\mathcal{L}_{\text{Yukawa}},
\label{eq:EWLAG}    
\end{eqnarray}
where each term is defined as follows:
\begin{itemize}
    \item $\mathcal{L}_{\text{Gauge}}\to$ Describes the kinetic terms
      for the gauge fields and their self-interactions.
    \item $\mathcal{L}_{\text{Fermion}}\to$ Represents the
      interactions between fermions and gauge bosons.
    \item $\mathcal{L}_{\text{Higgs}}\to$ Encodes the dynamics of the
      Higgs field and its interactions with itself and with gauge
      bosons, leading to spontaneous symmetry breaking and the generation
      of the gauge boson masses.
    \item $\mathcal{L}_{\text{Yukawa}}\to$ Describes the interaction
      between the Higgs field and fermions, leading to their mass
      generation.
\end{itemize}
\subsubsection{Gauge Field Lagrangian}
\begin{equation}
\mathcal{L}_{\text{Gauge}} = -\frac{1}{4} W_{\mu\nu}^a W^{a,\mu\nu} -
\frac{1}{4} B_{\mu\nu} B^{\mu\nu},
\label{gauge_boson_dyn}
\end{equation} with $W_{\mu\nu}^a$ and $B_{\mu\nu}$ are the field strength
tensors of $SU(2)_L$ and $U(1)_Y$, respectively.
These terms describe the dynamics of the gauge bosons.
\subsubsection{Fermion Lagrangian}
The fermions are organized into doublets under $SU(2)_L$ and singlets
under $U(1)_Y$ and there are three families of each fermion. The
interaction terms are given by:
\begin{equation}
\mathcal{L}_{\text{Fermion}}^\psi = \sum_{\text{generations}}
\bar{\psi}_L i \gamma^\mu \left( \partial_\mu - i g \frac{\tau^a}{2}
W_\mu^a - i g' \frac{Y}{2} B_\mu \right) \psi_L + \bar{\psi}_R i
\gamma^\mu \left( \partial_\mu - i g' \frac{Y}{2} B_\mu \right) \psi_R,
\label{eq:boson_fermion_lag}
\end{equation}
where $g$ and $g'$ are the gauge couplings and  $\tau^a/2$.

\subsubsection{Higgs Lagrangian and Spontaneous Symmetry Breaking}
The Higgs mechanism introduces a scalar doublet under $SU(2)_L$, which
will induce the breaking of  the gauge symmetry as
\[ SU(2)_L \times U(1)_Y \to U(1)_{em}, \]
where $U(1)_{em}$ represents the electromagnetic interaction. This
mechanism gives mass to the weak bosons ($W^\pm, Z^0$) while leaving
the photon massless, consistent with experimental observations.

The Higgs field $\phi$ is introduced as a complex scalar doublet under
$SU(2)_L$:
\begin{equation}
\label{eq:higgs_def}
\phi = \begin{pmatrix} \phi^+ \\ \phi^0
\end{pmatrix}.
\end{equation}
Its Lagrangian is:
\begin{equation}
\mathcal{L}_{\text{Higgs}} = (D_\mu \phi)^\dagger (D^\mu \phi) -
V(\phi),
\label{eq:lag_higgs}
\end{equation}
where $D_\mu$ is the covariant derivative:
\begin{equation}
    D_\mu = \partial_\mu - i g \frac{\tau^a}{2} W_\mu^a - i g'
    \frac{Y}{2} B_\mu,
\label{eq:covderiv}
\end{equation}
and the potential $V(\phi)$ is:
\begin{equation}
V(\phi) = \mu^2 \phi^\dagger \phi + \lambda (\phi^\dagger \phi)^2.
\label{eq:pot_higgs}
\end{equation}

When $\mu^2 < 0$, the Higgs field acquires a vacuum expectation value
(VEV):
\[
\langle \phi \rangle = \frac{1}{\sqrt{2}} \begin{pmatrix} 0 \\ v
\end{pmatrix}.
\]
Quantizing above this ground state induces the 
spontaneous breaking of  $SU(2)_L \times U(1)_Y \to U(1)_{em}$ and
it generates masses fo the gauge  bosons:
\[
M_W = \frac{gv}{2}, \quad\quad
M_Z = \frac{\sqrt{g^2 + g'^2} v}{2}, \quad\quad
M_\gamma = 0,
\]
where $W^\pm$, $Z^0$, and the photon ($\gamma$) are the resulting
physical mass eigenstate gauge bosons.
The photon remains massless, corresponding to
the unbroken $U(1)_{em}$ symmetry.

\subsubsection{Yukawa interaction}
The Yukawa interactions in the Standard Model describe the coupling
between fermions and the Higgs field and they are responsible for
generating their masses after spontaneous electroweak symmetry
breaking. The Yukawa Lagrangian is expressed as:
\begin{equation}
\mathcal{L}_{\text{Yukawa}} = - \sum_{\text{generations}} y^u
\bar{Q}_L \tilde{\phi} u_R + y^d \bar{Q}_L \phi d_R + y^e \bar{L}_L
\phi e_R + \text{h.c.}
\label{eq:yukawa_lag}
\end{equation}
where $ y^f $ represents the Yukawa coupling matrices, which are
generally complex and non-diagonal, and $\tilde{\phi}=i\sigma_2
\phi^*$ is the conjugate of the $\phi$ scalar doublet field.
After the Higgs field acquires a vacuum
expectation value (VEV), the Yukawa interactions lead to
the fermion mass terms:
\begin{equation}
    m_f = \frac{y^f v}{\sqrt{2}}
    \label{eq:fermion_mass}
\end{equation}
For the quark sector, the Higgs spontaneously generates masses for both
up-and down-quarks via their couplings to $\phi$ and $\tilde{\phi}$,
while for the leptonic sector, there is mass generated by the Higgs only
for the electrically charged leptons, while neutrinos are massless
as the SM does not contain the  $\nu_R$ field needed to construct
the required  Yukawa interaction which would generate a neutrino mass.
This was totally consistent with all experimental results at the time
when the model was constructed.  The first robust
experimental evidence for neutrino mass came from the observation of
neutrino oscillations by the Super-Kamiokande experiment in 1998
\cite{Fukuda:1998mi}. In the next section, we will discuss how to
extend the SM to include massive
neutrinos and their implications.

A final remark in this subsection concerns the structure of the Yukawa
matrices $ y^f $. The non-diagonal nature of the Yukawa matrices
enables flavour mixing between different generations of
fermions. This phenomenon was first observed in the oscillations of
neutral mesons, such as $ K^0 $-$ \bar{K}^0 $, $ B^0 $-$ \bar{B}^0 $,
and $ D^0 $-$ \bar{D}^0 $ systems.
However, this is not the case for leptons because no Yukawa term can
be built for the neutrinos. As a consequence, the SM with the particle contents
of Table \ref{tab:smcont} has a global accidental symmetry which implies
that together with the total baryon number, each of the lepton flavours
(and therefore the total lepton number) is conserved.

\subsection{Neutrino interactions}
Since neutrinos are neutral particles, in the SM they interact at tree-level
exclusively through the  weak interactions, which can be
classified into two main types:
\begin{itemize}
    \item \textbf{Charged current interactions (CC):} These
      interactions are mediated by the exchange of $W^\pm$ bosons and
      involve transitions between neutrinos and their corresponding
      charged lepton 
    \begin{equation}
    \label{eq:cc_womixing}
    \mathcal{L}_{\text{CC}} = -\frac{g}{\sqrt{2}} \left( \bar{\ell}_L
    \gamma^\mu \nu_L W_\mu^- + \text{h.c.} \right),
    \end{equation}
    where $g$ is the weak coupling constant, $\ell_L$ represents the
    left-handed charged lepton, and $\nu_L$ denotes the left-handed
    neutrino. Charged current interactions are
    allow for the identification of neutrino
    {\sl flavour} through the observation of the associated charged
    lepton. We label those neutrinos as ($\nu_e, \nu_\mu, \nu_\tau$).    
   CC interactions are essential in experiments involving neutrino
   scattering off nuclei, where the detection of electrons, muons, or
   tau leptons signals the presence of electron, muon, or tau
   neutrinos, respectively.   
    \item \textbf{Neutral current interactions (NC):} Neutral current
      interactions are mediated by the exchange of the $Z^0$ boson as
    \begin{equation}
    \label{eq:nc_womixing}
    \mathcal{L}_{\text{NC}} = -\frac{g}{2\cos\theta_W} \bar{\nu}_L
    \gamma^\mu \nu_L Z_\mu + \text{h.c.},
    \end{equation}
    where $\theta_W$ is the Weinberg angle. Unlike charged current
    interactions, NC processes are flavour-blind, meaning they provide
    limited information about the specific neutrino flavour
    involved. However, they contribute to neutrino-nucleon scattering
    and play a vital role in understanding the total neutrino flux in
    detectors.
\end{itemize}
Both types of interactions underpin the core mechanisms through which
neutrinos are detected in a variety of experimental setups, from large
underground detectors to particle accelerators and astrophysical
observatories. Their precise measurement offers valuable information
on neutrino masses, mixing angles, and potential physics beyond SM
(BSM).

\subsection{The number of neutrinos species}

The discovery of the $Z$ boson and subsequent precision measurements
at LEP (Large Electron-Positron Collider) in the 1990s provided
critical insights into the EW sector of the SM. The $Z$ boson mediates
neutral current interactions and can decay into fermion-antifermion
pairs, including neutrinos
\cite{ALEPH:2005ab,LEP:2006tt,Patrignani:2016xqp}.  The total decay
width of the $Z$ boson, $\Gamma_Z$, is determined by summing the
partial widths for its decay into all possible fermions:
\begin{eqnarray}
\Gamma_Z = \sum_f \Gamma(Z \to f\bar{f}).
\label{eq:Zdecay}
\end{eqnarray}

Each partial width depends on the number of fermion generations and
their couplings to the $Z$, which are specified by the SM:
\begin{eqnarray}
\label{eq:Znu}
\Gamma(Z \to f\bar{f}) \propto N_c^f \cdot \left( g_V^2 + g_A^2
\right),
\end{eqnarray}
where $N_c^f$ is the color factor ($N_c = 3$ for quarks, $N_c = 1$ for
leptons), and $g_V$ and $g_A$ are the vector and axial couplings of the
fermion to the $Z$ boson.

An important result of LEP was the precise determination of
 the invisible decay width of the $Z$ boson, $\Gamma_{\text{inv}}$.
In the SM  $\Gamma_{\text{inv}}$
is due to the decay of the $Z$ boson
into neutrinos. Since the SM
predicts that each generation of neutrinos contributes equally to the decay mode, one gets:
\begin{equation}
\label{eq:Zinv}    
\Gamma_{\text{inv}} = \Gamma(Z \to \nu\bar{\nu}) N_\nu= \frac{G_F
  M_Z^3}{12 \pi \sqrt{2}}N_\nu,
\end{equation}
where $N_\nu$ is the number of neutrino species. LEP measured
$\Gamma_{\text{inv}}$ precisely resulting in
\cite{ALEPH:2005ab,LEP:2006tt,Patrignani:2016xqp}:
\begin{equation}
    N_\nu = 2.984 \pm 0.008,
\label{eq:number_nu_act}
\end{equation}
consistent with three active neutrino flavours
($\nu_e, \nu_\mu,
\nu_\tau$) as predicted by the SM.

Hence, this measurement is fundamental for us to state that there are
only three {\sl active} neutrinos (those that participate in the weak
interactions). If there are more neutrinos, they must be {\sl sterile}
\textit{i.e.}, they do not participate in the weak interactions).

\section{Introducing Neutrino Masses}

As we will discuss in the next chapter in detail, neutrino oscillation
experiments have shown beyond doubt 
that neutrinos are massive and the SM as described above
needs to be extended to account for this. 
In this section we will introduce the simplest SM extensions with massive
neutrinos and the generation of lepton flavour mixing. 
\subsection{Dirac Neutrinos}
The {\sl apparently}  minimal way to extend the SM and generate
neutrino masses is by introducing three right-handed neutrinos (RHNs),
$\nu_{Ri}$. As mentioned above, they are singlets of the SM group and
for that reason they are often referred to as {\sl sterile} neutrinos.
The introduction of three RHNs allows for
a new Yukawa interaction, similar to those for up-type quarks:
\begin{equation}
\mathcal{L} \supset - y^\nu_{ij} \overline{L_i} \tilde{\phi} \nu_{Rj} +
\text{h.c.},
\label{eq:dirac_mass_0}
\end{equation}
After spontaneous symmetry breaking, this interaction
generates the following {\sl Dirac} neutrino mass term:
\begin{equation}
  \mathcal{L}_D = - M_D \overline{\nu_{L_i}}
\nu_{R_j} + \text{h.c.},
\label{eq:dirac_mass}
\end{equation}
with 
\begin{equation}
M_D = \frac{v}{\sqrt{2}} y_{ij}.
\label{eq:Dirac_mass_term}
\end{equation}
being a $3\times 3$ matrix.

The three neutrino mass eigenstates are obtained by diagonalizing $M_D$ via
unitary transformations:
\begin{equation}
\label{eq:Dirac_mix_chiral}
\nu^D_L = V_L^{\nu\dagger} \nu_L, \quad\quad
\nu^D_R = V_R^{\nu\dagger}
\nu_R,
\end{equation}
where $\nu^D_{R,L}$ represent the mass eigenstates and $\nu_{R,L}$ the
flavour eigenstates. The diagonalization condition is:
\begin{equation}
\label{eq:Dirac_mix}
V^{\nu\dagger}_L M_D V_R^\nu = M_D^{\text{diag}}.
\end{equation}
We call the massive neutrino states in this scenario {\sl Dirac} neutrinos.

Several comments are in order
\begin{itemize}
\item In this case total lepton number is still conserved since
 one can assign opposite lepton number to 
$\nu^D$ and its charge conjugate  
\begin{equation}
{\nu^D}^c \equiv C {\nu^D}\dagger,
\end{equation}
(here $C$ us the charge conjugate matrix for spinors) because they are
two distinct states.

But as we will argue below, in this scenario the conservation of total
lepton number is not longer accidental but it has been implicitly {\sl imposed}
by adding only the term in Eq.~\eqref{eq:dirac_mass_0} and not 
additional terms involving only the RHN's which, being purely singlets
of the gauge group, would be equally allowed by the gauge symmetry.
\item  As we will see in Chapter~\ref{cap:nufit} global analyses of neutrino
experimental data \cite{Esteban:2024eli}
robustly indicate sub-eV masses for neutrinos. For Dirac neutrinos,
the Yukawa couplings in eq.~\eqref{eq:Dirac_mass_term} must be of order
$|y^\nu| \lesssim 10^{-12}$, significantly smaller than those for
other fermions ($|y^{\text{other\,fermions}}| \gtrsim 10^{-6}$). This
raises a naturalness question: \textit{Why are neutrinos so light
compared to other fermions?} This question remains unresolved
in this minimal scenario.
 While this thesis will not dive deeply into extended 
scenarios, it is worth noting that introducing new particles into the
SM can generate Dirac neutrino masses with Yukawa couplings larger
than $10^{-6}$ \cite{mohapatra_massive_neutrinos} .
\item
Dirac neutrinos predict a non-zero magnetic moment proportional to their mass:
\begin{equation}
\mu_\nu = \frac{3 e G_F m_\nu}{8 \pi^2 m_e},
\end{equation}
where $m_e$ is the electron mass. This allows interactions with external
electromagnetic fields, though heavily suppressed by the small neutrino mass.
\end{itemize}  
\subsection{Majorana Neutrinos}
A Majorana neutrino is a fermion that is its own antiparticle,
implying that the neutrino field satisfies the condition:
\begin{equation}
\nu^M = {\nu^M}^c = C \overline{\nu^M}^T,
\end{equation}
In the minimal extension of the SM there are three Majorana neutrinos which are
mass eigenstates of a mass term built out of the three left-handed states and
their charge conjugate
\begin{equation}
\mathcal{L}_M = - \frac{1}{2} M_M \left( \overline{(\nu_L)^c} \nu_L +
\overline{\nu_L} (\nu_L)^c \right),
\label{eq:massmaj}
\end{equation}
such that $M_M$ represents a $3\times 3$ Majorana mass which is a symmetric
matrix by construction and the diagonalization condition now reads
\begin{equation}
\label{eq:majorana_mix}
V^{\nu\dagger} M_M V^{\nu*} = M_M^\text{diag}
\end{equation}
so the mass eigenstates verify
\begin{equation}
\label{eq:majorana_mix_chiral}
\nu^M_L = V^{\nu\dagger} \nu_L, \quad\quad
\nu^M_R = V^{*} (\nu_L)^C\,.
\end{equation}
We notice that in this case total lepton violated as the Majorana mass
term Eq.~\eqref{eq:massmaj} has total lepton number $\pm 2$.

But most importantly Eq.~\eqref{eq:massmaj} violates gauge invariance and
therefore can only be understood as a low energy effective operator generated
by some beyond the standard model physics as we will discuss next.

\subsection{Minimal gauge invariant neutrino mass}
Once we include $m$ right-handed neutrinos the most general form of the
renormalizable gauge invariant mass term which can be built is
\begin{equation}
\mathcal{L}_m = - \bar{\nu}_R M_D \nu_L - \frac{1}{2} \bar{\nu}_R^c M_R \nu_R + \text{h.c.},
\label{eq:neutrino_mass_general}
\end{equation}
where $M_D$ is $m\times 3$ matrix  and $M_R$ is a $m\times m$ symmetric
matrix. In general this scenario leads to $3+m$ massive Majorana neutrinos with
\begin{equation}
\nu^M = V^{\nu\dagger} \vec\nu, \quad\quad  
\end{equation}
where $\vec\nu=(\vec\nu_L,\vec \nu^C_R)^T$ contains the  3+m weak-eigenstate
neutrinos.

Two interesting limiting cases are:
\begin{itemize}
\item $M_R\gg M_D$. In this case diagonalization leads to three states
which are much lighter than the other $m$ with an effective 
Majorana mass 
\begin{equation}
  m_\nu \simeq - M_D M_R^{-1}
  M_D^T.
\end{equation}
The three light states are Majorana neutrinos
combination of the 3 active neutrinos
with projections over the right-handed states suppressed as
$M_D M_R^{-1}$.

This is the so-called see-saw mechanism ~\cite{seesaw1,seesaw2,seesaw3,seesaw4}
in which the light neutrino masses are suppressed by
the scale of $M_R$, which can be as high as the Grand Unified Theory
(GUT) scale, thus naturally explaining why neutrino masses are orders
of magnitude smaller than the masses of other fermions.
\item Some $s$ number of the eigenvalues of $M_R$ is not large compared to
those of $M_D$. In this case the spectrum contains $3+s$ light neutrino states
and $m-s$ heavy neutrinos. All of them are  Majorana.
These are the scenarios usually named in the literature of
ligh sterile neutrino models. 
\end{itemize}  

\subsection{Leptonic Mixing and CP Violation}
The charged current introduced in Eq.~\eqref{eq:cc_womixing} describes
the interaction between the gauge boson $W^\pm$ and the leptons in the
weak (or flavour) basis $\alpha=e,\mu,\tau$. In the absence of any
new physics in the charged lepton sector, it is always possible to
identify the three flavour charged lepton states with the states of
well defined mass. But in that basis, as we have seen, the massive neutrinos
will still be a superposition of the weak eigenstates so for massive leptons
the CC lagrangian reads
\begin{equation}
\mathcal{L}_{\text{CC}} = -\sum_\alpha \sum_k \frac{g}{\sqrt{2}}
\left(U_{\alpha k} \bar{\ell}_{\alpha L} \gamma^\mu \nu_{k L} W_\mu^-
+ \text{h.c.} \right),
\label{eq:cc_wmixing}
\end{equation}
where  $\nu_k$ is a neutrino with definite mass. In the most general case
with $N$ massive neutrinos the {\sl leptonic mixing matrix}
$U$ will be a $3\times N$ matrix which
verifies $U\times U^\dagger=I_{3\times 3}$ but in general
$U^\dagger U\neq I_{N\times N}$ and in this basis it can be identify with the
upper 3 lines of the $V^\nu$  ($V^\nu_L$) matrix diagonalizing the neutrino
mass matrix for the case of Majorana (Dirac) neutrinos after the non-physical
phases have been absorbed in the definition of the mass eigenstates
which depends if neutrinos are Dirac or Majorana particles.
Following Chapter 4 of \cite{Giunti:2007ry} one finds 
that for Majorana [Dirac] neutrinos the U matrix contains a
total of $6(N-2)$ [$5N-11$] real parameters, of which
$3(N-2)$ are angles and $3(N-2)$ [$2N-5$] can be interpreted as
physical phases. So for a given $N$ number of massive neutrinos
\begin{equation}
  U^\text{Maj}= U^\text{Dirac}\times P_\text{Maj}
  \label{eq:pmaj}
\end{equation}    
where $P_\text{Maj}$ is a diagonal matrix containing the  additional $N-1$ phases.

For most of this thesis we will be studying scenarios with three light
states; in this case, the $U$ matrix contains 3 mixing angles plus 1
phase if neutrinos are Dirac and 3 if they are Majorana.
It can be parametrized as
\begin{equation}
U \equiv
\begin{pmatrix}
U_{e1} & U_{e2} & U_{e3} \\ U_{\mu1} & U_{\mu2} & U_{\mu3}
\\ U_{\tau1} & U_{\tau2} & U_{\tau3}
\end{pmatrix} =
\begin{pmatrix}
1 & 0 & 0 \\ 0 & c_{23} & s_{23} \\ 0 & -s_{23} & c_{23}
\end{pmatrix}
\begin{pmatrix}
c_{13} & 0 & s_{13} e^{-i\delta_\text{CP}} \\ 0 & 1 & 0 \\ - s_{13}
e^{i\delta_\text{CP}} & 0 & c_{13}
\end{pmatrix}
\begin{pmatrix}
c_{12} & s_{12} & 0 \\ - s_{12} & c_{12} & 0 \\ 0 & 0 & 1
\end{pmatrix}\times  P_\text{Maj}
\label{eq:ULEP2}
\end{equation}
where $c_{ij} = \cos \theta_{ij}$ and $s_{ij} = \sin \theta_{ij}$ and the
three mixing angle $\theta_{ij}$
can be taken with generality to be in the first quadrant
$0\leq \theta_{ij}\leq \pi/2$ and the phases to be all between 0 and $2\pi$.
$P_\text{Maj}=I$ for Dirac neutrinos while for Majorana neutrinos contains
the additional two phases
\begin{equation}
P_\text{Maj}=\begin{pmatrix}
1 & 0 & 0 \\ 0 & e^{i\alpha_{21}/2} & 0 \\ 0 & 0 & e^{i\alpha_{31}/2}
\end{pmatrix},
\label{PMNS_MAJ}
\end{equation}

The presence of complex phases in the leptonic mixing matrix opens up
the possibility of CP violation in the leptonic sector. 
For three  Dirac or Majorana neutrinos one can build a basis independent
quantity that directly measures the presence of leptonic CP violation:
the Jarlskog invariant $J_{\text{CP}}$  defined as:
\begin{eqnarray}
\label{eq:Jarlskog_inv}    
J_{\text{CP}} = \text{Im}(U_{e1} U_{\mu 2} U_{e2}^* U_{\mu 1}^*).
\end{eqnarray}
which in the parametrization of the leptonic mixing matrix above reads
\begin{eqnarray}
\label{eq:Jarlskog_inv_2}    
J_{\text{CP}} = \frac{1}{8} \sin 2\theta_{12} \sin 2\theta_{23} \sin
2\theta_{13} \cos \theta_{13} \sin \delta_\text{CP}.
\end{eqnarray}
From the expression above, if $\delta=0$ or $\pi$, $J_{\text{CP}}=0$
and there is no CP violation.  A non-zero value of $J_{\text{CP}}$
implies CP violation in the neutrino sector.

As for the additional Majorana phases they cannot be measured in conventional
neutrino oscillation experiments because they do not affect oscillation
probabilities, as will be described below. However, they can be probed
through neutrinoless double-beta decay ($0\nu\beta\beta$), a rare
nuclear process that violates lepton number by two units.
In the minimal scenario with three massive Majorana neutrinos the
decay rate depends on the so-called {\sl effective Majorana mass}:
\begin{eqnarray}
\label{eq:doublebetadecay}
    m_{ee} = \left| \sum_{i} U_{ei}^2 m_i \right| = \left| m_1
    c_{12}^2 c_{13}^2 + m_2 s_{12}^2 c_{13}^2 e^{i\alpha_1} + m_3
    s_{13}^2 e^{i(\alpha_2 - 2\delta_\text{CP})} \right|,
\end{eqnarray}
where $m_i$ are the neutrino masses. The phases $\alpha_1$ and
$\alpha_2$ affect the interference terms in $m_{\beta\beta}$, making
it possible to constrain them experimentally.

\section{Neutrino flavour Oscillations}
\label{sec:vacosc}
Neutrino flavour oscillations is the phenomenon by which a neutrino
produced in  a flavour eigenstate $\alpha$ after traveling
some distance can be detected as a neutrino with flavour $\beta$
which could be different from $\alpha$. 

More quantiatively a neutrino with flavour $\alpha$ and momentum $\mathbf{p}$,
created in a vertex via a
charged current interaction process, is described by the flavour
state:
\begin{equation}
|\nu_\alpha\rangle = \sum_i U^*_{\alpha i} |\nu_i\rangle.
\label{eq:neutrino_states}
\end{equation}
expressed as a superposition of the mass eigenstates $|\nu_i\rangle.$ which
will have energy  $E_i = \sqrt{|\mathbf{p_i}|^2 + m_i^2}$. 

After a time $t$ the state will have evolved as 
\begin{equation}
|\nu_\alpha(t)\rangle = \sum_\beta \left( \sum_i U_{\alpha i}^*
\exp(-i E_i t) U_{\beta i} \right) |\nu_\beta\rangle.
\label{eq:flavour_time_evo}
\end{equation}

The probability of detecting it as  neutrino of flavour $\beta$ at time
$t$ is then
\begin{equation}
P_{\alpha \to \beta}(t) = \left| \langle \nu_\beta | \nu_\alpha(t)
\rangle \right|^2 = \left| \sum_i U_{\beta i} U_{\alpha i}^*
e^{-i E_i t} \right|^2.
\label{eq:probabi_eq}
\end{equation}
In the ultrarelativistic limit, the energy $E_i$ can be approximated
by:
\begin{equation}
E_i \approx p + \frac{m_i^2}{2E},
\label{eq:ultrar_lim}
\end{equation}
where $p$ is the neutrino momentum, and $E$ is its energy. This
approximation is known as the \textit{equal momentum assumption}
\cite{Giunti:2007ry}. Furthermore in this limit  $t=L/c$ where
$L$ is the distance traveled by the neutrino at time $t$
(called \textit{light-ray approximation}\cite{Giunti:2007ry}).
With this one can write 
\begin{eqnarray}
    \label{eq:prob_time}
    P_{\nu_\alpha\to \nu_\beta}(L,E)&=&
   \left| \sum_i U_{\beta i} U_{\alpha i}^*
   e^{-i \frac{m^2_i L}{2E}} \right|^2\nonumber 
   =
    \sum_{k,j}U^*_{\alpha k}U^*_{\beta
      j}U_{\alpha j}U_{\beta k}\exp{\Bigg(}-i \frac{\Delta m_{kj}^2t}{2E}{\Bigg)}
    \nonumber\\
&=&\sum_i |U_{\alpha i}|^2|U_{\beta i}|^2 +
2 \sum_{i < j} \text{Re}(U_{\alpha i} U_{\beta i}^* U_{\alpha j}^*
U_{\beta j}) \cos \left( \frac{\Delta m_{ij}^2 L}{2E} \right) 
\nonumber \\ &&
\hspace*{2.5cm}+2 \sum_{i < j} \text{Im}(U_{\alpha i} U_{\beta i}^*
U_{\alpha j}^* U_{\beta j}) \sin \left( \frac{\Delta m_{ij}^2 L}{2E}
\right)\nonumber\\
&=& \delta_{\alpha\beta} -
4 \sum_{i < j} \text{Re}(U_{\alpha i} U_{\beta i}^* U_{\alpha j}^*
U_{\beta j}) \sin^2 \left( \frac{\Delta m_{ij}^2 L}{4E} \right) 
\nonumber \\ &&
\hspace*{1cm}
+2 \sum_{i < j} \text{Im}(U_{\alpha i} U_{\beta i}^*
U_{\alpha j}^* U_{\beta j}) \sin \left( \frac{\Delta m_{ij}^2 L}{2E}
\right)
    \label{eq:prob_vac}
\end{eqnarray}
where $\Delta m^2_{ij}=m_i^2-m_j^2$.
The probability in Eq.~\eqref{eq:prob_vac} presents
an oscillatory behaviour with wavelengths
\begin{equation}
  \lambda^\text{vac}_{ij}=\frac{4\pi E}{|\Delta m^2_{ij}|}\simeq
  4.5\;{\rm km} \frac{E/{\rm GeV}}{|\Delta m_{ij}^2|/{\rm eV^2}}
\end{equation}
The probabilities described in Eq.~\eqref{eq:prob_vac} are categorized
into two types. When $\alpha = \beta$, we call it survival
probabilities (or disappearance probabilities). However, when $\alpha
\neq \beta$, we call it transition probabilities (or appearance
probabilities).

For the case of antineutrinos
the derivation is the same with the change $U\rightarrow U^*$ 
oscillation.
\begin{eqnarray}
  P_{\bar{\alpha} \to \bar{\beta}}(L,E) &= &
\delta_{\alpha\beta} -
4 \sum_{i < j} \text{Re}(U_{\alpha i} U_{\beta i}^* U_{\alpha j}^*
U_{\beta j}) \sin^2 \left( \frac{\Delta m_{ij}^2 L}{4E} \right) 
\nonumber \\ &&
\hspace*{1cm}
-2 \sum_{i < j} \text{Im}(U_{\alpha i} U_{\beta i}^*
U_{\alpha j}^* U_{\beta j}) \sin \left( \frac{\Delta m_{ij}^2 L}{2E}
\right)
    \label{eq:prob_vac_anti}
\end{eqnarray}
so if the last term is non zero it is  possible to have different
probabilities for neutrinos and antineutrinos and the leptonic CP
violation can be experimentally established.
It is also clear that for this to occur it is required that $\alpha\neq\beta$ and
to have at least three massive neutrinos involved in the oscillation
so that the $U$ matrix contains some complex phase.
For the case of three massive neutrinos unitarity relates
all the combinations in the last term of the probabilities
with the Jarlskog invariant defined in Eq.~\eqref{eq:Jarlskog_inv}    

From these expressions we also see that the oscillation probabilities
are the same for Dirac or Majorana neutrinos since the matrix $P_\text{Maj}$
(see Eq.~\eqref{eq:pmaj}) cancels in both
Eq.~\eqref{eq:prob_vac} and Eq.~\eqref{eq:prob_vac_anti}. 

In real experiments, neutrino beams are not monoenergetic and furthermore
detectors have finite energy resolution and sometimes the distance
traveled is also not fixed. Consequently
experiments measure  an average of the oscillation probability over some
range of energy and sometimes also distance. 
Therefore there are three different characteristic
regimes:
\begin{itemize}
\item $L \ll \lambda^\text{vac}_{i j}$: in this case, oscillations do not
  have enough time to develop and the experiment is basically not
  sensitive to either the mass splittings or the mixing angles.
\item $L \sim \lambda^\text{vac}_{i j}$: in this, case  the 
  experiment can be  sensitive to both $\Delta m_{ij}^2$ and the leptonic
  mixing matrix elements.
\item $L \gg \lambda^\text{vac}_{i j}$: in this case, the oscillation
  phase goes through many cycles. So
  when averaging over the energy or distance
  $\sin^2 \frac{\Delta m_{ij}^2 L}{4E}\sim \frac{1}{2}$, the experiment still provides information  on the leptonic
  mixing matrix elements but not on the mass splittings.
\end{itemize}

Most experimental set ups are restricted to some range of neutrino
energies and/or distances. Consequently they are mostly sensitive to
oscillations driven by one particular $\Delta m^2$ and the results can
be approximately understood in terms of oscillation probabilities derived
with two neutrino massive states. 
In that case, there is an angle
dominantly controlling the observable oscillation amplitude and a mass
difference dominantly controlling the observable frequency and
both Eqs.~\eqref{eq:prob_vac} and ~\eqref{eq:prob_vac_anti} become
\begin{equation}
P_{\alpha \beta} = \delta_{\alpha \beta} - \left( 2 \delta_{\alpha
  \beta} - 1\right) \sin^2 2 \theta \sin^2 \frac{\Delta m^2 L}{4E} \,
.
\label{eq:2fam}
\end{equation}
In this case, the oscillation probability
is symmetric under the exchange $\theta \leftrightarrow
\frac{\pi}{2}-\theta$ and/or $\Delta m^2\leftrightarrow - \Delta
m^2$. Furthermore, there is no physical CP violation in this
limit. More-than-two neutrino mixing as well as matter effects
break all these symmetries.

\begin{figure}[t]
\centering \includegraphics[width=0.75\textwidth]{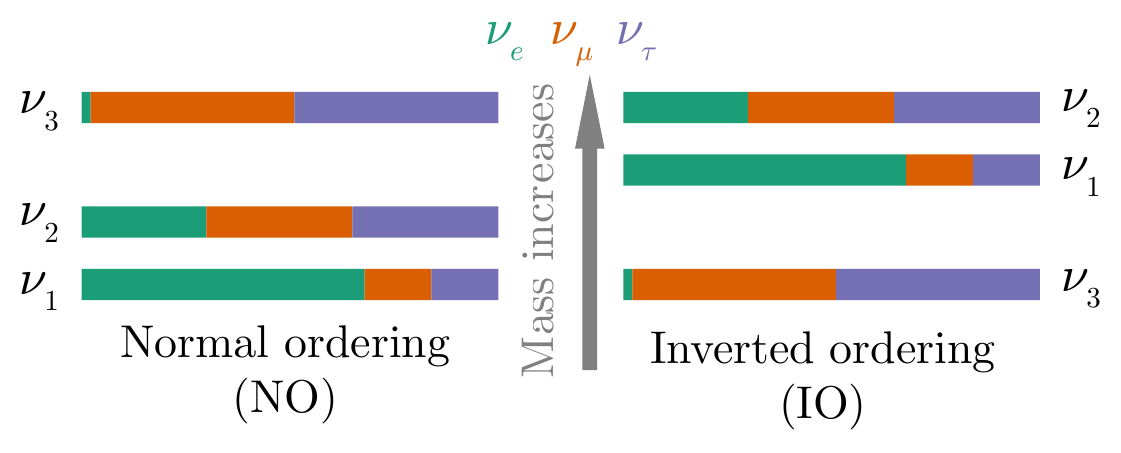}
\caption{Convention for the numbering of mass eigenstates and possible
orderings (NO in left, IO in right). The colours indicate the amount
of mixing between mass and flavour eigenstates as obtained from the
global analysis of neutrino oscillation experiments as we will 
describe and quantify in Chapters~\ref{chap:exp} and ~\ref{cap:nufit}}
\label{fig:ordering}
\end{figure}

Let us make a final comment about the signs of the mass
splittings which are physical in neutrino oscillations.
As seen above for the case of two massive neutrinos there is a total
degeneracy between the octact of the mixing angle and the sign of the
mass splitting. 
For three massive neutrinos and in the convention
that  we are using for the physical range of the mixing angles
($0\leq \theta_{ij}\leq \pi/2$) and the
CP phase $0\leq\delta\leq 2\pi$
one of the mass splittings, here $\Delta m^2_{21}$, is by
convention positive while the splitting $\Delta m^2_{31}$
can be positive or negative. We refer to these two signs as the
{\sl neutrino mass ordering}. So we have two possible orderings
\begin{itemize}
\item Normal   ordering: $m_1 < m_2 <m_3$ 
\item Inverted ordering: $m_3 < m_1 <m_2$  \;.
\end{itemize}
diagrammatically shown in Fig.\ref{fig:ordering}.

\section{Flavour Transitions in Matter}
\label{sec:matter}
When neutrinos propagate through matter, their oscillation behaviour
is modified by coherent forward elastic scattering and incoherent
scattering with the matter that is crossing by, described by the
Mikheev-Smirnov-Wolfenstein (MSW) effect. However, the incoherent
contribution is substantially smaller than the coherent one in regular
scenarios (not too dense $N\gg N_A/\mathrm{cm}^3$ or too energetic
$E_\nu \gg 1$ TeV)\cite{Concha1,Giunti:2007ry}. For the cases studied
in this thesis, none of them are considered.

Forward scattering occurs when a neutrino interacts with particles in
a medium but continues moving in nearly the same direction as before
the interaction. Unlike incoherent scattering, where the neutrino's
direction changes significantly, forward scattering preserves the
neutrino's momentum almost entirely.
This process is coherent because the medium remains unchanged after
the interaction. The scattered neutrino waves overlap and interfere
with the unscattered waves, creating a combined
effect~\cite{Concha1}. Since the medium doesn't "remember" the
interaction, the scattered waves add up constructively, enhancing the
overall effect. This coherence allows us to describe the medium's
influence on neutrinos using an effective potential, which depends on
the density and composition of the matter\cite{Wolfenstein:1978ue}.

It is not the goal of this thesis to develop the most general formalism
for the evolution of neutrinos in matter. 
We will just briefly comment about how one could derive the
matter potential from the QFT formalism (following
\cite{Concha1,Giunti:2007ry}) and from that, how the matter potential
could change the oscillation pattern of neutrinos due to cross matter
regions for the relevant physical cases studied in this thesis.
\subsection{Matter potentials}
Following \cite{Concha1, Giunti:2007ry}, we derive the
effective potential for electron neutrinos propagating in a medium
composed of electrons, protons, and neutrons starting with the
relevant low-energy Hamiltonian describing neutrino
interactions is given by:
\begin{equation}
H_W = \frac{G_F}{\sqrt{2}} \left[ J^{(+)\alpha}(x) J^{(-)}_\alpha(x) +
  \frac{1}{4} J^{(N)\alpha}(x) J^{(N)}_\alpha(x) \right] \; ,
\label{eq:hamil_matter}
\end{equation}
where the $J_\alpha$'s are the standard fermionic currents:
\begin{eqnarray}
\label{eq:currents}
J^{(+)}_\alpha(x) &=& \overline{\nu_e}(x) \gamma_\alpha (1 - \gamma_5)
e(x) \; , \\ J^{(-)}_\alpha(x) &=& \overline{e}(x) \gamma_\alpha (1 -
\gamma_5) \nu_e(x) \; , \\ J^{(N)}_\alpha(x) &=& \overline{\nu_e}(x)
\gamma_\alpha (1 - \gamma_5) \nu_e(x) - \overline{e}(x) \left[
  \gamma_\alpha (1 - \gamma_5) - 4 \sin^2 \theta_W \gamma_\alpha
  \right] e(x) \nonumber \\ && + \overline{p}(x) \left[ \gamma_\alpha
  (1 - g_A^{(p)} \gamma_5) - 4 \sin^2 \theta_W \gamma_\alpha \right]
p(x) - \overline{n}(x) \gamma_\alpha (1 - g_A^{(n)} \gamma_5) n(x) \;
.
\end{eqnarray}
Here, $g_A^{(n,p)}$ are the axial couplings for neutrons and protons,
respectively, and $\theta_W$ is the Weinberg angle. For simplicity, we
focus on the CC interactions, which dominate the effective potential
for $\nu_e$.

The effective CC Hamiltonian due to electrons in the medium is:
\begin{eqnarray}
H_{CC}^{(e)} &=& \frac{G_F}{\sqrt{2}} \int d^3p_e \, f(E_e, T)
\nonumber \\ && \times \langle \langle e(s, p_e) | \overline{e}(x)
\gamma^\alpha (1 - \gamma_5) \nu_e(x) \overline{\nu_e}(x)
\gamma_\alpha (1 - \gamma_5) e(x) | e(s, p_e) \rangle \rangle
\nonumber \\ &=& \frac{G_F}{\sqrt{2}} \overline{\nu_e}(x)
\gamma_\alpha (1 - \gamma_5) \nu_e(x) \nonumber\\
&&\times \int d^3p_e \, f(E_e, T) \langle
\langle e(s, p_e) | \overline{e}(x) \gamma_\alpha (1 - \gamma_5) e(x)
| e(s, p_e) \rangle \rangle \; ,
\label{eq:HCC}
\end{eqnarray}
where $s$ is the electron spin, $p_e$ is its momentum, and $f(E_e, T)$
is the energy distribution function of the electrons in the medium,
assumed to be homogeneous and isotropic. The distribution function is
normalized as:
\begin{equation}
\int d^3p_e \, f(E_e, T) = 1 \; .
\label{eq:density}
\end{equation}
The notation $\langle \langle \dots \rangle \rangle$ denotes averaging
over electron spinors and summing over all electrons in the
medium. Coherence implies that the spin and momentum of the initial
and final electrons are the same.

Expanding the electron fields $e(x)$ in plane waves, we find:
\begin{eqnarray}
&&\langle e(s, p_e) | \overline{e}(x) \gamma^\alpha (1 - \gamma_5) e(x)
| e(s, p_e) \rangle =\nonumber\\ &&\frac{1}{V} \langle e(s, p_e) |
\overline{u_s}(p_e) a_s^\dagger(p_e) \gamma^\alpha (1 - \gamma_5)
a_s(p_e) u_s(p_e) | e(s, p_e) \rangle, \; \nonumber\\
\label{eq:bracket_average}
\end{eqnarray}
such that $V$ is a volume normalization factor. The averaging gives:
\begin{equation}
\label{eq:aver_oper}
\frac{1}{V} \langle \langle e(s, p_e) | a_s^\dagger(p_e) a_s(p_e) |
e(s, p_e) \rangle \rangle = N_e(p_e) \frac{1}{2} \sum_s \; ,
\end{equation}
where $N_e(p_e)$ is the number density of electrons with momentum
$p_e$. Assuming the medium has equal numbers of spin-up and spin-down
electrons, we obtain:
\begin{eqnarray}
\langle \langle e(s, p_e) | \overline{e}(x) \gamma^\alpha (1 -
\gamma_5) e(x) | e(s, p_e) \rangle \rangle &=& N_e(p_e) \frac{1}{2}
\sum_s \overline{u_s}(p_e) \gamma^\alpha (1 - \gamma_5) u_s(p_e)
\nonumber\\
&&= N_e(p_e) \frac{p_e^\alpha}{E_e} \; .
\label{eq:ave}
\end{eqnarray}

Isotropy implies that $\int d^3p_e \, \vec{p_e} f(E_e, T) = 0$, so
only the $p^0$ term contributes upon integration. Substituting
Eq.~\eqref{eq:ave}) into Eq.~\eqref{eq:HCC}, we obtain:
\begin{equation}
\label{eq:ham_cc_fin}
H_{CC}^{(e)} = \frac{G_F N_e}{\sqrt{2}} \overline{\nu_e}(x) \gamma_0
(1 - \gamma_5) \nu_e(x) \; ,
\end{equation}
where $N_e$ is the electron number density at position $x$.
The effective potential for $\nu_e$ induced by its charged current
interactions with electrons in matter is then:
\begin{equation}
V_{CC} = \langle \nu_e | \int d^3x \, H_{CC}^{(e)} | \nu_e \rangle =
\sqrt{2} G_F N_e \; .
\label{eq:effV}
\end{equation}

For $\overline{\nu_e}$, the sign of $V_{CC}$ is reversed. This
potential can also be expressed in terms of the matter density $\rho$:
\begin{equation}
V_{CC} = \sqrt{2} G_F N_e \simeq 7.6 \, Y_e \, \frac{\rho}{10^{14} \,
  \text{g/cm}^3} \, \text{eV} \; ,
\end{equation}
where $Y_e = \frac{N_e}{N_p + N_n}$ is the relative number density of
electrons.

For $\nu_\mu$ and $\nu_\tau$, $V_{CC} = 0$ in these media, while the
neutral current potential $V_{NC}$ generated by the fermions $f$ in the medium
is the same for all neutrino  flavour and can be derived similary
\cite{Concha1,Giunti:2007ry} and reads:
\begin{equation}
    \label{eq:matter_potential}
    N_e,\,\,\,V_{\mathrm{NC}}^f=\sqrt{2}G_F N_f\, g_V^f,
\end{equation}
with $g_V^p=-g_V^e$ and $g_V^n=-\frac{1}{2}$.
For neutral matter, $N_p=N_e$, so the overall NC potential can be written
as:
\begin{equation}
    \label{eq:nc_pot}
    V_{\mathrm{NC}}^\mathrm{total}=-\frac{\sqrt{2}}{2}G_F N_n.
\end{equation}

Altogether the potential for NC and CC that an active neutrino 
$\nu_\alpha$ with $\alpha=e,\mu\tau$ feels in a dense medium can
be written as:
\begin{equation}
    \label{eq:pot_tot}
    V_\alpha=\sqrt{2}G_F\Bigg( N_e \delta_{\alpha e}
    -\frac{1}{2}N_n\Bigg),
\end{equation}
while for sterile neutrinos $V_s=0$.

\subsection{Modified Evolution}
Altogether the total Hamiltonian governing the neutrino evolution  in a dense
medium can be written as the sum  of its free Hamiltonian
(here call $H_{\rm vac}$) and an effective potential generated by the
matter
\begin{equation}
\label{eq:hamiltonian_matter}    
H_{\text{M}} = H_{\text{vac}} + V 
\end{equation}
where $V$ is the matter potential, given by Eq.\eqref{eq:pot_tot}.

The Schr\"odinger equation for a neutrino ensemble in the flavour basis
can be then written in matrix form as
\begin{equation}
    \label{eq:time_evolution_matter}
    i\dfrac{d}{dt}|\nu(t)\rangle =
    H_\mathrm{M}|\nu(t)\rangle
    = (U\Delta_M U^\dagger +V)
    |\nu(t) \rangle \;,
\end{equation}
where allowing in general for the presence of light sterile neutrinos
$\Delta_M = \diag(0, \Dmq_{21}, \Dmq_{31},$ $\Delta_{41}\dots) / 2E$
and $V = \sqrt{2} G_F \diag (N_e(x), 0, 0,N_n/2,\dots)$,
after extracting a  diagonal term $\Big(p+\frac{m_1^2}{2E_\nu} +
V_\mathrm{NC}\Big)\times I $, which does not contribute to
the oscillation probabilities.

In general Eq.~\eqref{eq:time_evolution_matter} is difficult to solve
analytically except for the case when the neutrino is traveling through
a region with constant matter density.
In this case the oscillation probabilities can be recasted as
Eqs.~\eqref{eq:prob_vac} and ~\eqref{eq:prob_vac_anti}
with the substitutions
\begin{equation}
  \Delta m^2_{ij} \rightarrow \Delta \mu^2_{ij}\;,
  \hspace*{1cm} U_{\alpha i}
  \rightarrow  \widetilde{U}_{\alpha i}
  \end{equation}
where
\begin{equation}
U\Delta_M U^\dagger \pm V =\widetilde{U} \Delta_\mu \widetilde{U}^\dagger 
\end{equation}
for neutrinos and antineutrinos respectively.

This is a valid approximation for neutrinos traveling through the crust
of the Earth as it is the case in accelerator neutrinos at long baselines.

When the density of the matter crossed by the neutrino changes along its
trajectory one can always solve the evolution equation numerically 
by discretizing the evolution in layers sufficiently thin so that the
potential can be assumed constant in each layer. Schematically,
within a layer of thickness $L_j$ the evolution matrix ${\cal U}_j$ ,
which relates the initial and final neutrino states, takes the form:
\begin{eqnarray}
\label{eq:const_matter1}    
{\cal U}(L_i)_j = \exp\left(-i {H_M}_j  L_i \right)\;,
\end{eqnarray}
and can be evaluated numerically. 
The total evolution matrix is given by the product of the
evolution matrices for each layer
\begin{eqnarray}
\label{eq:const_matter2}    
{\cal U}_{\text{total}} = {\cal U}_n(L_n) {\cal U}_{n-1}(L_{n-1}) \cdots {\cal U}_1(L_1),
\end{eqnarray}
and the transition probability between flavors $\alpha $ and $ \beta $ is then:
\begin{eqnarray}
\label{eq:prob_const_matt}    
P_{\nu_\alpha \rightarrow \nu_\beta} = \left| \langle \nu_\beta |
{\cal U}_{\text{total}} | \nu_\alpha \rangle \right|^2.
\end{eqnarray}

It is also possible to find  approximate analytical solutions for the
evolution equation in non-uniform matter density under some simplifying 
conditions. A particularly
illuminating example is the evolution in the Sun which we discuss next.
For the sake of concreteness we are focusing in the evolution of
three active neutrino ensemble.

\subsection{Solar neutrinos: MSW effect}
We are going to solve the evolution of the 3 neutrino ensemble in the
solar matter. The first
approximation which we will use is that the largest
mass splitting, here $|\Delta m_{31}|\gg \Delta m^2_{21}$ (as we mentioned
above $\Delta m^2_{21}>$ by convention)  and that
$|\Delta m^2_{31}|/(2E)$ is always much larger than the matter potential
in any point in the evolution. 
With these assumptions, only the first two mass eigenstates are
dynamical.  Since physical quantities have to be independent of the
parametrization of the mixing matrix, we choose a parametrization
that makes analytical expressions particularly simple in this case.
We write $U = O U_{12}$, where $O = R_{23} R_{13}$
is real while $U_{12}$ is a complex rotation by angle $\theta_{12}$
and phase $\delta$, such that $O$ is given by:
\begin{equation}
  \label{eq:OU12matrix}
  O = \begin{pmatrix} c_{13} & 0 & s_{13} \\ -s_{13} s_{23} & c_{23} &
    c_{13} s_{23} \\ -s_{13} c_{23} & -s_{23} & c_{13} c_{23}
  \end{pmatrix}\;,\hspace*{1cm}
    U_{12} = \begin{pmatrix} c_{12} & s_{12}e^{-i\delta} &0
    \\ -s_{12}e^{i\delta}& c_{12} & 0\\ 0 & 0 & 1
    \end{pmatrix}\;.
\end{equation}    
With this definitions it is possible to rewrite
Eq.~\eqref{eq:time_evolution_matter} as:
\begin{equation}
  H_\text{M} = O \tilde{H} O^\dagger \qquad\text{with}\qquad
  \tilde{H} = U_{12} \Delta U_{12}^\dagger + O^\dagger V O \,.
  \label{eq:new_hamilt_matter}
\end{equation}
Defining 
\begin{equation}
\label{eq:rephasing}
|\hat{\nu} \rangle=O^\dagger|\nu \rangle 
\end{equation}
and taking the limt $|\Dmq_{3l}|
\to \infty$ for $l=1,2$,  the matrix $\tilde{H}$ takes the effective
block-diagonal form:
\begin{equation}
  \tilde{H} \approx
  \begin{pmatrix}
    H^{(2)} & \boldsymbol{0} \\ \boldsymbol{0} & \frac{\Delta m_{31}^2}{2E_\nu}
  \end{pmatrix}
\end{equation}
and $H^{(2)} = H_\text{vac}^\text{(2)} + H_\text{mat}^\text{(2)}$ with

\begin{align}
  \label{eq:HvacSol}
  H_\text{vac}^\text{(2)} &= \frac{\Dmq_{21}}{4 E_\nu}
  \begin{pmatrix}
    -\cos2\theta_{12} \, \hphantom{e^{-i\delta}} & ~\sin2\theta_{12}
    \, e^{i\delta} \\ \hphantom{-}\sin2\theta_{12} \, e^{-i\delta} &
    ~\cos2\theta_{12} \, \hphantom{e^{i\delta}}
  \end{pmatrix} ,
  \\
  \label{eq:HmatSol}
  H_\text{mat}^\text{(2)} &= \frac{\sqrt{2}}{2} G_F \left[N_e(x)
    \begin{pmatrix}
      c_{13}^2 & 0 \\ 0 & -c_{13}^2
    \end{pmatrix}
    \right].
\end{align}
The state $|\hat{\nu}\rangle_3$
decouples from $|\hat{\nu}\rangle_1$ and $|\hat{\nu}\rangle_2$. Hence,
such a state will have the following differential equation:
\begin{equation}
    \label{eq:dif_eq_third}
    i\frac{d}{dx}|\hat{\nu}(x)\rangle_3= \frac{1}{2 E_\nu}\Delta
    m_{31}^2 |\hat{\nu}(x)\rangle_3,
\end{equation}
that has the simple solution:
\begin{equation}
    \label{eq:dif_eq_sol_third}
    |\hat{\nu}(x)\rangle_3=\exp\Big(-i\frac{\Delta
      m_{31}^2x}{2E_\nu}\Big) |\hat{\nu}(0)\rangle_3.
\end{equation}

Now, by combining Eqs. \eqref{eq:HvacSol} and \eqref{eq:HmatSol} it is
possible to write the evolution equation for the states
$|\hat{\nu}(x)\rangle_1$ and $|\hat{\nu}(x)\rangle_2$ as:
\begin{equation}
    \label{eq:dif_eq_12_0}
    i\frac{d}{dx}\begin{pmatrix}
      |\hat{\nu}(x)\rangle_1\\ |\hat{\nu}(x)\rangle_2
    \end{pmatrix}
    = \frac{1}{4 E_\nu}\begin{pmatrix} -\cos2\theta_{12}\Delta
      m^2_{12} + Ac_{13}^2 & \sin2\theta_{12}\Delta
      m^2_{12}e^{i\delta}\\ \sin2\theta_{12}\Delta
      m^2_{12}e^{-i\delta}& +\cos2\theta_{12}\Delta m^2_{12} -
      Ac_{13}^2
    \end{pmatrix}
    \begin{pmatrix}
    |\hat{\nu}(x)\rangle_1\\ \hat{\nu}(x)\rangle_2
    \end{pmatrix},
\end{equation}
with  $A=A(x)=2\sqrt{2}G_F N_e(x)
E_\nu=V E_\nu$.
Eq. \ref{eq:dif_eq_12_0}
can be rewritten as:
\begin{equation}
    \label{eq:dif_eq_12}
    i\frac{d}{dx}\begin{pmatrix}
      |\hat{\nu}(x)\rangle_1\\ |\hat{\nu}(x)\rangle_2
    \end{pmatrix}
    = \frac{1}{4 E_\nu}\begin{pmatrix} -\cos2\theta_{M}(x)\Delta
      m^2_{M}(x) & \sin2\theta_{M}(x)\Delta
      m^2_{M}(x)\\ \sin2\theta_{M}(x)\Delta m^2_{M}(x)&
      +\cos2\theta_{M}(x)\Delta m^2_{M}(x)
    \end{pmatrix}
    \begin{pmatrix}
    |\hat{\nu}(x)\rangle_1\\ \hat{\nu}(x)\rangle_2
    \end{pmatrix},
\end{equation}
where we have dropped the phase $\delta$ which produces no effects in a two-neutrino framework, and
\begin{eqnarray}
\label{eq:delta_M}
    \Delta m_M^2(x)&=&\sqrt{(\cos2\theta_{12}\Delta m^2_{12} -
      Ac_{13}^2)^2+(\sin2\theta_{12}\Delta m^2_{12} )^2}\\
\label{eq:thetaM}
    \cos 2\theta_M(x)&=&\frac{\cos2\theta_{12}\Delta m^2_{12} -
      Ac_{13}^2}{\Delta m_M^2(x)}\,\,\,\text{and}\,\,\,\sin
    2\theta_M(x)=\frac{\sin2\theta_{12}\Delta m^2_{12} }{\Delta
      m_M^2(x)}
\end{eqnarray}
are the effective mass splitting and mixing angle in matter at position $x$,
The instantaneous Hamiltonian eigenstates at position $x$ are
\begin{equation}
\label{eq:change_var_2}
     |\hat{\eta} \rangle=U_M(x)|\hat{\nu}
     \rangle,\,\,\,\text{such}\,\,\text{that}\,\,\,U_M(x)=\begin{pmatrix}
     \cos\theta_M(x)&\sin\theta_M(x)\\-\sin\theta_M(x)&\cos\theta_M(x)
     \end{pmatrix}
\end{equation}
and in terms of these states the equation  \ref{eq:dif_eq_12} reads
\begin{eqnarray}
        \label{eq:dif_eq_12_new}
    iU_M(x)\frac{d}{dx}\Bigg(U_M^\dagger(x)\begin{pmatrix}
      |\hat{\eta}(x)\rangle_1\\ |\hat{\eta}(x)\rangle_2
    \end{pmatrix}\Bigg)
    &=& \frac{1}{4 E_\nu}\begin{pmatrix} -\Delta m^2_{M}(x) & 0\\ 0&
      \Delta m^2_{M}(x)
    \end{pmatrix}
    \begin{pmatrix}
    |\hat{\eta}(x)\rangle_1\\ \hat{\eta}(x)\rangle_2
    \end{pmatrix},
\end{eqnarray}
which after some algebra results into the evolution equation for
the instantaneous mass eigenstates 
\begin{eqnarray}
        \label{eq:dif_eq_12_new_final}
    i\frac{d}{dx}\begin{pmatrix}
      |\hat{\eta}(x)\rangle_1\\ |\hat{\eta}(x)\rangle_2
    \end{pmatrix}
    &=& \frac{1}{4 E_\nu}\begin{pmatrix} -\Delta m^2_{M}(x) & -4E_\nu
      i \frac{d\theta_M(x)}{dx}\\ 4E_\nu i \frac{d\theta_M(x)}{dx}&
      \Delta m^2_{M}(x)
    \end{pmatrix}
    \begin{pmatrix}
    |\hat{\eta}(x)\rangle_1\\ \hat{\eta}(x)\rangle_2
    \end{pmatrix}.
\end{eqnarray}
This evolution equation is non-diagonal which means that the
instantaneous mass eigenstates are not the eigenstates of the
evolution, \textit{i.e.}, the evolution is a priori not adiabatic.  
The non-diagonal terms in Eq.~\eqref{eq:dif_eq_12_new_final} are
proportional to the change of density on the way of neutrinos. We 
define the adiabaticity parameter\footnote{The
    Standard Solar Model (SSM) describes the Sun's electron density as
\[
N_e(r) = N_e(0) \exp(-r/r_0),
\] 
for $0.1 \leq r/R_\odot \leq 0.9$, where $N_e(0) =
245~N_A/\text{cm}^3$ and $r_0 = 10.54~R_\odot$. This profile aids in
analyzing resonance adiabaticity, characterized by the parameter
$\gamma_R$.  } 
\begin{equation}
    \label{eq:adiabaticity}
    \gamma\equiv \frac{\Delta
      m_M^2(x)}{4E_\nu|\frac{d\theta_M(x)}{dx}|}=\frac{(\Delta
      m_M^2(x))^2}{4\sqrt{2}G_FE_\nu^2\sin2\theta_M(x)c_{13}^2|\frac{dN_e(x)}{dx}|}.
\end{equation}
When $\gamma\gg 1$ for all $x$, we can neglect the off-diagonal terms
in Eq.\eqref{eq:dif_eq_12_new_final}, and this is called the adiabatic
approximation~\cite{ms2, Bethe:1986ej}. The adiabaticity condition implies
that transitions between the eigenstates $|\hat{\eta}\rangle_1$ and
$|\hat{\eta}\rangle_2$ are suppressed, and the evolution of the system
is governed by the separate propagation of each eigenstates. This
condition is satisfied when the electron number density
$N_e(x)$ varies sufficiently slowly along the neutrino's path compared
to the characteristic oscillation wavelength in vacuum
as is the case for solar neutrinos propagating through the Sun.

The Sun produces $|\nu_e\rangle$'s which from Eq.~\eqref{eq:rephasing} are
\begin{equation}
  |\nu_e\rangle= \cos\theta_{13} |\hat\nu\rangle_1+\cos\theta_{13} |\hat \nu\rangle_3
\end{equation}
 and $|\hat\nu\rangle_1$ is a superposition of the matter eigenstates 
$|\hat{\eta}\rangle_1$ and $|\hat{\eta}\rangle_2$ at the production
point $x_0$:
\begin{eqnarray}
\label{eq:initial_nue}    
|\nu_1\rangle = \cos\theta_M(x_0) |\hat{\eta}(x_0)\rangle_1 +
\sin\theta_M(x_0) |\hat{\eta}(x_0)\rangle_2.
\end{eqnarray}
As the neutrino propagates adiabatically, the  $|\hat{\eta}(x)\rangle_i$
(and  $|\hat{\nu}\rangle_3$ ) evolve
independently, and the state at a later point $x$ can be written as:
\begin{eqnarray}
\label{eq:prop_nue}    
|\nu(x)\rangle_1 = \cos\theta_M(x_0) e^{-i\phi_1(x)}
|\hat{\eta}(x)\rangle_1 + \sin\theta_M(x_0) e^{-i\phi_2(x)}
|\hat{\eta}(x)\rangle_2,
\end{eqnarray}
where $\phi_1(x)$ and $\phi_2(x)$ are the accumulated phases of the
eigenstates $|\hat{\eta}\rangle_1$ and $|\hat{\eta}\rangle_2$,
respectively. These phases are given by:
\begin{eqnarray}
\label{eq:phase_nue}    
\phi_i(x) = (-1)^{i+1}\int_{x_0}^x \frac{\Dmq_M(x')}{4 E_\nu} \, dx',
\quad i = 1, 2.
\end{eqnarray}
At the detection point $x_f$, the neutrino state is projected back
onto the flavour eigenstate $|\nu_e\rangle$. The survival probability
$P_{\hat\nu_1 \rightarrow \hat\nu_1}$ is then given by the squared overlap of
the evolved state with the initial state:
\begin{eqnarray}
\label{eq:probnue_2}    
P_{\hat\nu_1 \rightarrow \hat\nu_1} &=& \left| \langle \hat\nu_1 | \nu(x_f) \rangle
\right|^2= \cos^2\theta_M(x_0)
\cos^2\theta_M(x_f) + \sin^2\theta_M(x_0) \sin^2\theta_M(x_f)
\nonumber \\ &+& \frac{1}{2} \sin 2\theta_M(x_0) \sin 2\theta_M(x_f)
\cos(\phi_2(x_f) - \phi_1(x_f)).\nonumber\\
\end{eqnarray}
For the case of solar neutrinos the detection occurs in vacuum (or
very low density), where the effective mixing angle $\theta_M(x_f)$
reduces to the vacuum mixing angle $\theta_{12}$. The distance between
the neutrino source and the detector is huge, the phase of the cosine
is very large averaging it to zero, $\cos(\phi_2(x_f) - \phi_1(x_f))
\to 0$, the survival probability simplifies to:
\begin{eqnarray}
\label{eq:probnue_3}    
\langle P_{\hat\nu_1 \rightarrow \hat\nu_1} \rangle
&\simeq&
\cos^2\theta_M(x_0) \cos^2\theta_{12} + \sin^2\theta_M(x_0)
\sin^2\theta_{12}=
\frac{1}{2} + \frac{1}{2}\cos 2\theta_M(x_0) \cos
2\theta_{12}\nonumber\\
&\equiv& \langle P_{\nu_e \rightarrow \nu_e} \rangle^\text{$2\nu$}
\end{eqnarray}
Including the projection over the third state one gets 
\begin{equation}
    \langle P_{\nu_e \rightarrow \nu_e} \rangle^\text{3$\nu$} =
    \cos^4\theta_{13} \langle P_{\nu_e \rightarrow \nu_e} \rangle^\text{$2\nu$}+\sin^4\theta_{13} \;.    
    \label{eq:three_family}
\end{equation}

\subsubsection{The MSW effect}
From Eqs.\eqref{eq:thetaM} and \eqref{eq:delta_M} we find that the mixing
angle in matter is 
\begin{equation}
  \tan 2\theta_M
  = \frac{\sin 2\theta_{12}}{\cos 2\theta_{12} -
      \frac{c_{13}^2 V(x) E_\nu}{\Delta m^2_{21}}}\;.
\label{eq:tan_matter}
\end{equation}
So when 
\begin{equation}
  V(x)=2\sqrt{2}G_F N_e(x)=
\frac{\Delta m^2_{21}}{2E_\nu c_{13}^2} \cos 2\theta_{12}\;, 
\end{equation}
$\tan 2\theta_M$ changes sign and  the mixing angle becomes $\theta_M=\pi/4$
irrespective of its value in vacuum.
We call this the resonant condition.

Since  $\Delta m^2_{21} > 0$ by convention, this condition can be satisfied
for electron neutrinos propagating through matter with a
sufficiently high electron density $N_e$ if $\cos 2\theta_{12}>0$.
This is, if the mixing angle $\theta_{12}$ lies in the first octant,
$\theta_{12}<\frac{\pi}{4}$ the resonance condition is met
for neutrinos (rather than antineutrinos), as the matter potential $V$ has the
opposite sign for antineutrinos. At resonance, the mixing angle in
matter is $\theta_M=\pi/4$.

For solar neutrinos, which are produced in the core of the Sun as
electron neutrinos ($\nu_e$), the electron density is very high,
causing the effective mixing angle $\theta_{M,0}$ to initially be close to
$\pi/2$ ($\sin 2\theta_{M}(x_0) \to -1$). As the neutrinos propagate outward
through the Sun, the electron density decreases and $\theta_M$
evolves toward its vacuum value $\theta_{12}$. If the evolution is
adiabatic  the resulting
survival probability becomes
\begin{equation}
\langle  P_{\nu_e \rightarrow \nu_e} \rangle^\text{$2\nu$}=
  \frac{1}{2} - \frac{1}{2} \cos2\theta_{12}=\sin^2\theta_{12} \;.
\end{equation}
This is the MSW effect.

\section{Neutrino Mass Scale and Neutrino Mass Ordering}
Because of its quantum-interference nature, mass-induced flavor
oscillations are sensitive to the phase differences induced by the
mass-squared splitting $\Dmq_{ij}$ and to misalignment between the
detection and propagation eigenstates, \textit{i.e.}, to the leptonic
mixing matrix elements $U_{\alpha j}$.  They are, however, insensitive
to overall shifts of the energy levels, and hence they cannot provide
information on the absolute mass scale of the neutrinos other than the
obvious lower bound on the masses of the heaviest states involved in
the oscillations.

The most model-independent information on the neutrino mass, rather
than on mass differences, is obtained from kinematic studies of
reactions in which a neutrino or an antineutrino is involved.  In the
presence of mixing, the most relevant constraint comes from the study
of the end point ($E \sim E_0$) of the electron spectrum in Tritium
beta decay $\Nuc[3]{H} \to \Nuc[3]{He} + e^- + \bar\nu_e$.  This
spectrum can be effectively described by a single parameter,
$m_{\nu_e}$, if for all neutrino states $E_0 - E \gg m_i$:
\begin{equation}
  \label{eq:mbeta}
  m_{\nu_e}^2 \equiv \frac{\sum_i m_i^2 |U_{ei}|^2}{\sum_i |U_{ei}|^2}
  = \sum_i m_i^2 |U_{ei}|^2 \,,
\end{equation}
where the second equality holds if unitarity is assumed.  The most
recent result on the kinematic search for neutrino mass in tritium
decay is from KATRIN~\cite{Katrin:2024tvg}, which sets an upper
limit $m_{\nu_e} < 0.45~\text{eV}$ at 90\% CL.

Direct information on neutrino masses can also be obtained from
neutrinoless double beta decay $(A,Z) \to (A,Z+2) + e^- + e^-$.  This
process violates lepton number by two units, hence in order to induce
the $0\nu\beta\beta$ decay, neutrinos must be Majorana particles.  In
particular, if the only effective lepton number violation at low
energies is induced by a Majorana mass term for neutrinos, the rate of
$0\nu\beta\beta$ decay is proportional to the \emph{effective Majorana
mass of $\nu_e$}:
\begin{equation}
  m_{ee} = \Big| \sum_i m_i U_{ei}^2 \Big| \,.
\end{equation}
Currently the strongest bound on $0\nu\beta\beta$ decay lifetimes are
obtained with Germanium ($T^{0\nu}_{1/2} > 1.8\times 10^{26}$ yr) by
GERDA~\cite{GERDA:2020xhi} and with Xenon ($T^{0\nu}_{1/2} > 3.8\times
10^{26}$ yr) by KamLAND-Zen~\cite{KamLAND-Zen:2024eml}.  Depending on
the assumed nuclear matrix elements, these correspond to 90\% CL
limits of $m_{ee} \lesssim 0.079$--$0.180$~eV~\cite{GERDA:2020xhi} and
$m_{ee} \lesssim 0.028$--$0.122$~eV~\cite{KamLAND-Zen:2024eml},
respectively.

Finally, neutrino masses also have effects in cosmology.  In general,
cosmological data mostly gives information on the sum of the neutrino
masses, $\SumNu$, while it has very little to say on their mixing
structure and on the ordering of the mass states.  At present, no
positive evidence of the cosmological effect of a non-zero neutrino
mass has been observed, which results into upper bounds on $\SumNu$ in
the range of $\SumNu \lesssim 0.04$--$0.3$ eV (see, \textit{e.g.},
Refs.~\cite{Jiang:2024viw, Naredo-Tuero:2024sgf} and references
therein for post-DESI~\cite{DESI:2024mwx} global analyses) depending
on, \textit{e.g.}, the cosmological data included in the analysis,
assumptions on the cosmological model, the statistical approach, the
treatment of systematics, or parameter priors.

Within the $3\nu$-mixing scenario, for each mass ordering, the values
of these observables can be directly predicted in terms of the
parameters determined in the global oscillation analysis and a single
mass scale, which is usually taken to be the lightest neutrino mass
$m_0$.  In addition, the prediction for $m_{ee}$ also depends on the
unknown Majorana phases:
\begin{align}
\label{eq:numass}
  m_{\nu_e} & = \sqrt{m_1^2 \, c_{13}^2 c_{12}^2 + m_2^2 \, c_{13}^2 s_{12}^2
  + m_3^2 \, s_{13}^2} \,,
  \\
  m_{ee} &= \left| m_1 \, c_{13}^2 c_{12}^2 \, e^{2i(\alpha_1 - \dCP)}
  + m_2 \, c_{13}^2 s_{12}^2 \, e^{2i(\alpha_2 - \dCP)} + m_3 \, s_{13}^2
  \right| \,,
  \\
  \SumNu &= m_1 + m_2 + m_3 \,,
  \\
  \text{with~}
  &\left\{
  \begin{aligned}
    &\text{NO}:
    &
    m_1 &= m_0 \,,
    &\enspace
    m_2 &= \sqrt{m_0^2 + \Dmq_{21}} \,,
    &\enspace
    m_3 &= \sqrt{m_0^2 + \Dmq_{3\ell}} \,,
    \\
    &\text{IO}:
    &
    m_3 &= m_0 \,
    &\enspace
    m_2 &= \sqrt{m_0^2 - \Dmq_{3\ell}} \,,
    &\enspace
    m_1 &= \sqrt{m_0^2 - \Dmq_{3\ell} - \Dmq_{21}} \,.
  \end{aligned}\right.
\end{align}

where, following the convention adopted by the NuFIT group, we denote by $\Dmq_{3\ell}$ the mass-squared splitting with the largest absolute value for the given mass ordering (that is, 
$\Dmq_{3\ell}\equiv\Dmq_{31} >0$ for NO, and $\Dmq_{3\ell}\equiv\Dmq_{32} <0$ for IO. 

\chapter{Experimental Inputs}
\label{chap:exp}

%\section{Introduction}
%\label{sec:intro}

Building on the theoretical foundation of neutrino
oscillations—including the leptonic mixing matrix, matter effects, and
oscillation probabilities derived in Chapter~\ref{chap:theo}--- this chapter  
examines the experimental methodologies that have empirically
validated and refined these concepts. Neutrino oscillations are
dominantly probed with neutrinos that can be classified according to their
sources as solar, atmospheric, reactor, and accelerator neutrinos. Each
source exploits specific baseline-to-energy ratios ($L/E$) and
detection techniques to isolate different terms in the oscillation
probability, enabling precision measurements of some  mass splittings
and mixing angles.

In this chapter we will briefly describe the main features of
these experiments and their results, which will be employed in the
coming chapters to test different forms of new physics. 
Most quantitative details are provided for the two experiments for which
this thesis has contributed most in their simulation and
analysis: Borexino and NOvA.

In Section~\ref{sec:solar_exp}, we present the results obtained with
solar neutrinos. Solar neutrino experiments Homestake,
Gallex-GNO, Super-Kamiokande, SNO, and  Borexino have shown that 
$\nu_e$'s produced in the Sun's interior undergo matter-enhanced flavour
transitions via the MSW mechanism. In the framework of 3$\nu$ mixing
they mostly provide information on  $\Delta m^2_{21}$ and $\theta_{12}$.
At the end of Sec.~\ref{sec:solar_exp} we present the details of Borexino
Phase II and III data analyses, illustrating the interplay between
spectral decomposition, systematic uncertainty modeling (e.g., energy
scale corrections and background correlations), and statistical
methods (e.g., $\chi^2$ minimization and data analysis).

Section~\ref{sec:atm_exp} focuses on atmospheric neutrinos, generated
by cosmic ray interactions on Earth’s atmosphere. The experiments 
Super-Kamiokande and IceCube utilize their wide energy range (MeV–TeV)
and baselines (10–12,700 km) to observe  oscillations of $\nu_\mu$'s which  
are dominated by  the 3$\nu$ mixing parameters
$|\Delta m^2_{32}|$ and $\theta_{23}$.
Section~\ref{sec:reac_exp} describes reactor neutrino experiments, which
leverage controlled $\bar{\nu}_e$ fluxes from nuclear fission
to measure $\theta_{13}$ and $|\Delta m^2_{32}|$ (Daya Bay) and
$\Delta m^2_{21}$ and $\theta_{12}$  (KamLAND).
Section~\ref{sec:accel_exp} explores accelerator-based
long-baseline experiments (NOvA, T2K), where high-intensity
$\nu_\mu$ and $\overline\nu_\mu$
beams enable studies of both $\nu_\mu$ disappearance as well as 
$\nu_\mu \to \nu_e$ 
and $\overline{\nu}_\mu \to \overline{\nu}_e$ transitions and offer
a direct test to leptonic CP violation. In the end
of Section~\ref{sec:accel_exp}, we present detailed studies of NOvA
systematic uncertainty modeling (e.g., energy scale corrections and
background correlations), and statistical methods (e.g., $\chi^2$
minimization and data analysis).

We conclude this chapter by examining two types of experiments that
challenge our current understanding of the standard neutrino
paradigm. Section~\ref{sec:gallium} addresses the persistent
discrepancy observed in gallium-based neutrino detection experiments,
where measured electron neutrino ($\nu_e$) interaction rates from
radioactive sources systematically fall below SM predictions—a
phenomenon known as the gallium anomaly.  In Section~\ref{sec:CEvNS}
we explore another important class of experiments that bounds BSM physics:
Coherent Elastic Neutrino-Nucleus Scattering (CE$\nu$NS). Recent
precision measurements of this neutral-current process have
established new constraints on neutrino-quark coupling parameters,
bounding new physics that could add any contribution to the SM
predictions.

\section{Solar Neutrinos}
\label{sec:solar_exp}
The Sun is a powerful source of neutrinos, originating from nuclear
fusion processes within its core. As we will see in the coming
chapters, studying these neutrinos provides information on both particle
physics and solar physics. The fusion proceeds through two primary pathways:
\begin{itemize}
\item \textbf{Proton-Proton (pp) Chain:} This is the dominant fusion
  process in the Sun. It consists of several branches 
  \begin{itemize}
 \item \textbf{pp}: This is the dominant branch and proceeds as follows 
\begin{align*}
1. & \quad p + p \rightarrow d + e^+ + \nu_e \quad (E_{\nu} \leq 0.42
\ \text{MeV}) \\ 2. & \quad p + d \rightarrow \ ^3\text{He} + \gamma
\\ 3. & \quad ^3\text{He} + ^3\text{He} \rightarrow \ ^4\text{He} + 2p
\\
\end{align*}
\item \textbf{$^7\text{Be}$}: Second in energy production 
\begin{align*}
1.\,\,\, & ^3\text{He} + ^4\text{He} \rightarrow ^7\text{Be} + \gamma
\\ 2.\,\,\, & ^7\text{Be} + e^- \rightarrow ^7\text{Li} + \nu_e \quad
(E_{\nu} = 0.861 \ \text{MeV}) \\ 3. \,\,\,& ^7\text{Li} + p
\rightarrow 2\ ^4\text{He}.
\end{align*}
\item \textbf{pep}:
\begin{align*}
p + e^- + p \rightarrow d + \nu_e \quad (E_{\nu} = 1.44 \ \text{MeV}).
\end{align*}
\item  \textbf{$^8\text{B}$}: It is responsible for producing most of the  high-energy solar neutrinos:
\begin{align*}
1.&\,\,\,^3\text{He} + ^4\text{He} \rightarrow ^7\text{Be} + \gamma,
\\ 2.&\,\,\,^7\text{Be} + p \rightarrow ^8\text{B} + \gamma,
\\ 3.&\,\,\,^8\text{B} \rightarrow ^8\text{Be}^* + e^+ + \nu_e,
\\ 4.&\,\,\,^8\text{Be}^* \rightarrow 2\ ^4\text{He}.
\end{align*}
\item \textbf{hep} : It produces the neutrinos of highest energy via
\begin{align*}  
 &\,\,\,^3\text{He} + p \rightarrow ^4\text{He} + e^+ + \nu_e,
\end{align*}
but has the lowest flux.
\end{itemize}
\item \textbf{Carbon-Nitrogen-Oxygen (CNO) Cycle:} This process is
  less significant in the Sun but dominates in heavier stars. It uses
  carbon, nitrogen, and oxygen as catalysts to fuse hydrogen into
  helium, producing neutrinos in the process. Measuring precisely the
  flux of these neutrinos is essential to determine the abundance of
  heavy elements inside the Sun. The main reactions are:
\begin{align*}
1. & \quad ^{12}\text{C} + p \rightarrow ^{13}\text{N} + \gamma
\\ 2. & \quad ^{13}\text{N} \rightarrow ^{13}\text{C} + e^+ + \nu_e
\quad (E_{\nu} \leq 1.2 \ \text{MeV}) \\ 3. & \quad ^{13}\text{C} + p
\rightarrow ^{14}\text{N} + \gamma \\ 4. & \quad ^{14}\text{N} + p
\rightarrow ^{15}\text{O} + \gamma \\ 5. & \quad ^{15}\text{O}
\rightarrow ^{15}\text{N} + e^+ + \nu_e \quad (E_{\nu} \leq 1.7
\ \text{MeV}) \\ 6. & \quad ^{15}\text{N} + p \rightarrow
^{12}\text{C} + ^4\text{He}.
\end{align*}
\end{itemize}

\begin{figure}[t]
    \centering \includegraphics[width=0.6\linewidth,
      angle=-90]{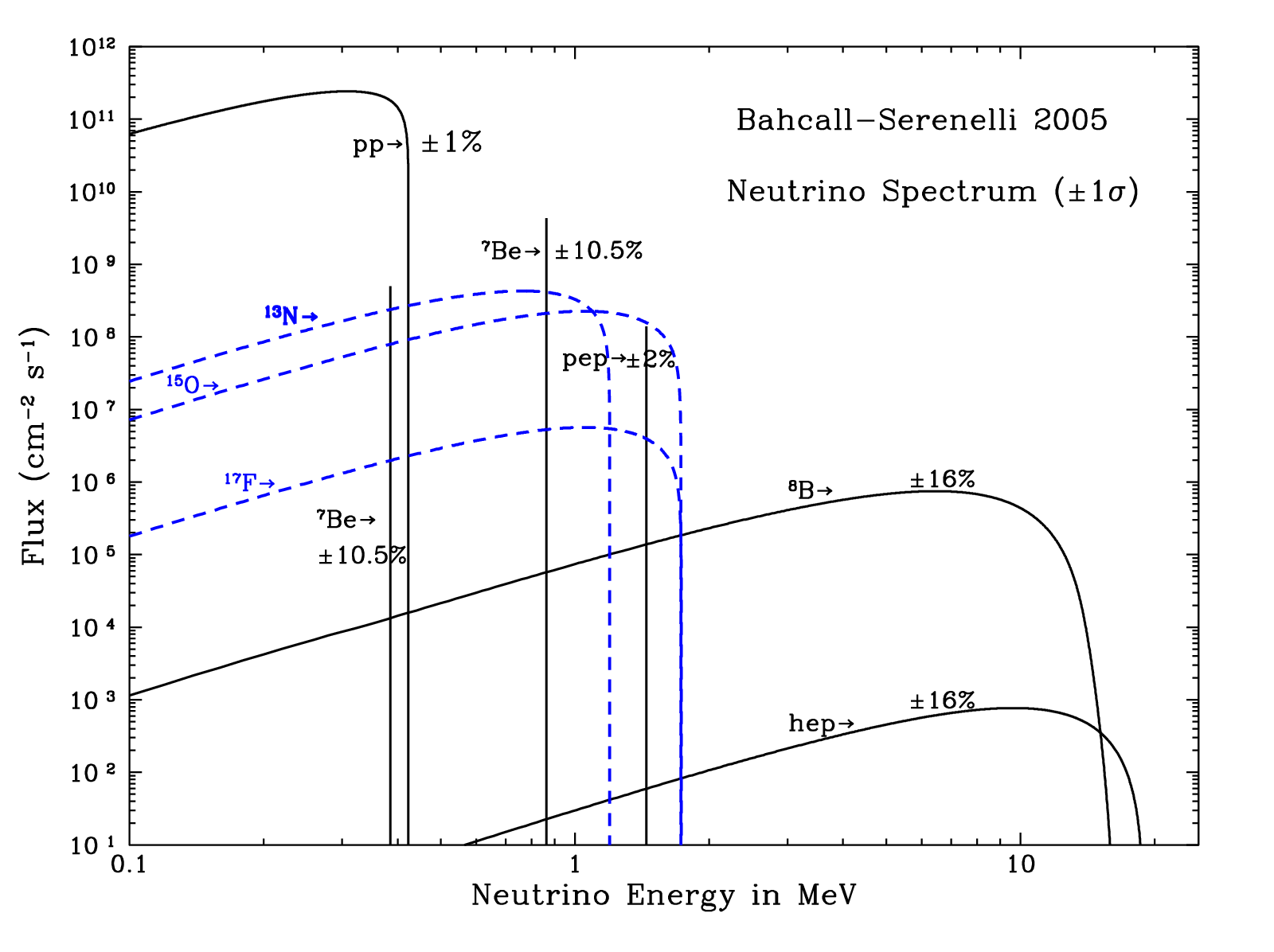}
    \caption{Neutrino fluxes predicted by the SSM\cite{Bahcall:2004pz}
      as a function of the neutrino energy. Figure extracted from
      \cite{Concha2}.}
    \label{fig:nuflux_sun}
\end{figure}

The resulting neutrino spectrum exhibits typical energies on the order
of $ \mathcal{O}(\text{MeV})$ as seen in Fig.~\ref{fig:nuflux_sun}.
The precise computation of this
spectrum requires detailed knowledge of the Sun's structure and
evolution \cite{Bahcall:2004pz,Si}. Standard Solar Models (SSM), are
mathematical representations used in astrophysics to describe the
structure, composition, and evolution of the Sun. Over the years several SSM, able to describe
the properties of the Sun and its evolution after entering the main
sequence, have been constructed with increasing level of
refinement~\cite{Bahcall:1987jc, TurckChieze:1988tj, Bahcall:1992hn,
  Bahcall:1995bt, Bahcall:2000nu, Bahcall:2004pz, PenaGaray:2008qe,
  Serenelli:2011py, Vinyoles:2016djt}.  Such models are numerical
calculations calibrated to match present-day surface properties of the
Sun, and developed under the assumption that the Sun was initially
chemically homogeneous and that mass loss is negligible at all moments
during its evolution up to the present solar age $\tau_\odot =
4.57$~Gyr.  The calibration is done in order to satisfy the
constraints imposed by the current solar luminosity $L_\odot$, radius
$R_\odot$, and surface metal to hydrogen abundance ratio
$(Z/X)_\odot$.  Refinements introduced over the years include more
precise observational and experimental information about the input
parameters (such as nuclear reaction rates and the surface abundances
of different elements), more accurate calculations of constituent
quantities (such as radiative opacity and equation of state), the
inclusion of new physical effects (such as element diffusion), and the
development of faster computers and more precise stellar evolution
codes. The
predictions of SSMs can be tested against the neutrino
observations as we will quantify in Chapter~\ref{chap:SSM_BX}.

The detection of solar neutrinos began in 1970 with the Homestake
experiment~\cite{Homestake}, which employed radiochemical methods to
observe neutrinos via
inverse beta decay interactions in a
chlorine-based target with a detection threshold of  0.814 MeV
(so mostly sensitive to $^7$Be and $^8$B fluxes)  
and integrating the interaction of the neutrinos over time.
This experiment provided the first direct evidence of solar
neutrinos but revealed a significant discrepancy: the observed
neutrino flux was approximately one-third of the flux predicted by the
SSM. This inconsistency was termed the "solar neutrino problem".
The second generation of radiochemical detectors
SAGE~\cite{Abdurashitov:2009tn} and Gallex/GNO\cite{Kaether:2010ag}
detect inverse beta decay interactions in a
gallium-based target which had a much lower  detection threshold of 0.233 MeV
and was able to detect the most abundant flux of pp neutrinos. They also
observed a deficit but it was $\sim$ 50\%.

The Kamiokande experiment~\cite{Kamioka1,Kamioka2,Kamioka3}, a
precursor to Super-Kamiokande, was a water Cherenkov detector located
in Japan. Unlike the radiochemical experiments, Kamiokande
was able to detect solar neutrinos in \textbf{real time} through elastic
scattering of neutrinos with electrons ($\nu + e^- \rightarrow \nu +
e^-$). This process produced Cherenkov light, enabling directional
information that confirmed neutrinos originated from the
Sun. Kamiokande observed a solar neutrino flux deficit of
approximately 60$\%$ compared to SSM predictions. However, its higher
energy threshold ($\sim 7.5$
MeV) limited sensitivity to only $^8$B neutrinos. While Kamiokande
could not resolve the solar neutrino problem alone, its real-time
detection capability and directional verification were essential for
subsequent experiments like Super-Kamiokande and SNO, which identified
neutrino flavour transitions as the solution.
The Super-Kamiokande experiment~\cite{Hosaka:2005um, Cravens:2008aa,Abe:2010hy, Super-Kamiokande:2023jbt}, a water Cherenkov detector,
measured solar neutrinos with improved precision and spectral
resolution. By detecting Cherenkov radiation from relativistic
electrons produced in neutrino interactions, Super-Kamiokande
confirmed the deficit of high-energy solar neutrinos and
determined  the  energy (in)dependence of the  suppression.
This spectral information provided critical evidence to establish the
MSW effect as the mechanism behind the observations.
Still because the radiochemical  experiments detected only $\nu_e$ and
the rate of $\nu_e$ interactions in water Cherenkov detectors is larger than that of $\nu_\mu$ and $\nu_\tau$,
these deficits could only be established through comparison with predictions of
some  SSM.  

The Sudbury Neutrino Observatory (SNO)~\cite{Aharmim:2011vm} resolved
the solar neutrino problem conclusively and independently of the SSM.
Using heavy water $(D_2O)$,
SNO simultaneously measured the flux of electron neutrinos (via
charged-current interactions) and the total flux of all active
neutrino flavours (via neutral-current interactions). The results
showed that the total neutrino flux matched SSM predictions, while the
electron neutrino flux remained suppressed. This confirmed that solar
neutrinos undergo flavour transitions, with electron neutrinos
transitioning to muon or tau neutrinos during propagation.

Ultimately all these results validated the neutrino flavour oscillation
hypothesis with  averaged oscillation in vacuum
and MSW matter transition as dominant mechanisms depending on the
neutrino energy. In a nut-shell in the Sun's high-density core ($\sim
150~\mathrm{g/cm^3}$), the higher-energy solar neutrinos ($E \gtrsim
10~\mathrm{MeV}$) undergo flavor conversion due to
matter-enhanced oscillations as described Sec.~\ref{sec:matter}.
In a two-neutrino system the adiabatic transition aligns the
produced $\nu_e$ with the heaviest matter eigenstate ($\nu_2^m$),
which exits the Sun as the vacuum mass eigenstate $\nu_2$, resulting
in $P_{ee} \approx \sin^2\theta\sim 0.31$ ($\theta\sim
34^\circ$), consistent with measurements from Homestake
and SK/SNO. At low energies ($E \ll
1~\mathrm{MeV}$), vacuum oscillations dominate: neutrinos propagate
incoherently as mass eigenstates ($\nu_1, \nu_2$), and $P_{ee}$
approaches $\cos^4\theta + \sin^4\theta =1-\frac{1}{2}\sin^22\theta
\approx 0.57$ (in agreement with the Gallium experiments),
reflecting the incoherent sum of probabilities for
$\nu_1$ and $\nu_2$ to be detected
as $\nu_e$. The transition between these regimes is governed by the
interplay of the solar matter potential ($V \sim
10^{-11}~\mathrm{eV}$) and vacuum oscillation wavelength and requires
$\Delta m^2_{21} \sim 7.5 \times 10^{-5}~\mathrm{eV^2}$.
In the framework of 3$\nu$ mixing these effect is controlled by
$\Delta m^2_{21}$ and $\theta_{12}$  with small corrections from $\theta_{13}$
as seen in Eq.~\eqref{eq:three_family}. In summary
\begin{equation}
    P_{ee} \approx
    \begin{cases} 
 \cos^4\theta_{13}\sin^2\theta_{12} +\sin^4\theta_{13}& \text{(high energy, adiabatic limit)},
  \\ \cos^4\theta_{13}\left(1-\frac{1}{2}\sin^22\theta_{12}\right)+\sin^4\theta_{13} & \text{(low energy, vacuum limit)}.
    \end{cases}
    \label{eq:Pee_limits}
\end{equation}

In the last decade Borexino~\cite{Borexino:2008fkj}, has significantly
advanced solar neutrino research by utilizing liquid scintillator
technology. It measures neutrinos through the process of elastic
scattering with electrons, a method that allows for the detection of
low-energy neutrinos with high sensitivity.  This capability has been
essential in observing solar neutrinos from all branches of the
pp-chain. Most recently, Borexino reported also the detection of CNO
neutrinos with a confidence level exceeding
6$\sigma$\cite{BOREXINO:2020aww}. This measurement provides new
insights into solar metallicity and confirms the dominant energy
production mechanism in heavy stars. During my PhD, together with
collaborators, we contributed important results on solar neutrinos. In
the next section, we will describe our analysis and in next chapters
we will present the main results obtained from this work.

\subsection{Simulation of Borexino Phase II and III spectra}
\label{sec:bxsimul}
The Borexino experiment, located in Gran Sasso (Italy), played a
crucial role in the precise measurement of solar neutrinos due to its
remarkably low background noise (it's still playing; however, the
experiment was discontinued recently). Phase I of the experiment was
successful in the detection of $^7$Be, adding significantly to our
understanding of the sun's energy generation \cite{Bellini:2011rx}.
Phase II of Borexino was characterized by strongly reduced
internal $^{85}$Kr and $^{210}$Bi backgrounds, leading to the most precise measurement
to date of $^7$Be solar neutrinos and the first real-time detection of pp and pep
neutrinos.  In Phase III, the Borexino succeeded in measuring for the first time the
CNO cycle by thermally stabilizing the detector, thus reducing the convection currents responsible for the diffusion of \Nuc[210]{Bi} into the scintillator. A detailed description of our analysis of the full spectrum
of the Phase-I~\cite{Bellini:2011rx, Borexino:2008fkj} and
Phase-II~\cite{Borexino:2017rsf} of Borexino can be found in
Ref.~\cite{Gonzalez-Garcia:2009dpj} and Ref.~\cite{Coloma:2022umy}
respectively.
Here we document the details of our analysis of the Borexino Phase II
(data collected between December 2011 and May 2016, corresponding
  to an exposure of $\text{1291.51 days}\times\text{71.3 tons}$) and
Phase-III (data collected from January 2017 to October
2021, corresponding to a total exposure of $\text{1431.6 days} \times
\text{71.3 tons}$), which we perform following closely the
details presented by the collaboration in
Refs.~\cite{Borexino:2017rsf} and~\cite{BOREXINO:2020aww, BOREXINO:2022abl}, respectively.

As discussed in a previous section, Borexino detects interactions of
solar neutrinos produced in the pp, pep, $^7$Be, and $^8$B reactions
of the pp-chain, as well as from all reactions in the CNO-cycle.  The
produced total flux is described by the differential spectrum:
\begin{equation}
  \label{eq:prod}
  \frac{\dd \phi_\nu^\text{prod}}{\dd E_\nu} = \sum_f \frac{\dd
    \phi_\nu^{f}}{\dd E_\nu}(E_\nu) ,
\end{equation}
where $E_\nu$ is the neutrino energy and $f$ refers to each of the
main components of the solar neutrino flux (pp, pep, CNO, $^7$Be, and
$^8$B).  In their journey to the Earth the produced $\nu_e$ will
change flavour and interact upon arrival via elastic scattering with
the $e^-$ of the detector.  Thus, the expected number of solar
neutrino events corresponding to $N_h$ hits is given by the
convolution of the oscillated solar neutrino spectrum with the
interaction cross section and the energy resolution function.
Quantitatively we compute the solar neutrino signal $S_i$ in the
$i$-th bin (\textit{i.e.}, with $N_h \in [N_{h,\text{min}}^i,
  N_{h,\text{max}}^i]$) as:
\begin{equation}
  \label{eq:binning}
  S_i = \int_{N_{h,\text{min}}^i}^{N_{h,\text{max}}^i} \int \frac{\dd
    S}{\dd T_e}(T_e)\, \frac{\dd\mathcal{R}}{\dd N_h}(T_e, N_h)\, \dd
  T_e\, \dd N_h \,.
\end{equation}
Here, $\dd S \big/ \dd T_e$ is the differential distribution of
neutrino-induced events as a function of recoil energy of the
scattered electrons ($T_e$)
\begin{equation}
  \label{eq:dSdTe}
  \frac{\dd S}{\dd T_e}(T_e) = \mathcal{F}_s \mathcal{N}_\text{tgt}\,
  \mathcal{T}_\text{run}\, \mathcal{E}_\text{cut} \int \frac{\dd
    \phi_\nu^\text{prod}}{\dd E_\nu}(E_\nu) \left[
    P_{e\alpha}(E_\nu)\, \frac{\dd \sigma^{\text{det}}}{\dd
      T_e}(\nu_\alpha) \right] \dd E_{\nu} 
\end{equation}
where $\mathcal{F}_s = 0.3572$ ($0.6359$) is the fraction for $s =
\text{``tagged''}$ (``subtracted'') signal events,
$\mathcal{N}_\text{tgt}$ is the number of $e^-$ targets
(\textit{i.e.}, the total number of electrons inside the fiducial
volume of the detector, corresponding to 71.3~ton of scintillator for
both phases), while $\mathcal{T}_\text{run}^{II} =
1291.51~\text{days}$ and $\mathcal{T}_\text{run}^{III} =
1431.6~\text{days}$ is the data taking time for phase II and III,
respectively, and $\mathcal{E}_\text{cut} = 98.5\%$ is the overall
efficiency for both phases.
$P_{e\alpha}(E_\nu)$ is the transition probability between the
flavours $e$ and $\alpha$, and $\dd\sigma^{\text{det}}(\nu_\alpha)/\dd
T_e$ is the flavour dependent $\nu_\alpha - e^-$ elastic scattering
detection cross section.

In addition, Eq.~\eqref{eq:binning} includes the energy resolution
function $\dd\mathcal{R} \big/ \dd T_e$ for the detector, which gives
the probability that an event with electron recoil energy $T_e$ yields
an observed number of hits $N_h$.  We assume it follows a Gaussian
distribution:
\begin{equation}
  \label{eq:dist-diff}
  \frac{\dd\mathcal{R}}{\dd N_h}(T_e, N_h) = \frac{1}{\sqrt{2\pi} \,
    \sigma_h(T_e)} \exp\bigg[ -\frac{1}{2} \bigg( \frac{N_h -
      \bar{N}_h(T_e)}{\sigma_h(T_e)} \bigg)^2 \bigg] \,,
\end{equation}
where $\bar{N}_h$ is the expected value of $N_h$ for a given true
recoil energy $T_e$.  We determine $\bar{N}_h(T_e)$ and
$\sigma_h(T_e)$ via the calibration procedure described below.

The data from BXII contains background contributions from a number of
sources.  The main backgrounds come from radioactive isotopes in the
scintillator \Nuc[14]{C}, \Nuc[11]{C}, \Nuc[10]{C}, \Nuc[10]{Po},
\Nuc[10]{Bi}, \Nuc[85]{Kr}, and \Nuc[6]{He}.  The collaboration
identifies two additional backgrounds due to pile-up of uncorrelated
events, and residual external backgrounds (see
Ref.~\cite{Borexino:2017rsf} for a complete description of these
backgrounds).

Among the considered backgrounds, the one coming from \Nuc[11]{C} is
particularly relevant for the analysis.  This isotope is produced in
the detector by muons through spallation on \Nuc[12]{C}.  In
Ref.~\cite{Borexino:2017rsf} the collaboration uses a Three-Fold
Coincidence (TFC) method to tag the \Nuc[11]{C} events, which are
correlated in space and time with a muon and a neutron.  Thus, they
divide the Phase II data set in two samples: one enriched (tagged) and
one depleted (subtracted) in \Nuc[11]{C} events.  The separation of
the solar neutrino signal (as well as most of the other backgrounds)
into these two samples is uncorrelated from the number of hits
$N_h$. Concretely, the tagged sample picks up $35.72\%$ of the solar
neutrino events, while the subtracted sample accounts for the
remaining $64.28\%$.

 Similarly to Phase-II, the main backgrounds of Phase-III come from
 radioactive isotopes in the scintillator \Nuc[11]{C}, \Nuc[210]{Bi},
 \Nuc[10]{C}, \Nuc[210]{Po} and \Nuc[85]{Kr}, extracted from Fig.~2(a)
 of Ref.~\cite{BOREXINO:2022abl} as well as the plot provided to us by
 the collaboration~\cite{borextag}.  These figures show the best-fit
 normalization of the different background components as obtained by
 the collaboration, and we take them as our nominal background
 predictions.  As for Phase II, the tagged sample of Phase III picks
 up about $35.72\%$ of the solar neutrino events, while the subtracted
 sample accounts for the remaining $64.28\%$.  The data and best fit
 components for the spectrum of the subtracted sample are shown in
 Fig.~2(a) of Ref.~\cite{BOREXINO:2022abl}.  The data points for this
 sample can also be found in the data release material in
 Ref.~\cite{borexdata}. The corresponding information for the tagged
 sample was kindly provided to us by the Borexino
 collaboration~\cite{borextag}.

In our analysis we include both sets of data, denoted in what follows
by $s = \text{``tagged''}$ or ``subtracted''.  As for the background,
we have read the contribution $B_{s,i}^c$ for each component $c$, in
each bin $i$, and for each data set $s$ from the corresponding lines
in the two panels in Fig.~7 of Ref.~\cite{Borexino:2017rsf}.  
Altogether the nominal number of expected events $T_{s,i}^0$ in some
bin $i$ of data sample $s$ is the sum of the neutrino-induced signal
and the background contributions:
\begin{equation}
  T_{s,i}^0 = \sum_f S_{s,i}^f + \sum_c B_{s,i}^c \,,
\end{equation}
where the index $f \in \{ \text{pp, $^7$Be, pep, CNO, $^8$B} \}$ runs
over the solar fluxes, while the index $c \in \{\text{\Nuc[14]{C},
  \Nuc[11]{C}, \Nuc[10]{C},}$ $\text{ \Nuc[210]{Po},
  \Nuc[210]{Bi},\Nuc[85]{Kr}, \Nuc[6]{He}, pile-up, ext} \}$ runs over
the background components.  In our calculation of $S_{s,i}^f$ we use
the high-metallicity (HZ) solar model for simplicity.\footnote{It
should be noted that this is the model currently favoured by the CNO
measurement at Borexino~\cite{BOREXINO:2020aww}, albeit with a modest
significance.}

\subsubsection{Energy Calibration}

The Borexino data fit is performed using the observed number of hits,
$N_h$, as an estimator for the electron recoil energy, $T_e$. The
energy resolution is modelled with a Gaussian function, parameterized
by the mean number of hits $\bar{N}_h(T_e)$ and its width
$\sigma_h(T_e)$. These parameters were derived in two steps:

1. The relation between $\bar{N}_h$ and $\sigma_h$ was calibrated
using $\gamma$-ray source data. During our investigations, we
discovered that each phase of Borexino has a particular way to relate
$\bar{N}_h$ and $\sigma_h$, particularly for the last two phases of
the experiment.

For \textbf{Borexino Phase II}:
   \begin{equation}
   \sigma_h(\bar{N}_h) = 1.21974 + 1.31394 \sqrt{\bar{N}_h} -
   0.0148585 \bar{N}_h \,.
   \end{equation}

And for the \textbf{Borexino Phase III} \begin{equation}
  \sigma_h(\bar{N}_h) = 1.21974 + 1.60121 \sqrt{\bar{N}_h} - 0.014859
  \bar{N}_h \,.
   \end{equation}

2. A third-degree polynomial was used to model $\bar{N}_h(T_e)$ for
reproducing solar neutrino spectra:
   \begin{equation}
   \bar{N}_h(T_e) = -8.065244 + 493.2560 \frac{T_e}{\text{MeV}} -
   64.09629 \left(\frac{T_e}{\text{MeV}}\right)^2 + 4.146102
   \left(\frac{T_e}{\text{MeV}}\right)^3 \,.
   \end{equation}

\begin{figure}[t]\centering
  \includegraphics[width=\linewidth]{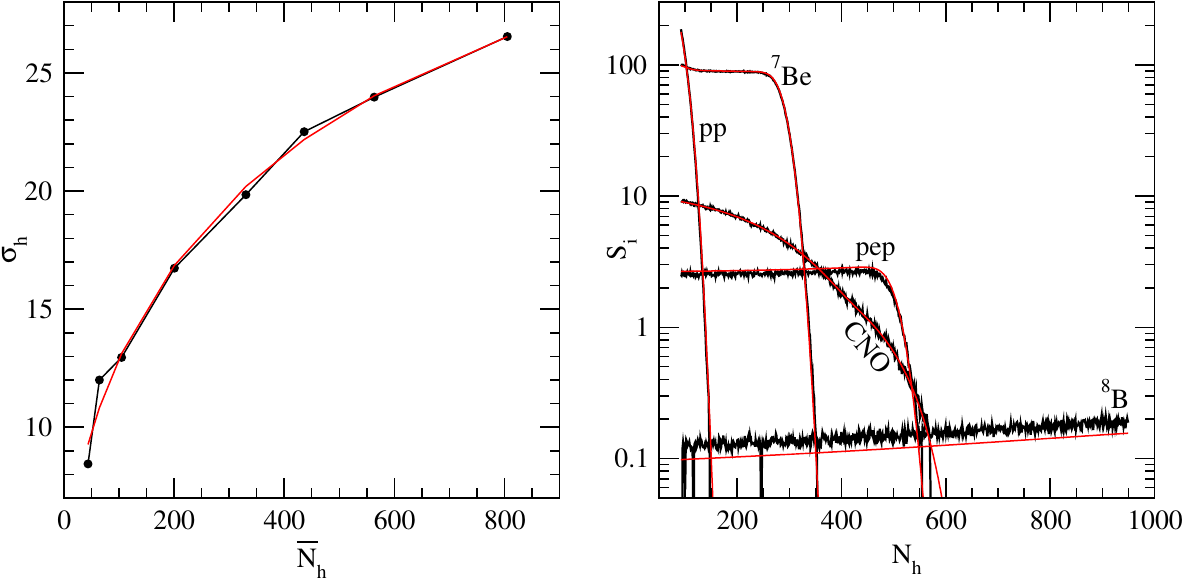}
  \caption{Left: our choice of the energy resolution function, based
    on the relation between $\sigma_h$ and $\bar{N}_h$ inferred from
    the upper panel of Fig.~22 of Ref.\cite{Borexino:2017rsf}.  Right:
    our reconstruction of solar spectra after the optimization of the
    energy scale function.}
  \label{fig:calibra}
\end{figure}

\begin{figure}[ht!]\centering
\includegraphics[width=\linewidth]{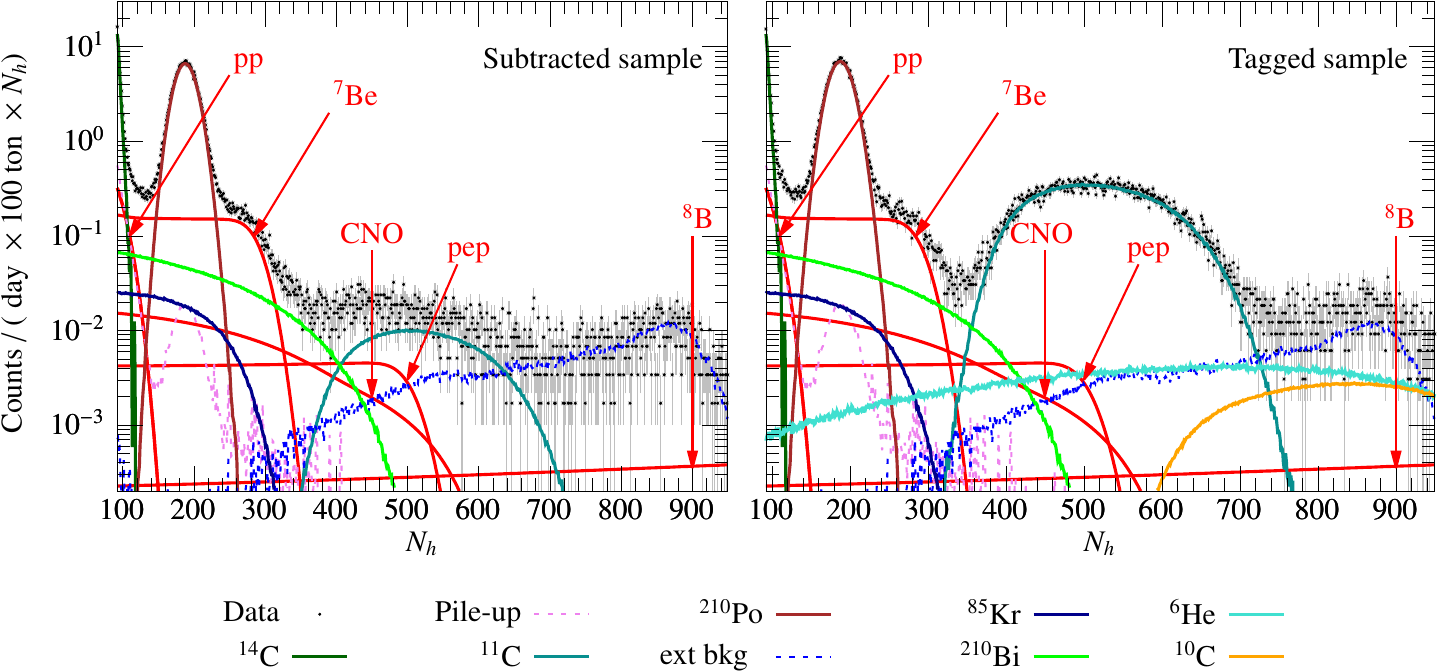}
\includegraphics[width=\linewidth]{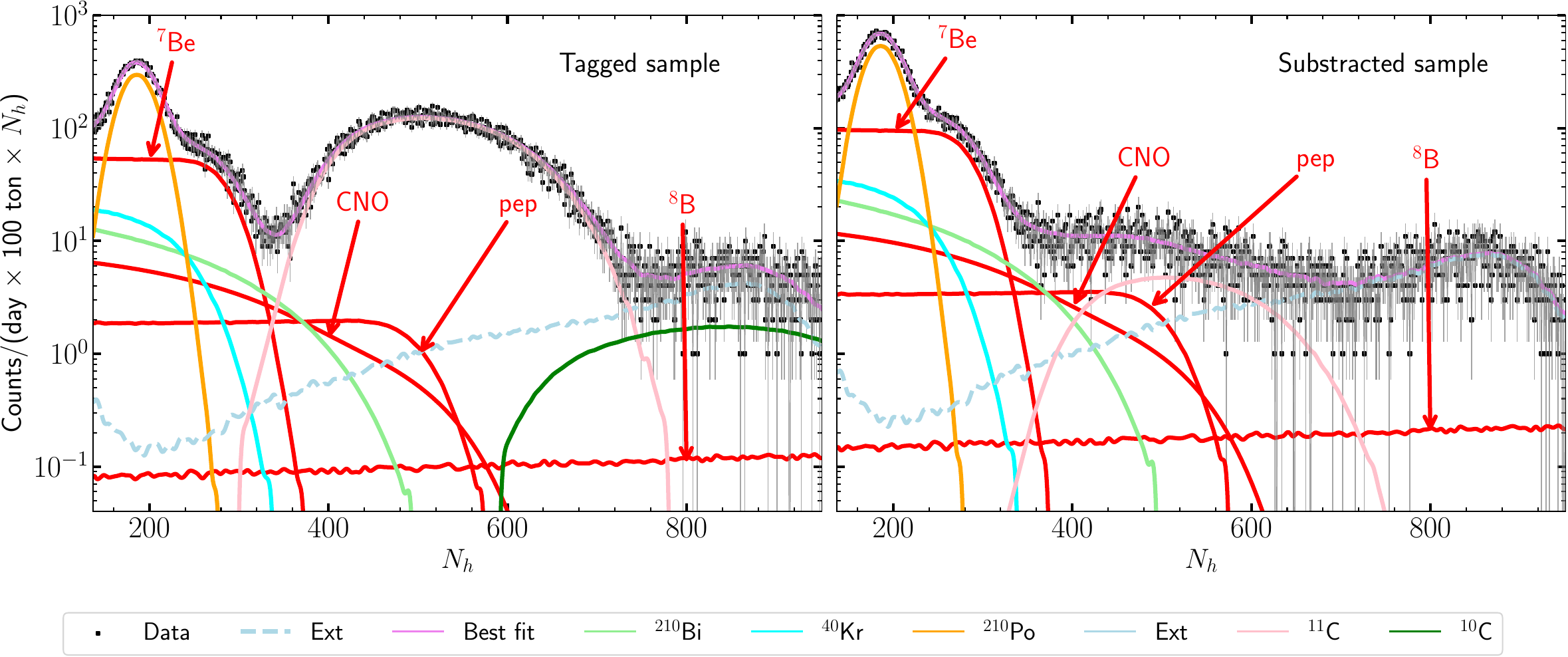}
  \caption{Spectrum for the best-fit normalizations of the different
    components obtained from our fit to the \textbf{(upper panel)}
    Borexino Phase II data for TFC-subtracted (left) and TFC-tagged
    events and \textbf{(lower panel)} Borexino Phase III data for
    TFC-subtracted (left) and TFC-tagged events.}
  \label{fig:borexcompa1}
\end{figure}

\begin{figure}[ht!]\centering
  \includegraphics[width=0.85\linewidth]{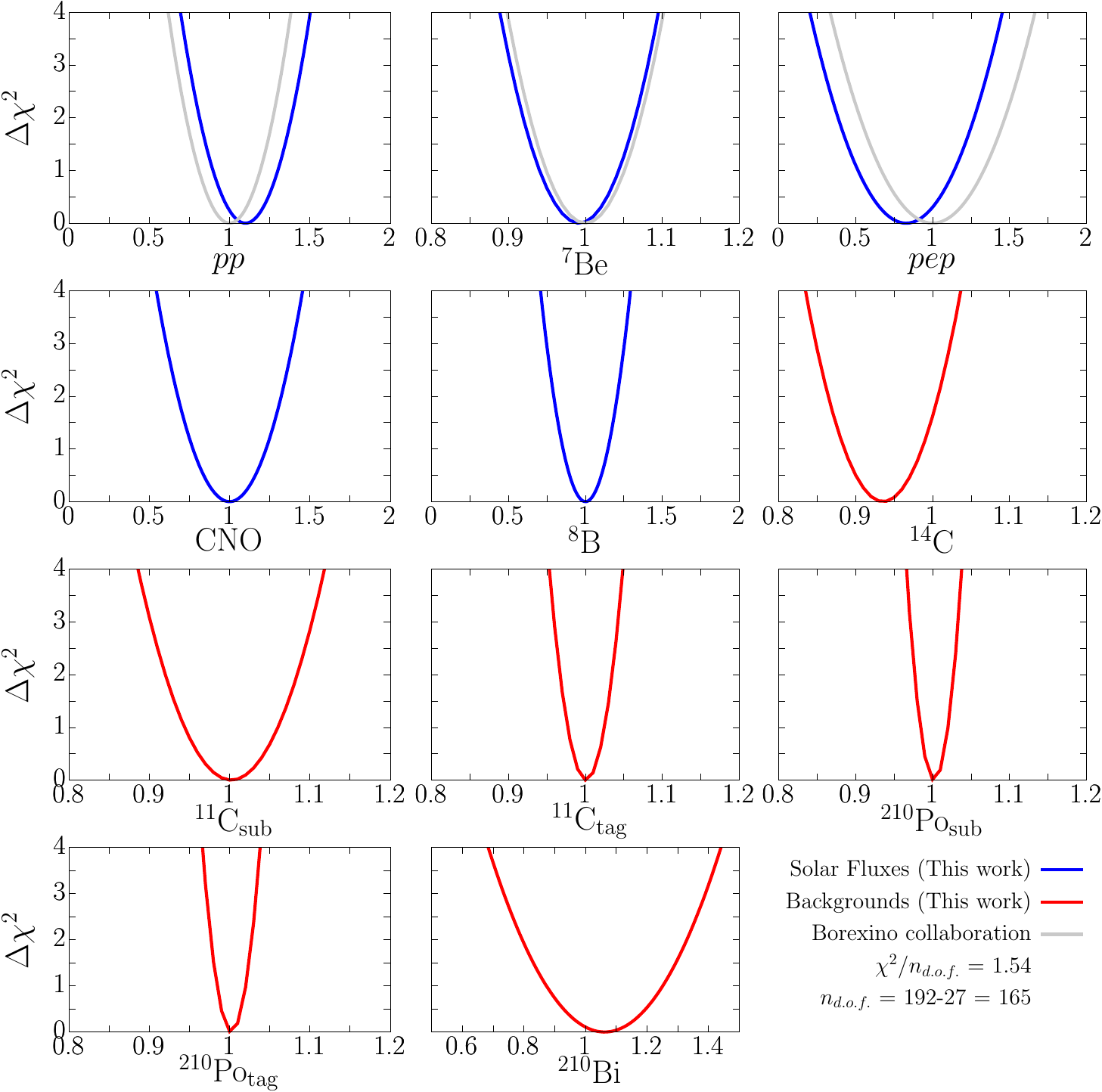}
  \caption{Dependence of our $\Delta\chi^2$ for the  fit to the Borexino
    phase-II spectra on the normalization of the solar fluxes and the
    dominant backgrounds (normalized to the corresponding best fit
    normalizations of the fit of the Borexino collaboration in
    Ref.\cite{Borexino:2017rsf}).  For comparison we show as light
    blue curves the corresponding results of the determination of
    solar fluxes in Ref.\cite{Borexino:2017rsf}) (see text for
    details).}
  \label{fig:borexcompa2}
\end{figure}

\subsubsection{Analysis of Borexino Phase II}

Our statistical analysis is based on the construction of a $\chi^2$
function based on the described experimental data, neutrino signal
expectations and sources of backgrounds, as well as the effect of
systematic uncertainties.

The analysis incorporates the concept of pulls, denoted as $\xi_r$,
which are varied to minimize the $\chi^2$. A total of 27 pulls are
included to account for uncertainties in background normalizations,
detector performance, priors on solar fluxes, and energy
shifts. Assuming a linear dependence of event rates on the pulls, the
$\chi^2$ minimization can be performed analytically if the $\chi^2$ is
Gaussian.

The Borexino collaboration, for Phase II, bins their data into 858
bins per sample, ranging from $N_h=92$ to $N_h=950$. However, this
fine binning does not provide sufficient events per bin to ensure
Gaussian statistics. To address this, we rebin the data using coarser
bins. Specifically:

\begin{itemize}
  \item The first 42 bins are grouped into sets of 2.
  \item The next 210 bins are grouped into sets of 5.
  \item The subsequent 260 bins are grouped into sets of 10.
  \item The final 346 bins are grouped into sets of 50.
\end{itemize}

This rebinning results in a total of 96 bins ($k$) for each data
sample ($s$), ensuring sufficient events in each bin to guarantee
Gaussian statistics. The resulting $\chi^2$ function is given by:

\begin{equation}
  \label{eq:chi2_0}
  \chi^2 = \min_{\vec{\xi}} \Bigg[ \sum_{s,k} \frac{\big[
        T_{s,k}(\vec{\xi}) - O_{s,k} \big]^2}{O_{s,k}} + \sum_r
    (\xi_r^{\text{unc}})^2 + \sum_{r,r'} (\Sigma^{-1})_{rr'}
    \xi_r^{\text{corr}} \xi_{r'}^{\text{corr}} \Bigg] \, ,
\end{equation}

where $O_{s,k}$ represents the observed number of events in bin $k$ of
sample $s$, and $T_{s,k}$ is the predicted number of events,
incorporating systematic uncertainties:

\begin{equation}
  \label{eq:tsys}
  T_{s,k}(\vec{\xi}) = T_{s,k}^0 + \sum_r D_{s,k}^r \xi_r \, .
\end{equation}
Here $D_{s,k}^r$ represents the derivatives of the total event rates
with respect to each pull.

The last two terms in Eq.~\eqref{eq:chi2} introduce Gaussian bias
factors to account for prior constraints on specific pulls. The first
term addresses pulls associated with uncorrelated (``unc'')
uncertainties, while the second term addresses correlated (``corr'')
uncertainties. In this context, $\Sigma$ represents the covariance
matrix for correlated uncertainties, detailed in the Appendix of
\cite{Coloma:2022umy}.

\subsubsection{Analysis of Borexino Phase-III spectrum}
\label{sec:BXIIIspec}

\begin{figure}[ht!]\centering
  \includegraphics[width=0.95\textwidth]{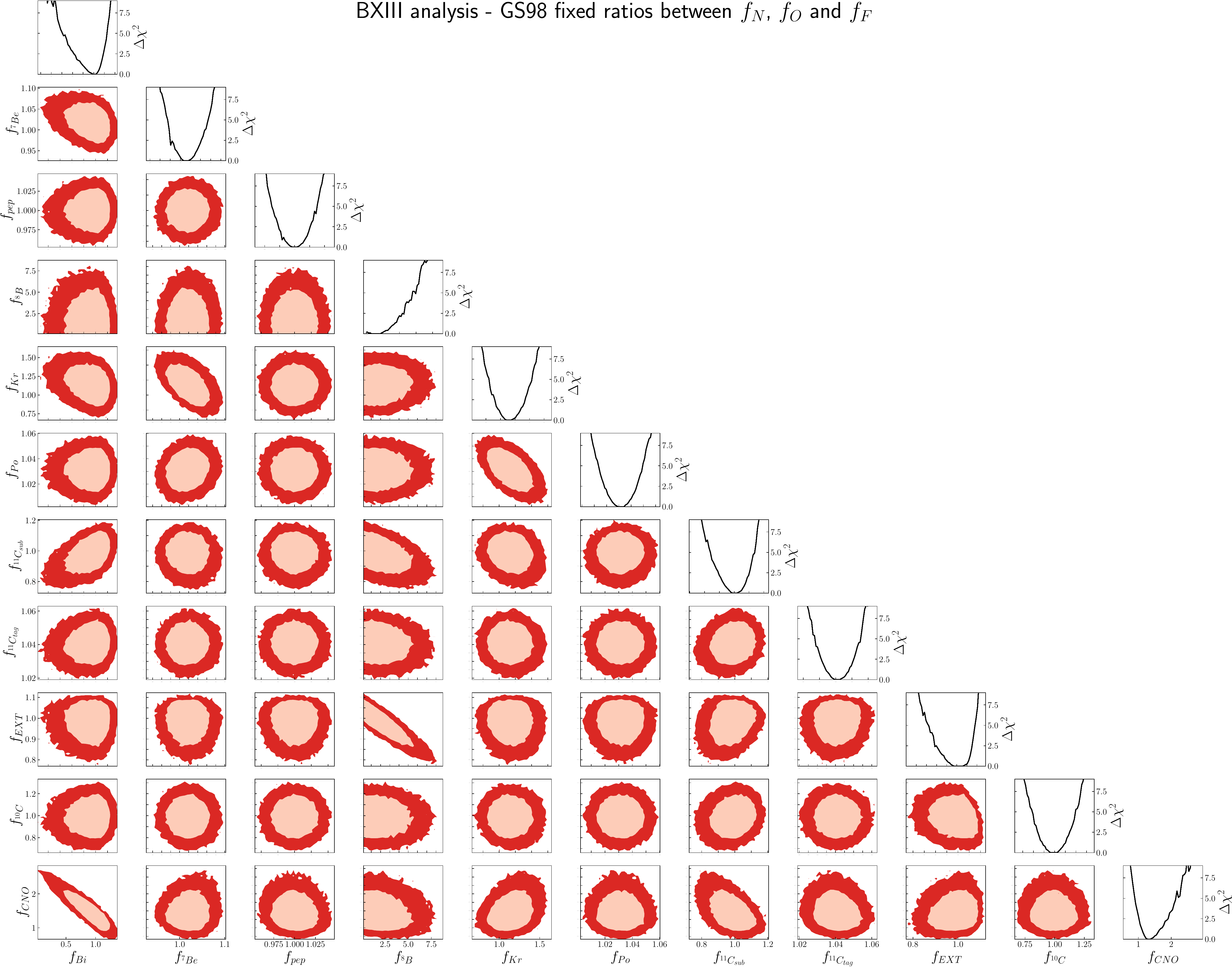}
  \caption{Constraints from our analysis of Borexino Phase-III spectra
    obtained with $\chi^2_\text{BXIII,test}$ in Eq.~\eqref{eq:bxtest}.
    Each panel shows a two-dimensional projection of the allowed
    multi-dimensional parameter space after minimization with respect
    to the undisplayed parameters.  The regions correspond to 90\% and
    99\% CL (2 d.o.f.).  The curves in the right-most panels show the
    marginalized one-dimensional $\Delta\chi^2_\text{BXIII,test}$ for
    each of the parameters.}
  \label{fig:BXIIIfit}
  \vskip 0.3cm
  \includegraphics[width=0.5\textwidth]{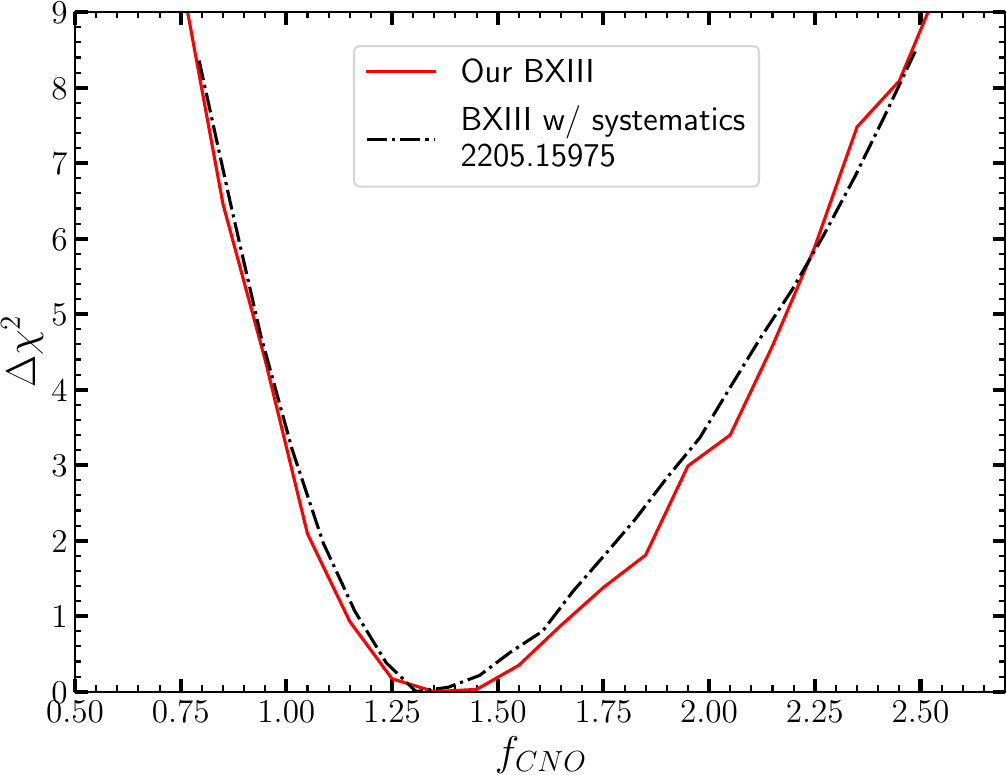}
  \caption{Dependence of $\Delta\chi^2$ of our fit to the Borexino
    Phase-III spectra on the common normalization of the CNO fluxes
    (red line).  For comparison we plot (black dot-dashed line) the
    corresponding results in figure 2(b) of~\cite{BOREXINO:2022abl}
    for their ``Fit w/ Systematics'', divided by the central value of
    the predicted CNO-$\nu$ rate of the B16-GS98 mode.}
  \label{fig:BXIIICNO}
\end{figure}

Our statistical analysis is based on the construction of a $\chi^2$
function built with the described experimental data, neutrino signal
expectations and sources of backgrounds.  Following
Refs.~\cite{BOREXINO:2020aww, BOREXINO:2022abl} we leave the
normalization of all the backgrounds as free parameters with the
exception of \Nuc[210]{Bi}.  The treatment of this background is
paramount to the positive evidence of CNO neutrinos.
As described in~\cite{BOREXINO:2020aww}, the extraction of the CNO
neutrino signal from the Borexino data faces two significant
challenges: the resemblance between spectra of CNO-$\nu$ recoil
electrons and the \Nuc[210]{Bi} $\beta^-$ spectra, and their
pronounced correlation with the \Nuc{pep}-$\nu$ recoil energy
spectrum.  In order to surpass the first challenge, the collaboration
restricted the rate of \Nuc[210]{Bi} for which it sets and upper
limit~\cite{BOREXINO:2022abl}:
\begin{equation}
  R(\Nuc[210]{Bi})\leq (10.8\pm 1.0)~\text{cpd} \big/ \text{100\,t}
  \,,
\end{equation}
while no constraint is imposed on its minimum value which is free to
be as low as allowed by the fit (as long as it remains
non-negative). In our analysis we implement this upper limit by
constraining the corresponding normalization factor
$f_{\Nuc[210]{Bi}}$ as
\begin{equation}
  f_{\Nuc[210]{Bi}}\leq \bigg( 1 \pm \frac{1.0}{10.8} \bigg),
\end{equation}
With this we construct the $\chi^2_\text{BXIII}$ as
\begin{equation}
  \label{eq:chi2BXIII}
  \chi^2_\text{BXIII} = \sum_{s,i} 2 \bigg[ T^0_{s,i} - O_{s,i} +
    O_{s,i} \log\bigg(\frac{O_{s,i}}{T^0_{s,i}} \bigg) \bigg] + \bigg(
  \frac{f_{\Nuc[210]{Bi}} - 1}{\sigma_{\Nuc[210]{Bi}}} \bigg)^2 \,
  \Theta(f_{\Nuc[210]{Bi}} -1)\,,
\end{equation}
where $O_{s,i}$ is the observed number of events in bin $i$ of sample
$s$, and $\sigma_{\Nuc[210]{Bi}} = 1.0 / 10.8$, and $\Theta(x)$ is the
Heaviside step function.

Constructed this way, $\chi^2_\text{BXIII}$ depends on 16 parameters:
the 3 oscillation parameters ($\Dmq_{21}$, $\theta_{12}$,
$\theta_{13}$), 6 solar flux normalizations ($f_{\Nuc[7]{Be}}$,
$f_{\Nuc{pep}}$, $f_{\Nuc[13]{N}}$, $f_{\Nuc[15]{O}}$,
$f_{\Nuc[17]{F}}$, $f_{\Nuc[8]{B}}$) and 7 background normalizations
($f_{\Nuc[210]{Po}}$, $f_{\Nuc[210]{Bi}}$, $f_{\Nuc[85]{Kr}}$,
$f_{\Nuc[10]{C}}$, $f_\text{ext}$ and two different factors
$f_{\Nuc[11]{C}}^\text{tag}$ and $f_{\Nuc[11]{C}}^\text{sub}$ for the
tagged and subtracted samples).

As a first validation of our $\chi^2$ function we perform an analysis
focused at reproducing the results on the solar neutrino fluxes found
by the Borexino collaboration in Ref.~\cite{BOREXINO:2022abl}, and in
particular the positive evidence of CNO neutrinos.  In this test fit
we fix the three oscillation parameters to their best fit value
($\sin^2\theta_{13} = 0.023$, $\sin^2\theta_{12} = 0.307$, $\Dmq_{21}
= 7.5\times 10^{-5}$), and following the procedure of the
collaboration we assume a common normalization factor for the three
CNO fluxes with respect to the SSM ($f_{\Nuc[13]{N}} = f_{\Nuc[15]{O}}
= f_{\Nuc[17]{F}} \equiv f_{\Nuc{CNO}})$.  Furthermore, in order to
break the pronounced correlation with the \Nuc{pep}-$\nu$ recoil
energy spectrum mentioned above, the collaboration introduced a prior
for the \Nuc{pep} neutrino signal flux following the SSM.  Thus we
define
\begin{equation}
  \label{eq:bxtest}
  \chi^2_\text{BXIII,test}=\chi^2_\text{BXIII} +
  \bigg(\frac{f_\text{pep} - 1}{\sigma_{\Nuc{pep}}} \bigg)^2
\end{equation}
with $\sigma_{\Nuc{pep}} = 0.04/2.74$ (for concreteness we choose the
B16-GS98 model for this prior).  The \Nuc[7]{Be} and \Nuc[8]{B} fluxes
are left completely free.

The results of this 11-parameter fit are shown in
Figs.~\ref{fig:BXIIIfit} and~\ref{fig:BXIIICNO}.  In
Fig.~\ref{fig:BXIIIfit} we plot the allowed ranges and correlations
for the parameters.  Notice that in this figure all parameters are
normalized to the best fit values obtained by the corresponding
analysis of the Borexino collaboration, hence a value of ``1'' means
perfect agreement.  We observe a strong correlation between the
normalization of the CNO fluxes $f_{\Nuc{CNO}}$ and the \Nuc[210]{Bi}
background.  This is expected because, as mentioned before, the
spectrum of CNO neutrinos and that of the \Nuc[210]{Bi} background are
similar.  Still, the two spectra are different enough so that, under
the assumption of the upper bound on the \Nuc[210]{Bi} background, the
degeneracy gets broken enough to lead to a positive evidence of CNO
neutrinos in an amount compatible with the prediction of the SSMs.

A quantitative comparison with the results of the collaboration is
shown in Fig.~\ref{fig:BXIIICNO} where we plot the dependence of our
marginalized $\Delta\chi^2$ on the common CNO flux normalization,
$f_{\Nuc{CNO}}$, together with that obtained by the collaboration as
extracted from Figure 2(b) of Ref.~\cite{BOREXINO:2022abl} (labeled
``Fit w/ Systematics'' in that figure).\footnote{Figure 2(b) of
Ref.~\cite{BOREXINO:2022abl} shows their $\Delta\chi^2$ as a function
of the CNO-$\nu$ event rate which we divide by the central value of
the expected rate in the B16-GS98 model to obtain the black dot-dashed
curve in Fig.~\ref{fig:BXIIICNO}.}
Altogether, these figures show that our constructed event rates and
the best-fit normalization of the CNO flux reproduce with very good
accuracy those of the fit performed by the collaboration.

\subsubsection{Analysis with Correlated Integrated Directionality Method}

In a very recent work~\cite{Borexino:2023puw} the Borexino
collaboration has presented a combined analysis of their three phases
making use of the Correlated and Integrated Directionality (CID)
method, which aims to enhance the precision of the determination of
the flux of CNO neutrinos.  In a nut-shell, the CID method exploits
the sub-dominant Cherenkov light in the liquid scintillator produced
by the electrons scattered in the neutrino interaction.  These
Cherenkov photons retain information of the original direction of the
incident neutrino, hence they can be used to enhance the
discrimination between the solar neutrino signal and the radioactive
backgrounds.

\begin{figure}[ht!]\centering
  \includegraphics[width=0.554\linewidth]{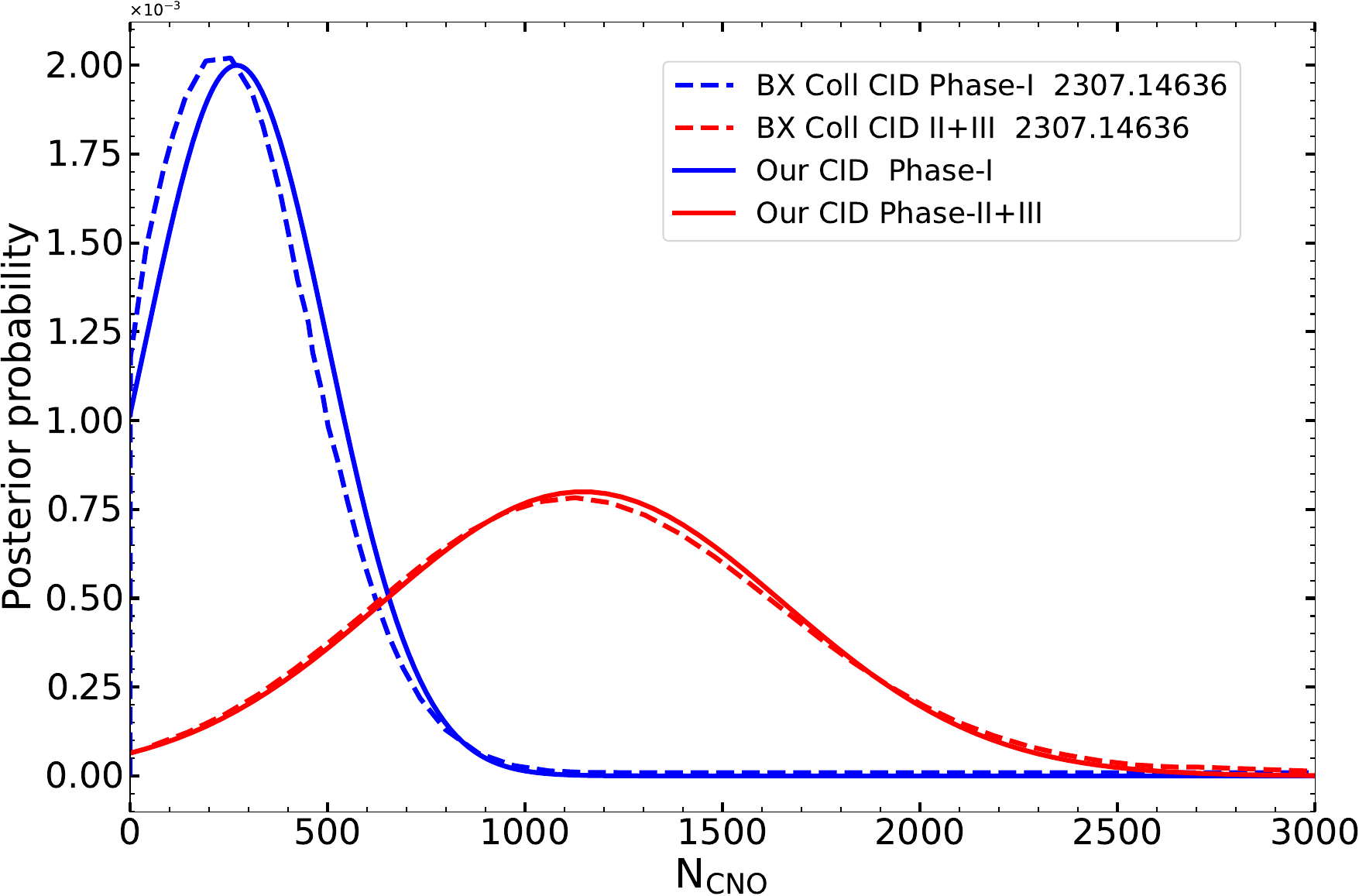}
  \hfill
  \includegraphics[width=0.426\linewidth]{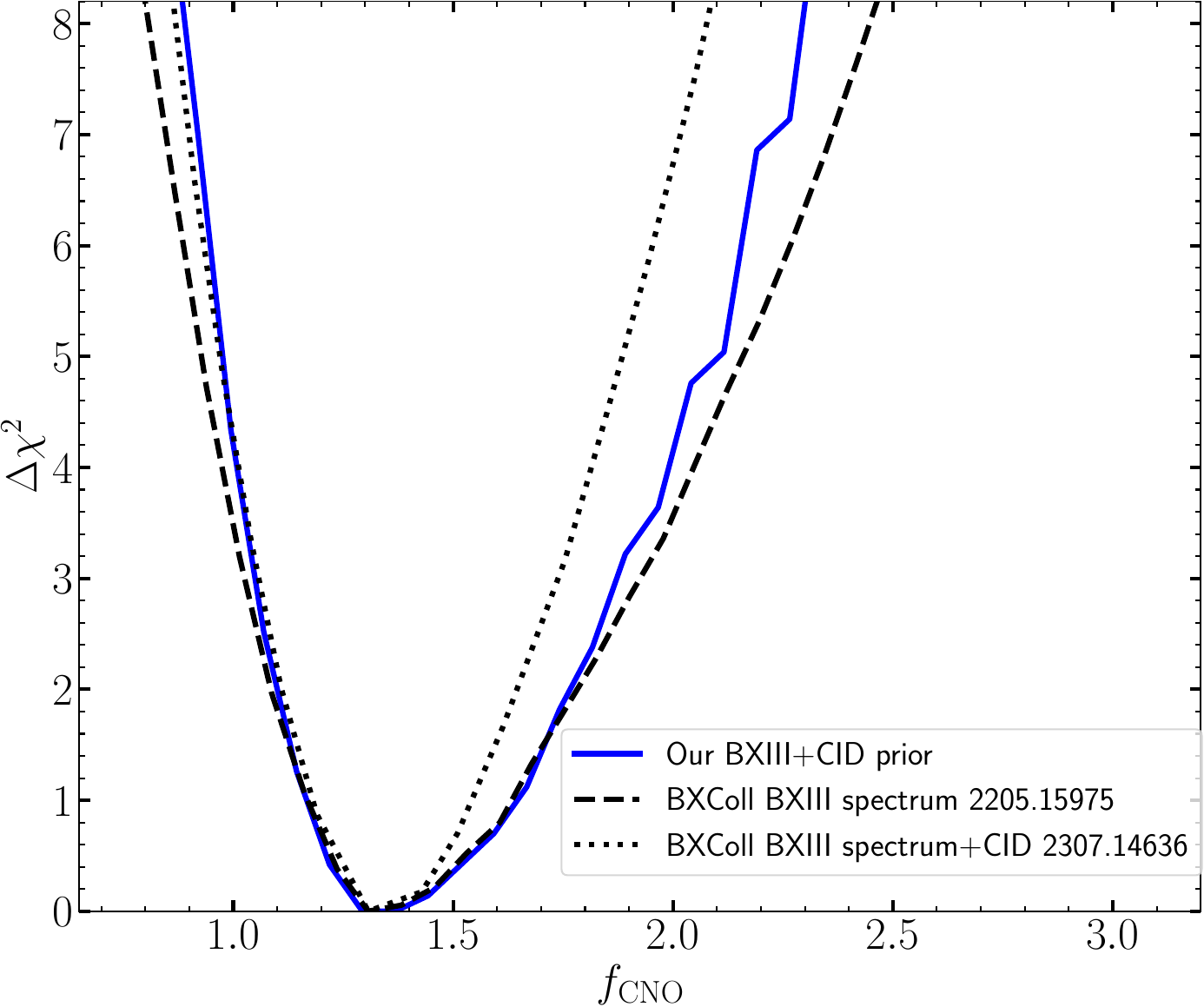}
  \caption{Left: CID posterior probabilities for the number for
    Phase-I and Phase-II+III of CNO-$\nu$ events after constraining
    \Nuc{pep} and \Nuc[8]{B} neutrino events to their SSM expectation.
    Right: Dependence of $\Delta\chi^2$ on the CNO flux normalization
    from our fit to the Borexino Phase-III spectra combined with the
    CID constraint (blue line) compared to that obtained by the
    Borexino collaboration (dotted black line).  For comparison we
    also show the result without CID information (dashed black line).}
  \label{fig:CIDcomp}
\end{figure}

Effectively, the CID analysis results into a determination of the
total number of solar neutrinos detected within a restricted range of
$N_h$ which corresponds to $0.85~\text{MeV} < T_e< 1.3~\text{MeV}$ for
Phase-I and $0.85~\text{MeV} < T_e< 1.29~\text{MeV}$ for Phase-II+III.
In this range the dominant contribution comes from \Nuc{CNO},
\Nuc{pep} and some \Nuc[8]{B}.  The increased fiducial volume for this
analysis brings the exposures to $740.7~\text{days} \times
104.3~\text{ton} \times 55.77\%$ for Phase-I and $2888.0~\text{days}
\times 94.4~\text{ton} \times 63.97\%$ for Phase-II+III.  The
resulting number of solar neutrinos detected is
$N^\text{P-I}_\text{obs} =
643^{+235}_{-224}\text{(stat)}^{+37}_{-30}\text{(sys)}$ for Phase-I
and $N^\text{P-II+III}_\text{obs} =
2719^{+518}_{-494}\text{(stat)}^{+85}_{-83}\text{(sys)}$ for
Phase-II+III.  After subtracting the expected SSM contribution from
\Nuc{pep} and \Nuc[8]{B} the Borexino collaboration obtains the
posterior probability distributions for the number of CNO neutrinos
shown in Fig.~9 of Ref.~\cite{Borexino:2023puw} (which we reproduce in
the left panel in Fig.~\ref{fig:CIDcomp}).  Furthermore, since this
new directional information is independent of the spectral
information, the collaboration proceeded to combine these two priors
on $N_{\Nuc{CNO}}$ with the their likelihood for the Borexino
Phase-III spectral analysis.  This resulted in a slightly stronger
dependence of the combined likelihood on the CNO-$\nu$ rate shown in
their Fig.~12 (which we reproduce in the right panel in
Fig.~\ref{fig:CIDcomp}).

In order to account for the CID information in our analysis we try to
follow as closely as possible the procedure of the collaboration.
With the information provided on the covered energy range and
exposures for the CID analysis, we integrate our computed spectra of
solar neutrino events in each phase to derive the corresponding total
number of expected events in Phase-I and Phase-II+III.  We then
subtract the SSM predictions for \Nuc{pep} and \Nuc[8]{B} neutrinos
from the observed number of events to derive an estimate for \Nuc{CNO}
neutrinos in in Phase-I and Phase-II+III, and construct a simple
Gaussian $\chi^2(N_{\Nuc{CNO}})$ for Phase-Y (Y=I or II+III)
\begin{equation}
  \label{eq:chi2CID}
  \chi^2_\text{CID,P-Y}(N_{\Nuc{CNO}}) = \bigg(
  \frac{N_\text{obs}^\text{P-Y} - N^\text{P-Y}_\text{SSM,pep} -
    N^\text{P-Y}_\text{SSM,\Nuc[8]{B}} -
    N_{\Nuc{CNO}}}{\sigma_\text{P-Y}} \bigg)^2
\end{equation}
where in $\sigma_{P-Y}$ we add in quadrature the symmetrized
statistical and systematic uncertainties in the number of observed
events.

We plot in the left panel in Fig.~\ref{fig:CIDcomp} our inferred
probability distributions $P_\text{P-Y}(N_{\Nuc{CNO}})\propto
\exp[-\chi^2_\text{P-Y}/2]$ compared to those from Borexino in Fig.~9
of Ref.~\cite{Borexino:2023puw}.  As seen in the figure our simple
procedure reproduces rather well the results of the collaboration for
the Phase-II+III but only reasonably for Phase-I.  This may be due to
differences in the reanalysis of the Phase-I data by the collaboration
in the CID analysis compared to their spectral analysis of 2011.  Our
simulations of the Phase-I event rates are tuned to their 2011 and
there is not enough information in Ref.~\cite{Borexino:2023puw} to
deduce what may have changed.  Thus we decide to introduce in our
analysis the CID prior for the Phase-II+III data but not for Phase-I.

We then combine the CID from Phase-II+III and Phase-III spectral
information as
\begin{equation}
  \chi^2_\text{CID+BXIII,test} = \chi^2_\text{BX-III} + \bigg(
  \frac{f_{\Nuc{pep}}-1}{\sigma_{\Nuc{pep}}} \bigg)^2 +
  \chi^2_\text{CID,P-II+III} \,.
\end{equation}
A quantitative comparison with the results of the collaboration for
this combined $\text{CID} + \text{Phase-III}$ spectrum analysis is
shown in the right panel of Fig.~\ref{fig:CIDcomp} where we plot the
dependence of our marginalized $\Delta\chi^2$ on the CNO flux
normalization after including the CID information compared to that
obtained by the collaboration in Fig.~12 of
Ref.~\cite{Borexino:2023puw}.  As seen in the figure, we reproduce
well the improved sensitivity for the lower range of the CNO flux
normalization but our constraints are more conservative in the higher
range, though they still represent an improvement over the
spectrum-only analysis.

Altogether, after all these tests and validations we define the
$\chi^2$ for the full Borexino analysis as
\begin{multline}
  \label{eq:chi2bxtot}
  \chi^2_\text{BX}(\vec\omega_\text{osc},\, \vec\omega_\text{flux}) =
  \chi^2_\text{BXI}(\vec\omega_\text{osc},\, \vec\omega_\text{flux}) +
  \chi^2_\text{BXII}(\vec\omega_\text{osc},\, \vec\omega_\text{flux})
  \\ + \chi^2_\text{BXIII}(\vec\omega_\text{osc},\,
  \vec\omega_\text{flux}) + \chi^2_\text{CID,P-II+III}
  (\vec\omega_\text{osc},\, \vec\omega_\text{flux}) \,.
\end{multline}
with $\chi^2_\text{BXIII}(\vec\omega_\text{osc}\,
\vec\omega_\text{flux})$ and $\chi^2_\text{CID,P-II+III}
(\vec\omega_\text{osc}, \vec\omega_\text{flux})$ in
Eqs.~\eqref{eq:chi2BXIII} and~\eqref{eq:chi2CID}, respectively.  We
finish by noticing that the inclusion of the CID information is not
enough to break the large degeneracy between the \Nuc[13]{N} and
\Nuc[210]{Bi} contributions to the spectra discussed in the previous
section.

\section{Atmospheric Neutrinos}
\label{sec:atm_exp}

\begin{figure}[t] \centering
\includegraphics[width=0.5\textwidth]{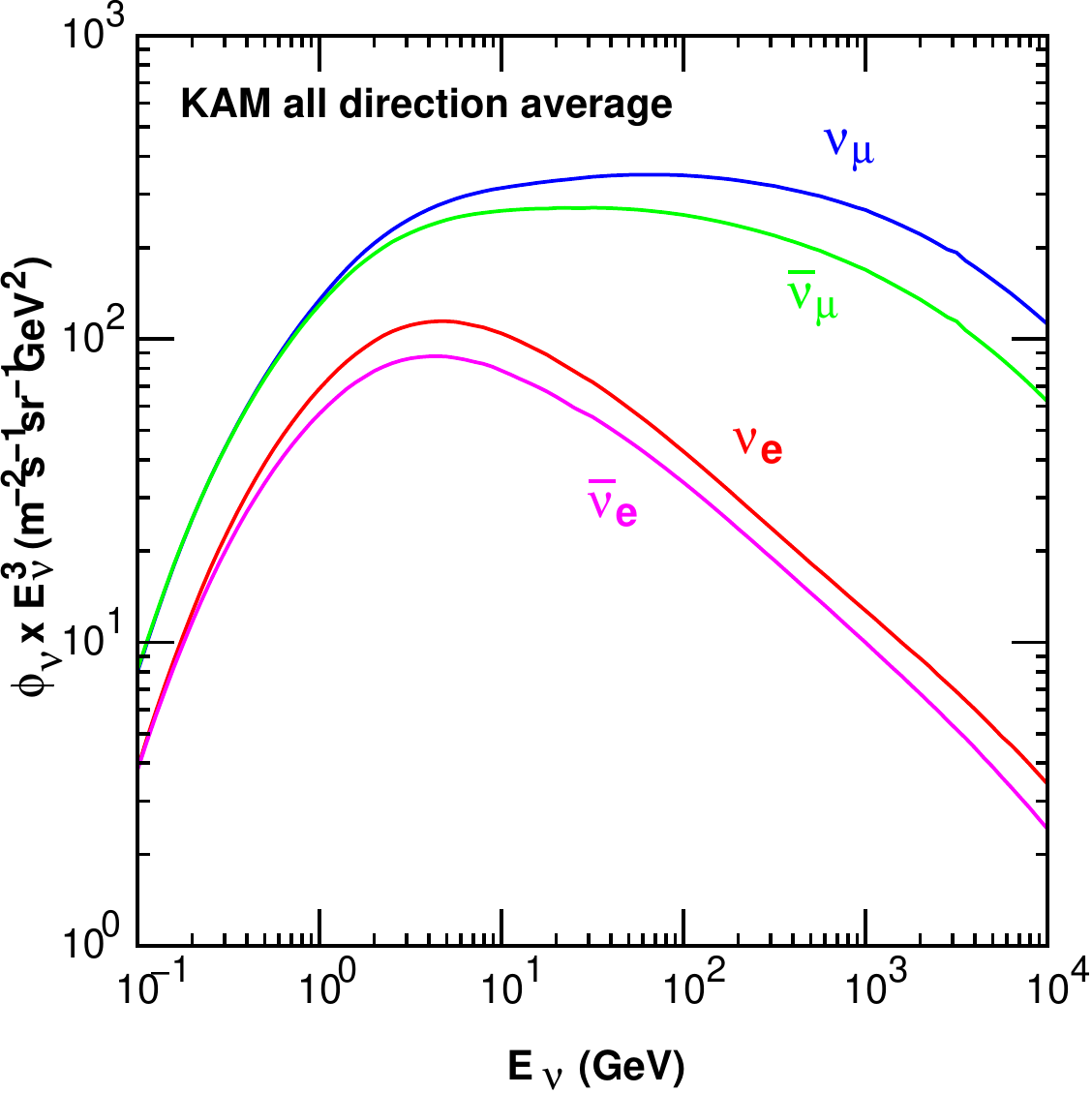}
\caption{Atmospheric neutrino flux at the Super-Kamiokande site,
  averaged over all directions and over one year. Extracted from
  Ref.~\cite{Honda:2015fha}.}
\label{fig:atmosFlux}
\end{figure}

Atmospheric neutrinos are produced by the interactions of cosmic
rays with atomic nuclei in the Earth's atmosphere. Cosmic rays, which
are primarily composed of high-energy protons and heavier nuclei,
originate from astrophysical sources such as supernovae, active
galactic nuclei, and other high-energy phenomena in the universe. When
these cosmic rays collide with atmospheric nuclei (mainly nitrogen and
oxygen), they produce a cascade of secondary particles, including
pions ($\pi$), kaons ($K$), and other mesons. The subsequent decays of
these unstable particles generate atmospheric neutrinos. For example,  
the charged pions then decay into muons and muon neutrinos:
\begin{eqnarray}
\label{eq:atm2}
\pi^\pm &\rightarrow& \mu^\pm +\begin{array}{l} \nu_{\mu}\\[-0.cm]
   {\overline\nu}_\mu\end{array}\nonumber \\
&&\mu^\pm \rightarrow e^\pm +\begin{array}{l}
\overline{\nu}_{\mu} + \nu_{e}\\[-0.cm]
\nu_{\mu} + \overline{\nu}_{e}\end{array}
\end{eqnarray}
Similar decay chains occur for kaons and heavier mesons contributing
to the flux of atmospheric neutrinos. As a result, atmospheric
neutrinos consist of a mixture of electron neutrinos, muon
neutrinos, and their corresponding antineutrinos.
The relative abundance of these
neutrino flavors depends on the energy of the primary cosmic rays and
the altitude at which the interactions occur. Atmospheric neutrinos
span a wide range of energies, from MeV to TeV scales, making them a
valuable probe for studying neutrino oscillations and other
fundamental properties of neutrinos.  A realistic representation of
the atmospheric neutrinos is provided in Fig. \ref{fig:atmosFlux}.

The study of atmospheric neutrinos has been realized by large-scale
detectors such as Super-Kamiokande~\cite{Super-Kamiokande:2023ahc,SKatm:data2024} and most recently IceCube~\cite{IceCube:2019dqi, IceCube:2019dyb}.
Super-Kamiokande  is a water Cherenkov detector
located in Japan. It played a pivotal role in the study of the discovery
of neutrino oscillations. By measuring the zenith angle dependence of the
atmospheric neutrino flux, Super-Kamiokande provided conclusive
evidence that muon neutrinos oscillate into tau neutrinos as they
travel through the Earth. 
\begin{figure}[ht!] \centering
\includegraphics[width=\textwidth]{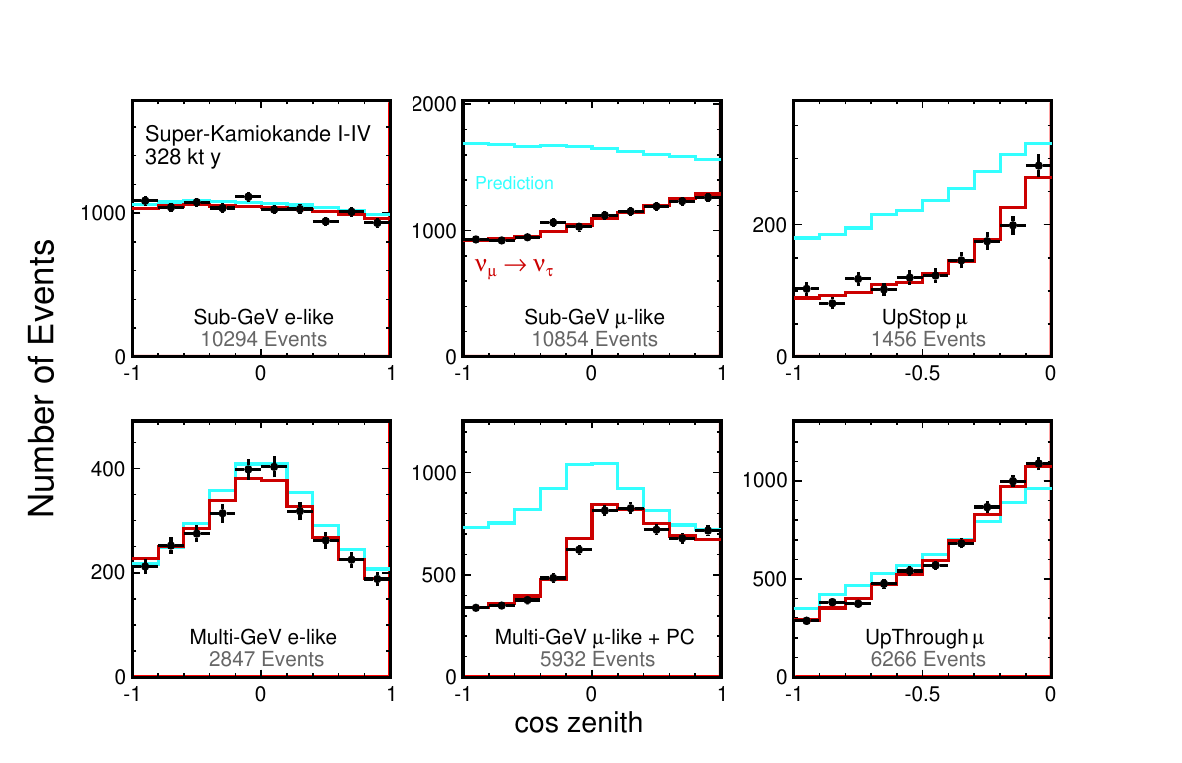}
\caption{Zenith-angle distribution of Super-Kamiokande neutrino events, comparing data (points) to expectations without (blue) and with (red) oscillations. A clear $\nu_\mu/\bar{\nu}_\mu$ disappearance is observed, enhanced for upward-going neutrinos. Events are categorized as Sub-GeV ($<1.33\;\text{GeV}$), Multi-GeV ($>1.33\;\text{GeV}$), or up-going muons (``UpStop''/``UpThrough'' from rock interactions, the latter involving higher-energy neutrinos). Figure extracted from
  \cite{PDG}.}
\label{fig:SK-zenith}
\end{figure}
As seen in Fig.~\ref{fig:SK-zenith}, Super-Kamiokande data reveals a
zenith-angle-dependent
$\nu_\mu/\bar{\nu}_\mu$ deficit, increasing with baseline $L$ (e.g.,
upward-going neutrinos traversing Earth), consistent with the survival
probability:
\begin{equation}
P_{\mu\mu} = 1 - \sin^2 2\theta \sin^2\left(\frac{\Delta m^2
  L}{4E}\right),
\label{eq:Pmumuatm}
\end{equation}
where $\theta$ and $\Delta m^2$ parametrize $\nu_\mu \to \nu_\tau$
oscillations. No $\nu_e/\bar{\nu}_e$ appearance or disappearance is
observed, and matter effects are negligible (no sensitivity to
$\theta$'s octant or $\Delta m^2$'s sign). Higher-energy neutrinos
(e.g., ``UpThrough'' or ``Multi-GeV'' samples) show reduced
disappearance due to the $L/E$ dependence. In the three-flavor framework,
this corresponds to $\theta_{23}\sim 45^\circ$ and
$|\Delta m^2_{32}|\sim 2.5 \times 10^{-3}$ eV$^2$, which is much
larger than the mass splitting observed in solar neutrinos.

IceCube, located at the South Pole, utilizes a
cubic kilometer of Antarctic ice as a detection medium and it also
detects neutrinos through the Cherenkov radiation emitted by secondary
particles produced in neutrino interactions.
These two experiments still collect data, most recent results
are the IceCube's IC24 configuration corresponding to 9.3-year dataset
~\cite{IceCube:2024xjj, IC:data2024} and the combined results from
Super-Kamiokande's SK1-5 phases~\cite{Super-Kamiokande:2023ahc,SKatm:data2024}.

\section{Reactor Neutrinos}
\label{sec:reac_exp}

\begin{figure}[ht!] \centering
\includegraphics[width=0.6\textwidth]{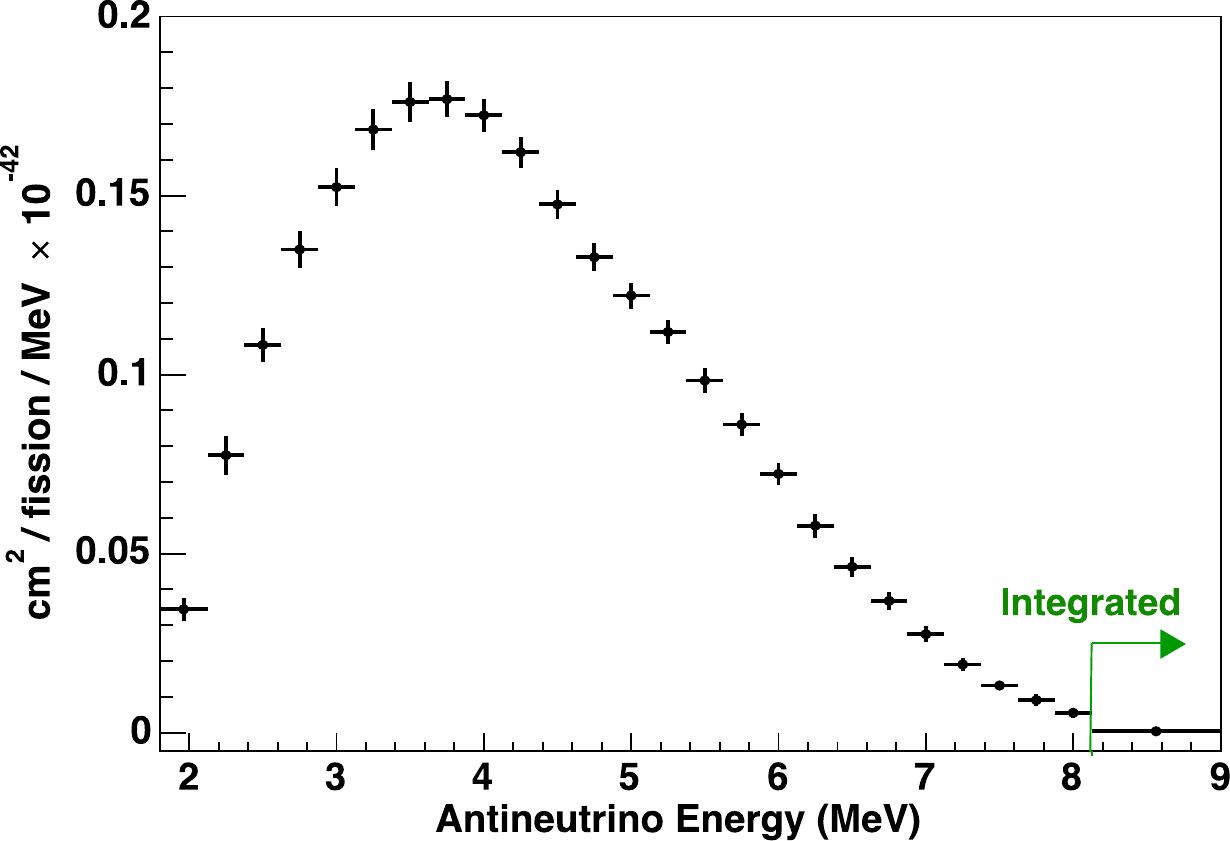}
\caption{Reactor $\bar{\nu}_e$ spectrum, as measured by the Daya Bay
  collaboration~\cite{DayaBay:2015lja}.}
\label{fig:DB-spectrum}
\end{figure}

Reactor neutrinos are electron antineutrinos ($\bar{\nu}_e$) produced
by power reactors. Nuclear reactors are powerful sources of neutrinos
generated in  the $\beta$ decay processes in the
fission of $^{235}$U, $^{238}$U, $^{239}$Pu, and $^{241}$Pu
\begin{equation}
\label{eq:betadecay}
    n \to p + e^- + \overline{\nu}_e,
\end{equation}
resulting in $\overline{\nu}_e$ with energies in the range
$E_{\overline{\nu}_e} \approx 1 \text{--} 10\,\text{MeV}$, with flux
represented in Fig.\ref{fig:DB-spectrum}. These antineutrinos are
emitted isotropically and can be  detected in liquid
scintillator detectors through the inverse beta decay process:
\begin{equation}
\label{eq:inversebetadecay}
    \overline{\nu}_e + p \to n + e^+.
\end{equation}

From  Eq~\eqref{eq:prob_vac_anti} we find that  the survival probability of
$\overline{\nu}_e$ in 3$\nu$ mixing reads:
\begin{equation}
\label{eq:prob_antinu}
    P_{\overline{\nu}_e \to \overline{\nu}_e} = 1 - c_{13}^4 \sin^2
    2\theta_{12} \sin^2\left(\frac{\Delta m_{21}^2 L}{4E}\right) -
    \sin^2 2\theta_{13} \left[c_{12}^2 \sin^2\left(\frac{\Delta
        m_{31}^2 L}{4E}\right) + s_{12}^2 \sin^2\left(\frac{\Delta
        m_{32}^2 L}{4E}\right)\right].
\end{equation}
where the dominant terms depend on the baseline of the experiment.
According to that they can be classified as follows:
\subsubsection{Short-Baseline (SBL) Reactor Experiments}
For baselines $L = \mathcal{O}(10 \text{--} 100\,\text{m})$,
these
experiments are not sensitive to the standard three-neutrino
oscillation framework. Oscillations at such short baselines would
indicate the existence of a new, sterile neutrino species. Examples
include:
\begin{itemize}
    \item Early experiments: Bugey-3~\cite{Declais:1994su},
      Goesgen~\cite{Zacek:1986cu}, ILL~\cite{Kwon:1981ua},
    \item Recent experiments: NEOS~\cite{Ko:2016owz},
      DANSS~\cite{Alekseev:2018efk},
      PROSPECT~\cite{Ashenfelter:2018iov},
      STEREO~\cite{AlmazanMolina:2019qul}.
\end{itemize}
\subsubsection{Medium-Baseline (MBL) Reactor Experiments}
For baselines $L = \mathcal{O}(1\,\text{km})$, these experiments are
sensitive to $\Delta m_{31}^2$ and $\theta_{13}$. Examples include:
\begin{itemize}
    \item Early experiments: CHOOZ~\cite{Apollonio:1999ae}, Palo
      Verde~\cite{Boehm:2001ik},
    \item Recent experiments: Daya
      Bay~\cite{An:2012eh,DayaBay:2022orm, DayaBay:2021dqj},
      RENO~\cite{Ahn:2012nd}, Double Chooz~\cite{Abe:2011fz}.
\end{itemize}

Among the experiments listed, Daya Bay experiment was the most
important in the determination of $\theta_{13}$ by observing
$\bar{\nu}_e$ disappearances at different baselines. Located near the
Daya Bay Nuclear Power Plant in China, the experiment has six 20-ton
gadolinium-doped liquid scintillator detectors that registered the IBD
reactions resulting from the interaction of reactor antineutrinos with
protons.  In 2012, the Daya Bay collaboration announced the discovery
of a non-zero value of $\theta_{13}$ with more than 5 $\sigma$
significance, a significant milestone in neutrino physics.  The results
from the Daya Bay experiment ~\cite{DayaBay:2022orm, DayaBay:2021dqj}
are critical because a non-zero $\theta_{13}$ is a necessary condition
for the investigation of CP violation in the lepton sector by future
LBL neutrino experiments. Their measurement  of
$|\Delta m^2_{31}|$, complementary to that of atmospheric (and long-baseline)
experiments is also an essential input for the determination of the neutrino
mass ordering as we will see in the next chapter.

\subsubsection{Long-Baseline (LBL) Reactor Experiments}
For baselines $L = \mathcal{O}(100\,\text{km})$, these experiments are
sensitive to $\Delta m_{21}^2$, $\theta_{12}$, and $\theta_{13}$,
similar to solar neutrino experiments. Until now, there is only one
experiment of this type, called Kamioka Liquid
Scintillator Antineutrino Detector (KamLAND)~\cite{Gando:2013nba},
located in the Kamioka mine. KamLAND detects antineutrinos from
multiple reactors across Japan and South Korea, with baselines ranging
from $80$ km to $800$ km and an effective baseline of $180$ km. Data
from KamLAND confirm the oscillation of neutrinos in vacuum with
parameters as observed in solar neutrino experiments. In terms of
precision it outperforms solar neutrino
measurements in constraining $\Delta m_{21}^2$.  In the very near future the forthcoming experiment JUNO (presently under advanced
  construction) will observe neutrinos at about $L =
  \mathcal{O}(50\,\text{km})$ baseline, measuring the solar parameters
  $\Dmq_{21}$ and $\theta_{12}$ with unprecedented precision while
  also aiming at a complete determination of $\Dmq_{31}$ in both
  magnitude and sign.

\section{Accelerator Neutrinos }
\label{sec:accel_exp}
In accelerator-based experiments, a high-energy proton beam collides
with a target composed of a light material. These collisions produce
mesons such as $\pi$, $K$, and heavier mesons (depending on the energy
of the incoming proton beam) which then decay producing neutrinos and antineutrinos
in a way similar to the chain in Eq.\eqref{eq:atm2}.
In the accelerator beam line after the target, there is a
vacuum space known as the decay pipe, allowing these mesons to decay
in-flight into neutrinos. Following the decay pipe, a heavy material,
called the dump, is used to stop all remaining particles. In the dump,
mesons decay predominantly at rest, ensuring that only neutrinos
survive beyond this point.
Furthermore, a magnetic field can be used to discard negative or positive pions,
generating a beam of muon neutrinos or antineutrinos,
respectively. This allows exploring matter effects and CP violation, 
that affect differently neutrinos and antineutrinos.

Accelerator-produced neutrinos have typical energies $\mathcal{O}
(\sim\text{GeV})$. The orientation of the detector with respect to the beam
can be exploited to enhance the beam in a narrow range of energies.

Based on the distance between the neutrino source and the detector,
accelerator experiments are categorized as Short-Baseline (SBL) or
Long-Baseline (LBL):
\begin{itemize}
    \item \textbf{Short-Baseline (SBL):} Baselines on the order of
      $0.01-1$ km .
    \item \textbf{Long-Baseline (LBL):} Baselines in the range
      $100-1300$ km.
\end{itemize}
SBL accelerator  experiments are not sensitive to the standard three-neutrino
oscillation framework. Oscillations at such short baselines would
indicate the existence of a new, sterile neutrino species.  In the
context of 3$\nu$ analysis they play a critical role in constraining
systematic uncertainties in neutrino-nucleus interactions.

In brief, generically accelerator neutrino experiments typically employ
two detectors:
\begin{itemize}
    \item \textbf{Near Detector (ND):} Positioned a few hundred meters
      from the target to measure the unoscillated flux and reduce
      energy-dependent systematic uncertainties. These detectors are also
      themselves capable of studies of SBL physics.
    \item \textbf{Far Detector (FD):} Placed hundreds of kilometers
      from the target to measure neutrino oscillations. Combined with
      the ND, they form the usual structure of LBL experiments.
\end{itemize}

The number of neutrino events of flavour $\alpha$, $N_{\nu_\alpha}$,
observed at each detector can be approximated as:
\begin{equation}
    N_{\nu_\beta} \sim \phi_{\nu_\mu} \times P_{\nu_\mu \to
      \nu_\alpha} \times \sigma_{\nu_\beta},
\end{equation}
where $\phi_{\nu_\mu}$ is the neutrino flux, $P_{\nu_\mu \to
  \nu_\alpha}$ is the oscillation probability, and
$\sigma_{\nu_\beta}$ is the cross-section for neutrinos of flavour
$\beta$. The oscillation probability can be extracted using:
\begin{equation}
    P_{\nu_\mu \to \nu_\beta} \simeq
    \frac{N^\mathrm{far}_\beta}{N^\mathrm{near}_\mu} \cdot
    \frac{\phi^\mathrm{near}_{\nu_\mu}}{\phi^\mathrm{far}_{\nu_\mu}}
    \cdot \frac{\sigma_{\nu_\mu}}{\sigma_{\nu_\beta}}.
\end{equation}

This method reduces systematic uncertainties related to flux and
cross-sections. As mentioned above, far Detectors can be placed
on-axis (aligned with the
beam) or off-axis (displaced from the beam line). Off-axis
configurations reduce background but also decrease neutrino flux.
Another method that accelerators use to reduce systematic effects is
by studying separately two types of events: Neutral Current (NC) and
Charged Current (CC) interactions. CC events, where the neutrino
transfers its flavor (e.g., $\nu_\mu \rightarrow \mu^-$), are used to
directly measure oscillations. NC events, which are flavor-blind
(e.g., $\nu_x + N \rightarrow \nu_x + X$), provide a flux
normalization sample, as their rate is unaffected by oscillations. By
comparing NC rates between near and far detectors, experiments
constrain flux and cross-section uncertainties, isolating true
oscillation signals in CC channels.

\subsection{Long-Baseline (LBL) Experiments}
As mentioned above, in an LBL experiment neutrinos are detected  after crossing distances of $\cal
O$(100 km). The most important experiments in
this field are the NuMI Off-Axis $\nu_e$ Appearance (NOvA) experiment
\cite{NOvA:nu24, NOvA:nu2020}, the Tokai-to-Kamioka (T2K) experiment~
\cite{T2K:nu24, T2K:2023mcm, T2K:2023smv} and MINOS
~\cite{Adamson:2013whj, Adamson:2013ue}.
\begin{itemize}
    \item \textbf{MINOS~\cite{Adamson:2013whj, Adamson:2013ue}:} Used
      the 120 GeV NUMI beam at Fermilab. The far detector was located
      735 km away, on-axis and underground to mitigate cosmic ray
      backgrounds.
      \item \textbf{NOvA~\cite{NOvA:nu24, NOvA:nu2020}:} NOvA is a
       second-generation long-baseline neutrino experiment at Fermilab
       \cite{NOvA:nu2020} with a detection baseline of 810 km.
       Below we give more details of our simulation and analysis
       of  NOvA. 
     \item \textbf{T2K~\cite{T2K:nu24, T2K:2023mcm, T2K:2023smv}:}
       Utilizes the  muon neutrino beam at
       the J-PARC facility in Tokai, which is then directed (off-axis)
       towards the Super-Kamiokande detector located 295 km away.
\end{itemize}

\begin{figure}[t] \centering
\includegraphics[width=0.7\textwidth]{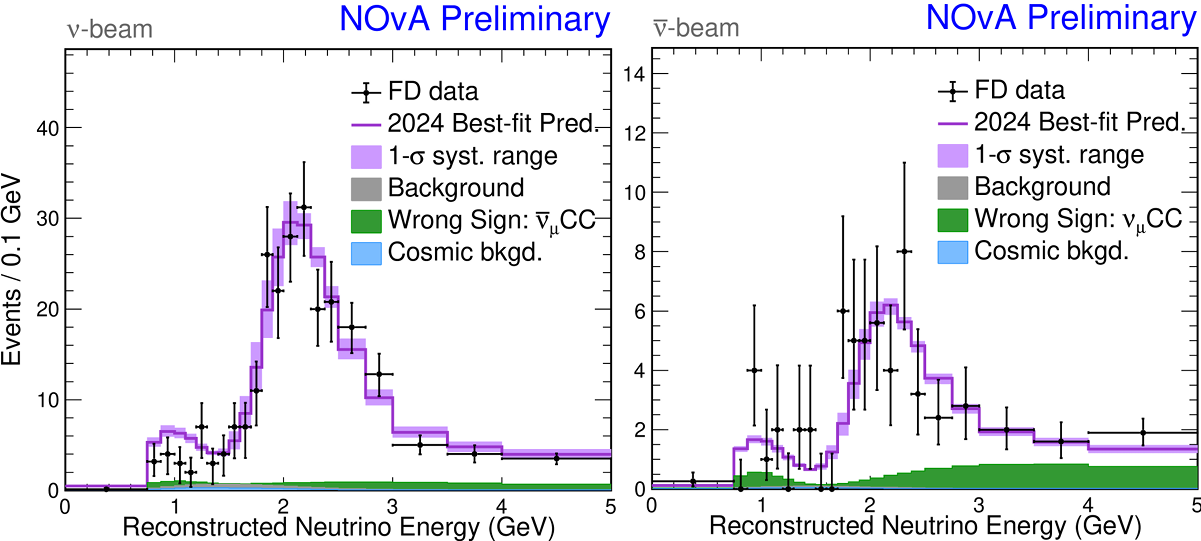}
\includegraphics[width=0.7\textwidth]{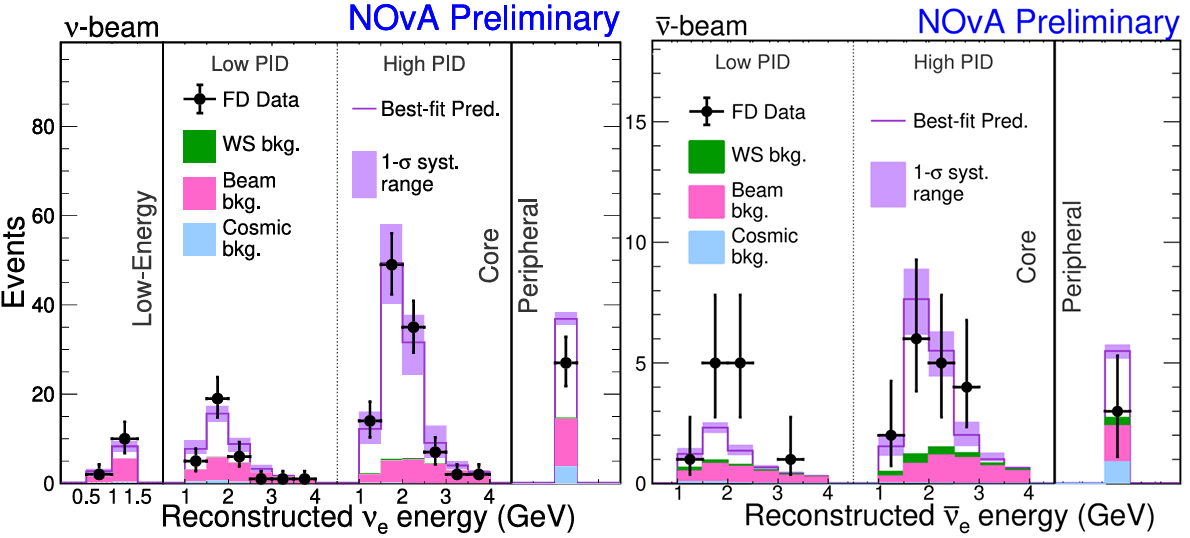}
\caption{NOvA results as of June 2024 on
  $\nu_\mu$ and $\overline{\nu}_\mu$ disappearance
      (upper panels) and $\nu_e$ and $\overline{\nu}_e$ appearance.
  The appearance events are
  classified in three bins from lowest to highest purity:
  “Peripheral”, “Low PID”, and “High PID”. Extracted from
  Ref.\cite{NOvA:nu24}}
\label{fig:nova_events}
\end{figure}
In these experiments, neutrinos traverse a nearly
constant layer of matter, enabling the study of matter effects. The
dominant contributions to the $\nu_\mu$/$\bar{\nu}_\mu$ (disappearance
probability-\textbf{dis}) $P_{\mu\mu}$ in 3$\nu$ scenario is
\begin{equation}  
P_{\mu\mu} \approx 1 - \sin^2 2\theta_{23} \sin^2\left(\frac{\Delta
  m^2_{32} L}{4E}\right) + \text{(subdominant $\Dmq_{21}$ terms)}.
\label{eq:dis_acc}
\end{equation}  
For $\nu_e$/$\bar{\nu}_e$ (appearance probability-\textbf{app})
$P_{\mu e}$:
\begin{eqnarray}
P_{\mu e} &=& \sin^2 \theta_{23} \sin^2 2\theta_{13} \frac{\sin^2
  \Delta (1 - A)}{(1 - A)^2} \pm \alpha \cos \theta_{13}\sin
2\theta_{13} \sin 2\theta_{23} \sin 2\theta_{12} \nonumber
\\ &&\times\cos(\Delta + \delta_{\text{CP}}) \frac{\sin \Delta A}{A}
\frac{\sin \Delta (1 - A)}{(1 - A)} + \text{other terms}\;{\cal
  O}(\alpha^2)
\label{eq:app_acc}
\end{eqnarray}
where $\Delta=\Delta m_{31}^2\frac{L}{4 E_\nu}$, $\alpha \equiv \Delta
m^2_{21}/\Delta m^2_{31}$, and the $+$ ($-$) sign correspond to $\nu$
($\bar{\nu}$).  Matter effects ($A \propto \pm \sqrt{2} G_F n_e$, $+$
for $\nu$, $-$ for $\bar{\nu}$) induce a $\sim$5–10$\%$ correction in
NOvA due to its longer baseline. They enhance (suppress) $P_{\mu e}$
for normal (inverted) mass ordering. T2K, on the contrary,
is largely vacuum-dominated  making it more sensitive to $\delta_{CP}$-driven
$\nu/\bar{\nu}$ asymmetry.

\subsection{Simulation of Nova}
\label{sec:novasimul}
Compared to the case of solar neutrinos, the analysis of NOvA is
relatively simpler. Using the formalism for LBL described above for CC
interactions, we will perform a $\chi^2$-analysis in order to obtain
the confidence level regions for the oscillation parameters. In
general, we define the number of expected events in a given energy bin
$i$ and for a given channel $\alpha$ ($\alpha \in \{\nu_\mu, \nu_e,
\bar\nu_\mu, \bar\nu_e\}$). This can generically be calculated as
\begin{equation} N_i^\alpha = N_\text{bkg,i} + \int^{E_{i+1}}_{E_i} \,
\mathrm{d}E_\text{rec} \int_0^\infty \, \mathrm{d}E_\nu
R(E_\text{rec}, E_\nu) \varepsilon (E_\nu) \sum_\beta \frac{\mathrm{d}
  \Phi^\beta}{\mathrm{d} E_\nu } P_{\nu_\beta \rightarrow
  \nu_\alpha}(E_\nu) \sigma_\alpha (E_\nu) \, ,
\label{eq:events_nova}
\end{equation} where
\begin{itemize}
\item $N_\text{bkg,i}$ is the number of background events in that bin.
  If there is a neutrino component in the background, its oscillation
  has to be consistently included.
\item $[E_i, E_{i+1}]$ are the bin limits.
\item $E_\text{rec}$ is the reconstructed neutrino energy.
\item $E_\nu$ is the true neutrino energy.
\item $R(E_\text{rec}, E_\nu)$ is the energy reconstruction function:
  the probability to observe a reconstructed energy $E_\text{rec}$ if
  the true neutrino energy is $E_\nu$. We usually take it to be
  Gaussian, i.e.,
\begin{equation} R(E_\text{rec}, E_\nu) = \frac{1}{\sqrt{2 \pi}
\sigma_E E_\nu} \exp\left[-\frac{1}{2 \sigma_E^2}\left(\frac{E_\nu -
      E_\text{rec}} {E_\nu}\right)^2\right] \, .
\end{equation} That is, $\frac{E_\nu - E_\text{rec}}{E_\nu}$ is
Gaussian-distributed around zero with standard deviation
$\sigma_E$. The $\frac{1}{E_\nu}$ prefactor is for normalisation
purposes.
\item $\varepsilon$ is the detection efficiency.
\item $\frac{\mathrm{d} \Phi^\beta}{\mathrm{d} E_\nu }$ is the
  incident neutrino flux with flavour $\beta$.
\item $P_{\nu_\beta \rightarrow \nu_\alpha} (E_\nu)$ is the $\nu_\beta
  \rightarrow \nu_\alpha$ transition probability, with the Earth's
  crust density assumed as constant at $2.76~\text{g/cm}^3$, with
  matter composed of equal proton and neutron densities.
\item $\sigma_\alpha$ is the $\nu_\alpha$ detector cross section.
\end{itemize}
The antineutrino channels are obtained by switching $\nu$ to $\bar{\nu}$.
The incident flux of muon (anti)neutrinos, background events and
energy resolutions are provided by the collaboration, extracting
information from Ref. \cite{NOvA:nu24}. The scattering cross section
between the neutrinos and nuclear targets (for the case of NOvA,
Hydrocarbons) is provided by the collaboration, too.
\begin{figure}[ht!]
    \centering
    \begin{subfigure}[ht!]{0.48\textwidth}
        \includegraphics[width=\textwidth]{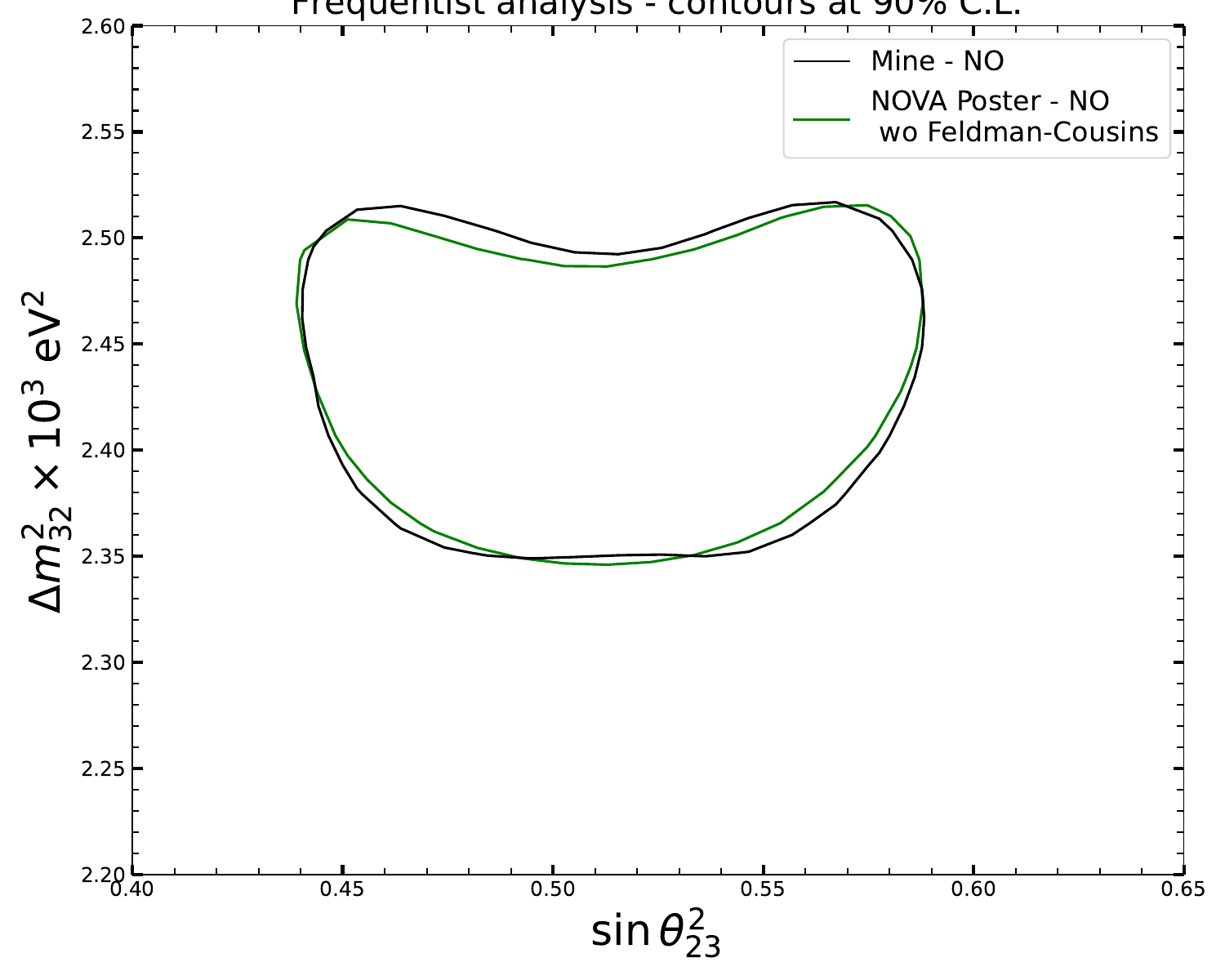}
    \end{subfigure}\\
%    \hfill
%    \begin{subfigure}[ht!]{0.48\textwidth}
%        \includegraphics[width=\textwidth]{Figuras/pics/NOVA_6.pdf}
%    \end{subfigure}
%    \vspace{1em}
    \begin{subfigure}[ht!]{0.49\textwidth}
        \includegraphics[width=\textwidth]{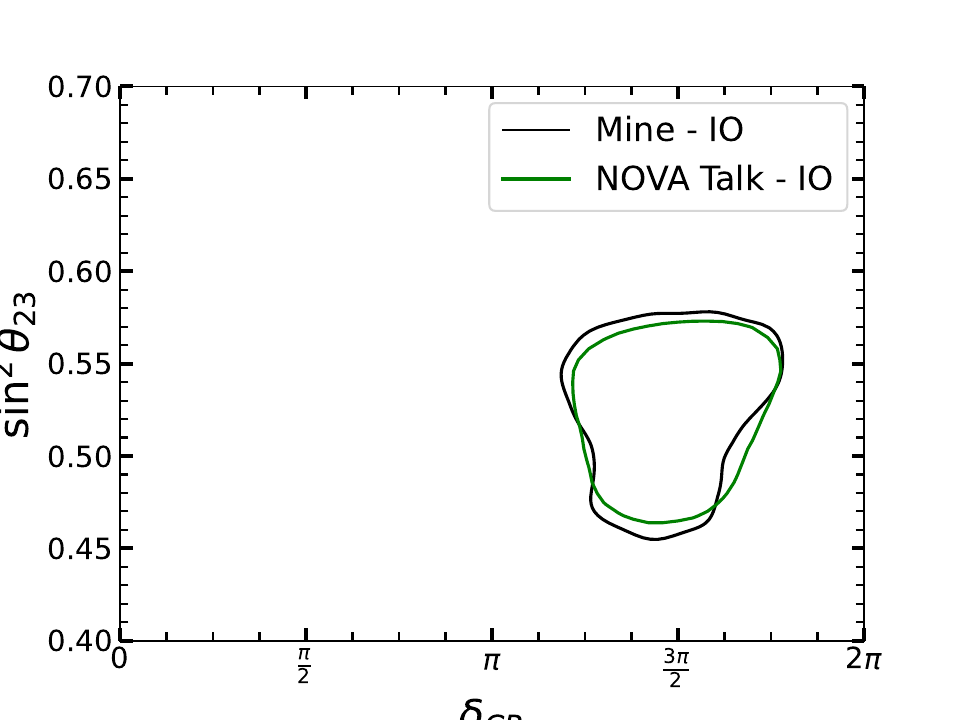}
    \end{subfigure}
    \hfill
    \begin{subfigure}[ht!]{0.49\textwidth}
        \includegraphics[width=\textwidth]{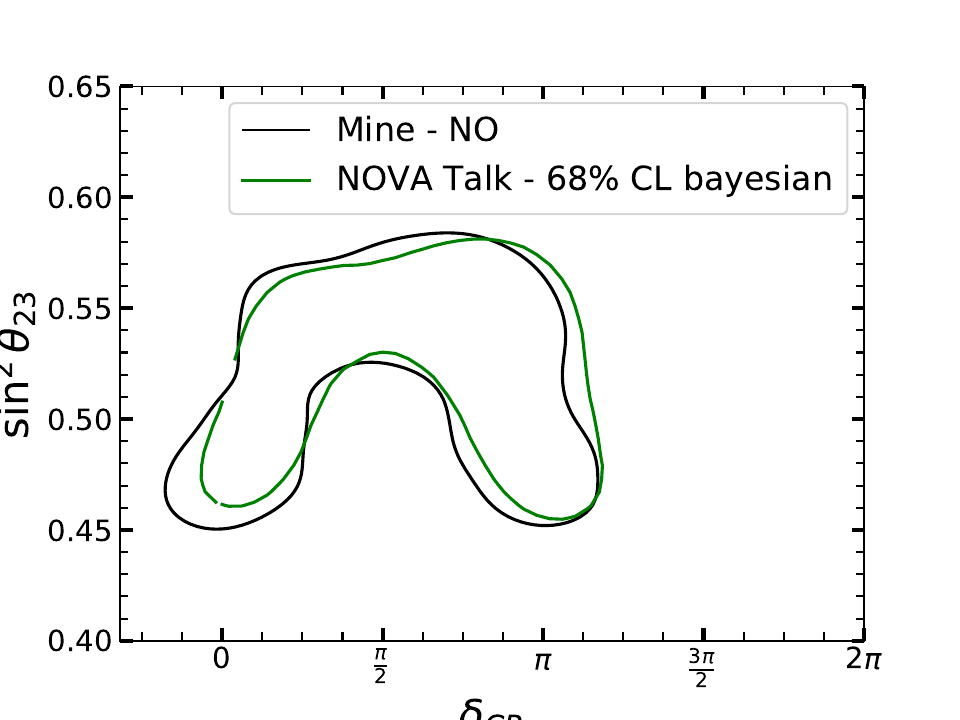}
    \end{subfigure} \\ [-0.2cm]  
    \begin{subfigure}[ht!]{0.45\textwidth}
  \hspace*{-0.7cm}      \includegraphics[width=\textwidth]{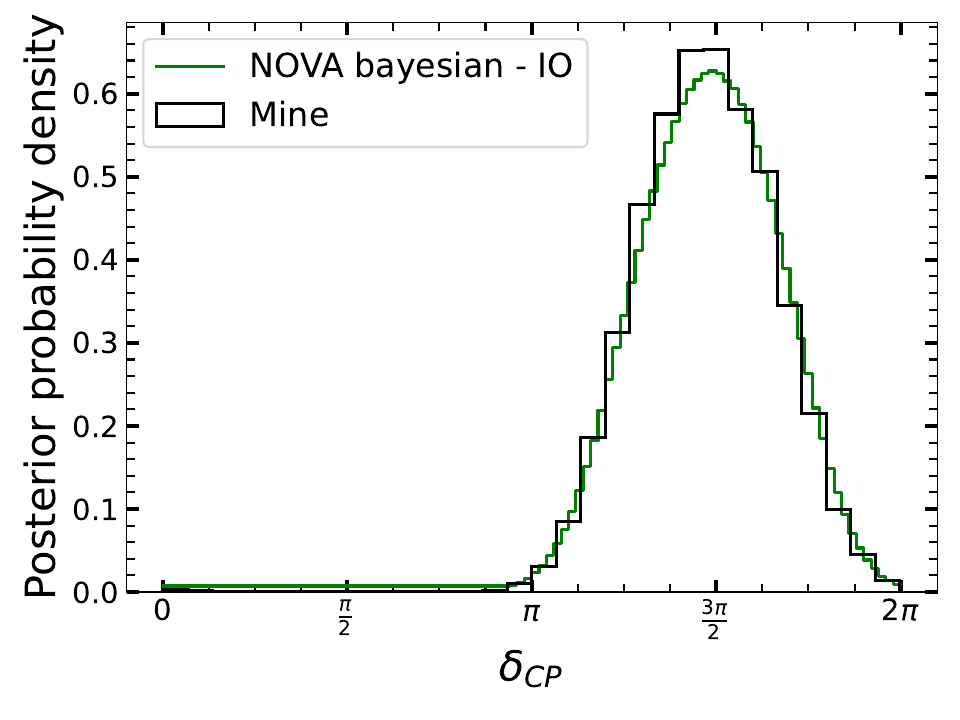}
    \end{subfigure}
%    \hfill
    \begin{subfigure}[ht!]{0.45\textwidth}
        \includegraphics[width=\textwidth]{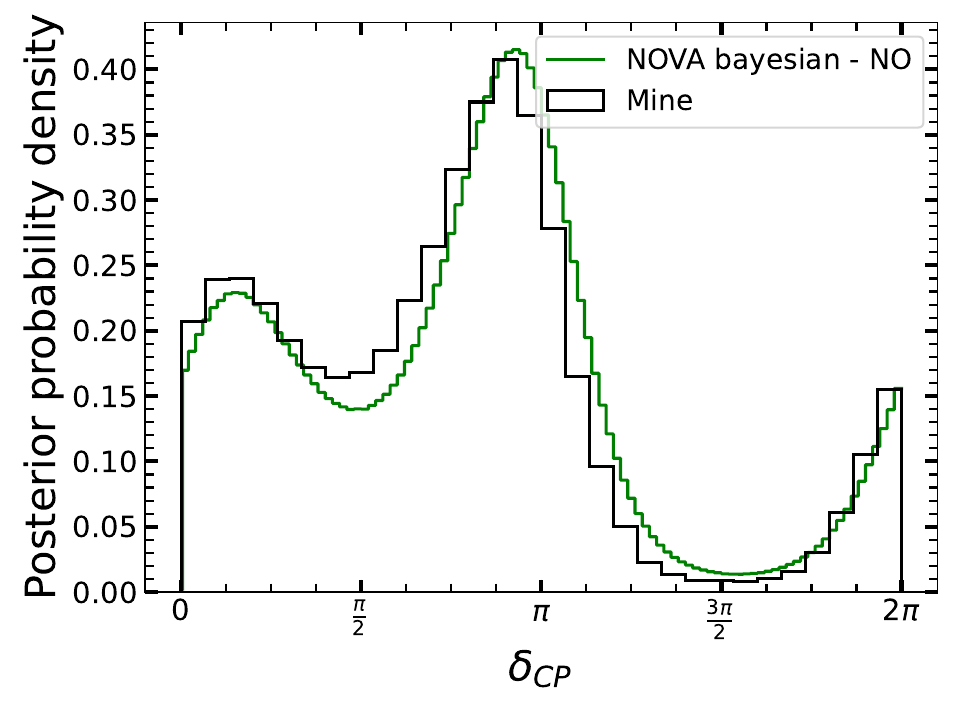}
    \end{subfigure}
    \caption{Validation of our simulation of NOvA. 
      Each figure represents
      in green the information released by the collaboration in
      Ref. \cite{NOvA:nu24} while in black  the corresponding results 
      obtained with our simulations.}
    \label{fig:nova_grid}
\end{figure}

The analysis procedure commenced with the unoscillated event
distribution from a collaboration-provided benchmark
point. Predictions were subsequently reweighted to match the official
collaboration spectra through iterative adjustments of detection
efficiencies, ensuring consistency with the expected oscillated event
rates.

For the analyses, a priori there are three nuisance parameters
associated to systematic uncertainties for each channel (\textbf{DIS}
and \textbf{APP}): a signal normalisation factor $\eta_1\equiv
\eta_S$, a background normalisation factor $\eta_2\equiv \eta_b$, and
an absolute energy scale uncertainty $\eta_3\equiv \rho$, all of them
centred around 1. In that case,the prediction $N_i^\alpha$ for a given
energy bin $[E^i_\nu,E^{i+1}_\nu]$ is given by

\begin{equation} N_i^\alpha = \eta_b N_\text{bkg,i} + \eta_s \int^{\rho E_{i+1}}_{\rho E_i} \,
\mathrm{d}E_\text{rec} \int_0^\infty \, \mathrm{d}E_\nu
R(E_\text{rec}, E_\nu) \varepsilon (E_\nu) \sum_\beta \frac{\mathrm{d}
  \Phi^\beta}{\mathrm{d} E_\nu } P_{\nu_\beta \rightarrow
  \nu_\alpha}(E_\nu) \sigma_\alpha (E_\nu) \, ,
\label{eq:events_nova_2}
\end{equation}

The observed number of events per bin $i$ and neutrino flavor
$\alpha$, $n_i^\alpha$, as well as the nuisance parameters $\theta_j$,
are assumed to be independent. The $\chi^2$ statistic can be written
as (for each channel):
\begin{equation}
\chi^2_\alpha = 2 \sum_i \left( n_i^\alpha + N_i^\alpha \ln
\frac{n_i^\alpha}{N_i^\alpha} \right) + \sum_j \frac{\left(\eta_j -
  1\right)^2}{\sigma_j^2},
\label{eq:chi2}
\end{equation}
where the logarithmic term is taken to be zero if $n_i =
0$. $\sigma_j$ are the systematic uncertainties of the nuisance
parameters. These parameters were calibrated to replicate reported
confidence intervals through $\chi^2$ minimization. The statistical
minimization employed a dual-methodology approach: Bayesian parameter
estimation and evidence calculation were performed using the
\textsc{MultiNest} algorithm \cite{Feroz2008,Feroz2013}, while
frequentist optimization utilized the Nelder-Mead simplex
implementation from the GNU Scientific Library (GSL) \cite{GSL}. This
hybrid strategy ensured robust convergence across parameter space
landscapes while maintaining computational efficiency.

Using the $\nu/\bar{\nu}$ beam data and background estimates from
Fig.~\ref{fig:nova_events}, the framework simultaneously analysed
appearance and disappearance channels. All oscillation parameters
except $\Delta m^2_{32}$, $\sin^2\theta_{23}$, and $\delta_\text{CP}$
were fixed to central values from Ref.~\cite{nufit}. The resulting
confidence contours (Fig.~\ref{fig:nova_grid}) demonstrate qualitative
agreement with NOvA's official results, with residual quantitative
discrepancies attributable to approximations in energy scale
systematics and efficiency modeling. The framework's robustness was
further verified through spectrum morphology checks at best-fit
parameter values.

\section{CEvNS}
\label{sec:CEvNS}

Coherent Elastic Neutrino-Nucleus Scattering (CE$\nu$NS) is a
neutral-current weak interaction process where low-energy neutrinos
scatter elastically off entire atomic nuclei. The coherence condition,
satisfied when the momentum transfer $|q| \ll 1/R$ (with $R$ being the
nuclear radius), enables constructive interference of scattering
amplitudes from individual nucleons. This results in a cross-section
enhancement proportional to the square of the neutron number $N^2$,
making heavy nuclei ideal targets. SM predicts the differential
cross-section as:

\begin{equation}
\frac{d\sigma_{\nu N}}{dT} \approx \frac{G_F^2 M}{4\pi} \left[ N - (1 - 4\sin^2\theta_W)Z \right]^2 \left(1 - \frac{MT}{2E_\nu^2}\right) |F_{\text{weak}}(q)|^2,
\end{equation}
where $G_F$ is the Fermi coupling constant, $M$ the nuclear mass, $T$
the nuclear recoil energy, $E_\nu$ the neutrino energy, and $N$ ($Z$)
the neutron (proton) number.  $F_{\text{weak}}(q)$ is the weak nuclear form factor and is
dominated by neutron contributions. The recoil energies (sub-keV to
keV) and low neutrino fluxes at suitable sources posed significant
detection challenges until the COHERENT collaboration's landmark 2017
observation using a 14.6 kg CsI[Na] scintillator at the Spallation
Neutron Source (SNS) \cite{Akimov:2017ade}, forty-three years after
Freedman's theoretical proposal \cite{Freedman1974}.

Spallation sources like SNS provide pulsed $\pi^+$-decay-at-rest neutrinos with energies $\lesssim 50$ MeV, satisfying the coherence condition while enabling background suppression via timing cuts. In addition to spallation sources, CE$\nu$NS is also searched for in experiments using electron antineutrinos from nuclear reactors, including TEXONO~\cite{TEXONO}, GeN~\cite{GeN}, CONNIE~\cite{CONNIE}, MINER~\cite{MINER}, Ricochet~\cite{Ricochet}, $\nu$-cleus~\cite{nucleus}, RED-100~\cite{RED100}, NEON~\cite{NEON}, CONUS~\cite{CONUS}, and NCC-1701 at Dresden-II~\cite{DresdenII:2021}. To date, most reactor-based experiments have not yet achieved definitive CE$\nu$NS detection. The exception is the NCC-1701 experiment at the Dresden-II reactor, which reported an event spectrum with a low-energy excess in its first published data~\cite{DresdenII:2021}, consistent with SM CE$\nu$NS predictions. Recently, the collaboration released updated results with increased exposure, observing CE$\nu$NS with \textit{strong} to \textit{very strong} statistical preference (relative to a background-only hypothesis) in a Bayesian framework, depending on the assumed quenching factor~\cite{DresdenII:2023}. 

CE$\nu$NS processes provide unique nuclear structure insights through sensitivity to neutron distributions, inaccessible via electromagnetic probes \cite{Donnelly:1984rg}, constraints on non-standard neutrino interactions (NSI) \cite{Coloma:2019mbs}, neutrino electromagnetic properties \cite{Papoulias:2019txv}, sterile neutrinos \cite{Giunti:2019xpr}, and hidden-sector particles \cite{Baxter:2019mcx}.

\section{Gallium Source Experiments}
\label{sec:gallium}

It is almost two decades since the so-called \emph{Gallium
Anomaly}~\cite{Giunti:2006bj, Laveder:2007zz} became a standing puzzle
in neutrino physics.  In general terms, the anomaly accounts for the
deficit of the event rate measured in Gallium source experiments with
respect to the expectation.  It was originally observed in the
calibration of the gallium solar-neutrino detectors
GALLEX~\cite{GALLEX:1997lja, Kaether:2010ag} and
SAGE~\cite{SAGE:1998fvr, Abdurashitov:2005tb} with radioactive
\Nuc[51]{Cr} and \Nuc[37]{Ar} sources:
\begin{equation}
  \label{eq:detproc}
  \nu_e + \Nuc[71]{Ga} \to \Nuc[71]{Ge} + e^- .
\end{equation}
Using the detection cross section as predicted by
Bahcall~\cite{Bahcall:1997eg}, the average ratio of observed vs
predicted rates was found to be ${R}_{\text{GALLEX+SAGE}} = 0.88 \pm
0.05$~\cite{Abdurashitov:2005tb}, which represented a $2.4\sigma$
statistically significant deficit.
Most interestingly the Gallium Anomaly has been recently rechecked by
the BEST experiment~\cite{Barinov:2021asz, Barinov:2022wfh}, which
placed the \Nuc[51]{Cr} radioactive source at the center of a
concentric two-zone gallium target (thus effectively probing two
distinctive neutrino path lengths, of about 0.5~m and 1.1~m).  In both
zones they observe consistent deficits of $R_{\text{in}}= 0.79
\pm0.05$ and $R_{\text{out}}= 0.77 \pm0.05$~\cite{Barinov:2021asz,
  Barinov:2022wfh}, so the current combined level of the deficit is
$R_{\text{GALLEX+SAGE+BEST}} = 0.80 \pm0.05$~\cite{Barinov:2021asz,
  Barinov:2022wfh} promoting the statistical significance of the
anomaly beyond $4\sigma$.

Careful scrutiny~\cite{Elliott:2023xkb, Giunti:2022xat} of the
neutrino capture cross sections and its uncertainties does not provide
an explanation of the deficit, leaving open a possible effect in the
neutrino propagation.  The idea that $\nu_e$ may disappear during
propagation from source to detector is no surprise nowadays, as the
phenomenon of mass-induced neutrino flavour oscillations has been
established beyond doubt (see for example the review in
Ref.~\cite{ParticleDataGroup:2024cfk}) and the involved masses and
mixing are being determined with increasing accuracy by the combined
results of solar, reactor, atmospheric and long-baseline neutrino
experiments (see Ref.~\cite{Esteban:2024eli} for the latest global
analysis).  Unfortunately, it is precisely such accuracy which puts in
jeopardy the possible interpretation of the Gallium anomaly in terms
of neutrino oscillations.  Given the characteristic baseline
$\mathcal{O}(\text{meter})$ of the GALLEX, SAGE, and BEST radioactive
source experiments and their average neutrino energy
$\mathcal{O}(\text{MeV})$, a $\Dmq\gtrsim \mathcal{O}(\eVq)$ is
required to produce visible effects, and this is more than two orders
of magnitude larger than what is indicated by the global analysis.  Hence
at least a fourth massive state must be involved in the propagation of
the neutrino ensemble with mass $m_4\sim\mathcal{O}(\text{eV})$.  This
in turn requires the introduction of a fourth neutrino weak
eigenstate, which must be an $SU(2)$ singlet to comply with the bounds
from the $Z$ invisible width~\cite{ParticleDataGroup:2024cfk}.  This
is how the light sterile neutrino scenario makes its entrance, but in
order to explain the Gallium anomaly in such a way, the fourth state
must significantly mix with the three standard neutrinos,
$\sin^2\theta\sim \mathcal{O}(10)\%$.

The problem is that such large mixing would impact heavily the
oscillation signals included in the global analysis.  This results in
a strong tension between the sterile-neutrino interpretation of the
Gallium anomaly and other neutrino data~\cite{Berryman:2021yan,
  Goldhagen:2021kxe, Giunti:2022btk}.  In particular solar neutrinos
and reactor antineutrinos provide a clean test of the possible
projection of $\nu_e$ and $\bar \nu_e$ on a $\mathcal{O}(\text{eV})$
massive state.  Given the long baselines and the energies involved,
the oscillations driven by $\Dmq \sim \mathcal{O}(\eVq)$ are averaged
in both solar and KamLAND experiments so they directly test the mixing
relevant for the interpretation of the Gallium anomaly.  The analyses
presented in Refs.~\cite{Giunti:2022btk, Goldhagen:2021kxe} lead to
$2\sigma$ bounds $\sin^2\theta\lesssim 0.025$--$0.045$ of the
corresponding mixing angle, clearly disfavoring the sterile
oscillation interpretation of the Gallium anomaly.  Alternative
non-conventional scenarios have also been considered (see for example
Refs.~\cite{Brdar:2023cms, Farzan:2023fqa, Arguelles:2022bvt,
  Hardin:2022muu, Banks:2023qgd}), but they are nevertheless not free
from severe tension with other data~\cite{Giunti:2023kyo}.

\section{Summary}
In this chapter we have briefly presented the neutrino oscillation  
experiments performed with neutrinos produced  in the  Sun
(Sec.~\ref{sec:solar_exp}), in the  
atmosphere (Sec.~\ref{sec:atm_exp}),  
at nuclear reactors (Sec.~\ref{sec:reac_exp})  
and at dedicated accelerator beams (Sec.~\ref{sec:accel_exp}).
They have  robustly confirmed the three-neutrino mixing scenario.
For convenience we compile in Table~\ref{tab:expsum} the
different experiments which contribute to the determination of
the 3$\nu$ oscillation parameters.

The global interpretation of this bulk of data is in the hands
of phenomenological groups and in the next chapter we will present
the results of the last analysis of NuFIT. Analysis of this data
has also been used to constrain other forms of new physics as
we will present in the subsequent chapters.
Paramount to this type of work is the precise simulation of the
experiments and the consistent statistical analysis of their data.
On that front, the work in this thesis contains the original contributions
to the analysis of the solar neutrino results from the phases-II
and III of Borexino
and of the accelerator neutrino experiment NOvA 
which have been described in detail in Secs.~\ref{sec:bxsimul} and
~\ref{sec:novasimul} respectively.

\begin{table}[ht!]
\begin{tabular}{l|l|l}
Experiment & Dominant &  Important  \\
\hline
Solar Experiments &  {$\theta_{12}$} 
&  {$\Delta m^2_{21}$}  , {$\theta_{13}$} 
\\
Reactor LBL (KamLAND)  &  
{$\Delta m^2_{21}$}  
& {$\theta_{12}$}   , {$\theta_{13}$} 
\\
Reactor MBL (Daya-Bay, Reno, D-Chooz)   
&  {$\theta_{13}$}, {$|\Delta m^2_{31,32 }|$} &
\\
Atmospheric Experiments (SK, IC-DC)  
&   &
{$\theta_{23}$},{$|\Delta m^2_{31,32}|$}, 
{$\theta_{13}$},{$\delta_{\rm CP}$}
\\
Accel LBL $\nu_\mu$,$\bar\nu_\mu$,
Disapp (K2K, MINOS, T2K, NO$\nu$A) 
&  {$|\Delta m^2_{31,32 }|$}, {$\theta_{23}$} &
\\
Accel LBL $\nu_e$,$\bar\nu_e$ App (MINOS, T2K, NO$\nu$A)  
&  {$\delta_{\rm CP}$}   &  
$\theta_{13}$  ,  {$\theta_{23}$}\\\hline
\end{tabular}
\caption{Summary of experiments contributing to the present determination of  
the oscillation parameters in the 3$\nu$ framework
(table taken from Ref.~\cite{PDG}).}
\label{tab:expsum}
\end{table}

We close this chapter with a short description of two  additional
experimental inputs  which are also relevant to the work presented in the
following chapters.
 In Section~\ref{sec:CEvNS}, we introduced  CE$\nu$NS measurements which we later on will employ in Sections~\ref{sec:nuCS_nsi} and~\ref{sec:results_glob_bsm}  to bound  new physics that could add any contribution to the SM predictions at the scattering level via neutral-current processes. In Section~\ref{sec:gallium} we added the results from source experiments in gallium detectors which are a persistent puzzle that we will address in Section~\ref{sec:frame_gal}.

\chapter{Global analysis of Three neutrino oscillations}
\label{cap:nufit}

In Chapter \ref{chap:exp}, we discussed the role of the neutrino
experiments in the determination of the oscillation parameters. Each
experiment, including solar, atmospheric, reactor, and accelerator
sources, is able to constrain a specific region of the parameter
space. When performing a joint fit of all main experiments, it is
expected that such analysis will yield the most stringent determination of
the oscillation parameters, as free of degeneracies as presently possible. This is the main goal of a combined fit. 
The global analysis of neutrino oscillation data provides a
comprehensive understanding of the flavor structure of leptons. While
experimental collaborations like T2K and NOvA have made strides in
combining their data~\cite{NOvA:nu24}, the primary responsibility for
global phenomenological analyses still lies with groups like
NuFIT\cite{nufit}. Over the past decade, these studies have
consistently shown that mass-driven oscillations between three
neutrino states, characterized by different masses and mixing angles,
explain the majority of experimental results. Subdominant effects,
such as the mass ordering (MO), the maximality of $\theta_{23}$, and
leptonic CP violation, remain open questions and are the focus of
upcoming experiments like JUNO, DUNE, and Hyper-Kamiokande.

The following chapter is dedicated to presenting the main results of
the NuFIT collaboration after the Neutrino 2024 conference. It
describes the joint analysis of the main experiments up to this
moment. For solar neutrinos, we include constraints from Standard
Solar Models~\cite{B23Fluxes} and experimental results from the
chlorine experiment~\cite{Cleveland:1998nv}, Gallex/GNO~\cite{Kaether:2010ag} and SAGE~\cite{Abdurashitov:2009tn}, as
well as spectral measurements from Super-Kamiokande (SK) phases 1 to
4~\cite{Hosaka:2005um, Cravens:2008aa, Abe:2010hy,
  Super-Kamiokande:2023jbt}, SNO~\cite{Aharmim:2011vm}, and Borexino
Phases I-III~\cite{Bellini:2011rx, Bellini:2008mr, Borexino:2017rsf,
  BOREXINO:2022abl}. Atmospheric neutrino data include
IceCube/DeepCore measurements from IC19
(3-year)~\cite{IceCube:2019dqi, IceCube:2019dyb} and IC24
(9.3-year)~\cite{IceCube:2024xjj, IC:data2024}, along with
Super-Kamiokande (SK1-5) results~\cite{Super-Kamiokande:2023ahc,
  SKatm:data2024}. Reactor neutrino data incorporate KamLAND
spectra~\cite{Gando:2013nba}, SNO+ results~\cite{SNO:2024wzq,
  SNO+:nu24, SNO+poster:nu24}, Double Chooz spectral
ratios~\cite{DoubleC:nu2020}, as well as extensive measurements from
Daya Bay~\cite{DayaBay:2022orm, DayaBay:2021dqj} and
Reno~\cite{RENO:nu2020}. Finally, accelerator neutrino data include
MINOS disappearance and appearance channels~\cite{Adamson:2013whj,
  Adamson:2013ue}, T2K $\nu_\mu$ and $\nu_e$ appearance and
disappearance measurements~\cite{T2K:nu24, T2K:2023mcm, T2K:2023smv},
and NOvA results for both neutrino and antineutrino
modes~\cite{NOvA:nu24, NOvA:nu2020}. These datasets collectively
provide a comprehensive picture of neutrino oscillation parameters.

\section{Global analysis}
\label{sec:global24}

We start by presenting the results of the NuFIT~6.0 global
fit\cite{Esteban:2024eli}, which includes all data available up
to September 2024.  As is customary in the NuFIT releases since
v4.0~\cite{Esteban:2018azc}, there are two versions of the analysis.
They differ in the inclusion of some atmospheric neutrino
data, for which there is not enough information for us to make an
independent fit comparable to that performed by the
collaborations.  In NuFIT~6.0, this is the case for the atmospheric
data from Super-Kamiokande phases 1-5 (SK-atm) and from the
latest 9.3-year results from IceCube/DeepCore (IC24).  For those, we
use their tabulated $\chi^2$ maps provided in
Refs.~\cite{SKatm:data2024} and~\cite{IC:data2024}, respectively,
which we can combine with our global analysis of solar, reactor
and LBL experiments.  We note that for IceCube/DeepCore we have
performed an independent fit of their previous 3-year atmospheric
neutrino data sample~\cite{IceCube:2019dyb, IceCube:2019dqi}, which we
do include in the version of the analysis without tabulated $\chi^2$
maps.  In what follows, we refer as <<IC19 w/o SK-atm>> to the
analysis variant without tabulated $\chi^2$ maps, and as <<IC24 with
SK-atm>> to the one that includes the tabulated SK-atm
and IC24 $\chi^2$ maps instead of our 3-year IceCube/DeepCore
analysis.

\begin{figure}[ht!]\centering
 \includegraphics[width=0.86\textwidth]{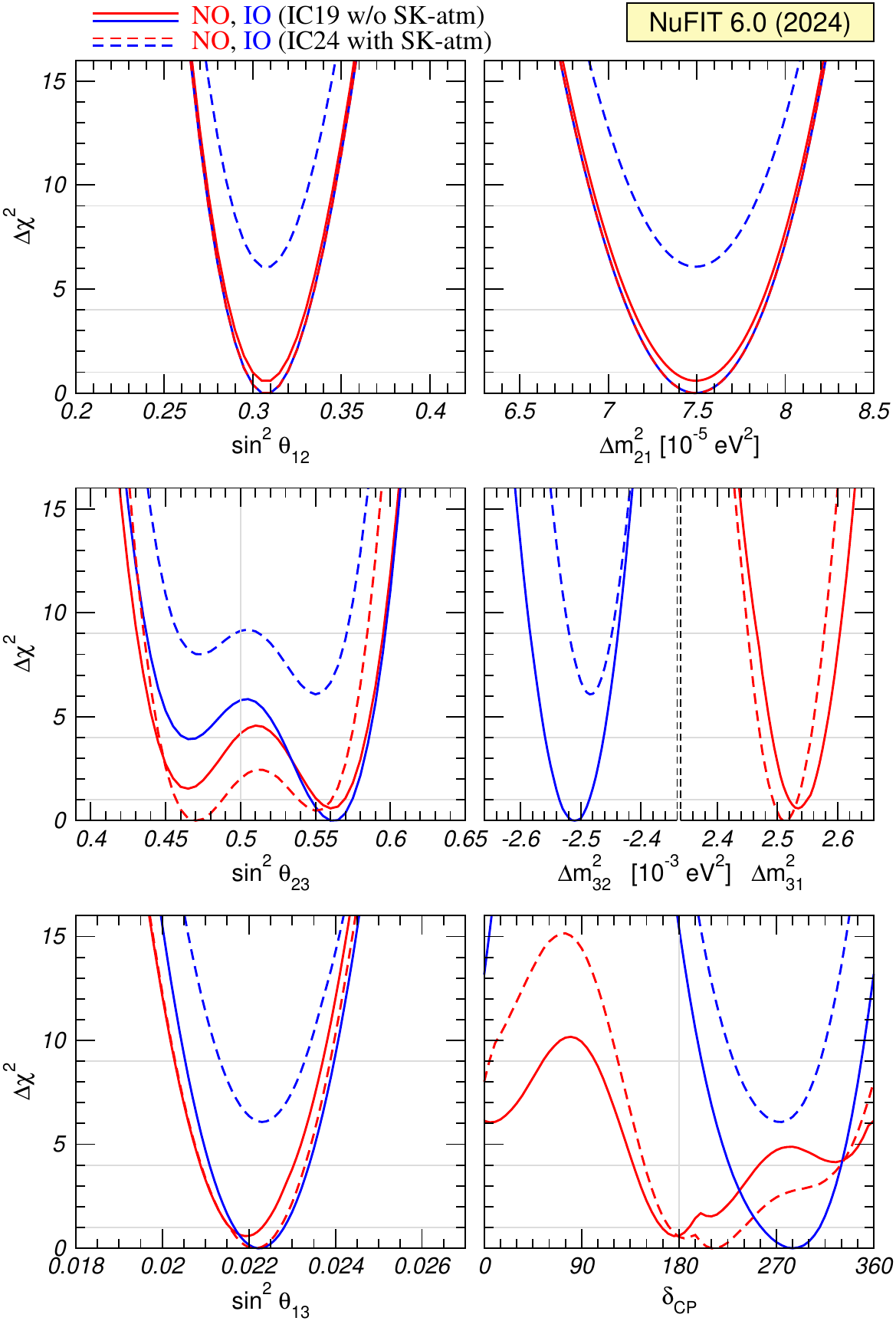}
  \caption{Global $3\nu$ oscillation analysis.  We show $\Delta\chi^2$
    profiles minimized with respect to all undisplayed parameters.
    The red (blue) curves correspond to Normal (Inverted) Ordering.
    Solid and dashed curves correspond to the two variants of the
    analysis as described in the labels.}
  \label{fig:chisq-glob24}
\end{figure}

\begin{figure}[ht!]\centering
 \includegraphics[width=0.81\textwidth]{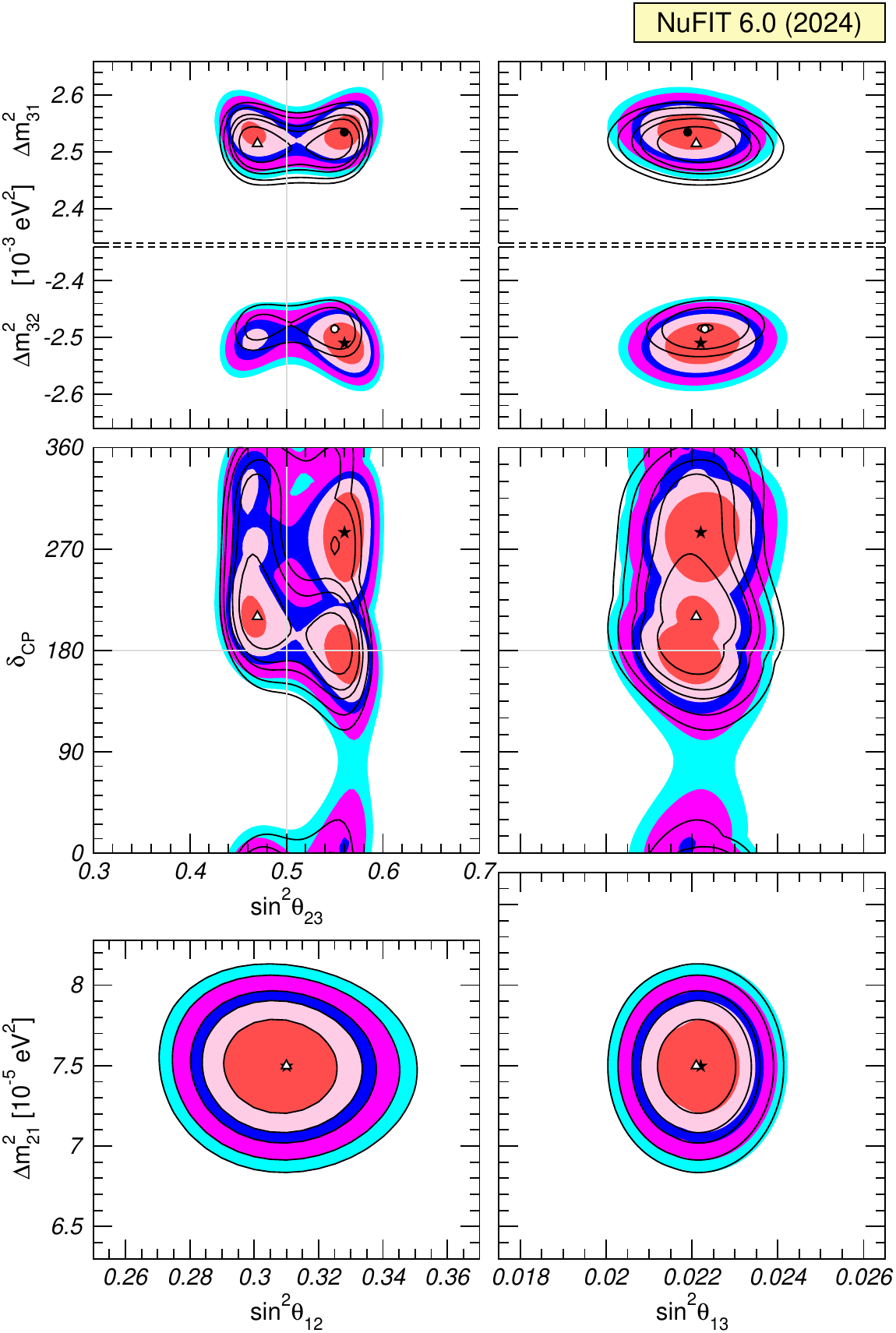}
  \caption{Global $3\nu$ oscillation analysis.  Each panel shows the
    two-dimensional projection of the allowed six-dimensional region
    after minimization with respect to the undisplayed parameters.
    The regions in the four lower panels are obtained from
    $\Delta\chi^2$ minimized with respect to the mass ordering.  The
    different contours correspond to $1\sigma$, 90\%, $2\sigma$, 99\%,
    $3\sigma$ CL (2 dof).  Colored regions (black contours) correspond
    to the variant with IC19 and without SK-atm (with IC24 and with
    SK-atm).}
  \label{fig:region-glob24}
\end{figure}

A selection of the results of our global fit are displayed in
Fig.~\ref{fig:chisq-glob24} (one-dimensional $\Delta\chi^2$ curves)
and Fig.~\ref{fig:region-glob24} (two-dimensional projections of
confidence regions).  In Table~\ref{tab:bfranges24} we give the
best-fit values as well as $1\sigma$ and $3\sigma$ confidence
intervals for the oscillation parameters in both mass orderings,
relative to the local best-fit points in each ordering.  Additional
figures and tables corresponding to this global analysis can be found
in the NuFIT webpage~\cite{nufit}.

\begin{table}\centering
  \begin{footnotesize}
    \begin{tabular}{c|l|cc|cc}
      \hline\hline
      \multirow{2}{*}{} & 
      & \multicolumn{2}{c|}{Normal Ordering (NO)}
      & \multicolumn{2}{c}{Inverted Ordering (IO)}
      \\
      \cline{3-6}
      & & bfp $\pm 1\sigma$ & $3\sigma$ range
      & bfp $\pm 1\sigma$ & $3\sigma$ range
      \\
      \hline\hline
      \multicolumn{6}{c}{\textbf{IC19 wo SK atmospheric data} ($\Delta\chi^2 = 0.6$ for NO)} \\
      \hline
      & $\sin^2\theta_{12}$ & $0.307_{-0.011}^{+0.012}$ & $0.275 \to 0.345$ & $0.308_{-0.011}^{+0.012}$ & $0.275 \to 0.345$ \\
      & $\theta_{12}/^\circ$ & $33.68_{-0.70}^{+0.73}$ & $31.63 \to 35.95$ & $33.68_{-0.70}^{+0.73}$ & $31.63 \to 35.95$ \\
      & $\sin^2\theta_{23}$ & $0.561_{-0.015}^{+0.012}$ & $0.430 \to 0.596$ & $0.562_{-0.015}^{+0.012}$ & $0.437 \to 0.597$ \\
      & $\theta_{23}/^\circ$ & $48.5_{-0.9}^{+0.7}$ & $41.0 \to 50.5$ & $48.6_{-0.9}^{+0.7}$ & $41.4 \to 50.6$ \\
      & $\sin^2\theta_{13}$ & $0.02195_{-0.00058}^{+0.00054}$ & $0.02023 \to 0.02376$ & $0.02224_{-0.00057}^{+0.00056}$ & $0.02053 \to 0.02397$ \\
      & $\theta_{13}/^\circ$ & $8.52_{-0.11}^{+0.11}$ & $8.18 \to 8.87$ & $8.58_{-0.11}^{+0.11}$ & $8.24 \to 8.91$ \\
      & $\dCP/^\circ$ & $177_{-20}^{+19}$ & $96 \to 422$ & $285_{-28}^{+25}$ & $201 \to 348$ \\
      & $\frac{\Dmq_{21}}{10^{-5}~\eVq}$ & $7.49_{-0.19}^{+0.19}$ & $6.92 \to 8.05$ & $7.49_{-0.19}^{+0.19}$ & $6.92 \to 8.05$ \\
      & $\frac{\Dmq_{3\ell}}{10^{-3}~\eVq}$ & $+2.534_{-0.023}^{+0.025}$ & $+2.463 \to +2.606$ & $-2.510_{-0.025}^{+0.024}$ & $-2.584 \to -2.438$ \\
      \hline\hline
      \multicolumn{6}{c}{\textbf{IC24 w SK atmospheric data} ($\Delta\chi^2 = 6.1$ for IO)} \\
      \hline
      & $\sin^2\theta_{12}$ & $0.308_{-0.011}^{+0.012}$ & $0.275 \to 0.345$ & $0.308_{-0.011}^{+0.012}$ & $0.275 \to 0.345$ \\
      & $\theta_{12}/^\circ$ & $33.68_{-0.70}^{+0.73}$ & $31.63 \to 35.95$ & $33.68_{-0.70}^{+0.73}$ & $31.63 \to 35.95$ \\
      & $\sin^2\theta_{23}$ & $0.470_{-0.013}^{+0.017}$ & $0.435 \to 0.585$ & $0.550_{-0.015}^{+0.012}$ & $0.440 \to 0.584$ \\
      & $\theta_{23}/^\circ$ & $43.3_{-0.8}^{+1.0}$ & $41.3 \to 49.9$ & $47.9_{-0.9}^{+0.7}$ & $41.5 \to 49.8$ \\
      & $\sin^2\theta_{13}$ & $0.02215_{-0.00058}^{+0.00056}$ & $0.02030 \to 0.02388$ & $0.02231_{-0.00056}^{+0.00056}$ & $0.02060 \to 0.02409$ \\
      & $\theta_{13}/^\circ$ & $8.56_{-0.11}^{+0.11}$ & $8.19 \to 8.89$ & $8.59_{-0.11}^{+0.11}$ & $8.25 \to 8.93$ \\
      & $\dCP/^\circ$ & $212_{-41}^{+26}$ & $124 \to 364$ & $274_{-25}^{+22}$ & $201 \to 335$ \\
      & $\frac{\Dmq_{21}}{10^{-5}~\eVq}$ & $7.49_{-0.19}^{+0.19}$ & $6.92 \to 8.05$ & $7.49_{-0.19}^{+0.19}$ & $6.92 \to 8.05$ \\
      & $\frac{\Dmq_{3\ell}}{10^{-3}~\eVq}$ & $+2.513_{-0.019}^{+0.021}$ & $+2.451 \to +2.578$ & $-2.484_{-0.020}^{+0.020}$ & $-2.547 \to -2.421$ \\
      \hline\hline
    \end{tabular}
  \end{footnotesize}
  \caption{Three-flavor oscillation parameters from our fit to global data for the two variants of
the analysis described in the text. The numbers in the 1st (2nd) column are obtained assuming
NO (IO), i.e., relative to the respective local minimum. Note that $\Dmq_{3\ell} \equiv \Dmq_{31} > 0$ for NO and
$\Dmq_{3\ell} \equiv \Dmq_{32} < 0$ for IO.}
  \label{tab:bfranges24}
\end{table}

With these results, we obtain the following $3\sigma$ relative
precision of each parameter $x$, defined as $2(x^\text{up} -
x^\text{low}) / (x^\text{up} + x^\text{low})$, where $x^\text{up}$
($x^\text{low}$) is the upper (lower) bound on $x$ at the $3\sigma$
level:
\begin{equation}
  \label{eq:precision24}
  \begin{aligned}
    \theta_{12} &: 13\% \,,
    &\quad
    \theta_{13} &: \left\{
    \begin{array}{lr}
      \text{NO} & 8.1\% ~[8.2\%] \,, \\
      \text{IO} & 7.8\% ~[7.9\%] \,,
    \end{array}\right.
    &\quad
    \theta_{23} &: \left\{
    \begin{array}{lr}
      \text{NO} & 21\% ~[19\%] \,, \\
      \text{IO} & 20\% ~[18\%] \,,
    \end{array}\right.
    \\
    \Dmq_{21} &: 15\% \,,
    &\quad
    |\Dmq_{3\ell}| &: \left\{
    \begin{array}{lr}
      \text{NO} & 5.6\% ~[5.1\%] \,, \\
      \text{IO} & 5.8\% ~[5.1\%] \,,
    \end{array}\right.
    &\quad
    \dCP &: \left\{
    \begin{array}{lr}
      \text{NO} & 100\% ~[98\%] \,, \\
      \text{IO} & 54\% ~[55\%] \,,
    \end{array}\right.
  \end{aligned}
\end{equation}
where the numbers between brackets show the impact of including IC24
and SK-atm.  We note that given the non-Gaussianity of
$\Delta\chi^2(\dCP)$, the above estimated precision for $\dCP$ can
only be taken as indicative, in particular for NO.

Altogether, we derive the following $3\sigma$ ranges on the magnitude
of the elements of the leptonic mixing matrix (see
Ref.~\cite{GonzalezGarcia:2003qf} for details on how we derive the
ranges and their correlations):
\begin{equation}
  \label{eq:umatrix24}
  \begin{aligned}
    |U|_{3\sigma}^\text{IC19 w/o SK-atm} &=
    \begin{pmatrix}
      0.801 \to 0.842 &\qquad
      0.519 \to 0.580 &\qquad
      0.142 \to 0.155
      \\
      0.248 \to 0.505 &\qquad
      0.473 \to 0.682 &\qquad
      0.649 \to 0.764
      \\
      0.270 \to 0.521 &\qquad
      0.483 \to 0.690 &\qquad
      0.628 \to 0.746
    \end{pmatrix}
    \\[1mm]
    |U|_{3\sigma}^\text{IC24 with SK-atm} &=
    \begin{pmatrix}
      0.801 \to 0.842 &\qquad
      0.519 \to 0.580 &\qquad
      0.142 \to 0.155
      \\
      0.252 \to 0.501 &\qquad
      0.496 \to 0.680 &\qquad
      0.652 \to 0.756
      \\
      0.276 \to 0.518 &\qquad
      0.485 \to 0.673 &\qquad
      0.637 \to 0.743
    \end{pmatrix}
  \end{aligned}
\end{equation}

\begin{figure}[ht!]\centering
  \includegraphics[width=0.9\textwidth]{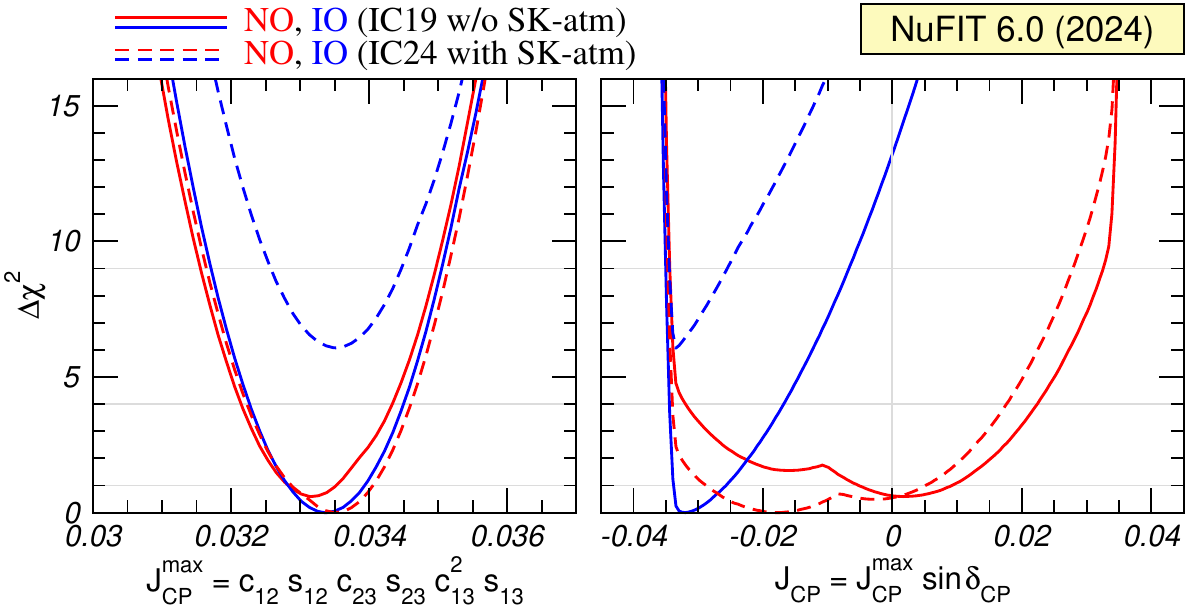}
  \caption{Dependence of the global $\Delta\chi^2$ function on the
    Jarlskog invariant.  The red (blue) curves are for NO (IO).  Solid
    (dashed) curves are for the <<IC19 w/o SK-atm>> (<<IC24 with
    SK-atm>>) $\Delta\chi^2$.}
  \label{fig:chisq-viola}
\end{figure}

We quantify the presence of leptonic CP violation  in vacuum in a convention-independent form in terms of
the leptonic Jarslkog invariant in EQs.\ref{eq:Jarlskog_inv} and
\ref{eq:Jarlskog_inv_2}. Its present determination is shown in
Fig.~\ref{fig:chisq-viola}. While the precise value of $J_\text{CP}$---and in
  particular its deviation from zero---is still highly uncertain, its
  maximum possible value $J_\text{CP}^\text{max}$ (which depends
  solely on the mixing angles and not on $\delta_\text{CP}$) is now
  well determined:
\begin{equation}
  \label{eq:jmax}
  J_\text{CP}^\text{max} = 0.0333 \pm 0.0007 \, (\pm 0.0017) 
\end{equation}
at $1\sigma$ ($3\sigma$) for both orderings. As for $J_\text{CP}$ itself, the right
  panel of figure~\ref{fig:chisq-viola} shows that in NO
the best-fit value $J_\text{CP}^\text{best} = 0.0017\,(-0.018)$ (where
the value in parenthesis corresponds to the analysis with IC24 and
SK-atm) is only favored over CP conservation $J_\text{CP} = 0$ with
$\Delta\chi^2=0.02\,(0.55)$.  In contrast, in IO CP conservation is
disfavored with respect to $J_\text{CP}^\text{best} = -0.032$ with
$\Delta\chi^2=13\,(16)$, which corresponds to $3.6\sigma$ ($4\sigma$)
when evaluated for 1~dof. Let us stress that $J_\text{CP}$ is
  totally analogous to the invariant introduced in
  Ref.~\cite{Jarlskog:1985ht} for the description of CP-violating
  effects in the quark sector, presently determined to be
  $J_\text{CP}^\text{quarks} = (3.12^{+0.13}_{-0.12}) \times
  10^{-5}$~\cite{PDG}.

\section{Status of neutrino mass ordering, leptonic CP violation, and \texorpdfstring{$\boldsymbol{\theta_{23}}$}{theta23}}
\label{sec:atm}

\subsection{Updates from T2K and NOvA}
\label{sec:lbl}

\begin{figure}[ht!]\centering
  \includegraphics[width=0.85\textwidth]{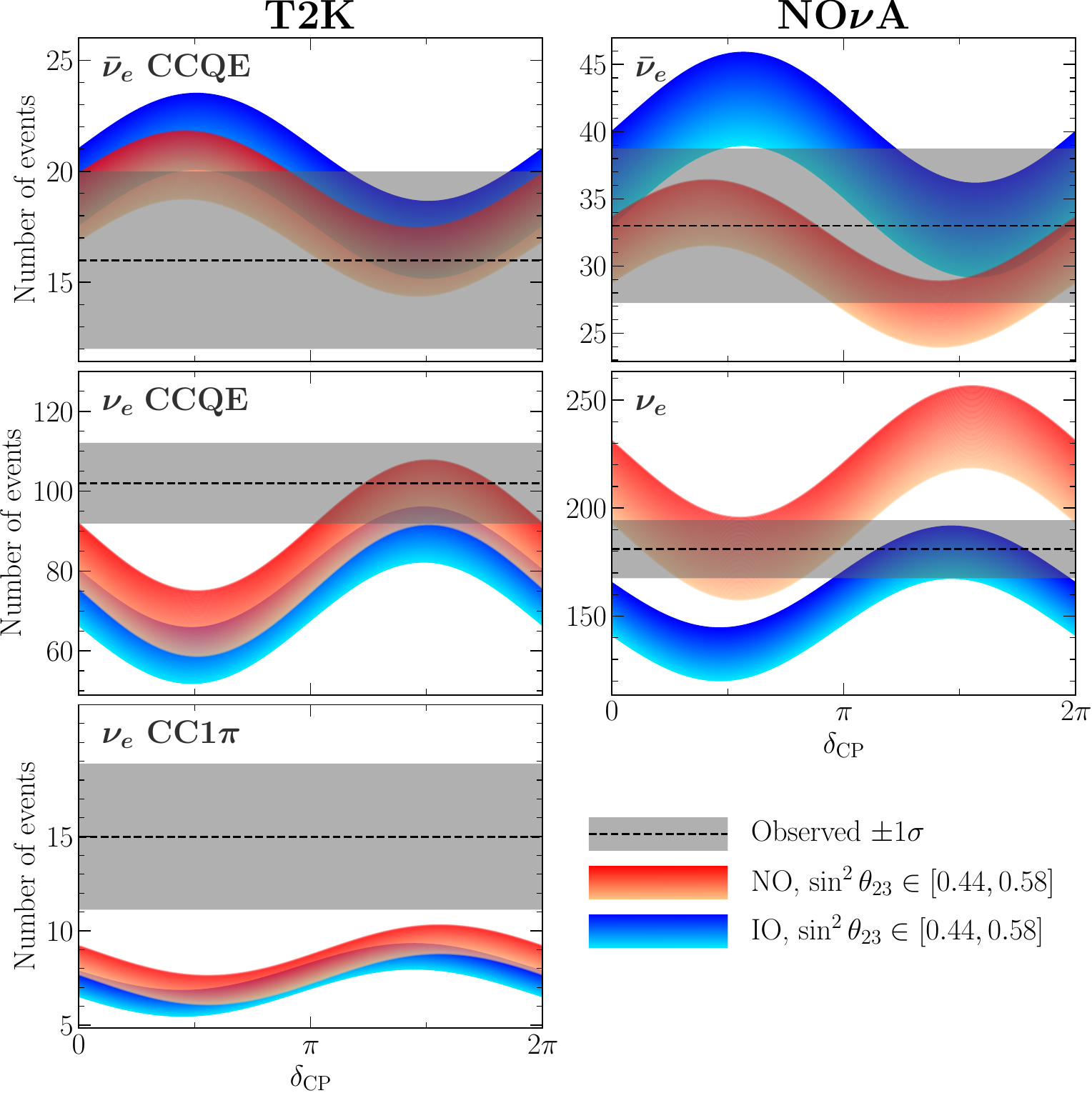}
  \caption{Predicted number of events as a function of $\dCP$ for the
    T2K (left) and NOvA (right) appearance data sets.
    $\sin^2\theta_{23}$ varies between 0.44 and 0.58, where the
    lower-light (upper-dark) bound of the colored bands corresponds to
    0.44 (0.58).  Red (blue) bands correspond to NO (IO).  For the
    other oscillation parameters we have adopted $\sin^2\theta_{13} =
    0.0222$, $|\Dmq_{3\ell}| = 2.5\times 10^{-3}~\eVq$,
    $\sin^2\theta_{12} = 0.32$, $\Dmq_{21} = 7.5\times 10^{-5}~\eVq$.
    The horizontal dashed lines show the observed number of events,
    with the $\pm 1\sigma$ statistical error indicated by the gray
    shaded band.}
  \label{fig:nevts24}
\end{figure}
%%%%%%%%%%%%%%%%%%%%%%%%%%%%%%%%%%%%%%%%

We analyze the latest data from T2K and NOvA presented at Neutrino
2024. Fig.~\ref{fig:nevts24} shows the predicted number of events as a
function of $\dCP$ for varying $\sin^2\theta_{23}$ and mass ordering,
compared to observations. Predictions are based on numerical
simulations, with approximate expressions from
Refs.~\cite{Elevant:2015ska, Esteban:2018azc} describing the general
behavior:

\begin{align}
  N_{\nu_e} &\approx \mathcal{N}_\nu \left[ 2 s_{23}^2(1+2oA) - C' \sin\dCP(1+oA) \right], \label{eq:Nnu} \\
  N_{\bar\nu_e} &\approx \mathcal{N}_{\bar\nu} \left[ 2 s_{23}^2(1-2oA) + C' \sin\dCP(1-oA) \right], \label{eq:Nan}
\end{align}

where $o \equiv \text{sgn}(\Dmq_{3\ell})$. For T2K, $A \approx 0.05$,
while for NOvA, $A=0.1$ empirically. Setting the oscillation parameters to their global
  best-fit values, we get $C' \approx 0.28$. normalization
constants $\mathcal{N}_{\nu,\bar\nu}$ are given in
Table~\ref{tab:app24}, along with observed events and backgrounds. The
ratio $r \equiv N_{\nu_e,\bar\nu_e} / \mathcal{N} = {}
(N_\text{obs}-N_\text{bck})/\mathcal{N}$ provides insights:

\begin{itemize}
  \item T2K data shows $r > 1$ for neutrinos and $r < 1$ for antineutrinos, favoring $\dCP \simeq 3\pi/2$ in NO. This preference, present since early T2K results, remains significant in NuFIT~6.0, strongest in the CC$1\pi$ sample.
  
  \item NOvA antineutrino results ($r<1$) are consistent with NO and $\dCP \simeq \pi/2$ or IO and $\dCP \simeq 3\pi/2$, slightly favoring NO.
  
  \item NOvA neutrino results, with 60\% more statistics, show $r\sim 1$, compatible with NO and $\dCP \simeq \pi/2$ or IO and $\dCP \simeq 3\pi/2$, aligning with antineutrino results. NOvA shows only a mild NO preference.
  
  \item T2K and NOvA favor different $\dCP$ values in NO, making their combination better described by IO with $\dCP \simeq 3\pi/2$, a trend strengthened in NuFIT~6.0.
\end{itemize}

\begin{table}\centering
  \catcode`?=\active\def?{\hphantom{0}}
  \catcode`!=\active\def!{\hphantom{.}}
  \begin{tabular}{c|cccccc}
    \hline\hline
    & \multicolumn{3}{c}{T2K ($\nu$)} & T2K ($\bar\nu$) & NOvA ($\nu$) & NOvA ($\bar\nu$) \\
    & CCQE & CC1$\pi$ & Sum & \\
    \hline
    $\mathcal{N}$              & ?54!? & ?5!? & ?59!? & 13 & 104!? & 23 \\
    $N_\text{obs}$              & 102!? & 15!? & 117!? & 16 & 181!? & 33 \\
    $N_\text{obs}-N_\text{bck}$  & ?81.6 & 12.5 & ?94.2 & 10 & 117.3 & 19 \\
    $r=\frac{N_\text{obs}-N_\text{bck}}{\mathcal{N}}$ & 1.5 (1.6) & 2.5 (2.4) & 1.59 (1.65) & 0.77 (0.61)& 1.13 (1.14) & 0.83 (0.83) \\
    \hline\hline
  \end{tabular}
  \caption{Normalization coefficients $\mathcal{N}_\nu$ and $\mathcal{N}_{\bar\nu}$ for T2K and NOvA appearance samples. Numbers in parentheses correspond to NuFIT~5.0.}
  \label{tab:app24}
\end{table}

\begin{figure}[ht!]\centering
  \includegraphics[width=0.7\textwidth]{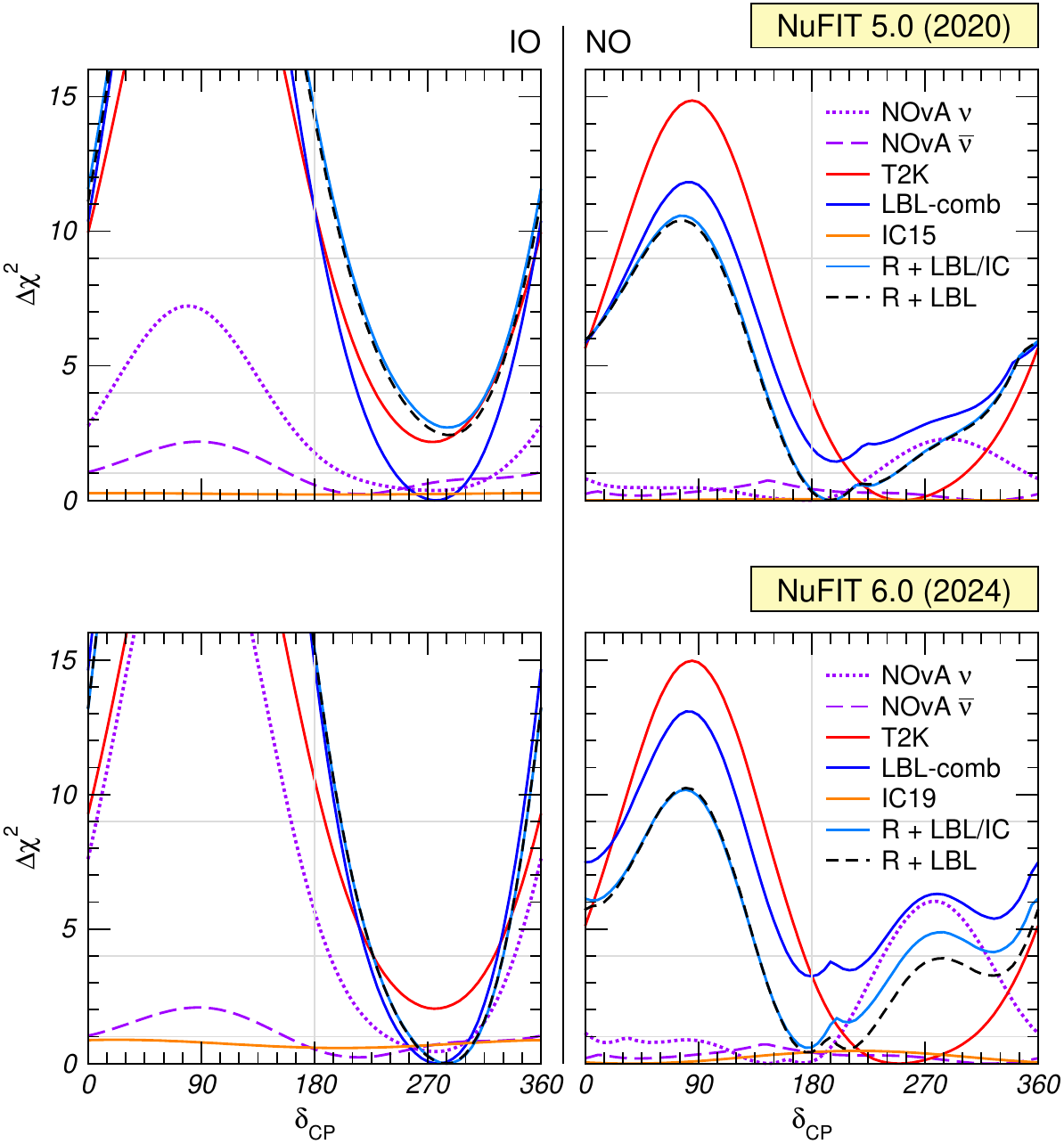}
  \caption{$\Delta\chi^2$ profiles as a function of $\dCP$ for
    different data sets and combinations as labeled in the figure.  In
    the curves where the reactors $R$ are not included in the
    combination we have fixed $\sin^2\theta_{13} = 0.0222$ as well as
    the solar parameters and minimized with respect to $\theta_{23}$
    and $|\Dmq_{3\ell}|$.  When the reactors are included
    $\theta_{13}$ is also marginalized.  Left (right) panels are for
    IO (NO) and $\Delta\chi^2$ is shown with respect to the global
    best-fit point for each curve.  Upper panels are for the NuFIT~5.0
    data set, whereas lower panels correspond to the current update.}
  \label{fig:compare-dcp24}
\end{figure}

%%%%%%%%%%%%%%%%%%%%%%%%%%%%%%%%%%5
Fig.~\ref{fig:compare-dcp24} shows the $\Delta\chi^2$ profiles as a function of $\dCP$ for T2K, NOvA, and their combination, including MBL reactor and IC atmospheric data analyzed in NuFIT~5.0 and NuFIT~6.0 (discussed in Sec.~\ref{sec:reacatm}). The NOvA neutrino results disfavor NO and $\dCP\simeq 3\pi/2$ by $\Delta\chi^2 \approx 6$ (versus 3 in NuFIT~5.0). For IO, T2K and NOvA show a statistically stronger preference for $\dCP\simeq 3\pi/2$. This drives the LBL combination's preference for IO, with $\Delta\chi^2(\text{NO}-\text{IO}) \approx 3.2$ (versus 1.5 in NuFIT~5.0).

\begin{figure}[ht!]\centering
  \includegraphics[width=\textwidth]{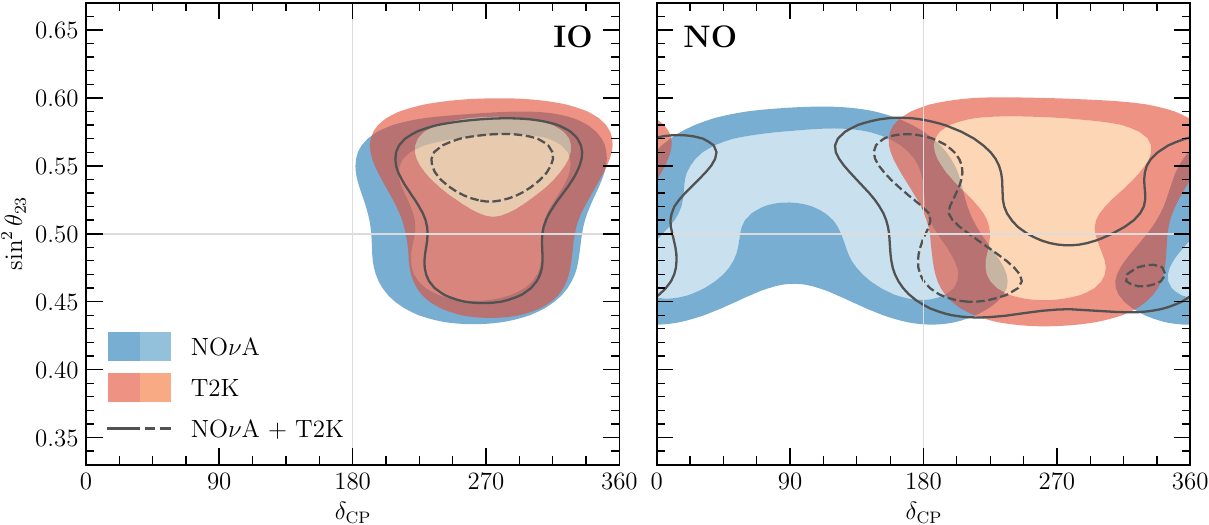}
  \caption{$1\sigma$ and $2\sigma$ allowed regions (2 dof) for T2K
    (red shading), NOvA (blue shading) and their combination (black
    curves).  Contours are defined with respect to the local minimum
    for IO (left) or NO (right).  We fix $\sin^2\theta_{13}=0.0222$,
    $\sin^2\theta_{12}=0.31$, $\Dmq_{21}=7.5\times 10^{-5}~\eVq$ and
    minimize with respect to $|\Dmq_{3\ell}|$.}
  \label{fig:sq23-dCP24}
\end{figure}

\begin{figure}[ht!]\centering
  \includegraphics[width=0.9\textwidth]{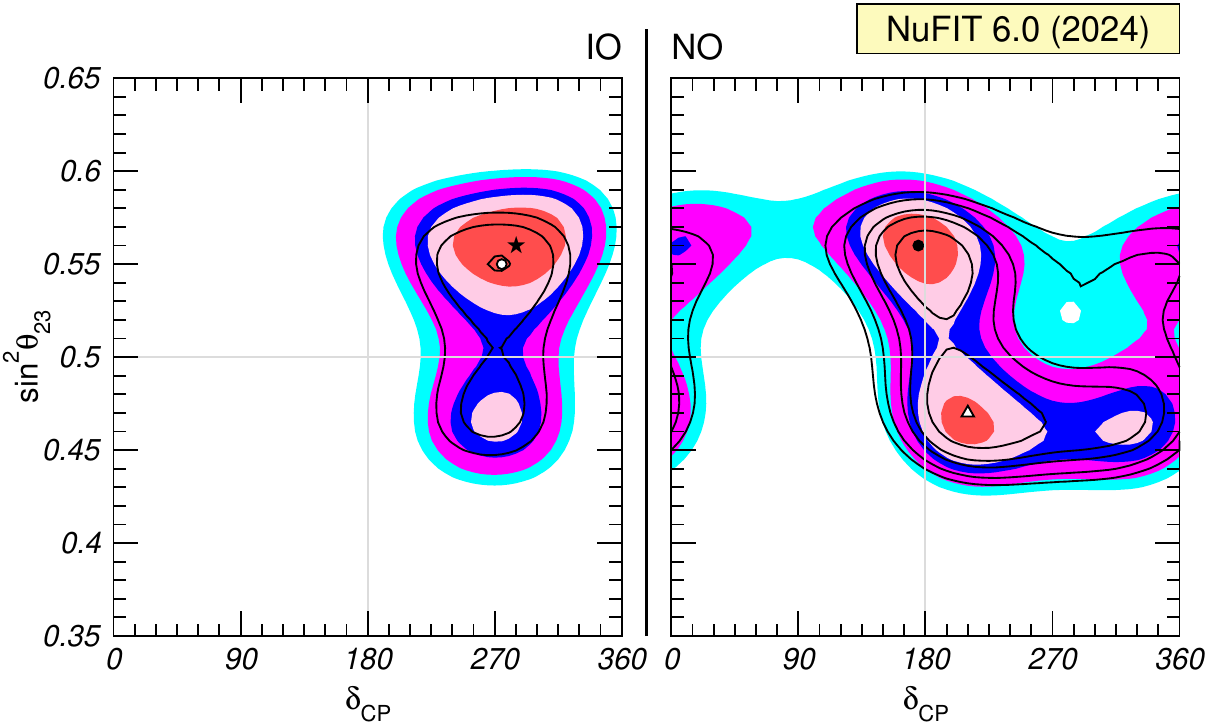}
  \caption{Two-dimensional projection of the allowed six-dimensional
    region from global data in the plane of ($\dCP,\sin^2\theta_{23}$)
    for IO (left) and NO (right) after minimization with respect to
    the undisplayed parameters.  Regions for both orderings are
    defined with respect to the global best-fit point.  The different
    contours correspond to $1\sigma$, 90\%, $2\sigma$, 99\%, $3\sigma$
    CL (2 dof).  Colored regions (black contours) correspond to the
    variant with IC19 and without SK-atm (with IC24 and with SK-atm).}
  \label{fig:region-cp23}
\end{figure}

The two-dimensional regions for T2K and NOvA in the $(\dCP, \sin^2\theta_{23})$ plane (Fig.~\ref{fig:sq23-dCP24}) show better consistency for IO. Unlike NuFIT~5.0, the $1\sigma$ regions for NO do not overlap. The global fit projections (Fig.~\ref{fig:region-cp23}) closely resemble the T2K+NOvA combination, exhibiting correlations with the MO. For IO, $\dCP \simeq 270^\circ$ is highly preferred, whereas NO presents a more complex structure with multiple local minima. The $\theta_{23}$ octant degeneracy persists with $\Delta\chi^2 < 4$ for both MOs and data variants, with minima near $\sin^2\theta_{23} \approx 0.56$ and $0.47$.

Then, after the discussion of the absence of overlapping regions for NO, a natural question  is how large is the tension between T2K and NOvA.  Consistency among different data sets can be quantified with the parameter goodness-of-fit
(PG)~\cite{Maltoni:2003cu}.  For a number $N$ of different data sets
$i$, each depending on $n_i$ model parameters, and globally depending
on $n_\text{glob}$ parameters, it can be shown that the test statistic
\begin{equation}
  \chi^2_\text{PG} \equiv \chi^2_\text{min,glob} - \sum_i^N
  \chi^2_{\text{min}, i} = \min\bigg[ \sum_i^N\chi^2_i \bigg]
  - \sum_i \chi^2_{\text{min}, i} \,,
\end{equation}
follows a $\chi^2$ distribution with $n \equiv \sum_i n_i -
n_\text{glob}$ degrees of freedom~\cite{Maltoni:2003cu}.

Applying this test to the full NOvA and T2K samples (including both
appearance and disappearance data for neutrinos and antineutrinos) we
obtain the values in Table~\ref{tab:PG24}.  We carry out the analysis
separately for each mass ordering, in all cases fixing $\Dmq_{21}$ and
$\theta_{12}$ to their best fit.  In the results reported in the upper
part of the table $\theta_{13}$ is varied in the minimization, so
$n_\text{T2K} = n_\text{NOvA} = n_\text{glob=T2K+NOvA} = 4$
(\textit{i.e.}, $\Dmq_{3\ell}$, $\theta_{23}$, $\dCP$, and
$\theta_{13}$).  In the lower part $\theta_{13}$ is kept fixed to its
best fit so $n_\text{T2K} = n_\text{NOvA} = n_\text{glob=T2K+NOvA} =
3$.  From the table we read that, as expected, agreement is better in
IO, where irrespective on $\theta_{13}$ the samples are compatible at
the $0.5\sigma$ level or better.  In NO, compatibility arises at
$1.7\sigma$ ($2.0\sigma$) for free (fixed) $\theta_{13}$.  This is to
be compared with the NuFIT~5.0 results of $1.4\sigma$ ($1.7\sigma$)
respectively.  We conclude that the tension between T2K and NOvA in NO
has slightly strengthened with the new results, reaching at most the
$2\sigma$ level.

\begin{table}\centering
  \catcode`?=\active\def?{\hphantom{0}}
  \catcode`!=\active\def!{\hphantom{.}}
  \begin{tabular}{l|ccc|ccc}
  \hline\hline
  Data sets &  \multicolumn{3}{c|}{Normal Ordering}
  &  \multicolumn{3}{c}{Inverted Ordering} \\
  & $\chi^2_\text{PG} / n$ & $p$-value & $\#\sigma$
  & $\chi^2_\text{PG} / n$ & $p$-value & $\#\sigma$ \\
  \hline
  T2K vs NOvA                  & ?7.9/4? & 0.093 & 1.7? & ?2.3/4? & 0.67 & 0.42 \\
  T2K vs React                 & 0.23/2? & 0.89? & 0.14 & ?1.7/2? & 0.43 & 0.79 \\
  NOvA vs React                & ?1.1/2? & 0.58? & 0.56 & ?4.3/2? & 0.12 & 1.6? \\
  T2K vs NOvA vs React         & ?8.6/6? & 0.20? & 1.3? & ?6.0/6? & 0.42 & 0.80 \\
  (T2K \& NOvA) vs React       & 0.76/2? & 0.68? & 0.41 & ?3.4/2? & 0.18 & 1.3? \\
  T2K vs IC19                  & ?2.7/4? & 0.61? & 0.51 & ?1.2/4? & 0.88 & 0.15 \\
  NOvA vs IC19                 & ?3.3/4? & 0.51? & 0.66 & ?2.3/4? & 0.68 & 0.41 \\
  Reac vs IC19                 & ?2.1/2? & 0.35? & 0.93 & 0.88/2? & 0.64 & 0.84 \\
  NOvA vs T2K vs IC19          & !?11/8? & 0.20? & 1.3? & ?4.3/8? & 0.83 & 0.21 \\
  NOvA vs T2K vs React vs IC19 & 11.5/10 & 0.33? & 0.96 & ?7.2/10 & 0.71 & 0.38 \\
  \hline
  T2K vs NOvA                  & ?8.0/3 & 0.045 & 2.0? & 1.8/3 & 0.61? & 0.50? \\
  T2K vs NOvA vs React         & ?8.3/4 & 0.081 & 1.7? & 4.1/4 & 0.39? & 0.85? \\
  (T2K \& NOvA) vs React       & 0.25/1 & 0.62? & 0.50 & 2.0/1 & 0.16? & 1.4?? \\
  T2K vs IC19                  & 0.72/3 & 0.86? & 0.16 & 0.2/3 & 0.98? & 0.028 \\
  NOvA vs IC19                 & ?1.5/3 & 0.68? & 0.41 & 1.0/3 & 0.80? & 0.25? \\
  NOvA vs T2K vs IC19          & ?9.3/6 & 0.16? & 1.4? & 2.4/6 & 0.88? & 0.15? \\
  NOvA vs T2K vs React vs IC19 & ?9.4/7 & 0.22? & 1.2? & 4.5/7 & 0.72? & 0.36? \\
  NOvA vs T2K vs IC24          & ?9.5/6 & 0.15? & 1.4? & 4.4/6 & 0.62? & 0.49? \\
  NOvA vs T2K vs React vs IC24 & !?10/7 & 0.19? & 1.3? & 8.2/7 & 0.27? & 1.1?? \\
  \hline\hline
  \end{tabular}
  \caption{Consistency test among different data sets, shown in the
    first column, assuming either normal or inverted ordering.
    ``React'' includes Daya-Bay, RENO and Double-Chooz.  In the
    analyses above the horizontal line, $\theta_{13}$ is a free
    parameter, whereas below the line we have fixed $\sin^2\theta_{13}
    = 0.0222$.  See text for more details.}
    \label{tab:PG24}
\end{table}

\subsection{Effects from \texorpdfstring{$\nu_\mu/\bar\nu_\mu$}{numu/antinumu} versus \texorpdfstring{$\bar\nu_e$}{antinue} disappearance}
\label{sec:reacatm}

\begin{figure}[ht!]\centering
  \includegraphics[width=0.9\textwidth]{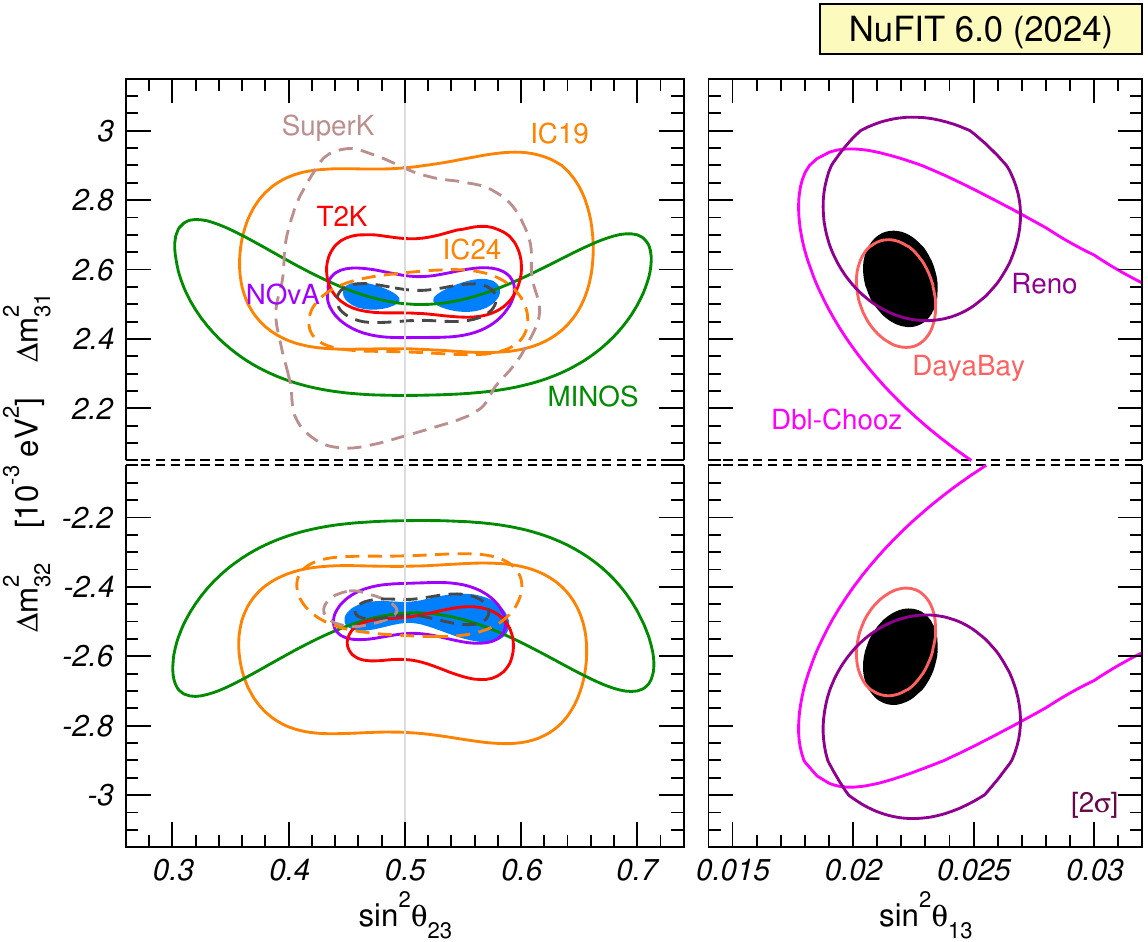}
  \caption{Confidence regions at 95.45\% CL (2~dof) in the plane of
    $\sin^2\theta_{23}$ ($\sin^2\theta_{13}$) and $\Dmq_{3\ell})$ in
    the left (right) panels.  For the left panels we use both
    appearance and disappearance data from MINOS (green), NOvA
    (purple) and T2K (red), as well as atmospheric data from IC
    (orange) and Super-Kamiokande (light-brown); the colored region
    corresponds to the combination of these accelerator data with
    IC19, whereas the black-dashed contour corresponds to the
    combination with IC24 and Super-Kamiokande.  A prior on
    $\theta_{13}$ is included to account for the reactor constraint.
    The right panels show regions using data from Daya-Bay (pink),
    Double-Chooz (magenta), RENO (violet), and their combination
    (black regions).  In all panels solar, KamLAND and SNO+ data are
    included to constrain $\Dmq_{21}$ and $\theta_{12}$.  Contours are
    defined with respect to the global minimum of the two orderings
    for each data set.}
  \label{fig:regions-dis}
\end{figure}

Figure~\ref{fig:regions-dis} illustrates the determination of $\sin^2\theta_{23}$, $\sin^2\theta_{13}$, and $\Dmq_{3\ell}$ using different neutrino disappearance channels: $\nu_\mu/\bar\nu_\mu$ disappearance from long-baseline (LBL) accelerator and atmospheric data (left panel) and $\bar\nu_e$ disappearance from reactor experiments (right panel). The combination of these datasets demonstrates a strong synergy, as the global constraints are significantly tighter than those from individual experiments, indicating a high level of consistency (Table~\ref{tab:PG24}). The interplay between LBL and reactor data is particularly crucial for mass ordering (MO) determination, as these experiments provide complementary constraints~\cite{Nunokawa:2005nx, Minakata:2006gq}.  

\begin{figure}[ht!]\centering
  \includegraphics[width=0.49\textwidth]{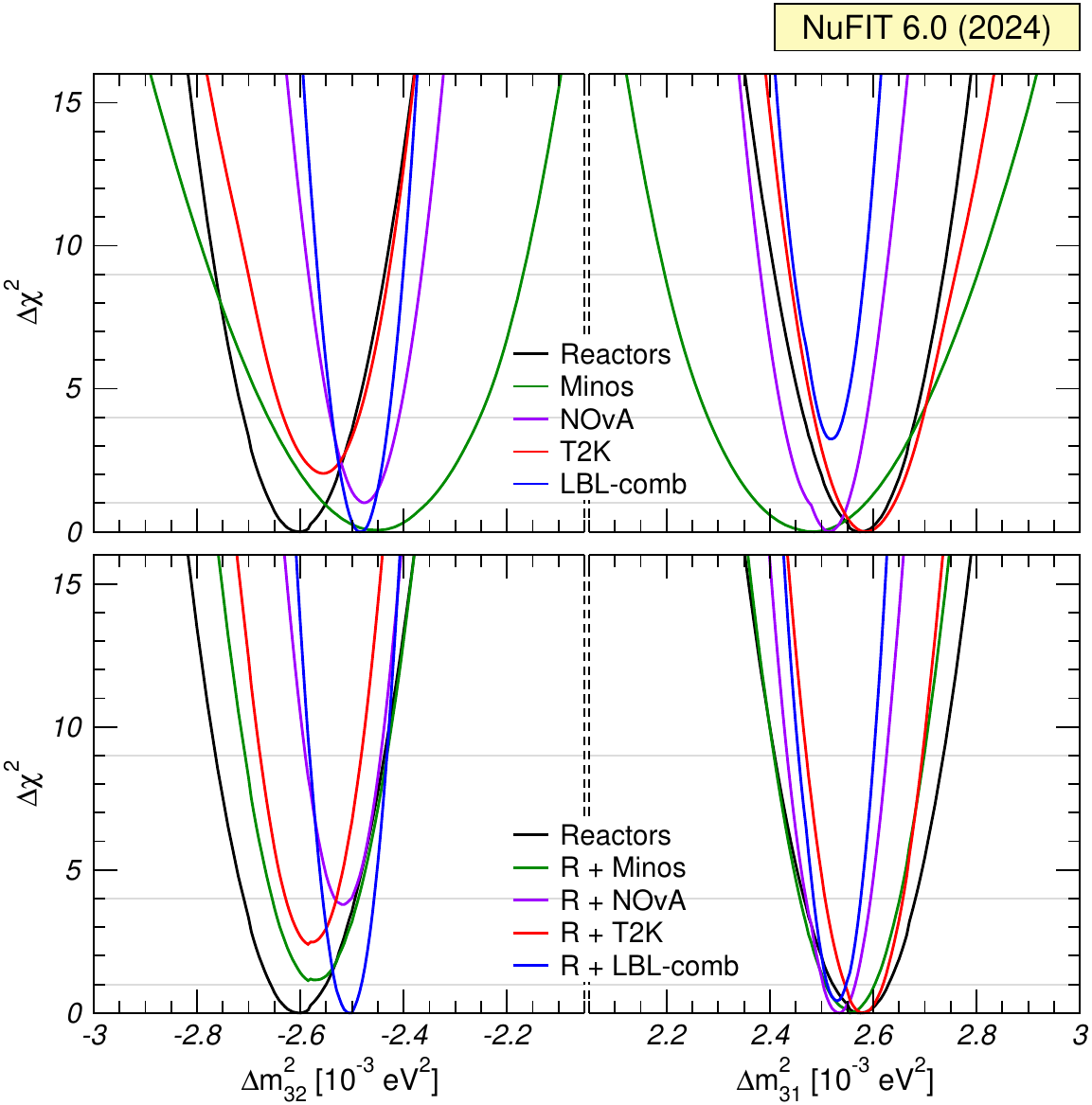}\hfill
  \includegraphics[width=0.49\textwidth]{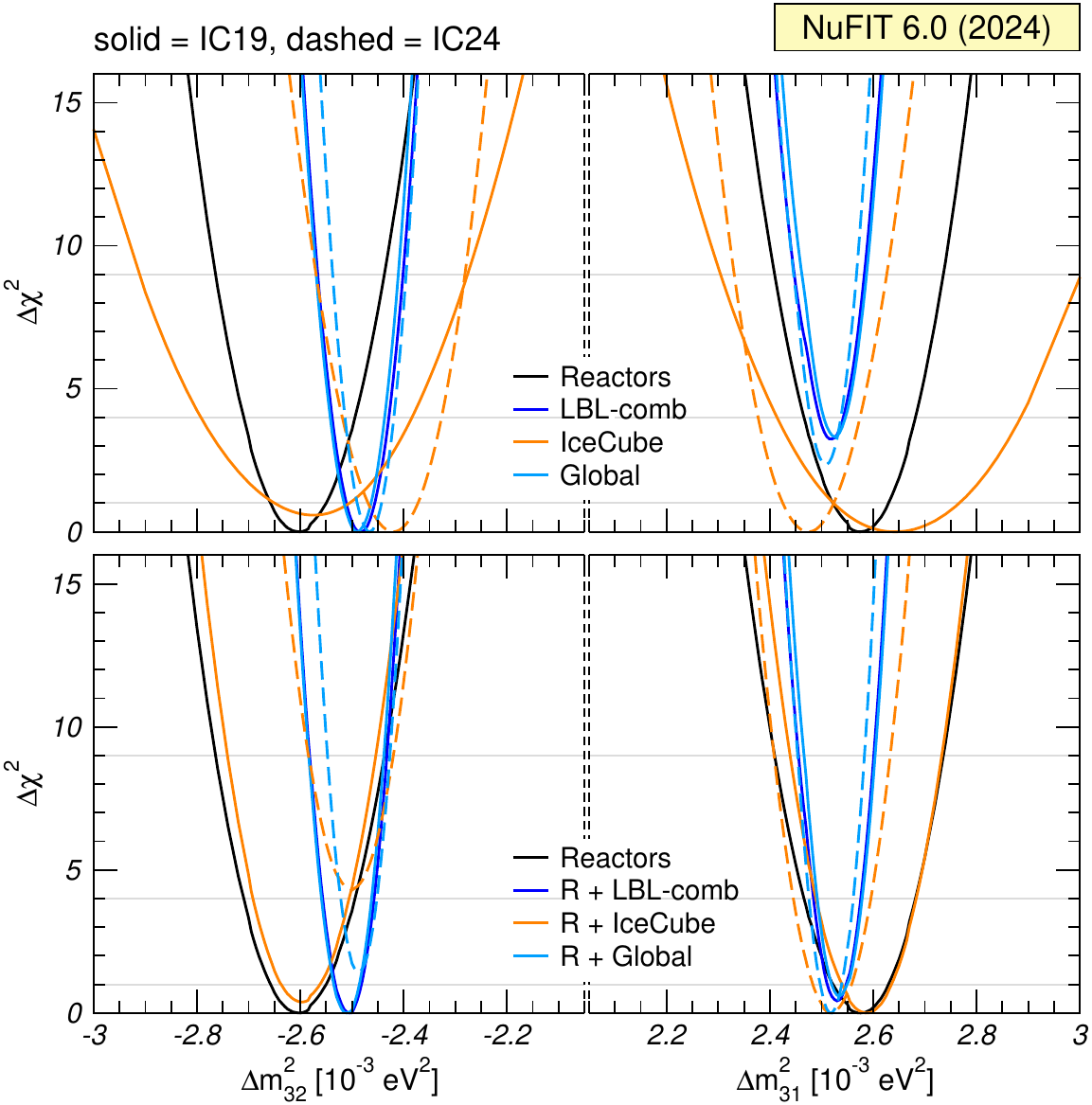}
  \caption{$\Delta\chi^2$ profiles as a function of $\Dmq_{3\ell}$ for
    different data sets and combinations as labeled in the figures.
    In the curves where the reactors $R$ are not included in the
    combination we have fixed $\sin^2\theta_{13}=0.0222$ as well as
    the solar parameters and minimized with respect to $\theta_{23}$
    and $\dCP$.  When the reactors are included $\theta_{13}$ is also
    marginalized.  $\Delta\chi^2$ is shown with respect to the global
    best-fit point (IO or NO) for each curve.  The left set of panels
    visualizes the reactor/LBL combination, whereas in the right set
    of panels we are illustrating the impact of the IC19 or IC24 data
    sets.}
  \label{fig:chisq-dma}
\end{figure}

The $\chi^2$ projections for $\Dmq_{3\ell}$ (Fig.~\ref{fig:chisq-dma}) reveal different sensitivities to the mass ordering across datasets. While most individual datasets either show little sensitivity or slightly prefer normal ordering (NO), the combination of LBL accelerator data with IceCube atmospheric neutrinos (LBL/IC) favors inverted ordering (IO), primarily due to the tension between T2K and NOvA results. However, when LBL and reactor data are combined, they yield a preference for NO with $\Delta\chi^2_\text{IO,NO} = -0.6$, effectively canceling out the preference for IO from T2K and NOvA.  

Compatibility tests using the parameter goodness-of-fit (PG) method (Table~\ref{tab:PG24}) confirm a better agreement with NO when comparing the combined T2K and NOvA dataset with reactor data. However, this preference diminishes when T2K and NOvA are tested separately, reflecting their internal tension. Figure~\ref{fig:chisq-dma} also shows that the older IceCube dataset (IC19) has a negligible impact on MO determination due to its weak constraint on $|\Dmq_{3\ell}|$. In contrast, the more recent IceCube dataset (IC24), which includes 9.3 years of data~\cite{IceCube:2024xjj, IC:data2024}, significantly strengthens the preference for NO when combined with reactor data, leading to $\Delta\chi^2_\text{IO,NO} \approx 4.5$. However, this result does not simply add to the previous $\Delta\chi^2_\text{IO,NO} = -0.6$ from LBL and reactor data, as their interplay shifts the best-fit region, ultimately leading to an overall preference of $\Delta\chi^2_\text{IO,NO} \approx 1.5$.  

Super-Kamiokande atmospheric neutrino data~\cite{SKatm:data2024} alone also favors NO with $\Delta\chi^2_\text{IO,NO} \approx 5.7$. However, this preference appears to stem from a statistical fluctuation~\cite{Super-Kamiokande:2023ahc}, as the data is not strongly conclusive for either ordering. Finally, when combining the IC24, Super-Kamiokande, and global fit data, the overall preference for NO increases to $\Delta\chi^2_\text{IO,NO} \approx 6.1$ (Sec.~\ref{sec:global24}). This suggests that, despite the conflicting trends among individual datasets, the global combination increasingly supports normal ordering.

\subsection{Sensitivity to the neutrino mass ordering}
\label{sec:MO}

To better understand how current global data constrains the neutrino mass ordering (MO), we analyze its sensitivity following the methodology in Ref.~\cite{Blennow:2013oma}. A widely used test statistic for this purpose is the $\chi^2$ difference between the best-fit points for NO and IO, defined as  
\begin{equation}
  T \equiv \Delta\chi^2_\text{IO,NO} = \chi^2_\text{min,IO} - \chi^2_\text{min,NO} \,.
\end{equation}
A positive value of $T$ favors NO, while a negative value favors IO. As demonstrated in Refs.~\cite{Blennow:2013oma, Qian:2012zn}, under certain conditions $T$ follows a Gaussian distribution with mean $\pm T_0$ and standard deviation $2\sqrt{T_0}$. The value of $T_0$ is obtained by considering the expected event rates for the opposite MO:  
\begin{equation}
  \begin{split}
    T_0^\text{NO} = \min_{\theta} \chi^2\big[ p_i(\text{IO},\theta);\, p_i(\text{NO},\theta^\text{true}) \big] \,, \\
    T_0^\text{IO} = \min_{\theta} \chi^2\big[ p_i(\text{NO},\theta);\, p_i(\text{IO},\theta^\text{true}) \big] \,.
  \end{split}
\end{equation}
Here, $p_i(o, \theta)$ represents the expected event number in bin $i$ for ordering $o \in \{\text{NO}, \text{IO}\}$, and $d_i$ is the observed event number. This approach ensures that $T_0$ is always positive, with its magnitude depending on the true values of the oscillation parameters $\theta^\text{true}$, which we take as the best-fit points from Table~\ref{tab:bfranges24}.  

\begin{figure}[ht!]\centering
  \includegraphics[width=0.9\textwidth]{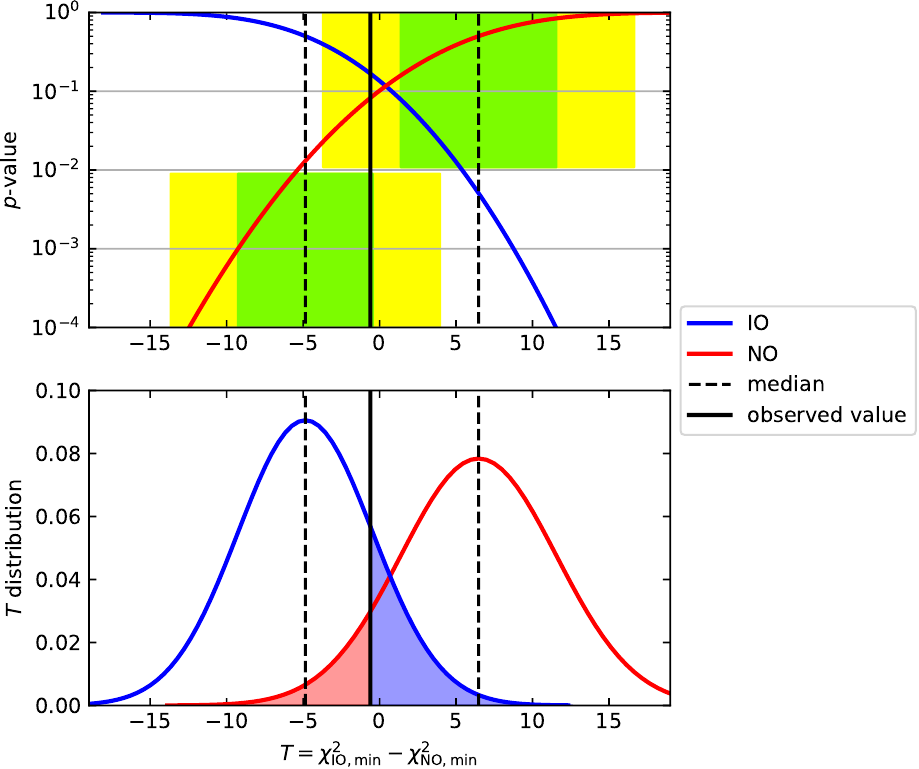}
  \caption{$p$-values (top) and distributions (bottom) for the test
    statistic $T = \chi^2_\text{IO,min} - \chi^2_\text{NO,min}$
    corresponding to the <<IC19 w/o SK-atm>> analysis, assuming true
    NO (red) or true IO (blue).  The observed value $T_\text{obs} =
    -0.6$ is shown by the solid vertical black line.  The
    corresponding median values are shown by the dashed vertical
    lines.  The green and yellow bands in the top panel ---vertically
    displaced to avoid graphical overlap--- correspond to the
    $1\sigma$ and $2\sigma$ intervals for $T$ assuming NO (upwards
    displaced bands) and IO (downwards displaced bands).}
  \label{fig:T-test}
\end{figure}

Since IC24 and Super-Kamiokande data are provided as numerical $\chi^2$ tables, we cannot apply this test directly to them. Thus, we perform the analysis using the dataset labeled <<IC19 w/o SK-atm>>. The results for the global combination are  
\begin{equation}
  T_0^\text{NO} = 6.47 \,, \qquad T_0^\text{IO} = 4.85 \,.
\end{equation}
Figure~\ref{fig:T-test} (bottom panel) shows the corresponding Gaussian distributions. Given the observed value $T_\text{obs} = -0.6$, we compute the probability ($p$-value) of obtaining a more extreme result under each hypothesis:
\begin{equation}
  \begin{aligned}
    &\text{NO:} & p_\text{NO} &= 8.2\% \,, && 91.8\%~\text{CL} \,, && 1.7\sigma \,, \\
    &\text{IO:} & p_\text{IO} &= 16.7\% \,, && 83.3\%~\text{CL} \,, && 1.4\sigma \,.
  \end{aligned}
\end{equation}
As $|T_\text{obs}| < 1$, the $p$-values for both orderings are similar, preventing a strong preference for either. Neither NO nor IO can be rejected at high significance, with both $p$-values remaining below $2\sigma$. The slightly lower $p_\text{NO}$ reflects the negative $T_\text{obs}$ value.

The median sensitivity of the dataset, assuming $T_\text{obs}$ follows the expected mean, gives  
\begin{equation}
  \begin{aligned}
    &\text{NO:} & p_\text{NO}^\text{med} &= 1.3\% \,, && 98.7\%~\text{CL} \,, && 2.5\sigma \,, \\
    &\text{IO:} & p_\text{IO}^\text{med} &= 0.51\% \,, && 99.49\%~\text{CL} \,, && 2.8\sigma \,.
  \end{aligned}
\end{equation}
This implies that, under ideal conditions, the dataset has a nominal sensitivity above $2.5\sigma$ to the MO. However, the observed $T_\text{obs}$ lies between the peaks of the distributions due to competing trends in the data, weakening the rejection of either ordering.  

To assess the likelihood of this result, the top panel of Fig.~\ref{fig:T-test} shows the expected $T_\text{obs}$ range with 68.27\% (green) and 95.45\% (yellow) probability for each ordering. The observed value $T_\text{obs} = -0.6$ lies well within the $1\sigma$ ($2\sigma$) range for IO (NO), confirming that it is not an unusual outcome for either scenario.

\section{Updates in the ``12'' sector}
\label{sec:solar}

\begin{figure}[ht!]\centering
  \includegraphics[width=0.9\textwidth]{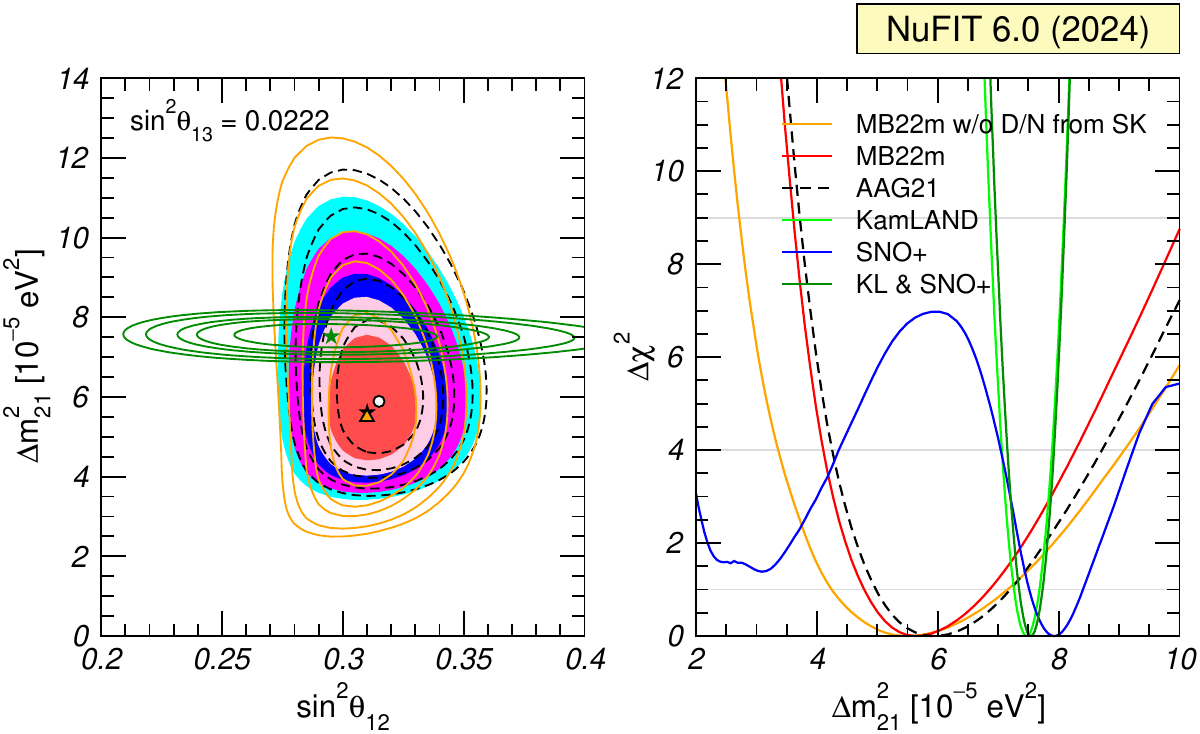}
  \caption{Left: Allowed parameter regions (at $1\sigma$, 90\%,
    $2\sigma$, 99\%, and $3\sigma$ CL for 2 dof) from the combined
    analysis of solar data for MB22-met model (full regions with best
    fit marked by black star) and AAG21 model (dashed void contours
    with best fit marked by a white dot), and for the analysis of the
    combination of KamLAND and SNO+ data (solid green contours with
    best fit marked by a green star) for fixed
    $\sin^2{\theta_{13}}=0.0222$.  For comparison we also show as
    orange contours the results obtained with the MB22-met model
    without including the results of the day-night variation in SK.
    Right: $\Delta\chi^2$ dependence on $\Dmq_{21}$ for the same four
    analyses after marginalizing over $\theta_{12}$.  In addition we
    show separately the results from KamLAND and SNO+.}
  \label{fig:sun-tension24}
\end{figure}

The dominant constraints on the solar neutrino oscillation parameters, $\Dmq_{21}$ and $\theta_{12}$, come from solar neutrino experiments and LBL reactor experiments at distances of $\mathcal{O}(100~\text{km})$. Figure~\ref{fig:sun-tension24} compares the current determination of these parameters from the global solar analysis with that from LBL reactor data.

In the solar neutrino sector, several new data sets have been incorporated since NuFIT~5.0, including the full day-night spectrum from Super-Kamiokande phase IV~\cite{Cravens:2008aa}, as well as the final spectra from Borexino phases II~\cite{Borexino:2017rsf} and III~\cite{BOREXINO:2022abl}. Additionally, the solar neutrino flux predictions have been updated using the latest generation of Standard Solar Models (SSMs)~\cite{B23Fluxes}.  

For the past two decades, solar modeling has faced the so-called solar composition problem, which stems from differences in heavy element abundances. Earlier models used abundances from Ref.~\cite{Grevesse1998} (GS98), which predicted a higher metallicity and provided excellent agreement with helioseismology. However, more modern analyses~\cite{Asplund2009} (AGSS09) yielded lower metallicity values, leading to discrepancies with helioseismology. Consequently, two versions of the Standard Solar Model were developed, one based on GS98 abundances and another on AGSS09~\cite{Serenelli:2009yc, Serenelli:2011py, Vinyoles:2016djt}. More recently, an updated AGSS09-based model (AAG21)~\cite{Asplund2021} slightly increased the metallicity, while a separate study (MB22)~\cite{Magg:2022rxb} found metallicity values closer to GS98. Importantly, MB22-based models align well with helioseismology observations, reinforcing their credibility.

Figure~\ref{fig:sun-tension24} compares the current determinations of $\Dmq_{21}$ and $\theta_{12}$ using solar data with two extreme solar model versions: one based on AAG21 and another on MB22-met abundances. While the overall determination of these parameters remains robust across models, the allowed parameter ranges—especially at higher confidence levels—vary slightly. Notably, a recent model-independent solar flux determination~\cite{Gonzalez-Garcia:2023kva} shows better agreement with MB22 predictions, leading us to adopt MB22 as the reference model for NuFIT~6.0.

For reactor neutrino data, we have updated the predicted antineutrino
fluxes at the KamLAND location to incorporate the latest Daya
Bay measurements~\cite{DayaBay:2021dqj}. Furthermore, our analysis now
includes the first results from the SNO+ experiment, combining 114
ton-years of data from its partial-fill phase~\cite{SNO:2024wzq} with
the first 286 ton-years from its full-fill phase~\cite{SNO+:nu24,
  SNO+poster:nu24}. The right panel of Fig.~\ref{fig:sun-tension24}
displays the $\Delta\chi^2$ dependence on $\Dmq_{21}$ after
marginalizing over $\theta_{12}$ (with $\sin^2\theta_{13} = 0.0222$
fixed). Although SNO+ is currently less precise than KamLAND, its
best-fit value for $\Dmq_{21}$ is slightly higher. However, its impact
on the global determination remains minor at this stage.

Looking ahead, as SNO+ accumulates more data, it will provide valuable insights into the potential tension between solar and reactor determinations of $\Dmq_{21}$. Presently, the best-fit $\Dmq_{21}$ from reactor experiments corresponds to $\Delta\chi^2_\text{solar,MB22} = 2.5$, representing a slight increase over the $\Delta\chi^2_\text{solar,GS98} = 1.3$ reported in NuFIT~5.0. To illustrate the effect of different data inputs, Fig.~\ref{fig:sun-tension24} also shows the solar analysis results without including Super-Kamiokande’s day-night variation data (orange curve). Removing this information reduces the tension, bringing the agreement to $\Delta\chi^2 \sim 1.5$.

Overall, the latest updates lead to only minor shifts in the determination of solar neutrino oscillation parameters—a $\sim 1\%$ increase in the best-fit value and a $\sim 10\%$ improvement in precision. These small variations reaffirm the robustness of the current results.

\section{Projections on neutrino mass scale observables}
\label{sec:absmass}

\begin{figure}[ht!]\centering
  \includegraphics[width=\textwidth]{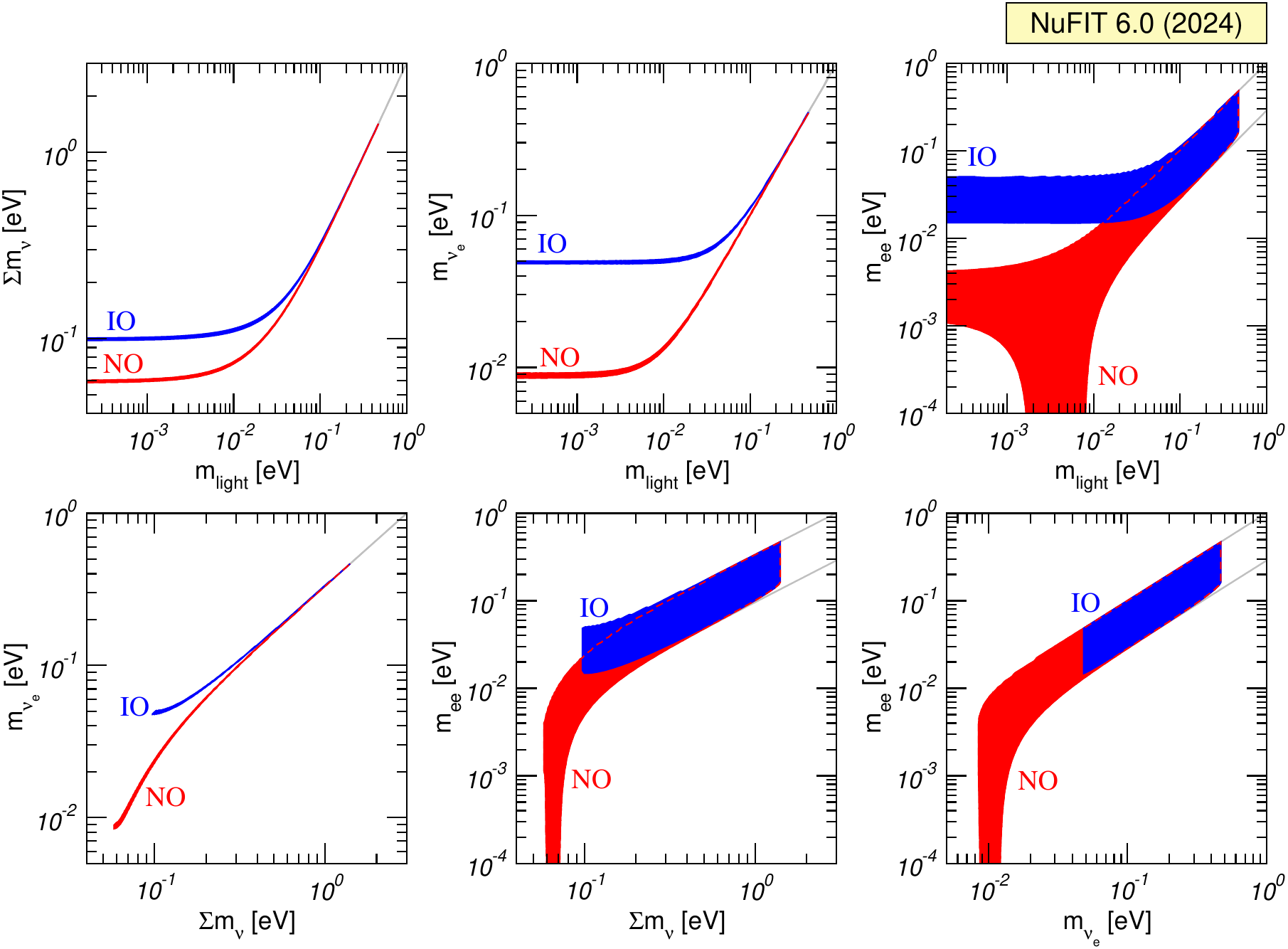}
  \caption{Upper: 95\% CL allowed ranges of the three probes of the
    absolute neutrino mass $\SumNu$, $m_{\nu_e}$, $m_{ee}$ as a
    function of the mass of the lightest neutrino obtained from
    projecting the results of the global analysis of oscillation data.
    The regions are defined with respect to the minimum for each
    ordering.  Lower: Corresponding 95\% CL allowed regions (for 2
    dof) in the planes ($m_{\nu_e}$, $\SumNu$), ($m_{ee}$, $\SumNu$),
    and ($m_{\nu_e}$, $m_{ee}$).}
  \label{fig:mprobes24}
\end{figure}

Neutrino oscillations arise from quantum interference effects, making
them sensitive to mass-squared differences $\Dmq_{ij}$ and to the
structure of the leptonic mixing matrix $U_{\alpha j}$. However,
oscillations do not reveal the absolute neutrino mass scale, only
setting a lower bound on the mass of the heaviest participating
neutrino.

The most direct information on absolute neutrino masses comes from
kinematic studies of processes involving neutrinos as discussed in
Section \ref{eq:numass}.  The most recent result from the KATRIN
experiment~\cite{Katrin:2024tvg} sets an upper limit of $m_{\nu_e} <
0.45$~eV at 90$\%$ CL. The most stringent limits on neutrinoless
double beta decay come from Germanium-based
GERDA~\cite{GERDA:2020xhi}, which sets $T^{0\nu}_{1/2} > 1.8\times
10^{26}$ yr, and Xenon-based KamLAND-Zen~\cite{KamLAND-Zen:2024eml},
with $T^{0\nu}_{1/2} > 3.8\times 10^{26}$ yr. Depending on the assumed
nuclear matrix elements, these translate into limits of $m_{ee}
\lesssim 0.079$--$0.180$~eV for GERDA and $m_{ee} \lesssim
0.028$--$0.122$~eV for KamLAND-Zen.

As already commented in Chapter~\ref{chap:theo}, neutrino masses also leave an imprint in cosmology, primarily through
their impact on the large-scale structure of the
Universe. Cosmological constraints typically provide information on
the sum of neutrino masses provided in EQ. \ref{eq:numass}, while
being largely insensitive to the mass ordering or mixing
angles. Current cosmological observations set upper limits in the
range $\SumNu \lesssim 0.04$--$0.3$ eV~\cite{Jiang:2024viw,
  Naredo-Tuero:2024sgf}, depending on the dataset used, the assumed
cosmological model, and the statistical approach.

Figure~\ref{fig:mprobes24} (upper panels) shows the 95\% CL allowed
regions for $m_{\nu_e}$, $m_{ee}$, and $\SumNu$ obtained from the
NuFIT~6.0 global fit, plotted as a function of $m_0$. The spread in
$m_{ee}$ predictions is due to the unknown Majorana phases. The figure
compares constraints from oscillations alone (void regions) and from
oscillations combined with the KATRIN limit (filled regions). The
KATRIN result is included using a simple $\chi^2$ function based on
their measurement, $m_{\nu_e}^2 = -0.14^{+0.13}_{-0.15}$~eV, assuming
Gaussian errors and Wilks' theorem.\footnote{This gives a 90\% CL
upper limit of $m_{\nu_e} \leq 0.35$~eV, which lies between the limits
obtained with the Lokhov-Tkachov ($0.45$~eV) and Feldman-Cousins
($0.31$~eV) methods.}

Since these observables are linked through the neutrino mass spectrum,
knowing two of them allows in principle for constraints on the
Majorana phases and/or the mass ordering~\cite{Fogli:2004as,
  Pascoli:2005zb}. Figure~\ref{fig:mprobes24} (lower panels)
illustrates these correlations, showing that a strong constraint on
any one parameter can provide valuable insights into the
others~\cite{Gariazzo:2022ahe}.

The combined constraints from oscillation data and KATRIN at 95\% CL are:
\begin{align}
  0.00085~\text{eV} \leq m_{\nu_e} \leq 0.4~\text{eV} & \quad \text{(NO)}, \quad
  0.048~\text{eV} \leq m_{\nu_e} \leq 0.4~\text{eV} & \quad \text{(IO)}, \\
  0.058~\text{eV} \leq \SumNu \leq 1.2~\text{eV} & \quad \text{(NO)}, \quad
  0.098~\text{eV} \leq \SumNu \leq 1.2~\text{eV} & \quad \text{(IO)}.
\end{align}
For Majorana neutrinos, the corresponding $m_{ee}$ limits are:
\begin{equation}
  0 \leq m_{ee} \leq 0.41~\text{eV} \quad \text{(NO)}, \qquad
  0.015~\text{eV} \leq m_{ee} \leq 0.41~\text{eV} \quad \text{(IO)}.
\end{equation}
These results illustrate how complementary approaches—oscillations, direct kinematic searches, neutrinoless double beta decay, and cosmology—collectively constrain the absolute neutrino mass scale and mass ordering.

\section{Summary}

We have presented an updated global analysis of neutrino oscillation data available up to September 2024, as outlined in the introduction of this chapter. Our results are provided in two versions:  
\begin{itemize}
    \item \textbf{IC19 w/o SK-atm}: This dataset includes all data for which enough information is available to perform an independent and detailed fit.  
    \item \textbf{IC24 with SK-atm}: This dataset incorporates numerical $\chi^2$ tables provided by the IceCube and Super-Kamiokande collaborations, which are added to our own $\chi^2$ analysis.  
\end{itemize}  
The global best-fit values, along with their $1\sigma$ and $3\sigma$ confidence ranges, are presented in Table~\ref{tab:bfranges24}. Our main findings can be summarized as follows:  

\begin{itemize}
    \item The parameters $\theta_{12}$, $\theta_{13}$, $\Dmq_{21}$, and $|\Dmq_{3\ell}|$ remain very well constrained, with nearly Gaussian $\chi^2$ distributions up to high confidence levels. Their relative uncertainties at $3\sigma$ are approximately 13\%, 8\%, 16\%, and (5--6)\%, respectively, showing strong stability across datasets.  

    \item The mixing angle $\theta_{23}$ remains the least precisely determined, with a relative uncertainty of about 20\% at $3\sigma$. The well-known octant ambiguity persists, as both solutions ($\theta_{23} < 45^\circ$ and $\theta_{23} > 45^\circ$) have nearly degenerate $\chi^2$ minima. There is a slight overall preference for the second octant ($\theta_{23} > 45^\circ$), except in NO for the <<IC24 with SK-atm>> dataset. However, for all dataset combinations and mass orderings, the alternative octant remains viable, with $\Delta\chi^2 < 4$ between the two minima.  

    \item The determination of the CP-violating phase $\dCP$ strongly depends on the assumed mass ordering. For NO, the best-fit value is close to the CP-conserving scenario $\dCP \approx 180^\circ$, with $\Delta\chi^2 < 1$, making its profile highly non-Gaussian and dependent on the dataset variant and the $\theta_{23}$ octant. In contrast, for IO, the best-fit values for both data variants favor maximal CP violation at $\dCP \approx 270^\circ$, disfavoring CP conservation at $3.6\sigma$ ($4\sigma$) in the <<IC19 w/o SK-atm>> (<<IC24 with SK-atm>>) analysis.  

    \item The mass ordering preference remains a subject of debate. The difference in $\chi^2$ between IO and NO is found to be:
    \begin{equation}
        \Delta\chi^2_\text{IO,NO} = -0.6 \quad \text{(IC19 w/o SK-atm)}, \qquad 6.1 \quad \text{(IC24 with SK-atm)}.
    \end{equation}
    The nearly indecisive result in the <<IC19 w/o SK-atm>> dataset arises from opposing trends in different experiments. The appearance data from T2K and NOvA exhibit a $2\sigma$ tension for NO, while they remain fully consistent for IO. However, the disappearance channels from accelerator and reactor experiments show better agreement with NO than IO, particularly in their determination of $|\Dmq_{3\ell}|$.  

    Despite a sensitivity estimate indicating that the <<IC19 w/o SK-atm>> dataset should have a median ability to reject NO (IO) at $2.5\sigma$ ($2.8\sigma$), the actual observed constraints are much weaker at only $1.7\sigma$ ($1.4\sigma$) due to competing trends in the data. The inclusion of IceCube (IC24) and Super-Kamiokande atmospheric data enhances the preference for NO, leading to the more decisive result of $\Delta\chi^2_\text{IO,NO} = 6.1$ in the <<IC24 with SK-atm>> dataset.  
\end{itemize}  

Additionally, we provide updated constraints and correlations for parameters sensitive to the absolute neutrino mass, obtained from beta decay, neutrinoless double beta decay, and cosmological observations. These results, along with supplementary materials such as detailed figures and data tables, can be accessed on the NuFit webpage~\cite{nufit}.  

\chapter{ Three neutrino oscillations with new interactions: Formalism}
\label{chap:theo_bsm}

As established in Chapter~\ref{chap:exp}, modern neutrino physics exploits four primary neutrino sources—solar, atmospheric, reactor, and accelerator-generated—to probe oscillation phenomena through both charged-current (CC) and neutral-current (NC) interaction channels. While Chapter~\ref{cap:nufit} demonstrated the precision achievable in standard three-flavor oscillation parameters through global analyses of multi-experiment data, the following chapters address a critical open question: \textit{How do current experimental constraints limit beyond-Standard Model (BSM) scenarios affecting both neutrino propagation and detection?}

This chapter develops the theoretical formalism for Non-Standard Neutrino Interactions (NSI) and related BSM frameworks that modify neutrino dynamics. We focus on three complementary approaches: 
\begin{enumerate}
    \item Generic NSI-induced modifications to matter effects in neutrino propagation and detection
    (Section~\ref{sec:formalism_NSI});
    \item Heavy mediator scenarios altering detection cross-sections (Section~\ref{sec:formCS});
    \item Long-range monopole-dipole interactions mediated by ultra-light scalars (Section~\ref{sec:MD_forces}).
\end{enumerate}

\section{Non-Standard Interactions - Formalism }
\label{sec:formalism_NSI}

Generic BSM physics can be systematically incorporated via an
effective field theory (EFT) approach. This involves constructing a
hierarchy of gauge-invariant effective operators, ordered by their
mass dimension and suppressed by powers of a high-energy NP scale
$\Lambda$. Remarkably, neutrino masses emerge naturally in this
construction through the sole dimension-five operator compatible with
the SM gauge symmetry and field content. This observation
places neutrino masses as the leading-order
phenomenological signature of BSM physics. At the next order in the
EFT expansion, dimension-six four-fermion operators dominate. Among
these, operators involving neutrino fields induce non-standard
modifications to neutrino interactions—referred to as NSI—which alter
neutrino production, propagation, and detection processes. In this
section, we specifically analyze NSI arising from dimension-six
operators that generate purely vector or axial-vector couplings
between neutrinos and matter fermions, as these interactions are most
relevant for neutrino oscillation and scattering phenomenology (for
recent works including additional operators with different Lorentz
structures see for example Refs.~\cite{Falkowski:2021bkq,
  Breso-Pla:2023tnz, Falkowski:2019kfn, Falkowski:2019xoe}).
Generically these can be classified in charged-current (CC) NSI
\begin{equation}
  \label{eq:nsi-cc}
  \mathcal{L}_\text{NSI,CC} = -2\sqrt{2}\, G_F
  \sum_{f,f^\prime,\alpha,\beta}
  \Eps_{\alpha\beta}^{ff^\prime,P}
  (\bar\ell_\alpha \gamma_\mu P_L \nu_\beta)
  (\bar f \gamma^\mu P f^\prime)
  + \text{h.c.},
\end{equation}
and neutral-current (NC) NSI
\begin{equation}
  \label{eq:nsi-nc}
  \mathcal{L}_\text{NSI,NC} = -2\sqrt{2}\, G_F
  \sum_{f,P,\alpha,\beta} \Eps_{\alpha\beta}^{f,P}
  (\bar\nu_\alpha\gamma^\mu P_L\nu_\beta)
  (\bar f\gamma_\mu P f) \,.
\end{equation}
In Eqs.~\eqref{eq:nsi-cc} and~\eqref{eq:nsi-nc}, $f$ and $f^\prime$
refer to SM charged fermions, $\ell$ denotes a SM charged lepton and
$P$ can be either a left-handed or a right-handed projection operator
($P_L$ or $P_R$, respectively).  Moreover, the normalization of the
couplings is deliberately chosen to match that of the weak currents in
the SM, so the values of $\Eps_{\alpha\beta}^{f,P}$ indicate the
strength of the new interaction with respect to the Fermi constant,
$G_F$.
The corresponding vector and axial-vector combinations of NSI
coefficients are defined as:
\begin{equation}
  \label{eq:vect-axial}
  \Eps_{\alpha\beta}^{f,V} \equiv
  \Eps_{\alpha\beta}^{f,L} + \Eps_{\alpha\beta}^{f,R}
  \quad\text{and}\quad
  \Eps_{\alpha\beta}^{f,A} \equiv
  \Eps_{\alpha\beta}^{f,L} - \Eps_{\alpha\beta}^{f,R} \,.
\end{equation}

Precise measurements of meson and muon decays place severe constraints
on the possible strength of CC NSI (see for example
Refs.~\cite{Davidson:2003ha, Biggio:2009nt, Biggio:2009kv,
  Falkowski:2021bkq}).  Conversely, NC NSI are much more difficult to
probe directly, given the intrinsic difficulties associated with  
neutrino detection via neutral currents.  A priori, the requirement of
gauge invariance would generate similar operators in the charged
lepton sector, in severe conflict with experimental
observables~\cite{Gavela:2008ra, Antusch:2008tz}.  However, such
bounds may be alleviated (or evaded) in NP models in which NC NSI are
generated by exchange of neutral mediators with masses well below the
EW scale (for a recent review on viable NSI models see, \textit{e.g.},
Ref.~\cite{Dev:2019anc}).  It is in these scenarios that one can
envision observable effects in present and future neutrino
experiments.
Several models have been proposed, involving new gauge symmetries and
light degrees of freedom, that would give rise to relatively large NC
NSI.  These include, for example, models where the NSI are generated
from the $Z'$ boson associated to a new $U(1)'$
symmetry~\cite{Babu:2017olk, Farzan:2015doa, Farzan:2016wym,
  Farzan:2015hkd, Greljo:2022dwn, Heeck:2018nzc, Farzan:2019xor,
  Bernal:2022qba}, radiative neutrino mass models involving new
scalars~\cite{Babu:2019mfe}, or models with
leptoquarks~\cite{Wise:2014oea, Greljo:2021npi, Babu:2019mfe}.  Many
of these extensions involve a gauge symmetry based on a combination of
baryon and lepton quantum numbers, and would therefore induce equal
NSI for up and down quarks.  However, exceptions to this rule arise,
for example, in models with leptoquarks where NSI may be only
generated for down-quarks~\cite{Wise:2014oea, Babu:2019mfe}.
Similarly it should be stressed that, while many of these models
typically lead to diagonal NSI in lepton flavor space, this is not
always the case and, depending on the particular extension, sizable
off-diagonal NSI may also be obtained (see, \textit{e.g.},
Refs.~\cite{Farzan:2015hkd, Farzan:2019xor, Farzan:2016wym}).  In
order to derive constraints to such a wide landscape of models, the
use of the effective operator approach advocated above is extremely
useful.  For bounds from oscillation data on $U(1)'$ models with
diagonal couplings in flavor space, see Ref.~\cite{Coloma:2020gfv}.

In fact, some of the best model-independent bounds on NC NSI are
obtained from global fits to oscillation data.  These are affected by
vector couplings involving fermions present in matter,
$\Eps_{\alpha\beta}^{f,V}$ with $f \in \{u,d,e \}$, since they modify
the effective matter potential~\cite{Wolfenstein:1977ue,
  Mikheev:1986gs} felt by neutrinos as they propagate in a medium.
In Ref.~\cite{Gonzalez-Garcia:2013usa} such global analysis was
performed in the context of vector NC NSI with either up or down
quarks.  In Ref.~\cite{Esteban:2018ppq} the study was extended to
account for the possibility vector NC NSI with up and down quarks
simultaneously, under the restriction that the neutrino flavour
structure of the NSI interactions is independent of the quark type.
However, NSI with electrons were not considered.

One important effect of the presence of NSI~\cite{Wolfenstein:1977ue,
  Valle:1987gv, Guzzo:1991hi} affecting the neutrino propagation in
oscillation experiments is the appearance of a degeneracy leading to a
qualitative change of the lepton mixing pattern.  This was first
observed in the context of solar neutrinos, for which the established
standard MSW
solution (discussed in Chapter \ref{chap:theo}), requires a mixing
angle $\theta_{12}$ in the first octant, while with suitable NSI the
data could be described by a mixing angle $\theta_{12}$ in the second
octant, the so-called LMA-Dark (LMA-D)~\cite{Miranda:2004nb} solution.
The origin of the LMA-D solution is a degeneracy in the oscillation
probabilities due to a symmetry of the Hamiltonian describing neutrino
evolution in the presence of NSI~\cite{GonzalezGarcia:2011my,
  Gonzalez-Garcia:2013usa, Bakhti:2014pva, Coloma:2016gei}.  Such
degeneracy makes it impossible to determine the neutrino mass ordering
by oscillation experiments alone~\cite{Coloma:2016gei} and therefore
jeopardizes one of the main goals of the upcoming neutrino oscillation
program.  Although the degeneracy is not exact when including
oscillation data from neutrino propagating in different environments,
quantitatively the breaking is small and global fits show that the
LMA-D solution is still pretty much allowed by oscillation data
alone~\cite{Gonzalez-Garcia:2013usa, Esteban:2018ppq}.  A second
important limitation of oscillation data is that it is only sensitive
to differences in the potential felt by different neutrino flavours.
Thus, oscillation data may be used to derive constraints on the
differences between diagonal NC NSI parameters, but not on the
individual parameters themselves.

To make the analysis feasible, the following
simplifications are introduced:
\begin{itemize}
\item we assume that the neutrino flavour structure of the interactions
  is independent of the charged fermion properties;

\item we further assume that the chiral structure of the charged
  fermion vertex is the same for all fermion types.
\end{itemize}
Under these hypotheses, we can factorize $\Eps_{\alpha\beta}^{f,P}$ as
the product of three terms:
\begin{equation}
  \label{eq:eps-fact}
  \Eps_{\alpha\beta}^{f,P}
  \equiv \Eps_{\alpha\beta} \, \xi^f \chi^P
\end{equation}
where the matrix $\Eps_{\alpha\beta}$ describes the dependence on the
neutrino flavour, the coefficients $\xi^f$ parametrize the coupling to
the charged fermions, and the terms $\chi^P$ account for the chiral
structure of such couplings, normalized so that
$\Eps_{\alpha\beta}^{f,L}$ ($\Eps_{\alpha\beta}^{f,R}$) corresponds to
$\chi^L = 1/2$ and $\chi^R = 0$ ($\chi^R = 1/2$ and $\chi^L = 0$).
With these assumptions the Lagrangian in Eq.~\eqref{eq:nsi-nc} takes
the form:
\begin{equation}
  \mathcal{L}_\text{NSI,NC} = -2\sqrt{2} G_F
  \bigg[ \sum_{\alpha,\beta} \Eps_{\alpha\beta}
  (\bar\nu_\alpha\gamma^\mu P_L\nu_\beta) \bigg]
  \bigg[ \sum_{f,P} \xi^f \chi^P (\bar f\gamma_\mu P f) \bigg] \,.
\end{equation}
Concerning the chiral structure of the charged fermion vertex, in this
work we will consider either vector couplings ($\chi^L = \chi^R =
1/2$), or axial-vector couplings ($\chi^L = -\chi^R = 1/2)$.  The
corresponding combinations of NSI coefficients are given in
Eq.~\eqref{eq:vect-axial}.

For what concerns the dependence of the NSI on the charged fermion
type, we notice that ordinary matter is composed of electrons ($e$),
up quarks ($u$) and down quarks ($d$), so that only the coefficients
$\xi^e$, $\xi^u$ and $\xi^d$ are experimentally accessible.  For
vector NSI, since quarks are always confined inside protons ($p$) and
neutrons ($n$), we may define (see, \textit{e.g.},
Ref.~\cite{Breso-Pla:2023tnz}):
\begin{equation}
  \label{eq:xi-nucleon}
  \xi^p = 2\xi^u + \xi^d \,,
  \qquad
  \xi^n = 2\xi^d + \xi^u
\end{equation}
so that $\Eps_{\alpha\beta}^{p,V} \equiv 2\Eps_{\alpha\beta}^{u,V} +
\Eps_{\alpha\beta}^{d,V} = \Eps_{\alpha\beta}\, \xi^p\, (\chi^L +
\chi^R)$ and $\Eps_{\alpha\beta}^{n,V} \equiv
2\Eps_{\alpha\beta}^{d,V} + \Eps_{\alpha\beta}^{u,V} =
\Eps_{\alpha\beta}\, \xi^n\, (\chi^L + \chi^R)$.
For axial-vector NSI, the correspondence between quark NSI and nucleon
NSI is not that obvious: for example, for non-relativistic nucleons an
axial-vector hadronic current would induce a change in the spin of the
nucleon.  In any case, it is clear that a simultaneous re-scaling of all
$\{\xi^f\}$ by a common factor can be reabsorbed into a re-scaling of
$\Eps_{\alpha\beta}$, so that only the direction of $(\xi^e, \xi^u,
\xi^d)$ --~or, equivalently, $(\xi^e, \xi^p, \xi^n)$~-- is
phenomenologically non-trivial.  We parametrize such direction using
spherical coordinates, in terms of a ``latitude'' angle $\eta$ and a
``longitude'' angle $\zeta$, as an extension of the framework and
notation introduced in Ref.~\cite{Esteban:2018ppq} (see also
Ref.~\cite{Amaral:2023tbs}).  Concretely, we define:
\begin{equation}
  \label{eq:xi-eta}
  \xi^e = \sqrt{5} \cos\eta \sin\zeta \,,
  \qquad
  \xi^p = \sqrt{5} \cos\eta \cos\zeta \,,
  \qquad
  \xi^n = \sqrt{5} \sin\eta
\end{equation}
or, in terms of the ``quark'' couplings:
\begin{equation}
  \label{eq:xi-eta-quark}
  \xi^u = \frac{\sqrt{5}}{3} (2 \cos\eta \cos\zeta - \sin\eta) \,,
  \qquad
  \xi^d = \frac{\sqrt{5}}{3} (2 \sin\eta - \cos\eta \cos\zeta)
\end{equation}
Using this parametrization, the case of NSI with quarks analyzed in
Ref.~\cite{Esteban:2018ppq} correspond to the ``prime meridian''
$\zeta = 0$, with the pure up-quark and pure down-quark cases located
at $\eta = \arctan(1/2) \approx 26.6^\circ$ and $\eta = \arctan(2)
\approx 63.4^\circ$, respectively.  The ``poles'' ($\eta = \pm
90^\circ$) correspond to NSI with neutrons, while NSI with protons
lies on the ``equator'' ($\eta = \zeta = 0$).  Finally, the pure
electron case is also on the equator, at a right angle ($\zeta = \pm
90^\circ$) from the pure proton case.  Notice that an overall sign
flip of $(\xi^e, \xi^u, \xi^d)$ is just a special case of re-scaling
and produces no observable effect, hence it is sufficient to restrict
both $\eta$ and $\zeta$ to the $[-90^\circ, +90^\circ]$ range.

The presence of vector NC NSI will affect both neutrino propagation in
matter and neutrino scattering in the detector, while axial-vector NC
NSI only affect some of the interaction cross sections.  Both
propagation and interaction effects lead to a modification of the
expected number of events which can be described by the generic
expression~\cite{Coloma:2022umy}:
\begin{equation}
  \label{eq:ES-dens}
  N_\text{ev} \propto
  \Tr\Big[ \rho^\text{det}\, \sigma^\text{det} \Big]
\end{equation}
where $\rho^\text{det}$ is the density matrix characterizing the
flavour state of the neutrinos reaching the detector, while the
\emph{generalized} cross section $\sigma^\text{det}$ is a matrix in
flavour space containing enough information to describe the
interaction of \emph{any} neutrino configuration.  The form of
Eq.~\eqref{eq:ES-dens} is manifestly basis-independent and permits a
separate description of propagation effects (encoded into
$\rho^\text{det}$) and of the scattering process (contained in
$\sigma^\text{det}$), while at the same time properly taking into
account possible interference between them.  Notice that both
$\rho^\text{det}$ and $\sigma^\text{det}$ are hermitian matrices,
which ensures that $N_\text{ev}$ is real.  Actually,
Eq.~\eqref{eq:ES-dens} is invariant under the joint transformation:
\begin{equation}
  \label{eq:conjugate}
  \rho^\text{det} \to \big[ \rho^\text{det} \big]^*
  \quad\text{and}\quad
  \sigma^\text{det} \to \big[ \sigma^\text{det} \big]^*.
\end{equation}
whose implications will be discussed later on in this section.

\subsection{Neutrino oscillations in the presence of NSI}
\label{sec:formOSC}

In general, the evolution of the neutrino and antineutrino flavour
state during propagation is governed by the Hamiltonian:
\begin{equation}
  H^\nu = H_\text{vac} + H_\text{mat}
  \quad\text{and}\quad
  H^{\bar\nu} = ( H_\text{vac} - H_\text{mat} )^* \,,
\end{equation}
where $H_\text{vac}$ is the vacuum part which in the flavour basis
has been defined in Chapter \ref{chap:theo}.Here
$U_\text{vac}=O\cdot U_{12}$ denotes the three-lepton mixing
matrix in vacuum, as defined in Eq. \ref{eq:OU12matrix}. The advantage
of defining $U_\text{vac}$ in this way rather than as in Eq.~\eqref{eq:ULEP2} is that the
transformation $H_\text{vac} \to -H_\text{vac}^*$, whose relevance for
the present work will be discussed below, can be implemented exactly
(up to an irrelevant multiple of the identity) by the following
transformation of the parameters:
\begin{equation}
  \label{eq:osc-deg}
  \begin{aligned}
    \Dmq_{31} &\to -\Dmq_{31} + \Dmq_{21} = -\Dmq_{32} \,,
    \\
    \theta_{12} & \to \pi/2 - \theta_{12} \,,
    \\
    \delta_\text{CP} &\to \pi - \delta_\text{CP}
  \end{aligned}
\end{equation}
which does not spoil the commonly assumed restrictions on the range of
the vacuum parameters ($\Dmq_{21} > 0$ and $0 \leq \theta_{ij} \leq
\pi/2$).

Concerning the matter part $H_\text{mat}$ of the Hamiltonian which
governs neutrino oscillations, if all possible operators in
Eq.~\eqref{eq:nsi-nc} are added to the SM Lagrangian we get:
\begin{equation}
  \label{eq:Hmat}
  H_\text{mat} = \sqrt{2} G_F N_e(x)
  \begin{pmatrix}
    1+\Epx_{ee}(x) & \Epx_{e\mu}(x) & \Epx_{e\tau}(x) \\
    \Epx_{e\mu}^*(x) & \Epx_{\mu\mu}(x) & \Epx_{\mu\tau}(x) \\
    \Epx_{e\tau}^*(x) & \Epx_{\mu\tau}^*(x) & \Epx_{\tau\tau}(x)
  \end{pmatrix}
\end{equation}
where the ``$+1$'' term in the $ee$ entry accounts for the standard
contribution, and
\begin{equation}
  \label{eq:epx-nsi}
  \Epx_{\alpha\beta}(x) = \sum_{f=e,u,d}
  \frac{N_f(x)}{N_e(x)} \Eps_{\alpha\beta}^{f,V}
\end{equation}
describes the non-standard part.  Here $N_f(x)$ is the number density
of fermion $f$ as a function of the distance traveled by the neutrino
along its trajectory.  In Eq.~\eqref{eq:epx-nsi} we have limited the
sum to the charged fermions present in ordinary matter, $f=e,u,d$.
Taking into account that $N_u(x) = 2N_p(x) + N_n(x)$ and $N_d(x) =
N_p(x) + 2N_n(x)$, and also that matter neutrality implies $N_p(x) =
N_e(x)$, Eq.~\eqref{eq:epx-nsi} becomes:
\begin{equation}
  \label{eq:epx-nuc}
  \Epx_{\alpha\beta}(x) =
  \big( \Eps_{\alpha\beta}^{e,V} + \Eps_{\alpha\beta}^{p,V} \big)
  + Y_n(x)\, \Eps_{\alpha\beta}^{n,V}
  \quad\text{with}\quad
  Y_n(x) \equiv \frac{N_n(x)}{N_e(x)}
\end{equation}
which shows that, from the phenomenological point of view, the
propagation effects of NSI with electrons can be mimicked by NSI with
quarks by means of a suitable combination of up-quark and down-quark
contributions.

Since the matter potential can be determined by oscillation
experiments only up to an overall multiple of the identity, each
$\Eps_{\alpha\beta}^{f,V}$ matrix introduces 8 new parameters: two
differences of the three diagonal real parameters (\textit{e.g.},
$\Eps_{ee}^{f,V} - \Eps_{\mu\mu}^{f,V}$ and $\Eps_{\tau\tau}^{f,V} -
\Eps_{\mu\mu}^{f,V}$) and three off-diagonal complex parameters
(\textit{i.e.}, three additional moduli and three complex phases).
Under the assumption that the neutrino flavour structure of the
interactions is independent of the charged fermion type, as described
in Eq.~\eqref{eq:eps-fact}, we get:
\begin{equation}
  \label{eq:epx-eta}
  \begin{aligned}
    \Epx_{\alpha\beta}(x)
    &= \Eps_{\alpha\beta} \big[ \xi^e + \xi^p + Y_n(x) \xi^n \big]
    \big( \chi^L + \chi^R \big)
    \\
    &= \sqrt{5} \, \big[\! \cos\eta\, (\cos\zeta + \sin\zeta)
      + Y_n(x) \sin\eta \big] \big( \chi^L + \chi^R \big)\,
    \Eps_{\alpha\beta}
  \end{aligned}
\end{equation}
so that the phenomenological framework adopted here is characterized
by 10 matter parameters: eight related to the matrix
$\Eps_{\alpha\beta}$ plus two directions $(\eta, \zeta)$ in the
$(\xi^e, \xi^p, \xi^n)$ space.  Notice, however, that the dependence
on $\zeta$ in Eq.~\eqref{eq:epx-eta} can be reabsorbed into a
re-scaling of $\Eps_{\alpha\beta}$ by introducing a new effective angle
$\eta^\prime$:
\begin{multline}
  \label{eq:epx-etapr}
  \Epx_{\alpha\beta}(x)
  = \sqrt{5} \, \big[\! \cos\eta^\prime
    + Y_n(x) \sin\eta^\prime \big] \big( \chi^L + \chi^R \big)\,
  \Eps_{\alpha\beta}^\prime
  \\
  \text{with}\quad
  \tan\eta^\prime
  \equiv \tan\eta \mathbin{\big/} (\cos\zeta + \sin\zeta)
  \quad\text{and}\quad
  \Eps_{\alpha\beta}^\prime
  \equiv \Eps_{\alpha\beta} \sqrt{1 + \cos^2\eta \sin(2\zeta)} \,.
\end{multline}
This is a consequence of the fact that electron and proton NSI always
appear together in propagation, as explained after
Eq.~\eqref{eq:epx-nuc}.  Indeed, $\eta^\prime$ is just a practical way
to express the direction in the $(\xi^e + \xi^p,\, \xi^n)$ plane.  For
$\zeta = 0$, which is the case studied in Ref.~\cite{Esteban:2018ppq},
one trivially recovers $\eta^\prime = \eta$ and
$\Eps_{\alpha\beta}^\prime = \Eps_{\alpha\beta}$.
Furthermore, for $\zeta = -45^\circ$ one gets $\xi^e + \xi^p = 0$, so
that NSI effects in oscillations depend solely on the neutron coupling
$\xi^n$.
This would be the case, for example, in models where the $Z'$ does not
couple directly to matter fermions, and NSI are generated through
$Z-Z'$ mass mixing (see the related discussion in
Ref.~\cite{Heeck:2018nzc}).  In fact, in this case, the specific value
of $\eta$ becomes irrelevant as long as it is different from zero
(since Eq.~\eqref{eq:epx-etapr} always yields $\eta^\prime = \pm
90^\circ$ in this case), while for the special point where $\eta = 0$
and $\zeta = -45^\circ$ NSI completely cancel from oscillations.  It
should be stressed, however, that this only applies to
\emph{oscillations}: the implications of NSI for neutrino scattering
described in Sec.~\ref{sec:formCS} will still depend non-trivially on
$\eta$.

\subsubsection*{LMA-D solution}

An important piece of the discussion of NSI is about the LMA-Dark solution. Neutrino transition probabilities remain invariant --~and more generically the density
matrix $\rho^\text{det}$ undergoes complex conjugation, as described
in Eq.~\eqref{eq:conjugate}~-- if the Hamiltonian $H^\nu =
H_\text{vac} + H_\text{mat}$ is transformed as $H^\nu \to -(H^\nu)^*$.
This requires a simultaneous transformation of both the vacuum and the
matter terms.  The transformation of $H_\text{vac}$ is described in
Eq.~\eqref{eq:osc-deg} and involves a change in the octant of
$\theta_{12}$,  as well as a change in the neutrino mass ordering (\textit{i.e.}, the
sign of $\Dmq_{31}$), which is why it has been called ``generalized
mass-ordering degeneracy'' in Ref.~\cite{Coloma:2016gei}.  As for
$H_\text{mat}$ we need:
\begin{equation}
  \label{eq:NSI-deg}
  \begin{aligned}
    \big[ \Epx_{ee}(x) - \Epx_{\mu\mu}(x) \big]
    &\to - \big[ \Epx_{ee}(x) - \Epx_{\mu\mu}(x) \big] - 2 \,,
    \\
    \big[ \Epx_{\tau\tau}(x) - \Epx_{\mu\mu}(x) \big]
    &\to -\big[ \Epx_{\tau\tau}(x) - \Epx_{\mu\mu}(x) \big] \,,
    \\
    \Epx_{\alpha\beta}(x)
    &\to - \Epx_{\alpha\beta}^*(x) \qquad (\alpha \neq \beta) \,,
  \end{aligned}
\end{equation}
see Refs.~\cite{Gonzalez-Garcia:2013usa, Bakhti:2014pva,
  Coloma:2016gei}.  As seen in Eqs.~\eqref{eq:epx-nsi},
\eqref{eq:epx-nuc} and~\eqref{eq:epx-eta} the matrix
$\Epx_{\alpha\beta}(x)$ depends on the chemical composition of the
medium, which may vary along the neutrino trajectory, so that in
general the condition in Eq.~\eqref{eq:NSI-deg} is fulfilled only in
an approximate way.  The degeneracy becomes exact in the following two
cases:\footnote{Strictly speaking, Eq.~\eqref{eq:NSI-deg} can be
satisfied exactly for \emph{any} matter chemical profile $Y_n(x)$ if
$\Eps_{\alpha\beta}^{n,V}$ and $\Eps_{\alpha\beta}^{e,V} +
\Eps_{\alpha\beta}^{p,V}$ are allowed to transform independently of
each other.  This possibility, however, is incompatible with the
factorization constraint of Eq.~\eqref{eq:eps-fact}, so it will not be
discussed here.}
\begin{itemize}
\item if the effective NSI coupling to neutrons vanishes, so that
  $\Eps_{\alpha\beta}^{n,V} = 0$ in Eq.~\eqref{eq:epx-nuc}.  In terms
  of fundamental quantities this occurs when $\Eps_{\alpha\beta}^{u,V}
  = -2 \Eps_{\alpha\beta}^{d,V}$, \textit{i.e.}, the NSI couplings are
  proportional to the electric charge of quarks.  In our
  parametrization this corresponds to the ``equator'' $\eta=0$ for
  arbitrary $\zeta$, as shown in Eq.~\eqref{eq:epx-eta};

\item if the neutron/proton ratio $Y_n(x)$ is constant along the
  entire neutrino propagation path.  This is certainly the case for
  reactor and long-baseline experiments, where only the Earth's mantle
  is involved, and to a good approximation also for atmospheric
  neutrinos, since the differences in chemical composition between
  mantle and core can safely be neglected in the context of
  NSI~\cite{GonzalezGarcia:2011my}.  In this case the matrix
  $\Epx_{\alpha\beta}(x)$ becomes independent of $x$ and can be
  regarded as a new phenomenological parameter, as we will describe in
  Sec.~\ref{sec:formalism-earth}.
\end{itemize}

Further details on the implications of this degeneracy for different
classes of neutrino experiments (solar, atmospheric, \textit{etc.}) is
provided below in the corresponding section.

\subsubsection{Matter potential in atmospheric and long-baseline neutrinos}
\label{sec:formalism-earth}

As discussed in Ref.~\cite{GonzalezGarcia:2011my}, in the Earth the
neutron/proton ratio $Y_n(x)$ which characterizes the matter chemical
composition can be taken to be constant to very good approximation.
The PREM model~\cite{Dziewonski:1981xy} fixes $Y_n = 1.012$ in the
Mantle and $Y_n = 1.137$ in the Core, with an average value
$Y_n^\oplus = 1.051$ all over the Earth.  Setting therefore $Y_n(x)
\equiv Y_n^\oplus$ in Eqs.~\eqref{eq:epx-nsi} and~\eqref{eq:epx-nuc}
we get $\Epx_{\alpha\beta}(x) \equiv \Eps_{\alpha\beta}^\oplus$ with:
\begin{equation}
  \label{eq:eps-earth0}
  \Eps_{\alpha\beta}^\oplus
  = \Eps_{\alpha\beta}^{e,V} + \big( 2 + Y_n^\oplus \big) \Eps_{\alpha\beta}^{u,V}
  + \big( 1 + 2Y_n^\oplus \big) \Eps_{\alpha\beta}^{d,V}
  = \big( \Eps_{\alpha\beta}^{e,V} + \Eps_{\alpha\beta}^{p,V} \big)
  + Y_n^\oplus \Eps_{\alpha\beta}^{n,V} \,.
\end{equation}
If we impose quark-lepton factorization as in Eq.~\eqref{eq:epx-eta}
we get:
\begin{equation}
  \label{eq:eps-earth}
  \begin{aligned}
    \Eps_{\alpha\beta}^\oplus
    &= \sqrt{5} \, \big[\! \cos\eta\, (\cos\zeta + \sin\zeta)
      + Y_n^\oplus \sin\eta \big] \big( \chi^L + \chi^R \big)\,
    \Eps_{\alpha\beta}
    \\
    &= \sqrt{5} \, \big[\! \cos\eta^\prime
      + Y_n^\oplus \sin\eta^\prime \big]
    \big( \chi^L + \chi^R \big)\, \Eps_{\alpha\beta}^\prime \,.
  \end{aligned}
\end{equation}
In other words, within this approximation the analysis of atmospheric
and LBL neutrinos holds for any combination of NSI with up, down or
electrons and it can be performed in terms of the effective NSI
couplings $\Eps_{\alpha\beta}^\oplus$, which play the role of
phenomenological parameters.  In particular, the best-fit value and
allowed ranges of $\Eps_{\alpha\beta}^\oplus$ are independent of
$\eta$ and $\zeta$, while the bounds on $\Eps_{\alpha\beta}$ simply
scale as $[\cos\eta\, (\cos\zeta + \sin\zeta) + Y_n^\oplus \sin\eta]$.
Moreover, it is immediate to see that for $\eta^\prime =
\arctan(-1/Y_n^\oplus) \approx -43.6^\circ$, with $\eta^\prime$
defined in Eq.~\eqref{eq:epx-etapr}, the contribution of NSI to the
matter potential vanishes, so that no bound on $\Eps_{\alpha\beta}$
can be derived from atmospheric and LBL data in such case.
This would be approximately the case, for example, for $U(1)'$ models
associated to the combination $B - 2L_e + \alpha L_\mu - \beta L_\tau$
and, consequently, oscillation bounds are significantly relaxed for
this type of models~\cite{Greljo:2022dwn}.

Following the approach of Ref.~\cite{GonzalezGarcia:2011my}, the
matter Hamiltonian $H_\text{mat}$, given in Eq.~\eqref{eq:Hmat} after
setting $\Epx_{\alpha\beta}(x) \equiv \Eps_{\alpha\beta}^\oplus$, can
be parametrized in a way that mimics the structure of the vacuum
term:
\begin{equation}
  \label{eq:HmatGen}
  H_\text{mat} = Q_\text{rel} U_\text{mat} D_\text{mat}
  U_\text{mat}^\dagger Q_\text{rel}^\dagger
  \text{~~with~~}
  \left\lbrace
  \begin{aligned}
    Q_\text{rel} &= \diag\left(
    e^{i\alpha_1}, e^{i\alpha_2}, e^{-i\alpha_1 -i\alpha_2} \right),
    \\
    U_\text{mat} &= R_{12}(\varphi_{12}) R_{13}(\varphi_{13})
    \tilde{R}_{23}(\varphi_{23}, \delta_\text{NS}) \,,
    \\
    D_\text{mat} &= \sqrt{2} G_F N_e(x)
    \diag(\Eps_\oplus, \Eps_\oplus^\prime, 0)
  \end{aligned}\right.
\end{equation}
where $R_{ij}(\varphi_{ij})$ is a rotation of angle $\varphi_{ij}$ in
the $ij$ plane and $\tilde{R}_{23}(\varphi_{23},\delta_\text{NS})$ is
a complex rotation by angle $\varphi_{23}$ and phase
$\delta_\text{NS}$.
Note that the two phases $\alpha_1$ and $\alpha_2$ included in
$Q_\text{rel}$ are not a feature of neutrino-matter interactions, but
rather a relative feature of the vacuum and matter terms.
In order to simplify the analysis we impose that two eigenvalues of
$H_\text{mat}$ are equal, $\Eps_\oplus^\prime = 0$.  This assumption
is justified since, as shown in Ref.~\cite{Friedland:2004ah}, in this
case strong cancellations in the oscillation of atmospheric neutrinos
occur, and this is precisely the situation in which the weakest
constraints can be placed.
Setting $\Eps_\oplus^\prime \to 0$ implies that the $\varphi_{23}$
angle and the $\delta_\text{NS}$ phase disappear from neutrino
oscillations, so that the effective NSI couplings
$\Eps_{\alpha\beta}^\oplus$ can be parametrized as:
\begin{equation}
  \label{eq:eps_atm}
  \begin{aligned}
    \Eps_{ee}^\oplus - \Eps_{\mu\mu}^\oplus
    &= \hphantom{-} \Eps_\oplus \, (\cos^2\varphi_{12} - \sin^2\varphi_{12})
    \cos^2\varphi_{13} - 1\,,
    \\
    \Eps_{\tau\tau}^\oplus - \Eps_{\mu\mu}^\oplus
    &= \hphantom{-} \Eps_\oplus \, (\sin^2\varphi_{13}
    - \sin^2\varphi_{12} \, \cos^2\varphi_{13}) \,,
    \\
    \Eps_{e\mu}^\oplus
    &= -\Eps_\oplus \, \cos\varphi_{12} \, \sin\varphi_{12} \,
    \cos^2\varphi_{13} \, e^{i(\alpha_1 - \alpha_2)} \,,
    \\
    \Eps_{e\tau}^\oplus
    &= -\Eps_\oplus \, \cos\varphi_{12} \, \cos\varphi_{13} \,
    \sin\varphi_{13} \, e^{i(2\alpha_1 + \alpha_2)} \,,
    \\
    \Eps_{\mu\tau}^\oplus
    &= \hphantom{-} \Eps_\oplus \, \sin\varphi_{12} \, \cos\varphi_{13} \,
    \sin\varphi_{13} \, e^{i(\alpha_1 + 2\alpha_2)} \,.
  \end{aligned}
\end{equation}
As further simplification, in order to keep the fit manageable we
assume real NSI, which we implement by choosing $\alpha_1 = \alpha_2 =
0$ and $-\pi/2 \leq \varphi_{ij} \leq \pi/2$, and also restrict
$\delta_\text{CP} \in \{0, \pi\}$.  It is important to note that with
these approximations the formalism for atmospheric and long-baseline
data is CP-conserving; we will go back to this point when discussing
the experimental results included in the fit.  In addition to
atmospheric and LBL experiments, important information on neutrino
oscillation parameters is provided also by reactor experiments with a
baseline of about 1~km.  Due to the very small amount of matter
crossed, both standard and non-standard matter effects are completely
irrelevant for these experiments, so that neutrino propagation depends
only on the vacuum parameters.

\subsubsection{Matter potential for solar and KamLAND neutrinos}
\label{sec:formalism-solar}

For the study of propagation of solar and KamLAND neutrinos one can
work in the one mass dominance approximation, $\Dmq_{31} \to \infty$
(the same approach of Chapter \ref{chap:theo}).  In this limit the
neutrino evolution can be calculated in an effective $2\times 2$ model
described by the Hamiltonian $H_\text{eff} = H_\text{vac}^\text{eff} +
H_\text{mat}^\text{eff}$ (similarly to Eqs. \ref{eq:HvacSol} and \ref{eq:HmatSol} ), with:

\begin{align}
  \label{eq:HvacSol_0}
  H_\text{vac}^\text{eff}
  &= \frac{\Dmq_{21}}{4 E_\nu}
  \begin{pmatrix}
    -\cos2\theta_{12} \, \hphantom{e^{-i\delta_\text{CP}}}
    & ~\sin2\theta_{12} \, e^{i\delta_\text{CP}}
    \\
    \hphantom{-}\sin2\theta_{12} \, e^{-i\delta_\text{CP}}
    & ~\cos2\theta_{12} \, \hphantom{e^{i\delta_\text{CP}}}
  \end{pmatrix},
  \\
  \label{eq:HmatSol_2}
  H_\text{mat}^\text{eff}
  &= \sqrt{2} G_F N_e(x)
  \left[
    \frac{c_{13}^2}{2}
    \begin{pmatrix}
      1 & \hphantom{-}0 \\
      0 & -1
    \end{pmatrix}
    + \big[ \xi^e + \xi^p + Y_n(x) \xi^n \big]
    \big( \chi^L + \chi^R \big)\!
    \begin{pmatrix}
      -\Eps_D^{\hphantom{*}} & \Eps_N \\
      \hphantom{+} \Eps_N^* & \Eps_D
    \end{pmatrix}
    \right],
\end{align}
where we have imposed the quark-lepton factorization of
Eq.~\eqref{eq:epx-eta}.  The coefficients $\Eps_D$ and
$\Eps_N$ are related to the original parameters $\Eps_{\alpha\beta}$
by the following relations:
\begin{align}
  \label{eq:eps_D}
  \begin{split}
    \Eps_D
    &= c_{13} s_{13}\, \Re\!\big( s_{23} \, \Eps_{e\mu}
    + c_{23} \, \Eps_{e\tau} \big)
    - \big( 1 + s_{13}^2 \big)\, c_{23} s_{23}\,
    \Re\!\big( \Eps_{\mu\tau} \big)
    \\
    & \hphantom{={}}
    -\frac{c_{13}^2}{2} \big( \Eps_{ee} - \Eps_{\mu\mu} \big)
    + \frac{s_{23}^2 - s_{13}^2 c_{23}^2}{2}
    \big( \Eps_{\tau\tau} - \Eps_{\mu\mu} \big) \,,
  \end{split}
  \\[2mm]
  \label{eq:eps_N}
  \Eps_N &=
  c_{13} \big( c_{23} \, \Eps_{e\mu} - s_{23} \, \Eps_{e\tau} \big)
  + s_{13} \left[
    s_{23}^2 \, \Eps_{\mu\tau} - c_{23}^2 \, \Eps_{\mu\tau}^*
    + c_{23} s_{23} \big( \Eps_{\tau\tau} - \Eps_{\mu\mu} \big)
    \right].
\end{align}
Denoting by $S_\text{eff}$ the $2\times 2$ unitary matrix obtained
integrating $H_\text{eff}$ along the neutrino trajectory, the full
density matrix $\rho^\text{det}$ introduced in Eq.~\eqref{eq:ES-dens}
can be written as:
\begin{equation}
  \rho_{\alpha\beta}^\text{det} = c_{13}^2 \big[
    A_{\alpha\beta} P_\text{osc}
    + B_{\alpha\beta} P_\text{int}
    + i C_{\alpha\beta} P_\text{ext} \big]
  + D_{\alpha\beta}
\end{equation}
where the effective probabilities $P_\text{osc}$, $P_\text{int}$ and
$P_\text{ext}$ are given by
\begin{equation}
  P_\text{osc} \equiv |S_{21}^\text{eff}|^2 \,,
  \qquad
  P_\text{int} \equiv \Re\big(
  S_{11}^\text{eff} S_{21}^{\text{eff}\,*} \big) \,,
  \qquad
  P_\text{ext} \equiv \Im\big(
  S_{11}^\text{eff} S_{21}^{\text{eff}\,*} \big) \,,
\end{equation}
and the numerical coefficients $A_{\alpha\beta}$, $B_{\alpha\beta}$,
$C_{\alpha\beta}$ and $D_{\alpha\beta}$ are defined as
\begin{equation}
  \begin{aligned}
    A_{\alpha\beta} &\equiv
    O_{\alpha 2} O_{\beta 2} - O_{\alpha 1} O_{\beta 1} \,,
    &\qquad
    B_{\alpha\beta} &\equiv
    O_{\alpha 1} O_{\beta 2} + O_{\alpha 2} O_{\beta 1} \,,
    \\
    C_{\alpha\beta} &\equiv
    O_{\alpha 1} O_{\beta 2} - O_{\alpha 2} O_{\beta 1} \,,
    &\qquad
    D_{\alpha\beta} &\equiv
    \sum_{i=\text{all}} O_{\alpha i} O_{\beta i} |O_{ei}|^2 \,.
  \end{aligned}
\end{equation}
with $O = R_{23}(\theta_{23}) R_{13}(\theta_{13})$.  Unlike in
Ref.~\cite{Esteban:2018ppq} where NSI did \emph{not} affect the
scattering process and only the $\rho_{ee}^\text{det}$ entry (which
depends exclusively on $P_\text{osc}$) was required, a rephasing of
$S_\text{eff}$ now produces visible consequences as it affects
$P_\text{int}$ and $P_\text{ext}$.  Moreover, $\nu_\mu$ and $\nu_\tau$
are no longer indistinguishable as their scattering amplitude may be
different under NSI, so that the $\theta_{23}$ angle acquires
relevance.  Hence, for each fixed value of $\eta$ and $\zeta$ the
density matrix $\rho^\text{det}$ for solar and KamLAND neutrinos
depends effectively on eight quantities: the four real oscillation
parameters $\theta_{12}$, $\theta_{13}$, $\theta_{23}$ and
$\Dmq_{21}$, the real $\Eps_D$ and complex $\Eps_N$ matter parameters,
and the CP phase $\delta_\text{CP}$.
As stated in Sec.~\ref{sec:formalism-earth} in this work we will
assume real NSI, implemented here by setting $\delta_\text{CP} \in
\{0, \pi\}$ and considering only real values for $\Eps_N$.

Unlike in the Earth, the matter chemical composition of the Sun varies
substantially along the neutrino trajectory, and consequently the
potential depends non-trivially on the specific combinations of
$(\xi^e + \xi^p)$ and $\xi^n$ couplings --~\textit{i.e.}, on the value
of $\eta^\prime$ as determined by the $(\eta, \zeta)$ parameters.
This implies that the generalized mass-ordering degeneracy is not
exact, except for $\eta = 0$ (in which case the NSI potential is
proportional to the standard MSW potential and an exact inversion of
the matter sign is possible).  However the transformation described in
Eqs.~\eqref{eq:osc-deg} and~\eqref{eq:NSI-deg} still results in a good
fit to the global analysis of oscillation data for a wide range of
values of $\eta^\prime$, and non-oscillation data are needed to break
this degeneracy~\cite{Coloma:2016gei}.

\subsubsection{Departures from adiabaticity in presence of NSI}
\label{sec:formADIA}

When computing neutrino evolution in the Sun, it is often assumed that
it takes place in the adiabatic regime, as this considerably
simplifies the calculation, as discussed in Chapter \ref{chap:theo}.  While this is the case for neutrino
oscillations within the LMA solution with a standard matter potential,
it is worthwhile asking if the inclusion of NP effects (such as NSI)
can lead to non-adiabatic transitions.  If this were the case, it
would require special care and may lead to interesting new
phenomenological consequences.  This possibility has been largely
overlooked in the literature, where most studies of NSI in the Sun
assume adiabatic transitions.

Let us consider the two-neutrino case, with a matter potential that
depends on the position $x$ along the neutrino path inside the Sun.
In the instantaneous mass basis, the Hamiltonian can be written as:
\begin{equation}
  \label{eq:instant-mass}
  i \frac{\dd}{\dd x}
  \begin{pmatrix}
    \tilde \nu_1 \\ \tilde \nu_2
  \end{pmatrix}
  = \begin{pmatrix}
    -\Delta_m(x) \enspace & -i \theta_m^\prime(x)
    \\
    i \theta_m^\prime(x) & \Delta_m(x)
  \end{pmatrix}
  \begin{pmatrix}
    \tilde \nu_1 \\ \tilde \nu_2
  \end{pmatrix}
\end{equation}
where $\theta_m(x)$ and $\Delta_m(x)$ refer to the mixing angle and
oscillation frequency in the presence of matter effects, and
$\theta^\prime(x) \equiv \dd\theta(x) \big/ \dd x$.  In this basis,
transitions between $\tilde \nu_1 \leftrightarrow \tilde \nu_2$ are
negligible as long as the off-diagonal terms $\theta_m^\prime(x)$ are
small compared to the diagonal entries $\Delta_m(x)$ in
Eq.~\eqref{eq:instant-mass}.  This leads to the adiabaticity
condition discussed in Eq.~\eqref{eq:adiabaticity}.
Neutrino transitions will be adiabatic if this condition is satisfied
along all points in the neutrino trajectory.  Of course, this argument
is general and may be applied both in the standard case and in the
presence of NP.  In the present work neutrino propagation is described
by the effective $2\times 2$ Hamiltionian $H_\text{eff}$ introduced in
Eqs.~\eqref{eq:HvacSol_0} and~\eqref{eq:HmatSol_2}.  Assuming
$\delta_\text{CP} = 0$ and real NSI, so that $H_\text{eff}$ is
traceless and real with $H_{11}^\text{eff} = -H_{22}^\text{eff}$ and
$H_{12}^\text{eff} = H_{21}^\text{eff}$, we can write:
\begin{equation}
  \theta_m(x)
  \equiv \frac{1}{2} \arctan\!\big[
    H_{12}^\text{eff}(x) \mathbin{\big/} H_{22}^\text{eff}(x) \big]
  \quad\text{and}\quad
  \Delta_m(x)
  \equiv \sqrt{\big[ H_{12}^\text{eff}(x) \big]^2
    + \big[ H_{22}^\text{eff}(x) \big]^2} \,.
\end{equation}
In the presence of NSI, for a given matter density and neutrino
energy, it is possible to choose $\Eps_D$ and $\Eps_N$ so that the
contribution from NP cancels the standard one, resulting in
$\Delta_m(x) \to 0$.  It is easy to show analytically that, for such
values, the adiabatic condition in Eq.~\eqref{eq:adiabaticity} is no
longer satisfied.  Specifically, such cancellation takes place when:
\begin{equation}
  \label{eq:cancel}
  \begin{aligned}
    \big[ \xi^e + \xi^p + Y_n(x) \xi^n \big]
    \big(\chi^L + \chi^R \big)\, \Eps_D
    &\to -\frac{\Dmq_{12} \cos 2\theta_{12}}{4 E_\nu V(x)}
    + \frac{c_{13}^2}{2} \,,
    \\
    \big[ \xi^e + \xi^p + Y_n(x) \xi^n \big]
    \big(\chi^L + \chi^R \big)\, \Eps_N
    &\to -\frac{\Dmq_{12} \sin 2\theta_{12}}{4 E_\nu V(x)} \,,
  \end{aligned}
\end{equation}
where $V(x) \equiv \sqrt{2} G_F N_e(x)$ is the SM matter potential.
In other words, for a neutrino with a given energy $E_\nu$ at some
point $x$ along the trajectory, it is possible to find a pair of
values $(\Eps_D, \Eps_N)$ for which transitions are no longer
adiabatic.
Note that the cancellation condition for $\Eps_D$ depends on
$\cos2\theta_{12}$ and therefore will take different values for the
LMA and LMA-D regions, while the corresponding value for $\Eps_N$ will
remain invariant under a change of octant for $\theta_{12}$.  Also, as
the cancellation conditions in Eqs.~\eqref{eq:cancel} depend on the
values of $\xi^f$, the regions will depend on whether NSI take place
with electrons/protons, neutrons, or a combination of the two.

\begin{figure}[ht!]\centering
  \includegraphics[width=\textwidth]{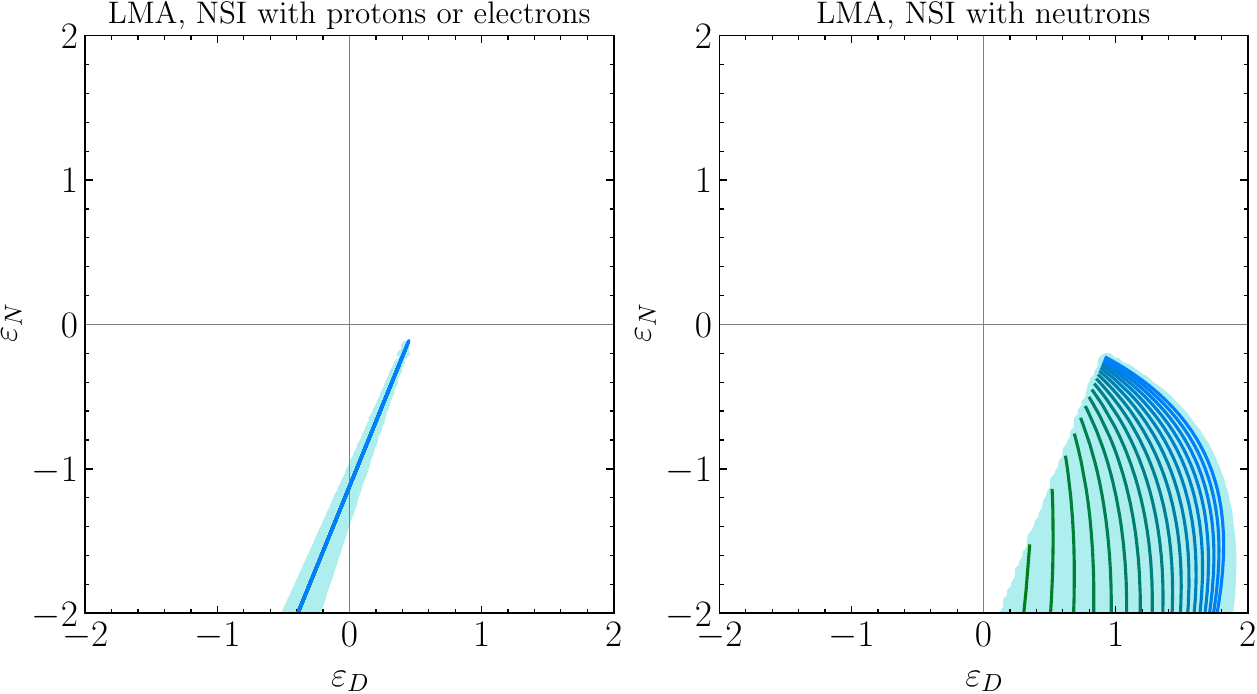}
  \caption{Departure from the adiabatic regime in presence of NSI, for
    a solar mixing angle in the LMA region.  The pale blue regions
    show the values of NSI parameters leading to $\gamma^{-1} > 1$,
    obtained from a numerical computation.  The coloured lines show
    the values of NSI parameters where Eqs.~\eqref{eq:cancel} are
    satisfied at some point along the neutrino trajectory, for a
    neutrino produced at the center of the Sun.  The different lines
    in the right panel correspond to neutrino energies between 1 MeV
    (green lines, at the left edge of the region) and 20~MeV (blue
    lines, towards the right edge).}
  \label{fig:adiab}
\end{figure}

Figure~\ref{fig:adiab} shows the regions where the transitions are not
adiabatic, for NSI with protons or electrons (left panel) and for NSI
with neutrons (right panel).  The shaded pale blue regions show the
results from a numerical computation.  In contrast, the coloured lines
show the points satisfying the analytic conditions in
Eqs.~\eqref{eq:cancel}, for neutrino energies between 1~MeV (green
lines, towards the left edge of the region) and 20~MeV (blue lines,
towards the right edge of the region), for $\theta_{12} = 33^\circ$
and $\Dmq_{12} = 7.5\times 10^{-5}~\eVq$.  As can be seen, the
agreement with the numerical computation is excellent.  Also, note the
very different shape of the regions in the two panels.  The reason
behind this is that for NSI with electrons or protons the dependence
with $E_\nu$ and $x$ comes in Eqs.~\eqref{eq:cancel} through the
product $E_\nu V(x)$, while for NSI with neutrons there is an extra
dependence on $Y_n(x)$.  Therefore, while the regions in the left
panel in Fig.~\ref{fig:adiab} span essentially a straight line, in the
right panel the dependence is more complex.

In our past study~\cite{Esteban:2018ppq}, where only NSI with quarks
were considered, neutrino evolution in the Sun was based on a fully
numerical approach.  In the present work, however, exploring the full
parameter space of the most general NSI with protons, electrons, and
neutrons without assuming adiabaticity is challenging from the
numerical point of view.  To overcome this problem in our calculations
we start by evaluating the adiabaticity index for the point in the NSI
and oscillation parameter space to be surveyed.  If for that point
$\gamma^{-1}(x) < 1$ along the whole neutrino trajectory in the Sun we
use the adiabatic approximation when computing the corresponding
flavour transition probabilities and the subsequent prediction for all
observables and $\chi^2$ value.  If, on the contrary, the adiabaticity
condition is violated somewhere inside the Sun such point is removed
from the parameter space to be surveyed.  We do so because we have
verified that for parameter values for which adiabaticity in the Sun
is violated the predicted observables with the properly computed
flavour transition probability without assuming adiabaticity never
lead to good description of the data.  But we also find that if one
evaluates the probabilities for those parameters wrongly using the
adiabatic approximation, one can find a ``fake'' good fit to the Solar
and KamLAND data, which would lead to wrong conclusions about their
acceptability.  In conclusion, the adiabatic approximation can be
safely used for the purposes of this work as long as one removes from
the parameter space those points for which adiabaticity in the Sun is
violated.  Furthermore, once the data of atmospheric neutrino
experiments is added to the fit, it totally disfavors the parameter
regions where transitions in the Sun are not adiabatic.

\subsection{Neutrino detection cross sections in the presence of NSI}
\label{sec:nuCS_nsi}

Neutrino scattering data is also sensitive to NC NSI as they would
affect the interaction rates directly~\cite{Davidson:2003ha,
  Scholberg:2005qs, Barranco:2005ps, Barranco:2005yy, Barranco:2007ej,
  Bolanos:2008km}.  Therefore the combination of oscillation and
scattering data may be used to break the LMA-D degeneracy (see for
example Refs.~\cite{Miranda:2004nb, Escrihuela:2009up} for early works
on this topic).  

In this
respect, it should be mentioned that all the atmospheric, reactor and
accelerator experiments included in our fit rely on CC processes in
order to detect neutrinos, so the corresponding cross-section is not
affected by the NC-like NSI considered here.\footnote{It should be
noted that a subleading component of NC interactions is present as
background in many of these experiments, so that NSI parameter may in
principle have an impact on the number of events.  During my PhD, I have been studying the effects of NCNSIs using NOvA background of NC events. See more in \cite{Gehrlein:2024vwz}. }
This implies that the $\sigma^\text{det}$ generalized cross-section
matrix entering Eq.~\eqref{eq:ES-dens} is diagonal in the SM flavour
basis, and its non-zero entries coincide with the usual SM
cross-section, that is,
\begin{equation}
  \sigma^\text{det} = \diag(
  \sigma_e^\text{SM},\, \sigma_\mu^\text{SM},\, \sigma_\tau^\text{SM}) \,
  \qquad\text{for CC processes.}
\end{equation}
Some solar neutrino experiments, on the contrary, are sensitive to NC
NSI in some of the detection processes involved.  This is the case of
Borexino and SK (and SNO, albeit with lower sensitivity) which observe
neutrino-electron ES, which may be affected by electron NSI.
Regarding NSI with nuclei SNO can also probe axial-vector NSI in NC
events, and CE$\nu$NS experiments are able to set important
constraints on vector interactions.  In the rest of this subsection we
review the phenomenological implications for these three cases
separately.

\subsubsection{Neutrino-electron elastic scattering}
\label{sec:formES}

The presence of flavour-changing effects in NSI implies that the SM
flavour basis no longer coincides with the interaction eigenstates of
the neutrino-electron scattering.  In such case the generalized cross
section $\sigma^\text{det}$ can be obtained as the integral over the
electron recoil kinetic energy $T_e$ of the following matrix
expression:
\begin{multline}
  \label{eq:nsi-elec}
  \dfrac{\dd\sigma^\text{ES}}{\dd T_e}(E_\nu, T_e)
  = \dfrac{2 G_F^2 m_e}{\pi}
  \bigg\lbrace
  C_L^2 \Big[ 1 + \dfrac{\alpha}{\pi} f_-(y) \Big]
  + C_R^2\, (1-y)^2 \Big[ 1 + \dfrac{\alpha}{\pi}f_+(y) \Big]
  \\
  - \big\{ C_L, C_R \big\}\,
  \dfrac{m_e y}{2E_\nu} \Big[ 1 + \dfrac{\alpha}{\pi} f_\pm(y) \Big]
  \bigg\rbrace
\end{multline}
where $y \equiv T_e / E_\nu$ and $f_+$, $f_-$, $f_\pm$ are loop
functions given in Ref.~\cite{Bahcall:1995mm}, while $\alpha$ stands
for the fine-structure constant and $m_e$ is the electron mass.  In
this formula $C_L$ and $C_R$ are $3\times 3$ hermitian matrices which
incorporate both SM and NSI contributions:
\begin{equation}
\label{eq:CL-CR}
  C_{\alpha\beta}^L
  \equiv c_{L\beta}\, \delta_{\alpha\beta} + \Eps_{\alpha\beta}^{e,L}
  \quad\text{and}\quad
  C_{\alpha\beta}^R
  \equiv c_{R\beta}\, \delta_{\alpha\beta} + \Eps_{\alpha\beta}^{e,R} \,.
\end{equation}
The effective couplings $c_{L\beta}$ and $c_{R\beta}$ account for the
SM part, and contain both the flavour-universal NC terms and the
$\nu_e$-only CC scattering:
\begin{equation}
  \label{eq:cm-coupl}
  \begin{aligned}
    c_{Le}
    &= \rho\,\Big[ \kappa_{e}(T_e)\sin^2\theta_w - \dfrac{1}{2} \Big] + 1 \,,
    &\quad
    c_{Re}
    &= \rho\, \kappa_{e}(T_e)\sin^2\theta_w \,,
    \\
    c_{L\tau} = c_{L\mu}
    &= \rho\, \Big[ \kappa_\mu(T_e)\sin^2\theta_w - \dfrac{1}{2} \Big] \,,
    &\quad
    c_{R\tau} = c_{R\mu}
    &= \rho\, \kappa_\mu(T_e)\sin^2\theta_w \,,
  \end{aligned}
\end{equation}
with $\theta_w$ being the weak mixing angle, and $\rho$ and
$\kappa_\beta(T_e)$ departing from $1$ due to radiative corrections of
the gauge boson self-energies and vertices~\cite{Bahcall:1995mm}.
It is immediate to see that, if the NSI terms
$\Eps_{\alpha\beta}^{e,L}$ and $\Eps_{\alpha\beta}^{e,R}$ are set to
zero, the matrix $\dd\sigma^\text{ES} \big/ \dd T_e$ becomes diagonal.
Imposing the factorization of Eq.~\eqref{eq:eps-fact} for the vector
($+$) and axial-vector ($-$) NSI we get:
\begin{equation}
  \Eps_{\alpha\beta}^{e,V(A)}=
  \Eps_{\alpha\beta}^{e,L}\pm \Eps_{\alpha\beta}^{e,R} =
    \Eps_{\alpha\beta} \, \xi^e\, (\chi^L \pm \chi^R)
    = \sqrt{5} \cos\eta \sin\zeta\,
    (\chi^L \pm \chi^R)\, \Eps_{\alpha\beta}\,.
\end{equation}

Let us finalize this section by discussing briefly the impact that the
inclusion of NSI effects on ES could have on the generalized
mass-ordering degeneracy discussed in Sec.~\ref{sec:formOSC}.  We have
seen that the parameter transformations~\eqref{eq:osc-deg}
and~\eqref{eq:NSI-deg} lead to a complex conjugation of the neutrino
density matrix, $\rho^\text{det} \to [\rho^\text{det}]^*$.  As shown
in Eq.~\eqref{eq:conjugate}, this does not affect the overall number
of events (thus resulting in the appearance of the degeneracy) as long
as it is accompanied by a similar transformation $\sigma^\text{det}
\to [\sigma^\text{det}]^*$.  The latter can be realized either as $C_L
\to C_L^*$ and $C_R \to C_R^*$, which occur when both
$\Eps_{\alpha\beta}^{e,L}$ and $\Eps_{\alpha\beta}^{e,R}$ undergo
simple complex conjugation, or as $C_L \to -C_L^*$ and $C_R \to
-C_R^*$, which require ad-hoc transformations of the diagonal entries
$\Eps_{\alpha\alpha}^{e,L}$ and $\Eps_{\alpha\alpha}^{e,R}$ to
compensate for the SM contribution of Eq.~\eqref{eq:cm-coupl}.  While
conceptually identical to the situation occurring in neutrino
oscillations, where the extra freedom introduced by NSI allows to
``flip the sign'' of the standard matter effects, the specific
transformations required to achieve a perfect symmetry of
$\sigma^\text{det}$ differ from those of Eq.~\eqref{eq:NSI-deg}.  In
principle one may first choose $\Eps_{\alpha\beta}^{e,L}$ and
$\Eps_{\alpha\beta}^{e,R}$ accounting for Eq.~\eqref{eq:cm-coupl} and
then tune $\Eps_{\alpha\beta}^{p,V}$ and $\Eps_{\alpha\beta}^{n,V}$ to
fulfill Eq.~\eqref{eq:NSI-deg}, but this procedure is incompatible
with the factorization constraint of Eq.~\eqref{eq:eps-fact}, which
assumes that all NSI have the same neutrino flavour structure
independently of their chirality and of the charged fermion type.  The
net conclusion is that, for NSI involving electrons (that is,
$\zeta\ne 0$) and assuming that the factorization in
Eq.~\eqref{eq:eps-fact} holds, ES effects \emph{break} the generalized
mass-ordering degeneracy.

\subsubsection{SNO neutral-current cross-section}
\label{sec:sno-nc}

The SNO experiment observed NC interactions of solar neutrinos on
deuterium, as discussed briefly in Section \ref{sec:solar_exp}.  At low energies, the corresponding cross section is
dominated by the Gamow-Teller transition and it scales as $g_A^2$
where $g_A$ is the coupling of the neutrino current to the axial
isovector hadronic current which in the SM is given by $g_A\equiv
g_A^u-g_A^d$~\cite{Bahcall:1988em, Bernabeu:1991sd, Chen:2002pv}.
Using that the nuclear corrections to $g_A$ are the same when the NSI
are added, we obtain that in the presence of the NC NSI we can write
\begin{equation}
\label{eq:sigma_axial}
  \sigma^\text{det} = \sigma_\text{SM} \bigg(\frac{G_A}{g_A}\bigg)^2
\end{equation}
where $G_A$ is an hermitian matrix in flavour space
\begin{equation}
  \frac{G_A}{g_A}
  = \delta_{\alpha\beta}
  + \Eps_{\alpha\beta}^{u,A} -\Eps_{\alpha\beta}^{d,A}
  = \delta_{\alpha\beta} +
  \Eps_{\alpha\beta} \, (\xi^u - \xi^d) \, (\chi^L - \chi^R) \,,
\end{equation}
Clearly for vector NSI ($\chi^L =\chi^R$), the NSI contributions
vanish and $\sigma^\text{det}$ takes just the SM value times the
identity in flavour space.  Conversely for axial-vector NSI one gets
$\chi^L - \chi^R = 1$ and the NSI term contributes.

\subsubsection{Coherent elastic neutrino-nucleus scattering}
\label{sec:formCNUES}

The generalized cross-section $\sigma^\text{det}$ describing CE$\nu$NS
in the presence of NSI can be obtained by integrating over the recoil
energy of the nucleus $E_R$ the following expression (see Section~\ref{sec:CEvNS}):
\begin{equation}
  \label{eq:xsec-SM}
  \frac{\dd\sigma^\text{coh} (E_R, E_\nu)}{\dd E_R}
  = \frac{G_F^2}{2\pi} \,
  \mathcal{Q}^2 \, F^2(q^2) \, m_A
  \bigg(2 - \frac{m_A E_R}{E_\nu^2} \bigg)
\end{equation}
where $m_A$ is the mass of the nucleus and $F(q^2)$ is its nuclear
form factor evaluated at the squared momentum transfer of the process,
$q^2 = 2 m_A E_R$.  In this formalism, the structure in flavour space
which characterizes $\sigma^\text{det}$ is encoded into the hermitian
matrix $\mathcal{Q}$, which is just the generalization of the weak
charge of the nucleus for this formalism.  For a nucleus with $Z$
protons and $N$ neutrons, it reads:
\begin{equation}
  \label{eq:Qweak}
  \mathcal{Q}_{\alpha\beta}
  = Z \big(g_p^V \delta_{\alpha\beta} + \Eps_{\alpha\beta}^{p,V} \big)
  + N \big(g_n^V \delta_{\alpha\beta} + \Eps_{\alpha\beta}^{n,V} \big)
\end{equation}
where $g_p^V = 1/2 - 2\sin^2\theta_w$ and $g_n^V = -1/2$ are the SM
vector couplings to protons and neutrons, respectively.  In
experiments with very short baselines such as those performed so far,
neutrinos have no time to oscillate and therefore the density matrix
at the detector $\rho^\text{det}$ is just the identity matrix.  Taking
this explicitly into account in Eq.~\eqref{eq:ES-dens} we get:
\begin{equation}
  \rho^\text{det} = I
  \quad\Rightarrow\quad
  N_\text{ev} \propto \mathcal{Q}_\alpha^2
  \quad\text{with}\quad
  \mathcal{Q}_\alpha^2 \equiv \big[ \mathcal{Q}^2 \big]_{\alpha\alpha}
  = (\mathcal{Q}_{\alpha\alpha})^2
  + \sum_{\beta\ne\alpha} |\mathcal{Q}_{\alpha\beta}|^2
\end{equation}
for incident neutrino flavour $\alpha$, thus recovering the
expressions for the ordinary weak charges $\mathcal{Q}_\alpha^2$ used
in our former publications.  Coming back to Eq.~\eqref{eq:Qweak}, let
us notice that it can be rewritten as:
\begin{equation}
  \mathcal{Q}_{\alpha\beta}
  = Z \big[ (g_p^V + Y_n^\text{coh} g_n^V)\, \delta_{\alpha\beta}
  + \Eps_{\alpha\beta}^\text{coh} \big]
  \quad\text{with}\quad
  \Eps_{\alpha\beta}^\text{coh}
  \equiv \Eps_{\alpha\beta}^{p,V} + Y_n^\text{coh} \Eps_{\alpha\beta}^{n,V}
\end{equation}
where $Y_n^\text{coh} \equiv N/Z$ is the neutron/proton ratio
characterizing the target of a given CE$\nu$NS experiment.  Imposing
the quark-lepton factorization of Eq.~\eqref{eq:eps-fact} we get:
\begin{equation}
  \Eps_{\alpha\beta}^\text{coh}
  = \Eps_{\alpha\beta} \big( \xi^p +Y_n^\text{coh} \xi^n \big)
  \big( \chi^L + \chi^R \big)
  = \sqrt{5} \, \big[\! \cos\eta\, \cos\zeta
    + Y_n^\text{coh} \sin\eta \big] \big( \chi^L + \chi^R \big)\,
  \Eps_{\alpha\beta}\,.
\end{equation}
Similarly to Eq.~\eqref{eq:eps-earth}, this expression suggests that
the analysis of coherent scattering data can be performed in terms of
the effective couplings $\Eps_{\alpha\beta}^\text{coh}$, whose
best-fit value and allowed ranges are independent of $(\eta, \zeta)$.
As a consequence, the bounds on $\Eps_{\alpha\beta}$ simply scale as
$[\cos\eta \cos\zeta + Y_n^\text{coh} \sin\eta]$.  In analogy to
Eq.~\eqref{eq:epx-etapr}, one can define an effective angle
$\tan\eta^{\prime\prime} \equiv \tan\eta\, \big/ \cos\zeta$
parametrizing the direction in the $(\xi^p, \xi^n)$ plane, such that
all CE$\nu$NS experiments depend on $(\eta, \zeta)$ only through the  
combination $\eta^{\prime\prime}$.  In particular, it is
straightforward to see that a coherent scattering experiment
characterized by a given $Y_n^\text{coh}$ ratio will yield no bound on
$\Eps_{\alpha\beta}$ for $\eta^{\prime\prime} =
\arctan(-1/Y_n^\text{coh})$, as for this value the effects of NSI on
protons and neutrons cancel exactly.
Such a cancellation can be obtained, for example, for models of the
type proposed in Ref.~\cite{Bernal:2022qba}, where the $Z'$ associated
to a new gauge symmetry also has a sizable kinetic mixing with the SM
photon, which allows for arbitrary relative size of NSI with up and
down quarks.

\section{New Physics scenarios of  neutrino interactions }
\label{sec:formCS}

As outlined in the preceding section, NSI can manifest in neutrino
experiments through two distinct mechanisms: modifications to flavor
propagation (governed by oscillation dynamics) and alterations to
scattering processes (affecting both production and detection
channels). In this section, we focus only on the NP effects on the
interaction cross-sections at the detection stage, neglecting for simplicity its possible impact on flavor
transitions during propagation. In this approach
  we preserve the standard oscillation framework but introduce new
couplings between neutrinos and first-generation fermions ($e,d,u$)
via the exchange of hypothetical mediators:

\begin{itemize}
    \item Non-zero neutrino magnetic moment $\mu_\nu$;
    \item Neutral scalar boson $\phi$;
    \item Neutral pseudoscalar boson $\varphi$;
    \item Neutral vector boson $Z'$ from $U(1)$ gauge symmetries;
\end{itemize}

A non-vanishing neutrino magnetic moment enables spin-flip
interactions with charged particles, producing characteristic
low-energy recoil spectra in scattering processes. In parallel, the
scalar ($\phi$) and pseudoscalar ($\varphi$) mediators generate
Yukawa-type interactions, which modify the differential cross-section
through distinct recoil energy dependencies. The vector mediator
$Z^\prime$ introduces new current-current interactions, analogous to
the SM weak force but with couplings dependent on the mediator's mass
and the charge of the new gauge symmetry.  Each scenario predicts
unique signatures in neutrino detectors, such as deviations in event
rates, spectral distortions, or anomalous scattering angles. By
analysing these phenomenological consequences, we establish
constraints on the parameter spaces of the mediators (coupling
strengths, masses) and the neutrino magnetic moment, while
distinguishing the interaction-centric aspects of
  this kind of NP from propagation-dependent effects. Following
  Ref.~\cite{Coloma:2022umy}, in the rest of this section we will
  focus on solar experiments sensitive to ES interactions, such as
  Borexino.

\subsection{Neutrino Magnetic-Moments}

In the presence of a neutrino magnetic moment ($\mu_\nu$) the
scattering cross sections on electrons get additional contributions
which do not interfere with the SM ones.  Neutrino magnetic moments
arise in a variety of BSM models and, in particular, they do not need
to be flavour-universal.  Therefore, in what follows we will allow
different magnetic moments for the different neutrino flavours --
albeit still imposing that they are flavour-diagonal.  Under this
assumption the ES differential cross section for either neutrinos or
antineutrinos, up to order $\mathcal{O}(y^2)$, takes the
form~\cite{Vogel:1989iv}
\begin{equation}
  \label{eq:mag-elec}
  \frac{\dd\sigma^{\mu_\nu}_\beta}{\dd T_e}
  = \frac{\dd\sigma^\text{SM}_\beta}{\dd T_e}
  + \left(\frac{\mu_{\nu_\beta}}{\mu_B}\right)^2
  \frac{\alpha^2\,\pi}{m_e^2}
  \left[\frac{1}{T_e} - \frac{1}{E_\nu} \right] .
\end{equation}
Since no flavour-changing contributions are present in this case, the
expression for the number of ES interactions will be just proportional to the probability weighted total cross
section.

\subsection{Models with light scalar, pseudoscalar, and vector mediators}

Concerning the
scalar, the Lagrangian we consider is~\cite{Cerdeno:2016sfi}
\begin{equation}
  \label{eq:lagscal}
  \mathcal{L}_\phi = g_\phi \, \phi
  \bigg( q_\phi^e\, \bar{e} e + \sum_\alpha
  q_\phi^{\nu_\alpha}\, \bar\nu_{\alpha,R}\, \nu_{\alpha,L}
  + \text{h.c.} \bigg)
  - \frac{1}{2} \, M^2_\phi \, \phi^2 \,,
\end{equation}
where $q_\phi^j$ are the individual scalar charges and $j =
\{\nu_\alpha, e\}$.  The corresponding cross section for
neutrino-electron (or antineutrino-electron) scattering is
\begin{equation}
  \label{eq:csscale}
  \frac{\dd \sigma_{\beta}^\phi}{\dd T_e}
  = \frac{\dd\sigma^\text{SM}_\beta}{\dd T_e}
  + \frac{g_\phi^4 \, (q^{\nu_\beta}_\phi)^2 \, (q^e_\phi)^2 \, m_e^2 \, T_e}
  {4\pi\, E_\nu^2\, (2\, m_e\, T_e + M_\phi^2)^2}
  \left( 1 + \frac{T_e}{2\,m_e} \right) ,
\end{equation}
where we have included all orders in $T_e/E_\nu$ and $T_e/m_e$ (beyond
the leading-order terms given in Ref.~\cite{Cerdeno:2016sfi}), since
these are non-negligible for solar event rates at Borexino.

For the pseudoscalar scenario, we consider the Lagrangian
\begin{equation}
  \label{eq:lagpscal}
  \mathcal{L}_\varphi = i\, g_\varphi \, \varphi
  \bigg( q_\varphi^e\, \bar{e}\gamma^5 e + \sum_\alpha
  q_\varphi^{\nu_\alpha}\, \bar\nu_{\alpha,R} \gamma^5 \nu_{\alpha,L}
  + \text{h.c.} \bigg)
  - \frac{1}{2} \, M^2_\varphi \, \varphi^2 \,,
\end{equation}
where $q_\varphi^j$ are the individual pseudoscalar charges and $j =
\{\nu_\alpha, e\}$.  The corresponding cross section for
neutrino-electron (or antineutrino-electron) scattering is
\begin{equation}
  \label{eq:cspscale}
  \frac{\dd \sigma_{\beta}^{\varphi}}{\dd T_e}
  = \frac{\dd\sigma^\text{SM}_\beta}{\dd T_e}
  + \frac{g_\varphi^4 \, (q^{\nu_\beta}_\varphi)^2 \, (q^e_\varphi)^2 \, m_e \, T_e^2}
  {8 \pi \, E_\nu^2 \, (2\, m_e\, T_e + M_\varphi^2)^2} \,.
  \end{equation}

As for the vector mediator, our assumed Lagrangian is:
\begin{equation}
  \label{eq:lagvec}
  \mathcal{L}_V = g_{Z'}\, Z'_\mu \bigg( q_{Z'}^e\, \bar{e} \gamma^\mu e
  + \sum_\alpha q_{Z'}^{\nu_\alpha}\, \bar\nu_{\alpha,L}\gamma^\mu \nu_{\alpha,L}
 \bigg) + \frac{1}{2} M^2_{Z'} {Z'}^\mu Z'_\mu\,,
\end{equation}
with the individual vector charges $q_{Z'}^j$.  Unlike for the scalar
and magnetic-moment cases, a neutral vector interaction interferes
with the SM contribution.  The differential cross section for
neutrino-electron scattering reads~\cite{Lindner:2018kjo}
\begin{multline}
  \label{eq:csvece}
  \frac{\dd \sigma_{\beta}^{Z'}}{\dd T_e}
  = \frac{\dd\sigma^\text{SM}_\beta}{\dd T_e}
  + \frac{g^2_{Z'} \, m_e}{2 \, \pi} \Bigg\{
  \frac{g^2_{Z'}\, (q_{Z'}^{\nu_\beta})^2\, (q^e_{Z'})^2}{\big(2 m_e T_e + M_{Z'}^2\big)^2}
  \left[ 1 - \frac{m_e\, T_e}{2\,E_\nu^2}
    + \frac{T_e}{2\,E_\nu} \left( \frac{T_e}{E_\nu} - 2 \right) \right]
  \\
  + \frac{2\sqrt{2}\, G_F\, q_{Z'}^{\nu_\beta}\, q_{Z'}^e}{\big(2 m_e T_e + M_{Z'}^2\big)}
  \left[ c_{V\beta} \left( 1 - \frac{m_e\, T_e}{2\,E_\nu^2} \right)
    + c_{R\beta}\, \frac{T_e}{E_\nu} \left( \frac{T_e}{E_\nu} - 2 \right)
    \right]
  \Bigg\} \,,
\end{multline}
where $c_{V\beta}$ is the SM effective vector coupling, defined as $c_{L\beta}+c_{R\beta}$, such that $c_{L\beta}$,$c_{R\beta}$ are defined in Eq. \ref{eq:cm-coupl}.  For the sake of simplicity, in our calculations describing the effects of light mediators, we have
neglected the small SM radiative corrections to the BSM contributions.
The corresponding cross section for $\bar\nu_\beta \, e^-$ ES can be
obtained from Eq. ~\eqref{eq:csvece} with the exchange
$c_{L\beta}\leftrightarrow c_{R\beta}$.

Similarly to the SM and the magnetic moment cases, no flavour-changing
contributions are considered for scalar, pseudoscalar, or vector
mediators, hence the number of interactions  will be just
proportional to the probability-weighted total cross section.

For scalar and pseudoscalar mediators we
will focus on a model in which the mediator couples universally to
fermions.  For vector mediators we will show the results for three
anomaly-free\footnote{We assume that three right-handed neutrinos are
added to the SM particle content which, in addition of generating
non-zero neutrino masses, cancel the anomalies in the $B-L$ case.}
models coupling to electrons, namely $B-L$, $L_e-L_\mu$ and
$L_e-L_\tau$.  For convenience Tab.~\ref{tab:charges} lists the
relevant charges.

\begin{table}[t]\centering
  \catcode`?=\active \def?{\hphantom{+}}
  \begin{tabular}{l|cccc}
    Model & $q^e$ & $q^{\nu_e}$ & $q^{\nu_\mu}$ & $q^{\nu_\tau}$
    \\
    \hline
    Universal/leptonic scalar (or pseudoscalar) & $?1$ & $?1$ & $?1$ & $?1$
    \\
    $B-L$ vector & $-1$ & $-1$ & $-1$ & $-1$
    \\
    $L_e-L_\mu$ vector & $?1$ & $?1$ & $-1$ & $?0$
    \\
    $L_e-L_\tau$ vector & $?1$ & $?1$ & $?0$ & $-1$
  \end{tabular}
  \caption{Charges for the vector, scalar, and pseudoscalar mediators
    considered in this work.}
  \label{tab:charges}
\end{table}

\section{New pseudoscalar interactions in neutrino propagation}
\label{sec:MD_forces}

A spin-zero bosonic mediator $\phi$ with ultralight mass ($m_\phi \ll \text{eV}$) can interact with SM fermions $\psi$ through two Lorentz-invariant structures: scalar couplings ($\bar{\psi}\psi\phi$), mediating spin-independent forces, and pseudoscalar couplings ($\bar{\psi}\gamma^5\psi\phi$), generating spin-dependent monopole-dipole and dipole-dipole interactions~\cite{Moody:1984ba}. Unlike NSIs that alter local interaction cross-sections (detection/production), the ultralight $\phi$ does not perturb local scattering amplitudes due to its feeble couplings. Instead, its macroscopic coherence length $\sim m_\phi^{-1}$ enables cumulative phase shifts over neutrino propagation baselines, imprinting non-local effective forces on flavor transitions. For pseudoscalar couplings, the gradient $\nabla \phi$ induces an effective background magnetic field $\mathbf{B}_{\text{eff}} \propto \nabla\phi$, driving quantum spin precession in non-relativistic regimes.

In neutrino physics, the coherent superposition of $\phi$-mediated long-range interactions modifies the MSW resonance mechanism (see Chapter~\ref{chap:theo}) through flavor-dependent potential terms integrated over macroscopic scales. This effective force, arising from the mediator's coherence, introduces oscillatory modulations in the neutrino Hamiltonian, distinct from conventional matter effects. These dynamics fundamentally alter oscillation phenomenology~\cite{Grifols:2003gy, Joshipura:2003jh, GonzalezGarcia:2006vp, Davoudiasl:2011sz, Wise:2018rnb, Smirnov:2019cae, Coloma:2020gfv}. Additionally, spin-coupled interactions enable spin-flavor precession (SFP)~\cite{Akhmedov:1987nc, Lim:1987tk}, resonantly enhanced in solar matter density gradients. For Dirac neutrinos, SFP converts left-handed states to undetectable right-handed counterparts, mimicking active-sterile oscillations. For Majorana neutrinos, this process generates right-handed antineutrinos, potentially producing observable solar $\bar{\nu}_e$ fluxes~\cite{Cisneros:1970nq, Okun:1986hi, Okun:1986na, Okun:1986uf, Voloshin:1986ty}. The helicity-dependent response to the coherent $\phi$ field provides a unique discriminant between Dirac and Majorana neutrino mass mechanisms through solar neutrino spectroscopy.
\subsection{Monopole-dipole potential: Mathematical formulation}

We consider the interactions of a field $\phi$ with scalar couplings
to the nucleons $f=$ neutrons, protons, and pseudoscalar couplings to neutrinos:
\begin{equation}
  \label{eq:lagran}
  \mathcal{L}_\phi = \sum_f g_s^f \phi \bar{f}f
  + i\, [g_p^\nu]^{\alpha\beta} \phi\,
  \overline{\nu}_\alpha \gamma^5 \nu_\beta \,,
\end{equation}
where $g_p^\nu$ parametrize the pseudoscalar couplings to the spin of
the neutrinos, which in its most generality is a $3\times 3$ hermitian
matrix in the neutrino flavour space.

Considering solar neutrinos, the interactions in Eq.~\eqref{eq:lagran}
will generate a flavour dependent potential sourced by the nucleons in
the Sun which will affect the flavour evolution of the neutrinos and
can lead to observable signatures.  We present in
Appendix~\ref{sec:MD_pot} the derivation of this potential.  In
brief, we find that the interactions in Eq.~\eqref{eq:lagran} generate
a spin-flip potential on a neutrino of energy $E_\nu$ at position
$\vec{x}$ of the form
\begin{equation}
  \label{eq:Vsp}
  V_\text{sp}(\vec{x})
  = -\frac{g_s^f g_p^\nu}{4E_\nu} \mathcal{E}_\perp(\vec{x})
\end{equation}
where $\mathcal{E}_\perp$ is the size of the transverse component
(with respect to the neutrino trajectory) of a vector field
$\vec{\mathcal{E}}$ defined as
\begin{equation}
  \label{eq:nhatdef}
  \vec{\mathcal{E}}(\vec{x})
  \equiv \frac{1}{2\pi} \vec{\nabla}_x
  \int N_f(\vec\rho)\,
    \frac{e^{-m_\phi |\vec\rho - \vec{x}|}}{|\vec\rho - \vec{x}|}\,
    d^3\vec\rho\,
  \equiv \frac{2}{m_\phi^2} \vec\nabla
  \hat{N}_f(\vec{x}, m_\phi) \,.
\end{equation}
In writing Eq.~\eqref{eq:nhatdef} we have introduced the quantity
$\hat{N}_f(\vec{x}, m_\phi)$ which represents the potential-weighted
matter density within a radius $\sim 1/m_\phi$ around the neutrino
location~\cite{Grifols:2003gy, Joshipura:2003jh,
  GonzalezGarcia:2006vp, Davoudiasl:2011sz, Wise:2018rnb}.  Its
normalization factor ensures that $\hat{N}_f(\vec{x}, m_\phi) \to
N_f(\vec{x})$ for $m_\phi \to \infty$, because that is the limit in
which one should recover the contact interaction expectation
(\textit{i.e.}, the neutrino wave packet localized at position
$\vec{x}$ should be directly sensitive to the number density
distribution $N_f(\vec{x})$ of fermion $f$ at such point).

\begin{figure}[t]\centering
  \includegraphics[width=0.85\textwidth]{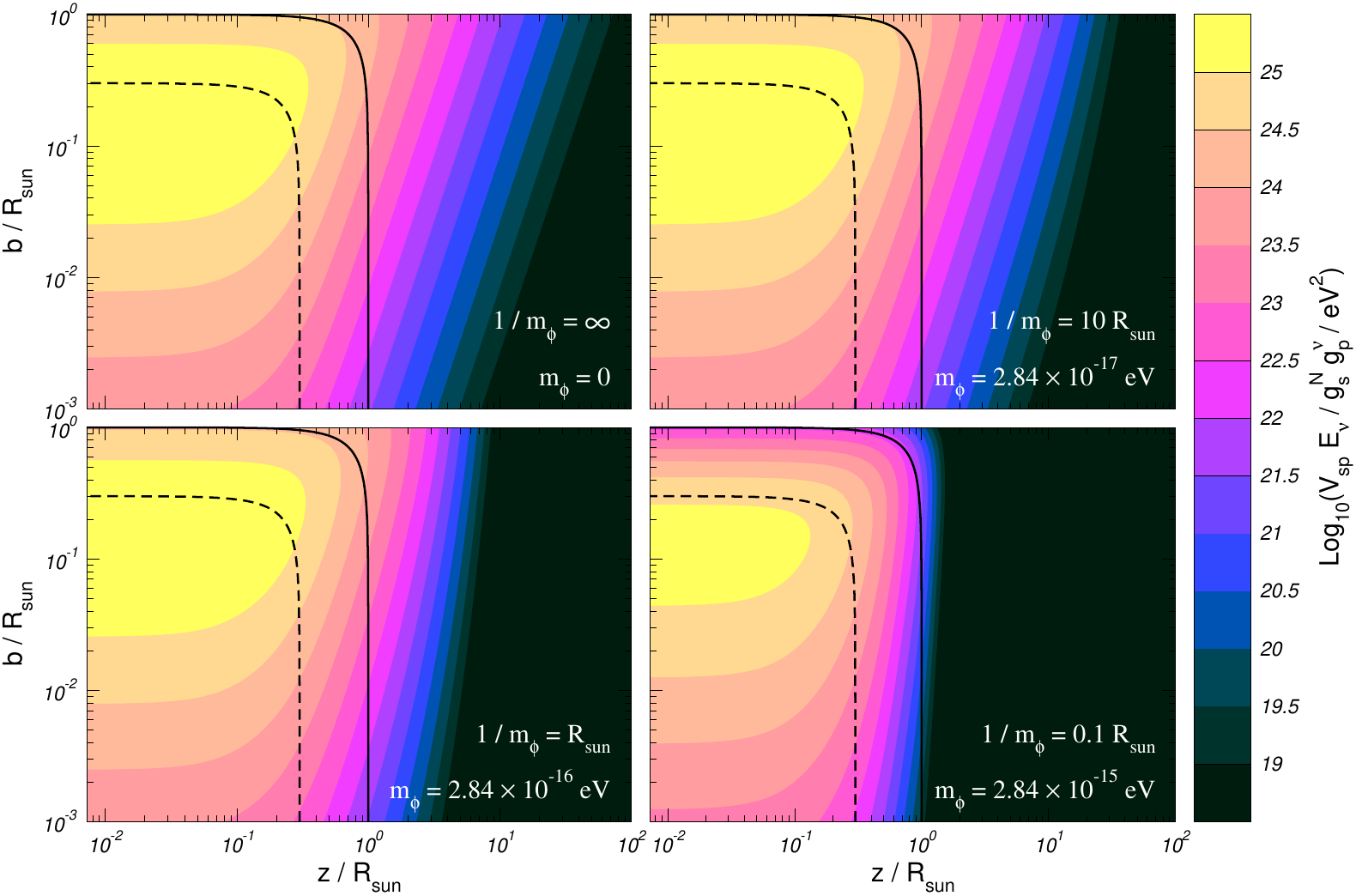}
  \caption{Isocontours of the potential $V_{sp}$ sourced by nucleons
    with coupling $g_s^N$ to $\phi$ felt by $\nu$ with coupling
    $g_p^\nu$ to $\phi$ as a function of $b$ (its transverse distance
    to the line connecting the Sun and Earth centers) and $z$ (the
    distance along the Sun-Earth direction characterizing its
    trajectory to the Earth) for several values of $m_\phi$.  For
    reference the black full and dashed lines represent the position of the edge of
    the Sun, $\sqrt{b^2+z^2} = R_\odot$, and the outer edge of the
    solar core, $\sqrt{b^2+z^2} = 0.3\, R_\odot$, respectively.}
  \label{fig:F3}
\end{figure}

Denoting by $r \equiv |\vec{x} - \vec{x}_\odot|$ the distance from the
center of the Sun $\vec{x}_\odot$ and taking into account the
spherical symmetry of the solar matter distribution $N_f^\odot(r)$, we
see that $\hat{N}_f(\vec{x}, m_\phi)$ simplifies to $\hat{N}_f(r,
m_\phi)$ with:
\begin{equation}
  \hat{N}_f(r, m_\phi)
  = \frac{m_\phi}{2\,r}
  \int_{0}^{R_\odot} \rho\, N_f^\odot(\rho)
  \big[ e^{-m_\phi|\rho-r|} - e^{-m_\phi(\rho+r)} \big] \, d\rho \,.
  \end{equation}
Notice that since $\hat{N}_f(r, m_\phi)$ only depends on the radial
distance to the center of the Sun, the direction of the vector
$\vec{\mathcal{E}}(\vec{x})$ is radial.  As the spin-flip potential is
proportional to the component of $\vec{\mathcal{E}}$ orthogonal to the
neutrino direction, it is clear that it will vanish for trajectories
along the line connecting the solar center and the center of the
Earth.  For neutrinos at a transverse distance $b$ from the line
connecting the Earth and Sun centers, and at a radial distance $r$
from the center of the Sun, the potential is
\begin{equation}
  \label{eq:Vsp2}
  V_\text{sp}(r, b)
  = -\frac{g_s^N g_p^\nu}{2E_\nu\, m_\phi^2}
  \cdot \frac{b}{r} \cdot
  \frac{d\hat{N}_f(r, m_\phi)}{dr} \,.
\end{equation}
In Fig.~\ref{fig:F3} we plot the value of this potential (for
different values of $m_\phi$) at a given position $(b,z)$, where $b$
is the transverse distance to the line connecting the Sun and Earth
centers, and $z=\sqrt{r^2-b^2}$ is the distance along the Sun-Earth
direction (which, as we will see below, is the variable parametrizing
the neutrino trajectory).  For the sake of concreteness we show the
results for $g_s^{\rm proton} = g_s^{\rm neutron} \equiv g_s^N$.

From the figure we see that, as expected, the potential decreases as
$b\to 0$.  However, it increases very rapidly with $b$ and it reaches
its maximum for values of $b$ well within the neutrino production
region $b\lesssim 0.3\, R_\odot$.  Quantitatively we see that for the
$m_\phi$ considered here the characteristic values of the potential
inside the solar core are comparable with the inverse of the
oscillation length of solar neutrinos with energy $E_\nu$, $E_\nu /
\lambda \sim \Dmq_{21} \sim 10^{-4}~\eVq$ for $g_s^N g_p^\nu \sim
10^{-30}$ within the range of values of the coupling to muons proposed
to account for the $(g-2)_\mu$ anomaly~\cite{Agrawal:2022wjm,
  Davoudiasl:2022gdg}.

To account for the effect of this potential in the solar neutrino
observables we need to solve the evolution equation for the neutrino
ensemble.  Since the potential flips the helicity of the neutrino,
such equation is different for Dirac and Majorana neutrinos.  For a
Dirac neutrino the evolution equation for its flavour components
$\vec\nu\equiv (\nu_e, \nu_\mu, \nu_\tau)^T$ reads
\begin{equation}
  \label{eq:dir}
  i\, \frac{d}{dt}
  \begin{pmatrix}
    \vec\nu_L(t)
    \\
    \vec\nu_R(t)
  \end{pmatrix}
  =
  \begin{pmatrix}
    H^\nu(r) & V_\text{sp}(r, b)
    \\
    V_\text{sp}(r, b)^\dagger & H_\text{vac}
  \end{pmatrix}
  \begin{pmatrix}
    \vec\nu_L(t)
    \\
    \vec\nu_R(t)
  \end{pmatrix},
\end{equation}
where, given the large distance between the Sun and the Earth, we can
assume with high precision that the neutrino travels along a
trajectory which runs parallel to the line connecting the Sun and
Earth centers, so that its distance $b$ from such line remains
constant during propagation.  Denoting by $z$ the longitudinal
position of the neutrino along its trajectory, we have that at a given
time $t$ the distance from the neutrino location to the center of the
Sun is given by $r=\sqrt{b^2+(z-z_0)^2}$ with $z = z_0 + c t$.

For a Majorana neutrino the evolution equation is
\begin{equation}
  \label{eq:maj}
  i\, \frac{d}{dt}
  \begin{pmatrix}
    \vec\nu_L(t)
    \\
    \vec{\overline\nu}_R(t)
  \end{pmatrix}
  =
  \begin{pmatrix}
    H^\nu(r) & V_\text{sp}(r, b)
    \\
    V_\text{sp}(r, b)^\dagger & H^{\overline{\nu}}(r)
  \end{pmatrix}
  \begin{pmatrix}
    \vec\nu_L(t)
    \\
    \vec{\overline\nu}_R(t)
  \end{pmatrix},
\end{equation}
where
\begin{equation}
  H^\nu(r) = H_\text{vac} + H_\text{mat} (r)
  \quad\text{and}\quad
  H^{\overline{\nu}}(r) = [H_\text{vac} - H_\text{mat}(r)]^* \,.
\end{equation}
Here $H_\text{vac}$ is the vacuum part which in the flavour basis is defined in Chapter \ref{chap:theo}. $H_\text{mat}(r)$ is the matter potential due to the SM weak
interactions, which in the flavour basis takes the diagonal form
\begin{equation}
  \label{eq:hmsw}
  H_\text{mat}(r) = \frac{G_F}{\sqrt{2}}
  \diag\Big[ 2N_e(r) -N_n(r),\, -N_n(r),\, -N_n(r) \Big]
\end{equation}
and $N_e(r)$ and $N_n(r)$ are the electron and neutron number
density at distance $r$ from the solar center.

We numerically solve the evolution equation from the production point
of the ${\nu_e}_L$ in the Sun to the Earth.  With our choice of
variables we can characterize the production point by its radial
distance $r_0$ from the center of the Sun, its transverse distance $b$
from the line connecting the Sun-Earth centers, and the angle $\phi$
in that transverse plane, and we are left with two discrete
possibilities $\vec{x}_0^\pm = (\pm \sqrt{r_0^2-b^2}, b\cos\phi_0,
b\sin\phi_0)$ where positive (negative) sign corresponds to neutrinos
produced in the hemisphere of the Sun which is closer (further) from
the Earth.  In this way we obtain the oscillation probability into
$\nu_\alpha$ (here denoting generically either a left-handed neutrino
state, or a right-handed neutrino for Dirac and a right-handed
antineutrino for Majorana), $P_{e\alpha}(\vec{x}_0^\pm, E_\nu)$.  In
the Standard Solar Models the probability distribution of the neutrino
production point only depends on the radial distance to the center of
the Sun, $r_0$ (with different distributions for the different
neutrino production reactions) so it is convenient to define
\begin{equation}
  P_{e\alpha}(r_0, E_\nu)
  \equiv
  \frac{1}{4\pi}\int d\Omega\, P_{e\alpha}(\vec{x}_0, E_\nu)
  =\frac{1}{4\pi r_0}\int_0^{r_0} db\, \frac{b}{\sqrt{r_0^2-b^2}}
    \int_0^{2\pi} d\phi\, \sum_{\pm}P_{e\alpha}(\vec{x}_0^\pm, E_\nu) \,.
\end{equation}
As mentioned above each of the eight reactions producing solar
neutrinos ---~labeled by $i=1\dots 8$ for \Nuc{pp}, \Nuc[7]{Be},
\Nuc{pep}, \Nuc[13]{N}, \Nuc[15]{O}, \Nuc[17]{F}, \Nuc[8]{B}, and
\Nuc{hep}~--- generates neutrinos with characteristic distributions
$\mathcal{R}_i(r_0)$ (normalized to one) which are energy independent.
Thus we can obtain the mean survival probability for a neutrino of
energy $E_\nu$ produced in reaction $i$ as
\begin{equation}
  P^{i}_{e\alpha}(E_\nu)
  = \int_0^{R_{\odot}} dr_0 \, \mathcal{R}_i(r_0)\, P_{e\alpha}(r_0, E_\nu)
\end{equation}
and with those, obtain the predictions for the solar observables in our
analysis.

\begin{figure}[t]\centering
  \includegraphics[width=\textwidth]{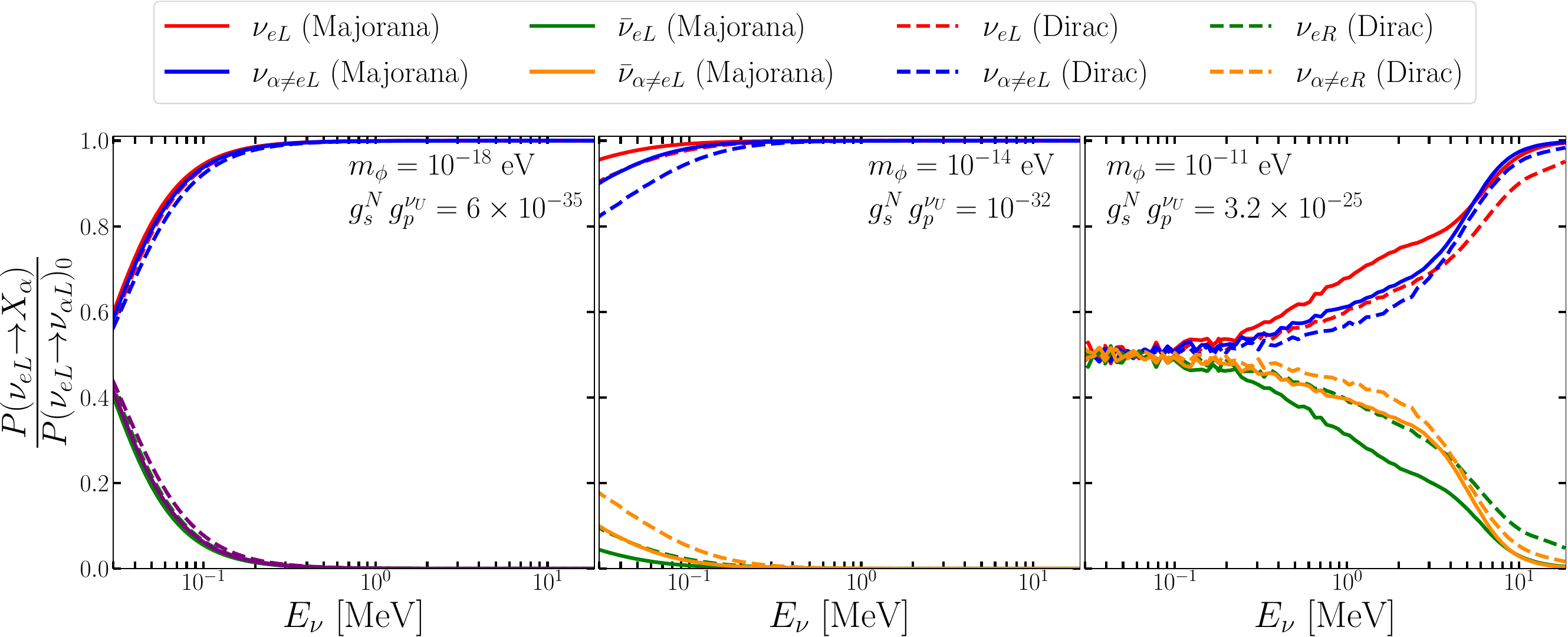}
  \caption{Relevant oscillation probabilities at the Earth surface
    of solar $\nu_{eL}$ into
    left and right neutrinos (antineutrinos) of flavour $\alpha$
    divided by $P(\nu_{e_L}\to \nu_{\alpha L})_0$ (the probability for
    zero coupling to the $\phi$ field) as a function of the neutrino
    energy for different values of model parameters as labeled in the
    figure.  Oscillation parameters have been fixed to $\Dmq_{21} =
    7.4\times 10^{-5}~\eVq$, $\sin^2\theta_{12} = 0.303$,
    $\sin^2\theta_{13} = 0.02203$, $\Dmq_{31} = 2.5\times
    10^{-3}~\eVq$, $\sin^2\theta_{23} = 0.5$, and $\delta_\text{CP} =
    0$.}
  \label{fig:probs}
\end{figure}

In what follows, for the sake of concreteness, we are going to focus
on four cases: three in which $\phi$ couples to a specific flavour
$\gamma$ and a fourth in which the coupling is universal in flavour
space
\begin{equation}
  \label{eq:nucoup}
  [g_p^\nu]^{\alpha\beta}
  = g_p^{\nu_\gamma} \, \delta^{\alpha\gamma}\delta^{\beta\gamma}
  \enspace\text{for}\enspace
  \gamma \in \{e, \mu, \tau \} \,,
  \quad\text{and}\quad
  [g_p^\nu]^{\alpha\beta}
  = g_p^{\nu_U} \, \delta^{\alpha\beta}
  \enspace\text{for universal}.
\end{equation}
As illustration we plot in Fig.~\ref{fig:probs} the relevant
oscillation probabilities for Dirac and Majorana cases as a function
of the neutrino energy.  The figure is shown for the case of coupling
to $\nu_U$ but the behaviour is qualitatively the same for couplings
to $\nu_e$, $\nu_\mu$ or $\nu_\tau$.\footnote{For concreteness the
probabilities shown here have been generated integrating over the
production point distribution $\mathcal{R}_\text{\Nuc[8]{B}}(r_0)$,
but the corresponding plots obtained with the distribution
probabilities for the other solar fluxes are very similar.}  We assume
fixed oscillation parameters $\Dmq_{21} = 7.4\times 10^{-5}~\eVq$,
$\sin^2\theta_{12} = 0.303$, $\sin^2\theta_{13} = 0.02203$, $\Dmq_{31}
= 2.5\times 10^{-3}~\eVq$, $\sin^2\theta_{23}=0.5$, and
$\delta_\text{CP}=0$, and display different values for the two model
parameters ($m_{\phi}$, $g_s^N g_p^{\nu_U}$).    As seen in the figure the relevant
probabilities are not very different for Dirac and Majorana neutrinos.
Most importantly the figure shows that for moderate values of the
model parameters the new interaction has the largest effect on the
disappearance of $\nu_{eL}$ and the appearance of $\nu_{\alpha\neq
  e,L}$ and $\nu_{\alpha R}$ ($\overline{\nu}_{\alpha R}$) for Dirac
(Majorana) neutrinos with the lowest energies.  This is expected as
the potential in Eq.~\eqref{eq:Vsp} is inversely proportional to the
neutrino energy, as characteristic of interactions with spin-zero
mediators~\cite{GonzalezGarcia:2006vp}.  This is at a difference with
the helicity-conserving MSW potential of Eq.~\eqref{eq:hmsw} (or with
any vector interaction in general~\cite{Coloma:2020gfv}), and with the
helicity-flip potential generated by a neutrino magnetic
moment~\cite{Akhmedov:1987nc, Lim:1987tk} which are independent of the
neutrino energy.  We also see that for these cases the relative effect  
on the disappearance of $\nu_{eL}$ with $\sim
\mathcal{O}(\text{10-100})$~KeV energy is larger than the
$\overline{\nu}_{eR}$ appearance at MeV.  These two facts are of
relevance in the derivation of the bounds from the analysis of the
solar neutrino and antineutrino data as we describe next.

\section{Summary}

To begin, Section \ref{sec:formalism_NSI} started by describing the
formalism of NSIs within the framework of neutrino oscillations and
interactions for different neutrino sources. After introducing NSIs
and exploring their impact on neutrino propagation by modifying matter
effects in various neutrino sources, we discussed the LMA-D
degeneracy.

Following this, Section \ref{sec:formCS} examined how non-standard
scenarios can influence neutrino detection in experiments and
discussed the diverse origins of New Physics (NP) scenarios. To
finish, Section \ref{sec:MD_forces} established the formalism of how
Monopole-Dipole interactions mediated by a very light scalar field
that interacts with neutrinos and SM fermions can alter the
propagation of neutrinos in the Sun.

Beyond NSI, we developed the formalism for several relevant scenarios
involving interactions between quarks/leptons and neutrinos through
BSM interactions, focusing in particular on the study of
vector, scalar, and pseudoscalar mediators and the neutrino magnetic
moment. The last section was concentrated on an exotic BSM, a very
light scalar that affects the oscillation pattern of neutrinos
crossing solar matter, and interacts with charged fermions
via monopole and with neutrinos via dipole interactions.

Now, we will utilize data from various experiments to analyse and
determine the constraints that can be derived from a global fit for
these different physical scenarios.

\chapter{ Three neutrino oscillations with new interactions: Analysis}
\label{chap:exp_bsm}

This chapter employs the experimental datasets outlined in
Chapter~\ref{chap:exp} to impose bounds over new interactions in the
three-neutrino (3$\nu$) framework, as described theoretically in
Chapter~\ref{chap:theo_bsm}. Section~\ref{sec:NSI_BX} initiates this
investigation with a detailed analysis of Borexino Phase-II solar
neutrino spectral data, aiming to constrain new physics scenarios that
alter neutrino matter potential or modify
elastic scattering processes. These include neutrino magnetic moments,
NC NSI, and simplified models with light scalar, pseudoscalar, or
vector mediators.

Section~\ref{sec:results_glob_bsm} extends this work through a global
analysis of neutrino data, updating and expanding upon previous
studies. This analysis simultaneously accounts for NC NSI couplings to
up quarks, down quarks, and electrons, treating vector and
axial-vector couplings independently. The combined fit of solar, atmospheric, reactor and
  accelerator neutrino data introduces computational challanges,
  particularly in modeling NSI effects on all these experiments at the
  same time. Additionally, CE$\nu$NS data—from COHERENT (CsI and Ar
targets) and the Dresden-II (both of them discussed in Section
\ref{sec:CEvNS}) reactor experiment—are incorporated to resolve
degeneracies in NSI parameter space, particularly those linked to the
LMA-D solution.

Finally, Section~\ref{sec:MD_section} derives constraints on an exotic
scenario involving a very light scalar field interacting with
neutrinos via pseudoscalar couplings and with nucleons via scalar
couplings. Building on the discussion in Section~\ref{sec:MD_forces},
which outlines how such interactions generate effective potentials
altering solar neutrino oscillations, this section presents a global
analysis of solar neutrino data (including KamLAND) and solar
antineutrino flux bounds. The analysis distinguishes between Dirac and
Majorana neutrino mass mechanisms, leveraging their distinct
helicity-dependent responses to the scalar-mediated
potential. Together, these studies demonstrate how multi-experiment
synergies and novel interaction frameworks refine the boundaries of
BSM physics in neutrino systems.

\section{Confronting new interactions with electrons with Borexino data}
\label{sec:NSI_BX}

\begin{figure}[t]\centering
  \includegraphics[width=0.65\textwidth]{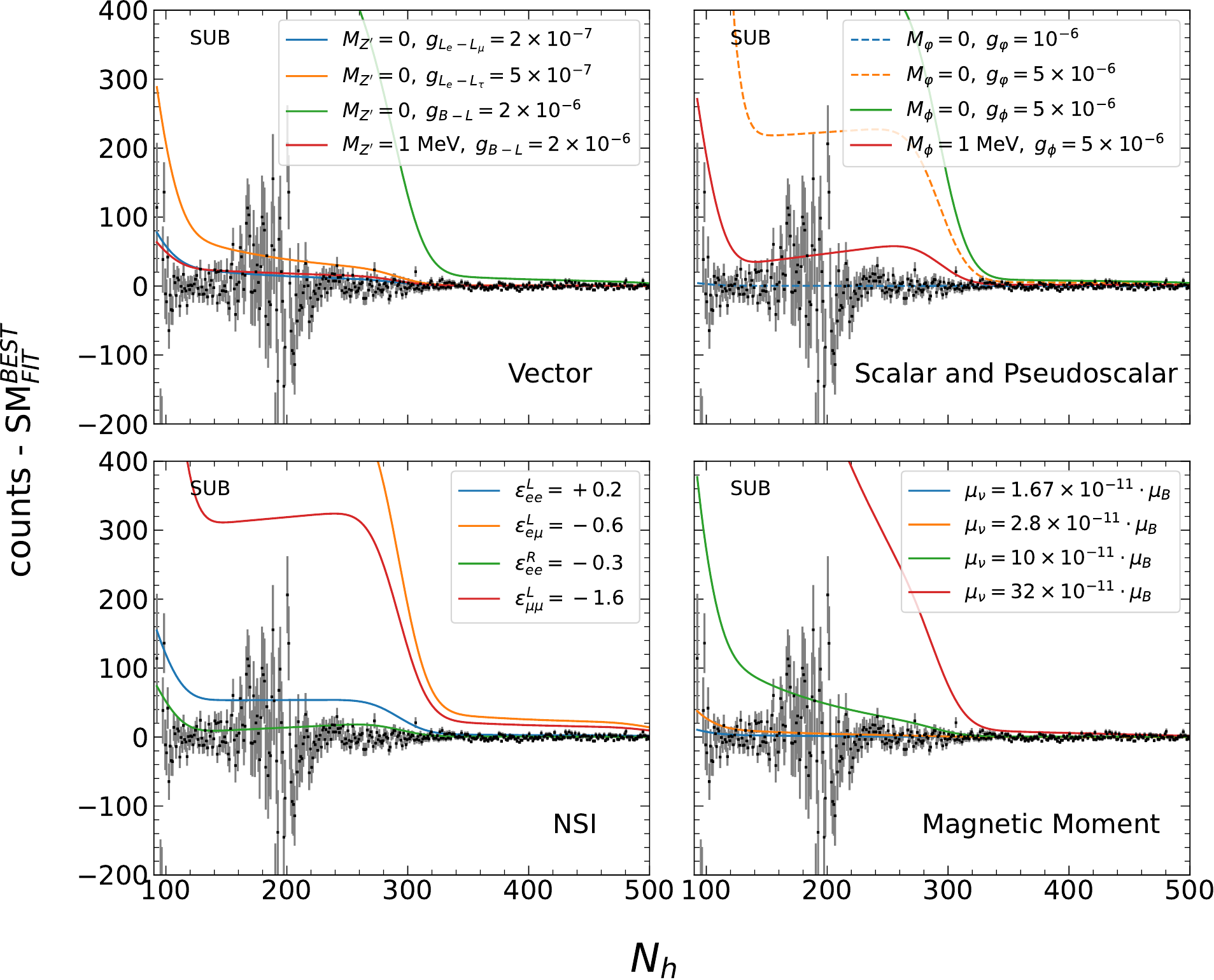}
  \caption{Difference between the number of events with respect to the
    SM expectation, as a function of $N_h$.  For reference, the black
    dots are obtained for the data in the ``subtracted'' sample, where
    the error bars indicate statistical errors only.  The coloured
    lines correspond to the predicted event rates (using the same
    exposure as that of the ``subtracted'' sample) for different BSM
    scenarios and parameter values, as indicated by the labels.}
  \label{fig:events}
\end{figure}

We start by confronting the high statistics data from Borexino described in~\ref{sec:bxsimul}, with several new physics scenarios: NSI, anomalous magnetic moment, vector, pseudoscalar and scalar mediators. For illustration, we show in Fig.~\ref{fig:events}
the predicted event distributions for each new physics scenario. We also include the Borexino data points, after subtracting the best-fit background model and SM signal. For the sake of concreteness, we show the results for the “subtracted” event sample. 

\subsection{Non-Standard Interactions with electrons - Results}
First, let us discuss the case of NSI with electrons.   In this scenario
Borexino is sensitive to a total of 12 operators: 6 involving left-handed electrons (with coefficients
$\Eps_{\alpha\beta}^L$) and 6  involving right-handed
electrons (with coefficients $\Eps_{\alpha\beta}^R$), see
Eq.~\eqref{eq:nsi-nc}.  
In all generality, \emph{a priori} all
operators should be considered at once in the analysis since all of
them enter at the same order in the effective theory.  However,
including such a large number of parameters makes the problem
numerically challenging, and the extraction of meaningful conclusions
is also jeopardized by the multiple interference effects between the
different coefficients.  Thus, here we focus on four representative
cases with a reduced set of operators, depending on the type of
interaction that could lead to the terms in Eq.~\eqref{eq:nsi-nc}: 
\begin{itemize}
\item \textbf{(1)}
the axial vector case, where both left-handed and right-handed NSI operators
are generated with equal strength but with opposite signs; \item \textbf{(2)} the
purely vector case, where both types of operators are generated with
equal strength and equal sign;  
\item \textbf{(3)} the purely left-handed or purely
right-handed cases, where only operators involving electrons of a
given chirality are considered.  
\end{itemize}

For each of these three benchmarks we
include all operators in neutrino flavour space (both flavour-diagonal
and off-diagonal operators) simultaneously in the analysis, allowing
for interference and correlation effects among them.

\begin{figure}[t]\centering
  \includegraphics[width=0.7\textwidth]{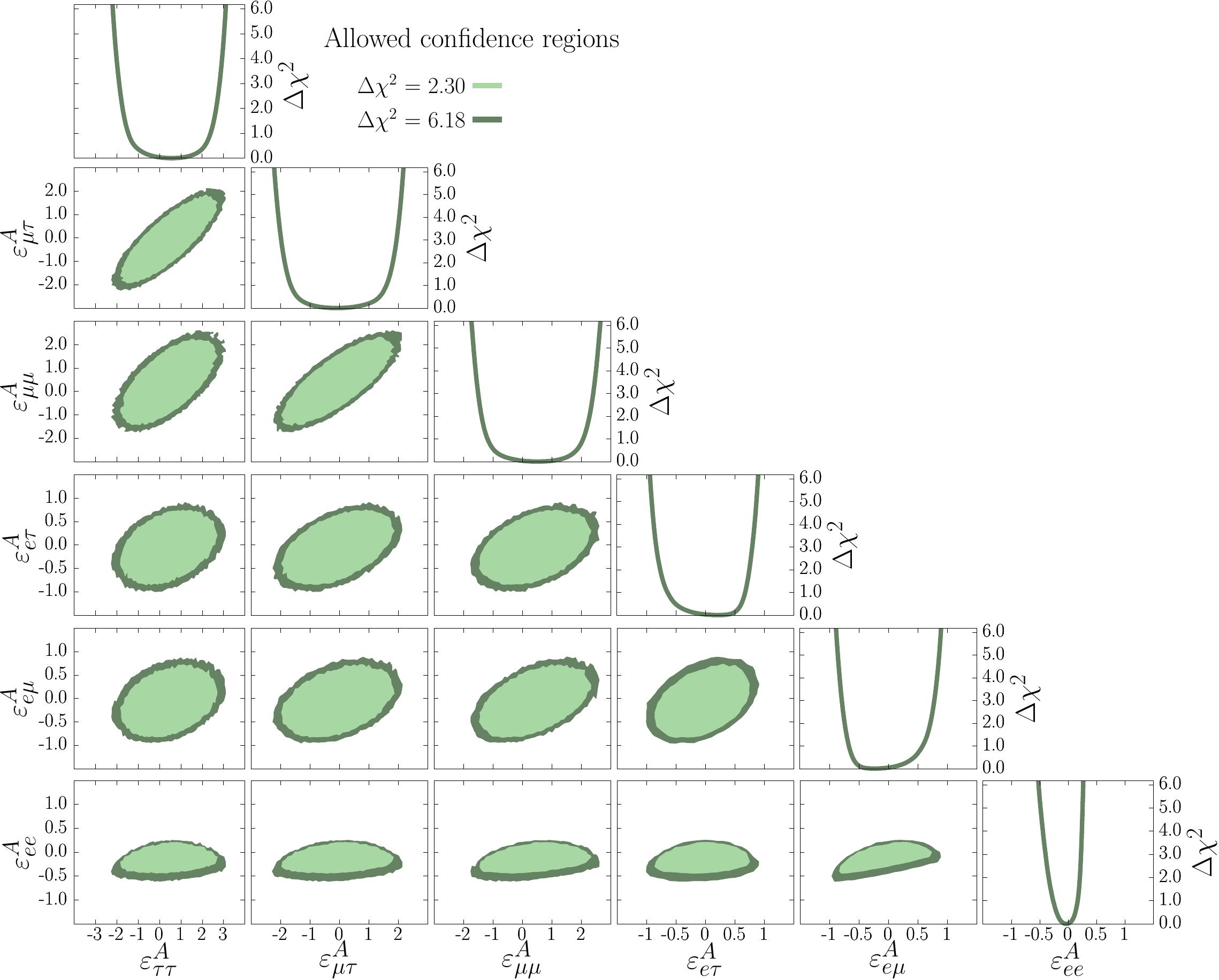}
  \caption{Results for axial NSI with electrons ($\Eps_{\alpha\beta}^L
    = -\Eps_{\alpha\beta}^R$).  The central panels show the
    two-dimensional allowed confidence regions at $1\sigma$ and
    $2\sigma$ (for 2 d.o.f., using two-sided intervals $\Delta\chi^2 =
    2.30$, $6.18$), while the remaining panels show the profile of the
    $\Delta\chi^2$ for each of the parameters individually.  In each
    panel, the results are obtained after minimization over the
    parameters not shown.}
  \label{fig:NSIcorner-a}
\end{figure}

We start discussing the axial NSI case.  For axial NSI, the effect
cancels out in the matter potential, rendering neutrino oscillations
insensitive to the new interactions.  However, Borexino can still
constrain these operators since they affect the interaction cross
section in the detector.  First, we notice that even allowing all 6
axial NSI coefficients to vary freely we get
$\chi^2_\mathrm{min,SM+NSI^A} = 255.7$, which is only one unit lower
that the SM.  So we proceed to derive the bounds on the full parameter
space.

The results are presented in Fig.~\ref{fig:NSIcorner-a} where we show
one-dimensional and two-dimensional projections of
$\Delta\chi^2_\mathrm{SM+NSI^A}$ after minimization over the
parameters not shown in each panel.  We notice that since the
cross-section of flavour-conserving interactions receives contributions
from both SM and BSM terms, which could lead to cancellations among
them, the corresponding regions are generically larger than those
involving only flavour-off-diagonal NSI, for which no SM term is
present.  From the figure we see that when all the parameters are
included in the fit, Borexino is able to constrain NSI roughly only at
the level of $|\Eps^A_{\alpha\alpha}| < \mathcal{O}(2)$ for
flavour-diagonal interactions, and $|\Eps^A_{\alpha\neq\beta}| <
\mathcal{O}(1)$ for the flavour-off-diagonal ones (see also
Tab.~\ref{tab:VA}).  The only exception to this is the parameter
$\Eps_{ee}^A$, for which tighter constraints are obtained, roughly at
the level $|\Eps_{ee}^A|< \mathcal{O}(0.5)$ as it can be seen from the
lowest row in the figure.  The main reason behind this is that the
$\Eps_{ee}^A$ parameter enters the effective cross section in
Eq.~\eqref{eq:nsi-nc} multiplied by $P_{ee}\sim 0.5$, while the
remaining NSI parameters are multiplied either by $P_{e\mu}, P_{e\tau}
\sim 0.2 - 0.25$ or by off-diagonal elements in the density matrix
(which are at most $\sim 0.2$).

Note also that the allowed regions for flavour-diagonal NSI are a bit
asymmetric with respect to zero.  In general, given the large
parameter space it is difficult to identify a single reason for the
specific shape of the allowed regions.  Still, the origin of the
asymmetry can be qualitatively understood if the generalized cross
section matrix in Eq.~\eqref{eq:nsi-elec} is written in terms of the
axial and vector coefficients, defined (in analogy to
Eq.~\eqref{eq:CL-CR}) as
\begin{equation}
  \begin{aligned}
    C^V_{\alpha\beta}
    &= C^L_{\alpha\beta} + C^R_{\alpha\beta} =
    c_{V\beta}\delta_{\alpha\beta} + \Eps^V_{\alpha\beta} \,,
    \\
    C^A_{\alpha\beta}
    &= C^L_{\alpha\beta} - C^R_{\alpha\beta} =
    c_{A\beta}\delta_{\alpha\beta} + \Eps^A_{\alpha\beta} \,.
  \end{aligned}
\end{equation}
In this case, the differential cross section for axial NSI, setting
$\alpha=\beta$, depends on the following combination of parameters:
\begin{equation}
  \label{eq:nsi-cv-ca}
  \frac{1}{2}\left(c_{V \alpha}^2 + c_{A \alpha}^2\right) + c_{A
    \alpha}\Eps_{\alpha\alpha}^A + \frac{(\Eps^A_{\alpha\alpha})^2}{2}
  + \mathcal{O}(y) \,.
\end{equation}
We notice that due to the interference between the SM and NSI
parameters, a second \emph{SM-like} solution appears in
Eq.~\eqref{eq:nsi-cv-ca} with $\Eps^A_{\alpha\alpha} \simeq -2\,
c_{A\alpha}$ (\textit{i.e.}, for which $C^A_{\alpha\alpha} =
-c_{A,\alpha}$).  It can actually be shown that for $\alpha = \mu$ and
$\alpha = \tau$ such \emph{SM-like} solution holds also after
including the full $y$ dependence of the cross section and so it
provides a fit of similar quality to the SM solution.  When only one
NSI parameter is considered at a time this leads to the two
disconnected allowed ranges seen for $\Eps^A_{\mu\mu}$ and
$\Eps^A_{\tau\tau}$ in Table~\ref{tab:VA}.  When all NSI parameters
are included both allowed regions merge in a unique range which is
asymmetric about zero.  In the case of $\Eps^A_{ee}$, on the contrary,
the $y$-dependent terms break such degeneracy and the second
\emph{SM-like} solution gets lifted.

From Eq.~\eqref{eq:nsi-cv-ca} we also see that for
$\Eps^A_{\alpha\alpha} \simeq -c_{A\alpha}$ it is possible to reduce
the cross section with respect to the SM case.  This leads to a
slightly better fit to the data (albeit mildly, at the level of
$\Delta\chi^2 = -1$ as mentioned above) and results into the preferred
solution to be at $\Eps_{ee}^A \lesssim 0$ (since $c_{Ae} \sim 0.5$).
Conversely, for $\Eps_{\mu\mu}^A$ and $\Eps_{\tau\tau}^A$ the fit
shows a mild preference for positive values since $c_{A\mu} =
c_{A\tau} \sim -0.5$.

\begin{table}[t]\centering
  \renewcommand{\arraystretch}{1.2}
  \begin{tabular}{|c || c | c ||c | c ||}
    \hline
    & \multicolumn{4}{|c||}{Allowed regions at 90\% CL $(\Delta\chi^2 = 2.71)$}
    \\
    \cline{2-5}
    & \multicolumn{2}{|c||}{Vector}
    & \multicolumn{2}{|c||}{Axial Vector}
    \\
    \cline{2-5}
    & 1 Parameter & Marginalized
    & 1 Parameter & Marginalized
    \\
    \hline
    $\Eps_{ee}$
    & $[-0.09, +0.14]$ & $[-1.05, +0.17]$
    & $[-0.05, +0.10]$ & $[-0.38, +0.24]$
    \\
    $\Eps_{\mu\mu}$
    & $[-0.51, +0.35]$ & $[-2.38, +1.54]$
    &  $[-0.29, +0.19] \oplus [+0.68, +1.45]$ & $[-1.47, +2.37]$
    \\ $\Eps_{\tau\tau}$
    & $[-0.66, +0.52]$ & $[-2.85, +2.04]$
    & $[-0.40, +0.36] \oplus [+0.69, +1.44]$ & $[-1.82, +2.81]$
    \\
    $\Eps_{e\mu}$
    & $[-0.34, +0.61]$ & $[-0.83, +0.84]$
    & $[-0.30, +0.43]$ & $[-0.79, +0.76]$
    \\
    $\Eps_{e\tau}$
    & $[-0.48, +0.47]$ & $[-0.90, +0.85]$
    & $[-0.40, +0.38]$ & $[-0.81, +0.78]$
    \\
    $\Eps_{\mu\tau}$
    & $[-0.25, +0.36]$ & $[-2.07, +2.06]$
    & $[-1.10, -0.75] \oplus [-0.13, +0.22]$ & $[-1.95, +1.91]$
    \\
    \hline
  \end{tabular}
  \caption{90\% CL bounds (1 d.o.f., 2-sided) on the coefficients of
    NSI operators with electrons, for the vector ($\Eps^V$) and axial
    vector ($\Eps^A$) scenarios.  Results are provided separately for
    two cases: when only one NSI operator is included at at time (``1
    Parameter'') or when the remaining NSI coefficients are allowed to
    float freely in the fit (``Marginalized'').}
  \label{tab:VA}
\end{table}

Let us now turn to the analysis of vector NSI.  In this case NSI
affect the matter potential felt by neutrinos in propagation and a
relevant question is how sensitive the analysis is to this effect, and
whether the sensitivity to NSI is still dominated by the impact on the
neutrino detection cross section.  We have numerically checked that
the results of the analysis are rather insensitive to NSI effects in
propagation.  This is partly so because in the energy window
relevant for Borexino the contribution from the higher-energy components
(mainly \Nuc{8}{B}, but also pep neutrinos) to the event rates is
subdominant, and moreover it is affected by relatively large
backgrounds uncertainties.  As a result we find that the impact of the
matter potential on the fit is very limited and the sensitivity comes
dominantly from NSI effects on the detection cross section.\footnote{A
similar finding was also reported in Ref.~\cite{Borexino:2019mhy},
although only on-diagonal operators were considered in that case.}  We
also find that, even allowing all 6 vector NSI coefficients to vary
freely in the analysis, $\chi^2_\mathrm{min,SM+NSI^V} -
\chi^2_\mathrm{min,SM}$ is only $-0.2$.

\begin{figure}[t]\centering
  \includegraphics[width=0.7\textwidth]{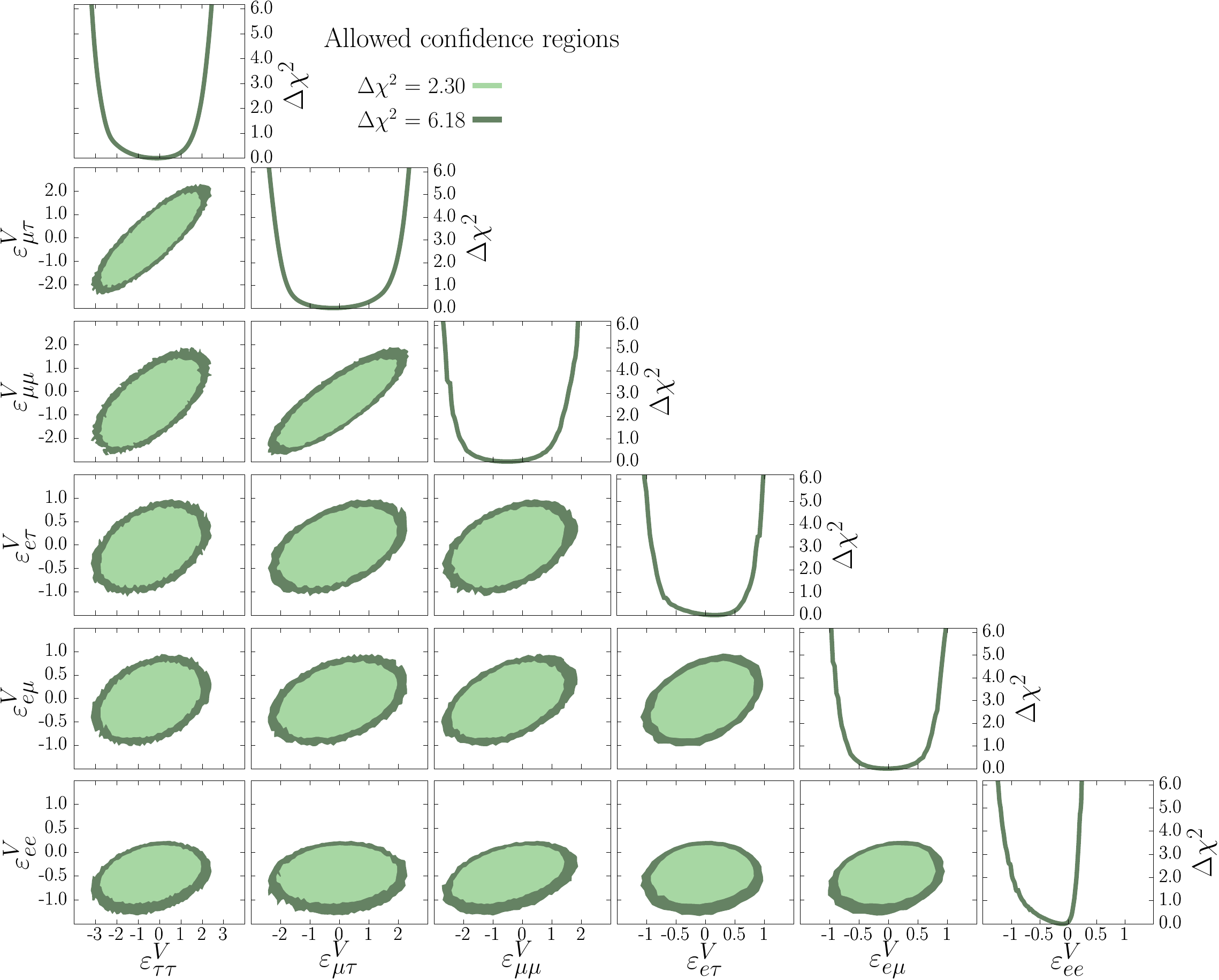}
  \caption{Same as Fig.~\ref{fig:NSIcorner-a}, but for vector NSI with
    electrons ($\Eps_{\alpha\beta}^L = \Eps_{\alpha\beta}^R$).}
  \label{fig:NSIcorner-v}
\end{figure}

Our allowed regions for vector NSI are presented in
Fig.~\ref{fig:NSIcorner-v}, where we see that the potential of
Borexino to test NSI with electrons is again limited in this
multi-parameter scenario: Borexino is only able to constrain NSI at
the level of $|\Eps_{\alpha\alpha}^V| < \mathcal{O}(2-3)$ for
flavour-diagonal interactions, and $|\Eps_{\alpha\beta}^V| <
\mathcal{O}(1)$ for the off-diagonal parameters.  Precise values for
the allowed regions at 90\% CL are given in Tab.~\ref{tab:VA}.

The features describing the allowed regions in the two-dimensional
projections of the $\Delta\chi^2$ are qualitatively similar to the
results found for the axial case, albeit with some differences.  In
particular the regions for $\Eps_{\mu\mu}^V$ and $\Eps_{\tau\tau}^V$
are slightly more symmetric than the axial case.  Again, this can be
qualitatively understood if the cross section in
Eq.~\eqref{eq:nsi-elec} is rewritten in terms of the axial and vector
couplings which results in Eq.~\eqref{eq:nsi-cv-ca} but replacing
$c_{A\alpha} \leftrightarrow c_{V\alpha}$ and $\Eps^A_{\alpha\alpha}
\leftrightarrow \Eps^V_{\alpha\alpha}$.  Thus, a similar interference
as in the axial case takes place here; however since $c_{V\mu} =
c_{V\tau} \sim -0.04$ the quasi-degenerate \emph{SM-like} solution
lies closer to the SM so the region is more symmetric about zero.

For completeness, we also provide our results for NSI involving only
left-handed and right-handed electrons in Tab.~\ref{tab:LR}, where we
again show the results obtained turning on only one operator at a
time, and allowing all NSI operators simultaneously in the fit.  We
have numerically checked that, if only a single operator is included,
we recover the results of the Borexino collaboration for both
left-handed and right-handed NSI (Fig.~5 in
Ref.~\cite{Borexino:2019mhy}) with a very good accuracy. The only
difference is the secondary minimum for $\Eps^L_{ee}\sim -1.3$ which
is outside of the range of parameters explored in
Ref.~\cite{Borexino:2019mhy}.

\begin{table}[t]\centering
  \renewcommand{\arraystretch}{1.2}
  \begin{tabular}{|c || c | c ||c | c ||}
    \hline
    & \multicolumn{4}{|c||}{Allowed regions at 90\% CL $(\Delta\chi^2 = 2.71)$}
    \\
    \cline{2-5} &
    \multicolumn{2}{|c||}{Left-handed} &
    \multicolumn{2}{|c||}{Right-handed}
    \\
    \cline{2-5}
    & 1 Parameter & Marginalized
    & 1 Parameter & Marginalized
    \\
    \hline
    $\Eps_{ee}$
    & $[-1.37, -1.29] \oplus [-0.03, +0.06]$ & $[-1.48, +0.09]$
    & $[-0.23, +0.07]$ & $[-0.53, +0.09]$
    \\
    $\Eps_{\mu\mu}$
    & $[-0.20, +0.13] \oplus [+0.58, +0.81]$ & $[-1.59, +2.21]$
    & $[-0.36, +0.37]$ & $[-1.26, +0.77]$
    \\
    $\Eps_{\tau\tau}$
    & $[-0.26, +0.26] \oplus [+0.45, +0.86]$ & $[-2.00, +2.60]$
    & $[-0.58, +0.47]$ & $[-1.45, +1.04]$
    \\
    $\Eps_{e\mu}$
    & $[-0.17, +0.29]$ & $[-0.78, +0.75]$ & $[-0.21, +0.41]$
    & $[-0.43, +0.42]$
    \\
    $\Eps_{e\tau}$
    & $[-0.26, +0.23]$ & $[-0.81, +0.79]$
    & $[-0.35, +0.31]$ & $[-0.39, +0.43]$
    \\
    $\Eps_{\mu\tau}$
    & $[-0.62, -0.52] \oplus [-0.09, +0.14]$ & $[-1.95, +1.91]$
    & $[-0.26, +0.23]$ & $[-1.10, +1.03]$
    \\
    \hline
  \end{tabular}
  \caption{Constraints at 90\% CL (for 1 d.o.f., 2-sided) on the
    coefficients of NSI operators involving only left-handed
    ($\Eps^L$) or right-handed ($\Eps^R$) electrons. Results are
    provided separately for two cases: when only one NSI operator is
    included at at time (``1 Parameter'') or when the remaining NSI
    coefficients are allowed to float freely in the fit
    (``Marginalized'').}
  \label{tab:LR}
\end{table}

\subsection{Neutrino magnetic moment}
\label{sec:magnetic}

\begin{figure}[t]\centering
  \includegraphics[width=0.65\textwidth]{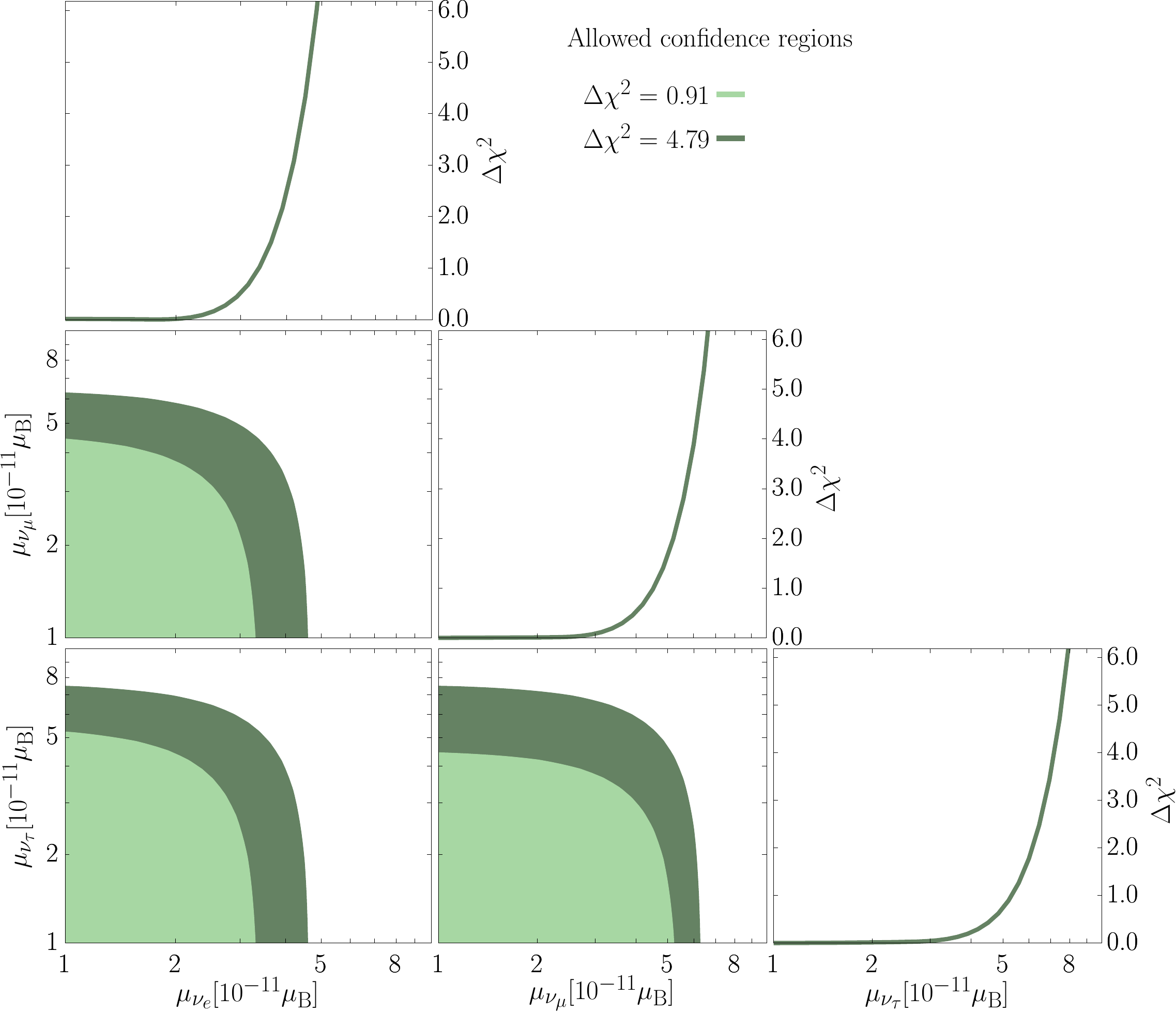}
  \caption{Results for the magnetic moment scenario, assuming three
    independent parameters (one for each of the SM neutrino flavours).
    Since the neutrino magnetic moment is positive definite, the cuts
    have been taken according to a one-sided $\chi^2$ distribution.
    In the central panels, we show the allowed confidence regions (at
    $1\sigma$ and $2\sigma$, for 2 d.o.f.\ using one-sided intervals)
    while in the right-most panels we show the profile of the
    $\Delta\chi^2$ for each of the parameters individually.  In each
    panel, the results are obtained after minimization over the
    undisplayed parameters.}
  \label{fig:magnetic}
\end{figure}

For the analysis with neutrino magnetic moment we find
$\chi^2_\mathrm{min,SM+\mu_\nu} = 256.5$.  In other words, the data do
not show a statistically significant preference for a non-vanishing
magnetic moment, which allows to set constraints on the parameter
space for this scenario.  The corresponding results are summarized in
Fig.~\ref{fig:magnetic}, assuming three independent parameters for the
$\nu_e$, $\nu_\mu$ and $\nu_\tau$ magnetic moments.  In each panel,
the $\Delta\chi^2$ is minimized over the undisplayed parameters.
From the one-dimensional projections we can read the upper bounds at
90\% CL (for 1 d.o.f., one-sided intervals, \textit{i.e.},
{$\Delta\chi^2 = 1.64$})
\begin{equation}
  \label{eq:mubounds}
  \mu_{\nu_e} < 3.7 \times 10^{-11} \mu_B \,,
  \qquad
  \mu_{\nu_\mu} < 5.0 \times 10^{-11} \mu_B \,,
  \qquad
  \mu_{\nu_\tau} < 5.9 \times 10^{-11} \mu_B \,.
\end{equation}
Our results are fully compatible with those in
Ref.~\cite{Borexino:2017fbd}, considering the different choice of
oscillation parameters assumed.  If we assume that all flavours carry
the same value of the neutrino magnetic moment then the dependence on
the oscillation parameters cancels out and we obtain the following
bound (at 90 \% CL for 1 d.o.f., one-sided)
\begin{equation}
  \label{eq:mubounds2}
  \mu_\nu < 2.8 \times 10^{-11} \mu_B \,.
\end{equation}
This coincides precisely with the result obtained by the collaboration
in Ref.~\cite{Borexino:2017fbd}, and serves as validation for the
$\chi^2$ implementation and the choice of systematic uncertainties
adopted in our analysis.

We notice that the Borexino bounds are stronger than those obtained by
other experiments under the same flavour assumptions and CL.  This is
the case for the (over a decade old) experimental results of
GEMMA~\cite{Beda:2010hk, Beda:2012zz} ($\mu_\nu < 2.9 \times 10^{-11}
\mu_B$) and TEXONO~\cite{TEXONO:2006xds} ($\mu_\nu < 7.4 \times
10^{-11} \mu_B$), as well as the bounds derived from recent results
such as CONUS~\cite{CONUS:2022qbb} ($\mu_\nu < 7.5 \times 10^{-11}
\mu_B$) and the combined analysis of Dresden-II and
COHERENT~\cite{Coloma:2022avw} ($\mu_\nu < 1.8 \times 10^{-10}
\mu_B$).

\subsection{Light vector mediators}
\label{sec:light-vec}

The potential of Borexino to probe simplified models with light vector
mediators has been demonstrated in the literature for the $B-L$
case~\cite{Harnik:2012ni, Bilmis:2015lja}, and the bound has been
later recast to other $U(1)'$ models in Refs.~\cite{Kaneta:2016uyt,
  Bauer:2018onh}.  However, Refs.~\cite{Harnik:2012ni, Bilmis:2015lja}
date from 2012 and 2015 and therefore did not include Phase-II data.
Furthermore, their bound was only approximate: it was derived
requiring that the new physics contribution to the event rates should
not exceed the SM expectation.  Here we report our limits for
simplified models, for our data analysis which includes full spectral
information for the Phase II data and a careful implementation of
systematic uncertainties.

\begin{figure}[t]\centering
  \includegraphics[width=0.75\textwidth]{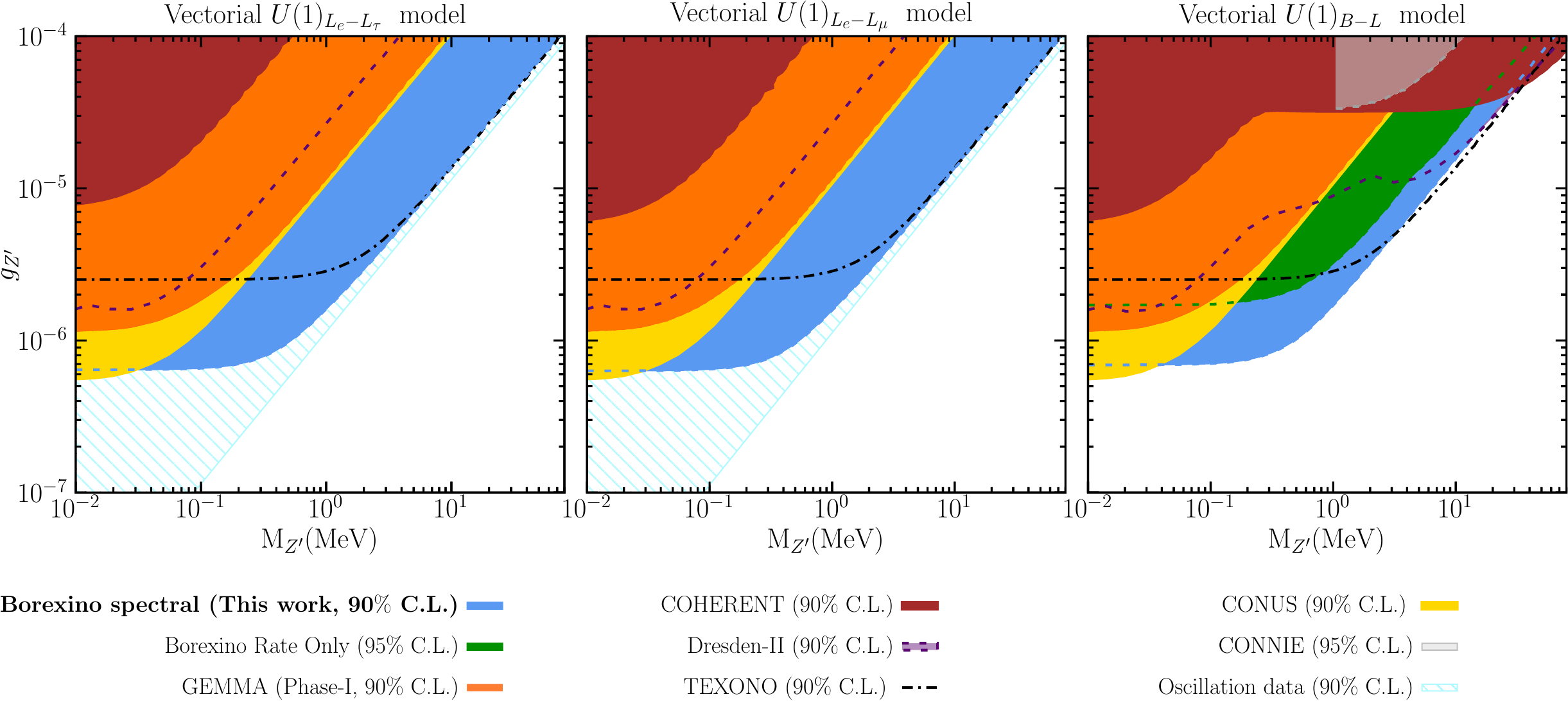}
  \caption{Bounds from our analysis of Borexino Phase II spectral data
    (at 90\% CL for 1 d.o.f., using two-sided intervals,
    \textit{i.e.}, $\Delta\chi^2=2.71$) on the vector mediators
    associated to a new $U(1)'$ symmetry, as indicated in the top
    label for each panel.  Our results are compared to our derived
    bounds from GEMMA~\cite{Beda:2010hk, Coloma:2020gfv},
    TEXONO~\cite{Deniz:2009mu, Coloma:2020gfv},
    COHERENT~\cite{COHERENT:2017ipa, COHERENT:2020iec,
      Coloma:2022avw}, and Dresden-II reactor
    experiment~\cite{Colaresi:2022obx, Coloma:2022avw}, as well as
    those derived from a global fit to oscillation
    data~\cite{Coloma:2020gfv}, all of them obtained with the same
    statistical criterion.  We also show the bounds from
    CONNIE~\cite{CONNIE:2019xid} (95\% CL) and
    CONUS~\cite{CONUS:2021dwh} (90\% CL) which we have obtained
    applying an overall rescaling factor to their published bounds on
    a universally coupled vector (see text for details).  For the
    $B-L$ model, the line labelled ``Borexino Rate Only'' shows the
    approximate bound derived (at 95\% CL, 1 d.o.f., for one-sided
    intervals) in Ref.~\cite{Bilmis:2015lja} (see also
    Ref.~\cite{Harnik:2012ni}).}
  \label{fig:vector-med}
\end{figure}

Our results for vector mediators are shown in Fig.~\ref{fig:vector-med} for three anomaly-free models with couplings to $B-L$, $L_e-L_\mu$, or $L_e-L_\tau$. The similarity between constraints arises from the dominant interference term in Eq. ~\eqref{eq:csvece}, where the event rate modification
\begin{equation}
  \Delta N_\text{ev} \propto P_{ee}\, q_{Z'}^{\nu_e}\, q_{Z'}^e\, c_{Ve}
\end{equation}
dominates due to $c_{Ve} \gg c_{V\mu(\tau)}$ and $P_{ee} > P_{e\mu(\tau)}$. Identical $q_{Z'}^{\nu_e} q_{Z'}^e = 1$ values yield comparable bounds across models.

We compare Borexino constraints with previous limits from neutrino-electron scattering (GEMMA Phase-I~\cite{Beda:2010hk}, TEXONO~\cite{Deniz:2009mu}), CE$\nu$NS measurements (COHERENT~\cite{COHERENT:2020iec,COHERENT:2017ipa}, Dresden-II~\cite{Colaresi:2022obx}), oscillation data~\cite{Coloma:2020gfv}, and CE$\nu$NS/ES searches (CONNIE~\cite{CONNIE:2019xid}, CONUS~\cite{CONUS:2021dwh}). For $B-L$, pre-2015 ``Rate only'' Borexino estimates~\cite{Harnik:2012ni,Bilmis:2015lja} are included. Technical details follow established methodologies for quenching factors~\cite{Coloma:2022avw} and confidence levels~\cite{Rink:2022rsx}.

Our Borexino Phase-II analysis improves constraints by $\sim$60\% for light mediators and $\sim$30\% on $g_{Z'}/M_{Z'}$ in contact-interaction regimes. While surpassing most scattering experiment bounds, oscillation data remain dominant for $L_e-L_{\mu/\tau}$ models.

\subsection{Light scalar and pseudoscalar mediators}
\label{sec:light-phi}

\begin{figure}[t]\centering
  \includegraphics[width=0.75\textwidth]{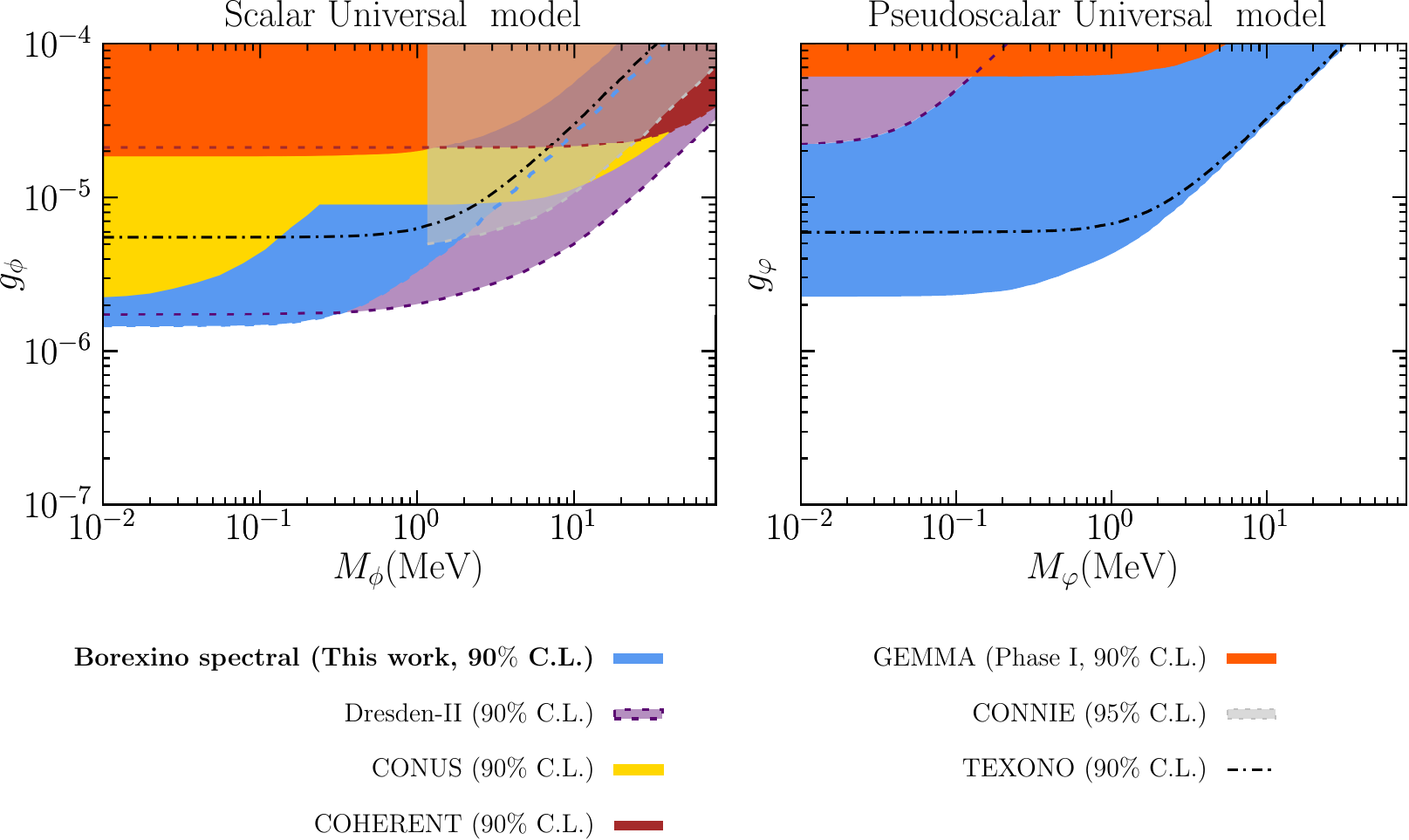}
  \caption{Bounds from our analysis of Borexino Phase-II spectral data
    on a scalar (left panel) and pseudoscalar (right panel) mediator
    (at 90\% CL for 1 d.o.f., using two-sided intervals,
    \textit{i.e.}, $\Delta\chi^2=2.71$) which couples universally to
    all fermions in the SM.  Our results are compared to our derived
    bounds from GEMMA~\cite{Beda:2010hk, Coloma:2020gfv},
    TEXONO~\cite{Deniz:2009mu, Coloma:2020gfv},
    COHERENT~\cite{COHERENT:2017ipa, COHERENT:2020iec,
      Coloma:2022avw}, and Dresden-II reactor
    experiment~\cite{Colaresi:2022obx, Coloma:2022avw}, obtained
    using the same statistical criterion.  For comparison we also show
    the bounds from CONNIE~\cite{CONNIE:2019xid} (95~\%CL) and
    CONUS~\cite{CONUS:2021dwh} (90\%~CL).}
  \label{fig:scalar-med}
\end{figure}

Next let us discuss our results for a scalar (pseudoscalar) mediator
coupled universally to the SM fermions, shown in the left (right)
panel of Fig.~\ref{fig:scalar-med}.   For comparison, we also show the bounds
derived from neutrino scattering measurements, computed following (or
obtained from) the same references as in Sec.~\ref{sec:light-vec}.Our results for scalar (pseudoscalar) mediators with universal SM couplings are shown in Fig.~\ref{fig:scalar-med} left (right) panel. Unlike vector mediators, scalar/pseudoscalar scenarios evade oscillation constraints. Following the methodology from Sec.~\ref{sec:light-vec}, we compare neutrino scattering bounds which are generally weaker than vector counterparts due to: (1)~quartic coupling dependence (no SM interference) versus quadratic for vectors, and (2)~cross-section suppression factors $T_e m_e/E_\nu^2$ (scalar) and $T_e^2/E_\nu^2$ (pseudoscalar) relative to vector interactions in Eqs.~\eqref{eq:csscale}-\eqref{eq:csvece}.

For scalar mediators (left panel), Borexino Phase II improves over reactor experiments (GEMMA/TEXONO) but Dresden-II provides dominant constraints except at $M_\phi \sim 0.5$~MeV. This stems from enhanced CE$\nu$NS sensitivity via scalar-quark couplings~\cite{Coloma:2022avw}. Conversely for pseudoscalars (right panel), Borexino leads for $M_\varphi \lesssim 5$~MeV as CE$\nu$NS contributions become negligible (spin-dependent coupling). Dresden-II limits here consider only ES contributions, while missing CONUS pseudoscalar bounds (not studied experimentally) would likely remain weaker than Borexino due to $T_e/m_e \sim 2\times10^{-3}$ suppression.

TEXONO provides comparable pseudoscalar constraints, though all scattering limits weaken significantly compared to vector case. The distinct suppression mechanisms and interaction topologies underscore the importance of combined spectral analyses across multiple interaction types.

\section{Constraining NSI with electrons and quarks with global oscillation data}
\label{sec:results_glob_bsm}

This section summarizes the main results of our study using global analysis to constrain NCNSIs.  In
Sec.~\ref{sec:sim} we first review the data included in the fit,
introduce our $\chi^2$ definition, and outline the details related to
the sampling of the multi-dimensional parameter space.  We then
proceed to present our results for NSI with electrons
(Sec.~\ref{sec:resule}), with quarks (Sec.~\ref{sec:resulq}), and for
simultaneous NSI with electrons and quarks (Sec.~\ref{sec:resulgen}).
The status of the LMA-D solution in this general case is discussed in
Sec.~\ref{sec:resulLMAD}.

\subsection{Simulation details}
\label{sec:sim}

The data samples included in our oscillation analysis mostly coincide
with those in NuFIT-5.2~\cite{nufit-5.2}\footnote{ In brief, in the analysis
of solar neutrino data we consider the total rates from the
radiochemical experiments Chlorine~\cite{Cleveland:1998nv},
Gallex/GNO~\cite{Kaether:2010ag}, and SAGE~\cite{Abdurashitov:2009tn},
the spectral data (including day-night information) from the four
phases of Super-Kamiokande in Refs.~\cite{Hosaka:2005um,
  Cravens:2008aa, Abe:2010hy,SK:nu2020}, the results of the three
phases of SNO in the form of the day-night spectrum data of
SNO-I~\cite{Aharmim:2007nv}, and SNO-II~\cite{Aharmim:2005gt} and the
three total rates of SNO-III~\cite{Aharmim:2008kc}, and the spectra from Borexino
Phase-I~\cite{Bellini:2011rx, Bellini:2008mr}, and
Phase-II~\cite{Borexino:2017rsf}.  For reactor neutrinos we include
the separate DS1, DS2, DS3 spectra from KamLAND~\cite{Gando:2013nba}
with Daya Bay reactor $\nu$ fluxes~\cite{An:2016srz}, the FD/ND
spectral ratio, with 1276-day (FD), 587-day (ND) exposures of
Double-Chooz~\cite{DoubleC:nu2020}, the 3158-day separate EH1, EH2,
EH3 spectra~\cite{DayaBay:2022orm} of Daya-Bay, and the 2908-day FD/ND
spectral ratio of RENO~\cite{RENO:nu2020}.  For atmospheric neutrinos
we use the four phases of Super-Kamiokande (up to 1775 days of
SK4~\cite{Wendell:2014dka}), the complete set of DeepCore 3-year
$\mu$-like events presented in Ref.~\cite{Aartsen:2014yll} and
publicly released in Ref.~\cite{deepcore:2016}, and the results on
$\nu_\mu$-induced upgoing muons reported by
IceCube~\cite{TheIceCube:2016oqi} based on one year of data taking.
Finally, for LBL experiments we include the final neutrino and
antineutrino spectral data on $\nu_e$-appearance and
$\nu_\mu$-disappearance in MINOS~\cite{Adamson:2013whj}, the
$19.7\times 10^{20}$ pot $\nu_\mu$-disappearance and $16.3\times
10^{20}$ pot $\bar\nu_\mu$-disappearance data in
T2K~\cite{T2K:nu2020}, and the $13.6\times 10^{20}$ pot
$\nu_\mu$-disappearance and $12.5\times 10^{20}$ pot
$\bar{\nu}_\mu$-disappearance data in NO$\nu$A~\cite{NOvA:nu2020}.
Notice that to ensure full consistency with our CP-conserving
parametrization we have chosen not to include in the present study the
data from the $\nu_e$ and $\bar\nu_e$ appearance channels in NO$\nu$A
and T2K.  With this data we construct
$\chi^2_\text{OSC}(\vec\omega,\vec\Eps)$ where we denote by
$\vec\omega$ the 3$\nu$ oscillation parameters and $\vec\Eps$ the NSI
parameters considered in the analysis.}. 
When combining with CE$\nu$NS we include the results from COHERENT
data on CsI~\cite{COHERENT:2017ipa, COHERENT:2018imc} and Ar
targets~\cite{COHERENT:2020iec, COHERENT:2020ybo} (see
Refs.~\cite{Coloma:2019mbs, Coloma:2022avw} for details).  In
particular, for the analysis of COHERENT CsI data we use the quenching
factor from Ref.~\cite{Collar:2019ihs} and the nuclear form factor
from Ref.~\cite{Klos:2013rwa}.  For the analysis of COHERENT Ar data
we use the quenching factor provided by the COHERENT collaboration in
Ref.~\cite{COHERENT:2020ybo} and Helm~\cite{Helm:1956zz} nuclear form
factor (the values of the parameters employed are the same as in
Ref.~\cite{Coloma:2022avw}).  For CE$\nu$NS searches using reactor
neutrinos at Dresden-II reactor experiment~\cite{Colaresi:2022obx} we
follow the analysis presented in Ref.~\cite{Coloma:2022avw} with YBe
quenching factor~\cite{Collar:2021fcl} (the characteristic
momentum-transfer in this experiment is very low so the nuclear form
factor can be taken to be 1).  With all this we construct the
corresponding $\chi^2_\text{COH,CsI}(\vec\Eps)$,
$\chi^2_\text{COH,Ar}(\vec\Eps)$, and
$\chi^2_\text{D-II,Ge}(\vec\Eps)$.

Our goal is to find the global minimum of the total $\chi^2$ which,
unless otherwise stated, is obtained adding the contributions from our
global analysis of oscillation data (<<GLOB-OSC w NSI in ES>>) and
CE$\nu$NS data:
\begin{equation}
  \chi^2 (\vec\omega, \vec\Eps)
  = \chi^2_\text{OSC}(\vec\omega, \vec \Eps)
  + \chi^2_\text{CE$\nu$NS} (\vec\Eps)
\end{equation}
where we have defined
\begin{equation}
  \chi^2_\text{CE$\nu$NS}
  = \chi^2_\text{COH,CsI} + \chi^2_\text{COH,Ar} + \chi^2_\text{D-II,Ge} \,.
\end{equation}
The minimum of the total $\chi^2$ is obtained after minimization over
all the nuisance parameters, which are included as pull terms in our
fit.\footnote{For details on the numerical implementation of
systematic uncertainties see Refs.~\cite{Coloma:2022avw,
  Coloma:2022umy, nufit-5.2}.}  Vector NSI effects in detection can be suppressed while maintaining oscillation impacts depending on mediator mass~\cite{Farzan:2015doa}. For oscillations, light mediators require $\Mmed \gtrsim \mathcal{O}(10^{-12})$ eV~\cite{Gonzalez-Garcia:2006vp,Coloma:2020gfv}. Detection effects require momentum transfer $q$: 

\begin{itemize}
    \item Elastic scattering (ES): Borexino ($q \sim 500$~keV) constrains $\Mmed \ll 500$~keV (<<GLOB-OSC w/o NSI in ES>>), while SNO/SK ($q \sim 5-10$~MeV) require $\Mmed \gtrsim 10$~MeV (<<GLOB-OSC w NSI in ES>>). Intermediate masses ($0.5-10$~MeV) primarily affect Borexino.
    
    \item CE$\nu$NS: COHERENT ($q \sim 30-50$~MeV) and Dresden-II ($q \sim 5$~MeV) constrain $\Mmed \gtrsim 50$~MeV (<<GLOB-OSC+CE$\nu$NS>>) versus $\Mmed \ll 5$~MeV (<<GLOB-OSC>>). Intermediate $5-50$~MeV ranges require case-by-case analysis.
\end{itemize}

Axial-vector NSI exclusively affect detection: ES bounds follow mediator thresholds (e.g., $\Mmed \gtrsim 3$~MeV for SNO NC interactions). 

Our global fit scans parameters are:
\begin{equation}
    \theta_{12},\, \theta_{23},\, \Dmq_{21},\, \Dmq_{31},\, \Eps_\oplus,\, \varphi_{12},\, \varphi_{23},\, \Eps_{\mu\mu}^\oplus,\, \eta,\, \zeta
\end{equation}
with $\delta_\text{CP} \in \{0,\pi\}$. Since the determination of $\theta_{13}$ is
  dominated by reactor data which are mostly insensitive to NSI
  (see Sec.~\ref{sec:formalism-earth}), we fix
  $\sin^2\theta_{13} = 0.022$. Atmospheric analyses are perfomed in the limit $\Dmq_{21}
\to 0$. Parameter space exploration uses
MultiNest~\cite{Feroz:2013hea,Feroz:2008xx} and
Diver~\cite{Martinez:2017lzg}, accounting for detection-induced
breaking of diagonal NSI degeneracies through $\Eps_{\mu\mu}^\oplus$.

\begin{figure}[t]\centering
  \includegraphics[width=0.7\textwidth]{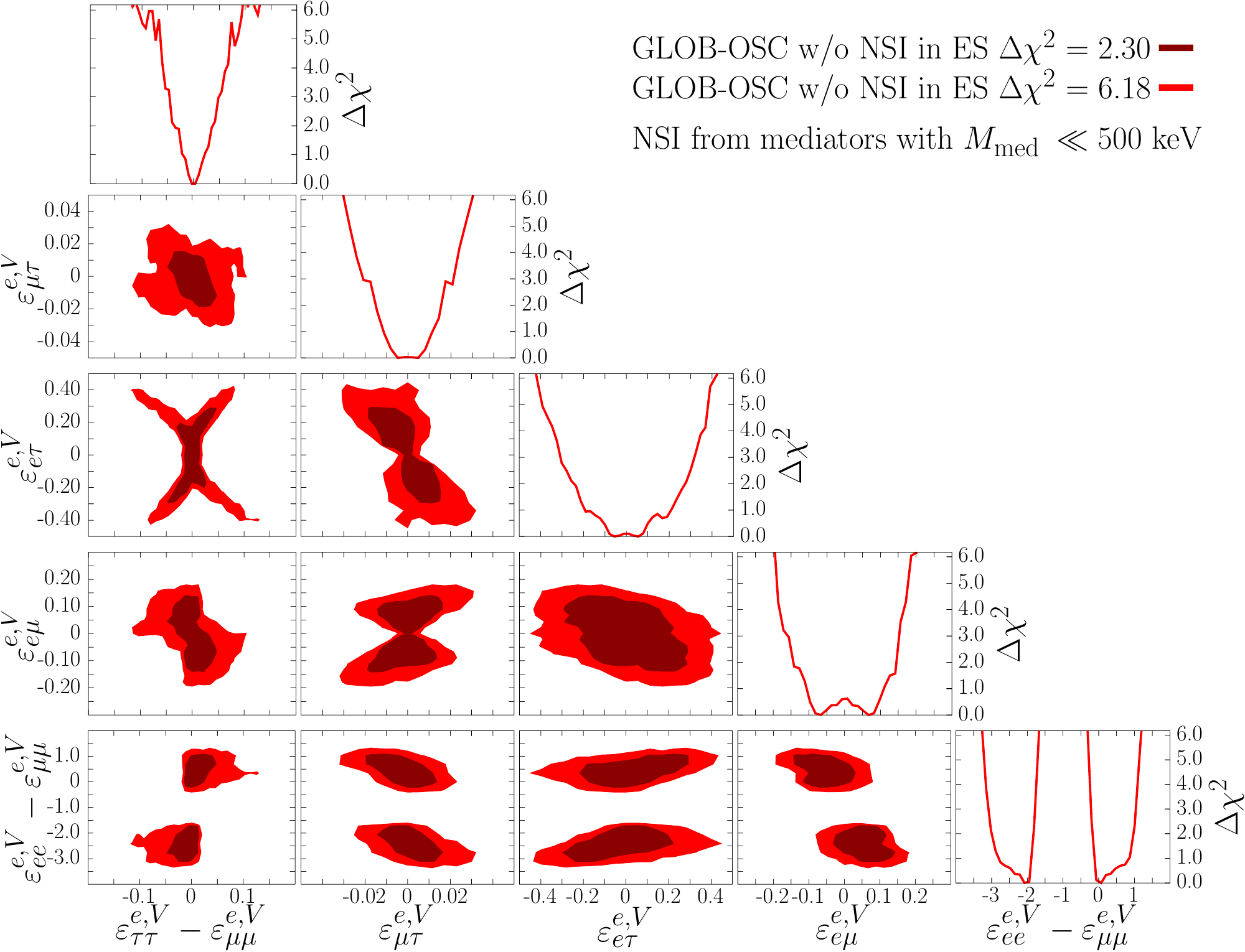}
  \caption{Constraints on the coefficients for vector NSI with
    electrons from the global analysis of oscillation data
    \textit{without including the effect of NSI in the detection cross
      section}.  Each panel shows a two-dimensional projection of the
    allowed multi-dimensional parameter space after minimization with
    respect to the undisplayed parameters.  The regions correspond to
    $1\sigma$ and $2\sigma$ (2 d.o.f.).}
  \label{fig:etriangVwoES}
\end{figure}

\begin{table}[t]\centering
  \catcode`?=\active\def?{\hphantom{0}}
  \begin{tabular}{|l||c|c|}
    \hline
    & \multicolumn{2}{|c|}{Allowed ranges at 90\% CL (marginalized)}
    \\
    \hline
    & \multicolumn{2}{|c|}{GLOB-OSC w/o NSI in ES}
    \\
    \hline
    & LMA & $\text{LMA}\oplus\text{LMA-D}$
    \\
    \hline
    $\Eps_{ee}^{e,V} - \Eps_{\mu\mu}^{e,V}$
    & $[-0.21, +1.0]?$
    & $[-3.0, -1.8] \oplus[-0.21, +1.0]$
    \\
    $\Eps_{\tau\tau}^{e,V} - \Eps_{\mu\mu}^{e,V}$
    & $[-0.015, +0.048]$
    & $[-0.040, +0.047]$
    \\
    $\Eps_{e\mu}^{e,V}$
    & $?[-0.15, +0.035]$
    & $[-0.15, +0.14]$
    \\
    $\Eps_{e\tau}^{e,V}$
    & $[-0.21, +0.31]$
    & $[-0.29, +0.31]$
    \\
    $\Eps_{\mu\tau}^{e,V}$
    & $[-0.020, +0.012]$
    & $[-0.020, +0.017]$
    \\
    \hline
    \end{tabular}
  \caption{90\% CL bounds (1 d.o.f., 2-sided) on the coefficients of
    NSI operators with electrons after marginalizing over all other
    NSI and oscillation parameters.  The bounds are derived from the
    global analysis of oscillation data \textit{without including the
      effect of NSI in the ES cross section} (NSI induced by mediators
    with mass $\Mmed \ll 500~\text{keV}$, see Sec.~\ref{sec:sim}).
    The ranges in the first column (labeled <<LMA>>) correspond to an
    analysis in which we restrict $\theta_{12} <45^\circ$.  In the
    second column (labeled <<$\text{LMA}\oplus\text{LMA-D}$>>), both
    $\theta_{12} <45^\circ$ and $\theta_{12} >45^\circ$ are allowed.
    The same bounds hold for vector NSI with protons.}
  \label{tab:nsieVwoES}
\end{table}

\subsection{New constraints on NSI with electrons}
\label{sec:resule}

\begin{figure}[t]\centering
  \includegraphics[width=0.7\textwidth]{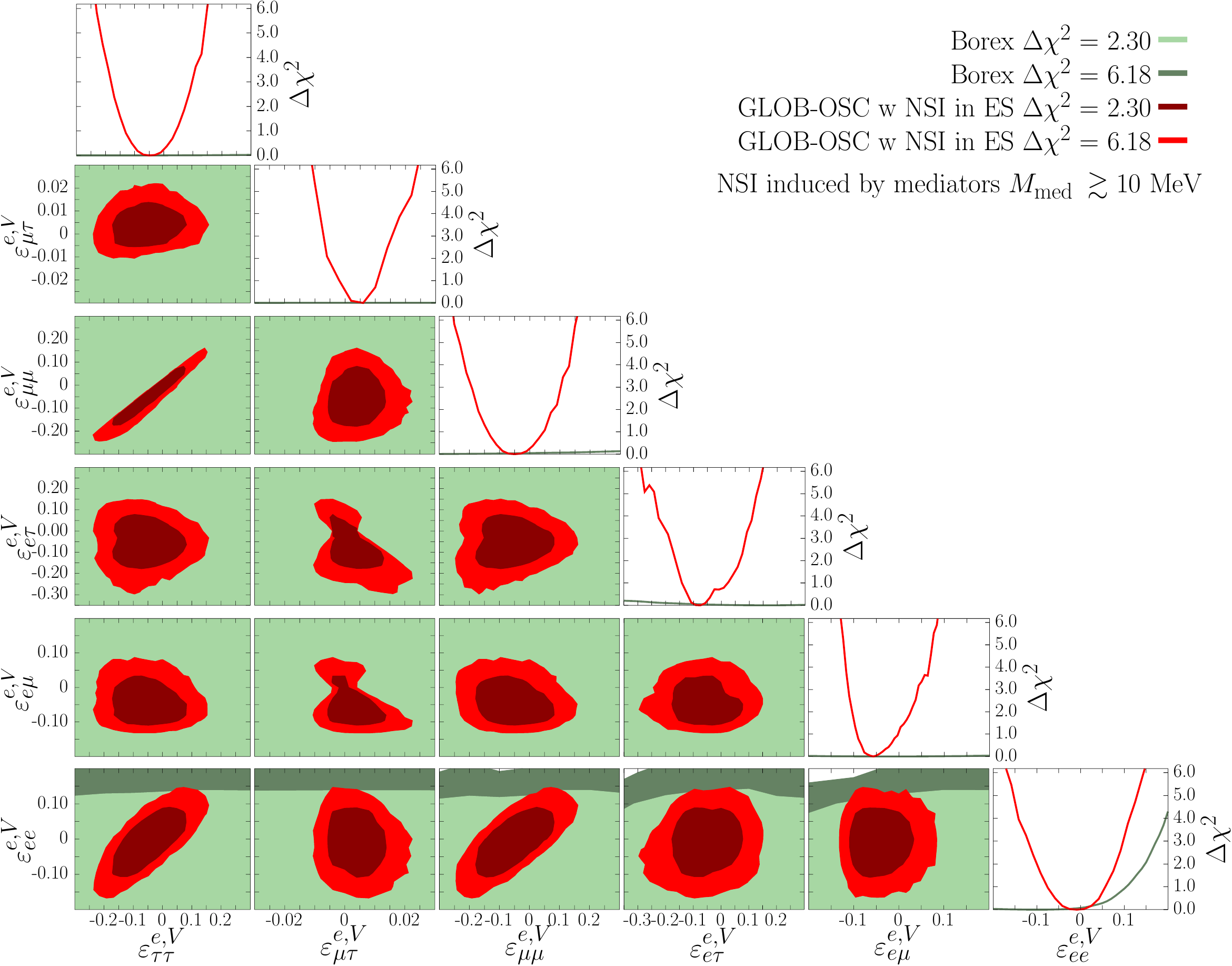}
  \caption{Constraints on the coefficients for vector NSI with
    electrons from the global analysis of oscillation data
    \emph{including the effect of NSI in the ES cross section}.  Each
    panel shows a two-dimensional projection of the allowed
    multi-dimensional parameter space after minimization with respect
    to the undisplayed parameters.  The contours correspond to
    $1\sigma$ and $2\sigma$ (2 d.o.f.).  The closed red regions
    correspond to the global oscillation analysis which involves the
    six NSI plus five oscillation parameters.  For the sake of
    comparison we also show as green regions the constraints obtained
    from the analysis of full Borexino Phase-II spectrum in
    Figure \ref{fig:NSIcorner-v}.}
  \label{fig:etriangVwES}
\end{figure}

The global analysis of neutrino data  reveals significant improvements in constraining NC NSI with electrons compared to the Borexino Phase-II spectral analysis presented in Sec.~\ref{sec:NSI_BX}.  We plot in
Fig.~\ref{fig:etriangVwoES} and~\ref{fig:etriangVwES} the constraints
on the different coefficients for the two scenarios outlined in the
previous section, without and with NSI effects in the detection cross
section, respectively.\footnote{Notice that from the point of view of
the data analysis the results from the <<GLOB-OSC w/o NSI in ES>> are
totally equivalent to those obtained for vector NSI which couple only
to protons.}  The corresponding 90\% CL allowed ranges are listed in
Table~\ref{tab:nsieVwoES} and on the left columns in
Table~\ref{tab:nsiewES} respectively.

\begin{table}[t]\centering
  \catcode`?=\active\def?{\hphantom{0}}
  \renewcommand{\arraystretch}{1.2}
  \begin{tabular}{|c || c | c || c | c ||}
    \hline
    & \multicolumn{4}{c||}{Allowed ranges at 90\% CL (marginalized)}
    \\
    \hline
    & \multicolumn{2}{c||}{Vector ($X=V$)}
    & \multicolumn{2}{c||}{Axial-vector ($X=A$)}
    \\
    \hline
    & Borexino & \small{GLOB-OSC w NSI in ES}
    & Borexino & \small{GLOB-OSC w NSI in ES}
    \\
    \hline
    $\Eps^{e,X}_{ee}$
    & $?[-1.1, +0.17]$
    & $[-0.13, +0.10]$
    & $[-0.38, +0.24]$
    & $[-0.13, +0.11]$
    \\
    $\Eps^{e,X}_{\mu\mu}$
    & $[-2.4, +1.5]$
    & $[-0.20, +0.10]$
    & $[-1.5, +2.4]$
    & $[-0.70, +1.2]?$
    \\ $\Eps^{e,X}_{\tau\tau}$
    & $[-2.8, +2.1]$
    & $?[-0.17, +0.093]$
    & $[-1.8, +2.8]$
    & $[-0.53, +1.0]?$
    \\
    $\Eps^{e,X}_{e\mu}$
    & $[-0.83, +0.84]$
    & $[-0.097, +0.011]$
    & $[-0.79, +0.76]$
    & $[-0.41, +0.40]$
    \\
    $\Eps^{e,X}_{e\tau}$
    & $[-0.90, +0.85]$
    & $?[-0.18, +0.080]$
    & $[-0.81, +0.78]$
    & $[-0.36, +0.36]$
    \\
    $\Eps^{e,X}_{\mu\tau}$
    & $[-2.1, +2.1]$
    & $[-0.0063, +0.016]?$
    & $[-1.9, +1.9]$
    & $[-0.79, +0.81]$
    \\
    \hline
  \end{tabular}
  \caption{90\% CL bounds (1 d.o.f., 2-sided) on the coefficients of
    vector NSI operators with electrons after marginalizing over all
    other NSI and oscillation parameters.  The bounds are derived from
    the global analysis of oscillation data \textit{including the
      effect of NSI in the ES cross section}.  For comparison, the
    results obtained from the analysis of Borexino Phase-II data in
    Ref.~\cite{Coloma:2022umy} are also shown for comparison.  Note
    that these bounds apply to interactions induced by mediators with
    masses $\Mmed \gtrsim 10~\text{MeV}$, see Sec.~\ref{sec:sim}.}
  \label{tab:nsiewES}
\end{table}

The first thing to notice is that in the scenario <<GLOB-OSC w/o NSI
in ES>>, for the flavour diagonal coefficients, only the combinations
$\Eps_{ee}^{e,V} - \Eps_{\mu\mu}^{e,V}$ and $\Eps_{\tau\tau}^{e,V} -
\Eps_{\mu\mu}^{e,V}$ can be constrained, and two separate allowed
ranges appear (see bottom row in Fig.~\ref{fig:etriangVwoES}): one
around $\Eps_{ee}^{e,V} - \Eps_{\mu\mu}^{e,V}\sim 0$ and another
around $\Eps_{ee}^{e,V} - \Eps_{\mu\mu}^{e,V}\sim -2$.  This is
nothing else than the result of the generalized mass-ordering
degeneracy of Eq.~\eqref{eq:NSI-deg}.  The two disjoint allowed
solutions correspond to regions of oscillation parameters with
$\theta_{12} <45^\circ$ (with ranges labeled <<LMA>> in
Table~\ref{tab:nsieVwoES}) and with $\theta_{12} >45^\circ$ (with
ranges labeled <<LMA-D>> in Table~\ref{tab:nsieVwoES}) respectively.
As discussed in Sec.~\ref{sec:formOSC}, the degeneracy is partly
broken by the variation of chemical composition of the matter along
the neutrino trajectory when NSI coupling to neutrons are involved.
But in this case, with only coupling to electrons, the degeneracy is
perfect as can be seen by the fact that $\Delta\chi^2=0$ in both
minima of $\Eps_{ee}^{e,V} - \Eps_{\mu\mu}^{e,V}$.
From the panels on the lower row we observe that the projection of the
allowed regions corresponding to each of the two solutions partly
overlap for $\Eps_{\tau\tau}^{e,V} - \Eps_{\mu\mu}^{e,V}$ and
$\Eps_{\alpha\neq\beta}^{e,V}$, which leads to the non-trivial shapes
of some of the corresponding two-dimensional regions for some pairs of
those paremeters.

\begin{figure}[t]\centering
  \includegraphics[width=0.7\textwidth]{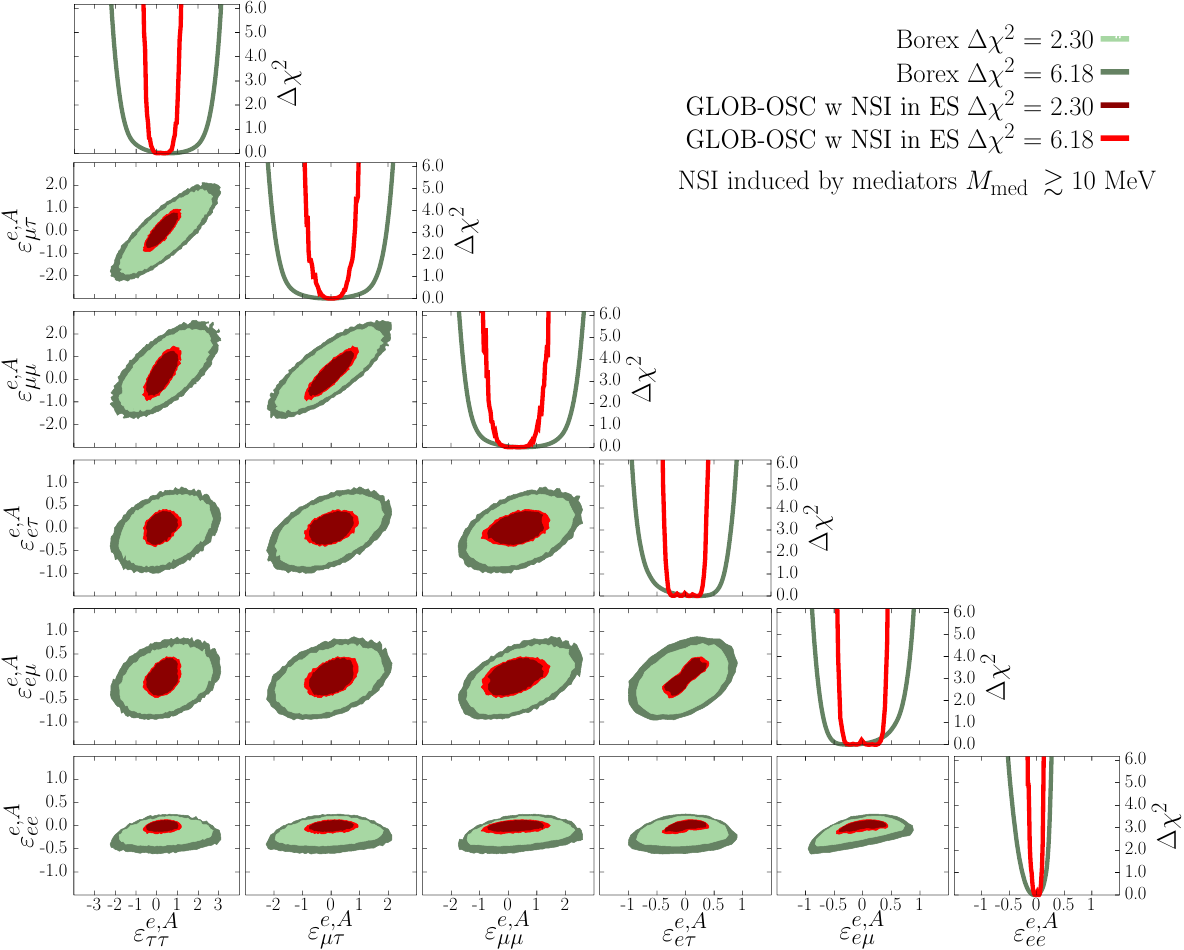}
  \caption{Same as Fig.~\ref{fig:etriangVwES} but for Axial-vector
    NSI, compared with Fig. ~\ref{fig:NSIcorner-a}.}
  \label{fig:etriangA}
\end{figure}

The results for the scenario <<GLOB-OSC w NSI in ES>> in
Fig.~\ref{fig:etriangVwES}, and on the left columns in
Table~\ref{tab:nsiewES}, show that the inclusion of the effect of the
vector NSI in the ES cross sections in Borexino, SNO, and SK totally
lifts the degeneracy.  In this scenario only the LMA solution is
allowed, and $\Eps_{ee}^{e,V}$, $\Eps_{\mu\mu}^{e,V}$, and
$\Eps_{\tau\tau}^{e,V}$ can be independently constrained.  Notice also
that in this case, the lifting of the LMA-D solution leads to allowed
regions with more standard (close-to-elliptical) shapes.  For the sake
of comparison we show for this scenario the corresponding bounds
derived from the analysis of Borexino spectra in
Section~\ref{sec:NSI_BX}.  The comparison shows that for vector NSI
with electrons the global analysis of the oscillation data reduces the
allowed ranges of the NSI coefficients by factors $\sim\text{4--200}$
with respect to those derived with Borexino spectrum only.  In other
words, the NSI contribution to ES is important to break the LMA-D
degeneracy and to impose independent bounds on the three
flavour-diagonal NSI coefficients, but within the LMA solution the
effect of the vector NSI on the matter potential also leads to
stronger constraints.  This is further illustrated by the results
obtained for the analysis with axial-vector NSI which are shown in
Fig.~\ref{fig:etriangA} and right columns in Table~\ref{tab:nsiewES}.
Axial-vector NSI do not contribute to the matter potential and
therefore the difference between the results of the global oscillation
and the Borexino-only analysis in this case arises solely from the
effect of the axial-vector NSI on the ES cross section in SNO and SK.
Comparing the red and green regions in Fig.~\ref{fig:etriangA} and the
two left columns in Table~\ref{tab:nsiewES} we see that for
axial-vector NSI the improvement over the bounds derived with
Borexino-only analysis is just a factor $\sim$ 2--3.

\subsection{Updated constraints on NSI with quarks}
\label{sec:resulq}

Next we briefly summarize the results of the analysis for the
scenarios of NSI with either up or down quarks (more general
combinations of couplings to quarks and electrons will be presented in
the next section).  For vector NSI this updates and complements the
results presented in Refs.~\cite{Esteban:2018ppq, Coloma:2019mbs} by
accounting for the effects of increased statistics in the oscillation
experiments, including the addition of new data from Borexino
Phase-II.  Notice also that in the present analysis, as mentioned
above, the treatment of the SNO data is different than in
Refs.~\cite{Esteban:2018ppq, Coloma:2019mbs}.  Furthermore when
combining with CE$\nu$NS we include here the results from COHERENT
both on CsI~\cite{COHERENT:2017ipa, COHERENT:2018imc} and Ar
targets~\cite{COHERENT:2020iec, COHERENT:2020ybo}, together with the
recent results from CE$\nu$NS searches using reactor neutrinos at
Dresden-II reactor experiment~\cite{Colaresi:2022obx, Coloma:2022avw}.

\begin{figure}[t]\centering
  \includegraphics[width=0.5\textwidth]{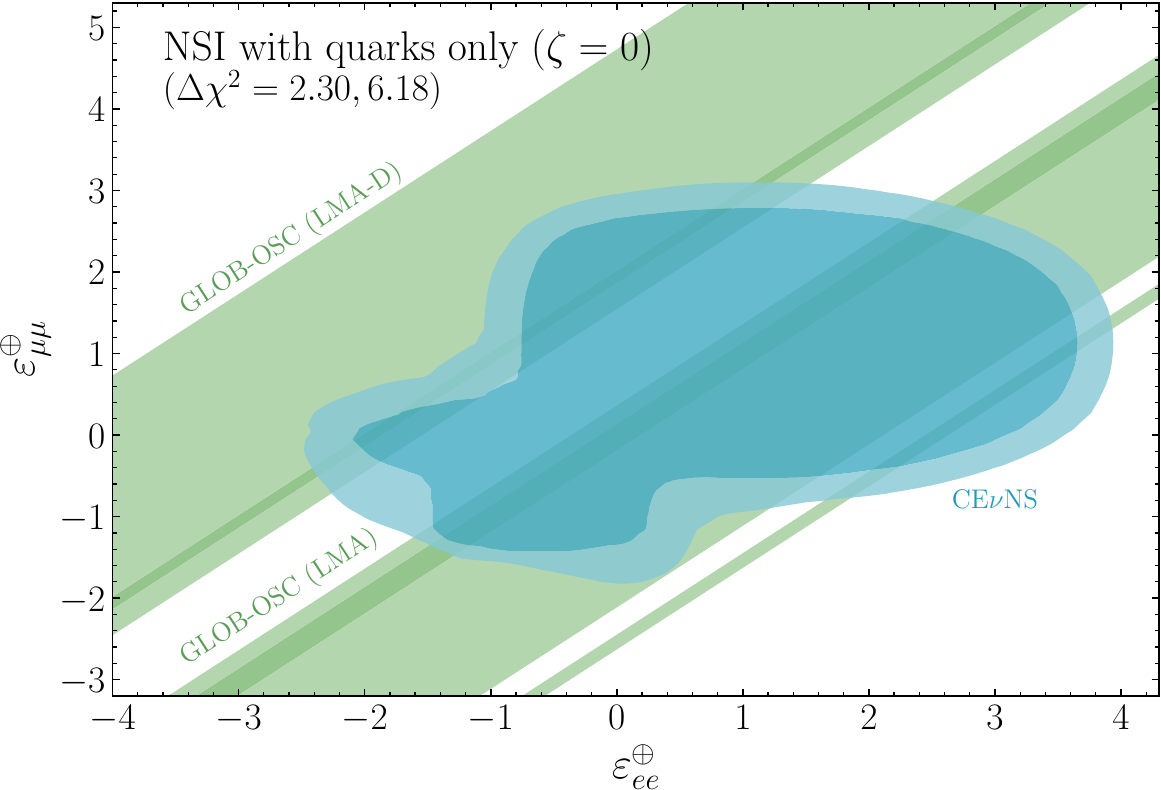}
  \caption{Allowed regions in the plane of $\Eps^\oplus_{ee}$ and
    $\Eps^\oplus_{\mu\mu}$ for vector NSI with quarks from the
    combinations of CE$\nu$NS data compared with the allowed regions
    from the global oscillation analysis which are the two diagonal
    shaded bands corresponding to the LMA and LMA-D solutions.  Both
    the green and blue regions are obtained after \emph{independently}
    marginalizing over all other relevant parameters: NSI couplings
    (including $\eta$) for the CE$\nu$NS region, and NSI couplings,
    $\eta$ and oscillation parameters for the <<GLOB-OSC>> regions.
    So the values of other NSI and $\eta$ in the blue and green
    regions are not forced to be the same.}
\label{fig:eemm}
\end{figure}

Let us start discussing the complementary sensitivity to vector-NSI
from the combined CE$\nu$NS results with that from present oscillation
data, for general models leading to NSI with quarks.  With this aim we
have first performed an analysis including only the effect of vector
NSI on the matter potential in the neutrino oscillation experiments.
The results of such analysis are given in the left column of
Table~\ref{tab:nsilblranges} in terms of the allowed ranges of the
effective NSI couplings to the Earth matter,
$\Eps^\oplus_{\alpha\beta}$, defined in Eq.~\eqref{eq:eps-earth0} (see
Sec.~\ref{sec:resulgen} for details).  Comparison with the constraints
from all available CE$\nu$NS data is illustrated in
Fig.~\ref{fig:eemm} where we plot the allowed regions in the plane
($\Eps^\oplus_{ee}$, $\Eps^\oplus_{\mu\mu}$).  The figure shows the
results obtained by the analysis of each of the two sets of
data \emph{independently}, after marginalization over the all other
(including off-diagonal) NSI parameters.\footnote{Technically the
regions for oscillations in Fig.~\ref{fig:eemm} are obtained by
performing each of the two analysis in terms of the 9 (out of the 10)
basic parameters  (fixing $\zeta=0$)
including the corresponding value of $Y_n$ to each data sample.
Therefore the output of each analysis is a $\chi^2$ function of the
basic 9 parameters, which is then marginalized with respect to all
parameters except for the two combinations shown in the figure.}  As
discussed in Sec.~\ref{sec:formCNUES}, even when considering NSI
coupling only to quarks, there is always a value of $\eta$ for which
the contribution of NSI to the CE$\nu$NS cross section cancels.
Consequently CE$\nu$NS data with a single nucleus does not lead to any
constraint on the $\Eps^\oplus_{\alpha\beta}$ parameters.  However,
the cancellation occurs at different values of $\eta$ for the three
nucleus considered (CsI, Ar, and Ge), and consequently, as seen in the
figure, the combination of CE$\nu$NS data with the three nuclear
targets does constrain the full space of effective
$\Eps^\oplus_{\alpha\beta}$, even after marginalization over $\eta$.
The constraints derived from the combined CE$\nu$NS data are
independent (and complementary) to those provided by the oscillation
analysis and, as seen in the figure, they are fully consistent.

The allowed ranges for vector NSI with up or down quarks
are compiled in Table~\ref{tab:nsiqranges} and show good qualitative
agreement with those of Refs.~\cite{Esteban:2018ppq, Coloma:2019mbs},
with the expected small deviations due to the differences in the
analysis, quoted CL, and included data.

 \begin{sidewaystable}\centering
  \catcode`?=\active\def?{\hphantom{0}}
  \begin{tabular}{|l|c|c||c|c|}
    \hline
    & \multicolumn{2}{|c||}{Allowed ranges at 90\% CL (marginalized)}
    && Allowed ranges at 90\% CL (marginalized)
    \\
    \hline
    & \multicolumn{2}{|c||}{GLOB-OSC}
    && GLOB-OSC+CE$\nu$NS
    \\ \hline
    & LMA & $\text{LMA}\oplus\text{LMA-D}$
    && $\text{LMA} = \text{LMA}\oplus\text{LMA-D}$
    \\
    \hline
    \begin{tabular}{@{}c@{}}
      $\Eps_{ee}^{u,V} - \Eps_{\mu\mu}^{u,V}$ \\
      $\Eps_{\tau\tau}^{u,V} - \Eps_{\mu\mu}^{u,V}$
    \end{tabular}
     &
    \begin{tabular}{@{}c@{}}
      $[-0.063, +0.36]?$ \\
      $[-0.0053, +0.017]?$
    \end{tabular}
    &
    \begin{tabular}{@{}c@{}}
      $[-1.1, -0.79] \oplus [-0.063, +0.36]$ \\
      $[-0.021, +0.018]$
    \end{tabular}
    &
    \begin{tabular}{@{}c@{}}
      $\Eps_{ee}^{u,V}$ \\
      $\Eps_{\mu\mu}^{u,V}$ \\
      $\Eps_{\tau\tau}^{u,V} $
    \end{tabular}
    &
    \begin{tabular}{@{}c@{}}
      $[-0.038, +0.034] \oplus [+0.34, +0.42]$ \\
      $[-0.046, +0.031] \oplus [+0.35, +0.42]$ \\
      $[-0.046, +0.033] \oplus [+0.35, +0.42]$
    \end{tabular}
    \\
    $\Eps_{e\mu}^{u,V}$
    & $[-0.057, +0.013]$
    & $[-0.057, +0.061]$
    & $\Eps_{e\mu}^{u,V}$
    & $?[-0.044, +0.0049] $
    \\
    $\Eps_{e\tau}^{u,V}$
    & $[-0.076, +0.11]?$
    & $[-0.12, +0.11]$
    & $\Eps_{e\tau}^{d,V}$
    & $[-0.079, +0.11]?$
    \\
    $\Eps_{\mu\tau}^{u,V}$
    & $[-0.0077, +0.0042]$
    & $[-0.0077, +0.0083]$
    & $\Eps_{\mu\tau}^{u,V}$
    & $[-0.0064, 0.0053]$
    \\ \hline
    \begin{tabular}{@{}c@{}}
      $\Eps_{ee}^{d,V} - \Eps_{\mu\mu}^{d,V}$ \\
      $\Eps_{\tau\tau}^{d,V} - \Eps_{\mu\mu}^{d,V}$
    \end{tabular}
    &
    \begin{tabular}{@{}c@{}}
      $[-0.069, +0.38]?$ \\
      $[-0.0058, +0.018]?$
    \end{tabular}
    &
    \begin{tabular}{@{}c@{}}
      $?[-1.3, -0.91] \oplus [-0.072, +0.38]?$ \\
      $[-0.029, +0.019]$
    \end{tabular}
    &
    \begin{tabular}{@{}l@{}}
      $\Eps_{ee}^{d,V}$ \\
      $\Eps_{\mu\mu}^{d,V}$ \\
      $\Eps_{\tau\tau}^{d,V} $
    \end{tabular}
    &
    \begin{tabular}{@{}r@{}}
      $[-0.036, +0.031] \oplus [+0.30, +0.39]$ \\
      $[-0.040, +0.038] \oplus [+0.31, +0.39]$ \\
      $[-0.041, +0.043] \oplus [+0.31, +0.39]$
    \end{tabular}
    \\
    $\Eps_{e\mu}^{d,V}$
    & $[-0.058, +0.014]$
    & $[-0.058, +0.098]$
    & $\Eps_{e\mu}^{d,V}$
    & $?[-0.054, +0.0045]$
    \\
    $\Eps_{e\tau}^{d,V}$
    & $[-0.079, +0.11]?$
    & $[-0.16, +0.11]$
    & $\Eps_{e\tau}^{d,V}$
    & $[-0.051, +0.11]?$
    \\
    $\Eps_{\mu\tau}^{d,V}$
    & $[-0.0087, +0.0051]$
    & $[-0.0087, +0.015]?$
    & $\Eps_{\mu\tau}^{d,V}$
    & $[-0.0075, +0.0046]$
    \\
    \hline
  \end{tabular}
  \caption{90\% allowed ranges for the vector NSI couplings
    $\Eps_{\alpha\beta}^{u,V}$ and $\Eps_{\alpha\beta}^{d,V}$ as
    obtained from the global analysis of oscillation data (left
    columns, applicable to NSI induced by mediators with $\Mmed \ll
    5~\text{MeV}$) and also including data from CE$\nu$NS experiments
    (right columns, applicable to NSI induced by mediators with $\Mmed
    \gtrsim 50~\text{MeV}$).  The results are obtained after
    marginalizing over oscillation and the other matter potential
    parameters either within the LMA only ($\theta_{12} < 45^\circ$)
    and within both LMA ($\theta_{12} < 45^\circ$) and LMA-D
    ($\theta_{12} > 45^\circ$) subspaces respectively (this second
    case is denoted as $\text{LMA} \oplus \text{LMA-D}$).  Notice that
    once CE$\nu$NS data is included the two columns become identical,
    since for NSI couplings with $f=u,d$ the LMA-D solution is only
    allowed well above 90\% CL.}
  \label{tab:nsiqranges}
\end{sidewaystable}

Finally for completeness we have also performed a new dedicated
analysis including axial-vector NSI with up or down quarks.
Axial-vector NSI with quarks do not contribute to matter effects nor
to CE$\nu$NS.  They only enter the global analysis via their
modification of the NC event rate in the SNO experiment (see
Sec.~\ref{sec:sno-nc}), which is not able to constrain the NSI
coefficients if all of them are included simultaneously due to
possible cancellations between their respective contributions to the
NC rate.  Thus in this case we derive the bounds \emph{assuming that
only one NSI coupling is different from zero at a time}.  The
corresponding allowed ranges can be found in
Table~\ref{tab:nsiqaxialranges}.

As seen in the table, for all the coefficients, the allowed range is
composed of several disjoint intervals.  They correspond to values of
the NSI couplings for which the SNO-NC event rate is approximately the
SM prediction.  This can occur for either flavour diagonal or flavour
non-diagonal coeficients because solar neutrinos reach the Earth as
mass eigenstates, so the density matrix describing the neutrino state
at the detector is diagonal in the mass basis, but not in the flavor
basis.  The presence of non-vanishing off-diagonal
$\rho_{\alpha\neq\beta}$ elements is responsible for the sensitivity
to off-diagonal $\Eps_{\alpha\neq\beta}$ coefficients.  More
quantitatively, the ranges correspond to coefficients verifying:
\begin{equation}
  0 \simeq
  \Tr\Bigg[ \rho^\text{SNO}\, \bigg(\frac{G_A}{g_A}\bigg)^2 \Bigg] - 1
  = \left\{
  \begin{array}{l}
    \rho_{\alpha\alpha}^\text{SNO} \big[ \big(\Eps^{q,A}_{\alpha\alpha}\big)^2
      \pm 2\,\Eps^{q,A}_{\alpha\alpha} \big] \,,
    \\[2mm]
    \big( \rho^\text{SNO}_{\alpha\alpha} + \rho^\text{SNO}_{\beta\beta} \big)\,
    \big( \Eps^{q,A}_{\alpha\neq\beta} \big)^2
    \pm 4\Re\!\big(\rho^\text{SNO}_{\alpha\neq\beta}\big)\,
    \Eps^{q,A}_{\alpha\neq\beta}
  \end{array}\right.
\end{equation}
where the upper line holds when the NSI coefficient included is
flavour diagonal, and the lower one when it is flavour-changing and
the $\pm$ sign correspond to $q=u$ and $q=d$ respectively.  Thus the
allowed range for flavour diagonal NSI is formed by two solutions
around $\Eps_{\alpha\alpha}^{q,A} = 0$ and $\Eps_{\alpha\alpha}^{q,A}
= -2$.  For flavour off-diagonal NSI it is formed by solutions around
$\Eps_{\alpha\neq\beta}^{q,A} = 0$ and $\Eps_{\alpha\neq\beta}^{q,A} =
\mp 4\Re(\rho^\text{SNO}_{\alpha\neq\beta}) \mathbin{\big/}
(\rho^\text{SNO}_{\alpha\alpha} + \rho^\text{SNO}_{\beta\beta})$.  For
$\Eps^{q,A}_{e\mu}$ (and similarly for $\Eps^{q,A}_{e\tau}$) this last
condition corresponds, in fact, to two distinct solutions, both around
$|\Eps^{q,A}_{e\mu}|\neq 0$ and two possible signs, due to the two
different signs of $\rho^\text{SNO}_{e\mu}$ for the two CP-conserving
values of $\delta_\text{CP} \in \{0, \pi\}$.  On the contrary,
$\rho^\text{SNO}_{\mu\tau}$ takes very similar values for
$\delta_\text{CP} \in \{0, \pi\}$, and consequently for
$\Eps^{q,A}_{\mu\tau}$ the two non-zero solutions closely overlap
around $\Eps^{u,A}_{\mu\tau}\sim 1.7$ ($\Eps^{d,A}_{\mu\tau}\sim
-1.7$).

\begin{table}[t]\centering
  \catcode`?=\active\def?{\hphantom{0}}
  \renewcommand{\arraystretch}{1.2}
  \begin{tabular}{|c || c || c |}\hline
    & {Allowed ranges at 90\% CL (1-parameter)}    &
    \\
    \hline
    & GLOB-OSC &
    \\
    \hline
    $\Eps^{u,A}_{ee}$
    & $??[-2.1, -1.8] \oplus [-0.19, +0.13]$
    & $-\Eps^{d,A}_{ee}$
    \\
    $\Eps^{u,A}_{\mu\mu}$
    & $??[-2.2, -1.7] \oplus [-0.26, +0.18]$
    & $-\Eps^{d,A}_{\mu\mu}$
    \\
    $\Eps^{u,A}_{\tau\tau}$
    & $??[-2.1, -1.8] \oplus [-0.20, +0.15]$
    & $-\Eps^{d,A}_{\tau\tau}$
    \\
    $\Eps^{u,A}_{e\mu}$
    & $[-1.5, -1.2] \oplus [-0.16, +0.12] \oplus [+1.4, +1.7]$
    & $-\Eps^{d,A}_{e\mu}$
    \\
    $\Eps^{u,A}_{e\tau}$
    & $[-1.5, -1.3] \oplus [-0.13, +0.10] \oplus [+1.4, +1.7]$
    & $-\Eps^{d,A}_{e\tau}$
    \\
    $\Eps^{u,A}_{\mu\tau}$
    & $[-0.085, +0.11] \oplus [+1.6, +1.9]???$
    & $-\Eps^{d,A}_{\mu\tau}$
    \\
    \hline
  \end{tabular}
  \caption{90\% CL bounds (1 d.o.f., 2-sided) on the effective
    axial-vector NSI couplings with quarks.  The bounds are derived
    from the global analysis of oscillation data including the effect
    of NSI in the SNO NC cross section and \emph{assuming only one NSI
    coupling different from zero at a time}.  As explained in
    Sec.~\ref{sec:sim}, these bounds apply to models with $\Mmed
    \gtrsim 3~\text{MeV}$.}
    \label{tab:nsiqaxialranges}
\end{table}

\subsection{Constraints on NSI with quarks and electrons: effective NSI in the Earth}
\label{sec:resulgen}

Let us now discuss the most general case in which NSI with quarks and
electrons are considered, parametrized by the angles $\eta$ and
$\zeta$ introduced in Eqs.~\eqref{eq:xi-eta}
and~\eqref{eq:xi-eta-quark}.  We focus on vector NSI because in this
case the interplay between matter and scattering effects, and
therefore the dependence on the couplings to charged fermions
involved, is expected to play a most relevant role.

\begin{figure}[t]\centering
  \includegraphics[width=0.7\textwidth]{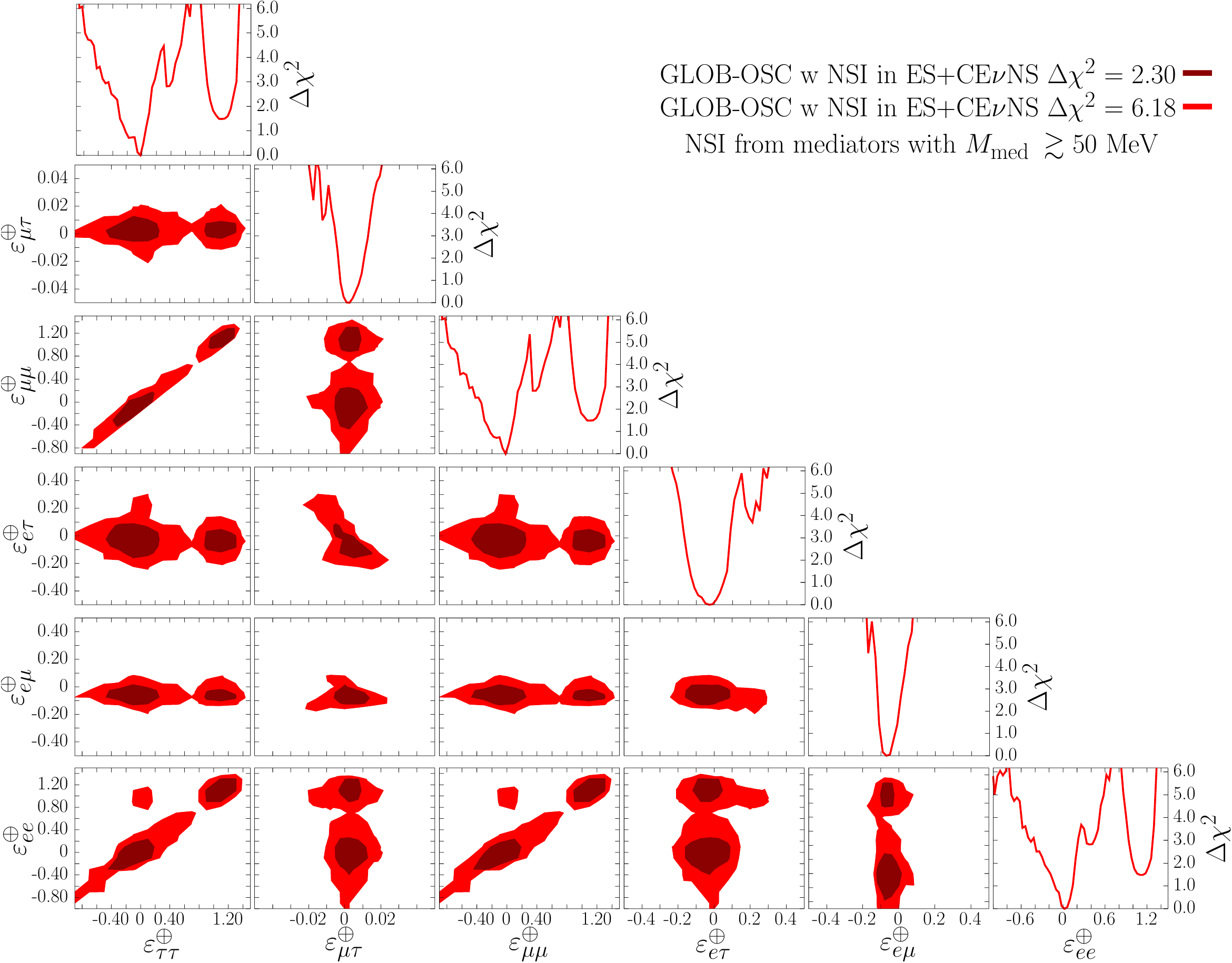}
  \caption{Constraints on the effective generalized NSI in the Earth
    matter (relevant for matter effects in LBL experiments) with
    arbitrary values of $\eta$ and $\zeta$.  Each panel shows a
    two-dimensional projection of the allowed multi-dimensional
    parameter space after minimization with respect to the undisplayed
    parameters.  The contours correspond to $1\sigma$ and $2\sigma$ (2
    d.o.f.).}
  \label{fig:earthtriang}
\end{figure}

In this framework, it is useful to quantify the results of our
analysis in terms of the effective NSI parameters which describe the
generalized Earth matter potential, which are in fact the relevant
quantities for the study of long-baseline and atmospheric oscillation
experiments.  The results are displayed in Fig.~\ref{fig:earthtriang}
and on the right column in Table~\ref{tab:nsilblranges} where we show
the allowed two-dimensional regions and one-dimensional ranges of the
effective NSI coefficients for the global analysis of oscillation and
CE$\nu$NS data, after marginalizing over all other parameters.
Therefore what we quantify in Fig.~\ref{fig:earthtriang} and the right
column in Table~\ref{tab:nsilblranges} is our present knowledge of the
matter potential for neutrino propagation in the Earth, for NSI
induced by mediators heavier than $\Mmed \gtrsim 50~\text{MeV}$ (as
discussed in Sec.~\ref{sec:sim}) and for \emph{any unknown value} of
$\eta$ and $\zeta$.  Technically this is obtained by marginalizing the
results of the global $\chi^2$ with respect to $\eta$ and $\zeta$ as
well, and the $\Delta\chi^2$ functions plotted in the figure are
defined with respect to the absolute minimum for any $\eta$ and
$\zeta$ which lies close to $\eta \sim -45^\circ$ and $\zeta\sim
10^\circ$.

\begin{sidewaystable}\centering
  \newcommand{\Twin}[2]{\begin{tabular}{@{}c@{}}$#1$\\[-2mm]$\big(#2\big)$\end{tabular}}
  \renewcommand{\arraystretch}{1.2}
  \begin{tabular}{|@{}c@{}||@{}c@{}|}
    \hline
    \multicolumn{2}{|c|}{Allowed ranges at \Twin{90\%~\text{CL}}{99\%~\text{CL}} marginalized}
    \\
    \hline\hline
    GLOB-OSC w/o NSI in ES
    & GLOB-OSC w NSI in ES + CE$\nu$NS
    \\
    \hline
    \begin{tabular}{c||c}
      \\[-1.15mm]
      $\Eps^\oplus_{ee}-\Eps^\oplus_{\mu\mu}$
      & \Twin{[-3.1, -2.8] \oplus [-2.1, -1.88] \oplus [-0.15, +0.17]}{[-4.8, -1.6] \oplus [-0.40, +2.6]}
      \\[+6mm]
      $\Eps^\oplus_{\tau\tau}-\Eps^\oplus_{\mu\mu}$
      & \Twin{[-0.0215, +0.0122]}{[-0.075, +0.080]}
      \\[+6mm]
      $\Eps^\oplus_{e\mu}$
      & \Twin{[-0.11, -0.021] \oplus[+0.045, +0.135]}{[-0.32, +0.40]}
      \\[+3mm]
      $\Eps^\oplus_{\mu\tau}$
      & \Twin{[-0.22, +0.088]}{[-0.49, +0.45]}
      \\[+3mm]
      $\Eps^\oplus_{\mu\tau}$
      & \Twin{[-0.0063, +0.013]}{[-0.043, +0.039]}
    \end{tabular}
    &
    \begin{tabular}{c||c}
      $\Eps^\oplus_{ee}$
      & \Twin{[-0.19, +0.20] \oplus [+0.95, +1.3]}{[-0.23, +0.25] \oplus [+0.81, +1.3]}
      \\[+3mm]
      $\Eps^\oplus_{\mu\mu}$
      & \Twin{[-0.43, +0.14]\oplus [+0.91, +1.3]}{[-0.29, +0.20] \oplus [+0.83, +1.4]}
      \\[+3mm]
      $\Eps^\oplus_{\tau\tau}$
      & \Twin{[-0.43, +0.14]\oplus [+0.91, +1.3]}{[-0.29,  +0.20]\oplus [+0.83, +1.4]}
      \\[+3mm]
      $\Eps_{e\mu}^\oplus$
      & \Twin{[-0.12, +0.011]}{[-0.18, +0.08]}
      \\[+3mm]
      $\Eps_{e\tau}^\oplus$
      & \Twin{[-0.16, +0.083]}{[-0.25, +0.33]}
      \\[+3mm]
      $\Eps_{\mu\tau}^\oplus$
      & \Twin{[-0.0047, +0.012]}{[-0.020, +0.021]}
    \end{tabular}
    \\
    \hline
  \end{tabular}
  \caption{90\% and 99\% CL bounds (1 d.o.f., 2-sided) on the
    effective NSI parameters relevant for matter effects in LBL
    experiments with arbitrary values of $\eta$ and $\zeta$, obtained
    after marginalizing over all other NSI and oscillation parameters.
    The bounds on the left (right) column are applicable to NSI
    induced by mediators with masses $\Mmed \ll 500~\text{keV}$
    ($\Mmed \gtrsim 50~\text{MeV}$).}
  \label{tab:nsilblranges}
\end{sidewaystable}

From these results we see that the allowed ranges for the diagonal
$\Eps^\oplus_{\alpha\alpha}$ parameters are composed of two disjoint
regions.  However let us stress that they both correspond to the LMA
solution since neither of them falls within the LMA-D region
(which requires $\Eps^\oplus_{ee} - \Eps^\oplus_{\mu\mu} = -2$).  In
fact, the LMA-D solution is ruled out beyond the CL shown in
Fig.~\ref{fig:earthtriang} and, as will be discussed in
Sec.~\ref{sec:resulLMAD} in more detail, the combination of
oscillation data with CE$\nu$NS results for different nuclear targets
is required to reach this sensitivity.

Conversely, for mediators with masses $\Mmed \ll 500~\text{keV}$,
effects on ES and CE$\nu$NS experiments would be suppressed even if
NSI involve couplings to both quarks and electrons, and the only
effect of NSI will be the modification of the matter potential in
neutrino oscillations.  The same holds for NSI with quarks only
(\textit{i.e.}, for $\zeta=0$) induced by mediators with masses $\Mmed
\ll 10~\text{MeV}$.  Since in the matter potential the effects for
protons or electrons are indistinguishable, the allowed ranges of the
effective $\Eps^\oplus_{\alpha\beta}$ are the same in both scenarios,
which are given on the left column in Table~\ref{tab:nsilblranges}
(and included in Fig.~\ref{fig:eemm}).  As discused in
Sec.~\ref{sec:formOSC}, in this case the analysis can only constrain
the differences between flavour-diagonal NSI coefficients.
Furthermore, the allowed range of $\Eps^\oplus_{ee} -
\Eps^\oplus_{\mu\mu}$ contains a disjoint interval around $-2$ (which
at 90\% CL further splits into two sub-intervals as seen in the table)
corresponding to the LMA-D solution, which is well allowed in these
scenarios.  We will discuss this in more detail in
Sec.~\ref{sec:resulLMAD}.

\begin{figure}[t]\centering
  \includegraphics[width=0.5\textwidth]{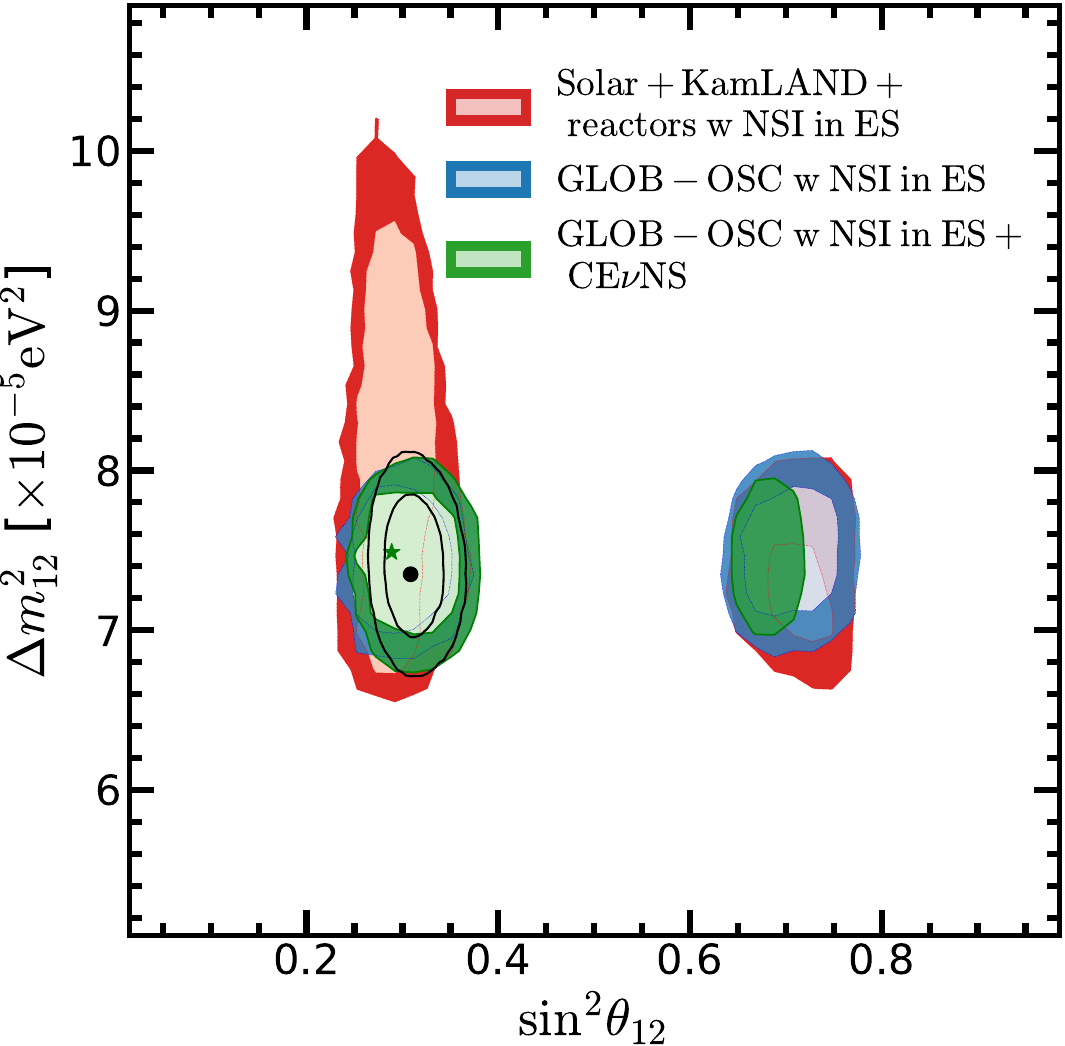}
  \caption{Two-dimensional projections of the allowed regions (at 90\%
    and $3\sigma$ CL) onto $\Dmq_{12}$ and $\theta_{12}$ parameters,
    after marginalizing over all other oscillation parameters and over
    NSI couplings to quarks and electrons.  Red regions correspond to
    the analysis of KamLAND and Solar data; blue regions include all
    oscillation data; and green regions include all oscillation and
    CE$\nu$NS data.  In all cases, NSI effects on ES are fully
    accounted for in the fit.  For comparison, the empty contours
    (solid black lines) show the corresponding regions for the global
    oscillation analysis without NSI.}
\label{fig:oscil}
\end{figure}

We finish this section by discussing the impact of NSI in this general
framework (where we allow couplings to quarks and electrons
simultaneously) on the determination of the oscillation parameters in
the solar sector.  This is shown in Fig.~\ref{fig:oscil}, where we see
that the determination of the oscillation parameters within the LMA
region, which is the region favored by the fit, is rather robust even
after the inclusion of NSI couplings to both quarks and electrons.
Comparing the blue and red regions in the figure we see that the
inclusion of data from atmospheric and LBL experiments is important to
reach such robustness.  This had been previously shown in
Refs.~\cite{Gonzalez-Garcia:2013usa, Esteban:2018ppq} for NSI with
quarks only; here we conclude that the same conclusions hold also in
presence of NSI with electrons, as long as their impact on ES is
accounted for in the fit.  Also, the allowed LMA regions are not very
much affected by the addition of the CE$\nu$NS data and it is very
close to that of the oscillation analysis without NSI.  As also shown
in the figure, the LMA-D is only allowed at 97\% CL or above (for
2~d.o.f.).  The current status of the LMA-D region is discussed in
more detail in the next section.

\subsection{Present status of the LMA-D solution}
\label{sec:resulLMAD}

\begin{figure}[t]\centering
  \includegraphics[width=\textwidth]{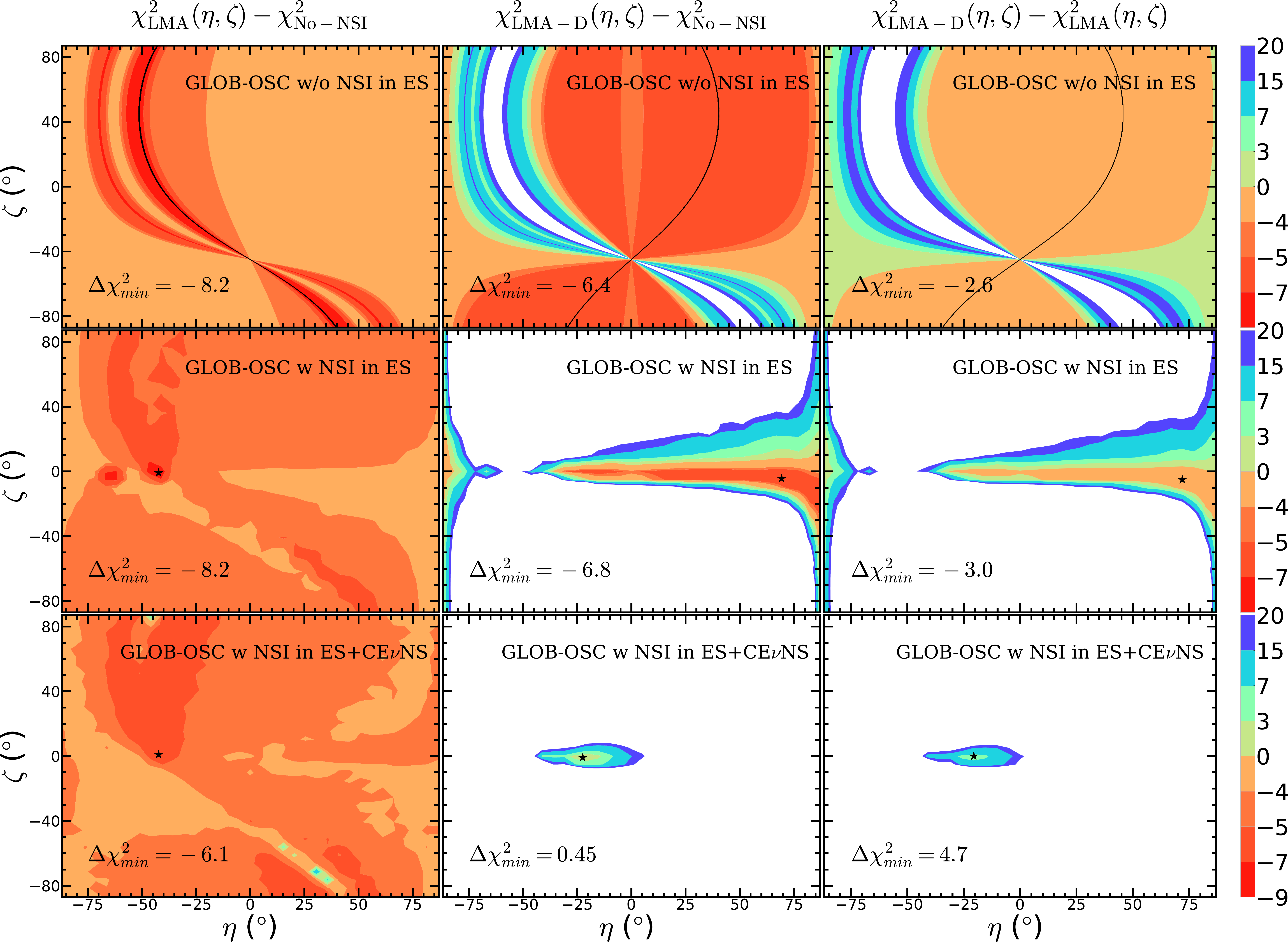}
  \caption{Isocontours of $\chi^2_\text{LMA}(\eta,\zeta) -
    \chi^2_\text{no-NSI}$, $\chi^2_\text{LMA-D}(\eta,\zeta) -
    \chi^2_\text{no-NSI}$ and $\chi^2_\text{LMA-D}(\eta,\zeta) -
    \chi^2_\text{LMA-D}(\eta,\zeta)$ of the global analysis of
    oscillation data without including NSI in the ES cross sections at
    Borexino, SNO, and SK (upper panels), and including the NSI in the
    ES cross sections at Borexino, SNO, and SK (middle panels).  The
    lower panels shows of the result adding the data from CE$\nu$NS
    experiments.  We show projections in the plane of angles ($\zeta$,
    $\eta$) (after marginalization of all other parameters) which
    parametrize the relative strength of the NSI couplings to
    up-quarks, down-quark, and electrons.  The levels corresponding to
    the different colours are given on the color bar on the right.
    Contours beyond 20 are white.  In each panel the best-fit point is
    marked with a star (middle and bottom rows) or by a solid black
    line (upper row).  Results in the upper row are applicable to NSI
    induced by mediators with $\Mmed \ll 500~\text{keV}$; in the
    middle row, for models with $\Mmed \gtrsim 10~\text{MeV}$; and in
    the lower row, for models with $\Mmed \gtrsim 50~\text{MeV}$, see
    Sec.~\ref{sec:sim} for details.}
  \label{fig:2D}
\end{figure}

In this section, we discuss in more detail the present status of the
LMA-D region in light of all available data, for models leading to NSI
couplings to quarks and electrons simultaneously.  We start by
exploring the dependence of the presence of the LMA-D solution on the
specific combination of couplings to the charged fermions considered.
In order to do so it is convenient to introduce the functions
$\chi^2_\text{LMA}(\eta,\zeta)$ and $\chi^2_\text{LMA-D}(\eta,\zeta)$
which are obtained by marginalizing the $\chi^2$ for a given value of
$\eta$ and $\zeta$ over both the oscillation and the matter potential
parameters within the regions $\theta_{12} < 45^\circ$ and
$\theta_{12} > 45^\circ$, respectively.  With this, in the left
(central) panels of Fig.~\ref{fig:2D} we plot isocontours of the
differences $\chi^2_\text{LMA}(\eta,\zeta) - \chi^2_\text{no-NSI}$
($\chi^2_\text{LMA-D}(\eta,\zeta) - \chi^2_\text{no-NSI}$) where
$\chi^2_\text{no-NSI}$ is the minimum $\chi^2$ for standard $3\nu$
oscillations (\textit{i.e.}, without NSI).  In the right panels we
plot $\chi^2_\text{LMA-D}(\eta,\zeta) - \chi^2_\text{LMA}(\eta,\zeta)$
which quantifies the relative quality of the LMA and LMA-D solutions.
In each row in this figure we include only a subset of the data as
indicated by the labels.

The upper panels of Fig.~\ref{fig:2D} show the results when only
oscillation data is analyzed accounting for the effects of NSI on the
matter potential, but without including the effect of NSI in the ES
cross sections in Borexino, SNO, and SK.  As outlined in
Sec.~\ref{sec:sim}, these results would therefore apply to NSI models
with very light mediators, $\Mmed \ll 500~\text{keV}$.  In this
scenario (as shown in Sec.~\ref{sec:formOSC}) only the combination of
NSI couplings to electrons, protons and neutrons, parametrized by the
effective angle $\eta^\prime$ in Eq.~\eqref{eq:epx-etapr}, is
relevant.  Therefore the $\Delta\chi^2$ isocontours are curves along
$\tan\eta^\prime = \tan\eta \mathbin{/} (\cos\zeta + \sin\zeta) =
\text{constant}$.  From the upper left panel we see that for most of
$(\eta,\zeta)$ values, the inclusion of NSI leads only to a mild
improvement of the global fit to oscillation data
($\chi^2_\text{LMA}(\eta,\zeta) - \chi^2_\text{no-NSI} > -4$).  As
discussed in Ref.~\cite{Esteban:2018ppq} (Addendum), with the updated
SK4 solar data the determination of $\Dmq_{21}$ in solar and in
KamLAND experiments are fully compatible at $\lesssim 2\sigma$ level.
The inclusion of NSI only leads to an overall better fit at a CL above
2$\sigma$ (but still not statistically significant in any of the
different data samples) for $(\zeta,\eta)$ along the darker band.  In
particular the best fit lies along $\tan\eta^\prime \sim -1 \big/
Y_n^\oplus \approx -0.95$ ($\eta^\prime \approx -43.6^\circ$) for
which NSI effects in the Earth matter cancel, so there is no
constraint from Atmospheric and LBL experiments on the NSI which can
lead to that slightly better fit to Solar + KamLAND data within the
LMA solution.  The central and right panels show the status of the
LMA-D solution in this scenario.  We find that LMA-D provides a good
solution in most of the $(\eta,\zeta)$ plane.  The LMA-D is only very
disfavoured for $(\zeta, \eta)$ along the band with $\-2.75\lesssim
\tan\eta^\prime \lesssim -1.75$.  For these values the NSI
contribution to the matter potential in the Sun cancels in some point
inside the neutrino production area and therefore the degeneracy
between NSI and the octant of $\theta_{12}$ cannot be realized.

The impact of including NSI in the ES cross section in Borexino, SNO,
and SK can be seen in the middle row panels in Fig.~\ref{fig:2D}.
Since these panels include the effect of NSI in ES, they apply to NSI
models with mediator masses $10~\text{MeV}\lesssim \Mmed \lesssim
50~\text{MeV}$.  The main effect is that the $\zeta$ values for which
the LMA-D solution is allowed become very restricted (in fact the best
fit point for all panels are always close to $\zeta=0$) as long as
$\eta$ does not approach $90^\circ$, and consequently the dependence
of $\zeta$ is heavily suppressed.  Thus, once the effect of NSI in the
ES cross section is included, the results are not very different from
those obtained for $\zeta=0$ (no coupling to electrons) in
Refs.~\cite{Esteban:2018ppq, Coloma:2019mbs}.  The middle and right
panels in this row also illustrate how, as long as only oscillation
data is included, the LMA-D solution is still allowed with a CL
comparable and even slightly better than LMA.

The lower panels in Fig.~\ref{fig:2D} include the combination of all
data available and are therefore applicable to NSI with mediators
above $\Mmed \gtrsim 50~\text{MeV}$.  When comparing the LMA (LMA-D)
to the SM hypothesis (\textit{i.e.,} no NSI) we find that the global
minimum of the fit is better (comparable) to that obtained in absence
of NSI, as shown in the middle and left panels.  However, we also see
that the inclusion of the CE$\nu$NS data in the analysis severely
constrains the LMA-D solution.  Quantitatively we find that LMA-D
becomes disfavoured with respect to LMA with $\Delta\chi^2>4.7$ for
any value of $(\eta,\zeta)$ (right panel) and it is only
allowed\footnote{Notice than this is different than what we show in
Fig.~\ref{fig:2D}, where marginalization over $\eta,\zeta$ has been
performed.} below $\Delta\chi^2=9$ for very specific combinations of
couplings to quarks and electrons, $-2.5^\circ\leq\zeta\leq 1.5^\circ$
and $-29^\circ\leq\eta\leq -13^\circ$.

\begin{figure}[t]\centering
  \includegraphics[width=0.5\textwidth]{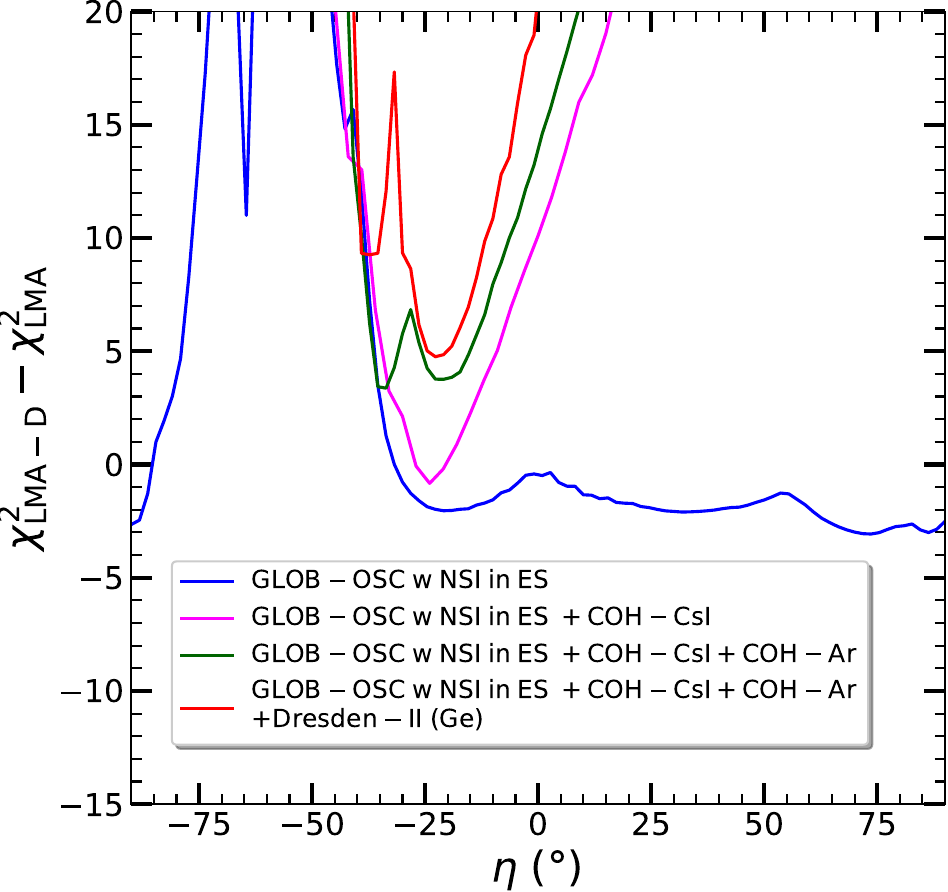}
  \caption{Dependence of $\chi^2_\text{LMA-D}(\eta,\zeta) -
    \chi^2_\text{LMA}(\eta,\zeta)$ on $\eta$ after marginalizing over
    $\zeta$ for different combination of experiments as labeled in the
    figure.}
  \label{fig:1Deta}
\end{figure}

To further illustrate the role of the different experiments in this
conclusion we show in Fig.~\ref{fig:1Deta} the projection of
$\chi^2_\text{LMA-D}(\eta,\zeta)-\chi^2_\text{LMA}(\eta,\zeta)$ on
$\eta$ after marginalizing over $\zeta$ (which, as discussed above, it
is effectively not very different from fixing $\zeta=0$).  The figure
illustrates the complementarity of the CE$\nu$NS data with different
targets.  As discussed in Sec.~\ref{sec:formCNUES}, the effects of NSI
in CE$\nu$NS on a given target, characterized by a value of $Y_n$,
cancel for $\eta^{\prime\prime} = \arctan(-1 / Y_n)$ (with
$\tan\eta^{\prime\prime} = \tan\eta / \cos\zeta \simeq \tan\eta$).
This corresponds to $\eta\simeq -35.4^\circ$ ($Y_n^\text{CsI} \approx
1.407$), $-39.3^\circ$ ($Y_n^\text{Ar} \approx 1.222$), and
$-38.4^\circ$ ($Y_n^\text{Ge} \approx 1.263$) for CsI, Ar, and Ge
respectively.  Thus, as seen in the figure, the combination of the
constraints from CE$\nu$NS with the different targets is important to
disfavour the LMA-D solution.

\section{Constraining Monopole-Dipole interactions with global oscillation data}
\label{sec:MD_section}

This section summarizes our results from the global fit to solar neutrino oscillation data in
the framework of three massive neutrinos including the new
neutrino-matter interactions generated by the Lagrangian in
Eq.~\eqref{eq:lagran} with the four choices of pseudoscalar couplings
to neutrinos in Eq.~\eqref{eq:nucoup}. 
\footnote{For the detailed description
of the methodology we refer to our published global analysis
NuFIT-5.0~\cite{Esteban:2020cvm} to which we have included the data
additions in NuFIT~\cite{nufit}; some technical aspects of our
treatment of neutrino propagation in the solar matter can be found in
Sec.~2.4 of Ref.~\cite{Maltoni:2023cpv}.  For solar neutrinos the
analysis includes the total rates from the radiochemical experiments
Chlorine~\cite{Cleveland:1998nv}, Gallex/GNO~\cite{Kaether:2010ag},
and SAGE~\cite{Abdurashitov:2009tn}, the spectral and zenith data from
the four phases of Super-Kamiokande  (SK) in Refs.~\cite{Hosaka:2005um,
  Cravens:2008aa, Abe:2010hy} including the latest SK4 2970-day energy
and day/night spectrum~\cite{Super-Kamiokande:2023jbt}, the results of
the three phases of SNO in the form of the day-night spectrum data of
SNO-I~\cite{Aharmim:2007nv} and SNO-II~\cite{Aharmim:2005gt} and the
three total rates of SNO-III~\cite{Aharmim:2008kc}, and the spectra
from Borexino Phase-I~\cite{Bellini:2011rx, Bellini:2008mr} (BX1),
Phase-II~\cite{Borexino:2017rsf} (BX2), and
Phase-III~\cite{BOREXINO:2022abl} (BX3).  In the framework of three-neutrino
mixing the oscillation probabilities for solar neutrinos dominantly
depend on $\theta_{12}$ and $\Dmq_{21}$, for which relevant
constraints arise from the analysis of the KamLAND reactor data, hence
we include in our fit the separate DS1, DS2, DS3 spectra from
KamLAND~\cite{Gando:2013nba}.  }
In what respect the relevant parameter space, Eqs.~\eqref{eq:dir}
and~\eqref{eq:maj} are a set of six coupled linear differential
equations whose solution depends on the two model parameters
($m_{\phi}$, $g_s^N g_p^{\nu_a}$) (for $a=e, \mu, \tau$ or $U$) and
the six oscillation parameters, $\theta_{12}$, $\theta_{13}$,
$\theta_{23}$, $\Dmq_{21}$, $\Dmq_{32}$, and $\delta_\text{CP}$.
However, as it is well-know, the present determination of $\Dmq_{32}$
---~derived dominantly from atmospheric, long-baseline accelerator,
and medium-baseline reactor neutrino experiments~--- implies that the
solar neutrino oscillations driven by $\Dmq_{32} / E_\nu$ are averaged
out.  Thus in our numerical evaluations we fix it to its present
best-fit value, but any other value would yield the same result as
long as we remain in the regime of averaged $\Dmq_{32}$
oscillations.\footnote{In fact this would allow for a reduction of the
evolution equations to an effective $4\times 4$ system.  However from
the computational point of view the reduction from a $6\times 6$ to a
$4\times 4$ set of equations is not substantial, so in our
calculations we stick to the complete $6\times 6$ system.}  In the
standard $3\nu$ oscillation scenario $\nu_\mu$ and $\nu_\tau$ are
totally indistinguishable at the energy scale of solar and reactor
experiments, hence also the dependence on $\theta_{23}$ and
$\delta_\text{CP}$ drops out in the evaluation of the solar
observables.  Including the flavour dependent scalar-pseudoscalar
potential reintroduces a mild dependence on $\theta_{23}$ and
$\delta_\text{CP}$ (except for the flavour-universal case $a=U$), but
quantitatively the phenomenology of solar neutrino data is still
dominated by $\Dmq_{21}$ and $\theta_{21}$.  The dependence on
$\theta_{13}$ is much weaker than its present precision from
medium-baseline reactor experiments (in particular from
Daya-Bay~\cite{DayaBay:2022orm}), while the impact of the CP phase is
very marginal.  So we can safely fix $\theta_{13}$ to its current
best-fit value from NuFIT $\sin^2\theta_{13} = 0.02203$ and set
$\delta_\text{CP}=0$.  In what respects $\theta_{23}$, for coupling to
specific flavours $a=e,\mu,\tau$ we marginalize it within the allowed
range from the global oscillation analysis by adding a prior for
$\sin^2\theta_{23}$ from NuFIT~\cite{nufit} to the $\chi^2$
function of the solar and KamLAND experiments.\footnote{Our results,
however, show that the difference in the bounds obtained marginalizing
over $\theta_{23}$ with this prior or fixing $\theta_{23}$ to its
best-fit value is very small.}  Altogether the parameter space to be
explored is five-dimensional.

We perform the analysis for both Dirac and Majorana neutrinos.
Besides the different $\nu_{eL}$ oscillation probabilities resulting
from the evolution equations Eqs.~\eqref{eq:dir} and~\eqref{eq:maj},
there are also differences in the evaluation of the neutrino-electron
elastic scattering event rates in Borexino, SNO and SK: while for the
Dirac case the produced $\nu_{\alpha R}$ does not interact in the
detector, for Majorana neutrinos the $\overline{\nu}_{\alpha R}$ do
interact (albeit with a different cross section than the corresponding
$\nu_{\alpha L}$ neutrinos) and such interactions must be included in
the evaluation of the event rates.  Furthermore the CC event rate at
SNO can also be different for Dirac and Majorana because the
$\overline{\nu}_{e R}$ interaction will produce a positron and two
neutrons which may or may not be discriminated.  Lacking the details
for a proper estimation, we have performed the analysis under two
extreme hypothesis: either assuming that the $e^+$ is recognized and
the whole event is rejected, or that the positron is misidentified as
an electron and it contributes as a CC event \emph{plus} two
additional NC events because of the generated neutrons.  We have
verified that the final bounds obtained under the two assumptions are
the same.

\begin{figure}[t]\centering
   \includegraphics[width=\textwidth]{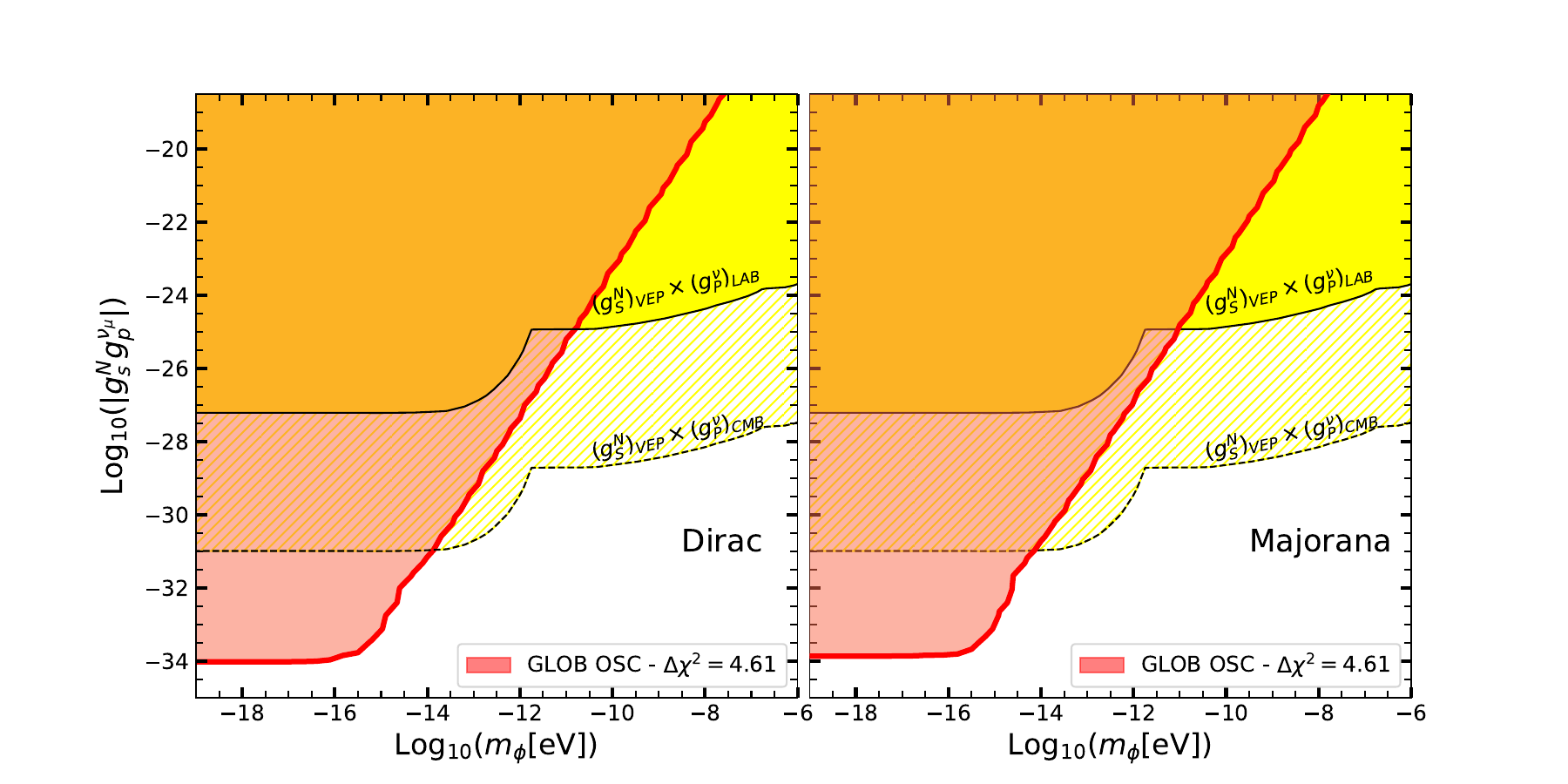}
   \caption{Excluded region at 90\% CL in the ($m_{\phi}$, $g_s^N
     g_p^{\nu_U}$) planeby the global analysis of solar neutrino data
     after marginalization over the relevant oscillation parameters
     ($\Dmq_{21}$, $\theta_{12}$) while keeping the other oscillation
     parameters fixed as described in the text.  Left (right) panel
     corresponds to Dirac (Majorana) neutrinos.
For comparison we
    show as yellow region the bounds on the product of couplings
     from the product  of the bound on $g_s^N$
     from  violation of weak equivalent (VEP)
     principle  and the bounds on $g_p^\nu$
     from either  kinematic and rare effects
     in weak decays
     in laboratory experiments  (LAB) or  
      cosmology bounds from cosmic microwave
     background (CMB).}
   \label{fig:limitsnu}
\end{figure}

Altogether we find that the inclusion of the new interaction does not
lead to any significant improvement on the description of the data,
and in fact the overall best-fit $\chi^2$ in the five-dimensional
parameter space is the same as in the standard $3\nu$ oscillation
scenario.  Consequently the statistical analysis results into bounds
on the allowed range of the scalar-pseudoscalar model parameters.

We show in Fig.~\ref{fig:limitsnu} the excluded region on the
parameter space for the case of flavour-universal coupling
($m_{\phi}$, $g_s^N g_p^{\nu_U}$) derived from the global analysis of
solar neutrino and KamLAND data for Dirac and Majorana cases after
marginalization over the relevant oscillation parameters ($\Dmq_{21}$,
$\theta_{12}$) while keeping the other oscillation parameters fixed to
their best-fit values as discussed above.  We show the regions at 90\%
CL (2~dof, two-sided).  Notice that when reporting the bounds on the
model parameters one has a choice on the statistical criteria to
derive the constraints.  The reason is that the experiment is in fact
only sensitive to $|g_s^N g_p^{\nu_U}|$: since the potential is
helicity-flipping its effect does not interfere with the
helicity-conserving vacuum oscillation and MSW matter potential, so
the probabilities only depend on the absolute value of the couplings
and on the square of the mediator mass.  Therefore, accounting for the
physical boundary, it is possible to report the limit on the absolute
value of the coupling (and mass) as a \emph{one-sided} limit, which at
90\% CL for 1~dof (2~dof) corresponds to $\Delta\chi^2 = 1.64$
($3.22$).  Conversely, if this restriction is not imposed, the result
obtained is what is denoted as a \emph{two-sided} limit, which at
90\%CL for 1~dof (2~dof) corresponds to $\Delta\chi^2 = 2.71$ ($4.61$)
and results into weaker constraints.  In what follows we list the
bounds obtained with the least constraining two-sided criterion.

As seen in Fig.~\ref{fig:limitsnu} the results are quantitatively
similar for Dirac or Majorana neutrinos.  In the figure we observe a
change in the slope of the exclusion region for masses $m_\phi\sim
10^{-15}$~eV.  For smaller mediator masses, the interaction length is
larger than the Sun radius and the corresponding potential becomes
saturated.  Conversely for $m_\phi\gtrsim 10^{-14}$~eV the interaction
range is short enough for the contact interaction approximation to
hold, in which case the analysis only depends on the combination
$|g_s^N g_p^{\nu_U}| \,\big/\, m_\phi^2$ and the region boundary
becomes a straight line of slope two in the log-log plane.

\begin{figure}[t]\centering
  \includegraphics[width=0.95\textwidth]{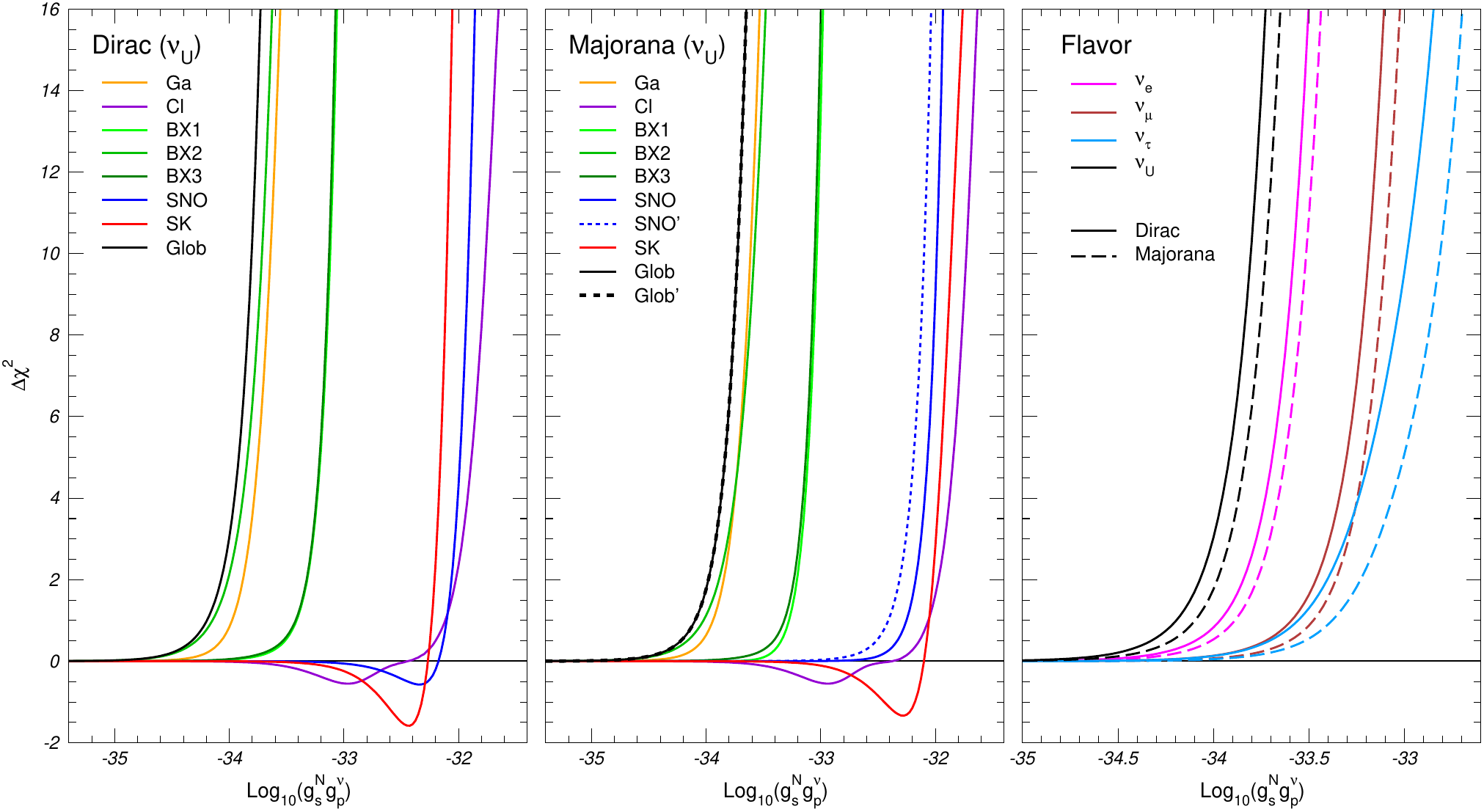}
  \caption{$\Delta\chi^2 \equiv \chi^2(g_s^N g_p^{\nu_U}) -
    \chi^2(g_s^N g_p^{\nu_U} = 0)$ as a function of $g_s^N
    g_p^{\nu_U}$ for $m_\phi\ll 10^{-15}$~eV fixing all oscillation
    parameters (in particular $\Dmq_{21}$, $\theta_{12}$ and
    $\theta_{23}$) to their best-fit values.  The left (central) panel
    show $\Delta\chi^2$ for each of the individual solar neutrino
    experiments for Dirac (Majorana) neutrinos and for the case of
    coupling to $\nu_U$.  In the central panel SNO and SNO' label the
    two variants of the SNO CC analysis for Majorana neutrinos (see
    text for details).  The right panel show the global $\Delta\chi^2$
    for four choices of the flavour dependence of the
    pseudoscalar-neutrino couplings; in particular, the black lines
    (coupling to $\nu_U$) for Dirac and Majorana cases coincide with
    the corresponding ``Glob'' ones in left and central panels,
    respectively.}
  \label{fig:solexp}
\end{figure}

To illustrate the relevance of the different solar neutrino
experiments on these results we plot in the left (central) panels in
Fig.~\ref{fig:solexp} the value of $\Delta\chi^2 \equiv \chi^2(g_s^N
g_p^{\nu_U}) - \chi^2(g_s^N g_p^{\nu_U} = 0)$ as a function of $g_s^N
g_p^{\nu_U}$ for the Dirac (Majorana) case in the effective
infinite-range interaction limit for each of the individual solar
neutrino experiments, fixing all oscillation parameters (in particular
$\Dmq_{21}$, $\theta_{12}$ and $\theta_{23}$) to their best-fit
values.  We see from the figure that the constraints are driven by the
experiments sensitive to the lowest energy part of the solar neutrino
spectrum (\textit{i.e.}, to the \Nuc{pp} flux), which correspond to
the spectral data of phase-II of Borexino (BX2 in the figure)
and the event rates in
Gallium experiments (Ga).  Conversely, those most sensitive to the higher
energy part of the solar spectrum (\textit{i.e.}, the \Nuc[8]{B}
flux), such as the spectral information of Super-Kamiokande and SNO
and the total rate in Chlorine (Cl), yield weaker sensitivity to the new
interaction.  As a consequence the combined constraints for Majorana
neutrinos are the same for both variants of the SNO CC analysis
(labeled as SNO and SNO' in the figure).  Furthermore, since the
oscillation parameters $\Dmq_{21}$ and $\theta_{12}$ are dominantly
determined by KamLAND reactor data as well as SNO and SK solar
neutrino data, their determination is very little affected by the
inclusion of the new interaction, and the allowed ranges that we found
in this five-parameter analysis are exactly the same as in the
standard $3\nu$ oscillation.

The right panel illustrates the dependence of the combined bounds on
the assumption for the neutrino flavour of the pseudoscalar coupling
for the four choices in Eq.~\eqref{eq:nucoup}.  As seen the variation
in the bounds is below $\mathcal{O}(10)$.  As expected the difference
is larger between coupling to $\nu_e$ and to $\nu_\mu$ (or $\nu_\tau$)
because the coupling to $\nu_e$ has larger effect on $P_{ee}$ which
gives the dominant contribution to the event rates in solar neutrino
experiments.  We notice in passing that the small variation between
the bounds for couplings to $\nu_\mu$ and $\nu_\tau$ is dominantly due
to the non-zero value of $\theta_{13}$.\footnote{In fact, it can be
shown that in the limit $\theta_{13} = 0$ the $\chi^2$ function
depends on $g_s^N g_p^{\nu_\mu}$, $g_s^N g_p^{\nu_\tau}$ and
$\theta_{23}$ only through the effective combination $g_s^N
g_p^{\nu_\mu} \cos^2\theta_{23} + g_s^N g_p^{\nu_\tau}
\sin^2\theta_{23}$.  In the general case $\theta_{13} \ne 0$ such
relation becomes only approximate.}

Finally we list in table~\ref{tab:bounds} the 90\% CL (1~dof) bounds
obtained from the global analysis of solar data in the two asymptotic
regimes and for the four choices of flavour dependence after
marginalization over the three relevant oscillation parameters
$\Dmq_{21}$, $\theta_{12}$ and $\theta_{23}$.

\begin{table}[t]\centering
  \begin{tabular}{|c|c||c|c||c|}
    \hline
    \multicolumn{2}{|c||}{}
    & \multicolumn{2}{c||}{Global Solar $\nu$} &
    Solar $\overline{\nu}_e$
    \\
    \cline{3-5}
    \multicolumn{2}{|c||}{}
    & Dirac & Majorana & Majorana
    \\
    \hline
    & $a=e$
    & $1.5\times 10^{-34}$
    & $1.7\times 10^{-34}$
    & $1.1\times 10^{-33}$
    \\
    $\big| g_s^N g_p^{\nu_a} \big|$
    & $a=\mu$
    & $3.6\times 10^{-34}$
    & $4.3\times 10^{-34}$
    & $7.6\times 10^{-34}$
    \\
    for $m_\phi< 10^{-16}$~eV
    & $a=\tau$
    & $3.6\times 10^{-34}$
    & $6.3\times 10^{-34}$
    & $7.9\times 10^{-34}$
    \\
    & $a=U$
    & $7.6\times 10^{-35}$
    & $9.6\times 10^{-35}$
    & $3.2\times 10^{-34}$
    \\\hline
    & $a=e$
    & $1.3\times 10^{-31}$
    & $2.5\times 10^{-31}$
    & $1.9\times 10^{-31}$
    \\
    $\frac{\left|g_s^N g_p^{\nu_a}\right|}{m_\phi \,/\, 10^{-14}~\eVq}$
    & $a=\mu$
    & $2.0\times 10^{-31}$
    & $5.0\times 10^{-31}$
    & $2.5\times 10^{-31}$
    \\
    for $m_\phi> 10^{-14}$~eV
    & $a=\tau$
    & $2.0\times 10^{-31}$
    & $7.9\times 10^{-31}$
    & $2.5\times 10^{-31}$
    \\
    & $a=U$
    & $6.3\times 10^{-32}$
    & $1.3\times 10^{-31}$
    & $7.0\times 10^{-32}$
    \\
    \hline
  \end{tabular}
  \caption{Bounds from the different analysis in the two asymptotic
    regimes, effective infinite interaction range and contact range
    respectively, and for the four choices of flavour dependence.  For
    the global solar analysis the bounds shown are obtained at 90\% CL
    (1~dof, two-sided), $\Delta\chi^2 = 2.71$.  The bounds from the
    antineutrino flux constraint are obtained from the 90\% CL
    experimental constraints in Eq.~\eqref{eq:panubound}.  See text
    for details.}
  \label{tab:bounds}
\end{table}
\subsection{Bounds from solar antineutrino constraints}
\begin{figure}[t]\centering
  \raisebox{0.2mm}{\includegraphics[width=0.48\textwidth]{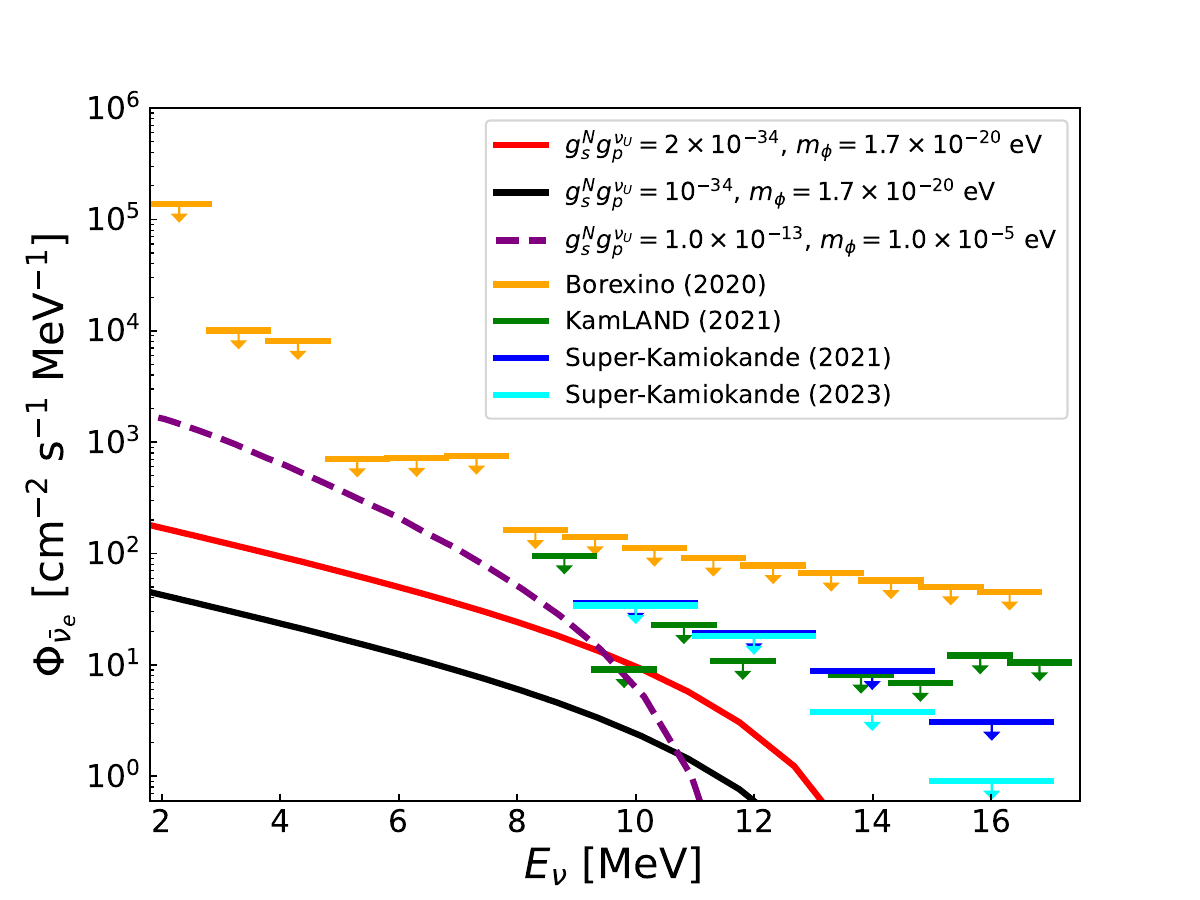}}
  \hfill\includegraphics[width=0.50\textwidth]{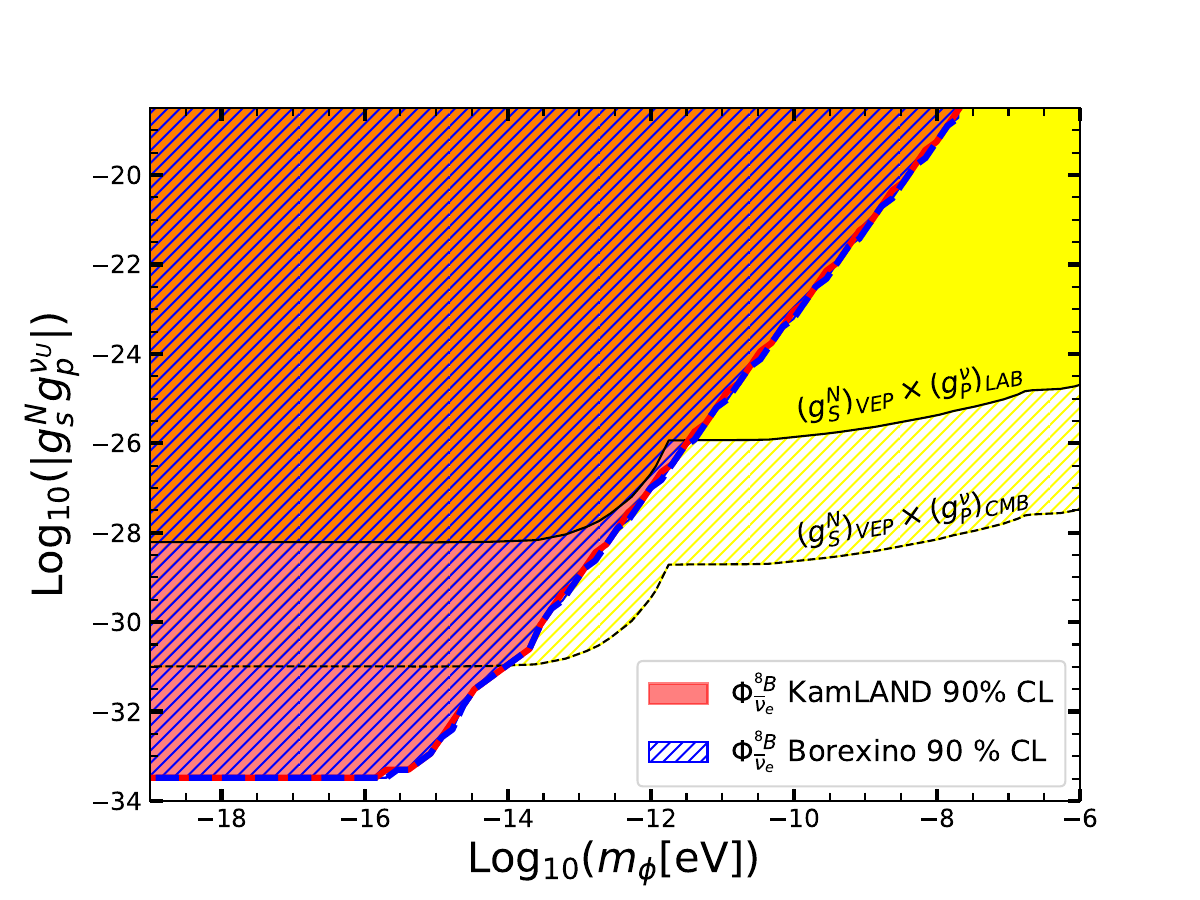}
  \caption{Left: Limits on $\bar{\nu}_e$ flux of astrophysical origin,
    as reported by KamLAND~\cite{KamLAND:2021gvi},
    Borexino~\cite{Borexino:2019wln} and
    Super-Kamiokande~\cite{Super-Kamiokande:2020frs,
      Super-Kamiokande:2021jaq,Super-Kamiokande:2023xup}.  For comparison, we show the expected
    solar $\bar{\nu}_e$ flux for several values of the model
    parameters as labeled in the figure.  Right: Excluded region on
    the ($m_{\phi}$, $g_s^N g_p^{\nu_U}$) plane obtained for Majorana
    neutrinos obtained from the 90\% CL constraint reported by
    KamLAND~\cite{KamLAND:2021gvi} and
    Borexino~\cite{Borexino:2019wln}, Eqs.~\eqref{eq:anubound1} and
    \eqref{eq:anubound2} (see text for details).  For comparison we
    show as yellow region the bounds on the product of couplings
        from the product of the bound on $g_s^N$
     from  violation of weak equivalent (VEP)
     principle  and the bounds on $g_p^\nu$
     from either  kinematic and rare effects
     in weak decays
     in laboratory experiments  (LAB) or  
      cosmology bounds from cosmic microwave background (CMB).}
  \label{fig:anures}
\end{figure}

Turning now to the bounds derived from the non observation of a flux
of solar antineutrinos, we plot in the left panel in
Fig.~\ref{fig:anures} a compilation of the model-independent limits on
$\bar{\nu}_e$ flux of astrophysical origin as reported by
KamLAND~\cite{KamLAND:2021gvi}, Borexino~\cite{Borexino:2019wln} and
Super-Kamiokande~\cite{Super-Kamiokande:2020frs,Super-Kamiokande:2021jaq,Super-Kamiokande:2023xup}.  In the same panel we also show the
predicted flux in the presence of the helicity-flip potential for
several model parameters.  From the figure we see that the most
stringent upper bounds on $\bar{\nu}_e$ come
KamLAND~\cite{KamLAND:2021gvi} and Borexino~\cite{Borexino:2019wln}
which constraint the $\bar{\nu}_e$ flux from the $\Nuc[8]{B}$ reaction
by
\begin{align}
  \label{eq:anubound1}
  &\text{KamLAND:}
  &\Phi^{\Nuc[8]{B}}_{\bar{\nu}_e} &< \hphantom{0}60~\text{cm}^{-2}\, \text{s}^{-1}
  ~\text{at 90\% CL for}~ E_\nu > 8.3~\text{MeV},
  \\
  \label{eq:anubound2}
  &\text{Borexino:}
  &\Phi^{\Nuc[8]{B}}_{\bar{\nu}_e} &< 138~\text{cm}^{-2}\, \text{s}^{-1}
  ~\text{at 90\% CL for}~ E_\nu > 7.8~\text{MeV}
\end{align}
which the experiments translated into a bound on the corresponding
integrated oscillation probabilities
\begin{align}
  \nonumber
  P(\nu_e\to \overline{\nu}_e)
  &\equiv
  \frac{\int_{E_\text{thres}} \Phi_{\Nuc[8]{B}}(E_\nu)\, \sigma(E_\nu)\,
    P_{\nu_{eL} \to \overline{\nu}_{eR}}(E_\nu)\, dE_\nu}{\int_{E_\text{thres}}
    \Phi_{\Nuc[8]{B}}(E_\nu)\, \sigma(E_\nu)\, dE_\nu}
  \\[1mm]
  \label{eq:panubound}
  &\leq
  \begin{cases}
    4.1\times 10^{-5} &\text{for}~ E_\text{thres} = 8.3~\text{MeV} \\
    7.4\times 10^{-5} &\text{for}~ E_\text{thres} = 7.8~\text{MeV}
  \end{cases}
\end{align}
where we have used the normalization of the $\Phi_{\Nuc[8]{B}}(E_\nu)$
of the latest version of the SSM~\cite{B23Fluxes, Magg:2022rxb}
which
results in the slight difference in the bounds on the probabilities in
Eq.~\eqref{eq:panubound} with respect to those quoted by the
experimental collaborations.

We show in the right panel of Fig.~\ref{fig:anures} the constraints on
the model parameters obtained by the projection of the bounds in
Eq.~\eqref{eq:panubound}.  For concreteness, in this case, we fix the
oscillation parameters to their best-fit values; however, as already  
noted, the dependence of the bounds on the oscillation parameters is  
very mild within their allowed ranges from global oscillation  
analysis~\cite{Esteban:2020cvm, nufit}.  We show the results for
pseudoscalar coupling to $\nu_U$, but similar bounds apply to the other
cases (see Table~\ref{tab:bounds} for the results obtained with the
different flavour assumptions in Eq.~\eqref{eq:nucoup}).  As seen from  
the figure, the bounds imposed with KamLAND and Borexino constraints
are very similar.  In addition, we notice the expected mass
independent asymptotic behaviour for effectively infinite range
interactions when $m_\phi\lesssim 10^{-16}$, as well as the dependence
on $|g_s^N g_p^{\nu_U}| \,/\, m_\phi^2$ in the contact interaction
regime for $m_\phi\gtrsim 10^{-14}$~eV.

\section{Summary}
\label{sec:summary_chap6}

Using Borexino Phase II data (Section~\ref{sec:NSI_BX}), we derived constraints on non-standard interactions (NSI), neutrino magnetic moments (NMM), and light mediators. For NSI, we accounted for flavor-diagonal and off-diagonal couplings using a density matrix formalism (Eq.~\eqref{eq:ES-dens}), resolving degeneracies between oscillation and interaction effects. Simultaneous inclusion of all vector/axial couplings relaxed bounds compared to single-parameter analyses (Figs.~\ref{fig:NSIcorner-a},~\ref{fig:NSIcorner-v}; Tables~\ref{tab:VA},~\ref{tab:LR}). Vector NSI with electrons eliminated the LMA-D solution, enabling independent determination of $\varepsilon_{\alpha\alpha}^{f,V}$.  

NMM bounds (Eqs.~\eqref{eq:mubounds},~\eqref{eq:mubounds2}) aligned with prior results, validating our $\chi^2$ implementation. Light mediators ($M \lesssim 0.1$ MeV) yielded mass-independent 90\% CL limits:
\begin{equation}
\begin{aligned}
|g_{Z'}^{B-L}| &\leq 6.3 \times 10^{-7},\; |g_{Z'}^{L_e-L_{\mu,\tau}}| \leq 5.8 \times 10^{-7}, \\
|g_\phi^\text{univ}| &\leq 1.4 \times 10^{-6},\; |g_\varphi^\text{univ}| \leq 2.2 \times 10^{-6}.
\end{aligned}
\end{equation}
For heavy mediators ($M \gtrsim$ MeV), contact-interaction bounds scaled as $(g/M)^n$:
\begin{equation}
|g_{Z'}|/M_{Z'} \leq 1.4 \times 10^{-6}\,\text{MeV}^{-1},\quad |g_{\phi,\varphi}|/M \leq 2.5 \times 10^{-6}\,\text{MeV}^{-1}.
\end{equation}

In Section \ref{sec:results_glob_bsm}, we combined neutrino oscillation data (excluding T2K/NO$\nu$A CPV results) with CE$\nu$NS datasets (COHERENT, Dresden-II) and constrained vector/axial NSI couplings to $e,u,d$ quarks. From this study, the main results can be resumed as:
\begin{itemize}

\item Vector NSI in matter potential allowed LMA-D solutions unless ES detection effects (Borexino, SK, SNO) were included (Fig.~\ref{fig:2D}). CE$\nu$NS data disfavored LMA-D beyond $2\sigma$ for quark-coupled mediators (Fig.~\ref{fig:1Deta}).
\item Oscillation parameters remained robust under NSI inclusion (Fig.~\ref{fig:oscil}).
\item Axial NSI bounds were weaker than vector counterparts (Fig.~\ref{fig:etriangA}; Table~\ref{tab:nsiewES}).
\end{itemize}
Earth matter potential projections for atmospheric and LBL experiments are shown in 
Fig.~\ref{fig:earthtriang} and Table~\ref{tab:nsilblranges}.

In Section \ref{sec:MD_section}, we discussed how global solar neutrino data provide stringent constraints on spin-zero mediators with pseudoscalar neutrino couplings ($g_p^\nu$) and scalar nucleon couplings ($g_s^N$). The induced SFP potential, dominant at low neutrino energies, is constrained by global solar+KamLAND analyses (Dirac/Majorana cases) and solar $\bar{\nu}_e$ flux limits (Majorana). Our bounds (Figs.~\ref{fig:limitsnu},~\ref{fig:anures}) surpass previous constraints for ultralight mediators ($m_\phi \lesssim 10^{-12}$ eV):

\begin{itemize}
\item For $m_\phi < 10^{-14}$ eV: Solar bounds exceed CMB limits ($g_p^\nu \leq 7 \times 10^{-7}$~\cite{Forastieri:2019cuf}).
\item For $10^{-14} < m_\phi/\text{eV} < 10^{-6}$: Exceed laboratory limits ($g_p^{\nu_e} \lesssim 3 \times 10^{-3}$~\cite{Lessa:2007up, Pasquini:2015fjv}) and equivalence principle tests~\cite{MICROSCOPE:2022doy, Schlamminger:2007ht}.
\end{itemize}

\chapter{Determination of the Solar Neutrino Fluxes and addressing Gallium Anomaly}
\label{chap:SSM_BX}

In Chapter \ref{chap:exp}, we discussed briefly solar neutrinos and
what their main mechanisms of production are, pp-chain and CNO
cycle. In this chapter, we will dig deep into the understanding of
solar neutrino fluxes and how we can use neutrino data to
determine them and incorporate the latest
results from both solar and non-solar neutrino experiments.  
Specifically, in Sec.~\ref{sec:SSM_full} we work in the context
  of three-neutrino oscillations and present the results of a SSM
  independent analysis where the normalization of the various
  neutrino-producing nuclear reactions is determined solely from
  experimental data, with no other theoretical input than the
  imposition of the luminosity constraint which links together the
  neutrino fluxes and the thermal energy produced by each nuclear
  reaction in the Sun (and accounts for the fact that the overall
  amount of generated thermal energy must match the observed solar
  radiated luminosity).  We then relax also such constraint and check
  it against neutrino data, finding that there is indeed excellent
  experimental agreement between the observed luminosity and its
  neutrino-inferred value.  In Sec.~\ref{sec:frame_gal} we extend our
  analysis to sterile neutrino models and address the issue of the
  \textit{Gallium anomaly} as described in Section~\ref{sec:gallium},
  performing consistency tests to assess the level of incompatibility
  between the 3+1 neutrino scenario invoked to explain the results
  from Gallium source experiments and the strong constraints for
  light-sterile neutrinos that come from solar data.  In the end of
  such section we also comment on the implications that assuming at
face value the sterile solution of the Gallium anomaly would have on
the mechanism for energy production in the Sun.

\section{Determination of solar neutrino fluxes}
\label{sec:SSM_full}

\subsection{Solar Neutrino Fluxes: State of the Art}

Following the discussion in Section \ref{sec:solar_exp}, the pp-chain five fusion reactions among
elements lighter than $A = 8$ produce neutrinos which are labeled by
the parent reaction as \Nuc{pp}, \Nuc[7]{Be}, \Nuc{pep}, \Nuc[8]{B},
and \Nuc{hep} neutrinos.  In the CNO-cycle the abundance of \Nuc{C}
and \Nuc{N} acts as a catalyst, and the \Nuc[13]{N} and \Nuc[15]{O}
beta decays provide the primary source of neutrinos, while \Nuc[17]{F}
beta decay produces a subdominant flux.  For each of these eight
processes the spectral energy shapes of the produced neutrinos is
known, but the calculation of the rate of neutrinos produced in each
reaction requires dedicated modeling of the Sun.

The detection of solar neutrinos, with their extremely small
interaction cross sections, enables us to see into the solar interior  
and verify directly our understanding of the
Sun~\cite{Bahcall:1964gx}~-- provided, of course, that one counts with
an established model of the physics effects relevant to their
production, interaction, and propagation.  The SM was thought to be such an established framework, but it
badly failed at the first attempt of this task giving rise to the
so-called ``solar neutrino problem''~\cite{Bahcall:1968hc,
  Bahcall:1976zz}.  Fortunately, we stand here, almost fifty years after
that first realization of the problem, with a different but well
established framework for the relevant effects in solar neutrino
propagation.  A framework in which the three flavour neutrinos
($\nu_e$, $\nu_\mu$, $\nu_\tau$) of the Standard Model are mixed
quantum superpositions of three massive states $\nu_i$ ($i=1,2,3$)
with different masses.  This allows for the flavour of the neutrino to
oscillate from production to detection~\cite{Pontecorvo:1967fh,
  Gribov:1968kq}, and for non-trivial effects (the so-called  
LMA-MSW~\cite{Wolfenstein:1977ue, Mikheev:1986gs} flavour transitions)
to take place when crossing dense regions of matter.  Furthermore, due
to the wealth of experiments exploring neutrino oscillations, the
value of the neutrino properties governing the propagation of solar  
neutrinos, mass differences, and mixing angles, are now precisely and
independently known.

Armed with this robust particle physics framework for neutrino
production, propagation, and detection, it is possible to turn to the
observation of solar neutrino experiments to test and refine the SSM.
Unfortunately soon after the particle physics side of the exercise was
clarified, the construction of the SSM run into a new puzzle: the so
called ``solar composition problem''.  In brief, SSMs built in the
1990's using the abundances of heavy elements on the surface of the
Sun from Ref.~\cite{Grevesse1998} (GS98) had notable successes in
predicting other observations, in particular helioseismology
measurements such as the radial distributions of sound speed and
density~\cite{Bahcall:1992hn, Bahcall:1995bt, Bahcall:2000nu,
  Bahcall:2004pz}.  But in the 2000's new determinations of these
abundances became available and pointed towards substantially lower
values, as summarized in Ref.~\cite{Asplund2009} (AGSS09).  The SSMs
built incorporating such lower metallicities failed at explaining the
helioseismic observations~\cite{Bahcall:2004yr}.
For almost two decades there was no successful solution of this puzzle
as changes in the modeling of the Sun did not seem able to account for
this discrepancy~\cite{Castro:2007, Guzik:2010, Serenelli:2011py}.
Consequently two different sets of SSMs were built, each based on the
corresponding set of solar abundances~\cite{Serenelli:2009yc,
  Serenelli:2011py, Vinyoles:2016djt}.

With this in mind, in Refs.~\cite{Gonzalez-Garcia:2009dpj,
  Bergstrom:2016cbh} we performed solar model independent analysis of
the solar and terrestrial neutrino data available at the time, in the
framework of three-neutrino masses and mixing, where the flavour
parameters and all the solar neutrino fluxes were simultaneously
determined with a minimum set of theoretical priors.  The results were
compared with the two variants of the SSM, but they were not precise
enough to provide a significant discrimination.

Since then there have been a number of developments.  First of all, a
substantial amount of relevant data has been accumulated, in
particular the full spectral information of the
Phase-II~\cite{Borexino:2017rsf} and Phase-III~\cite{BOREXINO:2022abl}
of Borexino and their results based on the correlated integrated
directionality (CID) method~\cite{Borexino:2023puw}.  All of them have
resulted into the first positive observation of the neutrino fluxes
produced in the CNO-cycle which are particularly relevant for
discrimination among the SSMs.

On the model front, an update of the AGSS09 results was recently
presented by the same group (AAG21)~\cite{Asplund2021}, though leading
only to a slight revision upwards of the solar metallicity.  Most
interestingly, almost simultaneously a new set of results
(MB22)~\cite{Magg:2022rxb}, based on similar methodologies and
techniques but with different atomic input data for the critical
oxygen lines among other differences, led to a substantial change in
solar elemental abundances with respect to AGSS09 (see the original
reference for details).  The outcome is a set of solar abundances
based on three-dimensional radiation hydrodynamic solar atmosphere
models and line formation treated under non-local thermodynamic
equilibrium that yields a total solar metallicity comparable to those
of the ``high-metallicity'' results by GS98.

Another issue which has come up in the interpretation of the solar
neutrino results is the appearance of the so-called ``gallium
anomaly''.  In brief, it accounts for the deficit of the rate of
events observed in Gallium source experiments with respect to the
expectation.  It was originally observed in the calibration of the
gallium solar-neutrino detectors GALLEX~\cite{GALLEX:1997lja,
  Kaether:2010ag} and SAGE~\cite{SAGE:1998fvr, Abdurashitov:2005tb}
with radioactive \Nuc[51]{Cr} and \Nuc[37]{Ar} sources, and it has
been recently confirmed by the BEST collaboration with a dedicated
source experiment using a \Nuc[51]{Cr} source with high statistical
significance~\cite{Barinov:2021asz, Barinov:2022wfh}.

The solution of this puzzle is an open question in neutrino physics
(see Ref.~\cite{Brdar:2023cms} and reference therein for a recent
discussion of --~mostly unsuccessful~-- attempts at explanations in
terms of standard and non-standard physics scenarios).  In particular,
in the framework of 3$\nu$ oscillations, the attempts at explanation
(or at least alleviation) of the anomaly invoke the uncertainties of
the capture cross section~\cite{Berryman:2021yan, Giunti:2022xat,
  Elliott:2023xkb}.  With this motivation, in this work we have
studied the (in)sensitivity of our results to the intrinsic
uncertainty on the observed neutrino rates in the Gallium source experiments
posed by possible modification of the capture cross section in
Gallium, or equivalently, of the detection efficiency of the Gallium
solar neutrino experiments.

All these developments motivate the new analysis which we present in
this section with the following outline.  In Sec.~\ref{sec:frame} we
describe the assumptions and methodology followed in our study of the
neutrino data.  As mentioned above, this work builds upon our previous
solar flux determination in Refs.~\cite{Gonzalez-Garcia:2009dpj,
  Bergstrom:2016cbh}.  Thus in Sec.~\ref{sec:frame}, for convenience,
we summarize the most prominent elements common to those analyses, but
most importantly, we detail the relevant points in which the present
analysis method deviates from them.  The new determination of the
solar fluxes is presented in Sec.~\ref{sec:res} where we also discuss
and quantify the role of the Gallium source experiments and address their
robustness with respect to the Gallium anomaly.  In Sec.~\ref{sec:CNO}
we have a closer look at the determination of the neutrino fluxes from
the CNO-cycle and its dependence on the assumptions on the relative
normalization of the fluxes produced in the three relevant reactions.
In Sec.~\ref{sec:compaSSM} we compare our determined fluxes with the
predictions of the SSMs in the form of a \emph{parameter goodness of
fit} test, and quantify the output of the test for the assumptions in
the analysis.  
\subsection{Analysis framework}
\label{sec:frame}

In the analysis of solar neutrino experiments we include the total
rates from the radiochemical experiments
Chlorine~\cite{Cleveland:1998nv}, Gallex/GNO~\cite{Kaether:2010ag},
and SAGE~\cite{Abdurashitov:2009tn}, the spectral and day-night data
from the four phases of Super-Kamiokande~\cite{Hosaka:2005um,
  Cravens:2008aa, Abe:2010hy, SK:nu2020}, the results of the three
phases of SNO in terms of the parametrization given in their combined
analysis~\cite{Aharmim:2011vm}, and the full spectra from Borexino
Phase-I~\cite{Bellini:2011rx}, Phase-II~\cite{Borexino:2017rsf}, and
Phase-III~\cite{BOREXINO:2022abl}, together with their latest results
based on the correlated integrated directionality (CID)
method~\cite{Borexino:2023puw}.  
In the framework of three neutrino masses and mixing the expected
values for these solar neutrino observables depend on the parameters
$\Dmq_{21}$, $\theta_{12}$, and $\theta_{13}$ as well as on the
normalizations of the eight solar fluxes.  Thus besides solar
experiments, we also include in the analysis the separate DS1, DS2,
DS3 spectra from KamLAND~\cite{Gando:2013nba} which in the framework
of three neutrino mixing also yield information on the parameters
$\Dmq_{21}$, $\theta_{12}$, and $\theta_{13}$.

In what follows we will use as normalization parameters for the solar
fluxes the reduced quantities:
\begin{equation}
  \label{eq:redflux}
  f_i = \frac{\Phi_i}{\Phi_i^\text{ref}}
\end{equation}
with $i = \Nuc{pp}$, \Nuc[7]{Be}, \Nuc{pep}, \Nuc[13]{N}, \Nuc[15]{O},
\Nuc[17]{F}, \Nuc[8]{B}, and \Nuc{hep}.  In this work the numerical
values of $\Phi_i^\text{ref}$ are set to the predictions of the latest
GS98 solar model, described in Ref.~\cite{Magg:2022rxb, B23Fluxes}.
They are listed in Table~\ref{tab:lumcoef}.
The methodology of the analysis presented in this work builds upon our
previous solar flux determination in
Refs.~\cite{Gonzalez-Garcia:2009dpj, Bergstrom:2016cbh}, which we
briefly summarize here for convenience, but it also exhibits a number
of differences besides the additional data included as described next.

The theoretical predictions for the solar and KamLAND observables
depend on eleven parameters: the eight reduced solar fluxes
$f_{\Nuc{pp}}$, \dots, $f_{\Nuc{hep}}$, and the three relevant
oscillation parameters $\Dmq_{21}$, $\theta_{12}$, $\theta_{13}$.  In
our analysis we include as well the complementary information on
$\theta_{13}$ obtained after marginalizing over $\Dmq_{3\ell}$,
$\theta_{23}$ and $\delta_\text{CP}$ the results of all the other
oscillation experiments considered in NuFIT-5.2~\cite{nufit-5.2}.
This results into a prior $\sin^2\theta_{13} = 0.0223\pm 0.0006$,
\textit{i.e.}, $\theta_{13} = 8.59^\circ\, (1\pm 0.014)$.  Given the
weak dependence of the solar and KamLAND observables on $\theta_{13}$,
including such prior yields results which are indistinguishable from
just fixing the value of $\bar\theta_{13}=8.59^\circ$.

Throughout this work, we follow a frequentist approach in order to
determine the allowed confidence regions for these parameters (unlike
in our former works~\cite{Gonzalez-Garcia:2009dpj, Bergstrom:2016cbh}
where we used instead a Bayesian analysis to reconstruct their
posterior probability distribution function).  To this end we make use
of the experimental data from the various solar and KamLAND samples
($D_\text{solar}$ and $D_\text{KamLAND}$, respectively) as well as the
corresponding theoretical predictions (which depends on ten free
parameters, as explained above) to build the $\chi^2$ statistical
function
\begin{equation}
  \label{eq:chi2g}
  \chi^2_\text{global}(\vec\omega_\text{flux},\, \vec\omega_\text{osc})
  \equiv \chi^2_\text{solar}(D_\text{solar} \,|\,
  \vec\omega_\text{flux},\, \vec\omega_\text{osc})
  + \chi^2_\text{KamLAND}(D_\text{KamLAND}\,|\, \vec\omega_\text{osc}) \,,
\end{equation}
with $\vec\omega_\text{flux} \equiv (f_{\Nuc{pp}},\, \dots,\,
f_{\Nuc{hep}})$ and $\vec\omega_\text{osc} \equiv (\Dmq_{21},\,
\theta_{12},\, \bar\theta_{13})$.  In order to scan this
multidimensional parameter space efficiently, we make use of the
MultiNest~\cite{Feroz:2013hea, Feroz:2008xx} and
Diver~\cite{Martinez:2017lzg} algorithms.

\begin{table}[t]\centering
  \catcode`?=\active\def?{\hphantom{0}}
  \begin{tabular}{r@{\hspace{20mm}}c@{\hspace{20mm}}c@{\hspace{20mm}}c}
    Flux & $\Phi_i^\text{ref}$ [$\text{cm}^{-2}\, \text{s}^{-1}$]
    & $\alpha_i$ [MeV] & $\beta_i$
    \\
    \hline
    \Nuc{pp}    & $5.960\times 10^{10}$ & $13.099?$ & $9.1864\times 10^{-1}$ \\
    \Nuc[7]{Be} & $4.854\times 10^{9?}$ & $12.552?$ & $7.1693\times 10^{-2}$ \\
    \Nuc{pep}   & $1.425\times 10^{8?}$ & $11.920?$ & $1.9987\times 10^{-3}$ \\
    \Nuc[13]{N} & $2.795\times 10^{8?}$ & $12.658?$ & $4.1630\times 10^{-3}$ \\
    \Nuc[15]{O} & $2.067\times 10^{8?}$ & $12.368?$ & $3.0082\times 10^{-3}$ \\
    \Nuc[17]{F} & $5.350\times 10^{6?}$ & $12.365?$ & $7.7841\times 10^{-5}$ \\
    \Nuc[8]{B}  & $5.025\times 10^{6?}$ & $?6.6305$ & $3.9205\times 10^{-5}$ \\
    \Nuc{hep}   & $7.950\times 10^{3?}$ & $?3.7355$ & $3.4944\times 10^{-8}$
  \end{tabular}
  \caption{The reference neutrino fluxes $\Phi_i^\text{ref}$ used for
    normalization (from Ref.~\cite{Magg:2022rxb, B23Fluxes}), the
    energy $\alpha_i$ provided to the star by nuclear fusion reactions
    associated with the $i^\text{th}$ neutrino flux (taken from
    Ref.~\cite{2021JPhG...48a5201V}), and the fractional
    contribution $\beta_i$ of the $i^\text{th}$ nuclear reaction to
    the total solar luminosity.}
  \label{tab:lumcoef}
\end{table}

The allowed range for the solar fluxes is further reduced when
imposing the so-called ``luminosity constraint'', \textit{i.e.}, the
requirement that the overall sum of the thermal energy generated
together with each solar neutrino flux coincides with the solar
luminosity~\cite{Spiro:1990vi}:
\begin{equation}
  \label{eq:lumsum1}
  \frac{L_\odot}{4\pi \, (\text{A.U.})^2}
  = \sum_{i=1}^8 \alpha_i \Phi_i \,.
\end{equation}
Here the constant $\alpha_i$ is the energy released into the star by
the nuclear fusion reactions associated with the $i^\text{th}$
neutrino flux; its numerical value is independent of details of the
solar model to an accuracy of one part in $10^4$ or
better~\cite{Bahcall:2001pf}.  A detailed derivation of this equation
and the numerical values of the coefficients $\alpha_i$ is presented
in Ref.~\cite{Bahcall:2001pf}, with some refinement and correction
following in~\cite{2021JPhG...48a5201V}.\footnote{We have explicitly
verified that the numerical differences between the results of the
analysis performed using the original $\alpha_i$ coefficients in
Ref.~\cite{Bahcall:2001pf} and those in
Ref.~\cite{2021JPhG...48a5201V} are below the quoted precision.}  The
coefficients employed in this work are listed in
Table~\ref{tab:lumcoef}.
In terms of the reduced fluxes Eq.~\eqref{eq:lumsum1} can be written
as:
\begin{equation}
  \label{eq:lumsum2}
  1 = \sum_{i=1}^8 \beta_i f_i
  \quad\text{with}\quad
  \beta_i \equiv
  \frac{\alpha_i \Phi_i^\text{ref}}{L_\odot \big/ [4\pi \, (\text{A.U.})^2]}
\end{equation}
where $\beta_i$ is the fractional contribution to the total solar
luminosity of the nuclear reactions responsible for the production of
the $\Phi_i^\text{ref}$ neutrino flux.  In
Refs.~\cite{Gonzalez-Garcia:2009dpj, Bergstrom:2016cbh} we adopted the
best-estimate value for the solar luminosity $L_\odot \big/ [4\pi \,
  (\text{A.U.})^2] = 8.5272 \times 10^{11}\, \text{MeV}\,
\text{cm}^{-2}\, \text{s}^{-1}$ given in Ref.~\cite{Bahcall:2001pf},
which was obtained from all the available satellite
data~\cite{Frohlich1998}.  This value was revised in
Ref.~\cite{kopp2011} using an updated catalog and calibration
methodology (see Ref.~\cite{Scafetta_2014} for a detailed comparative
discussion), yielding a slightly lower result which is now the
reference value listed by the PDG~\cite{Workman:2022ynf} and leads to
$L_\odot \big/ [4\pi \, (\text{A.U.})^2] = 8.4984 \times 10^{11} \,
\text{MeV} \, \text{cm}^{-2} \, \text{s}^{-1}$.  In this work we adopt
this new value when evaluating the $\beta_i$ coefficients listed in
Table~\ref{tab:lumcoef}.  Furthermore, in order to account for the
systematics in the extraction of the solar luminosity we now assign an
uncertainty of $0.34\%$ to the constraint in Eq.~\eqref{eq:lumsum2},
which we conservatively derive from the range of variation of the
estimates of $L_\odot$.
In what follows we will present results with and without imposing the
luminosity constraint.  For the analysis including the luminosity
constraint we add a prior
\begin{equation}
  \label{eq:priorLC}
  \chi^2_\text{LC}(\vec\omega_\text{flux})
  = \frac{\left( \displaystyle 1-\sum_{i=1}^8 \beta_i f_i\right)^2}{(0.0034)^2}
\end{equation}

Besides the imposition of the luminosity constraint in some of the
analysis, the flux normalizations are allowed to vary freely within a
set of physical constraints.  In particular:
\begin{itemize}
\item The fluxes must be positive:
  \begin{equation}
    \label{eq:fpos}
    \Phi_i \geq 0 \quad\Rightarrow\quad f_i \geq 0 \,.
  \end{equation}

\item Consistency in the pp-chain implies that the number of nuclear
  reactions terminating the pp-chain should not exceed the number of
  nuclear reactions which initiate it~\cite{Bahcall:1995rs,
    Bahcall:2001pf}:
  \begin{multline}
    \Phi_{\Nuc[7]{Be}} + \Phi_{\Nuc[8]{B}}
    \leq \Phi_{\Nuc{pp}} + \Phi_{\Nuc{pep}}
    \\
    \Rightarrow\quad
    8.12 \times 10^{-2} f_{\Nuc[7]{Be}}
    + 8.42 \times 10^{-5} f_{\Nuc[8]{B}}
    \leq f_{\Nuc{pp}} + 2.38 \times 10^{-3} f_{\Nuc{pep}} \,.
  \end{multline}

\item The ratio of the \Nuc{pep} neutrino flux to the \Nuc{pp}
  neutrino flux is fixed to high accuracy because they have the same
  nuclear matrix element.  We have constrained this ratio to match the
  average of the values in the five B23 SSMs
  (Sec.~\ref{sec:compaSSM}), with $1\sigma$ Gaussian uncertainty given
  by the difference between the values in the five models
  \begin{equation}
    \label{eq:pep-pp}
    \frac{f_{\Nuc{pep}}}{f_{\Nuc{pp}}} = 1.004 \pm 0.018 \,.
  \end{equation}
  Technically we implement this constraint by adding a Gaussian prior
  \begin{equation}
    \chi^2_\text{pep/pp}(f_{\Nuc{pp}},\, f_{\Nuc{pep}}) \equiv \left(
    \frac{f_{\Nuc{pep}} \big/ f_{\Nuc{pp}} - 1.004}{0.018}
    \right)^2.
  \end{equation}

\item For the CNO fluxes ($f\Phi_{\Nuc[13]{N}}$, $\Phi_{\Nuc[15]{O}}$,
  and $\Phi_{\Nuc[17]{F}}$) a minimum set of assumptions required by
  consistency are:
  \begin{itemize}
  \item The $\Nuc[14]{N}(p,\gamma) \Nuc[15]{O}$ reaction must be the
    slowest process in the main branch of the
    CNO-cycle~\cite{Bahcall:1995rs}:
    \begin{equation}
      \label{eq:CNOineq1}
      \Phi_{\Nuc[15]{O}} \leq \Phi_{\Nuc[13]{N}}
      \quad\Rightarrow\quad
      f_{\Nuc[15]{O}} \leq 1.35\, f_{\Nuc[13]{N}}
    \end{equation}

  \item the CNO-II branch must be subdominant:
    \begin{equation}
      \label{eq:CNOineq2}
      \Phi_{\Nuc[17]{F}} \leq \Phi_{\Nuc[15]{O}}
      \quad\Rightarrow\quad
      f_{\Nuc[17]{F}}\leq 40\, f_{\Nuc[15]{O}} \,.
    \end{equation}
  \end{itemize}
\end{itemize}
The conditions quoted above are all dictated by solar physics.
However, more practical reasons require that the CNO fluxes are
treated with a special care.  As discussed in detail in
Appendix~\ref{sec:bx3nfit}, the analysis of the Borexino Phase-III
spectra in Ref.~\cite{BOREXINO:2020aww} (which we closely reproduce)
has been optimized by the collaboration to maximize the sensitivity to
the overall CNO production rate, and therefore it may not be directly
applicable to a situation where the three \Nuc[13]{N}, \Nuc[15]{O} and
\Nuc[17]{F} flux normalizations are left totally free, subject only to
the conditions in Eqs.~\eqref{eq:CNOineq1} and~\eqref{eq:CNOineq2}.
Hence, following the approach of the Borexino collaboration in
Ref.~\cite{BOREXINO:2020aww}, we first perform an analysis where the
three CNO components are all scaled simultaneously by a unique
normalization parameter while their ratios are kept fixed as predicted
by the SSMs.  In order to avoid a bias towards one of the different
versions of the SSM we have constrained the two ratios to match the
average of the five B23 SSMs values
\begin{equation}
  \label{eq:CNOfix}
  \frac{\Phi_{\Nuc[15]{O}}}{\Phi_{\Nuc[13]{N}}} = 0.73
  \enspace\text{and}\enspace
  \frac{\Phi_{\Nuc[17]{F}}}{\Phi_{\Nuc[13]{N}}} = 0.016
  \quad\Rightarrow\quad
  \frac{f_{\Nuc[15]{O}}}{f_{\Nuc[13]{N}}} = 0.98
  \enspace\text{and}\enspace
  \frac{f_{\Nuc[17]{F}}}{f_{\Nuc[13]{N}}} = 0.85 \,.
\end{equation}
In these analysis, which we label <<CNO-Rfixed>>, the conditions in
Eq.~\eqref{eq:CNOfix} effectively reduces the number of free
parameters from ten to eight, namely the two oscillation parameters in
$\vec\omega_\text{osc}$ and six flux normalizations in
\begin{equation}
  \nonumber
  \vec\omega_\text{flux}^\text{CNO-Rfixed}
  \equiv (f_{\Nuc{pp}},\; f_{\Nuc[7]{Be}},\; f_{\Nuc{pep}},\; f_{\Nuc[13]{N}},\;
  f_{\Nuc[15]{O}} = 0.98\, f_{\Nuc[13]{N}},\;
  f_{\Nuc[17]{F}} = 0.85\, f_{\Nuc[13]{N}},\;
  f_{\Nuc[8]{B}},\; f_{\Nuc{hep}}) \,.
\end{equation}
In Sec.~\ref{sec:CNO} we will discuss and quantify the effect of
relaxing the condition of fixed CNO ratios.

\subsection{New determination of solar neutrino fluxes}
\label{sec:res}

\begin{figure}[t]\centering
  \includegraphics[width=0.95\textwidth]{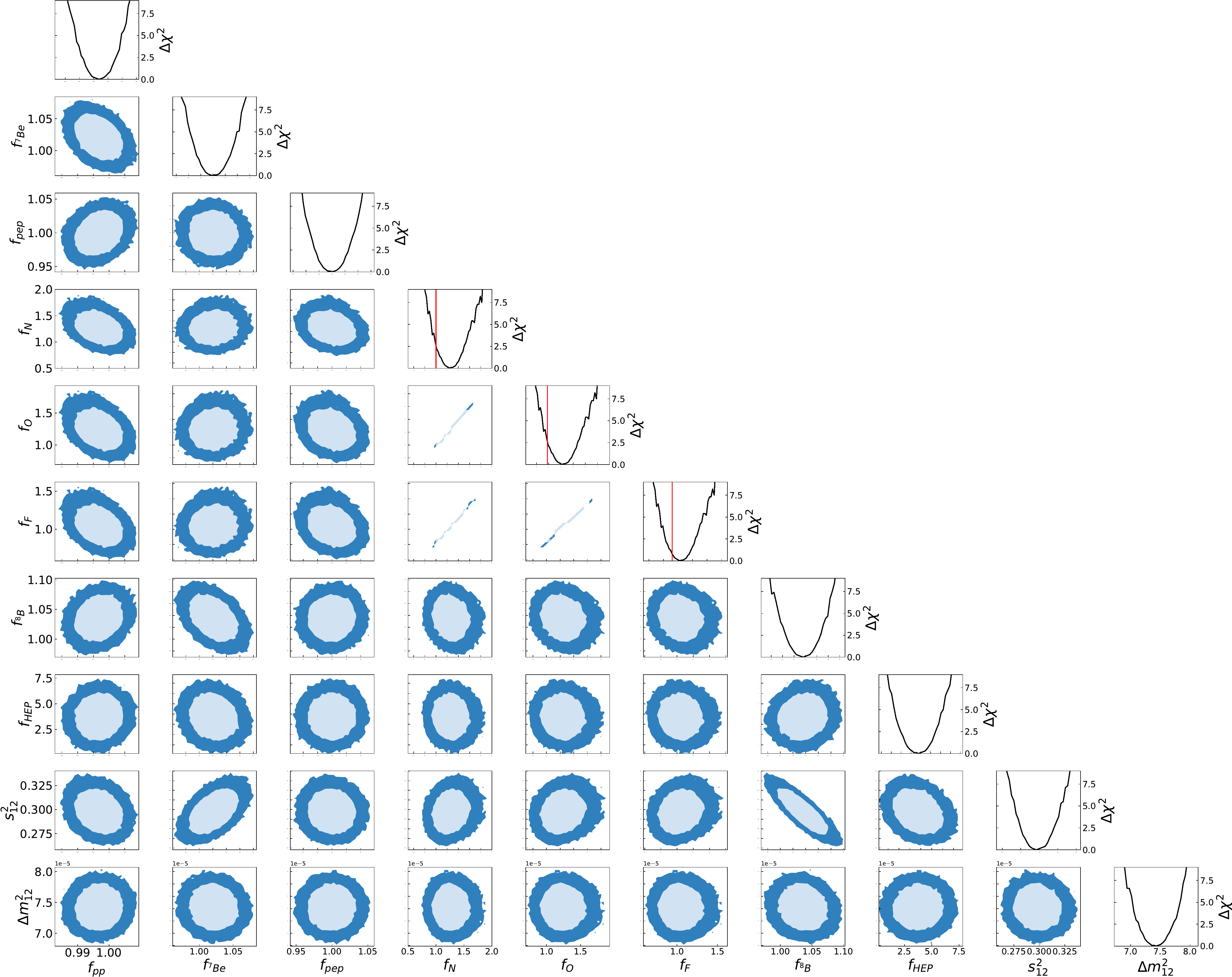}
  \caption{Constraints from our global analysis on the solar neutrino
    fluxes for the analysis with luminosity constraint and fixed
    ratios of the CNO fluxes (see Eq.~\eqref{eq:chi2wlccno}).  Each
    panel shows a two-dimensional projection of the allowed
    multidimensional parameter space after minimization with respect
    to the undisplayed parameters.  The regions correspond to 90\% and
    99\% CL (2 d.o.f.).  The curves in the rightmost panels show the
    marginalized one-dimensional $\Delta\chi^2_\text{wLC,CNO-Rfixed}$
    for each of the parameters.}
  \label{fig:triCNOLCGA1}
\end{figure}

We present first the results of our analysis with the luminosity
constraint and the ratios of the CNO fluxes fixed by the relations in
Eq.~\eqref{eq:CNOfix}, so that altogether for this case we construct
the $\chi^2$ function
\begin{equation}
  \label{eq:chi2wlccno}
  \chi^2_\text{wLC,CNO-Rfixed} \equiv
  \chi^2_\text{global}(\vec\omega_\text{osc},\, \vec\omega_\text{flux}^\text{CNO-Rfixed})
  + \chi^2_\text{pep/pp}(f_{\Nuc{pp}},\, f_{\Nuc{pep}})
  + \chi^2_\text{LC}(\vec\omega_\text{flux}^\text{CNO-Rfixed}) \,.
\end{equation}
The results of this analysis are displayed in
Fig.~\ref{fig:triCNOLCGA1}, where we show the two- and one-dimensional
projections of $\Delta\chi^2_\text{wLC,CNO-Rfixed}$.  From these
results one reads that the ranges at $1\sigma$ (and at the $99\%$ CL
in square brackets) for the two oscillation parameters are:
\begin{equation}
  \label{eq:bestosc}
  \begin{aligned}
    \Dmq_{21}
    &= 7.43_{-0.30}^{+0.30}\, [{}_{-0.49}^{+0.44}]
    \times 10^{-5} \eVq \,,
    \\[2pt]
    \sin^2\theta_{12}
    &= 0.300_{-0.017}^{+0.020}\, [{}_{-0.027}^{+0.031}] \,,
  \end{aligned}
\end{equation}
which are very similar to the results of NuFIT-5.2~\cite{nufit-5.2}
with the expected slight enlargement of the allowed ranges.  In other
words, within the $3\nu$ scenario the data is precise enough to
simultaneously constraint the oscillation parameters and the
normalizations of the solar flux components without resulting into a
substantial degradation of the former.
As for the solar fluxes, the corresponding ranges read:
\begin{equation}
  \label{eq:bestlc}
  \begin{aligned}
    f_{\Nuc{pp}}
    & = 0.9969_{-0.0039}^{+0.0041}\, [{}_{-0.0092}^{+0.0095}]
    \,, \qquad
    & \Phi_{\Nuc{pp}}
    & = 5.941_{-0.023}^{+0.024}\, [{}_{-0.055}^{+0.057}]
    \times 10^{10}~\text{cm}^{-2}~\text{s}^{-1} \,,
    \\
    f_{\Nuc[7]{Be}}
    & = 1.019_{-0.017}^{+0.020}\, [{}_{-0.041}^{+0.047}]
    \,, \qquad
    & \Phi_{\Nuc[7]{Be}}
    & = 4.93_{-0.08}^{+0.10}\, [{}_{-0.20}^{+0.23}]
    \times 10^{9}~\text{cm}^{-2}~\text{s}^{-1} \,,
    \\
    f_{\Nuc{pep}}
    & = 1.000_{-0.018}^{+0.016}\, [{}_{-0.042}^{+0.041}]
    \,, \qquad
    & \Phi_{\Nuc{pep}}
    & = 1.421_{-0.026}^{+0.023}\, [{}_{-0.060}^{+0.058}]
    \times 10^{8}~\text{cm}^{-2}~\text{s}^{-1} \,,
    \\
    f_{\Nuc[13]{N}}
    & = 1.25_{-0.14}^{+0.17}\, [{}_{-0.40}^{+0.47}]
    \,, \qquad
    & \Phi_{\Nuc[13]{N}}
    & = 3.48_{-0.40}^{+0.47}\, [{}_{-1.10}^{+1.30}]
    \times 10^{8}~\text{cm}^{-2}~\text{s}^{-1} \,,
    \\
    f_{\Nuc[15]{O}}
    & = 1.22_{-0.14}^{+0.17}\, [{}_{-0.39}^{+0.46}]
    \qquad
    & \Phi_{\Nuc[15]{O}}
    & = 2.53_{-0.29}^{+0.34}\, [{}_{-0.80}^{+0.94}]
    \times 10^{8}~\text{cm}^{-2}~\text{s}^{-1} \,,
    \\
    f_{\Nuc[17]{F}}
    & = 1.03_{-0.20}^{+0.20}\, [{}_{-0.48}^{+0.47}]
    \,, \qquad
    & \Phi_{\Nuc[17]{F}}
    & = 5.51_{-0.63}^{+0.75}\, [{}_{-1.75}^{+2.06}]
    \times 10^{7}~\text{cm}^{-2}~\text{s}^{-1} \,,
    \\
    f_{\Nuc[8]{B}}
    & = 1.036_{-0.020}^{+0.020}\, [{}_{-0.048}^{+0.047}]
    \,, \qquad
    & \Phi_{\Nuc[8]{B}}
    & = 5.20_{-0.10}^{+0.10}\, [{}_{-0.24}^{+0.24}]
    \times 10^{6}~\text{cm}^{-2}~\text{s}^{-1} \,,
    \\
    f_{\Nuc{hep}}
    & = 3.8_{-1.2}^{+1.1}\, [{}_{-2.7}^{+2.7}]
    \,, \qquad
    & \Phi_{\Nuc{hep}}
    & = 3.0_{-1.0}^{+0.9}\, [{}_{-2.1}^{+2.2}]
    \times 10^{4}~\text{cm}^{-2}~\text{s}^{-1} \,.
  \end{aligned}
\end{equation}
Notice that in Fig.~\ref{fig:triCNOLCGA1} we separately plot the
ranges for the three CNO flux normalization parameters, however they
are fully correlated since their ratios are fixed, which explains the
thin straight-line shape of the regions as seen in the three
corresponding panels.  Compared to the results from our previous
analysis we now find that all the fluxes are clearly determined to be
non-zero, while in Refs.~\cite{Gonzalez-Garcia:2009dpj,
  Bergstrom:2016cbh} only an upper bound for the CNO fluxes was found.
This is a direct consequence of the positive evidence of neutrinos
produced in the CNO cycle provided by Borexino Phase-III spectral
data, which is here confirmed in a fully consistent global analysis.
We will discuss this point in more detail in Sec.~\ref{sec:CNO}.  We
also observe that the inclusion of the full statistics of Borexino has
improved the determination of $f_{\Nuc[7]{Be}}$ by a factor
$\mathcal{O}(3)$.

\begin{figure}[t]\centering
  \includegraphics[width=0.95\textwidth]{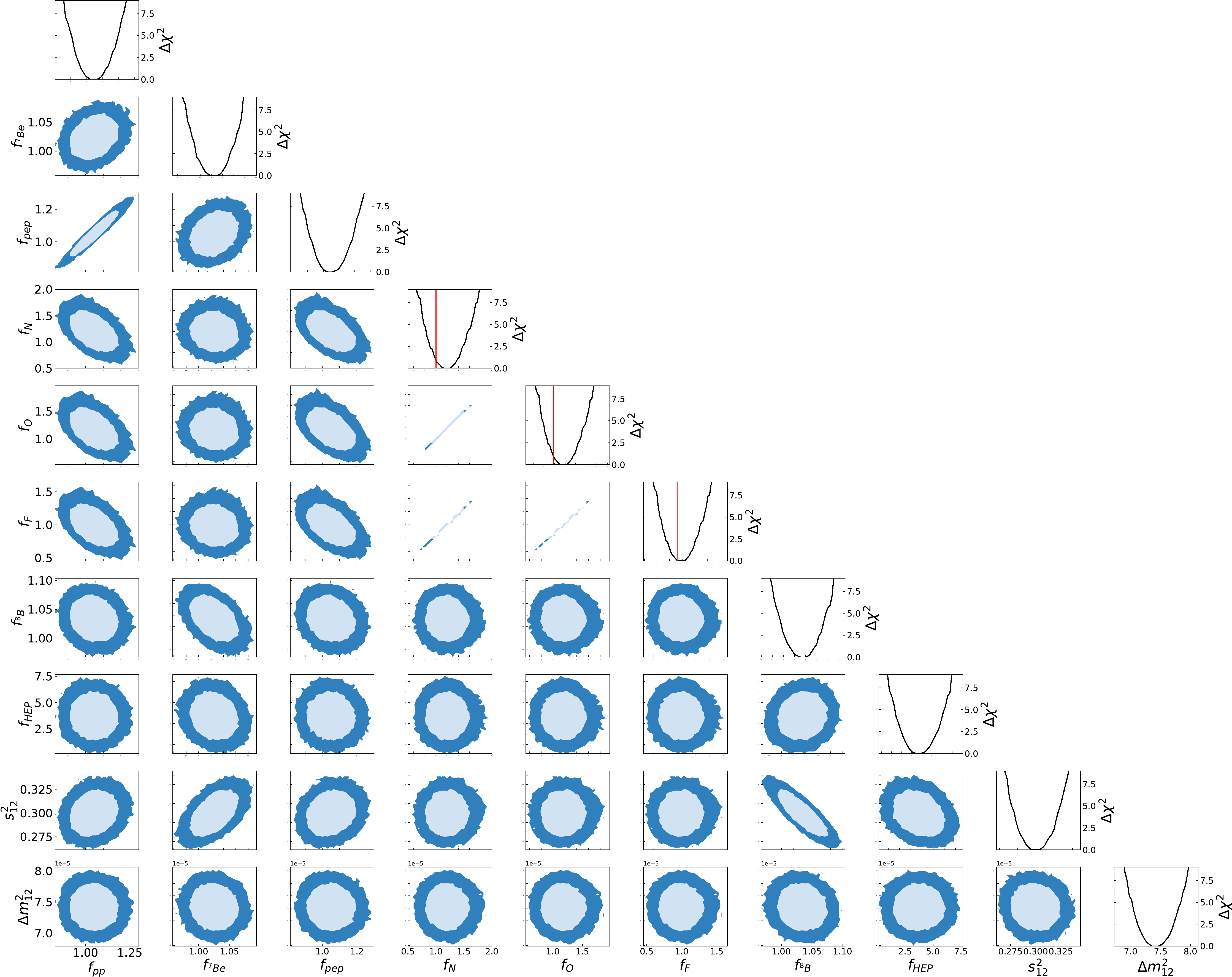}
  \caption{Same as Fig.~\ref{fig:triCNOLCGA1} but without imposing
    the luminosity constraint (see Eq.~\eqref{eq:chi2wolccno}).}
  \label{fig:triCNOwoLCGA1}
\end{figure}

Figure~\ref{fig:triCNOLCGA1} exhibits the expected correlation between
the allowed ranges of the \Nuc{pp} and \Nuc{pep} fluxes, which is a
consequence of the relation~\eqref{eq:pep-pp}.  This correlation is
somewhat weaker than what observed in the corresponding analysis in
Ref.~\cite{Bergstrom:2016cbh} because the spectral information from
Borexino Phase-II and Phase-III provides now some independent
information on $f_{\Nuc{pep}}$.  We also observe the presence of
anticorrelation between the allowed ranges of the two most intense
fluxes, \Nuc{pp} and \Nuc[7]{Be}, as dictated by the luminosity
constraint (see comparison with Fig.~\ref{fig:triCNOwoLCGA1}).
Finally we notice that the allowed ranges of $f_{\Nuc[7]{Be}}$ and
$f_{\Nuc[8]{B}}$ --~the two most precise directly determined flux
normalizations irrespective of the luminosity constraint (see
Fig.~\ref{fig:triCNOwoLCGA1})~-- are anticorrelated.  This is a direct
consequence of the different dependence of the survival probability
with $\sin^2\theta_{12}$ in their respective energy ranges.
\Nuc[8]{B} neutrinos have energies of the order of several MeV for
which the flavour transition occurs in the MSW regime and the survival
probability $P_{ee}\propto \sin^2\theta_{12}$.  Hence an increase in
$\sin^2\theta_{12}$ must be compensated by a decrease of
$f_{\Nuc[8]{B}}$ to get the correct number of events, which leads to
the anticorrelation between the $\sin^2\theta_{12}$ and
$f_{\Nuc[8]{B}}$ seen in the corresponding panel in
Fig.~\ref{fig:triCNOLCGA1}.  On the contrary, most \Nuc[7]{Be}
neutrinos have 0.86 MeV (some have 0.38 MeV) and for that energy the
flavour transition occurs in the transition regime between MSW and
vacuum average oscillations for which $P_{ee}$ decreases with
$\sin^2\theta_{12}$.  Hence the correlation between
$\sin^2\theta_{12}$ and $f_{\Nuc[7]{Be}}$ seen in the corresponding
panel.  Altogether, this leads to the anticorrelation observed between
$f_{\Nuc[7]{Be}}$ and $f_{\Nuc[8]{B}}$.  This was already mildly
present in the results in Ref.~\cite{Bergstrom:2016cbh} but it is now
a more prominent feature because of the most precise determination of
$f_{\Nuc[7]{Be}}$.

All these results imply the following share of the energy production
between the pp-chain and the CNO-cycle
\begin{equation}
  \label{eq:ppcnolum1}
  \frac{L_\text{pp-chain}}{L_\odot} =
  0.9919_{-0.0030}^{+0.0035}\, [{}_{-0.0077}^{+0.0082}]
  \quad\Longleftrightarrow\quad
  \frac{L_\text{CNO}}{L_\odot} =
  0.0079_{-0.0011}^{+0.0009}\, [{}_{-0.0026}^{+0.0028}]
\end{equation}
in perfect agreement with the SSMs which predict $L_\text{CNO} \big/
L_\odot \leq 1\%$ at the $3\sigma$ level.  Once again we notice that
in the present analysis the evidence for $L_\text{CNO}\neq 0$
clearly stands well above 99\% CL.

We next show in Fig.~\ref{fig:triCNOwoLCGA1} the results of the
analysis performed without imposing the luminosity constraint --~but
still with the ratios of the CNO fluxes fixed by the relations in
Eq.~\eqref{eq:CNOfix}~-- for which we employ
\begin{equation}
  \label{eq:chi2wolccno}
  \chi^2_\text{woLC,CNO-Rfixed} \equiv
  \chi^2_\text{global}(\vec\omega_\text{osc},\, \vec\omega_\text{flux}^\text{CNO-Rfixed})
  + \chi^2_\text{pep/pp}(f_{\Nuc{pp}},\, f_{\Nuc{pep}}) \,.
\end{equation}
The allowed ranges for the fluxes in this case are:
\begin{equation}
  \label{eq:bestwolc}
  \begin{aligned}
    f_{\Nuc{pp}}
    & = 1.038_{-0.066}^{+0.076}\, [{}_{-0.16}^{+0.18}]
    \,, \qquad
    & \Phi_{\Nuc{pp}}
    & = 6.19_{-0.39}^{+0.45}\, [{}_{-1.0}^{+1.1}]
    \times 10^{10}~\text{cm}^{-2}~\text{s}^{-1} \,,
    \\
    f_{\Nuc[7]{Be}}
    & = 1.022_{-0.018}^{+0.022}\, [{}_{-0.042}^{+0.051}]
    \,, \qquad
    & \Phi_{\Nuc[7]{Be}}
    & = 4.95_{-0.089}^{+0.11}\, [{}_{-0.22}^{+0.25}]
    \times 10^{9}~\text{cm}^{-2}~\text{s}^{-1} \,,
    \\
    f_{\Nuc{pep}}
    & = 1.039_{-0.065}^{+0.082}\, [{}_{-0.16}^{+0.19}]
    \,, \qquad
    & \Phi_{\Nuc{pep}}
    & = 1.48_{-0.09}^{+0.11}\, [{}_{-0.22}^{+0.26}]
    \times 10^{8}~\text{cm}^{-2}~\text{s}^{-1} \,,
    \\
    f_{\Nuc[13]{N}}
    & = 1.16_{-0.19}^{+0.19}\, [{}_{-0.45}^{+0.50}]
    \,, \qquad
    & \Phi_{\Nuc[13]{N}}
    & = 3.32_{-0.54}^{+0.53}\, [{}_{-1.24}^{+1.40}]
    \times 10^{8}~\text{cm}^{-2}~\text{s}^{-1} \,,
    \\
    f_{\Nuc[15]{O}}
    & = 1.16_{-0.19}^{+0.19}\, [{}_{-0.44}^{+0.49}]
    \qquad
    & \Phi_{\Nuc[15]{O}}
    & = 2.41_{-0.39}^{+0.38}\, [{}_{-0.90}^{+1.02}]
    \times 10^{8}~\text{cm}^{-2}~\text{s}^{-1} \,,
    \\
    f_{\Nuc[17]{F}}
    & = 1.01_{-0.16}^{+0.16}\, [{}_{-0.38}^{+0.45}]
    \,, \qquad
    & \Phi_{\Nuc[17]{F}}
    & = 5.25_{-0.85}^{+0.84}\, [{}_{-1.97}^{+2.21}]
    \times 10^{6}~\text{cm}^{-2}~\text{s}^{-1} \,,
    \\
    f_{\Nuc[8]{B}}
    & = 1.034_{-0.021}^{+0.020}\, [{}_{-0.051}^{+0.052}]
    \,, \qquad
    & \Phi_{\Nuc[8]{B}}
    & = 5.192_{-0.11}^{+0.10}\, [{}_{-0.26}^{+0.26}]
    \times 10^{6}~\text{cm}^{-2}~\text{s}^{-1} \,,
    \\
    f_{\Nuc{hep}}
    & = 3.6_{-1.1}^{+1.2}\, [{}_{-2.6}^{+3.0}]
    \,, \qquad
    & \Phi_{\Nuc{hep}}
    & = 2.9_{-0.9}^{+1.0}\, [{}_{-2.1}^{+2.4}]
    \times 10^{4}~\text{cm}^{-2}~\text{s}^{-1} \,.
  \end{aligned}
\end{equation}
As expected, the \Nuc{pp} flux is the most affected by the release of
the luminosity constraint as it is this reaction which gives the
largest contribution to the solar energy production and therefore its
associated neutrino flux is the one more strongly bounded when
imposing the luminosity constraint.  The \Nuc{pep} flux is also
affected due to its strong correlation with the \Nuc{pp} flux,
Eq.~\eqref{eq:pep-pp}.  The CNO fluxes are mildly affected in an
indirect way due to the modified contribution of the \Nuc{pep} fluxes
to the Borexino spectra.

Thus we find that the energy production in the pp-chain and the
CNO-cycle without imposing the luminosity constraint are given by:
\begin{equation}
  \label{eq:ppcnolum2}
  \frac{L_\text{pp-chain}}{L_\odot}
  = 1.030_{-0.061}^{+0.070}\, [{}_{-0.15}^{+0.17}]
  \qquad\text{and}\qquad
  \frac{L_\text{CNO}}{L_\odot}
  = 0.0075_{-0.0013}^{+0.0013}\, [{}_{-0.0029}^{+0.0030}] \,.
\end{equation}
Comparing Eqs.~\eqref{eq:ppcnolum1} and~\eqref{eq:ppcnolum2} we see
that while the amount of energy produced in the CNO cycle is about the
same in both analysis, releasing the luminosity constraint allows for
larger production of energy in the pp-chain.  So in this case we find
that the present value of the ratio of the neutrino-inferred solar
luminosity, $L_\odot\text{(neutrino-inferred)}$, to the photon
measured luminosity $L_\odot$ is:
\begin{equation}
  \label{eq:lnutot}
  \frac{L_\odot\text{(neutrino-inferred)}}{L_\odot}
  = 1.038_{-0.060}^{+0.069}\, [{}_{-0.15}^{+0.17}] \,.
\end{equation}
The neutrino-inferred luminosity is in good agreement with the one
measured in photons, with a $1\sigma$ uncertainty of $\sim 6\%$.  This
represents only a very small variation with respect to the previous
best determination~\cite{Bergstrom:2016cbh}.  Such result is expected
because the determination of the \Nuc{pp} flux, which, as mentioned
above gives the largest contribution to the neutrino-inferred solar
luminosity, has not improved sensibly with the inclusion of the full
statistics of the phases II and III of Borexino.

\begin{figure}[t]\centering
  \includegraphics[width=0.95\textwidth]{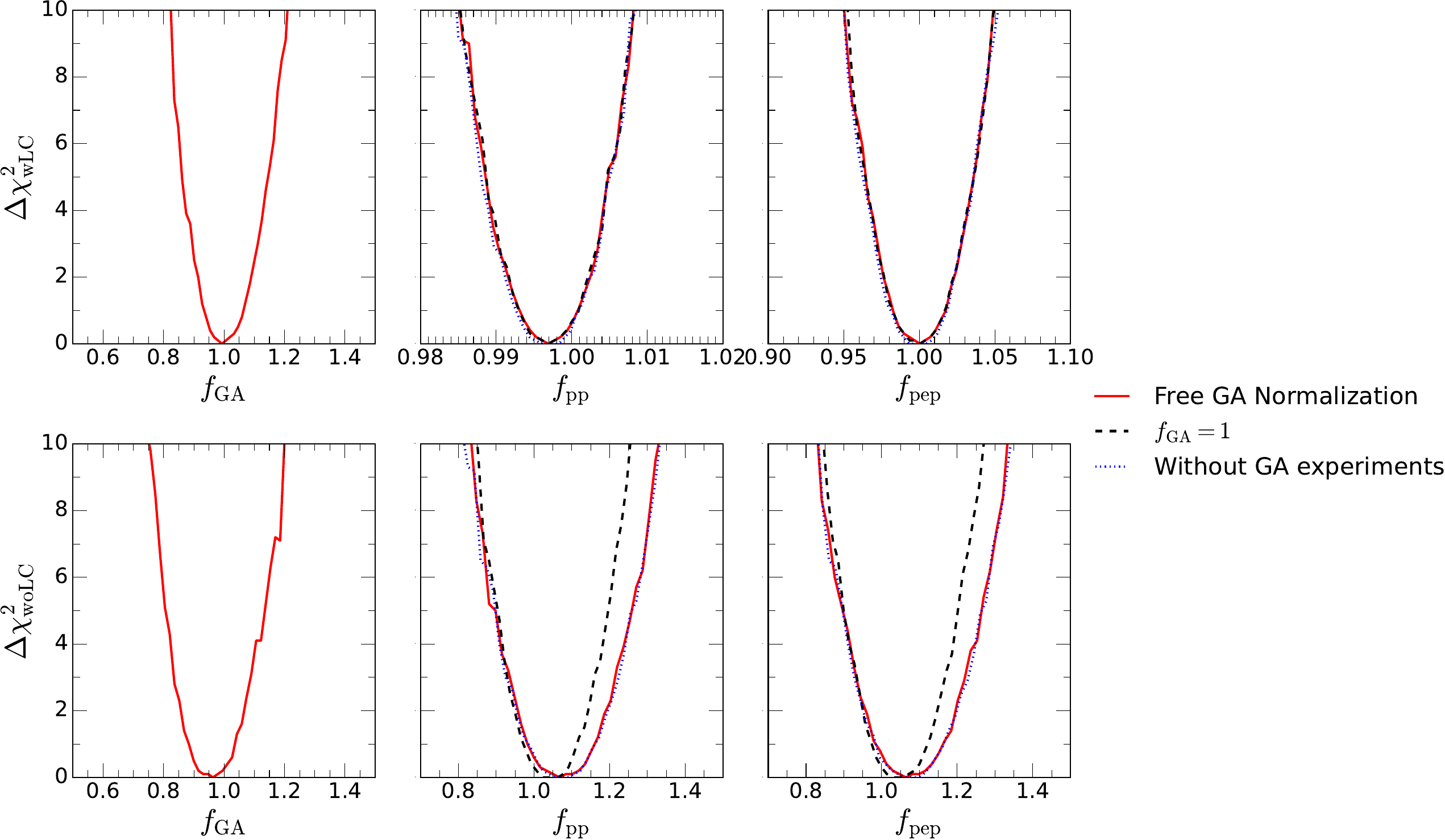}
  \caption{Dependence of the determination of the \Nuc{pp} and \Nuc{pep}
    fluxes on the assumptions about the Gallium source experiments included
    in the analysis.  The upper (lower) panels show
    the results of different variants of the analysis <<CNO-Rfixed>>
    with (without) luminosity constraint.  See text for details.}
  \label{fig:gallium}
\end{figure}

We finish this section by discussing the role of the Gallium
experiments in these results with the aim of addressing the possible
impact of the Gallium anomaly~\cite{Laveder:2007zz, Acero:2007su,
  Giunti:2010zu}.  As described in Section \ref{sec:gallium}, this anomaly
consists in a deficit of the event rate observed in Gallium source
experiments with respect to the expectation, which represents an
obvious puzzle for the interpretation of the results of the solar
neutrino Gallium measurements.  In this work we assume the well
established standard $3\nu$ oscillation scenario and in this context
the attempts at explanation (or at least alleviation) of the anomaly
invoke the uncertainties of the capture cross
section~\cite{Berryman:2021yan, Giunti:2022xat, Elliott:2023xkb}.
Thus the open question posed by the Gallium anomaly is the possible
impact of such modification of the cross section in the results of our
fit.

In order to quantify this we performed two additional variants of our
analysis.  In the first one we introduce an additional parameter,
$f_\text{GA}$, which multiplies the predicted event rates from all
solar fluxes in the Gallium solar experiments.  This parameter is left free
to vary in the fits and would mimic an energy independent modification
of the capture cross section (or equivalently of the detection
efficiency).  In the second variant we simply drop Gallium solar experiments
from our global fit.

The results of these explorations are shown in Fig.~\ref{fig:gallium}
where we plot the most relevant marginalized one-dimensional
projections of $\Delta\chi^2$ for these two variants.  The upper
(lower) panels correspond to analysis performed with (without) the
luminosity constraint.  The left panel shows the projection over the
normalization parameter $f_\text{GA}$ obtained in the variant of the
analysis which makes use of this parameter.  As seen from the figure,
the results of the fit favour $f_\text{GA}$ close to one, or, in other
words, the global analysis of the solar experiments do \emph{not}
support a modification of the neutrino capture cross section in
Gallium (or any other effect inducing an energy-independent reduction
of the detection efficiency in the Gallium source experiments).  This is so
because, within the $3\nu$ oscillation scenario, the global fit
implies a rate of \Nuc{pp} and \Nuc[7]{Be} neutrinos in the Gallium
experiment which is in good agreement with the luminosity constraint
as well as with the rates observed in Borexino.

On the central and right panel of the figure we show the corresponding
modification of marginalized one-dimensional projections of
$\Delta\chi^2$ on the \Nuc{pp} and \Nuc{pep} flux normalizations which
are those mostly affected in these variants.  For the sake of
comparison, in the upper and lower panels we also plot the results for
the $f_\text{GA} = 1$ analysis (also visible in the corresponding
windows in Figs.~\ref{fig:triCNOLCGA1} and~\ref{fig:triCNOwoLCGA1},
respectively).  The figure illustrates that once the luminosity
constraint is imposed, the determination of the solar fluxes is
totally unaffected by the assumptions about the capture rate in
Gallium.  As seen in the lower panels, even without the luminosity
constraint the impact on the \Nuc{pp} and \Nuc{pep} determination is
marginal, which emphasizes the robustness of the flux determination in
Eqs.~\eqref{eq:bestlc} and~\eqref{eq:bestwolc}.  This is the case
thanks to the independent precise determination of the \Nuc{pp} flux
in the phases I and II in Borexino.  Furthermore, the small
modification is the same irrespective of whether the Gallium capture
rate is left free or completely removed from the analysis; this is due
to the lack of spectral and day-night capabilities in Gallium
experiments, which prevents them from providing further information
beyond the overall normalization scale of the signal.

\subsection{Examination of the determination of the CNO fluxes}
\label{sec:CNO}

As mentioned above, one of the most important developments in the
experimental determination of the solar neutrino fluxes in the last
years have been the evidence of neutrinos produced in the CNO cycle
reported by Borexino~\cite{BOREXINO:2020aww, BOREXINO:2022abl,
  Borexino:2023puw}.  The detection was made possible thanks to a
novel method to constrain the rate of the \Nuc[210]{Bi} background.
In Ref.~\cite{BOREXINO:2020aww}, using a partial sample of their
Phase-III data, the collaboration found a $5.1\sigma$ significance of
the CNO flux observation, which increased to $7\sigma$ with the full
Phase-III statistics~\cite{BOREXINO:2022abl}, and to about $8\sigma$
when combined with the CID method~\cite{Borexino:2023puw}.

Key ingredients in the analysis performed by the collaboration in
Refs.~\cite{BOREXINO:2020aww, BOREXINO:2022abl, Borexino:2023puw} (and
therefore in the derivation of these results) are the assumptions
about the relative contribution of the three reactions producing
neutrinos in the CNO cycle, as well as those about other solar fluxes
in the same energy range, in particular the \Nuc{pep} neutrinos.  In a
nutshell, as mentioned above, the collaboration assumes a common shift
of the normalization of the CNO fluxes with respect to that of the
SSM, and it is the evidence of a non-zero value of such normalization
which is quantified in Refs.~\cite{BOREXINO:2020aww, BOREXINO:2022abl,
  Borexino:2023puw}.  In what respects the rate from the \Nuc{pep}
flux, the SSM expectation was assumed because the Phase-III data by
itself does not allow to constraint simultaneously the CNO and
\Nuc{pep} flux normalizations.

\begin{figure}[t]\centering
  \includegraphics[width=0.95\textwidth]{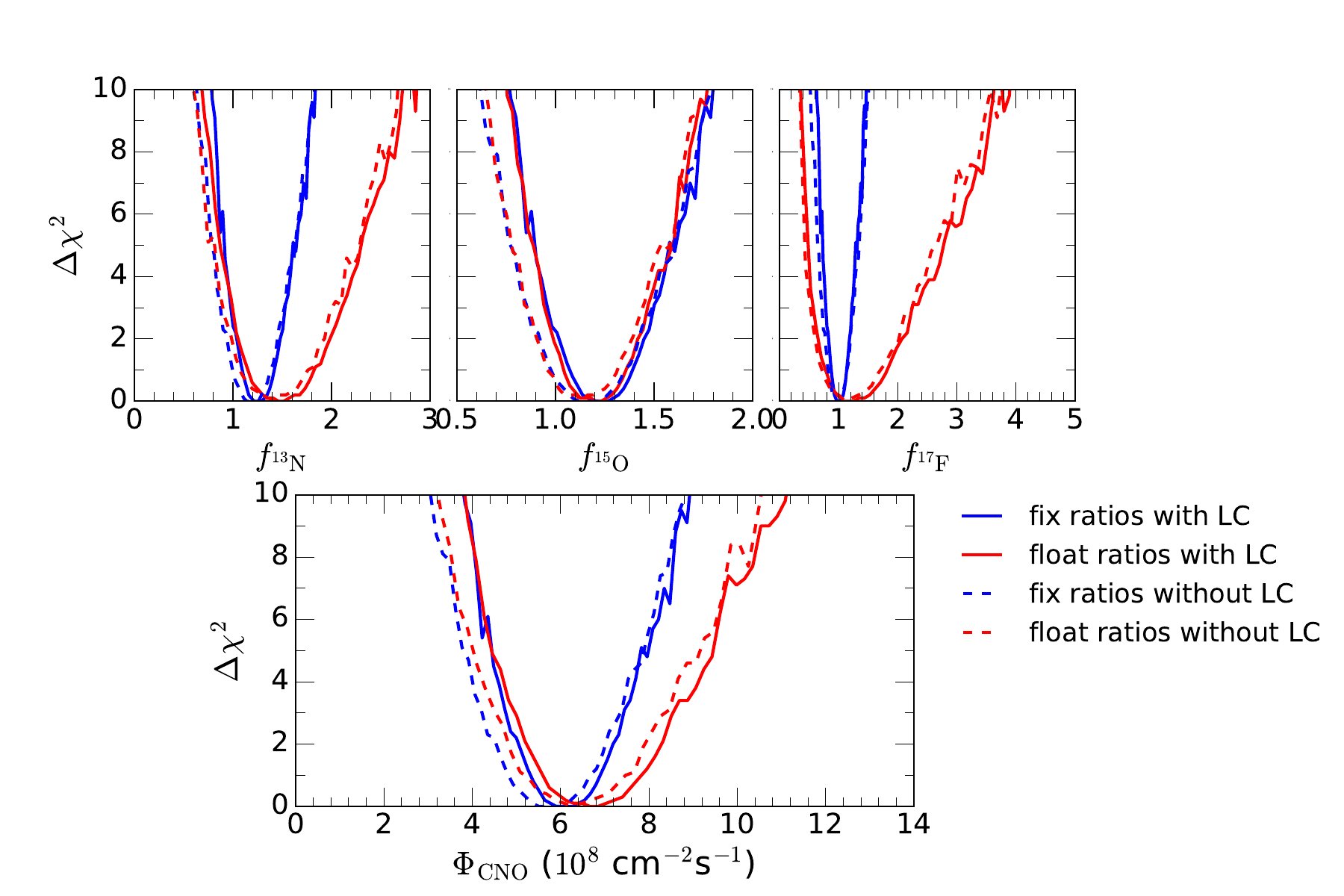}
  \caption{One dimensional projections of the global $\Delta\chi^2$ of
    the for the three neutrino fluxes produced in the CNO-cycle for
    different assumptions as labeled in the figure.  See text for
    details.}
  \label{fig:CNO}
\end{figure}

In this respect, the global analysis presented in the previous section
are performed under the same paradigm of a common shift normalization
of the CNO fluxes, but being global, the \Nuc{pep} flux normalization
is also simultaneously fitted.  For the sake of comparison we
reproduce in Fig.~\ref{fig:CNO} the projection of the marginalized
$\Delta\chi^2_\text{wLC,CNO-Rfixed}$~\eqref{eq:chi2wlccno} and
$\Delta\chi^2_\text{woLC,CNO-Rfixed}$~\eqref{eq:chi2wlccno} on the
normalization parameters for the three CNO fluxes.  For convenience we
also show the projections as a function of the total neutrino flux
produced in the CNO cycle.  As seen in the figure the results of the
analysis (either with or without luminosity constraint) yield a value
of $\Delta\chi^2$ well beyond $3\sigma$ for $\Phi_{\Nuc{CNO}}=0$.  A
dedicated run for this no-CNO scenario case gives $\Delta\chi^2 = 54$
($33$) for the analysis with (without) luminosity constraints, and it
is therefore excluded at $7.3\sigma$ ($5.7\sigma$) CL.

In order to study the dependence of the results on the assumption of a
unique common shift of the normalization of three CNO fluxes we
explored the possibility of making a global analysis in which the
three normalization parameters are varied independently.  As mentioned
above, a priori the three normalizations only have to be subject to a
minimum set of consistency relations in Eqs.~\eqref{eq:CNOineq1}
and~\eqref{eq:CNOineq2}.  However, as discussed in detail in
Sec.~\ref{sec:bx3nfit}, the background model in
Refs.~\cite{BOREXINO:2020aww, BOREXINO:2022abl, Borexino:2023puw} only
assumes an upper bound on the amount of \Nuc[210]{Bi} and cannot be
reliably employed to such general analysis because of the larger
degeneracy between the \Nuc[210]{Bi} background and the \Nuc[13]{N}
flux spectra.

With this limitation in mind, we proceed to perform two alternative
analysis (with and without imposing the luminosity constraints) in
which the normalization of the three CNO fluxes are left free to vary
independently but with ratios constrained within a range broad enough
to generously account for all variants of the B23 SSM, but not to
extend into regions of the parameter space where the assumptions on
the background model may not be applicable.
Conservatively neglecting correlations between their theoretical
uncertainties, the neutrino fluxes of SSMs presented in
Ref.~\cite{Magg:2022rxb} and available publicly through a public
repository \cite{B23Fluxes} verify
\begin{equation}
  \label{eq:cnoRanges}
  \frac{f_{\Nuc[15]{O}}}{f_{\Nuc[13]{N}}} =
  \begin{cases}
    1.00\,(1\pm 0.24) \\
    0.95\,(1\pm 0.22) \\
    0.96\,(1\pm 0.21) \\
    1.01\,(1\pm 0.23) \\
    1.00\,(1\pm 0.23)
  \end{cases}
  \qquad
  \frac{f_{\Nuc[17]{F}}}{f_{\Nuc[13]{N}}} =
  \begin{cases}
    1.00\,(1\pm 0.25) &\text{B23-GS98} \\
    0.84\,(1\pm 0.23) &\text{B23-AGSS09-met} \\
    0.80\,(1\pm 0.20) &\text{B23-AAG21} \\
    0.79\,(1\pm 0.22) &\text{B23-MB22-met} \\
    0.79\,(1\pm 0.22) &\text{B23-MB22-phot}
  \end{cases}
\end{equation}
Thus in these analyses, here onward labeled <<CNO-Rbound>>, we
introduce two pulls $\xi_1$ and $\xi_2$ for these two ratios.  Notice,
however, that we could have equally defined the priors with respect to
the reciprocal of the ratios in Eq.~\eqref{eq:cnoRanges}.  Hence, in
order to avoid a bias towards larger fluxes in the numerator versus
the denominator introduced by either choice, we resort instead to
logarithmic priors for the ratios:
\begin{multline}
  \vec\omega_\text{flux}^\text{CNO-Rbound} \equiv
  (f_{\Nuc{pp}},\; f_{\Nuc[7]{Be}},\; f_{\Nuc{pep}},\; f_{\Nuc[13]{N}},
  \\
  f_{\Nuc[15]{O}} = 0.98\, \exp(\xi_1) \, f_{\Nuc[13]{N}},\;
  f_{\Nuc[17]{F}} = 0.85\, \exp(\xi_2)\, f_{\Nuc[13]{N}},\;
  f_{\Nuc[8]{B}},\; f_{\Nuc{hep}})
\end{multline}
and add two Gaussian penalty factors for these pulls, so that the
corresponding $\chi^2$ function without the luminosity constraint is:
\begin{equation}
  \chi^2_\text{woLC,CNO-Rbound}
  \equiv \chi^2_\text{global}(\vec\omega_\text{osc},\, \vec\omega_\text{flux}^\text{CNO-Rbound})
  + \chi^2_\text{pep/pp}(f_{\Nuc{pp}},\, f_{\Nuc{pep}})
  + \frac{\xi_1^2}{\sigma_{\xi_1}^2} + \frac{\xi_2^2}{\sigma_{\xi_2}^2}
\end{equation}
with $\sigma_{\xi_1}=0.26$ and $\sigma_{\xi_2}=0.48$, chosen to cover
the ranges in Eq.~\eqref{eq:cnoRanges}.  In addition
$f_{\Nuc[13]{N}}$, $f_{\Nuc[15]{O}}$, and $f_{\Nuc[17]{F}}$ are
required to verify the consistency relations in
Eqs.~\eqref{eq:CNOineq1} and~\eqref{eq:CNOineq2}.  The $\chi^2$
function with the luminosity constraint is obtained by further
including the $\chi^2_\text{LC}$ prior of Eq.~\eqref{eq:priorLC}:
\begin{equation}
  \label{eq:chi2wlcnmod}
  \chi^2_\text{wLC,CNO-Rbound}
  \equiv \chi^2_\text{woLC,CNO-Rbound}
  + \chi^2_\text{LC}(\vec\omega_\text{flux}^\text{CNO-Rbound}) \,.
\end{equation}

We plot in Fig.~\ref{fig:CNO} the projection of the marginalized
$\Delta\chi^2_\text{wLC,CNO-Rbound}$~\eqref{eq:chi2wlcnmod} and
$\Delta\chi^2_\text{woLC,CNO-Rbound}$~\eqref{eq:chi2wlcnmod} on the
normalization parameters for the three CNO fluxes as well as on the
total neutrino flux produced in the CNO cycle.  As seen in the figure,
allowing for the ratios of the CNO normalizations to vary within the
intervals~\eqref{eq:cnoRanges} has little impact on the allowed range
of the \Nuc[15]{O} flux and on the lower limit of the \Nuc[13]{N} and
\Nuc[17]{F} fluxes.  As a consequence, the CL at which the no-CNO
scenario can be ruled out is unaffected.  On the contrary, we see in
Fig.~\ref{fig:CNO} that the upper bound on the \Nuc[13]{N} and
\Nuc[17]{F} fluxes, and therefore of the total neutrino flux produced
in the CNO-cycle, is  relaxed.\footnote{The allowed ranges
for the fluxes produced in the pp-chain are not substantially modified
with respect to the ones obtained from the <<CNO-Rfixed>> fits,
Eqs.~\eqref{eq:bestlc}~and~\eqref{eq:bestwolc}.}  This is a
consequence of the strong degeneracy between the spectrum of events
from these fluxes and those from the \Nuc[210]{Bi} background
mentioned above, see Fig.~\ref{fig:subspeccomp} and discussion in
Appendix~\ref{sec:bx3nfit}.  Conversely the fact that the range of the
\Nuc[15]{O} flux is robust under the relaxation of the constraints on
the CNO flux ratios, means that the high statistics spectral data of
the Phase-III of Borexino holds the potential to differentiate the
event rates from \Nuc[15]{O} $\nu$'s from those from \Nuc[13]{N} and
\Nuc[17]{F} $\nu$'s.  The reliable quantification of this possibility,
however, requires the knowledge of the minimum allowed value of the
\Nuc[210]{Bi} background which so far has not been presented by the
collaboration.

So, let us emphasize that our <<CNO-Rbound>> analysis have been
performed with the aim of testing the effect of relaxing the severe
constrains on the CNO fluxes in the studies of the Borexino
collaboration.  Our conclusion is that the statistical significance of
the evidence of detection of events produced by neutrinos from the
CNO-cycle is affected very little by the relaxation of the constraint
on their relative ratios.  However, their allowed range is, and this
can have an impact when confronting the results of the fit with the
predictions of the SSM as we discuss next.

\subsection{Comparison with Standard Solar Models}
\label{sec:compaSSM}

Next we compare the results of our determination of the solar fluxes
with the expectations from the five B23 solar models: SSMs computed
with the abundances compiled in table 5 of~\cite{Magg:2022rxb} based
on the photospheric and meteoritic solar mixtures (MB22-phot and
MB22-met models, respectively), and with the~\cite{Asplund2021}
(AAG21), the meteoritic scale from~\cite{Asplund2009} (AGSS09-met),
and~\cite{Grevesse1998} (GS98) compositions.  We use both MB22-phot
and MB22-met for completeness, although the abundances are very
similar in both scales, as clearly reflected by the results in this
section.  A similar agreement would be found using both the meteoritic
and photospheric scales from AAG21, and therefore we use only one
scale in this case.\footnote{The structures of these models, as well
as the total neutrino fluxes and internal distributions are available
at \cite{B23Fluxes}.}

\begin{figure}\centering
  \includegraphics[width=0.95\textwidth]{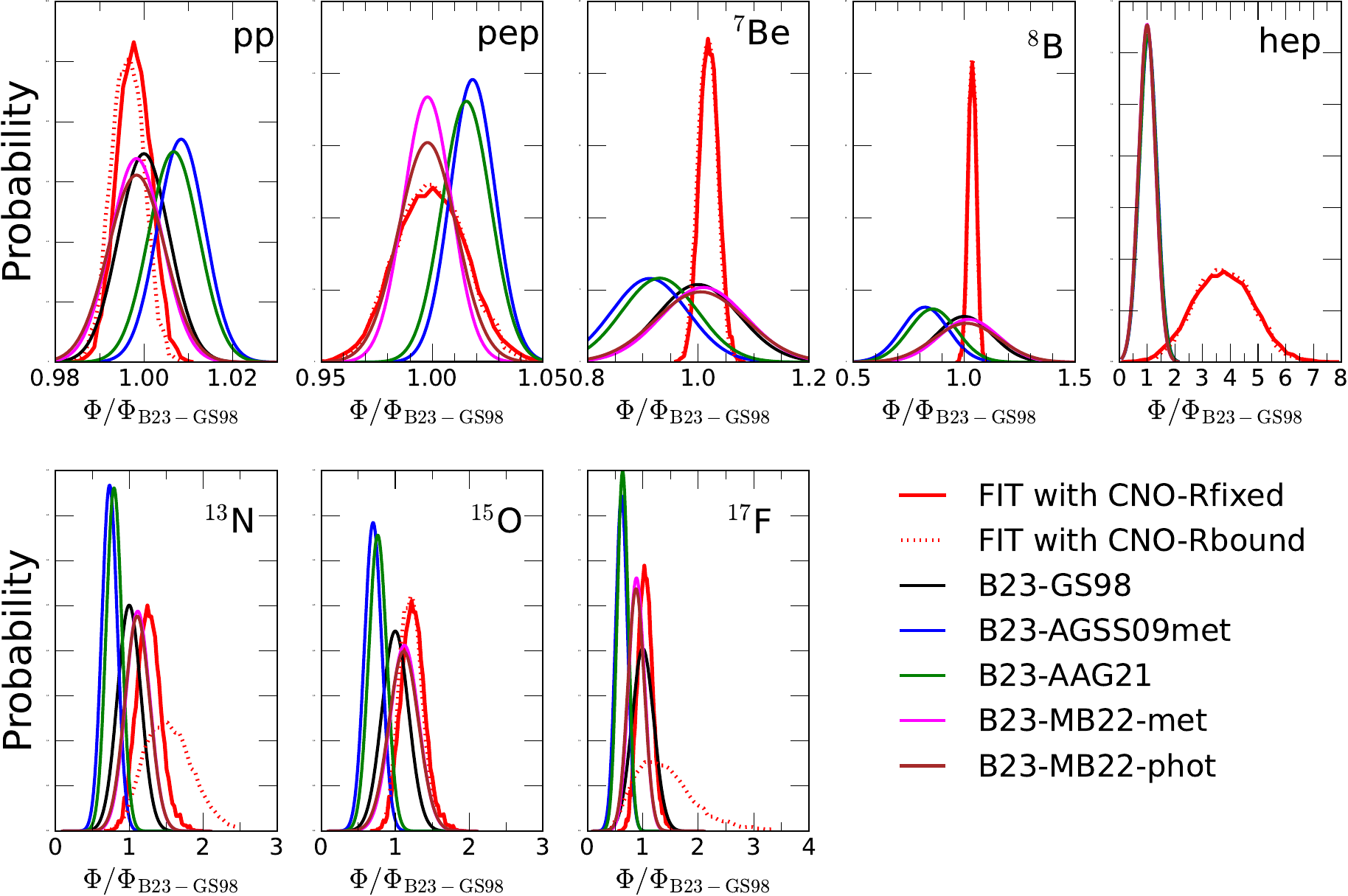}
  \caption{Marginalized one-dimensional probability distributions for
    the best determined solar fluxes in our analysis as compared to
    the predictions for the five SSMs in Ref.~\cite{Magg:2022rxb,B23Fluxes}.}
  \label{fig:compaSSM}
\end{figure}

SSMs predict that nuclear energy accounts for all the solar luminosity
(barred about a few parts in $10^4$ that are of gravothermal origin)
so for all practical matters the neutrino fluxes predicted by SSMs
satisfy the luminosity constraint.  Therefore we compare the
expectations of the various SSM models with the results of our
analysis performed with such constraint.  In what respects the
assumptions on the CNO fluxes, in order to explore the dependence of
our conclusions on the specific choice of flux ratios we quantify the
results obtained in both the <<FIT=CNO-Rfixed>> analysis (with
$\chi^2_\text{FIT}$ in Eq.~\eqref{eq:chi2wlccno}) and the
<<FIT=CNO-Rbounded>> one (with $\chi^2_\text{FIT}$ in
Eq.~\eqref{eq:chi2wlcnmod}).
For illustration we plot in Fig.~\ref{fig:compaSSM} the marginalized
one-dimensional probability distributions for the best determined
solar fluxes in such two cases as compared to the predictions for the
five B23 SSMs.  The probability distributions for our fits are
obtained from the one-dimensional marginalized
$\Delta\chi^2_\text{FIT}(f_i)$ of the corresponding analysis as
$P_\text{FIT}(f_i)\propto \exp[-\Delta\chi^2_\text{FIT}(f_i)/2]$
normalized to unity.  To construct the analogous distributions for
each of the SSMs we use the predictions $\langle f_i^\text{SSM}
\rangle$ for the fluxes, the relative uncertainties
$\sigma_i^\text{SSM}$ and their correlations $\rho_{ij}^\text{SSM}$ as
obtained from Refs.~\cite{B23Fluxes}, and also assume gaussianity so
to build the corresponding
$\chi^2_{\text{SSM}}(\vec\omega_\text{flux})$
\begin{equation}
  \label{eq:chi2mod}
  \chi^2_{\text{SSM}}(\vec\omega_\text{flux})=\sum_{i,j}
  (f_i-f_i^\text{SSM}) C^{-1}_{ij} (f_i-f_i^\text{SSM})
  \quad\text{with}\quad
  C_{ab} = \sigma_a^\text{SSM} \sigma_b^\text{SSM}\rho_{ab} \,,
\end{equation}
from which it is trivial to obtain the marginalized one-dimensional
$\Delta\chi^2_\text{SSM}(f_i)$ and construct the probability
$P_\text{SSM}(f_i)\propto \exp[-\Delta\chi^2_\text{SSM}(f_i)/2]$.

In the frequentist statistical approach, quantitative comparison of a
model prediction for a set of fluxes with the results from the data
analysis can be obtained using the \emph{parameter goodness of fit}
(PG) criterion introduced in Ref.~\cite{Maltoni:2003cu}, by comparing
the minimum value of $\chi^2$ function for the analysis of the data
with that obtained for the same analysis adding the prior imposed by
the model.\footnote{In this respect it is important to notice that, in
order to avoid any bias towards one of the models in the data
analysis, in both <<CNO-Rfixed>> and <<CNO-Rbound>> cases the
assumptions on the ratios of the three CNO fluxes have been chosen to
be ``model-democratic'', \textit{i.e.}, centered at the average of the
predictions of the models (and, in the case of <<CNO-Rbound>>, with
$1\sigma$ uncertainties covering the $1\sigma$ range allowed by all
SSM models).}  Thus, following Ref.~\cite{Maltoni:2003cu}, we
construct the test statistics
\begin{multline}
  \label{eq:dchifitmod}
  \Delta\chi^2_\text{FIT,SSM,SET}
  = \left.
  \big[\chi^2_\text{FIT}(\vec\omega_\text{osc},\, \vec\omega_\text{flux}^\text{FIT})
    + \chi^2_\text{SSM,SET}(\vec\omega_\text{flux}^\text{FIT}) \big]
  \right|_\text{min}
  \\
  - \left. \chi^2_\text{FIT}(\vec\omega_\text{osc},\, \vec\omega_\text{flux}^\text{FIT})
  \right|_\text{min}
  - \left. \chi^2_\text{SSM,SET}(\vec\omega_\text{flux}^\text{FIT})
  \right|_\text{min}
\end{multline}
where $\chi^2_\text{SSM,SET}(\vec\omega_\text{flux})$ is obtained as
Eq.~\eqref{eq:chi2mod} with $i,j$ (and $a,b$) fluxes restricted to a
specific subset as specified by ``SET''.  The minimization of each of
the terms in Eq.~\eqref{eq:dchifitmod} is performed independently in
the corresponding parameter space.  $\Delta\chi^2_\text{FIT,SSM,SET}$
follows a $\chi^2$ distribution with $n$ degrees of freedom, which, in
the present case, coincides with the number of free parameters in
common between $\chi^2_\text{FIT}(\vec\omega_\text{osc},\,
\vec\omega_\text{flux}^\text{FIT})$ and
$\chi^2_\text{SSM,SET}(\vec\omega_\text{flux}^\text{FIT})$.  Notice
that, by construction, the result of the test depends on the number of
fluxes to be compared, \textit{i.e.}, on the fluxes in ``SET'', both
because of the actual comparison between the measured and predicted
values for those specific fluxes, and because of the change in $n$
with which the $p$-value of the model is to be computed.  This is
illustrated in Table~\ref{tab:pgf} where we list the values of
$\Delta\chi^2_\text{FIT,SSM,SET}$ for different choices of ``SET''
which we have labeled as:
\begin{equation}
  \begin{tabular}{ll}
    SET  & constrained fluxes
    \\
    \hline
    FULL
    & ($f_{\Nuc{pp}}$, $f_{\Nuc[7]{Be}}$,\, $f_{\Nuc{pep}}$, $f_{\Nuc[13]{N}}$,
    $f_{\Nuc[15]{O}}$, $f_{\Nuc[17]{F}}$, $f_{\Nuc[8]{B}}$, $f_{\Nuc{hep}}$)
    \\
    Be+B+CNO
    & ($f_{\Nuc[7]{Be}}$, $f_{\Nuc[13]{N}}$, $f_{\Nuc[15]{O}}$,
    $f_{\Nuc[17]{F}}$, $f_{\Nuc[8]{B}}$)
    \\
    CNO & ($f_{\Nuc[13]{N}}$, $f_{\Nuc[15]{O}}$, $f_{\Nuc[17]{F}}$)
  \end{tabular}
\end{equation}

\begin{table}\centering
  \catcode`?=\active\def?{\hphantom{0}}
  \begin{footnotesize}
    \begin{tabular}{|c|l|ccc|ccc|ccc|}
      \hline
      FIT & B23-SSM & \multicolumn{3}{c|}{FULL}
      & \multicolumn{3}{c|}{Be+B+CNO} & \multicolumn{3}{c|}{CNO}
      \\
      \hline
      \multirow{6}{*}{\rotatebox{90}{CNO-Rfixed}}
      &
      & \multicolumn{3}{c|}{n=6}
      & \multicolumn{3}{c|}{n=3}
      & \multicolumn{3}{c|}{n=1}
      \\
      \cline{3-11}
      &
      & $\Delta\chi^2$ & $p_\text{GF}$ & CL [$\sigma$]
      & $\Delta\chi^2$ & $p_\text{GF}$ & CL [$\sigma$]
      & $\Delta\chi^2$ & $p_\text{GF}$ & CL [$\sigma$]
      \\
      \cline{2-11}
      & AGSS09-met    & 14.5 & 0.024 & 2.3 & 9.8 & 0.020 & 2.3? & 7.2 & 0.0073 & 2.7 \\
      & GS98          & ?8.1 & 0.24? & 1.2 & 3.0 & 0.39? & 0.86 & 2.4 & 0.12?? & 1.5 \\
      & AAG21         & 12.5 & 0.052 & 1.9 & 7.8 & 0.05? & 2.0? & 6.2 & 0.013? & 2.5 \\
      & MB22-met/phot & ?7.1 & 0.31? & 1.0 & 2.2 & 0.53? & 0.62 & 2.0 & 0.16?? & 1.4
      \\
      \hline
      \multirow{6}{*}{\rotatebox{90}{CNO-Rbound}}
      &
      & \multicolumn{3}{c|}{n=8}
      & \multicolumn{3}{c|}{n=5}
      & \multicolumn{3}{c|}{n=3}
      \\
      \cline{3-11}
      &
      & $\Delta\chi^2$ & $p_\text{GF}$ & CL [$\sigma$]
      & $\Delta\chi^2$ & $p_\text{GF}$ & CL [$\sigma$]
      & $\Delta\chi^2$ & $p_\text{GF}$ & CL [$\sigma$]
      \\
      \cline{2-11}
      & AGSS09-met    & 14.1 & 0.079 & 1.8? & 9.3 & 0.098 & 1.7? & 7.2 & 0.066 & 1.8? \\
      & GS98          & ?6.7 & 0.57? & 0.57 & 1.7 & 0.88? & 0.14 & 1.6 & 0.66? & 0.44 \\
      & AAG21         & 11.7 & 0.16? & 1.4? & 6.8 & 0.24? & 1.2? & 5.7 & 0.13? & 1.5? \\
      & MB22-met/phot & ?5.9 & 0.66? & 0.44 & 1.1 & 0.95? & 0.06 & 1.0 & 0.80? & 0.25 \\
      \hline
    \end{tabular}
  \end{footnotesize}
  \caption{Results of the PG test for the different models and data
    samples considered.  Within the given accuracy the results for
    MB22-met and MB22-phot models are the same.}
  \label{tab:pgf}
\end{table}

Upon analyzing the data in the Table~\ref{tab:pgf}, it becomes evident
that the B23-MB22 models (both the meteoritic and photospheric
variations) exhibit a significantly higher level of compatibility with
the observed data, even slightly better the B23-GS98 model.  On the
contrary the B23-AGSS09met and B23-AAG21 models exhibits a lower level
of compatibility with observations, with B23-AAG21 model slightly
better aligned with the data.
Maximum discrimination is provided by comparing mainly the CNO fluxes
for which the prediction of both models is mostly different.  On the
other hand, including all the fluxes from the pp-chain in the
comparison tends to dilute the discriminating power of the test.  The
table also illustrates how allowing for the three CNO fluxes
normalizations to vary in the fit tends to relax the CL at which the
models are compatible with the observations.

Let us remember that our previously determined fluxes in
Ref~\cite{Bergstrom:2016cbh} when confronted with the GS98 and AGSS09
models of the time~\cite{Serenelli:2011py} showed \emph{absolutely} no
preference for either model.  This was driven by the fact that the
most precisely measured \Nuc[8]{B} flux (and also of \Nuc[7]{Be}) laid
right in the middle of the prediction of both models.  The new
B16-GS98 model in Ref.~\cite{Vinyoles:2016djt} predicted a slightly
lower value for \Nuc[8]{B} flux in slightly better agreement with the
extracted fluxes of Ref~\cite{Bergstrom:2016cbh}, but the conclusion
was still that there was no significant preference for either model.
Compared to those results, both the most precisely determined
\Nuc[7]{Be} flux and, most importantly, the newly observed rate of CNO
events in Borexino have consistently moved towards the prediction of
the models with higher metallicity abundances.

Let us finish commenting on the relative weight of the experimental
precision versus the theoretical model uncertainties in the results in
Table~\ref{tab:pgf}.  To this end one can envision an ideal experiment
which measures $f_i$ to match precisely the values predicted by one of
the models with infinite accuracy.  Assuming the measurements to
coincide with the predictions of B23-GS98, one gets
$\Delta\chi^2_\text{SSM,SET} = 17.1$ and $16.7$ for SSM=B23-AGSS09-met
and SSM=B23-AAG21 with SET=FULL, which means that the maximum CL at
which these two SSM can be disfavoured is $2.2\sigma$ and $2.1\sigma$.
Choosing instead SET=CNO these numbers become
$\Delta\chi^2_\text{SSM,SET} = 15.0$ and $14.1$ for SSM=B23-AGSS09-met
and SSM=B23-AAG21, respectively, corresponding to a $3.1\sigma$ and
$3.0\sigma$ maximum rejection.
This stresses the importance of reducing the uncertainties in the
model predictions to boost the discrimination between the models.

\section{Confronting the gallium anomaly with solar neutrino data }
\label{sec:frame_gal}

The analysis presented in Section \ref{sec:SSM_full} was
performed in the framework of $3\nu$ oscillations, but the same methodology can be applied in the presence of mixing with a fourth
sterile neutrino state, thus determining the constraints on the
sterile interpretation of the Gallium anomaly in a way which is  completely independent of the modeling of the Sun.

As mentioned in Section \ref{sec:gallium}, an attempt to alleviate the Gallium
anomaly invokes the uncertainties of the capture cross
section~\cite{Berryman:2021yan, Giunti:2022xat, Elliott:2023xkb}.
This raises the issue of whether the estimated rate of solar neutrino
events in Gallium source experiments is really robust.  To quantify the
impact of such possibility, we performed two variants of our solar
neutrino data analysis.  In the first one we include the results of
the solar Gallium source experiments at their nominal values as provided by
the collaboration.  In the second one we introduce an additional
parameter, $f_\text{Ga}$, which is varied freely in the fits and
accounts for an overall scaling of the predicted event rates in the
solar Gallium source experiments.  Statistically this is equivalent to
removing the Gallium source experiments from the solar analysis, but done in
this way it allows us to quantify the range of $f_\text{Ga}$ favoured
by the solar data.  We will return to this point in the next section.

Concerning KamLAND, we include in the analysis the separate DS1, DS2,
DS3 spectral data~\cite{Gando:2013nba} (69 data points).  Since
KamLAND has no near detector, the theoretical uncertainties in the
calculations of the reactor neutrino spectra should be carefully taken
into account.  Here we will consider two limiting cases.  In the first
one we will assume some theoretically calculated reactor fluxes which
will be used as input in the analysis of the KamLAND data.  We will
label this analysis as <<reactor flux constrained>>, or
<<KamLAND-RFC>> in short.  For concreteness we use as
\emph{theoretical reactor fluxes} those predicted by an ad-hoc model
adjusted to perfectly reproduce the spectrum observed in Daya-Bay
experiment~\cite{DayaBay:2021dqj} (and their uncertainties) in the
absence of sterile oscillations.  Similar results would be obtained
with any of the reactor flux models in the literature as long as they
are consistent with the Daya-Bay measurements in the framework of
$3\nu$ mixing.  By construction this is the most limiting scenario.
Alternatively one can use the Daya-Bay reactor spectra as a truly
experimental input, taking into account that at the moment of their
detection the suppression induced by the $\Dmq\sim \mathcal{O}(\eVq)$
oscillations has already taken place (so that the neutrino flux
generated by the reactor cores must be proportionally larger).  This
is a more conservative scenario in which nothing is assumed about the
theoretical prediction of the reactor flux normalization.  We label
this analysis as <<reactor flux free>>, or <<KamLAND-RFF>> in short.

In what respects the relevant survival probabilities, we focus here on
a 3+1 scenario where $\{\nu_1, \nu_2, \nu_3\}$ (with mass-squared
splittings $\Dmq_{21}$ and $\Dmq_{31}$ as determined by the standard
$3\nu$ oscillation analysis) are dominantly admixtures of the
left--handed states $\{\nu_e, \nu_\mu, \nu_\tau \}$, and a fourth
massive state $\nu_4$ (with a mass-squared splitting $\Dmq_{41} \simeq
\Dmq_{42} \simeq \Dmq_{43} \sim \mathcal{O}(\eVq)$) is mostly sterile
(\textit{i.e.}, not coupled to the weak currents) but has some
non-vanishing projection over the left-handed states (see appendix C
of Ref.~\cite{Kopp:2013vaa} for details).  We obtain the oscillation
probabilities for solar neutrinos by numerically solving the evolution
equation for the neutrino ensemble from the neutrino production point
to the detector including matter effects both in the Sun and in the
Earth, with no other approximation than the assumption that the
evolution in the Sun is adiabatic.
We parametrize the mixing matrix as in Ref.~\cite{Kopp:2013vaa,
  Dentler:2018sju}:
\begin{equation}
  U = V_{34} V_{24} V_{14} V_{23} V_{13} V_{12} \,,
\end{equation}
where $V_{ij}$ is a rotation in the $ij$ plane by an angle
$\theta_{ij}$, which in general can also contain a complex phase (see
appendix~A of~\cite{Kopp:2013vaa} for a discussion).  Following
Ref.~\cite{Goldhagen:2021kxe} we make use of the fact that that bounds
from $\nu_\mu$ disappearance in atmospheric and long-baseline
neutral-current measurements render the solar neutrino data
effectively insensitive to $\theta_{34}$ and $\theta_{24}$, hence in
what follows we set $\theta_{34} = \theta_{24} = 0$ and obtain:
\begin{equation}
  \label{eq:U}
  U =
  \begin{pmatrix}
    c_{14} c_{13} c_{12} & c_{14} c_{13} s_{12} & c_{14} s_{13} & s_{14} \\
    \cdot & \cdot & \cdot & 0 \\
    \cdot & \cdot & \cdot & 0 \\
    -s_{14} c_{13} c_{12} & -s_{14} c_{13} s_{12} & -s_{14} s_{13} & c_{14}
  \end{pmatrix}.
\end{equation}
Under these approximations and taking into account that for the
distance and energies of solar and KamLAND neutrinos the oscillations
driven by $\Dmq_{31}$ and $\Dmq_{4i}$ are averaged out, the relevant
probabilities depend only on the three angles $\theta_{12}$,
$\theta_{13}$, $\theta_{14}$ as well as $\Dmq_{21}$.  Furthermore
in~\cite{Kopp:2013vaa} it has been shown that the determination of
$\theta_{13}$ is basically unaffected by the presence of a sterile
neutrino, so we can fix it to the best fit value $s_{13}^2 = 0.02224$
obtained in the $3\nu$ scenario and safely neglect its current
uncertainty.  Altogether the relevant probabilities depend on three
parameters: $\Dmq_{21}$, $\sin^2\theta_{12}$, and $\sin^2\theta_{14}$.

Finally, in the analysis of the Gallium source experiments one can set
$\Dmq_{21} = \Dmq_{31} = 0$, so that the corresponding $\nu_e$
survival probability reduces to the well-known $2\nu$ vacuum
oscillation formula and involves only two parameters, namely the large
mass-squared splitting $\Dmq_{41} = \Dmq_{42} = \Dmq_{43}$ and the
mixing angle $\theta_{14}$:
\begin{equation}
  P_{ee}^\text{Ga-source} =
  1 - \sin^2(2\theta_{14}) \sin^2\bigg( \frac{\Dmq_{41} L}{4E} \bigg) \,.
\end{equation}
Hence the only common parameter between the Gallium-source experiments
and the solar and KamLAND experiments is $\theta_{14}$.  In what
follows we will perform several compatibility tests of the oscillation
parameters allowed by solar data (both alone and in combination with
KamLAND) and those implied by the analysis of the Gallium source
experiments.  To this aim we will make use of a
$\Delta\chi^2_\text{Ga-source}(\theta_{14})$ function inferred from
the combined fit of the Gallium source experiments presented in
Ref.~\cite{Barinov:2022wfh}.

\subsection{Results}
\label{sec:res_gal}

\begin{figure*}[t]\centering
  \includegraphics[width=0.94\textwidth]{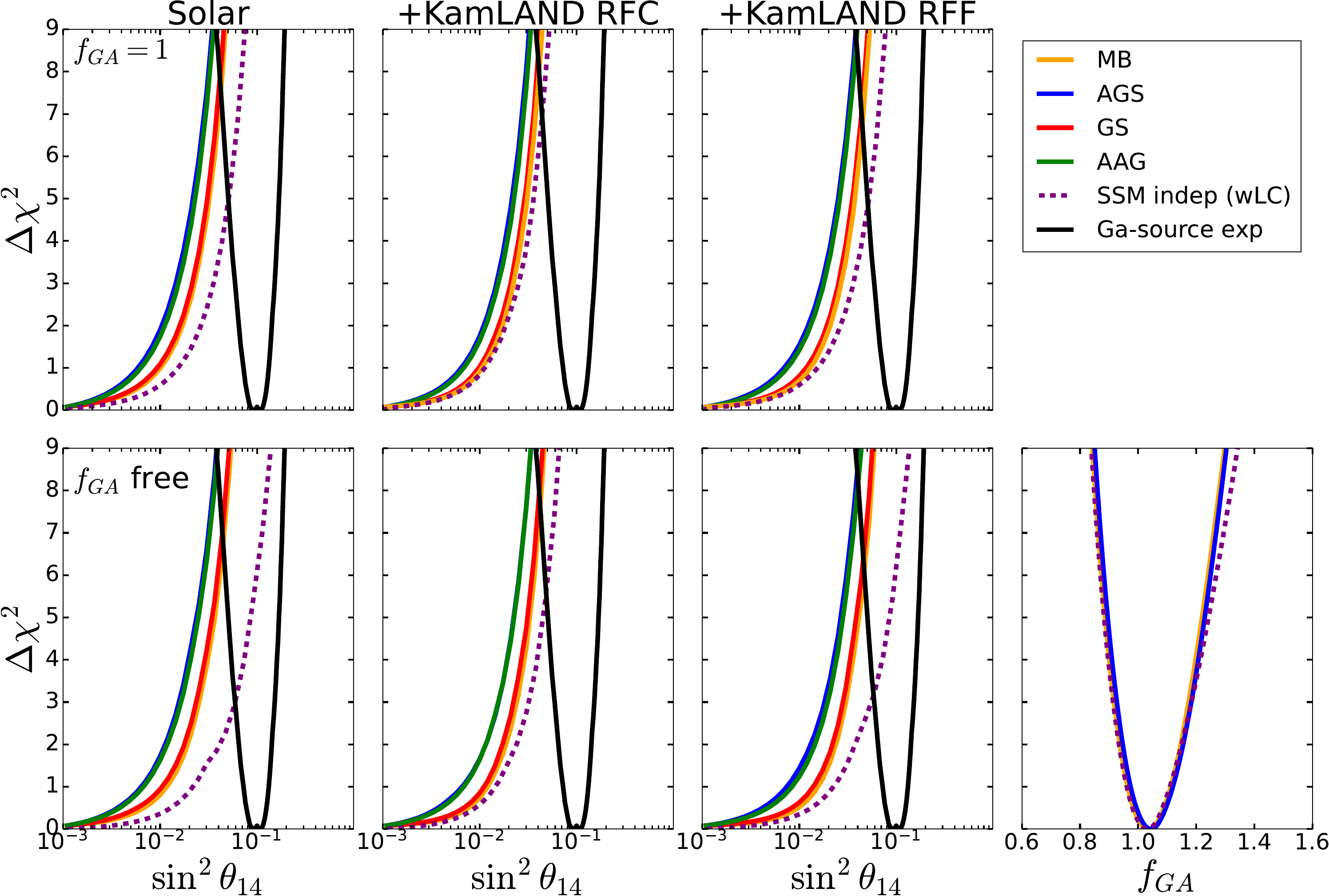}
  \caption{One-dimensional projection of the global $\Delta\chi^2$ on
    the mixing angle $\theta_{14}$ after marginalization over the
    undisplayed parameters, for different assumptions on the solar
    neutrino fluxes as labelled in the legend.  The first (second)
    [third] column corresponds to the analysis including solar-only
    (solar+KamLAND-RFC) [solar+KamLAND-RFF] data.  The upper (lower)
    panels show the results with $f_\text{Ga}=1$ (free $f_\text{Ga}$).
    In all panels the black parabola is
    $\Delta\chi^2_\text{Ga-source}(\theta_{14})$ as inferred from the
    combined analysis of the Gallium source experiments presented in
    Ref.~\cite{Barinov:2022wfh}.  In the rightmost lower panel we show
    the dependence of $\Delta\chi^2$ on the normalization parameter
    $f_\text{Ga}$.}
  \label{fig:ssmchi}
\end{figure*}

\subsubsection{Updated bounds with SSM fluxes}
\label{sec:SSM}

In the first round of analyses we present the results of our fits to
solar data in the framework of four different versions of the B23
standard solar models.  Concretely, we consider the SSMs computed with
the abundances compiled in table 5 of~\cite{Magg:2022rxb} based on the
photospheric solar mixtures (MB-phot; the results obtained with
meteoritic solar mixtures are totally equivalent), as well as models
with the solar composition taken from Ref.~\cite{Asplund2021} (AAG21),
from the meteoritic scale of Ref.~\cite{Asplund2009} (AGSS09-met), and
from Ref.~\cite{Grevesse1998} (GS98).

The results of the SSM constrained analysis are presented in
Fig.~\ref{fig:ssmchi} where we plot $\Delta\chi^2(\theta_{14})$ for
different choices of the solar fluxes and of the KamLAND analysis.
For sake of comparison we also show
$\Delta\chi^2_\text{Ga-source}(\theta_{14})$ as inferred from the
combined fit of the Gallium source experiments presented in
Ref.~\cite{Barinov:2022wfh}.
As can be seen all variants of the solar (+KamLAND) data analysis
favour $\theta_{14}=0$, so that the fit always results into an upper
bound on the allowed range of $\sin^2\theta_{14}$ which is in clear
tension with the values required to explain the Gallium anomaly.
Comparing the upper and lower panels we also see that relaxing the
constraint on the normalization parameter $f_\text{Ga}$ does not lead
to any significant difference in the outcome.  Let us notice that
$f_\text{Ga}$ is a factor introduced only in the fit of solar
neutrinos, without affecting the analysis of the Gallium source
experiments, because in here we are interested in testing if relaxing
some assumptions in the solar and KamLAND analysis can lead to a
better agreement with the sterile neutrino interpretation of the
results of Gallium source experiments.  Studies of correlated effects
affecting the Gallium capture rate in both solar and source
experiments, such as varitions of the capture cross sections, have
been presented in various works (see, \textit{e.g.},
Refs.~\cite{Bahcall:1997eg, Haxton:1998uc,
  Frekers:2015wga,Kostensalo:2019vmv, Semenov:2020xea,
  Giunti:2022xat,Elliott:2023xkb}).  Statistically, allowing
$f_\text{Ga}$ to be free in the solar and KamLAND analysis is
equivalent to removing the solar Gallium source experiments from the
fit, as clearly visible in Fig.~\ref{fig:gallium} where
both approches have been explicitly implemented. As explained at
  the end of Sec.~\ref{sec:res}, this is due to the lack of spectral
and day-night capabilities in Gallium data, which prevents them from
providing further information beyond the overall normalization scale
of the signal.  The advantage of performing the analysis introducing
the unconstrainted $f_\text{Ga}$ factor, is that one can quantify the
range of $f_\text{Ga}$ favoured by the fit.  Interestingly from the
bottom-right panel we see that the solar and KamLAND data always
favours $f_\text{Ga}$ close to one with $1\sigma$ uncertainty of about
$7\%$.  In comparison, the prediction of the neutrino capture cross
section in Gallium in the different models can vary up to about
$15$\%~\cite{Bahcall:1997eg, Haxton:1998uc, Frekers:2015wga,
  Kostensalo:2019vmv, Semenov:2020xea, Giunti:2022xat,
  Elliott:2023xkb}.  This means that the global analyses of solar
experiments do not support a significant modification of the neutrino
capture cross section in Gallium, or any other effect inducing an
energy independent reduction of the detection efficiency in the solar
experiments.

\begin{table*}[t]\centering
  \catcode`?=\active\def?{\hphantom{0}}
  \renewcommand{\arraystretch}{1.2}
  \begin{tabular}{|l|l|ccc|ccc|}
    \hline
    \multicolumn{2}{|c|}{}
    & \multicolumn{3}{c|}{$f_\text{Ga}=1$}
    & \multicolumn{3}{c|}{$f_\text{Ga}$ free}
    \\
    \hline
    & \hfil SSM
    & $\chi^2_\text{PG} / n$ & $p$-value ($\times 10^{-3}$) & $\#\sigma$
    & $\chi^2_\text{PG} / n$ & $p$-value ($\times 10^{-3}$) & $\#\sigma$
    \\
    \hline
    \multirow{3}{*}{Solar}
    & MB-phot/GS98 & 14.9 & 0.11 & 3.9 & 13.1 & 0.3? & 3.6
    \\
    & AAG21/AGSS09 & 18.7 & 0.2? & 4.3 & 17.3 & 0.03 & 4.2
    \\
    & SSM indep (wLC) & ?9.1 & 2.6? & 3.0 & ?4.9 & 27?  & 2.2
    \\
    \hline
    \multirow{3}{*}{\begin{tabular}{@{}l@{}}Solar +\\[2pt]KL-RFC\end{tabular}}
    & MB-phot/GS98 & 15.9 & 0.07 & 4.0 & 15.1 & 0.1? & 3.9
    \\
    & AAG21/AGSS09 & 19.4 & 0.1? & 4.4 & 18.7 & 0.01 & 4.3
    \\
    & SSM indep (wLC) & 13.5 & 0.23 & 3.7 & 10.5 & 1.2? & 3.2
    \\
    \hline
    \multirow{3}{*}{\begin{tabular}{@{}l@{}}Solar +\\[2pt]KL-RFF\end{tabular}}
    & MB-phot/GS98 & 13.2 & 0.28 & 3.6 & 11.7 & 0.64 & 3.4
    \\
    & AAG21/AGSS09 & 17.3 & 0.03 & 4.2 & 16.0 & 0.06 & 4.0
    \\
    & SSM indep (wLC) & ?8.7 & 3.1? & 2.9 & ?4.8 & 29?  & 2.2
    \\
    \hline
  \end{tabular}
  \caption{Results of the PG test for the different solar flux model
    assumptions and the different analysis variants.}
  \label{tab:ssmpg}
\end{table*}

The quantitative question to address is the level of (in)compatibility
of these results from the solar (+KamLAND) analysis with those from
the Gallium source experiments in the context of the 3+1 scenario.\footnote{Prof. Thomas Schwetz did a simple calculation assuming the same cross-section rescaling would hit both types of experiments equally, and found about a 2.8$\sigma$ disagreement between  solar gallium and gallium source results. }
Consistency among different data sets can be quantified with the
parameter goodness-of-fit (PG) test~\cite{Maltoni:2003cu}.  For a
number $N$ of uncorrelated data sets $i$, each one depending on $n_i$
model parameters and collectively depending on $n_\text{glob}$
parameters, it can be shown that the test statistic
\begin{equation}
  \label{eq:PGtest}
  \chi^2_\text{PG}
  \equiv \chi^2_\text{min,glob} - \sum_i^N \chi^2_{\text{min}, i}
  = \min\bigg[ \sum_i^N\chi^2_i \bigg] - \sum_i \min\chi^2_i
\end{equation}
follows a $\chi^2$ distribution with $n_\text{PG} \equiv \sum_i n_i -
n_\text{glob}$ degrees of freedom~\cite{Maltoni:2003cu}.  In this
section we have $N=2$, and the relevant number of parameters are
$n_\text{solar(+KamLAND)} = 3$ (or $4$ for analysis with $f_\text{Ga}$
free), $n_\text{Ga-source} = 2$ and $n_\text{global} = 4$ (or $5$ for
analysis with $f_\text{Ga}$ free).  So for all tests $n_\text{PG} =
1$, reflecting the fact that the only parameter in common between the
solar (+KamLAND) and the Gallium-source data sets is $\theta_{14}$.
We list in table~\ref{tab:ssmpg} the results of applying the PG test
to the different variants of the analysis.  As can be seen, from the
results in the lines labeled MB-phot/GS98 (AAG21/AGSS09) in the
``Solar'' case, the analysis of the solar neutrino data performed in
the framework of the SSM's with higher (lower) metallicity are
incompatible with the Gallium source experiments at the $3.6\sigma$
($4.2\sigma$) level even when allowing for a free $f_\text{Ga}$.
Looking at the corresponding lines for the ``Solar+KL-RFC'' and
``Solar+KL-RFF'' fits we see that combination with KamLAND data
results into a slight improvement or weakening of the incompatibility
depending on the assumption on the reactor fluxes.

To further illustrate the interplay between solar and KamLAND data we
show in Fig.~\ref{fig:triangle} the two-dimensional projection of the
global $\Delta\chi^2$ for the separate analysis of solar-only and
KamLAND-only results under different assumptions for the corresponding
input fluxes.  From the upper left panel, as expected, we see that the
determination of $\Dmq_{21}$ in KamLAND is robust irrespective of the
presence of sterile neutrinos or the assumptions on the reactor flux
normalization, since both occurrences only affect the overall scale of
the signal whereas $\Dmq_{21}$ is determined by the distortions of the
energy spectral shape.  As it is well known from the results of the
$3\nu$ analysis, the $\Dmq_{21}$ values favoured by KamLAND lie in the
upper $2\sigma$ allowed range of the solar neutrino fit.  From the
upper right panel we see that the dependence of
$\Delta\chi^2_\text{solar}$ on $\theta_{14}$ within the $\Dmq_{21}$
interval favoured by KamLAND is flatter than at the lower $\Dmq_{21}$
values preferred by the solar-only analysis, which means that the
solar-only bound on $\theta_{14}$ becomes weaker when $\Dmq_{21}$ is
constrained to the KamLAND range.  In addition, the lower panel shows
that in the <<KamLAND-RFF>> analysis (for which no information on the
absolute reactor flux normalization is included) there is a degeneracy
between $\theta_{12}$ and $\theta_{14}$.  Such degeneracy is expected
as the KamLAND survival probability in vacuum reads
\begin{multline}
  P_{ee}^\text{KamLAND} \simeq
  \cos^4\theta_{14} (\cos^4\theta_{13} + \sin^4\theta_{13})
  + \sin^4\theta_{14}
  \\
  - \cos^4\theta_{14}\, \cos^4\theta_{13} \,
  \sin^2(2\theta_{12})\, \sin^2(\Dmq_{21} L/E)
\end{multline}
so the spectral shape of the signal only provides information on the
ratio of the energy-dependent and energy-independent pieces
\begin{equation}
  \label{eq:ratio}
  \frac{\cos^4\theta_{14}\, \cos^4\theta_{13} \sin^2(2\theta_{12})}
       {\cos^4\theta_{14}(\cos^4\theta_{13} + \sin^4\theta_{13})
         + \sin^4\theta_{14}}
\end{equation}
whose isocontours precisely trace the magenta lines in the
($\sin^2\theta_{12}$, $\sin^2\theta_{14}$) plane observed in the lower
panel.  As a consequence of all this, the combination of
solar+KamLAND-RFF data leads to a slight weakening of the bounds on
$\theta_{14}$ compared to the solar-only analysis, as can be seen
comparing the left and right panels in Fig.~\ref{fig:ssmchi} as well
as the corresponding values of the PG test in Table~\ref{tab:ssmpg}.
On the contrary the analysis of KamLAND with constrained reactor
fluxes can independently bound both the numerator and denominator in
Eq.~\eqref{eq:ratio} and therefore provides an additional constraint
on $\theta_{14}$, as illustrated by the filled green regions in the
lower panel of Fig.~\ref{fig:triangle}.  In the end when combining
solar and KamLAND-RFC this second effect overcompensates the weakening
of the solar bound associated with the larger $\Dmq_{21}$ value, so
that the solar+KamLAND-RFC analysis results in a stronger
$\theta_{14}$ bound than the solar-only fit.

\subsubsection{Bounds from solar model independent analysis}
\label{sec:free_gal}

Let us now discuss the results of fits performed without the
assumption of standard solar model fluxes (but still retaining the
condition of consistency with the observed solar luminosity).
As mentioned in Section~\ref{sec:SSM_full}, the combined analysis of present
solar neutrino experiments and KamLAND reactor data allows for the
simultaneous determination of the relevant oscillation parameters
together with the normalizations $\Phi_i$ of the eight solar neutrino
fluxes~--- five produced in the reactions of the pp-chain, $i \in
\big\lbrace \Nuc{pp},\, \Nuc[7]{Be},\, \Nuc{pep},\, \Nuc[8]{B},\,
\Nuc{hep} \rbrace$, and three originating from the CNO-cycle, $i \in
\big\lbrace \Nuc[13]{N},\, \Nuc[15]{O},\, \Nuc[17]{F} \big\rbrace$.
In Section~\ref{sec:SSM_full} we presented such determination
in the framework of $3\nu$ oscillations (Eqs. (\ref{eq:lumsum1}-\ref{eq:CNOfix}) ).

\begin{figure}[t]\centering
  \includegraphics[width=0.7\linewidth]{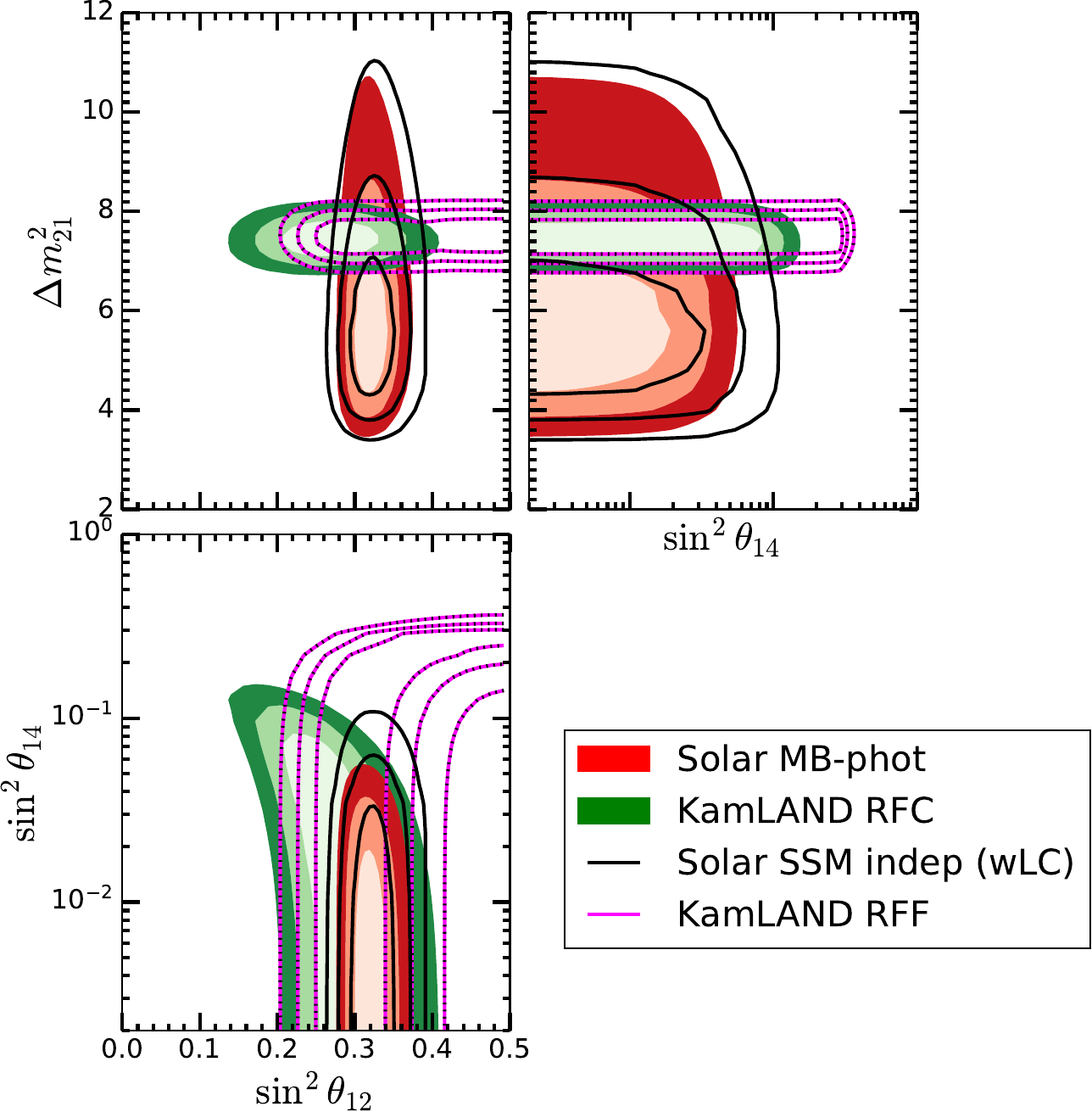}
  \caption{Two-dimensional projection of the global $\Delta\chi^2$
    (for $f_\text{GA}=1$) on
    the relevant oscillation parameters at $1\sigma$, $2\sigma$ and
    $3\sigma$ after marginalization over the undisplayed parameters.
    The full red regions (void black contours) correspond to the
    analysis of solar data with MB-phot (free) fluxes, while the full
    green regions (void magenta contours) correspond to the KamLAND-RFC
    (KamLAND-RFF) fit.}
  \label{fig:triangle}
\end{figure}

In Fig.~\ref{fig:triangle} we show as void black contours the
constraints on the oscillation parameters from our SSM independent
analysis of solar data (labeled as ``SSM indep (wLC)'' in short), as
obtained after marginalization over the various flux normalizations.
Comparing these contours with the full red regions (corresponding to
the analysis with solar fluxes as predicted by the MB-phot SSM) we see
that the determination of the oscillation parameters is only slightly
loosened in the SSM independent fit, and in particular the analysis
still yields a strong bound on $\theta_{14}$.
The corresponding one-dimensional projections
$\Delta\chi^2(\theta_{14})$ (both for solar-only and in combination
with the two variants of the KamLAND fit) are shown as dotted lines in
the various panels of Fig.~\ref{fig:ssmchi}, and the values of the PG
tests are given in Table~\ref{tab:ssmpg}.  From the table we read
that, as long as the normalization parameter $f_\text{Ga}$ is kept to
its nominal value $f_\text{Ga}=1$, the level of (in)compatibility
between the solar (+KamLAND) data and the Gallium source experiments
is at a level $\gtrsim 3\sigma$.  The tension can only be relaxed to
the $\sim 2.2\sigma$ level when allowing the value of $f_\text{Ga}$ to
float freely.

\subsubsection{Implications for the $\nu$-inferred Sun Luminosity}

\begin{figure*}[t]\centering
  \raisebox{1.2mm}{\includegraphics[width=0.48\textwidth]{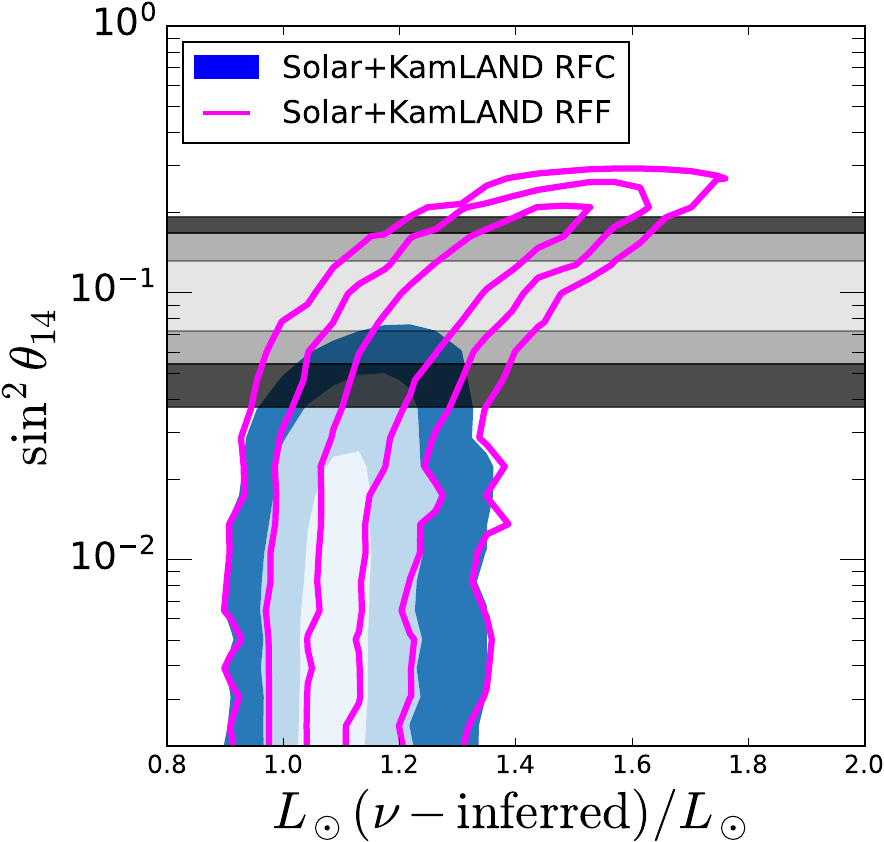}}
  \hfil\includegraphics[width=0.461\textwidth]{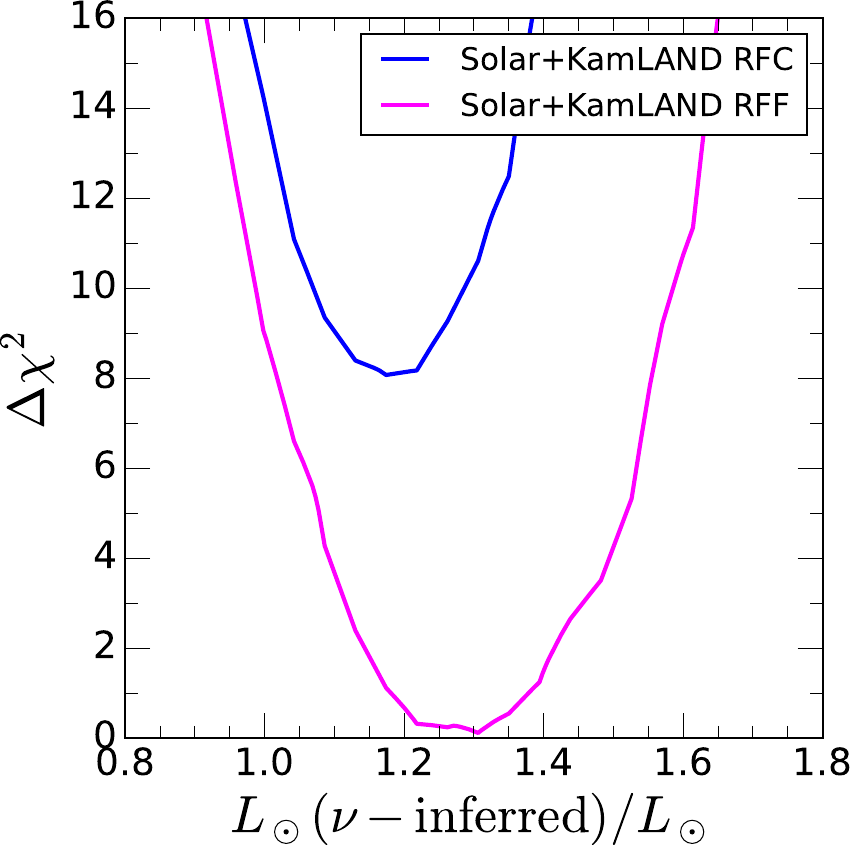}
  \caption{Left: dependence of the $1\sigma$, $2\sigma$, $3\sigma$
    ranges of $\sin^2\theta_{14}$ from the analysis of solar+KamLAND
    without imposing the constraint in Eq.~\eqref{eq:priorLC} on the
    resulting neutrino-inferred solar luminosity.  Fill (void) regions
    correspond to solar+KamLAND-RFC (solar+KamLAND-RFF) analysis.  The
    horizontal grey regions illustrate the $1\sigma$, $2\sigma$,
    $3\sigma$ ranges required to explain the Gallium source results.
    Right: value of $\Delta\chi^2$ from the joint analysis of
    solar+KamLAND and Ga-source data, defined with respect to the sum
    of the two separate best-fit $\chi^2_\text{min}$, as a function of
    the neutrino-inferred solar luminosity.  In these analysis we keep
    $f_\text{GA}=1$ (see text for details).}
  \label{fig:lumfree}
\end{figure*}

We finish by exploring the implications that assuming at face value
the sterile solution of the Gallium anomaly would have on the
mechanism for energy production in the Sun as inferred from neutrino
data.  As mentioned above, \Nuc{pp} neutrinos yield the largest
contribution both to $L_\odot(\text{$\nu$-inferred})$ ($\gtrsim 90\%$)
and to the event rate of the Gallium solar experiments ($\gtrsim
55\%$), which poses the question of what should be the deviation from
the relation in Eq.~\eqref{eq:lumsum1} required for solar observations
to be compatible with the Gallium source experimental results.  In
order to quantitatively answer this question we have performed a SSM
independent analysis similar to the one described above, but without
imposing the prior in Eq.~\eqref{eq:priorLC}.  This allows us to
determine the level of compatibility between solar+KamLAND data and
Gallium source experiments as a function of the $\nu$-inferred solar
luminosity, by comparing the $\sin^2\theta_{14}$ range preferred by
each data set.  The results are shown in Fig.~\ref{fig:lumfree}.
In this analysis we have kept $f_\text{Ga}=1$ because of the strong
correlation (induced by $\Phi_{\Nuc{pp}}$) between the solar
luminosity and the event rate in the Gallium source experiments, which
results in an almost complete degeneracy if both
$L_\odot(\text{$\nu$-inferred})$ and $f_\text{Ga}$ are left free to
vary at the same time.
In the left panel we plot the $1\sigma$, $2\sigma$, $3\sigma$ (1~dof)
ranges for $\theta_{14}$ (defined with respect to the global minimum)
allowed by the combined fit of solar and KamLAND data (for both
variants of the KamLAND analysis) as a function of the $\nu$-inferred
solar luminosity in units of the directly observed $L_\odot$.  For
comparison we show as horizontal grey bands the corresponding required
ranges to explain the Gallium anomaly.  As seen in the figure, for
either analysis, compatibility requires a substantial deviation of the
luminosity inferred from the solar neutrino observations with respect
to its value as directly determined.  This is further quantified in
the right panel, which shows the increase in $\chi^2$ when combining
together solar+KamLAND and Ga-source data (as a function of the
aforesaid luminosity ratio) with respect to the sum of the two
separate best-fits (so that by construction the minimum of each curve
yields the $\chi^2_\text{PG}$ value defined in Eq.~\eqref{eq:PGtest}).
As seen in the figure, for the KamLAND-RFC case the level of
compatibility is always higher than $2.8\sigma$, while in the
KamLAND-RFF case the compatibility only drops below the $2\sigma$
level if one allows $L_\odot(\text{$\nu$-inferred})$ to deviate by
more than $10\%$ from the directly determined solar luminosity.  In
other words, to accommodate the sterile neutrino interpretation of the
Gallium anomaly within the present observation of solar and KamLAND
neutrinos at better than the $2\sigma$ level, it is necessary to ($a$)
make no assumption on the normalization of the reactor antineutrino
fluxes, and ($b$) accept that more than 10\% of the energy produced in
the nuclear reactions in the Sun does \emph{not} result into observed
radiation --- despite the fact that, as mentioned before, the solar
radiated luminosity is directly determined with 0.34\% precision.

\section{Summary}
\label{sec:summary}

In this chapter, we have updated our former determination of solar
neutrino fluxes from neutrino data as presented in
Refs.~\cite{Gonzalez-Garcia:2009dpj, Bergstrom:2016cbh}, by
incorporating into the analysis the latest results from both solar and
non-solar neutrino experiments.  In particular, this includes the full
data from the three phases of the Borexino experiments, which have
provided us with the first direct evidence of neutrinos produced in
the CNO-cycle.

In Section~\ref{sec:SSM_full}, we have derived the best neutrino oscillation parameters and solar
fluxes constraints using a frequentist analysis with and without
imposing nuclear physics as the only source of energy generation
(luminosity constraint).  Compared to the results from previous
analysis, we find that the determination of the \Nuc[7]{Be} flux has
improved by a factor $\mathcal{O}(3)$, but most importantly, we now
find that the three fluxes produced in the CNO-cycle are clearly
determined to be non-zero, with $1\sigma$ precision ranges between
20\% to $\sim 100\%$ depending on the assumptions in the analysis
about their relative normalization.  Conversely, in
Refs.~\cite{Gonzalez-Garcia:2009dpj, Bergstrom:2016cbh} only an upper
bound for the CNO fluxes was found.  This also implies that it is
solidly established that at 99\% CL, the solar energy produced in the
CNO-cycle is between $0.46\%$ and $1.05\%$ of the total solar
luminosity.

The observation of the CNO neutrinos is also paramount to discriminate
among the different versions of the SSMs built with different inputs
for the solar abundances, since the CNO fluxes are the most sensitive
to the solar composition.  In this work we confront for the first time
the neutrino fluxes determined on a purely experimental basis with the
predictions of the latest generation of SSM obtained in
Ref.~\cite{Magg:2022rxb,B23Fluxes}.  Our results show that the SSMs built incorporating
lower metallicities are less compatible with the solar neutrino
observations.

In Section~\ref{sec:frame_gal}, we have applied our global analyses of solar
neutrino data in the framework of the 3+1 neutrino mixing
scenario commonly invoked to explain the results from Gallium source
experiments.  With these fits at hand, we have performed consistency
tests to assess the level of (in)compatibility between the range of
the sterile neutrino mixing preferred by each data set, and we have
studied the dependence of the results on the different assumptions
entering in the analysis.
All the fits considered here have shown compatibility (as measured by
the parameter goodness-of-fit) only at the $3\sigma$ level or higher.
This conclusion holds for analyses assuming solar neutrino fluxes as
predicted by any of the last generation Standard Solar Models, and
irrespective of the assumptions on the KamLAND reactor flux
normalization and on the gallium capture rate.  Relaxing the SSM
constraints on the solar fluxes ---~while still enforcing the relation
between the observed value of the solar radiated luminosity and the
total amount of thermal energy generated by the various
neutrino-emitting nuclear reactions~--- only improves the
compatibility to levels below $3\sigma$ when no prior knowledge in
either the gallium capture rate or the normalization of reactor
neutrino fluxes is assumed.  If the luminosity constraint is also
dropped, then it is formally possible to achieve a compatibility level
below $2\sigma$ as long as the flux of reactor antineutrinos is left
free, but such solution unavoidably requires that more than 10\% of
the energy produced together with neutrinos in the nuclear reactions
of the Sun does not result into observable radiation.

\include{capitulo8_v1}

\chapter{Conclusion}
\label{chap:conc}

In the quest of understanding the fundamental nature of neutrinos,
this thesis has explored multidirectional approaches to address some
intriguing puzzles surrounding these ghostly particles. Neutrinos
represent one of the most compelling subjects in present particle
physics due to their observed properties diverging from SM predictions
(at least the original content of the SM). The established non-zero
neutrino mass necessitates modifications to the traditional particle
physics framework, as evidenced by experimental results discussed
throughout this work. This requirement has stimulated extensive
experimental and theoretical research aimed at uncovering the
mechanisms governing neutrino behavior and nature, leading to
significant insights into physics beyond the SM.

The experimental foundation of three-neutrino mixing has been robustly
confirmed through analyses of solar, atmospheric, reactor, and
accelerator-based neutrino experiments, as detailed in
Chapter~\ref{chap:exp}. Sections~\ref{sec:solar_exp},
\ref{sec:atm_exp}, \ref{sec:reac_exp}, and \ref{sec:accel_exp}
presented comprehensive examinations of these experimental setups,
solidifying the three-flavor oscillation paradigm. However, persistent
discrepancies in gallium-based experiments (Section~\ref{sec:gallium})
and novel measurements of coherent elastic neutrino-nucleus scattering
(CE$\nu$NS, Section~\ref{sec:CEvNS}) challenge the standard formalism,
prompting investigations into new physics scenarios.

Chapter~\ref{cap:nufit} presented an updated global analysis of
neutrino oscillation data through September 2024, employing two
dataset variants: <<IC19 w/o SK-atm>> (independent detailed fits)
and <<IC24 with SK-atm>> (incorporating collaboration-provided
$\chi^2$ tables). Key parameters $\theta_{12}$, $\theta_{13}$,
$\Dmq_{21}$, and $|\Dmq_{3\ell}|$ exhibit strong constraints with
relative uncertainties of 13\%, 8\%, 16\%, and 5--6\% at $3\sigma$,
respectively. The mixing angle $\theta_{23}$ remains least constrained
(20\% uncertainty at $3\sigma$), maintaining persistent octant
ambiguity with $\Delta\chi^2 < 4$ between solutions. CP-violation
analyses reveal $\dCP \approx 180^\circ$ (NO) favoring CP
conservation, while IO favors $\dCP \approx 270^\circ$ with
$3.6\sigma$--$4\sigma$ rejection of CP conservation. Mass ordering
preferences show $\Delta\chi^2_\text{IO,NO} = -0.6$ (IC19) versus
$6.1$ (IC24), reflecting competing experimental trends and atmospheric
data impacts.

Theoretical extensions beyond the SM were systematically explored in
Chapter~\ref{chap:theo_bsm}. Section~\ref{sec:formalism_NSI}
established Non-Standard Interaction (NSI) formalism for neutrino
oscillations and matter effects, including discussion of LMA-D
degeneracy. Section~\ref{sec:formCS} examined detection-level
modifications from new physics scenarios, while
Section~\ref{sec:MD_forces} developed monopole-dipole interaction
formalisms for light scalar mediators. These frameworks enable
comprehensive analysis of vector, scalar, and pseudoscalar mediators
affecting neutrino propagation and detection.

Experimental constraints on beyond-SM scenarios were rigorously
derived in Chapter~\ref{chap:exp_bsm}. Borexino Phase II data
(Section~\ref{sec:NSI_BX}) yielded NC NSI wiht electrons bounds
through density matrix formalism (Eq.~\eqref{eq:ES-dens}), resolving
oscillation-interaction degeneracies. In particular, the LMA-D solution arising in the presence
  of vector NSI was eliminated, enabling independent
$\varepsilon_{\alpha\alpha}^{f,V}$ determination. Neutrino magnetic
moment (NMM) constraints
(Eqs.~\eqref{eq:mubounds},~\eqref{eq:mubounds2}) aligned with prior
results, while light mediator analyses produced mass-dependent limits:
\begin{equation}
\begin{aligned}
|g_{Z'}^{B-L}| &\leq 6.3 \times 10^{-7},\; |g_{Z'}^{L_e-L_{\mu,\tau}}| \leq 5.8 \times 10^{-7}, \\
|g_\phi^\text{univ}| &\leq 1.4 \times 10^{-6},\; |g_\varphi^\text{univ}| \leq 2.2 \times 10^{-6}
\end{aligned}
\end{equation}
for $M \lesssim 0.1$ MeV. Heavy mediator constraints scaled as $(g/M)^n$, with $|g_{Z'}|/M_{Z'} \leq 1.4 \times 10^{-6}\,\text{MeV}^{-1}$. Global fits combining oscillation and CEvNS data (Section~\ref{sec:results_glob_bsm}) excluded LMA-D beyond $2\sigma$ for quark-coupled mediators, while maintaining oscillation parameter robustness. Axial NSI bounds proved weaker than vector counterparts (Fig.~\ref{fig:etriangA}), with Earth matter effects quantified for atmospheric/LBL experiments (Fig.~\ref{fig:earthtriang}). Solar neutrino analyses constrained spin-zero mediators (Section~\ref{sec:MD_section}), surpassing previous limits for $m_\phi \lesssim 10^{-12}$ eV through SFP potential constraints.

Chapter~\ref{chap:SSM_BX} advanced solar neutrino flux determinations
by incorporating Borexino's CNO-cycle neutrino
observations. Section~\ref{sec:SSM_full} achieved $\mathcal{O}(3)$
improvement in \Nuc[7]{Be} flux precision, establishing non-zero CNO
fluxes (20\%--100\% uncertainties) and constraining CNO energy
contribution to 0.46\%--1.05\% of solar luminosity at 99\%
CL. Confrontation with SSM predictions revealed incompatibilities with
low-metallicity models. Section~\ref{sec:frame_gal} analyzed 3+1
neutrino mixing scenarios for gallium anomalies, finding $>3\sigma$
incompatibility across SSM assumptions, only reaching $<2\sigma$
compatibility when relaxing luminosity constraints and reactor flux
normalizations.

In summary, this thesis has pursued a multidimensional investigation
into the present limits of the standard neutrino parameters,
reinforcing the robustness of the three-flavor oscillation framework
and the investigation of BSM scenarios over the most up-to-date
experimental data. Starting with a rigorous analysis of the
three-neutrino oscillation framework, this work systematically probed
current limits on beyond-SM physics while utilizing neutrino flux data
to test solar models and address the unresolved gallium anomaly.  From
Pauli's hypothetical particle to current precision experiments,
neutrino physics continues to illuminate BSM phenomena while guiding
our understanding of astronomical objects. The field's future promises
continued surprises, building upon the rigorous methodologies and
insights presented throughout this work.

\cleardoublepage
\chapter*{Acknowledgements}
\markboth{Acknowledgements}{} %Write "Acknowledgements" in right page

Agora que a tese está finalizada, tenho a sensação de dever cumprido. Eu não consigo me ver de outra forma, a não ser como um físico (espero ser ao menos um razoável agora, com certeza muito melhor que há 4 anos atrás...). Ainda me lembro do dia em que resolvi escolher a física como carreira. Num ato de impulsividade, uma escolha em 2014 que mudou o rumo da minha vida. A minha amada avó Eudália me ajudou desde o princípio. Fui estudar em outro estado, com a ajuda dos meus tios Tárcio e Áurea me cedendo por uns meses um espaço em seu apartamento, e contando com o suporte financeiro da minha avó. Os agradeço muito por todo o suporte, sem vocês eu não estaria aqui. Porém, tenho que agradecer principalmente aos meus pais, Ednólia e José que sempre mostraram apoio, independente das minhas escolhas, e me ensinaram o que \'e o certo e o errado, o que \'e moral e imoral, e o valor do trabalho duro e consistente. Meus pais me ensinaram (e continuam me ensinando) o valor da fam\'ilia e do amor. \`A minha irmã, Paula, com quem sempre compartilhamos o que sentíamos e que durante esse período de Doutorado me ajudou inúmeras vezes financeiramente e psicologicamente, e que quando eu fui morar na Espanha, cuidou da minha cadelinha Liana. Também, agradeço a todos os meus outros familiares que me ajudaram nesse processo. M\~ae Santana, Marcelo, Ariane, Adnorah e Netinho.

Gostaria de agradecer aos meus amigos de infância que continuaram a me apoiar a todo momento. Sávio, uma pessoa excepcional e com quem mantemos contato sempre. A Samuel, que sempre foi próximo a mim, e não importa o nosso contato diário, sempre parece que nos falamos ontem. Sem vocês para dividir conquistas e dores durante esse período, minha vida seria muito ruim. E a Raick, que nos motivamos a continuar a carreira acadêmica desde o ensino médio. Sem vocês, dificilmente eu teria concluído todo esse trajeto.

Sair de uma cidade interiorana como Petrolina, e ir para uma capital como João Pessoa, não foi uma transição fácil. Ainda mais sendo uma pessoa difícil como eu, num período de formação aos 18 anos de idade. Na Universidade Federal da Paraíba me formei físico, e carrego todos os seus pontos fortes e fracos tatuados em mim. Nesse período, duas pessoas foram essenciais: os professores Carlos Pires e Paulo Sérgio. Paulo me acolheu durante a graduação como seu estudante de Iniciação Científica. Carlos foi essencial para a minha formação e continua sendo. Ele foi o responsável por colocar a ideia de fazer um doutorado fora do Brasil na minha cabeça e depois disso segui esse objetivo. Carlos é e continua sendo peça chave no meu processo formativo.

Durante o meu período na universidade, tive amigos que me ajudaram a seguir em frente. Vinícius, que sempre foi uma referência para mim e um amigo com quem eu sempre pude compartilhar tudo. Foi a primeira pessoa a saber que eu era pai. Espero levar a nossa amizade para sempre. E também foi a única pessoa que compartilhou esse sonho louco comigo de ir fazer um doutorado fora do país. Ver seus passos a cada dia me enche de orgulho. Também quero agradecer a Jefferson, William, Jacinto e Leandro, que sempre foram bons amigos durante esse período no Brasil.

Porém, a pessoa que mais devo no meu processo formativo como um todo se chama Concha. Concha basicamente pegou a minha mão e me ensinou como se eu fosse um estudante do jardim de infância. Te dei muita dor de cabeça nesses 4 anos. Coisas básicas que somente consegui entender ao adquirir mais maturidade, depois de refletir melhor sobre as coisas. Conceitos de ética, rigor, análise estatística, programação, fenomenologia... Muito do que sei foi apenas um fragmento que extraí da senhora, que é um mar de conhecimento vivo.

Mi período en España ha sido rico en todos los sentidos. He conocido una lengua distinta, con costumbres distintas. He conocido la comunidad de neutrinos de Europa, he conocido el mundo. Sin Concha y la Universitat de Barcelona, esto sería imposible para mí. En estos 4 años aquí he madurado mucho, y también vivido muchas experiencias. He conocido excelentes personas, como el grande Michele, una persona extremadamente agradable e inteligente. Me ha enseñado todo de programación, con sus consejos, códigos y longos correos. Muchas gracias por siempre ayudarme. Ahora, siempre que escribo un código pienso antes: ¿Cómo Michele lo haría? Muchas gracias por todo. A Jordi, por siempre estar con sus puertas abiertas para hablar de todo, desde física hasta sobre la vida en general. También a Salva, mi gran amigo que conocí antes de llegar a España y ha sido un gran profesor. Siempre me enseña algo nuevo. Muchas gracias. A Toni y David, que empezamos juntos y compartimos muchos momentos, nuestras angustias, dolores, dudas y momentos de felicidad. A Stefan, Pol e Iñigo, por siempre poder hablar, beber un café, almorzar, beber unas cañas, hablar de todo lo que sea. Muchas gracias. Agradezco a todos los que han pasado por mi vida en este momento de formación, muchas gracias.

During my PhD, I had the privilege of traveling to many different places around the world. I am deeply grateful to all those who welcomed me so warmly during my secondments and research visits. Throughout this journey, I met an incredible number of wonderful people: Jaime, Jay, Xavi, Stefan, Rasmi, Federica, Miguel, Federico, Sandra, Antonio, Gustavo, Alejandro, Androniki, Ushak, Giacomo, Gioaccino, Mario, Ana, Daniel, Joachim, Salva, Maria, Arturo, Christoph, and Michele (my apologies if I have forgotten anyone).

I extend my gratitude to Olcyr, Damir, and Asmaa for all their help and support and a very nice reception in Orsay. Pedro, você é um cara excepcional e espero sempre poder trabalhar contigo, me recepcionou muito bem durante a minha estadia curta no Fermilab. Tao, thank you for your exceptional hospitality in Pittsburgh. I would like to express my sincere appreciation to Renata and Oscar for their warm reception in São Paulo. I am also grateful to all my research collaborators: Pilar, Thomas, Ivan M, Saeed, Julia, Aldo, Adriano, and Patrício. Special thanks go to Ivan E, who provided tremendous assistance in helping me understand and simulate NOvA, as discussed in Chapter~\ref{cap:nufit}. I look forward to continued fruitful collaboration in the years ahead.

Finally, I would like to acknowledge the Hidden Network, which provided me with the financial support necessary to travel to different continents, disseminate my research, present my findings, and engage with a diverse community of outstanding researchers. I must give special recognition to Sheila, a dedicated staff member at the Universitat de Barcelona working with the Bosch i Gimpera Foundation, who handled all the bureaucratic processes related to reimbursements. Her diligent work and patience in managing these administrative tasks were invaluable throughout the first three years of my PhD journey. 

I am also grateful to the Universitat de Barcelona, which, like any institution, has both its strengths and challenges. The university is home to exceptional students and faculty, and Barcelona's favorable climate creates an environment where people tend to be more friendly and relaxed, fostering genuine human connections that made my experience truly memorable. However, I must admit that navigating the bureaucratic processes was not always straightforward. Despite these administrative challenges, the overall atmosphere and the quality of people I encountered made this journey worthwhile. Only a place like Barcelona could offer such a unique blend of academic rigor and  Mediterranean charm.

My deepest thanks go to everyone who has enriched my life and contributed positively to my journey thus far.

%Supress underfull warnings in list of figures, list of tables, and bibliography
\hbadness=11000 
\vbadness=11000
\hfuzz=1.11pt
%List of figures and tables

%-----------APENDICES--------------------

%\begin{appendices}
\appendix
% Appendix A
\chapter{} 
\label{sec:app-T}

\section{Assumed true values for the MO test}

As mentioned in Sec.~\ref{sec:MO}, the values of $T_0^o$ ---~and
therefore the distribution of $T$~--- depend on the unknown true value
of the oscillation parameters.  In
principle, one needs to consider the distribution of $T$ for all
possible values of $\theta^\text{true}$ and the final $p$-value of a
MO hypothesis will be given by the largest one among all choices of
$\theta^\text{true}$, \textit{i.e.}, by the weakest rejection (see the
discussion in Ref.~\cite{Blennow:2013oma}).  In the main text, we have
assumed that the best fit of the real data is representative of the
$T$-distribution at the unknown true value of $\theta$.  Indeed, given
the allowed regions of the oscillation parameters, we do not expect
$T_0$ to change significantly if we vary $\theta^\text{true}$ within
the allowed regions at reasonable confidence level.  The only
exception may be the sensitivity of LBL data as a function of $\dCP$,
which is known to affect the MO sensitivity, especially for NOvA as
shown in Fig.~\ref{fig:nevts24}.

For IO, the global fit constrains $\dCP$ reasonably well, so that we
do not expect strong variations of $T_0^\text{IO}$ for true values of
$\dCP$ in its allowed range for IO.  However, for NO, a significantly
larger range of $\dCP$ is allowed, in particular in correlation with
$\theta_{23}$, see Fig.~\ref{fig:region-cp23}.  Therefore, the
question arises, whether the MO test would give largely different
results when considering true values for $\dCP$ and $\theta_{23}$
within the allowed region for NO.

From Fig.~\ref{fig:nevts24}, we see that for $\dCP \approx 270^\circ$
and NO, the number of $\nu_e$ events in NOvA is maximal, and its value
cannot be obtained by any parameter choice in IO.  Therefore, we
expect best sensitivity to NO for this value of $\dCP$.  In contrast,
values around 0 or $180^\circ$ can be easily accommodated within IO,
and we expect that the sensitivity to NO is weakest around
CP-conserving values.  We have confirmed this expectation, as we
obtain
\begin{equation}
  \label{eq:T0max}
  T_0^\text{NO}\big( \dCP^\text{true}=270^\circ,
  \sin^2\theta_{23}^\text{true} = 0.46 \big) = 12.38 \,,
\end{equation}
with all other oscillation parameters kept at their best-fit values.
  Actually, for the value in
Eq.~\eqref{eq:T0max} the observed value $T_\text{obs} = -0.6$ would
imply a $p$-value for NO of 3.2\%.  Since values of $\dCP$ around
$90^\circ$ are significantly disfavored also for NO, we do not
consider them relevant for the MO test.

To summarize, for relevant choices of the oscillation parameters, the
sensitivity to the NO is weakest for values of $\dCP$ around
$180^\circ$.  In turn, for IO, $\dCP$ is sufficiently constrained, and
we expect only minor variations of $T_0$ within the relevant range
around $270^\circ$.  Therefore, it is appropriate to consider the
current best fit points to quote the final $p$-values for both
orderings.

\label{apendixA}

% Appendix B
\chapter{} 
\label{sec:bx3nfit}

\section{Allowing free normalizations for the three CNO fluxes}

\begin{figure}\centering
  \includegraphics[width=\textwidth]{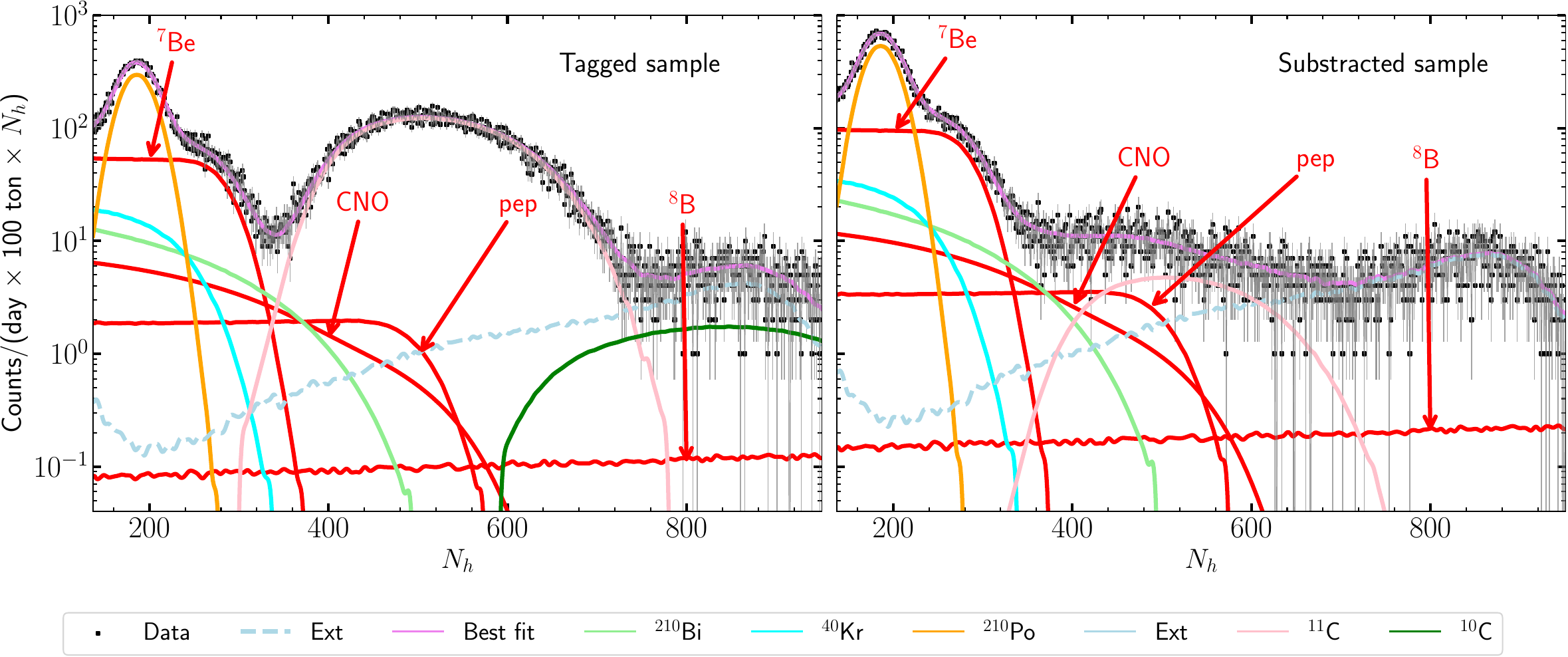}
  \caption{Spectrum for the best-fit normalizations of the different
    components obtained from our fit to the Borexino Phase-III data
    for TFC-tagged (left) and TFC-subtracted (right) events.  In this
    figure ``CNO'' labels the events produced by sum of the three
    fluxes produced in the CNO-cycle, $\Phi_{\Nuc[13]{N}}+
    \Phi_{\Nuc[15]{O}}+ \Phi_{\Nuc[13]{F}}$.}
  \label{fig:BXIIIspec}
\end{figure}

In their analysis of the different phases, the Borexino collaboration
always considers a common shift in the normalization for the three
fluxes of neutrinos produced in the CNO cycle with respect to their
values in the SSM.  On the contrary the normalization of the fluxes
produced in the pp-chain are fitted independently.

\begin{figure}\centering
  \includegraphics[width=0.95\textwidth]{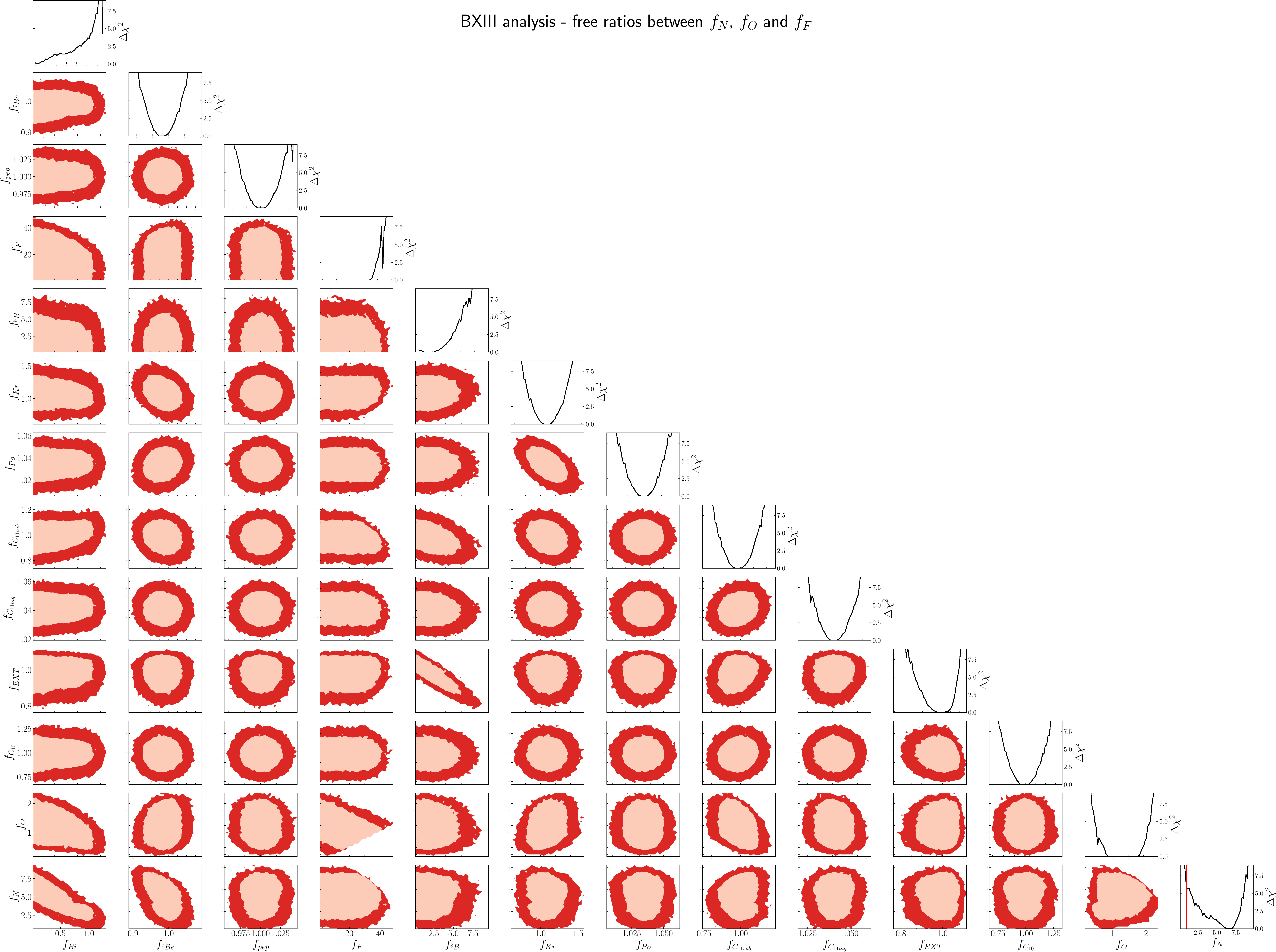}
  \caption{Same as Fig.~\ref{fig:BXIIIfit} but but allowing
    independent variation of the normalizations of the three CNO
    fluxes, only subject to the consistency conditions in
    Eqs.~\eqref{eq:CNOineq1} and~\eqref{eq:CNOineq2}.}
  \label{fig:BXIIINfit}
\end{figure}

In principle, once one departs from the constraints imposed by the
SSM, the normalization of the three CNO fluxes could be shifted
independently, subject only to the minimum set of consistency
relations in Eqs.~\eqref{eq:CNOineq1} and~\eqref{eq:CNOineq2}.  In
fact in our previous works~\cite{Gonzalez-Garcia:2009dpj,
  Bergstrom:2016cbh} we could perform such general analysis.  At the
time there was no evidence of CNO neutrinos and therefore those
analysis resulted into a more general set of upper bounds on their
allowed values compared to those obtained assuming a common shift.
With this as motivation, one can attempt to perform an analysis of the
present BXIII spectra under the same assumption of free normalization.
However, within the present modelling of the backgrounds in the
Borexino analysis, optimized to provide maximum sensitivity to a
positive evidence of CNO neutrinos, such generalized analysis runs
into trouble as we illustrate in Fig.~\ref{fig:BXIIINfit}.  As
expected, allowing three free CNO flux normalizations results in a
weaker constraints on each of the three parameters.  This is
particularly the case for the smaller \Nuc[17]{F} flux which is
allowed to take values as large as $\sim 40$ times the value predicted
by the SSM without however yielding substantial $\chi^2$ improvements
over the standard $f_{\Nuc[17]{F}} = 1$ value.  In the same way
$f_{\Nuc[15]{O}} $ is compatible with the prediction of the SSM,
$\Delta\chi^2(f_{\Nuc[15]{O}} = 1) \simeq 0$, with an upper bound
$f_{\Nuc[15]{O}} \lesssim 2$.\footnote{It is interesting to notice
that the Borexino bound on $\Phi_{\Nuc[17]{F}}$ is about one-half that
on $\Phi_{\Nuc[15]{O}}$.  This is no surprise since the energy spectra
of \Nuc[17]{F} and \Nuc[15]{O} neutrinos are extremely similar hence
neither Borexino nor any other experiment can separate them and what
is actually constrained is their \emph{sum}.  This is reflected in the
clear anticorrelation visible in Fig.~\ref{fig:BXIIINfit}, while the
factor of two stems from the consistency condition in
Eq.~\eqref{eq:CNOineq2}.}  On the contrary the fit results into a
favoured range for \Nuc[13]{N} which, it taken at face value, would
imply an incompatibility with the SSM at large CL:
$\Delta\chi^2(f_{\Nuc[13]{N}}=1) \gtrsim 6$.  This large \Nuc[13]{N}
flux comes at a price of a very low value of the \Nuc[210]{Bi}
normalization, which as seen in the figure is more strongly correlated
with \Nuc[13]{N} than with \Nuc[15]{O} and \Nuc[17]{F}.

\begin{figure}\centering
  \includegraphics[width=\textwidth]{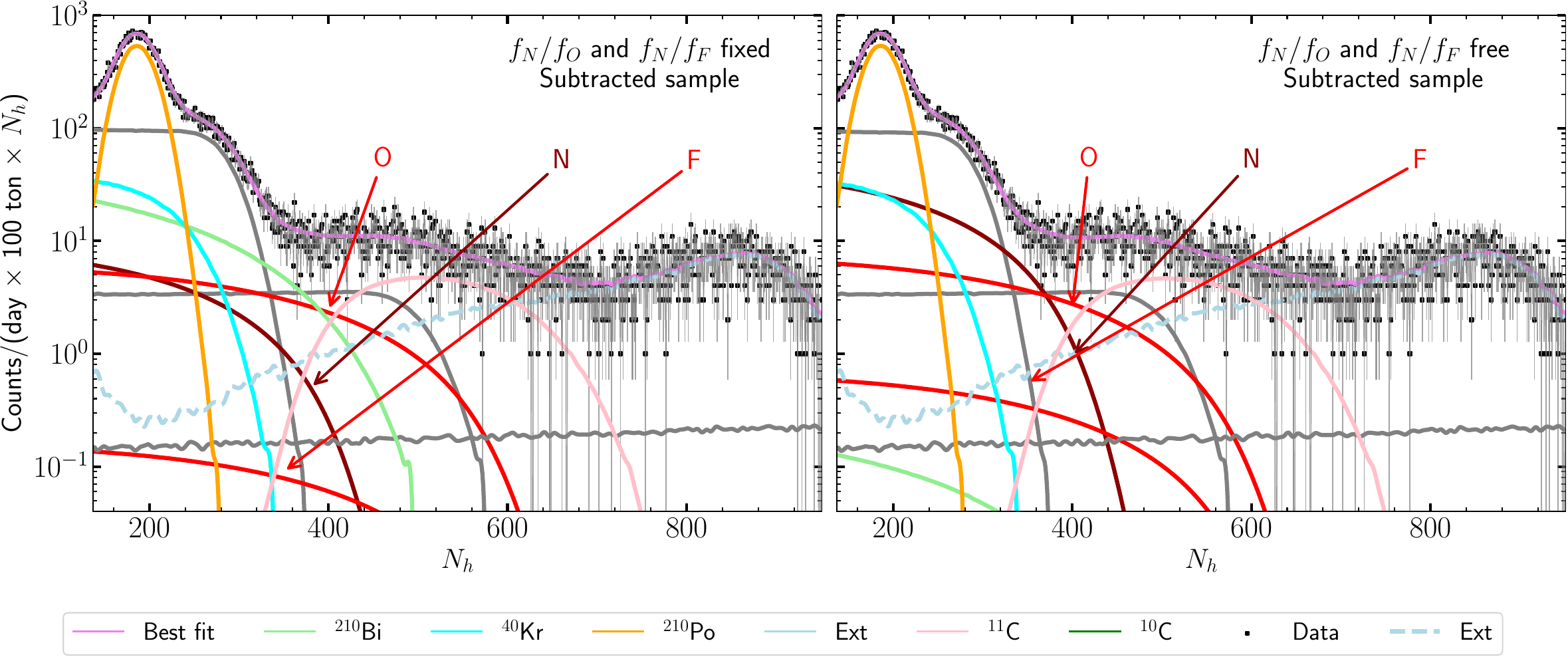}
  \caption{Spectrum of events in the TFC-subtracted sample for the
    best-fit normalizations of the different components obtained from
    two fits to the Borexino Phase-III data.  In the left panel the
    fit is performed assuming a common normalization shift for the 3
    CNO fluxes, while on the right panel the three normalizations are
    allowed to very free subject only the conditions
    Eqs.~\eqref{eq:CNOineq1} and~\eqref{eq:CNOineq2}.}
  \label{fig:subspeccomp}
\end{figure}

To illustrate further this point we show in Fig.~\ref{fig:subspeccomp}
our best fitted spectra of the ``subtracted'' sample for the analysis
where one common normalization for the three CNO fluxes is used (left,
in what follows ``CNO'' fit) and the one where all the three
normalizations are varied independently (right, in what follows ``N''
fit).  Thus the spectra in the left panel of
Fig.~\ref{fig:subspeccomp} are the same as the right panel of
Fig.~\ref{fig:BXIIIspec}, except that now, for convenience, we plot
separately the events from each of the CNO fluxes.  This highlights
clearly the different shape of the spectra of \Nuc[15]{O} and
\Nuc[13]{N}, with \Nuc[15]{O} extending to larger energies.  It is
also evident that \Nuc[13]{N} is the one mostly affected by
degeneracies with the \Nuc[210]{Bi} background.  Comparing the two
panels we see by naked eye that both spectra describe well the data:
in fact, the event rates for \Nuc[15]{O} are comparable in both
panels.  But in the right panel the normalization of the \Nuc[13]{N}
events is considerably enhanced while the \Nuc[210]{Bi} background is
suppressed: this is the option favoured by the fit.

\begin{figure}\centering
  \includegraphics[width=0.59\textwidth]{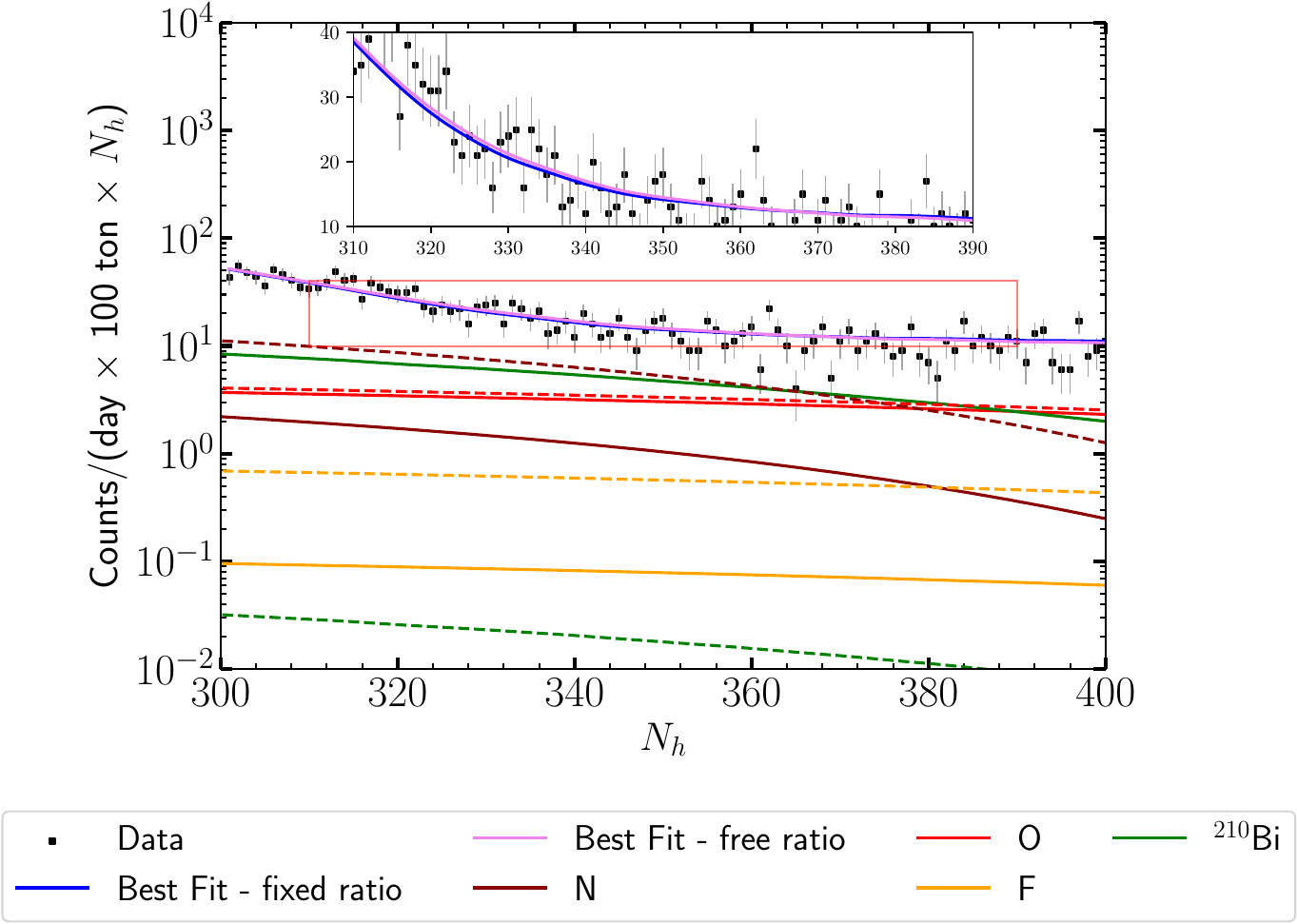}
  \hspace*{-30mm}\hfill
  \raisebox{12mm}{\includegraphics[width=0.455\textwidth]{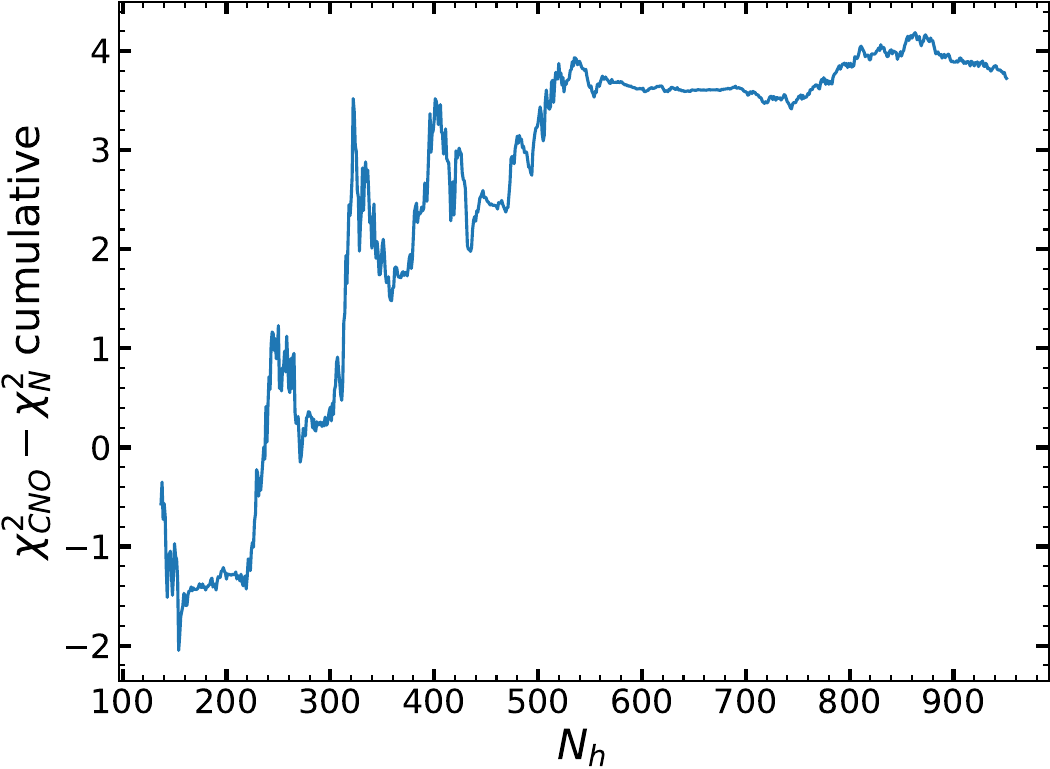}}
  \caption{Left: Spectra of subtracted event rates from best fit CNO
    fluxes and \Nuc[210]{Bi} background in the range of $300 \leq N_h
    \leq 400$ for the ``CNO'' fit (fit with a common normalization
    factor for the three CNO fluxes, full lines) and the ``N'' fit
    (fit with three independent normalizations, dashed lines).  We
    also show the best fit spectra for both fits compared to the data
    as labeled in the figure.  Right: Difference of the value of
    $\chi^2$ in both fits as a function of the maximum $N_h$ bin
    included in the fit.}
  \label{fig:compa2}
\end{figure}

Upon closer examination we find that in the range of $N_h$ spanning
between 300 and 400 photon hits, an increase in the value of
$f_{\Nuc[13]{N}}$ better fits the data while driving
$f_{\Nuc[210]{Bi}}$ towards 0.  In Fig.~\ref{fig:compa2} we show a
blow-up of the spectra in this $N_h$ window.  To quantify the
difference in the quality of the fit for those two solutions and the
relevant range of $N_h$ we plot in the right panel the cumulative
difference of $\chi^2_\text{BXIII,test}$ for the best fits of the
``CNO'' and ``N'' fits as a function of the maximum $N_h$ bin included
in the fit.

Clearly, this anomalously large \Nuc[13]{N} solution is possible only
because the sole information included in the fit for the \Nuc[210]{Bi}
background is the upper bound provided by the collaboration.  Such
upper bound is enough to ensure a lower bound on the amount of CNO
neutrinos, and indeed it results in a positive evidence of CNO fluxes
(in good agreement with at least some of the SSMs) when a common
normalization for the three CNO fluxes is enforced, as reported by the
collaboration in Refs.~\cite{BOREXINO:2020aww, BOREXINO:2022abl} (and
properly reproduced by us, as described in the previous section).  Our
results show that this is the case because the spectrum of
\Nuc[210]{Bi} and \Nuc[15]{O} are sufficiently different.  However,
once the normalization of the three CNO fluxes are not linked
together, the degeneracy between the spectral shape of \Nuc[13]{N} and
\Nuc[210]{Bi} --~together with the lack of a proper estimate for a
\emph{lower} bound on \Nuc[210]{Bi} which is not quantified in
Refs.~\cite{BOREXINO:2020aww, BOREXINO:2022abl}~-- pushes the best-fit
of \Nuc[13]{N} towards unnaturally large values.  In other words, the
background model proposed in Refs.~\cite{BOREXINO:2020aww,
  BOREXINO:2022abl} cannot be reliably employed for fits with
independent \Nuc[13]{N} and \Nuc[15]{O} normalizations.

We finish by noticing that this also implies that the high quality data
of Borexino Phase-III, besides having been able to yield the first
evidence of the presence of the CNO neutrinos, also holds the
potential to discriminate between the contributions from \Nuc[13]{N}
and \Nuc[15]{O}, a potential which may be interesting to explore by
the collaboration.

\chapter{} 
\label{sec:MD_pot}

\section{Derivation of the scalar-pseudoscalar potential}
\label{sec:appendix1}

Our starting point is the well-known relation between the static
potential generated by a non-relativistic source particle ---~here a
nucleon $f$ located at $\vec\rho$~--- felt by a probe particle
---~here a neutrino $\nu$ at position $\vec{x}$~--- to the elastic
scattering amplitude $\mathcal{M}$ of the process $\nu + f \to \nu + f
$ as computed in QFT in momentum space
\begin{equation}
  \label{eq:potential}
  V(\vec{r} \equiv \vec{x} - \vec\rho)
  = - \int \frac{d^3\vec{q}}{(2\pi)^3} \,
  e^{i\vec{q}\cdot\vec{r}}\, \mathcal{M}(\vec{q})
\end{equation}
with four-momentum transfer between the source particle and the
neutrino given by $q\equiv(0, \vec{q})\equiv p'-p$ where $p$ and $p'$
are the neutrino four-momenta before and after the elastic scattering,
which for a scalar-pseudoscalar interaction of a neutrino with mass
$m$
\begin{equation}
  \label{eq:lagran2}
  \mathcal{L} = g_s^f \phi \bar{f}f + ig_p^\nu \phi\bar{\nu}\gamma^5\nu
\end{equation}
reads
\begin{equation}
  \label{eq:amp}
  i\mathcal{M} = \frac{i(ig_s^f)(-g_p^\nu)}{q^2 - m_\phi^2}
  \Big[ \bar{u}_{r'}^f(k) u_{r}^f(k) \Big]
  \Big[ \bar{u}_{\lambda'}^\nu(p')\gamma^5 u_{\lambda}^\nu(p) \Big] \,.
\end{equation}
In writing Eq.~\eqref{eq:potential} we are implicitly assuming the
normalization of the spinors to be one particle per unit volume for
both nucleons and neutrinos.  Explicitly, in the Dirac representation
of the gamma matrices our choice for the spinor of fermion with
four-momentum $(E, \vec{p})$ and spin state $s$ is
\begin{equation}
  u_s(p) = \sqrt{\frac{E+m}{2E}}
  \begin{pmatrix}
    \chi_s
    \\
    \frac{\vec\sigma \cdot \vec{p}}{E+m} \chi_s
  \end{pmatrix}
\end{equation}
where $\chi_s$ are 2-spinors normalized as
$\chi^\dagger_{s'}\chi_s=\delta_{ss'}$.
With this choice, for the non-relativistic nucleons
$\bar{u}_{r'}^f(k) u_r^f(k) = \delta_{rr'}$, while for the neutrinos
we can write
\begin{equation}
  \bar{u}_{\lambda'}^\nu(p')\gamma^5 u_{\lambda}^\nu(p)
  = \sqrt{\frac{E'+m}{2E'}}\sqrt{\frac{E+m}{2E}} \,
  \Big( \chi^{\dag}_{\lambda'} \vec{\sigma} \chi_{\lambda} \Big) \cdot
  \bigg( \frac{\vec{p}'}{E' + m} - \frac{\vec{p}}{E + m} \bigg) \,.
\end{equation}
Expanding to the lowest non-vanishing order in $\vec{q}$ we find
\begin{equation}
  \label{eq:nuline}
  \bar{u}_{\lambda'}^\nu(p')\gamma^5 u_{\lambda}^\nu(p) =
  \frac{\chi^{\dag}_{\lambda'} \vec{\sigma} \chi_{\lambda}}{2E} \cdot
  \Big[ \vec{q} - \Big(1 - \frac{m}{E} \Big) (\hat{n} \cdot \vec{q})\,
    \hat{n} \Big],
\end{equation}
where $\hat{n}$ denotes the neutrino direction.  Introducing
Eq.~\eqref{eq:nuline} in Eq.~\eqref{eq:amp} and evaluating the
$\vec{q}$ integral in Eq.~\eqref{eq:potential} one finds
\begin{equation}
  \label{eq:vpoint}
  V(\vec{r}) = - \frac{g_s^f g_p^\nu}{8\pi E}
  \bigg[ \vec{S}_{\lambda\lambda'} - \Big( 1 - \dfrac{m}{E} \Big)
    (\vec{S}_{\lambda\lambda'} \cdot \hat{n})\, \hat{n} \bigg] \cdot
  \vec{\nabla}_r \bigg(\frac{e^{-m_\phi r}}{r} \bigg)\,,
\end{equation}
where we have defined $\vec{S}_{\lambda\lambda'} \equiv
\chi^{\dag}_{\lambda'}\, \vec{\sigma}\, \chi_{\lambda}$ as a spin-like
vector quantity.  
In the non-relativistic limit $\vec{S}_{\lambda\lambda'}$ corresponds
to the spin of the neutrino, and Eq.~\eqref{eq:vpoint} reduces to the
well-known monopole-dipole potential~\cite{Moody:1984ba}
\begin{equation}
  V(r)
  = g_s^f g_p^\nu\, \frac{\hat\sigma_\nu \cdot \hat{r}}{8\pi\, m_\nu}\,
  \bigg[ \frac{m_\phi}{r} + \frac{1}{r^2} \bigg]\, e^{-m_\phi r}
\end{equation}
where $\hat\sigma_\nu$ is the direction of the neutrino spin.
Conversely for relativistic neutrinos one can identify $\lambda$ with
the helicity of the neutrino and using the explicit form of the
spinors:
\begin{equation}
  \vec{S}_{++} = -\vec{S}_{--} = \hat{n} \,,
  \qquad
  \vec{S}_{+-} = \vec{S}_{-+}^* = \hat{u} + i \hat{v} \,,
  \qquad
  (\text{with~} \hat{n} \perp \hat{u} \perp \hat{v})
\end{equation}
we get
\begin{equation}
  \label{eq:potpoint}
  V(\vec{r}) = -\frac{g_s^f\, g_p^\nu}{8\pi E}\,
  \bigg[ \delta_{\lambda\neq \lambda'} (\vec{\nabla}_r)_\perp
    \pm \delta_{\lambda\lambda'}\, \frac{m}{E} (\vec{\nabla}_r)_\parallel
    \bigg] \, \bigg( \frac{e^{-m_\phi r}}{r} \bigg)
\end{equation}
where
\begin{equation}
  (\vec{\nabla}_r)_\parallel \equiv \hat{n}\cdot \vec{\nabla}_r
  \quad\text{and}\quad
  (\vec{\nabla}_r)_\perp \equiv (\hat{u} + i\hat{v})\cdot \vec{\nabla}_r
\end{equation}
are the gradient operators along and perpendicular to the neutrino
direction respectively.  We notice that Eq.~\eqref{eq:potpoint}
displays the relativistic $1/\gamma = m/E$ factor suppression of the
helicity-conserving potential in analogy with the well-known results
for the potential generated by a magnetic field in the presence of a
neutrino magnetic moment~\cite{Akhmedov:1988hd, Akhmedov:1988kih}.
Conversely the scalar-pseudoscalar interaction produces a
helicity-flip potential which is proportional to the variation of the
potential in the direction perpendicular to that of the neutrino
propagation.

We are interested in taking the Sun as a source of the potential and
for that we integrate Eq.~\eqref{eq:potpoint} with the number density
of nucleon $f$ at position $\vec\rho$ to obtain the potential at
neutrino position $\vec{x}$ to be
\begin{equation}
  V(\vec{x}) =
  -\frac{g_s^f\, g_p^\nu}{8\pi\, E}\
  \bigg[ \delta_{\lambda\neq\lambda'} (\vec{\nabla}_x)_\perp
    \pm \delta_{\lambda\lambda'}\, \frac{m}{E} (\vec{\nabla}_x)_\parallel \bigg]
  \int N_f(\vec\rho)\,
    \frac{e^{-m_\phi |\vec\rho - \vec{x}|}}{|\vec\rho - \vec{x}|} \,
    d^3\vec\rho \,.
\end{equation}
where we have replaced $\vec{\nabla}_r $ with $-\vec{\nabla}_\rho =
\vec{\nabla}_x$ because the position of the neutrino is fixed in the
integral.

%\end{appendices}

\cleardoublepage
\listoffigures
\cleardoublepage
\listoftables

\cleardoublepage
\bibliographystyle{unsrt}
\bibliography{references,references_2}

\providecommand{\href}[2]{#2}\begingroup\begin{thebibliography}{100}

\bibitem{Esteban:2016qun}
I.~Esteban, M.~C. Gonzalez-Garcia, M.~Maltoni, I.~Martinez-Soler, and
  T.~Schwetz, ``{Updated fit to three neutrino mixing: exploring the
  accelerator-reactor complementarity}'',
  \href{http://dx.doi.org/10.1007/JHEP01(2017)087}{{\em JHEP} {\bfseries 01}
  (2017) 087},
\href{http://arxiv.org/abs/1611.01514}{{\ttfamily arXiv:1611.01514 [hep-ph]}}.
%%CITATION = ARXIV:1611.01514;%%.

\bibitem{Esteban:2018ppq}
I.~Esteban, M.~C. Gonzalez-Garcia, M.~Maltoni, I.~Martinez-Soler, and
  J.~Salvado, ``{Updated Constraints on Non-Standard Interactions from Global
  Analysis of Oscillation Data}'',
  \href{http://dx.doi.org/10.1007/JHEP08(2018)180}{{\em JHEP} {\bfseries 08}
  (2018) 180},
\href{http://arxiv.org/abs/1805.04530}{{\ttfamily arXiv:1805.04530 [hep-ph]}}.
%%CITATION = ARXIV:1805.04530;%%.

\bibitem{Esteban:2018azc}
I.~Esteban, M.~C. Gonzalez-Garcia, A.~Hernandez-Cabezudo, M.~Maltoni, and
  T.~Schwetz, ``{Global analysis of three-flavour neutrino oscillations:
  synergies and tensions in the determination of $\theta_{23}$, $\delta_{CP}$,
  and the mass ordering}'',
  \href{http://dx.doi.org/10.1007/JHEP01(2019)106}{{\em JHEP} {\bfseries 01}
  (2019) 106},
\href{http://arxiv.org/abs/1811.05487}{{\ttfamily arXiv:1811.05487 [hep-ph]}}.
%%CITATION = ARXIV:1811.05487;%%.

\bibitem{Esteban:2019lfo}
I.~Esteban, M.~C. Gonzalez-Garcia, and M.~Maltoni, ``{On the Determination of
  Leptonic CP Violation and Neutrino Mass Ordering in Presence of Non-Standard
  Interactions: Present Status}'',
  \href{http://dx.doi.org/10.1007/JHEP06(2019)055}{{\em JHEP} {\bfseries 06}
  (2019) 055},
\href{http://arxiv.org/abs/1905.05203}{{\ttfamily arXiv:1905.05203 [hep-ph]}}.
%%CITATION = ARXIV:1905.05203;%%.

\bibitem{Coloma:2019mbs}
P.~Coloma, I.~Esteban, M.~C. Gonzalez-Garcia, and M.~Maltoni, ``{Improved
  global fit to Non-Standard neutrino Interactions using COHERENT energy and
  timing data}'', \href{http://dx.doi.org/10.1007/JHEP02(2020)023}{{\em JHEP}
  {\bfseries 02} (2020) 023},
\href{http://arxiv.org/abs/1911.09109}{{\ttfamily arXiv:1911.09109 [hep-ph]}}.
%%CITATION = ARXIV:1911.09109;%%.

\bibitem{Baxter:2019mcx}
D.~Baxter {\em et~al.}, ``{Coherent Elastic Neutrino-Nucleus Scattering at the
  European Spallation Source}'',
  \href{http://dx.doi.org/10.1007/JHEP02(2020)123}{{\em JHEP} {\bfseries 02}
  (2020) 123},
\href{http://arxiv.org/abs/1911.00762}{{\ttfamily arXiv:1911.00762
  [physics.ins-det]}}.
%%CITATION = ARXIV:1911.00762;%%.

\bibitem{Danby:1962nd}
G.~Danby, J.~M. Gaillard, K.~A. Goulianos, L.~M. Lederman, N.~B. Mistry,
  M.~Schwartz, and J.~Steinberger, ``{Observation of High-Energy Neutrino
  Reactions and the Existence of Two Kinds of Neutrinos}'',
\href{http://dx.doi.org/10.1103/PhysRevLett.9.36}{{\em Phys. Rev. Lett.}
  {\bfseries 9} (1962) 36--44}.
%%CITATION = PRLTA,9,36;%%.

\bibitem{Hasert:1973cr}
F.~J. Hasert {\em et~al.}, ``{Search for Elastic $\nu_\mu$ Electron
  Scattering}'', \href{http://dx.doi.org/10.1016/0370-2693(73)90494-2}{{\em
  Phys. Lett.} {\bfseries 46B} (1973) 121--124}.
[,5.11(1973); ,5.11(1973)].
%%CITATION = PHLTA,46B,121;%%.

\bibitem{Eichten:1973cs}
T.~Eichten {\em et~al.}, ``{Measurement of the Neutrino - Nucleon Anti-neutrino
  - Nucleon Total Cross-sections}'',
\href{http://dx.doi.org/10.1016/0370-2693(73)90702-8}{{\em Phys. Lett.}
  {\bfseries 46B} (1973) 274--280}.
%%CITATION = PHLTA,46B,274;%%.

\bibitem{Benvenuti:1975ru}
A.~C. Benvenuti {\em et~al.}, ``{Observation of New Particle Production by
  High-Energy Neutrinos and anti-neutrinos}'',
\href{http://dx.doi.org/10.1103/PhysRevLett.34.419}{{\em Phys. Rev. Lett.}
  {\bfseries 34} (1975) 419}.
%%CITATION = PRLTA,34,419;%%.

\bibitem{Pauli:83282}
W.~Pauli, ``{Pauli letter collection: letter to Lise Meitner}.'' Typed copy,
  Dec., 1930. \url{http://cds.cern.ch/record/83282}.

\bibitem{Bethe:1934qn}
H.~Bethe and R.~Peierls, ``{The 'neutrino'}'',
\href{http://dx.doi.org/10.1038/133532a0}{{\em Nature} {\bfseries 133} (1934)
  532}.
%%CITATION = NATUA,133,532;%%.

\bibitem{Cowan:1992xc}
C.~L. Cowan, F.~Reines, F.~B. Harrison, H.~W. Kruse, and A.~D. McGuire,
  ``{Detection of the free neutrino: A Confirmation}'',
\href{http://dx.doi.org/10.1126/science.124.3212.103}{{\em Science} {\bfseries
  124} (1956) 103--104}.
%%CITATION = SCIEA,124,103;%%.

\bibitem{Aker:2019uuj}
{ KATRIN} Collaboration, M.~Aker {\em et~al.}, ``{An improved upper limit on
  the neutrino mass from a direct kinematic method by KATRIN}'',
  \href{http://dx.doi.org/10.1103/PhysRevLett.123.221802}{{\em Phys. Rev.
  Lett.} {\bfseries 123} no.~22, (2019) 221802},
\href{http://arxiv.org/abs/1909.06048}{{\ttfamily arXiv:1909.06048 [hep-ex]}}.
%%CITATION = ARXIV:1909.06048;%%.

\bibitem{Ahmad:2002jz}
{ SNO} Collaboration, Q.~R. Ahmad {\em et~al.}, ``{Direct evidence for neutrino
  flavor transformation from neutral current interactions in the Sudbury
  Neutrino Observatory}'',
  \href{http://dx.doi.org/10.1103/PhysRevLett.89.011301}{{\em Phys. Rev. Lett.}
  {\bfseries 89} (2002) 011301},
\href{http://arxiv.org/abs/nucl-ex/0204008}{{\ttfamily arXiv:nucl-ex/0204008
  [nucl-ex]}}.
%%CITATION = NUCL-EX/0204008;%%.

\bibitem{Ahmad:2002ka}
{ SNO} Collaboration, Q.~R. Ahmad {\em et~al.}, ``{Measurement of day and night
  neutrino energy spectra at SNO and constraints on neutrino mixing
  parameters}'', \href{http://dx.doi.org/10.1103/PhysRevLett.89.011302}{{\em
  Phys. Rev. Lett.} {\bfseries 89} (2002) 011302},
\href{http://arxiv.org/abs/nucl-ex/0204009}{{\ttfamily arXiv:nucl-ex/0204009
  [nucl-ex]}}.
%%CITATION = NUCL-EX/0204009;%%.

\bibitem{Ahmed:2003kj}
{ SNO} Collaboration, S.~N. Ahmed {\em et~al.}, ``{Measurement of the total
  active B-8 solar neutrino flux at the Sudbury Neutrino Observatory with
  enhanced neutral current sensitivity}'',
  \href{http://dx.doi.org/10.1103/PhysRevLett.92.181301}{{\em Phys. Rev. Lett.}
  {\bfseries 92} (2004) 181301},
\href{http://arxiv.org/abs/nucl-ex/0309004}{{\ttfamily arXiv:nucl-ex/0309004
  [nucl-ex]}}.
%%CITATION = NUCL-EX/0309004;%%.

\bibitem{Aharmim:2005gt}
{ SNO} Collaboration, B.~Aharmim {\em et~al.}, ``{Electron energy spectra,
  fluxes, and day-night asymmetries of B-8 solar neutrinos from measurements
  with NaCl dissolved in the heavy-water detector at the Sudbury Neutrino
  Observatory}'', \href{http://dx.doi.org/10.1103/PhysRevC.72.055502}{{\em
  Phys. Rev.} {\bfseries C72} (2005) 055502},
\href{http://arxiv.org/abs/nucl-ex/0502021}{{\ttfamily arXiv:nucl-ex/0502021
  [nucl-ex]}}.
%%CITATION = NUCL-EX/0502021;%%.

\bibitem{BeckerSzendy:1992hq}
R.~Becker-Szendy {\em et~al.}, ``{The Electron-neutrino and muon-neutrino
  content of the atmospheric flux}'',
\href{http://dx.doi.org/10.1103/PhysRevD.46.3720}{{\em Phys. Rev.} {\bfseries
  D46} (1992) 3720--3724}.
%%CITATION = PHRVA,D46,3720;%%.

\bibitem{Fukuda:1994mc}
{ Kamiokande} Collaboration, Y.~Fukuda {\em et~al.}, ``{Atmospheric
  muon-neutrino / electron-neutrino ratio in the multiGeV energy range}'',
\href{http://dx.doi.org/10.1016/0370-2693(94)91420-6}{{\em Phys. Lett.}
  {\bfseries B335} (1994) 237--245}.
%%CITATION = PHLTA,B335,237;%%.

\bibitem{Fukuda:1998mi}
{ Super-Kamiokande} Collaboration, Y.~Fukuda {\em et~al.}, ``{Evidence for
  oscillation of atmospheric neutrinos}'',
  \href{http://dx.doi.org/10.1103/PhysRevLett.81.1562}{{\em Phys. Rev. Lett.}
  {\bfseries 81} (1998) 1562--1567},
\href{http://arxiv.org/abs/hep-ex/9807003}{{\ttfamily arXiv:hep-ex/9807003
  [hep-ex]}}.
%%CITATION = HEP-EX/9807003;%%.

\bibitem{t2k:ichep2016}
K.~Iwamoto, ``{Recent Results from T2K and Future Prospects}.'' Talk given at
  the {\it 38th International Conference on High Energy Physics}, Chicago, USA,
  August 3--10, 2016.

\bibitem{t2k:susy2016}
A.~Cervera, ``{Latest Results from Neutrino Oscillation Experiments}.'' Talk
  given at the SUSY~2016 Conference, Melbourne, Australia, July 3--8, 2016.

\bibitem{nova:nu2016}
P.~Vahle, ``{New results from NOvA}.'' Talk given at the {\it XXVII
  International Conference on Neutrino Physics and Astrophysics}, London, UK,
  July 4--9, 2016.

\bibitem{Cohen:1997mt}
A.~G. Cohen and A.~De~Rujula, ``{Scars on the Cosmic Background Radiation?}'',
  {\em Astrophys. J. Lett.} {\bfseries 496} no.~2, (Apr, 1998) L63--L65,
\href{http://arxiv.org/abs/astro-ph/9709132}{{\ttfamily arXiv:astro-ph/9709132
  [astro-ph]}}.
%%CITATION = ASTRO-PH/9709132;%%.

\bibitem{Kinney:1997ic}
W.~H. Kinney, E.~W. Kolb, and M.~S. Turner, ``{Ribbons on the CBR sky: A
  Powerful test of a baryon symmetric universe}'',
  \href{http://dx.doi.org/10.1103/PhysRevLett.79.2620}{{\em Phys. Rev. Lett.}
  {\bfseries 79} (1997) 2620--2623},
\href{http://arxiv.org/abs/astro-ph/9704070}{{\ttfamily arXiv:astro-ph/9704070
  [astro-ph]}}.
%%CITATION = ASTRO-PH/9704070;%%.

\bibitem{Sakharov:1967dj}
A.~D. Sakharov, ``{Violation of CP Invariance, C asymmetry, and baryon
  asymmetry of the universe}'',
  \href{http://dx.doi.org/10.1070/PU1991v034n05ABEH002497}{{\em Pisma Zh. Eksp.
  Teor. Fiz.} {\bfseries 5} (1967) 32--35}.
[JETP Lett.5,24(1967); Sov. Phys. Usp.34,no.5,392(1991); Usp. Fiz.
  Nauk161,no.5,61(1991)].
%%CITATION = ZFPRA,5,32;%%.

\bibitem{Lesgourgues:2018ncw}
J.~Lesgourgues, G.~Mangano, G.~Miele, and S.~Pastor, {\em {Neutrino
  Cosmology}}.
\newblock Cambridge University Press,
2013.
\newblock
%%CITATION = INSPIRE-1705305;%%.

\bibitem{Cleveland:1998nv}
B.~T. Cleveland, T.~Daily, R.~Davis, Jr., J.~R. Distel, K.~Lande, C.~K. Lee,
  P.~S. Wildenhain, and J.~Ullman, ``{Measurement of the solar electron
  neutrino flux with the Homestake chlorine detector}'',
\href{http://dx.doi.org/10.1086/305343}{{\em Astrophys. J.} {\bfseries 496}
  (1998) 505--526}.
%%CITATION = ASJOA,496,505;%%.

\bibitem{Abdurashitov:2002nt}
{ SAGE} Collaboration, J.~N. Abdurashitov {\em et~al.}, ``{Solar neutrino flux
  measurements by the Soviet-American Gallium Experiment (SAGE) for half the 22
  year solar cycle}'', \href{http://dx.doi.org/10.1134/1.1506424}{{\em J. Exp.
  Theor. Phys.} {\bfseries 95} (2002) 181--193},
  \href{http://arxiv.org/abs/astro-ph/0204245}{{\ttfamily
  arXiv:astro-ph/0204245 [astro-ph]}}.
[Zh. Eksp. Teor. Fiz.122,211(2002)].
%%CITATION = ASTRO-PH/0204245;%%.

\bibitem{Hampel:1998xg}
{ GALLEX} Collaboration, W.~Hampel {\em et~al.}, ``{GALLEX solar neutrino
  observations: Results for GALLEX IV}'',
\href{http://dx.doi.org/10.1016/S0370-2693(98)01579-2}{{\em Phys. Lett.}
  {\bfseries B447} (1999) 127--133}.
%%CITATION = PHLTA,B447,127;%%.

\bibitem{Altmann:2005ix}
{ GNO} Collaboration, M.~Altmann {\em et~al.}, ``{Complete results for five
  years of GNO solar neutrino observations}'',
  \href{http://dx.doi.org/10.1016/j.physletb.2005.04.068}{{\em Phys. Lett.}
  {\bfseries B616} (2005) 174--190},
\href{http://arxiv.org/abs/hep-ex/0504037}{{\ttfamily arXiv:hep-ex/0504037
  [hep-ex]}}.
%%CITATION = HEP-EX/0504037;%%.

\bibitem{Hirata:1991ub}
{ Kamiokande-II} Collaboration, K.~S. Hirata {\em et~al.}, ``{Real time,
  directional measurement of B-8 solar neutrinos in the Kamiokande-II
  detector}'', \href{http://dx.doi.org/10.1103/PhysRevD.44.2241,
  10.1103/PhysRevD.45.2170}{{\em Phys. Rev.} {\bfseries D44} (1991) 2241}.
[Erratum: Phys. Rev.D45,2170(1992)].
%%CITATION = PHRVA,D44,2241;%%.

\bibitem{Hosaka:2005um}
{ Super-Kamiokande} Collaboration, J.~Hosaka {\em et~al.}, ``{Solar neutrino
  measurements in Super-Kamiokande-I}'',
  \href{http://dx.doi.org/10.1103/PhysRevD.73.112001}{{\em Phys. Rev.}
  {\bfseries D73} (2006) 112001},
\href{http://arxiv.org/abs/hep-ex/0508053}{{\ttfamily arXiv:hep-ex/0508053
  [hep-ex]}}.
%%CITATION = HEP-EX/0508053;%%.

\bibitem{Aharmim:2011vm}
{ SNO} Collaboration, B.~Aharmim {\em et~al.}, ``{Combined Analysis of all
  Three Phases of Solar Neutrino Data from the Sudbury Neutrino Observatory}'',
  \href{http://dx.doi.org/10.1103/PhysRevC.88.025501}{{\em Phys. Rev.}
  {\bfseries C88} (2013) 025501},
\href{http://arxiv.org/abs/1109.0763}{{\ttfamily arXiv:1109.0763 [nucl-ex]}}.
%%CITATION = ARXIV:1109.0763;%%.

\bibitem{Gando:2014wjd}
{ KamLAND} Collaboration, A.~Gando {\em et~al.}, ``{$^7$Be Solar Neutrino
  Measurement with KamLAND}'',
  \href{http://dx.doi.org/10.1103/PhysRevC.92.055808}{{\em Phys. Rev.}
  {\bfseries C92} no.~5, (2015) 055808},
\href{http://arxiv.org/abs/1405.6190}{{\ttfamily arXiv:1405.6190 [hep-ex]}}.
%%CITATION = ARXIV:1405.6190;%%.

\bibitem{Bellini:2011rx}
G.~Bellini {\em et~al.}, ``{Precision measurement of the 7Be solar neutrino
  interaction rate in Borexino}'',
  \href{http://dx.doi.org/10.1103/PhysRevLett.107.141302}{{\em Phys. Rev.
  Lett.} {\bfseries 107} (2011) 141302},
\href{http://arxiv.org/abs/1104.1816}{{\ttfamily arXiv:1104.1816 [hep-ex]}}.
%%CITATION = ARXIV:1104.1816;%%.

\bibitem{Ambrosio:2001je}
{ MACRO} Collaboration, M.~Ambrosio {\em et~al.}, ``{Matter effects in upward
  going muons and sterile neutrino oscillations}'',
  \href{http://dx.doi.org/10.1016/S0370-2693(01)00992-3}{{\em Phys. Lett.}
  {\bfseries B517} (2001) 59--66},
\href{http://arxiv.org/abs/hep-ex/0106049}{{\ttfamily arXiv:hep-ex/0106049
  [hep-ex]}}.
%%CITATION = HEP-EX/0106049;%%.

\bibitem{Sanchez:2003rb}
{ Soudan 2} Collaboration, M.~C. Sanchez {\em et~al.}, ``{Measurement of the
  L/E distributions of atmospheric neutrinos in Soudan 2 and their
  interpretation as neutrino oscillations}'',
  \href{http://dx.doi.org/10.1103/PhysRevD.68.113004}{{\em Phys. Rev.}
  {\bfseries D68} (2003) 113004},
\href{http://arxiv.org/abs/hep-ex/0307069}{{\ttfamily arXiv:hep-ex/0307069
  [hep-ex]}}.
%%CITATION = HEP-EX/0307069;%%.

\bibitem{Abe:2008aa}
{ KamLAND} Collaboration, S.~Abe {\em et~al.}, ``{Precision Measurement of
  Neutrino Oscillation Parameters with KamLAND}'',
  \href{http://dx.doi.org/10.1103/PhysRevLett.100.221803}{{\em Phys. Rev.
  Lett.} {\bfseries 100} (2008) 221803},
\href{http://arxiv.org/abs/0801.4589}{{\ttfamily arXiv:0801.4589 [hep-ex]}}.
%%CITATION = ARXIV:0801.4589;%%.

\bibitem{Ahn:2001cq}
{ K2K} Collaboration, S.~H. Ahn {\em et~al.}, ``{Detection of accelerator
  produced neutrinos at a distance of 250-km}'',
  \href{http://dx.doi.org/10.1016/S0370-2693(01)00647-5}{{\em Phys. Lett.}
  {\bfseries B511} (2001) 178--184},
\href{http://arxiv.org/abs/hep-ex/0103001}{{\ttfamily arXiv:hep-ex/0103001
  [hep-ex]}}.
%%CITATION = HEP-EX/0103001;%%.

\bibitem{Adamson:2013whj}
{ MINOS} Collaboration, P.~Adamson {\em et~al.}, ``{Measurement of Neutrino and
  Antineutrino Oscillations Using Beam and Atmospheric Data in MINOS}'',
  \href{http://dx.doi.org/10.1103/PhysRevLett.110.251801}{{\em Phys. Rev.
  Lett.} {\bfseries 110} no.~25, (2013) 251801},
\href{http://arxiv.org/abs/1304.6335}{{\ttfamily arXiv:1304.6335 [hep-ex]}}.
%%CITATION = ARXIV:1304.6335;%%.

\bibitem{DoubleChooz:2019qbj}
{ Double Chooz} Collaboration, H.~de~Kerret {\em et~al.}, ``{First Double Chooz
  $\mathbf{\theta_{13}}$ Measurement via Total Neutron Capture Detection}'',
\href{http://arxiv.org/abs/1901.09445}{{\ttfamily arXiv:1901.09445 [hep-ex]}}.
%%CITATION = ARXIV:1901.09445;%%.

\bibitem{Adey:2018zwh}
{ Daya Bay} Collaboration, D.~Adey {\em et~al.}, ``{Measurement of the Electron
  Antineutrino Oscillation with 1958 Days of Operation at Daya Bay}'',
  \href{http://dx.doi.org/10.1103/PhysRevLett.121.241805}{{\em Phys. Rev.
  Lett.} {\bfseries 121} no.~24, (2018) 241805},
\href{http://arxiv.org/abs/1809.02261}{{\ttfamily arXiv:1809.02261 [hep-ex]}}.
%%CITATION = ARXIV:1809.02261;%%.

\bibitem{Bak:2018ydk}
{ RENO} Collaboration, G.~Bak {\em et~al.}, ``{Measurement of Reactor
  Antineutrino Oscillation Amplitude and Frequency at RENO}'',
  \href{http://dx.doi.org/10.1103/PhysRevLett.121.201801}{{\em Phys. Rev.
  Lett.} {\bfseries 121} no.~20, (2018) 201801},
\href{http://arxiv.org/abs/1806.00248}{{\ttfamily arXiv:1806.00248 [hep-ex]}}.
%%CITATION = ARXIV:1806.00248;%%.

\bibitem{Ayres:2004js}
{ NOvA} Collaboration, D.~S. Ayres {\em et~al.}, ``{NOvA: Proposal to Build a
  30 Kiloton Off-Axis Detector to Study $\nu_{\mu} \to \nu_e$ Oscillations in
  the NuMI Beamline}'',
\href{http://arxiv.org/abs/hep-ex/0503053}{{\ttfamily arXiv:hep-ex/0503053
  [hep-ex]}}.
%%CITATION = HEP-EX/0503053;%%.

\bibitem{Abe:2011ks}
{ T2K} Collaboration, K.~Abe {\em et~al.}, ``{The T2K Experiment}'',
  \href{http://dx.doi.org/10.1016/j.nima.2011.06.067}{{\em Nucl. Instrum.
  Meth.} {\bfseries A659} (2011) 106--135},
\href{http://arxiv.org/abs/1106.1238}{{\ttfamily arXiv:1106.1238
  [physics.ins-det]}}.
%%CITATION = ARXIV:1106.1238;%%.

\bibitem{Acciarri:2016ooe}
{ DUNE} Collaboration, R.~Acciarri {\em et~al.}, ``{Long-Baseline Neutrino
  Facility (LBNF) and Deep Underground Neutrino Experiment (DUNE)}'',
\href{http://arxiv.org/abs/1601.02984}{{\ttfamily arXiv:1601.02984
  [physics.ins-det]}}.
%%CITATION = ARXIV:1601.02984;%%.

\bibitem{Abe:2015zbg}
{ Hyper-Kamiokande Proto-Collaboration} Collaboration, K.~Abe {\em et~al.},
  ``{Physics potential of a long-baseline neutrino oscillation experiment using
  a J-PARC neutrino beam and Hyper-Kamiokande}'',
  \href{http://dx.doi.org/10.1093/ptep/ptv061}{{\em PTEP} {\bfseries 2015}
  (2015) 053C02},
\href{http://arxiv.org/abs/1502.05199}{{\ttfamily arXiv:1502.05199 [hep-ex]}}.
%%CITATION = ARXIV:1502.05199;%%.

\bibitem{nufit-3.1}
I.~Esteban, M.~Gonzalez-Garcia, A.~Hernandez-Cabezudo, M.~Maltoni,
  I.~Martinez-Soler, and T.~Schwetz, ``{NuFit 3.1 (2017)}.''.
  \href{http://www.nu-fit.org}{\tt http://www.nu-fit.org}.

\bibitem{nufit-3.2}
I.~Esteban, M.~Gonzalez-Garcia, A.~Hernandez-Cabezudo, M.~Maltoni,
  I.~Martinez-Soler, and T.~Schwetz, ``{NuFit 3.2 (2018)}.''.
  \href{http://www.nu-fit.org}{\tt http://www.nu-fit.org}.

\bibitem{nufit-4.1}
I.~Esteban, M.~Gonzalez-Garcia, A.~Hernandez-Cabezudo, M.~Maltoni, and
  T.~Schwetz, ``{NuFit 4.1 (2019)}.'' \url{http://www.nu-fit.org}.

\bibitem{Farzan:2015doa}
Y.~Farzan, ``{A model for large non-standard interactions of neutrinos leading
  to the LMA-Dark solution}'',
  \href{http://dx.doi.org/10.1016/j.physletb.2015.07.015}{{\em Phys. Lett.}
  {\bfseries B748} (2015) 311--315},
\href{http://arxiv.org/abs/1505.06906}{{\ttfamily arXiv:1505.06906 [hep-ph]}}.
%%CITATION = ARXIV:1505.06906;%%.

\bibitem{Farzan:2015hkd}
Y.~Farzan and I.~M. Shoemaker, ``{Lepton Flavor Violating Non-Standard
  Interactions via Light Mediators}'',
  \href{http://dx.doi.org/10.1007/JHEP07(2016)033}{{\em JHEP} {\bfseries 07}
  (2016) 033},
\href{http://arxiv.org/abs/1512.09147}{{\ttfamily arXiv:1512.09147 [hep-ph]}}.
%%CITATION = ARXIV:1512.09147;%%.

\bibitem{Babu:2017olk}
K.~S. Babu, A.~Friedland, P.~A.~N. Machado, and I.~Mocioiu, ``{Flavor Gauge
  Models Below the Fermi Scale}'',
\href{http://arxiv.org/abs/1705.01822}{{\ttfamily arXiv:1705.01822 [hep-ph]}}.
%%CITATION = ARXIV:1705.01822;%%.

\bibitem{Farzan:2017xzy}
Y.~Farzan and M.~Tortola, ``{Neutrino oscillations and Non-Standard
  Interactions}'', \href{http://dx.doi.org/10.3389/fphy.2018.00010}{{\em
  Front.in Phys.} {\bfseries 6} (2018) 10},
\href{http://arxiv.org/abs/1710.09360}{{\ttfamily arXiv:1710.09360 [hep-ph]}}.
%%CITATION = ARXIV:1710.09360;%%.

\bibitem{Denton:2018xmq}
P.~B. Denton, Y.~Farzan, and I.~M. Shoemaker, ``{A Plan to Rule out Large
  Non-Standard Neutrino Interactions After COHERENT Data}'',
\href{http://arxiv.org/abs/1804.03660}{{\ttfamily arXiv:1804.03660 [hep-ph]}}.
%%CITATION = ARXIV:1804.03660;%%.

\bibitem{Miranda:2015dra}
O.~G. Miranda and H.~Nunokawa, ``{Non standard neutrino interactions: current
  status and future prospects}'',
  \href{http://dx.doi.org/10.1088/1367-2630/17/9/095002}{{\em New J. Phys.}
  {\bfseries 17} no.~9, (2015) 095002},
\href{http://arxiv.org/abs/1505.06254}{{\ttfamily arXiv:1505.06254 [hep-ph]}}.
%%CITATION = ARXIV:1505.06254;%%.

\bibitem{Heeck:2018nzc}
J.~Heeck, M.~Lindner, W.~Rodejohann, and S.~Vogl, ``{Non-Standard Neutrino
  Interactions and Neutral Gauge Bosons}'',
  \href{http://dx.doi.org/10.21468/SciPostPhys.6.3.038}{{\em SciPost Phys.}
  {\bfseries 6} no.~3, (2019) 038},
\href{http://arxiv.org/abs/1812.04067}{{\ttfamily arXiv:1812.04067 [hep-ph]}}.
%%CITATION = ARXIV:1812.04067;%%.

\bibitem{Akimov:2015nza}
{ COHERENT} Collaboration, D.~Akimov {\em et~al.}, ``{The COHERENT Experiment
  at the Spallation Neutron Source}'',
\href{http://arxiv.org/abs/1509.08702}{{\ttfamily arXiv:1509.08702
  [physics.ins-det]}}.
%%CITATION = ARXIV:1509.08702;%%.

\bibitem{Dalton:1808}
J.~Dalton, {\em A new system of chemical philosophy}.
\newblock R. Bickerstaff, London, 1808.

\bibitem{Geiger:1909}
H.~Geiger, E.~Marsden, and E.~Rutherford, ``On a diffuse reflection of the
  $\alpha$-particles'',
  \href{http://dx.doi.org/https://doi.org/10.1098/rspa.1909.0054}{{\em Proc. R.
  Soc. Lond. A} {\bfseries 82} (1909) 495--500}.

\bibitem{Geiger:1910}
H.~Geiger and E.~Rutherford, ``The scattering of $\alpha$-particles by
  matter'',
  \href{http://dx.doi.org/https://doi.org/10.1098/rspa.1910.0038}{{\em Proc. R.
  Soc. Lond. A} {\bfseries 83} (1910) 492--504}.

\bibitem{Rutherford:1911}
E.~Rutherford, ``{The scattering of alpha and beta particles by matter and the
  structure of the atom}'',
\href{http://dx.doi.org/10.1080/14786440508637080}{{\em Phil. Mag. Ser.6}
  {\bfseries 21} (1911) 669--688}.
%%CITATION = INSPIRE-45404;%%.

\bibitem{Glashow:1961tr}
S.~L. Glashow, ``{Partial Symmetries of Weak Interactions}'',
\href{http://dx.doi.org/10.1016/0029-5582(61)90469-2}{{\em Nucl. Phys.}
  {\bfseries 22} (1961) 579--588}.
%%CITATION = NUPHA,22,579;%%.

\bibitem{Weinberg:1967tq}
S.~Weinberg, ``{A Model of Leptons}'',
\href{http://dx.doi.org/10.1103/PhysRevLett.19.1264}{{\em Phys. Rev. Lett.}
  {\bfseries 19} (1967) 1264--1266}.
%%CITATION = PRLTA,19,1264;%%.

\bibitem{Salam1196}
A.~Salam, ``{Elementary Particle Theory. Relativistic Groups and Analyticity.
  Proceedings of the eighth Nobel Symposium}'',
  \href{http://dx.doi.org/10.1126/science.168.3936.1196-a}{{\em Science}
  {\bfseries 168} no.~3936, (1970) 1196--1197}.

\bibitem{Fritzsch:1973pi}
H.~Fritzsch, M.~Gell-Mann, and H.~Leutwyler, ``{Advantages of the Color Octet
  Gluon Picture}'',
\href{http://dx.doi.org/10.1016/0370-2693(73)90625-4}{{\em Phys. Lett.}
  {\bfseries 47B} (1973) 365--368}.
%%CITATION = PHLTA,47B,365;%%.

\bibitem{Gross:1973ju}
D.~J. Gross and F.~Wilczek, ``{Asymptotically Free Gauge Theories - I}'',
\href{http://dx.doi.org/10.1103/PhysRevD.8.3633}{{\em Phys. Rev.} {\bfseries
  D8} (1973) 3633--3652}.
%%CITATION = PHRVA,D8,3633;%%.

\bibitem{Weinberg:1973un}
S.~Weinberg, ``{Nonabelian Gauge Theories of the Strong Interactions}'',
\href{http://dx.doi.org/10.1103/PhysRevLett.31.494}{{\em Phys. Rev. Lett.}
  {\bfseries 31} (1973) 494--497}.
%%CITATION = PRLTA,31,494;%%.

\bibitem{Higgs:1964pj}
P.~W. Higgs, ``{Broken Symmetries and the Masses of Gauge Bosons}'',
  \href{http://dx.doi.org/10.1103/PhysRevLett.13.508}{{\em Phys. Rev. Lett.}
  {\bfseries 13} (1964) 508--509}.
[,160(1964)].
%%CITATION = PRLTA,13,508;%%.

\bibitem{Englert:1964et}
F.~Englert and R.~Brout, ``{Broken Symmetry and the Mass of Gauge Vector
  Mesons}'', \href{http://dx.doi.org/10.1103/PhysRevLett.13.321}{{\em Phys.
  Rev. Lett.} {\bfseries 13} (1964) 321--323}.
[,157(1964)].
%%CITATION = PRLTA,13,321;%%.

\bibitem{Guralnik:1964eu}
G.~S. Guralnik, C.~R. Hagen, and T.~W.~B. Kibble, ``{Global Conservation Laws
  and Massless Particles}'',
  \href{http://dx.doi.org/10.1103/PhysRevLett.13.585}{{\em Phys. Rev. Lett.}
  {\bfseries 13} (1964) 585--587}.
[,162(1964)].
%%CITATION = PRLTA,13,585;%%.

\bibitem{Chatrchyan:2012xdj}
{ CMS} Collaboration, S.~Chatrchyan {\em et~al.}, ``{Observation of a New Boson
  at a Mass of 125 GeV with the CMS Experiment at the LHC}'',
  \href{http://dx.doi.org/10.1016/j.physletb.2012.08.021}{{\em Phys. Lett.}
  {\bfseries B716} (2012) 30--61},
\href{http://arxiv.org/abs/1207.7235}{{\ttfamily arXiv:1207.7235 [hep-ex]}}.
%%CITATION = ARXIV:1207.7235;%%.

\bibitem{Aad:2012tfa}
{ ATLAS} Collaboration, G.~Aad {\em et~al.}, ``{Observation of a new particle
  in the search for the Standard Model Higgs boson with the ATLAS detector at
  the LHC}'', \href{http://dx.doi.org/10.1016/j.physletb.2012.08.020}{{\em
  Phys. Lett.} {\bfseries B716} (2012) 1--29},
\href{http://arxiv.org/abs/1207.7214}{{\ttfamily arXiv:1207.7214 [hep-ex]}}.
%%CITATION = ARXIV:1207.7214;%%.

\bibitem{Weinberg:1973ew}
S.~Weinberg, ``{General Theory of Broken Local Symmetries}'',
\href{http://dx.doi.org/10.1103/PhysRevD.7.1068}{{\em Phys. Rev.} {\bfseries
  D7} (1973) 1068--1082}.
%%CITATION = PHRVA,D7,1068;%%.

\bibitem{Cabibbo:1963yz}
N.~Cabibbo, ``{Unitary Symmetry and Leptonic Decays}'',
  \href{http://dx.doi.org/10.1103/PhysRevLett.10.531}{{\em Phys. Rev. Lett.}
  {\bfseries 10} (1963) 531--533}.
[,648(1963)].
%%CITATION = PRLTA,10,531;%%.

\bibitem{Kobayashi:1973fv}
M.~Kobayashi and T.~Maskawa, ``{CP Violation in the Renormalizable Theory of
  Weak Interaction}'',
\href{http://dx.doi.org/10.1143/PTP.49.652}{{\em Prog. Theor. Phys.} {\bfseries
  49} (1973) 652--657}.
%%CITATION = PTPKA,49,652;%%.

\bibitem{Bernabeu:1986fc}
J.~Bernabeu, G.~C. Branco, and M.~Gronau, ``{CP Restrictions on Quark Mass
  Matrices}'',
\href{http://dx.doi.org/10.1016/0370-2693(86)90659-3}{{\em Phys. Lett.}
  {\bfseries 169B} (1986) 243--247}.
%%CITATION = PHLTA,169B,243;%%.

\bibitem{Gronau:1986xb}
M.~Gronau, A.~Kfir, and R.~Loewy, ``{Basis Independent Tests of {CP} Violation
  in Fermion Mass Matrices}'',
\href{http://dx.doi.org/10.1103/PhysRevLett.56.1538}{{\em Phys. Rev. Lett.}
  {\bfseries 56} (1986) 1538}.
%%CITATION = PRLTA,56,1538;%%.

\bibitem{Jarlskog:1985cw}
C.~Jarlskog, ``{A Basis Independent Formulation of the Connection Between Quark
  Mass Matrices, CP Violation and Experiment}'',
\href{http://dx.doi.org/10.1007/BF01565198}{{\em Z. Phys.} {\bfseries C29}
  (1985) 491--497}.
%%CITATION = ZEPYA,C29,491;%%.

\bibitem{Jarlskog:1985ht}
C.~Jarlskog, ``{Commutator of the Quark Mass Matrices in the Standard
  Electroweak Model and a Measure of Maximal CP Violation}'',
\href{http://dx.doi.org/10.1103/PhysRevLett.55.1039}{{\em Phys. Rev. Lett.}
  {\bfseries 55} (1985) 1039}.
%%CITATION = PRLTA,55,1039;%%.

\bibitem{Chivukula:1987py}
R.~S. Chivukula and H.~Georgi, ``{Composite Technicolor Standard Model}'',
\href{http://dx.doi.org/10.1016/0370-2693(87)90713-1}{{\em Phys. Lett.}
  {\bfseries B188} (1987) 99--104}.
%%CITATION = PHLTA,B188,99;%%.

\bibitem{DAmbrosio:2002vsn}
G.~D'Ambrosio, G.~F. Giudice, G.~Isidori, and A.~Strumia, ``{Minimal flavor
  violation: An Effective field theory approach}'',
  \href{http://dx.doi.org/10.1016/S0550-3213(02)00836-2}{{\em Nucl. Phys.}
  {\bfseries B645} (2002) 155--187},
\href{http://arxiv.org/abs/hep-ph/0207036}{{\ttfamily arXiv:hep-ph/0207036
  [hep-ph]}}.
%%CITATION = HEP-PH/0207036;%%.

\bibitem{tHooft:1976rip}
G.~'t~Hooft, ``{Symmetry Breaking Through Bell-Jackiw Anomalies}'',
  \href{http://dx.doi.org/10.1103/PhysRevLett.37.8}{{\em Phys. Rev. Lett.}
  {\bfseries 37} (1976) 8--11}.

\bibitem{Tanabashi:2018oca}
{ Particle Data Group} Collaboration, M.~Tanabashi {\em et~al.}, ``{Review of
  Particle Physics}'', \href{http://dx.doi.org/10.1103/PhysRevD.98.030001}{{\em
  Phys. Rev.} {\bfseries D98} no.~3, (2018) 030001}.
2019 update.
%%CITATION = PHRVA,D98,030001;%%.

\bibitem{Noether:1918zz}
E.~Noether, ``{Invariant Variation Problems}'',
  \href{http://dx.doi.org/10.1080/00411457108231446}{{\em Gott. Nachr.}
  {\bfseries 1918} (1918) 235--257},
  \href{http://arxiv.org/abs/physics/0503066}{{\ttfamily arXiv:physics/0503066
  [physics]}}.
[Transp. Theory Statist. Phys.1,186(1971)].
%%CITATION = PHYSICS/0503066;%%.

\bibitem{Fowler:1958zz}
W.~A. Fowler, ``{Completion of the Proton-Proton Reaction Chain and the
  Possibility of Energetic Neutrino Emission by Hot Stars}'',
\href{http://dx.doi.org/10.1086/146487}{{\em Astrophys. J.} {\bfseries 127}
  (1958) 551--556}.
%%CITATION = ASJOA,127,551;%%.

\bibitem{Cameron:1958vx}
A.~G.~W. Cameron, ``{Nuclear astrophysics}'',
\href{http://dx.doi.org/10.1146/annurev.ns.08.120158.001503}{{\em Ann. Rev.
  Nucl. Part. Sci.} {\bfseries 8} (1958) 299--326}.
%%CITATION = ARNUA,8,299;%%.

\bibitem{Bahcall:1997ha}
J.~N. Bahcall, W.~A. Fowler, I.~Iben, Jr., and R.~L. Sears, ``{Solar neutrino
  flux}'',
\href{http://dx.doi.org/10.1086/147513}{{\em Astrophys. J.} {\bfseries 137}
  (1963) 344--346}.
%%CITATION = ASJOA,137,344;%%.

\bibitem{Davis:1968cp}
R.~Davis, Jr., D.~S. Harmer, and K.~C. Hoffman, ``{Search for neutrinos from
  the sun}'',
\href{http://dx.doi.org/10.1103/PhysRevLett.20.1205}{{\em Phys. Rev. Lett.}
  {\bfseries 20} (1968) 1205--1209}.
%%CITATION = PRLTA,20,1205;%%.

\bibitem{Fukuda:1996sz}
{ Kamiokande} Collaboration, Y.~Fukuda {\em et~al.}, ``{Solar neutrino data
  covering solar cycle 22}'',
\href{http://dx.doi.org/10.1103/PhysRevLett.77.1683}{{\em Phys. Rev. Lett.}
  {\bfseries 77} (1996) 1683--1686}.
%%CITATION = PRLTA,77,1683;%%.

\bibitem{Smy:2003jf}
{ Super-Kamiokande} Collaboration, M.~B. Smy {\em et~al.}, ``{Precise
  measurement of the solar neutrino day / night and seasonal variation in
  Super-Kamiokande-1}'',
  \href{http://dx.doi.org/10.1103/PhysRevD.69.011104}{{\em Phys. Rev.}
  {\bfseries D69} (2004) 011104},
\href{http://arxiv.org/abs/hep-ex/0309011}{{\ttfamily arXiv:hep-ex/0309011
  [hep-ex]}}.
%%CITATION = HEP-EX/0309011;%%.

\bibitem{Ahmad:2001an}
{ SNO} Collaboration, Q.~R. Ahmad {\em et~al.}, ``{Measurement of the rate of
  $\nu_e+d \to p+p+e^-$ interactions produced by $^8B$ solar neutrinos at the
  Sudbury Neutrino Observatory}'',
  \href{http://dx.doi.org/10.1103/PhysRevLett.87.071301}{{\em Phys. Rev. Lett.}
  {\bfseries 87} (2001) 071301},
\href{http://arxiv.org/abs/nucl-ex/0106015}{{\ttfamily arXiv:nucl-ex/0106015
  [nucl-ex]}}.
%%CITATION = NUCL-EX/0106015;%%.

\bibitem{Kaether:2010ag}
F.~Kaether, W.~Hampel, G.~Heusser, J.~Kiko, and T.~Kirsten, ``{Reanalysis of
  the GALLEX solar neutrino flux and source experiments}'',
  \href{http://dx.doi.org/10.1016/j.physletb.2010.01.030}{{\em Phys. Lett.}
  {\bfseries B685} (2010) 47--54},
\href{http://arxiv.org/abs/1001.2731}{{\ttfamily arXiv:1001.2731 [hep-ex]}}.
%%CITATION = ARXIV:1001.2731;%%.

\bibitem{Abdurashitov:2009tn}
{ SAGE} Collaboration, J.~N. Abdurashitov {\em et~al.}, ``{Measurement of the
  solar neutrino capture rate with gallium metal. III: Results for the
  2002--2007 data-taking period}'',
  \href{http://dx.doi.org/10.1103/PhysRevC.80.015807}{{\em Phys. Rev.}
  {\bfseries C80} (2009) 015807},
\href{http://arxiv.org/abs/0901.2200}{{\ttfamily arXiv:0901.2200 [nucl-ex]}}.
%%CITATION = ARXIV:0901.2200;%%.

\bibitem{Cravens:2008aa}
{ Super-Kamiokande} Collaboration, J.~P. Cravens {\em et~al.}, ``{Solar
  neutrino measurements in Super-Kamiokande-II}'',
  \href{http://dx.doi.org/10.1103/PhysRevD.78.032002}{{\em Phys. Rev.}
  {\bfseries D78} (2008) 032002},
\href{http://arxiv.org/abs/0803.4312}{{\ttfamily arXiv:0803.4312 [hep-ex]}}.
%%CITATION = ARXIV:0803.4312;%%.

\bibitem{Abe:2010hy}
{ Super-Kamiokande} Collaboration, K.~Abe {\em et~al.}, ``{Solar neutrino
  results in Super-Kamiokande-III}'',
  \href{http://dx.doi.org/10.1103/PhysRevD.83.052010}{{\em Phys. Rev.}
  {\bfseries D83} (2011) 052010},
\href{http://arxiv.org/abs/1010.0118}{{\ttfamily arXiv:1010.0118 [hep-ex]}}.
%%CITATION = ARXIV:1010.0118;%%.

\bibitem{Nakano:th}
Y.~Nakano, {\em {8B solar neutrino spectrum measurement using Super-Kamiokande
  IV}}.
\newblock PhD thesis, U. Tokyo (main), 2016.
\newblock
\url{http://www-sk.icrr.u-tokyo.ac.jp/sk/_pdf/articles/2016/doc_thesis_naknao.pdf}.
\newblock
%%CITATION = INSPIRE-1667166;%%.

\bibitem{ikeda_motoyasu_2018_1286858}
M.~Ikeda, ``{Solar neutrino measurements with Super-Kamiokande}.'' Talk given
  at the \emph{XXVIII International Conference on Neutrino Physics and
  Astrophysics}, Heidelberg, Germany, June 4--9, 2018.

\bibitem{Bellini:2008mr}
{ Borexino} Collaboration, G.~Bellini {\em et~al.}, ``{Measurement of the solar
  8B neutrino rate with a liquid scintillator target and 3 MeV energy threshold
  in the Borexino detector}'',
  \href{http://dx.doi.org/10.1103/PhysRevD.82.033006}{{\em Phys. Rev.}
  {\bfseries D82} (2010) 033006},
\href{http://arxiv.org/abs/0808.2868}{{\ttfamily arXiv:0808.2868 [astro-ph]}}.
%%CITATION = ARXIV:0808.2868;%%.

\bibitem{Bellini:2014uqa}
{ BOREXINO} Collaboration, G.~Bellini {\em et~al.}, ``{Neutrinos from the
  primary proton–proton fusion process in the Sun}'',
\href{http://dx.doi.org/10.1038/nature13702}{{\em Nature} {\bfseries 512}
  no.~7515, (2014) 383--386}.
%%CITATION = NATUA,512,383;%%.

\bibitem{Aartsen:2014yll}
{ IceCube} Collaboration, M.~G. Aartsen {\em et~al.}, ``{Determining neutrino
  oscillation parameters from atmospheric muon neutrino disappearance with
  three years of IceCube DeepCore data}'',
  \href{http://dx.doi.org/10.1103/PhysRevD.91.072004}{{\em Phys. Rev.}
  {\bfseries D91} no.~7, (2015) 072004},
\href{http://arxiv.org/abs/1410.7227}{{\ttfamily arXiv:1410.7227 [hep-ex]}}.
%%CITATION = ARXIV:1410.7227;%%.

\bibitem{Abe:2017aap}
{ Super-Kamiokande} Collaboration, K.~Abe {\em et~al.}, ``{Atmospheric neutrino
  oscillation analysis with external constraints in Super-Kamiokande I-IV}'',
  \href{http://dx.doi.org/10.1103/PhysRevD.97.072001}{{\em Phys. Rev.}
  {\bfseries D97} no.~7, (2018) 072001},
\href{http://arxiv.org/abs/1710.09126}{{\ttfamily arXiv:1710.09126 [hep-ex]}}.
%%CITATION = ARXIV:1710.09126;%%.

\bibitem{Gando:2013nba}
{ KamLAND} Collaboration, A.~Gando {\em et~al.}, ``{Reactor On-Off Antineutrino
  Measurement with KamLAND}'',
  \href{http://dx.doi.org/10.1103/PhysRevD.88.033001}{{\em Phys. Rev.}
  {\bfseries D88} no.~3, (2013) 033001},
\href{http://arxiv.org/abs/1303.4667}{{\ttfamily arXiv:1303.4667 [hep-ex]}}.
%%CITATION = ARXIV:1303.4667;%%.

\bibitem{An:2016srz}
{ Daya Bay} Collaboration, F.~P. An {\em et~al.}, ``{Improved Measurement of
  the Reactor Antineutrino Flux and Spectrum at Daya Bay}'',
  \href{http://dx.doi.org/10.1088/1674-1137/41/1/013002}{{\em Chin. Phys.}
  {\bfseries C41} no.~1, (2017) 013002},
\href{http://arxiv.org/abs/1607.05378}{{\ttfamily arXiv:1607.05378 [hep-ex]}}.
%%CITATION = ARXIV:1607.05378;%%.

\bibitem{dc:cabrera2016}
A.~Cabrera~Serra, ``{Double Chooz Improved Multi-Detector Measurements}.'' Talk
  given at the \emph{CERN EP colloquium}, CERN, Switzerland, September 20,
  2016.

\bibitem{Adamson:2013ue}
{ MINOS} Collaboration, P.~Adamson {\em et~al.}, ``{Electron neutrino and
  antineutrino appearance in the full MINOS data sample}'',
  \href{http://dx.doi.org/10.1103/PhysRevLett.110.171801}{{\em Phys. Rev.
  Lett.} {\bfseries 110} no.~17, (2013) 171801},
\href{http://arxiv.org/abs/1301.4581}{{\ttfamily arXiv:1301.4581 [hep-ex]}}.
%%CITATION = ARXIV:1301.4581;%%.

\bibitem{t2k:vietnam2016}
A.~Izmaylov, ``{T2K Neutrino Experiment. Recent Results and Plans}.'' Talk
  given at the \emph{Flavour Physics Conference}, Quy Nhon, Vietnam, August
  13--19, 2017.

\bibitem{t2k:kek2019}
M.~Friend, ``{Updated Results from the T2K Experiment with $3.13 \times
  10^{21}$ Protons on Target}.'' KEK seminar, January 10, 2019.

\bibitem{Acero:2019ksn}
{ NOvA} Collaboration, M.~A. Acero {\em et~al.}, ``{First Measurement of
  Neutrino Oscillation Parameters using Neutrinos and Antineutrinos by NOvA}'',
  \href{http://dx.doi.org/10.1103/PhysRevLett.123.151803}{{\em Phys. Rev.
  Lett.} {\bfseries 123} no.~15, (2019) 151803},
\href{http://arxiv.org/abs/1906.04907}{{\ttfamily arXiv:1906.04907 [hep-ex]}}.
%%CITATION = ARXIV:1906.04907;%%.

\bibitem{Weinberg:1979sa}
S.~Weinberg, ``{Baryon and Lepton Nonconserving Processes}'',
\href{http://dx.doi.org/10.1103/PhysRevLett.43.1566}{{\em Phys. Rev. Lett.}
  {\bfseries 43} (1979) 1566--1570}.
%%CITATION = PRLTA,43,1566;%%.

\bibitem{GonzalezGarcia:2002dz}
M.~C. Gonzalez-Garcia and Y.~Nir, ``{Neutrino Masses and Mixing: Evidence and
  Implications}'', \href{http://dx.doi.org/10.1103/RevModPhys.75.345}{{\em Rev.
  Mod. Phys.} {\bfseries 75} (2003) 345--402},
\href{http://arxiv.org/abs/hep-ph/0202058}{{\ttfamily arXiv:hep-ph/0202058
  [hep-ph]}}.
%%CITATION = HEP-PH/0202058;%%.

\bibitem{GonzalezGarcia:2007ib}
M.~C. Gonzalez-Garcia and M.~Maltoni, ``{Phenomenology with Massive
  Neutrinos}'', \href{http://dx.doi.org/10.1016/j.physrep.2007.12.004}{{\em
  Phys. Rept.} {\bfseries 460} (2008) 1--129},
\href{http://arxiv.org/abs/0704.1800}{{\ttfamily arXiv:0704.1800 [hep-ph]}}.
%%CITATION = ARXIV:0704.1800;%%.

\bibitem{Kniehl:1996bd}
B.~A. Kniehl and A.~Pilaftsis, ``{Mixing renormalization in Majorana neutrino
  theories}'', \href{http://dx.doi.org/10.1016/0550-3213(96)00280-5}{{\em Nucl.
  Phys.} {\bfseries B474} (1996) 286--308},
\href{http://arxiv.org/abs/hep-ph/9601390}{{\ttfamily arXiv:hep-ph/9601390
  [hep-ph]}}.
%%CITATION = HEP-PH/9601390;%%.

\bibitem{Ramond:1979py}
P.~Ramond, ``{The Family Group in Grand Unified Theories}'', in {\em
  {International Symposium on Fundamentals of Quantum Theory and Quantum Field
  Theory Palm Coast, Florida, February 25-March 2, 1979}}, pp.~265--280.
\newblock 1979.
\newblock
\href{http://arxiv.org/abs/hep-ph/9809459}{{\ttfamily arXiv:hep-ph/9809459
  [hep-ph]}}.
\newblock
%%CITATION = HEP-PH/9809459;%%.

\bibitem{GellMann:1980vs}
M.~Gell-Mann, P.~Ramond, and R.~Slansky, ``{Complex Spinors and Unified
  Theories}'', {\em Conf. Proc.} {\bfseries C790927} (1979) 315--321,
\href{http://arxiv.org/abs/1306.4669}{{\ttfamily arXiv:1306.4669 [hep-th]}}.
%%CITATION = ARXIV:1306.4669;%%.

\bibitem{Yanagida:1979as}
T.~Yanagida, ``{Horizontal gauge symmetry and masses of neutrinos}'',
{\em Conf. Proc.} {\bfseries C7902131} (1979) 95--99.
%%CITATION = CONFP,C7902131,95;%%.

\bibitem{Mohapatra:1979ia}
R.~N. Mohapatra and G.~Senjanovic, ``{Neutrino Mass and Spontaneous Parity
  Nonconservation}'', \href{http://dx.doi.org/10.1103/PhysRevLett.44.912}{{\em
  Phys. Rev. Lett.} {\bfseries 44} (1980) 912}.
[,231(1979)].
%%CITATION = PRLTA,44,912;%%.

\bibitem{Pontecorvo:1967fh}
B.~Pontecorvo, ``{Neutrino Experiments and the Problem of Conservation of
  Leptonic Charge}'', {\em Sov. Phys. JETP} {\bfseries 26} (1968) 984--988.
[Zh. Eksp. Teor. Fiz.53,1717(1967)].
%%CITATION = SPHJA,26,984;%%.

\bibitem{Maki:1962mu}
Z.~Maki, M.~Nakagawa, and S.~Sakata, ``{Remarks on the unified model of
  elementary particles}'', \href{http://dx.doi.org/10.1143/PTP.28.870}{{\em
  Prog. Theor. Phys.} {\bfseries 28} (1962) 870--880}.
[,34(1962)].
%%CITATION = PTPKA,28,870;%%.

\bibitem{Coloma:2016gei}
P.~Coloma and T.~Schwetz, ``{Generalized mass ordering degeneracy in neutrino
  oscillation experiments}'',
  \href{http://dx.doi.org/10.1103/PhysRevD.95.079903,
  10.1103/PhysRevD.94.055005}{{\em Phys. Rev.} {\bfseries D94} no.~5, (2016)
  055005}, \href{http://arxiv.org/abs/1604.05772}{{\ttfamily arXiv:1604.05772
  [hep-ph]}}.
[Erratum: Phys. Rev.D95,no.7,079903(2017)].
%%CITATION = ARXIV:1604.05772;%%.

\bibitem{Akhmedov:2009rb}
E.~K. Akhmedov and A.~{\relax Yu}. Smirnov, ``{Paradoxes of neutrino
  oscillations}'', \href{http://dx.doi.org/10.1134/S1063778809080122}{{\em
  Phys. Atom. Nucl.} {\bfseries 72} (2009) 1363--1381},
\href{http://arxiv.org/abs/0905.1903}{{\ttfamily arXiv:0905.1903 [hep-ph]}}.
%%CITATION = ARXIV:0905.1903;%%.

\bibitem{Cohen:2008qb}
A.~G. Cohen, S.~L. Glashow, and Z.~Ligeti, ``{Disentangling Neutrino
  Oscillations}'', \href{http://dx.doi.org/10.1016/j.physletb.2009.06.020}{{\em
  Phys. Lett.} {\bfseries B678} (2009) 191--196},
\href{http://arxiv.org/abs/0810.4602}{{\ttfamily arXiv:0810.4602 [hep-ph]}}.
%%CITATION = ARXIV:0810.4602;%%.

\bibitem{Esteban:NuFIT41}
I.~Esteban, M.~C. Gonzalez-Garcia, A.~Hernandez-Cabezudo, M.~Maltoni, and
  T.~Schwetz, ``{Global analysis of three-flavour neutrino oscillations:
  synergies and tensions in the determination of $\theta_{23}$, $\delta_{CP}$,
  and the mass ordering}'',
  \href{http://dx.doi.org/10.1007/JHEP01(2019)106}{{\em JHEP} {\bfseries 01}
  (2019) 106}, \href{http://arxiv.org/abs/1811.05487}{{\ttfamily
  arXiv:1811.05487 [hep-ph]}}.
NuFIT 4.1 (2019), \url{www.nu-fit.org}.
%%CITATION = ARXIV:1811.05487;%%.

\bibitem{Halprin:1986pn}
A.~Halprin, ``{Neutrino Oscillations in Nonuniform Matter}'',
\href{http://dx.doi.org/10.1103/PhysRevD.34.3462}{{\em Phys. Rev.} {\bfseries
  D34} (1986) 3462--3466}.
%%CITATION = PHRVA,D34,3462;%%.

\bibitem{Baltz:1988sv}
A.~J. Baltz and J.~Weneser, ``{Matter Oscillations: Neutrino Transformation in
  the Sun and Regeneration in the Earth}'',
\href{http://dx.doi.org/10.1103/PhysRevD.37.3364}{{\em Phys. Rev.} {\bfseries
  D37} (1988) 3364}.
%%CITATION = PHRVA,D37,3364;%%.

\bibitem{Mannheim:1987ef}
P.~D. Mannheim, ``{Derivation of the Formalism for Neutrino Matter Oscillations
  From the Neutrino Relativistic Field Equations}'',
\href{http://dx.doi.org/10.1103/PhysRevD.37.1935}{{\em Phys. Rev.} {\bfseries
  D37} (1988) 1935}.
%%CITATION = PHRVA,D37,1935;%%.

\bibitem{Kim:1994dy}
C.~W. Kim and A.~Pevsner, {\em {Neutrinos in physics and astrophysics}},
  vol.~8.
\newblock Harwood Academic Press, 1993.

\bibitem{Wolfenstein:1977ue}
L.~Wolfenstein, ``{Neutrino Oscillations in Matter}'',
  \href{http://dx.doi.org/10.1103/PhysRevD.17.2369}{{\em Phys. Rev.} {\bfseries
  D17} (1978) 2369--2374}.
[,294(1977)].
%%CITATION = PHRVA,D17,2369;%%.

\bibitem{Mikheev:1986gs}
S.~P. Mikheyev and A.~{\relax Yu}. Smirnov, ``{Resonance Amplification of
  Oscillations in Matter and Spectroscopy of Solar Neutrinos}'', {\em Sov. J.
  Nucl. Phys.} {\bfseries 42} (1985) 913--917.
[Yad. Fiz.42,1441(1985); ,305(1986)].
%%CITATION = SJNCA,42,913;%%.

\bibitem{GonzalezGarcia:2004wg}
M.~C. Gonzalez-Garcia and M.~Maltoni, ``{Atmospheric neutrino oscillations and
  new physics}'', \href{http://dx.doi.org/10.1103/PhysRevD.70.033010}{{\em
  Phys. Rev.} {\bfseries D70} (2004) 033010},
\href{http://arxiv.org/abs/hep-ph/0404085}{{\ttfamily arXiv:hep-ph/0404085
  [hep-ph]}}.
%%CITATION = HEP-PH/0404085;%%.

\bibitem{Barger:1998xk}
V.~D. Barger, J.~G. Learned, S.~Pakvasa, and T.~J. Weiler, ``{Neutrino decay as
  an explanation of atmospheric neutrino observations}'',
  \href{http://dx.doi.org/10.1103/PhysRevLett.82.2640}{{\em Phys. Rev. Lett.}
  {\bfseries 82} (1999) 2640--2643},
\href{http://arxiv.org/abs/astro-ph/9810121}{{\ttfamily arXiv:astro-ph/9810121
  [astro-ph]}}.
%%CITATION = ASTRO-PH/9810121;%%.

\bibitem{Lisi:2000zt}
E.~Lisi, A.~Marrone, and D.~Montanino, ``{Probing possible decoherence effects
  in atmospheric neutrino oscillations}'',
  \href{http://dx.doi.org/10.1103/PhysRevLett.85.1166}{{\em Phys. Rev. Lett.}
  {\bfseries 85} (2000) 1166--1169},
\href{http://arxiv.org/abs/hep-ph/0002053}{{\ttfamily arXiv:hep-ph/0002053
  [hep-ph]}}.
%%CITATION = HEP-PH/0002053;%%.

\bibitem{Coleman:1997xq}
S.~R. Coleman and S.~L. Glashow, ``{Cosmic ray and neutrino tests of special
  relativity}'', \href{http://dx.doi.org/10.1016/S0370-2693(97)00638-2}{{\em
  Phys. Lett.} {\bfseries B405} (1997) 249--252},
\href{http://arxiv.org/abs/hep-ph/9703240}{{\ttfamily arXiv:hep-ph/9703240
  [hep-ph]}}.
%%CITATION = HEP-PH/9703240;%%.

\bibitem{Glashow:1997gx}
S.~L. Glashow, A.~Halprin, P.~I. Krastev, C.~N. Leung, and J.~T. Pantaleone,
  ``{Comments on neutrino tests of special relativity}'',
  \href{http://dx.doi.org/10.1103/PhysRevD.56.2433}{{\em Phys. Rev.} {\bfseries
  D56} (1997) 2433--2434},
  \href{http://arxiv.org/abs/hep-ph/9703454}{{\ttfamily arXiv:hep-ph/9703454
  [hep-ph]}}.
[,966(1997)].
%%CITATION = HEP-PH/9703454;%%.

\bibitem{GonzalezGarcia:2008ru}
M.~C. Gonzalez-Garcia and M.~Maltoni, ``{Status of Oscillation plus Decay of
  Atmospheric and Long-Baseline Neutrinos}'',
  \href{http://dx.doi.org/10.1016/j.physletb.2008.04.041}{{\em Phys. Lett.}
  {\bfseries B663} (2008) 405--409},
\href{http://arxiv.org/abs/0802.3699}{{\ttfamily arXiv:0802.3699 [hep-ph]}}.
%%CITATION = ARXIV:0802.3699;%%.

\bibitem{Mueller:2011nm}
T.~A. Mueller {\em et~al.}, ``{Improved Predictions of Reactor Antineutrino
  Spectra}'', \href{http://dx.doi.org/10.1103/PhysRevC.83.054615}{{\em Phys.
  Rev.} {\bfseries C83} (2011) 054615},
\href{http://arxiv.org/abs/1101.2663}{{\ttfamily arXiv:1101.2663 [hep-ex]}}.
%%CITATION = ARXIV:1101.2663;%%.

\bibitem{Huber:2011wv}
P.~Huber, ``{On the determination of anti-neutrino spectra from nuclear
  reactors}'', \href{http://dx.doi.org/10.1103/PhysRevC.85.029901,
  10.1103/PhysRevC.84.024617}{{\em Phys. Rev.} {\bfseries C84} (2011) 024617},
  \href{http://arxiv.org/abs/1106.0687}{{\ttfamily arXiv:1106.0687 [hep-ph]}}.
[Erratum: Phys. Rev.C85,029901(2012)].
%%CITATION = ARXIV:1106.0687;%%.

\bibitem{Mention:2011rk}
G.~Mention, M.~Fechner, T.~Lasserre, T.~A. Mueller, D.~Lhuillier, M.~Cribier,
  and A.~Letourneau, ``{The Reactor Antineutrino Anomaly}'',
  \href{http://dx.doi.org/10.1103/PhysRevD.83.073006}{{\em Phys. Rev.}
  {\bfseries D83} (2011) 073006},
\href{http://arxiv.org/abs/1101.2755}{{\ttfamily arXiv:1101.2755 [hep-ex]}}.
%%CITATION = ARXIV:1101.2755;%%.

\bibitem{Hayes:2013wra}
A.~C. Hayes, J.~L. Friar, G.~T. Garvey, G.~Jungman, and G.~Jonkmans,
  ``{Systematic Uncertainties in the Analysis of the Reactor Neutrino
  Anomaly}'', \href{http://dx.doi.org/10.1103/PhysRevLett.112.202501}{{\em
  Phys. Rev. Lett.} {\bfseries 112} (2014) 202501},
\href{http://arxiv.org/abs/1309.4146}{{\ttfamily arXiv:1309.4146 [nucl-th]}}.
%%CITATION = ARXIV:1309.4146;%%.

\bibitem{Fang:2015cma}
D.-L. Fang and B.~A. Brown, ``{Effect of first forbidden decays on the shape of
  neutrino spectra}'', \href{http://dx.doi.org/10.1103/PhysRevC.93.049903,
  10.1103/PhysRevC.91.025503}{{\em Phys. Rev.} {\bfseries C91} no.~2, (2015)
  025503}, \href{http://arxiv.org/abs/1502.02246}{{\ttfamily arXiv:1502.02246
  [nucl-th]}}.
[Erratum: Phys. Rev.C93,no.4,049903(2016)].
%%CITATION = ARXIV:1502.02246;%%.

\bibitem{Hayes:2016qnu}
A.~C. Hayes and P.~Vogel, ``{Reactor Neutrino Spectra}'',
  \href{http://dx.doi.org/10.1146/annurev-nucl-102115-044826}{{\em Ann. Rev.
  Nucl. Part. Sci.} {\bfseries 66} (2016) 219--244},
\href{http://arxiv.org/abs/1605.02047}{{\ttfamily arXiv:1605.02047 [hep-ph]}}.
%%CITATION = ARXIV:1605.02047;%%.

\bibitem{Acero:2007su}
M.~A. Acero, C.~Giunti, and M.~Laveder, ``{Limits on nu(e) and anti-nu(e)
  disappearance from Gallium and reactor experiments}'',
  \href{http://dx.doi.org/10.1103/PhysRevD.78.073009}{{\em Phys. Rev.}
  {\bfseries D78} (2008) 073009},
\href{http://arxiv.org/abs/0711.4222}{{\ttfamily arXiv:0711.4222 [hep-ph]}}.
%%CITATION = ARXIV:0711.4222;%%.

\bibitem{Giunti:2010zu}
C.~Giunti and M.~Laveder, ``{Statistical Significance of the Gallium
  Anomaly}'', \href{http://dx.doi.org/10.1103/PhysRevC.83.065504}{{\em Phys.
  Rev.} {\bfseries C83} (2011) 065504},
\href{http://arxiv.org/abs/1006.3244}{{\ttfamily arXiv:1006.3244 [hep-ph]}}.
%%CITATION = ARXIV:1006.3244;%%.

\bibitem{Aguilar:2001ty}
{ LSND} Collaboration, A.~Aguilar-Arevalo {\em et~al.}, ``{Evidence for
  neutrino oscillations from the observation of anti-neutrino(electron)
  appearance in a anti-neutrino(muon) beam}'',
  \href{http://dx.doi.org/10.1103/PhysRevD.64.112007}{{\em Phys. Rev.}
  {\bfseries D64} (2001) 112007},
\href{http://arxiv.org/abs/hep-ex/0104049}{{\ttfamily arXiv:hep-ex/0104049
  [hep-ex]}}.
%%CITATION = HEP-EX/0104049;%%.

\bibitem{Aguilar-Arevalo:2018gpe}
{ MiniBooNE} Collaboration, A.~A. Aguilar-Arevalo {\em et~al.}, ``{Significant
  Excess of ElectronLike Events in the MiniBooNE Short-Baseline Neutrino
  Experiment}'', \href{http://dx.doi.org/10.1103/PhysRevLett.121.221801}{{\em
  Phys. Rev. Lett.} {\bfseries 121} no.~22, (2018) 221801},
\href{http://arxiv.org/abs/1805.12028}{{\ttfamily arXiv:1805.12028 [hep-ex]}}.
%%CITATION = ARXIV:1805.12028;%%.

\bibitem{Dentler:2018sju}
M.~Dentler, A.~Hernandez-Cabezudo, J.~Kopp, P.~A.~N. Machado, M.~Maltoni,
  I.~Martinez-Soler, and T.~Schwetz, ``{Updated Global Analysis of Neutrino
  Oscillations in the Presence of eV-Scale Sterile Neutrinos}'',
  \href{http://dx.doi.org/10.1007/JHEP08(2018)010}{{\em JHEP} {\bfseries 08}
  (2018) 010},
\href{http://arxiv.org/abs/1803.10661}{{\ttfamily arXiv:1803.10661 [hep-ph]}}.
%%CITATION = ARXIV:1803.10661;%%.

\bibitem{Boser:2019rta}
S.~Böser, C.~Buck, C.~Giunti, J.~Lesgourgues, L.~Ludhova, S.~Mertens,
  A.~Schukraft, and M.~Wurm, ``{Status of Light Sterile Neutrino Searches}'',
\href{http://arxiv.org/abs/1906.01739}{{\ttfamily arXiv:1906.01739 [hep-ex]}}.
%%CITATION = ARXIV:1906.01739;%%.

\bibitem{Serenelli:2011py}
A.~M. Serenelli, W.~C. Haxton, and C.~Pena-Garay, ``{Solar models with
  accretion. I. Application to the solar abundance problem}'',
  \href{http://dx.doi.org/10.1088/0004-637X/743/1/24}{{\em Astrophys. J.}
  {\bfseries 743} (2011) 24},
\href{http://arxiv.org/abs/1104.1639}{{\ttfamily arXiv:1104.1639
  [astro-ph.SR]}}.
%%CITATION = ARXIV:1104.1639;%%.

\bibitem{Villante:2014txa}
F.~L. Villante, ``{ecCNO Solar Neutrinos: A Challenge for Gigantic Ultra-Pure
  Liquid Scintillator Detectors}'',
  \href{http://dx.doi.org/10.1016/j.physletb.2015.01.043}{{\em Phys. Lett.}
  {\bfseries B742} (2015) 279--284},
\href{http://arxiv.org/abs/1410.2796}{{\ttfamily arXiv:1410.2796 [hep-ph]}}.
%%CITATION = ARXIV:1410.2796;%%.

\bibitem{wiki:pp}
{Wikimedia Commons}, ``{Fusion in the Sun}.''
  \url{https://commons.wikimedia.org/w/index.php?title=File:Fusion_in_the_Sun.svg&oldid=345769195},
  2019.
\newblock [Online; accessed 14-February-2020].

\bibitem{wiki:CNO}
{Wikimedia Commons, User:Borb}, ``{CNO Cycle}.''
  \url{https://commons.wikimedia.org/w/index.php?title=File:CNO_Cycle.svg&oldid=272861408},
  2017.
\newblock [Online; accessed 14-February-2020].

\bibitem{Serenelli:2016dgz}
A.~Serenelli, ``{Alive and well: a short review about standard solar models}'',
  \href{http://dx.doi.org/10.1140/epja/i2016-16078-1}{{\em Eur. Phys. J.}
  {\bfseries A52} no.~4, (2016) 78},
\href{http://arxiv.org/abs/1601.07179}{{\ttfamily arXiv:1601.07179
  [astro-ph.SR]}}.
%%CITATION = ARXIV:1601.07179;%%.

\bibitem{Bahcall:2004pz}
J.~N. Bahcall, A.~M. Serenelli, and S.~Basu, ``{New solar opacities,
  abundances, helioseismology, and neutrino fluxes}'',
  \href{http://dx.doi.org/10.1086/428929}{{\em Astrophys. J.} {\bfseries 621}
  (2005) L85--L88},
\href{http://arxiv.org/abs/astro-ph/0412440}{{\ttfamily arXiv:astro-ph/0412440
  [astro-ph]}}.
%%CITATION = ASTRO-PH/0412440;%%.

\bibitem{Fukuda:2002pe}
{ Super-Kamiokande} Collaboration, S.~Fukuda {\em et~al.}, ``{Determination of
  solar neutrino oscillation parameters using 1496 days of Super-Kamiokande I
  data}'', \href{http://dx.doi.org/10.1016/S0370-2693(02)02090-7}{{\em Phys.
  Lett.} {\bfseries B539} (2002) 179--187},
\href{http://arxiv.org/abs/hep-ex/0205075}{{\ttfamily arXiv:hep-ex/0205075
  [hep-ex]}}.
%%CITATION = HEP-EX/0205075;%%.

\bibitem{Maltoni:2015kca}
M.~Maltoni and A.~{\relax Yu}. Smirnov, ``{Solar neutrinos and neutrino
  physics}'', \href{http://dx.doi.org/10.1140/epja/i2016-16087-0}{{\em Eur.
  Phys. J.} {\bfseries A52} no.~4, (2016) 87},
\href{http://arxiv.org/abs/1507.05287}{{\ttfamily arXiv:1507.05287 [hep-ph]}}.
%%CITATION = ARXIV:1507.05287;%%.

\bibitem{Honda:2015fha}
M.~Honda, M.~Sajjad~Athar, T.~Kajita, K.~Kasahara, and S.~Midorikawa,
  ``{Atmospheric neutrino flux calculation using the NRLMSISE-00 atmospheric
  model}'', \href{http://dx.doi.org/10.1103/PhysRevD.92.023004}{{\em Phys.
  Rev.} {\bfseries D92} no.~2, (2015) 023004},
\href{http://arxiv.org/abs/1502.03916}{{\ttfamily arXiv:1502.03916
  [astro-ph.HE]}}.
%%CITATION = ARXIV:1502.03916;%%.

\bibitem{Achar:1965ova}
C.~V. Achar {\em et~al.}, ``{Detection of muons produced by cosmic ray
  neutrinos deep underground}'',
\href{http://dx.doi.org/10.1016/0031-9163(65)90712-2}{{\em Phys. Lett.}
  {\bfseries 18} (1965) 196--199}.
%%CITATION = PHLTA,18,196;%%.

\bibitem{Reines:1965qk}
F.~Reines, M.~F. Crouch, T.~L. Jenkins, W.~R. Kropp, H.~S. Gurr, G.~R. Smith,
  J.~P.~F. Sellschop, and B.~Meyer, ``{Evidence for high-energy cosmic ray
  neutrino interactions}'',
\href{http://dx.doi.org/10.1103/PhysRevLett.15.429}{{\em Phys. Rev. Lett.}
  {\bfseries 15} (1965) 429--433}.
%%CITATION = PRLTA,15,429;%%.

\bibitem{Casper:1990ac}
D.~Casper {\em et~al.}, ``{Measurement of atmospheric neutrino composition with
  IMB-3}'',
\href{http://dx.doi.org/10.1103/PhysRevLett.66.2561}{{\em Phys. Rev. Lett.}
  {\bfseries 66} (1991) 2561--2564}.
%%CITATION = PRLTA,66,2561;%%.

\bibitem{Daum:1994bf}
{ Frejus} Collaboration, K.~Daum {\em et~al.}, ``{Determination of the
  atmospheric neutrino spectra with the Frejus detector}'',
\href{http://dx.doi.org/10.1007/BF01556368}{{\em Z. Phys.} {\bfseries C66}
  (1995) 417--428}.
%%CITATION = ZEPYA,C66,417;%%.

\bibitem{Aglietta:1988be}
{ NUSEX} Collaboration, M.~Aglietta {\em et~al.}, ``{Experimental study of
  atmospheric neutrino flux in the NUSEX experiment}'',
\href{http://dx.doi.org/10.1209/0295-5075/8/7/005}{{\em Europhys. Lett.}
  {\bfseries 8} (1989) 611--614}.
%%CITATION = EULEE,8,611;%%.

\bibitem{An:2015nua}
{ Daya Bay} Collaboration, F.~P. An {\em et~al.}, ``{Measurement of the Reactor
  Antineutrino Flux and Spectrum at Daya Bay}'',
  \href{http://dx.doi.org/10.1103/PhysRevLett.116.061801,
  10.1103/PhysRevLett.118.099902}{{\em Phys. Rev. Lett.} {\bfseries 116} no.~6,
  (2016) 061801}, \href{http://arxiv.org/abs/1508.04233}{{\ttfamily
  arXiv:1508.04233 [hep-ex]}}.
[Erratum: Phys. Rev. Lett.118,no.9,099902(2017)].
%%CITATION = ARXIV:1508.04233;%%.

\bibitem{Apollonio:2002gd}
{ CHOOZ} Collaboration, M.~Apollonio {\em et~al.}, ``{Search for neutrino
  oscillations on a long baseline at the CHOOZ nuclear power station}'',
  \href{http://dx.doi.org/10.1140/epjc/s2002-01127-9}{{\em Eur. Phys. J.}
  {\bfseries C27} (2003) 331--374},
\href{http://arxiv.org/abs/hep-ex/0301017}{{\ttfamily arXiv:hep-ex/0301017
  [hep-ex]}}.
%%CITATION = HEP-EX/0301017;%%.

\bibitem{Boehm:2001ik}
F.~Boehm {\em et~al.}, ``{Final results from the Palo Verde neutrino
  oscillation experiment}'',
  \href{http://dx.doi.org/10.1103/PhysRevD.64.112001}{{\em Phys. Rev.}
  {\bfseries D64} (2001) 112001},
\href{http://arxiv.org/abs/hep-ex/0107009}{{\ttfamily arXiv:hep-ex/0107009
  [hep-ex]}}.
%%CITATION = HEP-EX/0107009;%%.

\bibitem{Pontecorvo:1959sn}
B.~Pontecorvo, ``{Electron and Muon Neutrinos}'', {\em Sov. Phys. JETP}
  {\bfseries 10} (1960) 1236--1240.
[Zh. Eksp. Teor. Fiz.37,1751(1959)].
%%CITATION = SPHJA,10,1236;%%.

\bibitem{Schwartz:1960hg}
M.~Schwartz, ``{Feasibility of using high-energy neutrinos to study the weak
  interactions}'',
\href{http://dx.doi.org/10.1103/PhysRevLett.4.306}{{\em Phys. Rev. Lett.}
  {\bfseries 4} (1960) 306--307}.
%%CITATION = PRLTA,4,306;%%.

\bibitem{Dore:2018ldz}
U.~Dore, P.~Loverre, and L.~Ludovici, ``{History of accelerator neutrino
  beams}'', \href{http://dx.doi.org/10.1140/epjh/e2019-90032-x}{{\em Eur. Phys.
  J.} {\bfseries H44} no.~4-5, (2019) 271--305},
\href{http://arxiv.org/abs/1805.01373}{{\ttfamily arXiv:1805.01373
  [physics.acc-ph]}}.
%%CITATION = ARXIV:1805.01373;%%.

\bibitem{Abe:2012av}
{ T2K} Collaboration, K.~Abe {\em et~al.}, ``{T2K neutrino flux prediction}'',
  \href{http://dx.doi.org/10.1103/PhysRevD.87.012001,
  10.1103/PhysRevD.87.019902}{{\em Phys. Rev.} {\bfseries D87} no.~1, (2013)
  012001}, \href{http://arxiv.org/abs/1211.0469}{{\ttfamily arXiv:1211.0469
  [hep-ex]}}.
[Addendum: Phys. Rev.D87,no.1,019902(2013)].
%%CITATION = ARXIV:1211.0469;%%.

\bibitem{Aliu:2004sq}
{ K2K} Collaboration, E.~Aliu {\em et~al.}, ``{Evidence for muon neutrino
  oscillation in an accelerator-based experiment}'',
  \href{http://dx.doi.org/10.1103/PhysRevLett.94.081802}{{\em Phys. Rev. Lett.}
  {\bfseries 94} (2005) 081802},
\href{http://arxiv.org/abs/hep-ex/0411038}{{\ttfamily arXiv:hep-ex/0411038
  [hep-ex]}}.
%%CITATION = HEP-EX/0411038;%%.

\bibitem{Michael:2006rx}
{ MINOS} Collaboration, D.~G. Michael {\em et~al.}, ``{Observation of muon
  neutrino disappearance with the MINOS detectors and the NuMI neutrino
  beam}'', \href{http://dx.doi.org/10.1103/PhysRevLett.97.191801}{{\em Phys.
  Rev. Lett.} {\bfseries 97} (2006) 191801},
\href{http://arxiv.org/abs/hep-ex/0607088}{{\ttfamily arXiv:hep-ex/0607088
  [hep-ex]}}.
%%CITATION = HEP-EX/0607088;%%.

\bibitem{Wilks:1938dza}
S.~S. Wilks, ``{The Large-Sample Distribution of the Likelihood Ratio for
  Testing Composite Hypotheses}'',
\href{http://dx.doi.org/10.1214/aoms/1177732360}{{\em Annals Math. Statist.}
  {\bfseries 9} no.~1, (1938) 60--62}.
%%CITATION = AASTA,9,60;%%.

\bibitem{sksol:nakano2016}
Y.~Nakano, {\em {$^8$B solar neutrino spectrum measurement using
  Super-Kamiokande IV}}.
\newblock PhD thesis, Tokyo U., 2016-02.
\newblock
  \url{http://www-sk.icrr.u-tokyo.ac.jp/sk/_pdf/articles/2016/doc_thesis_naknao.pdf}.

\bibitem{sksol:ichep2016}
Y.~Nakano, ``{Solar neutrino results from Super-Kamiokande}.'' Talk given at
  the {\it 38th International Conference on High Energy Physics}, Chicago, USA,
  August 3--10, 2016.

\bibitem{Gando:2010aa}
{ KamLAND} Collaboration, A.~Gando {\em et~al.}, ``{Constraints on
  $\theta_{13}$ from A Three-Flavor Oscillation Analysis of Reactor
  Antineutrinos at KamLAND}'',
  \href{http://dx.doi.org/10.1103/PhysRevD.83.052002}{{\em Phys. Rev.}
  {\bfseries D83} (2011) 052002},
\href{http://arxiv.org/abs/1009.4771}{{\ttfamily arXiv:1009.4771 [hep-ex]}}.
%%CITATION = ARXIV:1009.4771;%%.

\bibitem{Apollonio:1999ae}
{ CHOOZ} Collaboration, M.~Apollonio {\em et~al.}, ``{Limits on Neutrino
  Oscillations from the CHOOZ Experiment}'',
  \href{http://dx.doi.org/10.1016/S0370-2693(99)01072-2}{{\em Phys. Lett.}
  {\bfseries B466} (1999) 415--430},
\href{http://arxiv.org/abs/hep-ex/9907037}{{\ttfamily arXiv:hep-ex/9907037}}.
%%CITATION = HEP-EX/9907037;%%.

\bibitem{Piepke:2002ju}
{ Palo Verde} Collaboration, A.~Piepke, ``{Final results from the Palo Verde
  neutrino oscillation experiment}'',
\href{http://dx.doi.org/10.1016/S0146-6410(02)00117-5}{{\em Prog. Part. Nucl.
  Phys.} {\bfseries 48} (2002) 113--121}.
%%CITATION = PPNPD,48,113;%%.

\bibitem{dc:moriond2016}
M.~Ishitsuka, ``{New results of Double Chooz}.'' {Talk given at the Conference
  {\it Rencontres de Moriond EW 2016}, La Thuile, Italy, March 12--19, 2016}.

\bibitem{db:nu2016}
Z.~Yu, ``{Recent Results from the Daya Bay Experiment}.'' Talk given at the
  {\it XXVII International Conference on Neutrino Physics and Astrophysics},
  London, UK, July 4--9, 2016.

\bibitem{reno:nu2014}
S.-H. Seo, ``{New Results from RENO}.'' Talk given at the {\it XXVI
  International Conference on Neutrino Physics and Astrophysics}, Boston, USA,
  June 2--7, 2014.

\bibitem{Kopp:2013vaa}
J.~Kopp, P.~A.~N. Machado, M.~Maltoni, and T.~Schwetz, ``{Sterile Neutrino
  Oscillations: The Global Picture}'',
  \href{http://dx.doi.org/10.1007/JHEP05(2013)050}{{\em JHEP} {\bfseries 1305}
  (2013) 050},
\href{http://arxiv.org/abs/1303.3011}{{\ttfamily arXiv:1303.3011 [hep-ph]}}.
%%CITATION = ARXIV:1303.3011;%%.

\bibitem{Kwon:1981ua}
H.~Kwon, F.~Boehm, A.~Hahn, H.~Henrikson, J.~Vuilleumier, {\em et~al.},
  ``Search for neutrino oscillations at a fission reactor'',
\href{http://dx.doi.org/10.1103/PhysRevD.24.1097}{{\em Phys.Rev.} {\bfseries
  D24} (1981) 1097--1111}.
%%CITATION = PHRVA,D24,1097;%%.

\bibitem{Zacek:1986cu}
{ CALTECH-SIN-TUM} Collaboration, G.~Zacek {\em et~al.}, ``{Neutrino
  Oscillation Experiments at the Gosgen Nuclear Power Reactor}'',
\href{http://dx.doi.org/10.1103/PhysRevD.34.2621}{{\em Phys.Rev.} {\bfseries
  D34} (1986) 2621--2636}.
%%CITATION = PHRVA,D34,2621;%%.

\bibitem{Vidyakin:1987ue}
G.~Vidyakin, V.~Vyrodov, I.~Gurevich, Y.~Kozlov, V.~Martemyanov, {\em et~al.},
  ``Detection of anti-neutrinos in the flux from two reactors'',
{\em Sov.Phys.JETP} {\bfseries 66} (1987) 243--247.
%%CITATION = SPHJA,66,243;%%.

\bibitem{Vidyakin:1994ut}
G.~Vidyakin, V.~Vyrodov, Y.~Kozlov, A.~Martemyanov, V.~Martemyanov, {\em
  et~al.}, ``{Limitations on the characteristics of neutrino oscillations}'',
{\em JETP Lett.} {\bfseries 59} (1994) 390--393.
%%CITATION = JTPLA,59,390;%%.

\bibitem{Afonin:1988gx}
A.~Afonin, S.~Ketov, V.~Kopeikin, L.~Mikaelyan, M.~Skorokhvatov, {\em et~al.},
  ``{A study of the reaction $\bar\nu_e + p \to e^+ + n$ on a nuclear
  reactor}'',
{\em Sov.Phys.JETP} {\bfseries 67} (1988) 213--221.
%%CITATION = SPHJA,67,213;%%.

\bibitem{Kuvshinnikov:1990ry}
A.~Kuvshinnikov, L.~Mikaelyan, S.~Nikolaev, M.~Skorokhvatov, and A.~Etenko,
  ``{Measuring the $\bar\nu_e + p \to n + e^+$ cross-section and beta decay
  axial constant in a new experiment at Rovno NPP reactor. (In Russian)}'',
{\em JETP Lett.} {\bfseries 54} (1991) 253--257.
%%CITATION = JTPLA,54,253;%%.

\bibitem{Declais:1994su}
Y.~Declais, J.~Favier, A.~Metref, H.~Pessard, B.~Achkar, {\em et~al.},
  ``{Search for neutrino oscillations at 15-meters, 40-meters, and 95-meters
  from a nuclear power reactor at Bugey}'',
\href{http://dx.doi.org/10.1016/0550-3213(94)00513-E}{{\em Nucl.Phys.}
  {\bfseries B434} (1995) 503--534}.
%%CITATION = NUPHA,B434,503;%%.

\bibitem{Declais:1994ma}
Y.~Declais, H.~de~Kerret, B.~Lefievre, M.~Obolensky, A.~Etenko, {\em et~al.},
  ``{Study of reactor anti-neutrino interaction with proton at Bugey nuclear
  power plant}'',
\href{http://dx.doi.org/10.1016/0370-2693(94)91394-3}{{\em Phys.Lett.}
  {\bfseries B338} (1994) 383--389}.
%%CITATION = PHLTA,B338,383;%%.

\bibitem{Greenwood:1996pb}
Z.~D. Greenwood {\em et~al.}, ``{Results of a two position reactor neutrino
  oscillation experiment}'',
\href{http://dx.doi.org/10.1103/PhysRevD.53.6054}{{\em Phys. Rev.} {\bfseries
  D53} (1996) 6054--6064}.
%%CITATION = PHRVA,D53,6054;%%.

\bibitem{nufit}
NuFIT webpage, \href{http://www.nu-fit.org}{\tt http://www.nu-fit.org}.

\bibitem{Gonzalez-Garcia:2014bfa}
M.~C. Gonzalez-Garcia, M.~Maltoni, and T.~Schwetz, ``{Updated fit to three
  neutrino mixing: status of leptonic CP violation}'',
  \href{http://dx.doi.org/10.1007/JHEP11(2014)052}{{\em JHEP} {\bfseries 11}
  (2014) 052},
\href{http://arxiv.org/abs/1409.5439}{{\ttfamily arXiv:1409.5439 [hep-ph]}}.
%%CITATION = ARXIV:1409.5439;%%.

\bibitem{GonzalezGarcia:2012sz}
M.~Gonzalez-Garcia, M.~Maltoni, J.~Salvado, and T.~Schwetz, ``{Global fit to
  three neutrino mixing: critical look at present precision}'',
  \href{http://dx.doi.org/10.1007/JHEP12(2012)123}{{\em JHEP} {\bfseries 1212}
  (2012) 123},
\href{http://arxiv.org/abs/1209.3023}{{\ttfamily arXiv:1209.3023 [hep-ph]}}.
%%CITATION = ARXIV:1209.3023;%%.

\bibitem{Schwetz:2006md}
T.~Schwetz, ``{What is the probability that theta(13) and CP violation will be
  discovered in future neutrino oscillation experiments?}'',
  \href{http://dx.doi.org/10.1016/j.physletb.2007.02.053}{{\em Phys. Lett.}
  {\bfseries B648} (2007) 54--59},
\href{http://arxiv.org/abs/hep-ph/0612223}{{\ttfamily arXiv:hep-ph/0612223
  [hep-ph]}}.
%%CITATION = HEP-PH/0612223;%%.

\bibitem{Blennow:2014sja}
M.~Blennow, P.~Coloma, and E.~Fernandez-Martinez, ``{Reassessing the
  sensitivity to leptonic CP violation}'',
  \href{http://dx.doi.org/10.1007/JHEP03(2015)005}{{\em JHEP} {\bfseries 03}
  (2015) 005},
\href{http://arxiv.org/abs/1407.3274}{{\ttfamily arXiv:1407.3274 [hep-ph]}}.
%%CITATION = ARXIV:1407.3274;%%.

\bibitem{GonzalezGarcia:2003qf}
M.~C. Gonzalez-Garcia and C.~Pena-Garay, ``{Three neutrino mixing after the
  first results from K2K and KamLAND}'',
  \href{http://dx.doi.org/10.1103/PhysRevD.68.093003}{{\em Phys. Rev.}
  {\bfseries D68} (2003) 093003},
\href{http://arxiv.org/abs/hep-ph/0306001}{{\ttfamily arXiv:hep-ph/0306001
  [hep-ph]}}.
%%CITATION = HEP-PH/0306001;%%.

\bibitem{PDB2016}
{ Particle Data Group} Collaboration, C.~Patrignani {\em et~al.}, ``{Review of
  Particle Physics}'',
\href{http://dx.doi.org/10.1088/1674-1137/40/10/100001}{{\em Chin. Phys.}
  {\bfseries C40} no.~10, (2016) 100001}.
%%CITATION = CHPHD,C40,100001;%%.

\bibitem{Bergstrom:2016cbh}
J.~Bergstrom, M.~C. Gonzalez-Garcia, M.~Maltoni, C.~Pena-Garay, A.~M.
  Serenelli, and N.~Song, ``{Updated determination of the solar neutrino fluxes
  from solar neutrino data}'',
  \href{http://dx.doi.org/10.1007/JHEP03(2016)132}{{\em JHEP} {\bfseries 03}
  (2016) 132},
\href{http://arxiv.org/abs/1601.00972}{{\ttfamily arXiv:1601.00972 [hep-ph]}}.
%%CITATION = ARXIV:1601.00972;%%.

\bibitem{Vinyoles:2016djt}
N.~Vinyoles, A.~M. Serenelli, F.~L. Villante, S.~Basu, J.~Bergström, M.~C.
  Gonzalez-Garcia, M.~Maltoni, C.~Peña-Garay, and N.~Song, ``{A new Generation
  of Standard Solar Models}'',
  \href{http://dx.doi.org/10.3847/1538-4357/835/2/202}{{\em Astrophys. J.}
  {\bfseries 835} no.~2, (2017) 202},
\href{http://arxiv.org/abs/1611.09867}{{\ttfamily arXiv:1611.09867
  [astro-ph.SR]}}.
%%CITATION = ARXIV:1611.09867;%%.

\bibitem{Bezerra:2012at}
T.~J.~C. Bezerra, H.~Furuta, and F.~Suekane, ``{Measurement of Effective
  $\Delta m_{31}^2$ using Baseline Differences of Daya Bay, RENO and Double
  Chooz Reactor Neutrino Experiments}'',
\href{http://arxiv.org/abs/1206.6017}{{\ttfamily arXiv:1206.6017 [hep-ex]}}.
%%CITATION = ARXIV:1206.6017;%%.

\bibitem{Seo:2016uom}
H.~Seo {\em et~al.}, ``{Spectral Measurement of the Electron Antineutrino
  Oscillation Amplitude and Frequency using 500 Live Days of RENO Data}'',
\href{http://arxiv.org/abs/1610.04326}{{\ttfamily arXiv:1610.04326 [hep-ex]}}.
%%CITATION = ARXIV:1610.04326;%%.

\bibitem{Wendell:2014dka}
{ Super-Kamiokande} Collaboration, R.~Wendell, ``{Atmospheric Results from
  Super-Kamiokande}'', \href{http://dx.doi.org/10.1063/1.4915569}{{\em AIP
  Conf. Proc.} {\bfseries 1666} (2015) 100001},
\href{http://arxiv.org/abs/1412.5234}{{\ttfamily arXiv:1412.5234 [hep-ex]}}.
%%CITATION = ARXIV:1412.5234;%%.

\bibitem{skatm:nufact2016}
J.~Kameda, ``{Recent results from Super-Kamokande on atmospheric neutrinos and
  next project: Hyper‐Kamioande}.'' Talk given at the {\it XII Rencontres de
  Vietnam: NuFact 2016}, Qui Nhon, Vietnam, August 21--27, 2016.

\bibitem{skatm:thesis}
K.~P. Lee, ``{Study of the neutrino mass hierarchy with the atmospheric
  neutrino data observed in SuperKamiokande}.'' {Ph.D. thesis, The University
  of Tokyo}, 2012.

\bibitem{Elevant:2015ska}
J.~Elevant and T.~Schwetz, ``{On the determination of the leptonic CP phase}'',
  \href{http://dx.doi.org/10.1007/JHEP09(2015)016}{{\em JHEP} {\bfseries 09}
  (2015) 016},
\href{http://arxiv.org/abs/1506.07685}{{\ttfamily arXiv:1506.07685 [hep-ph]}}.
%%CITATION = ARXIV:1506.07685;%%.

\bibitem{Blennow:2013oma}
M.~Blennow, P.~Coloma, P.~Huber, and T.~Schwetz, ``{Quantifying the sensitivity
  of oscillation experiments to the neutrino mass ordering}'',
  \href{http://dx.doi.org/10.1007/JHEP03(2014)028}{{\em JHEP} {\bfseries 03}
  (2014) 028},
\href{http://arxiv.org/abs/1311.1822}{{\ttfamily arXiv:1311.1822 [hep-ph]}}.
%%CITATION = ARXIV:1311.1822;%%.

\bibitem{Abe:2018wpn}
{ T2K} Collaboration, K.~Abe {\em et~al.}, ``{Search for CP Violation in
  Neutrino and Antineutrino Oscillations by the T2K Experiment with
  $2.2\times10^{21}$ Protons on Target}'',
  \href{http://dx.doi.org/10.1103/PhysRevLett.121.171802}{{\em Phys. Rev.
  Lett.} {\bfseries 121} no.~17, (2018) 171802},
\href{http://arxiv.org/abs/1807.07891}{{\ttfamily arXiv:1807.07891 [hep-ex]}}.
%%CITATION = ARXIV:1807.07891;%%.

\bibitem{reno:eps2017}
H.~Seo, ``{New Results from RENO}.''. Talk given at the {\it EPS Conference on
  High Energy Physics}, Venice, Italy, July 5--12, 2017.

\bibitem{Cervera:2000kp}
A.~Cervera, A.~Donini, M.~B. Gavela, J.~J. Gomez~Cadenas, P.~Hernandez,
  O.~Mena, and S.~Rigolin, ``{Golden measurements at a neutrino factory}'',
  \href{http://dx.doi.org/10.1016/S0550-3213(00)00606-4,
  10.1016/S0550-3213(00)00221-2}{{\em Nucl. Phys.} {\bfseries B579} (2000)
  17--55}, \href{http://arxiv.org/abs/hep-ph/0002108}{{\ttfamily
  arXiv:hep-ph/0002108 [hep-ph]}}.
[Erratum: Nucl. Phys.B593,731(2001)].
%%CITATION = HEP-PH/0002108;%%.

\bibitem{Freund:2001pn}
M.~Freund, ``{Analytic approximations for three neutrino oscillation parameters
  and probabilities in matter}'',
  \href{http://dx.doi.org/10.1103/PhysRevD.64.053003}{{\em Phys. Rev.}
  {\bfseries D64} (2001) 053003},
\href{http://arxiv.org/abs/hep-ph/0103300}{{\ttfamily arXiv:hep-ph/0103300
  [hep-ph]}}.
%%CITATION = HEP-PH/0103300;%%.

\bibitem{Akhmedov:2004ny}
E.~K. Akhmedov, R.~Johansson, M.~Lindner, T.~Ohlsson, and T.~Schwetz, ``{Series
  expansions for three flavor neutrino oscillation probabilities in matter}'',
  \href{http://dx.doi.org/10.1088/1126-6708/2004/04/078}{{\em JHEP} {\bfseries
  04} (2004) 078},
\href{http://arxiv.org/abs/hep-ph/0402175}{{\ttfamily arXiv:hep-ph/0402175
  [hep-ph]}}.
%%CITATION = HEP-PH/0402175;%%.

\bibitem{Abe:2017uxa}
{ T2K} Collaboration, K.~Abe {\em et~al.}, ``{Combined Analysis of Neutrino and
  Antineutrino Oscillations at T2K}'',
  \href{http://dx.doi.org/10.1103/PhysRevLett.118.151801}{{\em Phys. Rev.
  Lett.} {\bfseries 118} no.~15, (2017) 151801},
\href{http://arxiv.org/abs/1701.00432}{{\ttfamily arXiv:1701.00432 [hep-ex]}}.
%%CITATION = ARXIV:1701.00432;%%.

\bibitem{NOvA:2018gge}
{ NOvA} Collaboration, M.~A. Acero {\em et~al.}, ``{New constraints on
  oscillation parameters from $\nu_e$ appearance and $\nu_\mu$ disappearance in
  the NOvA experiment}'',
  \href{http://dx.doi.org/10.1103/PhysRevD.98.032012}{{\em Phys. Rev.}
  {\bfseries D98} (2018) 032012},
\href{http://arxiv.org/abs/1806.00096}{{\ttfamily arXiv:1806.00096 [hep-ex]}}.
%%CITATION = ARXIV:1806.00096;%%.

\bibitem{t2k:nu2018}
M.~Wascko, ``{T2K Status, Results, and Plans}.'' Talk given at the {\it XXVIII
  International Conference on Neutrino Physics and Astrophysics}, Heidelberg,
  Germany, June 4--9, 2018.

\bibitem{nova:nu2018}
M.~Sanchez, ``{NOvA Results and Prospects}.'' Talk given at the {\it XXVIII
  International Conference on Neutrino Physics and Astrophysics}, Heidelberg,
  Germany, June 4--9, 2018.

\bibitem{nova:nuphys2019}
J.~Hewes, ``{Neutrino Oscillation Results from NOvA}.'' Talk given at {\it
  NuPhys2019: Prospects in Neutrino Physics}, London, UK, December 16--18,
  2019.

\bibitem{Abi:2020wmh}
{ DUNE} Collaboration, S.~Jones {\em et~al.}, ``{Deep Underground Neutrino
  Experiment (DUNE), Far Detector Technical Design Report, Volume 1
  Introduction to DUNE}'',
\href{http://arxiv.org/abs/2002.02967}{{\ttfamily arXiv:2002.02967
  [physics.ins-det]}}.
%%CITATION = ARXIV:2002.02967;%%.

\bibitem{Abi:2020evt}
{ DUNE} Collaboration, S.~Jones {\em et~al.}, ``{Deep Underground Neutrino
  Experiment (DUNE), Far Detector Technical Design Report, Volume II DUNE
  Physics}'',
\href{http://arxiv.org/abs/2002.03005}{{\ttfamily arXiv:2002.03005 [hep-ex]}}.
%%CITATION = ARXIV:2002.03005;%%.

\bibitem{Abi:2020oxb}
{ DUNE} Collaboration, S.~Jones {\em et~al.}, ``{Deep Underground Neutrino
  Experiment (DUNE), Far Detector Technical Design Report, Volume III DUNE Far
  Detector Technical Coordination}'',
\href{http://arxiv.org/abs/2002.03008}{{\ttfamily arXiv:2002.03008
  [physics.ins-det]}}.
%%CITATION = ARXIV:2002.03008;%%.

\bibitem{Abi:2020loh}
{ DUNE} Collaboration, S.~Jones {\em et~al.}, ``{Deep Underground Neutrino
  Experiment (DUNE), Far Detector Technical Design Report, Volume IV Far
  Detector Single-phase Technology}'',
\href{http://arxiv.org/abs/2002.03010}{{\ttfamily arXiv:2002.03010
  [physics.ins-det]}}.
%%CITATION = ARXIV:2002.03010;%%.

\bibitem{Abe:2018uyc}
{ Hyper-Kamiokande} Collaboration, K.~Abe {\em et~al.}, ``{Hyper-Kamiokande
  Design Report}'',
\href{http://arxiv.org/abs/1805.04163}{{\ttfamily arXiv:1805.04163
  [physics.ins-det]}}.
%%CITATION = ARXIV:1805.04163;%%.

\bibitem{Broncano:2002rw}
A.~Broncano, M.~B. Gavela, and E.~E. Jenkins, ``{The Effective Lagrangian for
  the seesaw model of neutrino mass and leptogenesis}'',
  \href{http://dx.doi.org/10.1016/j.physletb.2006.04.003,
  10.1016/S0370-2693(02)03130-1}{{\em Phys. Lett.} {\bfseries B552} (2003)
  177--184}, \href{http://arxiv.org/abs/hep-ph/0210271}{{\ttfamily
  arXiv:hep-ph/0210271 [hep-ph]}}.
[Erratum: Phys. Lett.B636,332(2006)].
%%CITATION = HEP-PH/0210271;%%.

\bibitem{Broncano:2003fq}
A.~Broncano, M.~B. Gavela, and E.~E. Jenkins, ``{Neutrino physics in the seesaw
  model}'', \href{http://dx.doi.org/10.1016/j.nuclphysb.2003.09.011}{{\em Nucl.
  Phys.} {\bfseries B672} (2003) 163--198},
\href{http://arxiv.org/abs/hep-ph/0307058}{{\ttfamily arXiv:hep-ph/0307058
  [hep-ph]}}.
%%CITATION = HEP-PH/0307058;%%.

\bibitem{Abada:2007ux}
A.~Abada, C.~Biggio, F.~Bonnet, M.~B. Gavela, and T.~Hambye, ``{Low energy
  effects of neutrino masses}'',
  \href{http://dx.doi.org/10.1088/1126-6708/2007/12/061}{{\em JHEP} {\bfseries
  12} (2007) 061},
\href{http://arxiv.org/abs/0707.4058}{{\ttfamily arXiv:0707.4058 [hep-ph]}}.
%%CITATION = ARXIV:0707.4058;%%.

\bibitem{Fernandez-Martinez:2016lgt}
E.~Fernandez-Martinez, J.~Hernandez-Garcia, and J.~Lopez-Pavon, ``{Global
  constraints on heavy neutrino mixing}'',
  \href{http://dx.doi.org/10.1007/JHEP08(2016)033}{{\em JHEP} {\bfseries 08}
  (2016) 033},
\href{http://arxiv.org/abs/1605.08774}{{\ttfamily arXiv:1605.08774 [hep-ph]}}.
%%CITATION = ARXIV:1605.08774;%%.

\bibitem{Blennow:2016jkn}
M.~Blennow, P.~Coloma, E.~Fernandez-Martinez, J.~Hernandez-Garcia, and
  J.~Lopez-Pavon, ``{Non-Unitarity, sterile neutrinos, and Non-Standard
  neutrino Interactions}'',
  \href{http://dx.doi.org/10.1007/JHEP04(2017)153}{{\em JHEP} {\bfseries 04}
  (2017) 153},
\href{http://arxiv.org/abs/1609.08637}{{\ttfamily arXiv:1609.08637 [hep-ph]}}.
%%CITATION = ARXIV:1609.08637;%%.

\bibitem{Valle:1987gv}
J.~W.~F. Valle, ``{Resonant Oscillations of Massless Neutrinos in Matter}'',
\href{http://dx.doi.org/10.1016/0370-2693(87)90947-6}{{\em Phys. Lett.}
  {\bfseries B199} (1987) 432--436}.
%%CITATION = PHLTA,B199,432;%%.

\bibitem{Guzzo:1991hi}
M.~M. Guzzo, A.~Masiero, and S.~T. Petcov, ``{On the MSW effect with massless
  neutrinos and no mixing in the vacuum}'',
  \href{http://dx.doi.org/10.1016/0370-2693(91)90984-X}{{\em Phys. Lett.}
  {\bfseries B260} (1991) 154--160}.
[,369(1991)].
%%CITATION = PHLTA,B260,154;%%.

\bibitem{Ohlsson:2012kf}
T.~Ohlsson, ``{Status of non-standard neutrino interactions}'',
  \href{http://dx.doi.org/10.1088/0034-4885/76/4/044201}{{\em Rept. Prog.
  Phys.} {\bfseries 76} (2013) 044201},
\href{http://arxiv.org/abs/1209.2710}{{\ttfamily arXiv:1209.2710 [hep-ph]}}.
%%CITATION = ARXIV:1209.2710;%%.

\bibitem{Gavela:2008ra}
M.~B. Gavela, D.~Hernandez, T.~Ota, and W.~Winter, ``{Large gauge invariant
  non-standard neutrino interactions}'',
  \href{http://dx.doi.org/10.1103/PhysRevD.79.013007}{{\em Phys. Rev.}
  {\bfseries D79} (2009) 013007},
\href{http://arxiv.org/abs/0809.3451}{{\ttfamily arXiv:0809.3451 [hep-ph]}}.
%%CITATION = ARXIV:0809.3451;%%.

\bibitem{Davidson:2003ha}
S.~Davidson, C.~Pena-Garay, N.~Rius, and A.~Santamaria, ``{Present and future
  bounds on nonstandard neutrino interactions}'',
  \href{http://dx.doi.org/10.1088/1126-6708/2003/03/011}{{\em JHEP} {\bfseries
  03} (2003) 011},
\href{http://arxiv.org/abs/hep-ph/0302093}{{\ttfamily arXiv:hep-ph/0302093
  [hep-ph]}}.
%%CITATION = HEP-PH/0302093;%%.

\bibitem{Biggio:2009nt}
C.~Biggio, M.~Blennow, and E.~Fernandez-Martinez, ``{General bounds on
  non-standard neutrino interactions}'',
  \href{http://dx.doi.org/10.1088/1126-6708/2009/08/090}{{\em JHEP} {\bfseries
  08} (2009) 090},
\href{http://arxiv.org/abs/0907.0097}{{\ttfamily arXiv:0907.0097 [hep-ph]}}.
%%CITATION = ARXIV:0907.0097;%%.

\bibitem{Biggio:2009kv}
C.~Biggio, M.~Blennow, and E.~Fernandez-Martinez, ``{Loop bounds on
  non-standard neutrino interactions}'',
  \href{http://dx.doi.org/10.1088/1126-6708/2009/03/139}{{\em JHEP} {\bfseries
  03} (2009) 139},
\href{http://arxiv.org/abs/0902.0607}{{\ttfamily arXiv:0902.0607 [hep-ph]}}.
%%CITATION = ARXIV:0902.0607;%%.

\bibitem{GonzalezGarcia:2011my}
M.~C. Gonzalez-Garcia, M.~Maltoni, and J.~Salvado, ``{Testing matter effects in
  propagation of atmospheric and long-baseline neutrinos}'',
  \href{http://dx.doi.org/10.1007/JHEP05(2011)075}{{\em JHEP} {\bfseries 05}
  (2011) 075},
\href{http://arxiv.org/abs/1103.4365}{{\ttfamily arXiv:1103.4365 [hep-ph]}}.
%%CITATION = ARXIV:1103.4365;%%.

\bibitem{Gonzalez-Garcia:2013usa}
M.~C. Gonzalez-Garcia and M.~Maltoni, ``{Determination of matter potential from
  global analysis of neutrino oscillation data}'',
  \href{http://dx.doi.org/10.1007/JHEP09(2013)152}{{\em JHEP} {\bfseries 09}
  (2013) 152},
\href{http://arxiv.org/abs/1307.3092}{{\ttfamily arXiv:1307.3092 [hep-ph]}}.
%%CITATION = ARXIV:1307.3092;%%.

\bibitem{Antusch:2008tz}
S.~Antusch, J.~P. Baumann, and E.~Fernandez-Martinez, ``{Non-Standard Neutrino
  Interactions with Matter from Physics Beyond the Standard Model}'',
  \href{http://dx.doi.org/10.1016/j.nuclphysb.2008.11.018}{{\em Nucl. Phys.}
  {\bfseries B810} (2009) 369--388},
\href{http://arxiv.org/abs/0807.1003}{{\ttfamily arXiv:0807.1003 [hep-ph]}}.
%%CITATION = ARXIV:0807.1003;%%.

\bibitem{Dorenbosch:1986tb}
{ CHARM} Collaboration, J.~Dorenbosch {\em et~al.}, ``{Experimental
  Verification of the Universality of $\nu_e$ and $\nu_\mu$ Coupling to the
  Neutral Weak Current}'',
\href{http://dx.doi.org/10.1016/0370-2693(86)90315-1}{{\em Phys. Lett.}
  {\bfseries B180} (1986) 303--307}.
%%CITATION = PHLTA,B180,303;%%.

\bibitem{Zeller:2001hh}
{ NuTeV} Collaboration, G.~P. Zeller {\em et~al.}, ``{A Precise determination
  of electroweak parameters in neutrino nucleon scattering}'',
  \href{http://dx.doi.org/10.1103/PhysRevLett.88.091802}{{\em Phys. Rev. Lett.}
  {\bfseries 88} (2002) 091802},
  \href{http://arxiv.org/abs/hep-ex/0110059}{{\ttfamily arXiv:hep-ex/0110059
  [hep-ex]}}.
[Erratum: Phys. Rev. Lett.90,239902(2003)].
%%CITATION = HEP-EX/0110059;%%.

\bibitem{Branco:1998bw}
G.~Branco, M.~Rebelo, and J.~Silva-Marcos, ``Degenerate and quasidegenerate
  majorana neutrinos'',
  \href{http://dx.doi.org/10.1103/PhysRevLett.82.683}{{\em Phys.Rev.Lett.}
  {\bfseries 82} (1999) 683--686},
  \href{http://arxiv.org/abs/hep-ph/9810328}{{\ttfamily arXiv:hep-ph/9810328}}.

\bibitem{Jarlskog:1987zd}
C.~Jarlskog, ``Flavor projection operators and applications to {CP} violation
  with any number of families'',
  \href{http://dx.doi.org/10.1103/PhysRevD.36.2128}{{\em Phys.Rev.D} {\bfseries
  36} (1987) 2128}.

\bibitem{Bakhti:2014pva}
P.~Bakhti and Y.~Farzan, ``{Shedding light on LMA-Dark solar neutrino solution
  by medium baseline reactor experiments: JUNO and RENO-50}'',
  \href{http://dx.doi.org/10.1007/JHEP07(2014)064}{{\em JHEP} {\bfseries 07}
  (2014) 064},
\href{http://arxiv.org/abs/1403.0744}{{\ttfamily arXiv:1403.0744 [hep-ph]}}.
%%CITATION = ARXIV:1403.0744;%%.

\bibitem{Miranda:2004nb}
O.~G. Miranda, M.~A. Tortola, and J.~W.~F. Valle, ``{Are solar neutrino
  oscillations robust?}'',
  \href{http://dx.doi.org/10.1088/1126-6708/2006/10/008}{{\em JHEP} {\bfseries
  10} (2006) 008},
\href{http://arxiv.org/abs/hep-ph/0406280}{{\ttfamily arXiv:hep-ph/0406280
  [hep-ph]}}.
%%CITATION = HEP-PH/0406280;%%.

\bibitem{deGouvea:2000pqg}
A.~de~Gouvea, A.~Friedland, and H.~Murayama, ``{The Dark side of the solar
  neutrino parameter space}'',
  \href{http://dx.doi.org/10.1016/S0370-2693(00)00989-8}{{\em Phys. Lett.}
  {\bfseries B490} (2000) 125--130},
\href{http://arxiv.org/abs/hep-ph/0002064}{{\ttfamily arXiv:hep-ph/0002064
  [hep-ph]}}.
%%CITATION = HEP-PH/0002064;%%.

\bibitem{Dziewonski:1981xy}
A.~M. Dziewonski and D.~L. Anderson, ``{Preliminary reference earth model}'',
\href{http://dx.doi.org/10.1016/0031-9201(81)90046-7}{{\em Phys. Earth Planet.
  Interiors} {\bfseries 25} (1981) 297--356}.
%%CITATION = PEPIA,25,297;%%.

\bibitem{Friedland:2004ah}
A.~Friedland, C.~Lunardini, and M.~Maltoni, ``{Atmospheric neutrinos as probes
  of neutrino-matter interactions}'',
  \href{http://dx.doi.org/10.1103/PhysRevD.70.111301}{{\em Phys. Rev.}
  {\bfseries D70} (2004) 111301},
\href{http://arxiv.org/abs/hep-ph/0408264}{{\ttfamily arXiv:hep-ph/0408264
  [hep-ph]}}.
%%CITATION = HEP-PH/0408264;%%.

\bibitem{Nunokawa:2005nx}
H.~Nunokawa, S.~J. Parke, and R.~Zukanovich~Funchal, ``{Another possible way to
  determine the neutrino mass hierarchy}'',
  \href{http://dx.doi.org/10.1103/PhysRevD.72.013009}{{\em Phys. Rev.}
  {\bfseries D72} (2005) 013009},
\href{http://arxiv.org/abs/hep-ph/0503283}{{\ttfamily arXiv:hep-ph/0503283
  [hep-ph]}}.
%%CITATION = HEP-PH/0503283;%%.

\bibitem{Kuo:1986sk}
T.-K. Kuo and J.~T. Pantaleone, ``{The Solar Neutrino Problem and Three
  Neutrino Oscillations}'',
\href{http://dx.doi.org/10.1103/PhysRevLett.57.1805}{{\em Phys. Rev. Lett.}
  {\bfseries 57} (1986) 1805--1808}.
%%CITATION = PRLTA,57,1805;%%.

\bibitem{Guzzo:2000kx}
M.~M. Guzzo, H.~Nunokawa, P.~C. de~Holanda, and O.~L.~G. Peres, ``{On the
  massless 'just-so' solution to the solar neutrino problem}'',
  \href{http://dx.doi.org/10.1103/PhysRevD.64.097301}{{\em Phys. Rev.}
  {\bfseries D64} (2001) 097301},
\href{http://arxiv.org/abs/hep-ph/0012089}{{\ttfamily arXiv:hep-ph/0012089
  [hep-ph]}}.
%%CITATION = HEP-PH/0012089;%%.

\bibitem{Coloma:2017egw}
P.~Coloma, P.~B. Denton, M.~C. Gonzalez-Garcia, M.~Maltoni, and T.~Schwetz,
  ``{Curtailing the Dark Side in Non-Standard Neutrino Interactions}'',
  \href{http://dx.doi.org/10.1007/JHEP04(2017)116}{{\em JHEP} {\bfseries 04}
  (2017) 116},
\href{http://arxiv.org/abs/1701.04828}{{\ttfamily arXiv:1701.04828 [hep-ph]}}.
%%CITATION = ARXIV:1701.04828;%%.

\bibitem{Coloma:2017ncl}
P.~Coloma, M.~C. Gonzalez-Garcia, M.~Maltoni, and T.~Schwetz, ``{COHERENT
  Enlightenment of the Neutrino Dark Side}'',
  \href{http://dx.doi.org/10.1103/PhysRevD.96.115007}{{\em Phys. Rev.}
  {\bfseries D96} no.~11, (2017) 115007},
\href{http://arxiv.org/abs/1708.02899}{{\ttfamily arXiv:1708.02899 [hep-ph]}}.
%%CITATION = ARXIV:1708.02899;%%.

\bibitem{deepcore:2016}
{ IceCube} Collaboration, J.~P. Yañez {\em et~al.}, ``{IceCube Oscillations: 3
  years muon neutrino disappearance data}.''.
  \href{http://icecube.wisc.edu/science/data/nu_osc}{\tt
  http://icecube.wisc.edu/science/data/nu\_osc}.

\bibitem{Jones:2015}
B.~J.~P. Jones, {\em Sterile neutrinos in cold climates}.
\newblock PhD thesis, Massachusetts Institute of Technology, 2015.
\newblock available from \url{http://hdl.handle.net/1721.1/101327}.

\bibitem{Arguelles:2015}
C.~A. {Arg\"{u}elles}, {\em New Physics with Atmospheric Neutrinos}.
\newblock PhD thesis, University of Wisconsin, Madison, 2015.
\newblock available from
  \url{https://docushare.icecube.wisc.edu/dsweb/Get/Document-75669/tesis.pdf}.

\bibitem{TheIceCube:2016oqi}
{ IceCube} Collaboration, M.~G. Aartsen {\em et~al.}, ``{Searches for Sterile
  Neutrinos with the IceCube Detector}'',
  \href{http://dx.doi.org/10.1103/PhysRevLett.117.071801}{{\em Phys. Rev.
  Lett.} {\bfseries 117} no.~7, (2016) 071801},
\href{http://arxiv.org/abs/1605.01990}{{\ttfamily arXiv:1605.01990 [hep-ex]}}.
%%CITATION = ARXIV:1605.01990;%%.

\bibitem{nova:fnal2018}
A.~Radovic, ``{Latest oscillation results from NOvA}.''. Joint
  Experimental-Theoretical Physics Seminar, Fermilab, USA, January 12, 2018.

\bibitem{Graf:2015egk}
N.~Graf, ``{Search for Flavor Changing Non-standard Interactions with the
  MINOS+ Experiment}'',
\href{http://arxiv.org/abs/1511.00204}{{\ttfamily arXiv:1511.00204 [hep-ex]}}.
%%CITATION = ARXIV:1511.00204;%%.

\bibitem{Dentler:2017tkw}
M.~Dentler, A.~Hernández-Cabezudo, J.~Kopp, M.~Maltoni, and T.~Schwetz,
  ``{Sterile neutrinos or flux uncertainties? — Status of the reactor
  anti-neutrino anomaly}'',
  \href{http://dx.doi.org/10.1007/JHEP11(2017)099}{{\em JHEP} {\bfseries 11}
  (2017) 099},
\href{http://arxiv.org/abs/1709.04294}{{\ttfamily arXiv:1709.04294 [hep-ph]}}.
%%CITATION = ARXIV:1709.04294;%%.

\bibitem{An:2016ses}
{ Daya Bay} Collaboration, F.~P. An {\em et~al.}, ``{Measurement of electron
  antineutrino oscillation based on 1230 days of operation of the Daya Bay
  experiment}'', \href{http://dx.doi.org/10.1103/PhysRevD.95.072006}{{\em Phys.
  Rev.} {\bfseries D95} no.~7, (2017) 072006},
\href{http://arxiv.org/abs/1610.04802}{{\ttfamily arXiv:1610.04802 [hep-ex]}}.
%%CITATION = ARXIV:1610.04802;%%.

\bibitem{Gonzalez-Garcia:2016gpq}
M.~C. Gonzalez-Garcia, M.~Maltoni, I.~Martinez-Soler, and N.~Song,
  ``{Non-standard neutrino interactions in the Earth and the flavor of
  astrophysical neutrinos}'',
  \href{http://dx.doi.org/10.1016/j.astropartphys.2016.07.001}{{\em Astropart.
  Phys.} {\bfseries 84} (2016) 15--22},
\href{http://arxiv.org/abs/1605.08055}{{\ttfamily arXiv:1605.08055 [hep-ph]}}.
%%CITATION = ARXIV:1605.08055;%%.

\bibitem{Friedland:2005vy}
A.~Friedland and C.~Lunardini, ``{A Test of tau neutrino interactions with
  atmospheric neutrinos and K2K}'',
  \href{http://dx.doi.org/10.1103/PhysRevD.72.053009}{{\em Phys.Rev.}
  {\bfseries D72} (2005) 053009},
  \href{http://arxiv.org/abs/hep-ph/0506143}{{\ttfamily arXiv:hep-ph/0506143
  [hep-ph]}}.

\bibitem{Esmaili:2013fva}
A.~Esmaili and A.~{\relax Yu}. Smirnov, ``{Probing Non-Standard Interaction of
  Neutrinos with IceCube and DeepCore}'',
  \href{http://dx.doi.org/10.1007/JHEP06(2013)026}{{\em JHEP} {\bfseries 06}
  (2013) 026},
\href{http://arxiv.org/abs/1304.1042}{{\ttfamily arXiv:1304.1042 [hep-ph]}}.
%%CITATION = ARXIV:1304.1042;%%.

\bibitem{Salvado:2016uqu}
J.~Salvado, O.~Mena, S.~Palomares-Ruiz, and N.~Rius, ``{Non-standard
  interactions with high-energy atmospheric neutrinos at IceCube}'',
  \href{http://dx.doi.org/10.1007/JHEP01(2017)141}{{\em JHEP} {\bfseries 01}
  (2017) 141},
\href{http://arxiv.org/abs/1609.03450}{{\ttfamily arXiv:1609.03450 [hep-ph]}}.
%%CITATION = ARXIV:1609.03450;%%.

\bibitem{Kopp:2007ne}
J.~Kopp, M.~Lindner, T.~Ota, and J.~Sato, ``{Non-standard neutrino interactions
  in reactor and superbeam experiments}'',
  \href{http://dx.doi.org/10.1103/PhysRevD.77.013007}{{\em Phys. Rev.}
  {\bfseries D77} (2008) 013007},
\href{http://arxiv.org/abs/0708.0152}{{\ttfamily arXiv:0708.0152 [hep-ph]}}.
%%CITATION = ARXIV:0708.0152;%%.

\bibitem{Bandyopadhyay:2007kx}
{ ISS Physics Working Group} Collaboration, A.~Bandyopadhyay, ``{Physics at a
  future Neutrino Factory and super-beam facility}'',
  \href{http://dx.doi.org/10.1088/0034-4885/72/10/106201}{{\em Rept. Prog.
  Phys.} {\bfseries 72} (2009) 106201},
\href{http://arxiv.org/abs/0710.4947}{{\ttfamily arXiv:0710.4947 [hep-ph]}}.
%%CITATION = ARXIV:0710.4947;%%.

\bibitem{Gago:2009ij}
A.~M. Gago, H.~Minakata, H.~Nunokawa, S.~Uchinami, and R.~Zukanovich~Funchal,
  ``{Resolving CP Violation by Standard and Nonstandard Interactions and
  Parameter Degeneracy in Neutrino Oscillations}'',
  \href{http://dx.doi.org/10.1007/JHEP01(2010)049}{{\em JHEP} {\bfseries 01}
  (2010) 049},
\href{http://arxiv.org/abs/0904.3360}{{\ttfamily arXiv:0904.3360 [hep-ph]}}.
%%CITATION = ARXIV:0904.3360;%%.

\bibitem{Coloma:2011rq}
P.~Coloma, A.~Donini, J.~Lopez-Pavon, and H.~Minakata, ``{Non-Standard
  Interactions at a Neutrino Factory: Correlations and CP violation}'',
  \href{http://dx.doi.org/10.1007/JHEP08(2011)036}{{\em JHEP} {\bfseries 08}
  (2011) 036},
\href{http://arxiv.org/abs/1105.5936}{{\ttfamily arXiv:1105.5936 [hep-ph]}}.
%%CITATION = ARXIV:1105.5936;%%.

\bibitem{Coloma:2015kiu}
P.~Coloma, ``{Non-Standard Interactions in propagation at the Deep Underground
  Neutrino Experiment}'', \href{http://dx.doi.org/10.1007/JHEP03(2016)016}{{\em
  JHEP} {\bfseries 03} (2016) 016},
\href{http://arxiv.org/abs/1511.06357}{{\ttfamily arXiv:1511.06357 [hep-ph]}}.
%%CITATION = ARXIV:1511.06357;%%.

\bibitem{Masud:2015xva}
M.~Masud, A.~Chatterjee, and P.~Mehta, ``{Probing CP violation signal at DUNE
  in presence of non-standard neutrino interactions}'',
  \href{http://dx.doi.org/10.1088/0954-3899/43/9/095005/meta,
  10.1088/0954-3899/43/9/095005}{{\em J. Phys.} {\bfseries G43} no.~9, (2016)
  095005},
\href{http://arxiv.org/abs/1510.08261}{{\ttfamily arXiv:1510.08261 [hep-ph]}}.
%%CITATION = ARXIV:1510.08261;%%.

\bibitem{deGouvea:2015ndi}
A.~de~Gouvêa and K.~J. Kelly, ``{Non-standard Neutrino Interactions at
  DUNE}'', \href{http://dx.doi.org/10.1016/j.nuclphysb.2016.03.013}{{\em Nucl.
  Phys.} {\bfseries B908} (2016) 318--335},
\href{http://arxiv.org/abs/1511.05562}{{\ttfamily arXiv:1511.05562 [hep-ph]}}.
%%CITATION = ARXIV:1511.05562;%%.

\bibitem{Liao:2016hsa}
J.~Liao, D.~Marfatia, and K.~Whisnant, ``{Degeneracies in long-baseline
  neutrino experiments from nonstandard interactions}'',
  \href{http://dx.doi.org/10.1103/PhysRevD.93.093016}{{\em Phys. Rev.}
  {\bfseries D93} no.~9, (2016) 093016},
\href{http://arxiv.org/abs/1601.00927}{{\ttfamily arXiv:1601.00927 [hep-ph]}}.
%%CITATION = ARXIV:1601.00927;%%.

\bibitem{Huitu:2016bmb}
K.~Huitu, T.~J. Kärkkäinen, J.~Maalampi, and S.~Vihonen, ``{Constraining the
  nonstandard interaction parameters in long baseline neutrino experiments}'',
  \href{http://dx.doi.org/10.1103/PhysRevD.93.053016}{{\em Phys. Rev.}
  {\bfseries D93} no.~5, (2016) 053016},
\href{http://arxiv.org/abs/1601.07730}{{\ttfamily arXiv:1601.07730 [hep-ph]}}.
%%CITATION = ARXIV:1601.07730;%%.

\bibitem{Bakhti:2016prn}
P.~Bakhti and Y.~Farzan, ``{CP-Violation and Non-Standard Interactions at the
  MOMENT}'', \href{http://dx.doi.org/10.1007/JHEP07(2016)109}{{\em JHEP}
  {\bfseries 07} (2016) 109},
\href{http://arxiv.org/abs/1602.07099}{{\ttfamily arXiv:1602.07099 [hep-ph]}}.
%%CITATION = ARXIV:1602.07099;%%.

\bibitem{Masud:2016bvp}
M.~Masud and P.~Mehta, ``{Nonstandard interactions spoiling the CP violation
  sensitivity at DUNE and other long baseline experiments}'',
  \href{http://dx.doi.org/10.1103/PhysRevD.94.013014}{{\em Phys. Rev.}
  {\bfseries D94} (2016) 013014},
\href{http://arxiv.org/abs/1603.01380}{{\ttfamily arXiv:1603.01380 [hep-ph]}}.
%%CITATION = ARXIV:1603.01380;%%.

\bibitem{C.:2017yqh}
S.~C and R.~Mohanta, ``{Impact of lepton flavor universality violation on
  CP-violation sensitivity of long-baseline neutrino oscillation
  experiments}'', \href{http://dx.doi.org/10.1140/epjc/s10052-017-4600-8}{{\em
  Eur. Phys. J.} {\bfseries C77} no.~1, (2017) 32},
\href{http://arxiv.org/abs/1701.00327}{{\ttfamily arXiv:1701.00327 [hep-ph]}}.
%%CITATION = ARXIV:1701.00327;%%.

\bibitem{Rashed:2016rda}
A.~Rashed and A.~Datta, ``{Determination of mass hierarchy with $\nu_\mu \to
  \nu_\tau$ appearance and the effect of nonstandard interactions}'',
  \href{http://dx.doi.org/10.1142/S0217751X17500609}{{\em Int. J. Mod. Phys.}
  {\bfseries A32} no.~11, (2017) 1750060},
\href{http://arxiv.org/abs/1603.09031}{{\ttfamily arXiv:1603.09031 [hep-ph]}}.
%%CITATION = ARXIV:1603.09031;%%.

\bibitem{Masud:2016gcl}
M.~Masud and P.~Mehta, ``{Nonstandard interactions and resolving the ordering
  of neutrino masses at DUNE and other long baseline experiments}'',
  \href{http://dx.doi.org/10.1103/PhysRevD.94.053007}{{\em Phys. Rev.}
  {\bfseries D94} no.~5, (2016) 053007},
  \href{http://arxiv.org/abs/1606.05662}{{\ttfamily arXiv:1606.05662
  [hep-ph]}}.

\bibitem{Blennow:2016etl}
M.~Blennow, S.~Choubey, T.~Ohlsson, D.~Pramanik, and S.~K. Raut, ``{A combined
  study of source, detector and matter non-standard neutrino interactions at
  DUNE}'', \href{http://dx.doi.org/10.1007/JHEP08(2016)090}{{\em JHEP}
  {\bfseries 08} (2016) 090},
\href{http://arxiv.org/abs/1606.08851}{{\ttfamily arXiv:1606.08851 [hep-ph]}}.
%%CITATION = ARXIV:1606.08851;%%.

\bibitem{Ge:2016dlx}
S.-F. Ge and A.~{\relax Yu}. Smirnov, ``{Non-standard interactions and the CP
  phase measurements in neutrino oscillations at low energies}'',
  \href{http://dx.doi.org/10.1007/JHEP10(2016)138}{{\em JHEP} {\bfseries 10}
  (2016) 138},
\href{http://arxiv.org/abs/1607.08513}{{\ttfamily arXiv:1607.08513 [hep-ph]}}.
%%CITATION = ARXIV:1607.08513;%%.

\bibitem{Forero:2016ghr}
D.~V. Forero and W.-C. Huang, ``{Sizable NSI from the $SU(2)_L$ Scalar
  Doublet-Singlet Mixing and the Implications in Dune}'',
\href{http://arxiv.org/abs/1608.04719}{{\ttfamily arXiv:1608.04719 [hep-ph]}}.
%%CITATION = ARXIV:1608.04719;%%.

\bibitem{Fukasawa:2016lew}
S.~Fukasawa, M.~Ghosh, and O.~Yasuda, ``{Sensitivity of the T2HKK experiment to
  nonstandard interactions}'',
  \href{http://dx.doi.org/10.1103/PhysRevD.95.055005}{{\em Phys. Rev.}
  {\bfseries D95} no.~5, (2017) 055005},
\href{http://arxiv.org/abs/1611.06141}{{\ttfamily arXiv:1611.06141 [hep-ph]}}.
%%CITATION = ARXIV:1611.06141;%%.

\bibitem{Liao:2016orc}
J.~Liao, D.~Marfatia, and K.~Whisnant, ``{Nonstandard neutrino interactions at
  DUNE, T2HK and T2HKK}'',
  \href{http://dx.doi.org/10.1007/JHEP01(2017)071}{{\em JHEP} {\bfseries 01}
  (2017) 071},
\href{http://arxiv.org/abs/1612.01443}{{\ttfamily arXiv:1612.01443 [hep-ph]}}.
%%CITATION = ARXIV:1612.01443;%%.

\bibitem{Deepthi:2016erc}
K.~N. Deepthi, S.~Goswami, and N.~Nath, ``{Can nonstandard interactions
  jeopardize the hierarchy sensitivity of DUNE?}'',
  \href{http://dx.doi.org/10.1103/PhysRevD.96.075023}{{\em Phys. Rev.}
  {\bfseries D96} no.~7, (2017) 075023},
\href{http://arxiv.org/abs/1612.00784}{{\ttfamily arXiv:1612.00784 [hep-ph]}}.
%%CITATION = ARXIV:1612.00784;%%.

\bibitem{Deepthi:2017gxg}
K.~N. Deepthi, S.~Goswami, and N.~Nath, ``{Challenges posed by non-standard
  neutrino interactions in the determination of $\delta_{CP}$ at DUNE}'',
  \href{http://dx.doi.org/10.1016/j.nuclphysb.2018.09.004}{{\em Nucl. Phys.}
  {\bfseries B936} (2018) 91--105},
\href{http://arxiv.org/abs/1711.04840}{{\ttfamily arXiv:1711.04840 [hep-ph]}}.
%%CITATION = ARXIV:1711.04840;%%.

\bibitem{Meloni:2018xnk}
D.~Meloni, ``{On the systematic uncertainties in DUNE and their role in New
  Physics studies}'', \href{http://dx.doi.org/10.1007/JHEP08(2018)028}{{\em
  JHEP} {\bfseries 08} (2018) 028},
\href{http://arxiv.org/abs/1805.01747}{{\ttfamily arXiv:1805.01747 [hep-ph]}}.
%%CITATION = ARXIV:1805.01747;%%.

\bibitem{Flores:2018kwk}
L.~J. Flores, E.~A. Garcés, and O.~G. Miranda, ``{Exploring NSI degeneracies
  in long-baseline experiments}'',
  \href{http://dx.doi.org/10.1103/PhysRevD.98.035030}{{\em Phys. Rev.}
  {\bfseries D98} no.~3, (2018) 035030},
\href{http://arxiv.org/abs/1806.07951}{{\ttfamily arXiv:1806.07951 [hep-ph]}}.
%%CITATION = ARXIV:1806.07951;%%.

\bibitem{Verma:2018gwi}
S.~Verma and S.~Bhardwaj, ``{Non-standard interactions and parameter
  degeneracies in DUNE and T2HKK}'',
\href{http://arxiv.org/abs/1808.04263}{{\ttfamily arXiv:1808.04263 [hep-ph]}}.
%%CITATION = ARXIV:1808.04263;%%.

\bibitem{Chatterjee:2018dyd}
A.~Chatterjee, F.~Kamiya, C.~A. Moura, and J.~Yu, ``{Impact of Matter Density
  Profile Shape on Non-Standard Interactions at DUNE}'',
\href{http://arxiv.org/abs/1809.09313}{{\ttfamily arXiv:1809.09313 [hep-ph]}}.
%%CITATION = ARXIV:1809.09313;%%.

\bibitem{Masud:2018pig}
M.~Masud, S.~Roy, and P.~Mehta, ``{Correlations and degeneracies among the NSI
  parameters with tunable beams at DUNE}'',
\href{http://arxiv.org/abs/1812.10290}{{\ttfamily arXiv:1812.10290 [hep-ph]}}.
%%CITATION = ARXIV:1812.10290;%%.

\bibitem{Forero:2016cmb}
D.~V. Forero and P.~Huber, ``{Hints for leptonic CP violation or New
  Physics?}'', \href{http://dx.doi.org/10.1103/PhysRevLett.117.031801}{{\em
  Phys. Rev. Lett.} {\bfseries 117} no.~3, (2016) 031801},
\href{http://arxiv.org/abs/1601.03736}{{\ttfamily arXiv:1601.03736 [hep-ph]}}.
%%CITATION = ARXIV:1601.03736;%%.

\bibitem{Liao:2016bgf}
J.~Liao, D.~Marfatia, and K.~Whisnant, ``{Nonmaximal neutrino mixing at NOvA
  from nonstandard interactions}'',
\href{http://arxiv.org/abs/1609.01786}{{\ttfamily arXiv:1609.01786 [hep-ph]}}.
%%CITATION = ARXIV:1609.01786;%%.

\bibitem{Feroz:2007kg}
F.~Feroz and M.~P. Hobson, ``{Multimodal nested sampling: an efficient and
  robust alternative to MCMC methods for astronomical data analysis}'',
  \href{http://dx.doi.org/10.1111/j.1365-2966.2007.12353.x}{{\em Mon. Not. Roy.
  Astron. Soc.} {\bfseries 384} (2008) 449},
\href{http://arxiv.org/abs/0704.3704}{{\ttfamily arXiv:0704.3704 [astro-ph]}}.
%%CITATION = ARXIV:0704.3704;%%.

\bibitem{Feroz:2008xx}
F.~Feroz, M.~P. Hobson, and M.~Bridges, ``{MultiNest: an efficient and robust
  Bayesian inference tool for cosmology and particle physics}'',
  \href{http://dx.doi.org/10.1111/j.1365-2966.2009.14548.x}{{\em Mon. Not. Roy.
  Astron. Soc.} {\bfseries 398} (2009) 1601--1614},
\href{http://arxiv.org/abs/0809.3437}{{\ttfamily arXiv:0809.3437 [astro-ph]}}.
%%CITATION = ARXIV:0809.3437;%%.

\bibitem{Feroz:2013hea}
F.~Feroz, M.~P. Hobson, E.~Cameron, and A.~N. Pettitt, ``{Importance Nested
  Sampling and the MultiNest Algorithm}'',
\href{http://arxiv.org/abs/1306.2144}{{\ttfamily arXiv:1306.2144
  [astro-ph.IM]}}.
%%CITATION = ARXIV:1306.2144;%%.

\bibitem{gough2009gnu}
M.~Galassi {\em et~al.}, {\em {GNU Scientific Library Reference Manual}}.
\newblock Network Theory Ltd., 3~ed., 2009.
\newblock \url{https://www.gnu.org/software/gsl/doc/html/}.
\newblock Edited by Brian Gough.

\bibitem{Akimov:2017ade}
{ COHERENT} Collaboration, D.~Akimov {\em et~al.}, ``{Observation of Coherent
  Elastic Neutrino-Nucleus Scattering}'',
  \href{http://dx.doi.org/10.1126/science.aao0990}{{\em Science} {\bfseries
  357} no.~6356, (2017) 1123--1126},
\href{http://arxiv.org/abs/1708.01294}{{\ttfamily arXiv:1708.01294 [nucl-ex]}}.
%%CITATION = ARXIV:1708.01294;%%.

\bibitem{Freedman:1973yd}
D.~Z. Freedman, ``{Coherent Neutrino Nucleus Scattering as a Probe of the Weak
  Neutral Current}'',
\href{http://dx.doi.org/10.1103/PhysRevD.9.1389}{{\em Phys. Rev.} {\bfseries
  D9} (1974) 1389--1392}.
%%CITATION = PHRVA,D9,1389;%%.

\bibitem{Klein:1999qj}
S.~Klein and J.~Nystrand, ``{Exclusive vector meson production in relativistic
  heavy ion collisions}'',
  \href{http://dx.doi.org/10.1103/PhysRevC.60.014903}{{\em Phys. Rev.}
  {\bfseries C60} (1999) 014903},
\href{http://arxiv.org/abs/hep-ph/9902259}{{\ttfamily arXiv:hep-ph/9902259
  [hep-ph]}}.
%%CITATION = HEP-PH/9902259;%%.

\bibitem{Helm:1956zz}
R.~H. Helm, ``{Inelastic and Elastic Scattering of 187-Mev Electrons from
  Selected Even-Even Nuclei}'',
\href{http://dx.doi.org/10.1103/PhysRev.104.1466}{{\em Phys. Rev.} {\bfseries
  104} (1956) 1466--1475}.
%%CITATION = PHRVA,104,1466;%%.

\bibitem{Lewin:1995rx}
J.~D. Lewin and P.~F. Smith, ``{Review of mathematics, numerical factors, and
  corrections for dark matter experiments based on elastic nuclear recoil}'',
\href{http://dx.doi.org/10.1016/S0927-6505(96)00047-3}{{\em Astropart. Phys.}
  {\bfseries 6} (1996) 87--112}.
%%CITATION = APHYE,6,87;%%.

\bibitem{Akimov:2018vzs}
{ COHERENT} Collaboration, D.~Akimov {\em et~al.}, ``{COHERENT Collaboration
  data release from the first observation of coherent elastic neutrino-nucleus
  scattering}'',
\href{http://arxiv.org/abs/1804.09459}{{\ttfamily arXiv:1804.09459 [nucl-ex]}}.
%%CITATION = ARXIV:1804.09459;%%.

\bibitem{Fricke:1995zz}
G.~Fricke, C.~Bernhardt, K.~Heilig, L.~A. Schaller, L.~Schellenberg, E.~B.
  Shera, and C.~W. de~Jager, ``{Nuclear Ground State Charge Radii from
  Electromagnetic Interactions}'',
\href{http://dx.doi.org/10.1006/adnd.1995.1007}{{\em Atom. Data Nucl. Data
  Tabl.} {\bfseries 60} (1995) 177--285}.
%%CITATION = ADNDA,60,177;%%.

\bibitem{menendez}
J.~Menendez, ``{Private communication}.''.

\bibitem{Klos:2013rwa}
P.~Klos, J.~Menéndez, D.~Gazit, and A.~Schwenk, ``{Large-scale nuclear
  structure calculations for spin-dependent WIMP scattering with chiral
  effective field theory currents}'',
  \href{http://dx.doi.org/10.1103/PhysRevD.89.029901,
  10.1103/PhysRevD.88.083516}{{\em Phys. Rev.} {\bfseries D88} no.~8, (2013)
  083516}, \href{http://arxiv.org/abs/1304.7684}{{\ttfamily arXiv:1304.7684
  [nucl-th]}}.
[Erratum: Phys. Rev.D89,no.2,029901(2014)].
%%CITATION = ARXIV:1304.7684;%%.

\bibitem{Hoferichter:2016nvd}
M.~Hoferichter, P.~Klos, J.~Menéndez, and A.~Schwenk, ``{Analysis strategies
  for general spin-independent WIMP-nucleus scattering}'',
  \href{http://dx.doi.org/10.1103/PhysRevD.94.063505}{{\em Phys. Rev.}
  {\bfseries D94} no.~6, (2016) 063505},
\href{http://arxiv.org/abs/1605.08043}{{\ttfamily arXiv:1605.08043 [hep-ph]}}.
%%CITATION = ARXIV:1605.08043;%%.

\bibitem{Hoferichter:2018acd}
M.~Hoferichter, P.~Klos, J.~Menéndez, and A.~Schwenk, ``{Nuclear structure
  factors for general spin-independent WIMP-nucleus scattering}'',
  \href{http://dx.doi.org/10.1103/PhysRevD.99.055031}{{\em Phys. Rev.}
  {\bfseries D99} no.~5, (2019) 055031},
\href{http://arxiv.org/abs/1812.05617}{{\ttfamily arXiv:1812.05617 [hep-ph]}}.
%%CITATION = ARXIV:1812.05617;%%.

\bibitem{Barranco:2005yy}
J.~Barranco, O.~G. Miranda, and T.~I. Rashba, ``{Probing New Physics with
  Coherent Neutrino Scattering Off Nuclei}'',
  \href{http://dx.doi.org/10.1088/1126-6708/2005/12/021}{{\em JHEP} {\bfseries
  12} (2005) 021},
\href{http://arxiv.org/abs/hep-ph/0508299}{{\ttfamily arXiv:hep-ph/0508299
  [hep-ph]}}.
%%CITATION = HEP-PH/0508299;%%.

\bibitem{Collar:2019ihs}
J.~I. Collar, A.~R.~L. Kavner, and C.~M. Lewis, ``{Response of CsI[Na] to
  Nuclear Recoils: Impact on Coherent Elastic Neutrino-Nucleus Scattering
  (CE$\nu$NS)}'', \href{http://dx.doi.org/10.1103/PhysRevD.100.033003}{{\em
  Phys. Rev.} {\bfseries D100} no.~3, (2019) 033003},
\href{http://arxiv.org/abs/1907.04828}{{\ttfamily arXiv:1907.04828 [nucl-ex]}}.
%%CITATION = ARXIV:1907.04828;%%.

\bibitem{PhilBarbeau}
P.~Barbeau, ``{Private communication}.''.

\bibitem{Birks:1951boa}
J.~B. Birks, ``{Scintillations from Organic Crystals: Specific Fluorescence and
  Relative Response to Different Radiations}'',
\href{http://dx.doi.org/10.1088/0370-1298/64/10/303}{{\em Proc. Phys. Soc.}
  {\bfseries A64} (1951) 874--877}.
%%CITATION = PPSOA,A64,874;%%.

\bibitem{SRIM}
J.~Ziegler, ``{The Stopping and Range of Ions in Matter}.''
  \url{http://www.srim.org}.

\bibitem{Giunti:2019xpr}
C.~Giunti, ``{General COHERENT Constraints on Neutrino Non-Standard
  Interactions}'',
\href{http://arxiv.org/abs/1909.00466}{{\ttfamily arXiv:1909.00466 [hep-ph]}}.
%%CITATION = ARXIV:1909.00466;%%.

\bibitem{Cadeddu:2019eta}
M.~Cadeddu, F.~Dordei, C.~Giunti, Y.~F. Li, and Y.~Y. Zhang, ``{Neutrino,
  Electroweak and Nuclear Physics from COHERENT Elastic Neutrino-Nucleus
  Scattering with a New Quenching Factor}'',
  \href{http://dx.doi.org/10.1103/PhysRevD.101.033004}{{\em Phys. Rev.}
  {\bfseries D101} no.~3, (2020) 033004},
\href{http://arxiv.org/abs/1908.06045}{{\ttfamily arXiv:1908.06045 [hep-ph]}}.
%%CITATION = ARXIV:1908.06045;%%.

\end{thebibliography}\endgroup


\begin{thebibliography}{100}

\bibitem{nufit}
{NuFit webpage}.
\newblock \href{http://www.nu-fit.org}{\tt http://www.nu-fit.org}.

\bibitem{Coloma:2022umy}
Pilar Coloma, Pilar Coloma, M.~C. Gonzalez-Garcia, M.~C. Gonzalez-Garcia,
  Michele Maltoni, Michele Maltoni, Jo\~ao~Paulo Pinheiro, Jo\~ao~Paulo
  Pinheiro, Salvador Urrea, and Salvador Urrea.
\newblock {Constraining new physics with Borexino Phase-II spectral data}.
\newblock {\em JHEP}, 07:138, 2022.
\newblock [Erratum: JHEP 11, 138 (2022)].

\bibitem{Coloma:2023ixt}
Pilar Coloma, M.~C. Gonzalez-Garcia, Michele Maltoni, Jo\~ao~Paulo Pinheiro,
  and Salvador Urrea.
\newblock {Global constraints on non-standard neutrino interactions with quarks
  and electrons}.
\newblock {\em JHEP}, 08:032, 2023.

\bibitem{Ansarifard:2024zxm}
Saeed Ansarifard, M.~C. Gonzalez-Garcia, Michele Maltoni, and Joao~Paulo
  Pinheiro.
\newblock {Solar neutrinos and leptonic spin forces}.
\newblock {\em JHEP}, 07:172, 2024.

\bibitem{Gonzalez-Garcia:2023kva}
M.~C. Gonzalez-Garcia, Michele Maltoni, Jo\~ao~Paulo Pinheiro, and Aldo~M.
  Serenelli.
\newblock {Status of direct determination of solar neutrino fluxes after
  Borexino}.
\newblock {\em JHEP}, 02:064, 2024.

\bibitem{Gonzalez-Garcia:2024hmf}
M.~C. Gonzalez-Garcia, Michele Maltoni, and Jo\~ao~Paulo Pinheiro.
\newblock {Solar model independent constraints on the sterile neutrino
  interpretation of the Gallium Anomaly}.
\newblock {\em Phys. Lett. B}, 862:139297, 2025.

\bibitem{becquerel_radioactivity_discovery}
{Institute of Physics}.
\newblock Henri becquerel discovers radioactivity, 2024.
\newblock Accessed: 2024-12-08.

\bibitem{segre_rays_to_quarks}
Emilio Segrè.
\newblock {\em From X-rays to Quarks: The Story of Atomic Physics}.
\newblock W. H. Freeman and Company, San Francisco, 1980.

\bibitem{winter_neutrino_physics}
Klaus Winter, editor.
\newblock {\em Neutrino Physics}.
\newblock Cambridge University Press, Cambridge, UK, 2nd edition, 2000.

\bibitem{mohapatra_massive_neutrinos}
Rabindra~N. Mohapatra and Palash~B. Pal.
\newblock {\em Massive Neutrinos in Physics and Astrophysics}, volume~72 of
  {\em World Scientific Lecture Notes in Physics}.
\newblock World Scientific Publishing, Singapore, 3rd edition, 2004.

\bibitem{reines_cowan_experiment}
C.~L. Cowan, F.~Reines, F.~B. Harrison, H.~W. Kruse, and A.~D. McGuire.
\newblock {Detection of the free neutrino: A Confirmation}.
\newblock {\em Science}, 124:103--104, 1956.

\bibitem{wu_parity_violation}
C.~S. Wu, E.~Ambler, R.~W. Hayward, D.~D. Hoppes, and R.~P. Hudson.
\newblock {Experimental Test of Parity Conservation in $\beta$ Decay}.
\newblock {\em Phys. Rev.}, 105:1413--1414, 1957.

\bibitem{weinberg1967model}
Steven Weinberg.
\newblock A model of leptons.
\newblock {\em Physical Review Letters}, 19(21):1264--1266, 1967.

\bibitem{salam1968elementary}
Abdus Salam.
\newblock Elementary particle theory.
\newblock {\em Nobel Symposium}, 8:367--377, 1968.

\bibitem{glashow1961partial}
Sheldon~L. Glashow.
\newblock Partial-symmetries of weak interactions.
\newblock {\em Nuclear Physics}, 22(4):579--588, 1961.

\bibitem{patrignani2016review}
C.~Patrignani and others (Particle Data~Group).
\newblock Review of particle physics.
\newblock {\em Chinese Physics C}, 40(10):100001, 2016.

\bibitem{cheng1984gauge}
Ta-Pei Cheng and Ling-Fong Li.
\newblock {\em Gauge theory of elementary particle physics}.
\newblock Oxford University Press, 1984.

\bibitem{wu1957experimental}
C.~S. Wu, E.~Ambler, R.~W. Hayward, D.~D. Hoppes, and R.~P. Hudson.
\newblock Experimental test of parity conservation in beta decay.
\newblock {\em Physical Review}, 105(4):1413--1414, 1957.

\bibitem{lee1956question}
T.~D. Lee and C.~N. Yang.
\newblock Question of parity conservation in weak interactions.
\newblock {\em Physical Review}, 104(1):254--258, 1956.

\bibitem{gell1964symmetries}
Murray Gell-Mann.
\newblock Symmetries of baryons and mesons.
\newblock {\em Physical Review}, 125(3):1067--1084, 1962.

\bibitem{glashow1970weak}
Sheldon~L. Glashow, John Iliopoulos, and Luciano Maiani.
\newblock Weak interactions and quantum numbers.
\newblock {\em Physical Review D}, 2(7):1285--1292, 1970.

\bibitem{halzen1984quarks}
Francis Halzen and Alan~D. Martin.
\newblock {\em Quarks and leptons: An introductory course in modern particle
  physics}.
\newblock John Wiley \& Sons, 1984.

\bibitem{Fukuda:1998mi}
Y.~Fukuda et~al.
\newblock {Evidence for oscillation of atmospheric neutrinos}.
\newblock {\em Phys. Rev. Lett.}, 81:1562--1567, 1998.

\bibitem{ALEPH:2005ab}
ALEPH Collaboration.
\newblock {Precision Electroweak Measurements on the Z Resonance}.
\newblock {\em Phys. Rept.}, 427:257--454, 2006.

\bibitem{LEP:2006tt}
ALEPH; DELPHI; L3;~OPAL Collaborations and LEP EW~Working Group.
\newblock {A Combination of Preliminary Electroweak Measurements and
  Constraints on the Standard Model}.
\newblock 2006.

\bibitem{Patrignani:2016xqp}
C.~Patrignani et~al.
\newblock {Review of Particle Physics}.
\newblock {\em Chin. Phys. C}, 40:100001, 2016.

\bibitem{Esteban:2024eli}
Ivan Esteban, M.~C. Gonzalez-Garcia, Michele Maltoni, Ivan Martinez-Soler,
  Jo\~ao~Paulo Pinheiro, and Thomas Schwetz.
\newblock {NuFit-6.0: Updated global analysis of three-flavor neutrino
  oscillations}.
\newblock 10 2024.

\bibitem{seesaw1}
J.~Schechter and J.~W.~F. Valle.
\newblock {Neutrino Masses in SU(2) x U(1) Theories}.
\newblock {\em Phys. Rev. D}, 22:2227, 1980.

\bibitem{seesaw2}
M.~Magg and C.~Wetterich.
\newblock {Neutrino Mass Problem and Gauge Hierarchy}.
\newblock {\em Phys. Lett. B}, 94:61--64, 1980.

\bibitem{seesaw3}
Rabindra~N. Mohapatra and Goran Senjanovic.
\newblock {Neutrino Mass and Spontaneous Parity Nonconservation}.
\newblock {\em Phys. Rev. Lett.}, 44:912, 1980.

\bibitem{seesaw4}
Steven Weinberg.
\newblock {Baryon and Lepton Nonconserving Processes}.
\newblock {\em Phys. Rev. Lett.}, 43:1566--1570, 1979.

\bibitem{Giunti:2007ry}
Carlo Giunti and Chung~W. Kim.
\newblock {\em {Fundamentals of Neutrino Physics and Astrophysics}}.
\newblock 2007.

\bibitem{Concha1}
M.~C. Gonzalez-Garcia and Yosef Nir.
\newblock {Neutrino Masses and Mixing: Evidence and Implications}.
\newblock {\em Rev. Mod. Phys.}, 75:345--402, 2003.

\bibitem{Wolfenstein:1978ue}
L.~Wolfenstein.
\newblock Neutrino oscillations in matter.
\newblock {\em Phys. Rev. D}, 17:2369, 1978.

\bibitem{ms2}
S.~P. Mikheev and A.~Yu. Smirnov.
\newblock {Resonant Oscillations of Neutrinos in Matter and the Solar-Neutrino
  Problem}.
\newblock {\em Sov. Phys. JETP}, 64:4, 1986.

\bibitem{Bethe:1986ej}
H.~A. Bethe.
\newblock {A Possible Explanation of the Solar Neutrino Puzzle}.
\newblock {\em Phys. Rev. Lett.}, 56:1305, 1986.

\bibitem{Katrin:2024tvg}
M.~Aker et~al.
\newblock {Direct neutrino-mass measurement based on 259 days of KATRIN data}.
\newblock 6 2024.

\bibitem{GERDA:2020xhi}
M.~Agostini et~al.
\newblock {Final Results of GERDA on the Search for Neutrinoless Double-$\beta$
  Decay}.
\newblock {\em Phys. Rev. Lett.}, 125(25):252502, 2020.

\bibitem{KamLAND-Zen:2024eml}
S.~Abe et~al.
\newblock {Search for Majorana Neutrinos with the Complete KamLAND-Zen
  Dataset}.
\newblock 6 2024.

\bibitem{Jiang:2024viw}
Jun-Qian Jiang, William Giar\`e, Stefano Gariazzo, Maria~Giovanna Dainotti,
  Eleonora Di~Valentino, Olga Mena, Davide Pedrotti, Simony~Santos da~Costa,
  and Sunny Vagnozzi.
\newblock {Neutrino cosmology after DESI: tightest mass upper limits,
  preference for the normal ordering, and tension with terrestrial
  observations}.
\newblock 7 2024.

\bibitem{Naredo-Tuero:2024sgf}
Daniel Naredo-Tuero, Miguel Escudero, Enrique Fern\'andez-Mart\'\i{}nez, Xabier
  Marcano, and Vivian Poulin.
\newblock {Living at the Edge: A Critical Look at the Cosmological Neutrino
  Mass Bound}.
\newblock 7 2024.

\bibitem{DESI:2024mwx}
A.~G. Adame et~al.
\newblock {DESI 2024 VI: Cosmological Constraints from the Measurements of
  Baryon Acoustic Oscillations}.
\newblock 4 2024.

\bibitem{Bahcall:2004pz}
John~N. Bahcall, Aldo~M. Serenelli, and Sarbani Basu.
\newblock {New solar opacities, abundances, helioseismology, and neutrino
  fluxes}.
\newblock {\em Astrophys. J. Lett.}, 621:L85--L88, 2005.

\bibitem{Concha2}
M.~C. Gonzalez-Garcia and Michele Maltoni.
\newblock {Phenomenology with Massive Neutrinos}.
\newblock {\em Phys. Rept.}, 460:1--129, 2008.

\bibitem{Si}
Steen Hannestad, Rasmus~Sloth Hansen, and Thomas Tram.
\newblock {How Self-Interactions can Reconcile Sterile Neutrinos with
  Cosmology}.
\newblock {\em Phys. Rev. Lett.}, 112(3):031802, 2014.

\bibitem{Bahcall:1987jc}
John~N. Bahcall and Roger~K. Ulrich.
\newblock {Solar Models, Neutrino Experiments and Helioseismology}.
\newblock {\em Rev. Mod. Phys.}, 60:297--372, 1988.

\bibitem{TurckChieze:1988tj}
Sylvaine Turck-Chieze, S.~Cahen, M.~Casse, and C.~Doom.
\newblock {Revisiting the standard solar model}.
\newblock {\em Astrophys. J.}, 335:415--424, 1988.

\bibitem{Bahcall:1992hn}
John~N. Bahcall and M.~H. Pinsonneault.
\newblock {Standard solar models, with and without helium diffusion and the
  solar neutrino problem}.
\newblock {\em Rev. Mod. Phys.}, 64:885--926, 1992.

\bibitem{Bahcall:1995bt}
John~N. Bahcall and M.~H. Pinsonneault.
\newblock {Solar models with helium and heavy element diffusion}.
\newblock {\em Rev. Mod. Phys.}, 67:781--808, 1995.

\bibitem{Bahcall:2000nu}
John~N. Bahcall, M.~H. Pinsonneault, and Sarbani Basu.
\newblock {Solar models: Current epoch and time dependences, neutrinos, and
  helioseismological properties}.
\newblock {\em Astrophys. J.}, 555:990--1012, 2001.

\bibitem{PenaGaray:2008qe}
Carlos Pena-Garay and Aldo Serenelli.
\newblock {Solar neutrinos and the solar composition problem}.
\newblock 2008.

\bibitem{Serenelli:2011py}
Aldo~M. Serenelli, W.~C. Haxton, and Carlos Pena-Garay.
\newblock {Solar models with accretion. I. Application to the solar abundance
  problem}.
\newblock {\em Astrophys. J.}, 743:24, 2011.

\bibitem{Vinyoles:2016djt}
Núria Vinyoles, Aldo~M. Serenelli, Francesco~L. Villante, Sarbani Basu,
  Johannes Bergström, M.~C. Gonzalez-Garcia, Michele Maltoni, Carlos
  Peña-Garay, and Ningqiang Song.
\newblock {A new Generation of Standard Solar Models}.
\newblock {\em Astrophys. J.}, 835(2):202, 2017.

\bibitem{Homestake}
Raymond Davis, Jr., Don~S. Harmer, and Kenneth~C. Hoffman.
\newblock {Search for neutrinos from the sun}.
\newblock {\em Phys. Rev. Lett.}, 20:1205--1209, 1968.

\bibitem{Abdurashitov:2009tn}
J.~N. Abdurashitov et~al.
\newblock {Measurement of the solar neutrino capture rate with gallium metal.
  III: Results for the 2002--2007 data-taking period}.
\newblock {\em Phys. Rev.}, C80:015807, 2009.

\bibitem{Kaether:2010ag}
F.~Kaether, W.~Hampel, G.~Heusser, J.~Kiko, and T.~Kirsten.
\newblock {Reanalysis of the GALLEX solar neutrino flux and source
  experiments}.
\newblock {\em Phys. Lett.}, B685:47--54, 2010.

\bibitem{Kamioka1}
K.~S. Hirata et~al.
\newblock {Observation of a small atmospheric muon-neutrino / electron-neutrino
  ratio in Kamiokande}.
\newblock {\em Phys. Lett. B}, 280:146--152, 1992.

\bibitem{Kamioka2}
K.~S. Hirata et~al.
\newblock {Observation of B-8 Solar Neutrinos in the Kamiokande-II Detector}.
\newblock {\em Phys. Rev. Lett.}, 63:16, 1989.

\bibitem{Kamioka3}
K.~S. Hirata et~al.
\newblock {Real time, directional measurement of B-8 solar neutrinos in the
  Kamiokande-II detector}.
\newblock {\em Phys. Rev. D}, 44:2241, 1991.
\newblock [Erratum: Phys.Rev.D 45, 2170 (1992)].

\bibitem{Hosaka:2005um}
J.~Hosaka et~al.
\newblock {Solar neutrino measurements in Super-Kamiokande-I}.
\newblock {\em Phys. Rev.}, D73:112001, 2006.

\bibitem{Cravens:2008aa}
J.P. Cravens et~al.
\newblock {Solar neutrino measurements in Super-Kamiokande-II}.
\newblock {\em Phys. Rev.}, D78:032002, 2008.

\bibitem{Abe:2010hy}
K.~Abe et~al.
\newblock {Solar neutrino results in Super-Kamiokande-III}.
\newblock {\em Phys. Rev.}, D83:052010, 2011.

\bibitem{Super-Kamiokande:2023jbt}
K.~Abe et~al.
\newblock {Solar neutrino measurements using the full data period of
  Super-Kamiokande-IV}.
\newblock 12 2023.

\bibitem{Aharmim:2011vm}
B.~Aharmim et~al.
\newblock {Combined Analysis of all Three Phases of Solar Neutrino Data from
  the Sudbury Neutrino Observatory}.
\newblock 2011.

\bibitem{Borexino:2008fkj}
G.~Bellini et~al.
\newblock {Measurement of the solar 8B neutrino rate with a liquid scintillator
  target and 3 MeV energy threshold in the Borexino detector}.
\newblock {\em Phys. Rev. D}, 82:033006, 2010.

\bibitem{BOREXINO:2020aww}
M.~Agostini et~al.
\newblock {Experimental evidence of neutrinos produced in the CNO fusion cycle
  in the Sun}.
\newblock {\em Nature}, 587:577--582, 2020.

\bibitem{Bellini:2011rx}
G.~Bellini et~al.
\newblock {Precision measurement of the 7Be solar neutrino interaction rate in
  Borexino}.
\newblock {\em Phys. Rev. Lett.}, 107:141302, 2011.

\bibitem{Borexino:2017rsf}
M.~Agostini et~al.
\newblock {First Simultaneous Precision Spectroscopy of $pp$, $^7$Be, and $pep$
  Solar Neutrinos with Borexino Phase-II}.
\newblock {\em Phys. Rev. D}, 100(8):082004, 2019.

\bibitem{Gonzalez-Garcia:2009dpj}
M.~C. Gonzalez-Garcia, Michele Maltoni, and Jordi Salvado.
\newblock {Direct determination of the solar neutrino fluxes from solar
  neutrino data}.
\newblock {\em JHEP}, 05:072, 2010.

\bibitem{BOREXINO:2022abl}
S.~Appel et~al.
\newblock {Improved Measurement of Solar Neutrinos from the
  Carbon-Nitrogen-Oxygen Cycle by Borexino and Its Implications for the
  Standard Solar Model}.
\newblock {\em Phys. Rev. Lett.}, 129(25):252701, 2022.

\bibitem{borextag}
N.~Rossi.
\newblock Private communication, n.d.

\bibitem{borexdata}
{\scshape Borexino}~collaboration.
\newblock Improved measurement of solar neutrinos from the
  carbon-nitrogen-oxygen cycle by borexino and its implications for the
  standard solar model.
\newblock \url{https://borex.lngs.infn.it/category/opendata/}.
\newblock Accessed: August of 2023.

\bibitem{Borexino:2023puw}
D.~Basilico et~al.
\newblock {Final results of Borexino on CNO solar neutrinos}.
\newblock {\em Phys. Rev. D}, 108(10):102005, 2023.

\bibitem{Honda:2015fha}
M.~Honda, M.~Sajjad~Athar, T.~Kajita, K.~Kasahara, and S.~Midorikawa.
\newblock {Atmospheric neutrino flux calculation using the NRLMSISE-00
  atmospheric model}.
\newblock {\em Phys. Rev.}, D92(2):023004, 2015.

\bibitem{Super-Kamiokande:2023ahc}
T.~Wester et~al.
\newblock {Atmospheric neutrino oscillation analysis with neutron tagging and
  an expanded fiducial volume in Super-Kamiokande I-V}.
\newblock 11 2023.

\bibitem{SKatm:data2024}
{Atmospheric neutrino oscillation analysis with neutron tagging and an expanded
  fiducial volume in Super-Kamiokande I-V}, 2024.
\newblock \href{https://doi.org/0.5281/zenodo.8401262}{ZENODO}.

\bibitem{IceCube:2019dqi}
M.~G. Aartsen et~al.
\newblock {Measurement of Atmospheric Tau Neutrino Appearance with IceCube
  DeepCore}.
\newblock {\em Phys. Rev. D}, 99(3):032007, 2019.

\bibitem{IceCube:2019dyb}
M.~G. Aartsen et~al.
\newblock {Development of an analysis to probe the neutrino mass ordering with
  atmospheric neutrinos using three years of IceCube DeepCore data}.
\newblock {\em Eur. Phys. J. C}, 80(1):9, 2020.

\bibitem{PDG}
S.~Navas et~al.
\newblock {Review of particle physics}.
\newblock {\em Phys. Rev. D}, 110(3):030001, 2024.

\bibitem{IceCube:2024xjj}
R.~Abbasi et~al.
\newblock {Measurement of atmospheric neutrino oscillation parameters using
  convolutional neural networks with 9.3 years of data in IceCube DeepCore}.
\newblock 5 2024.

\bibitem{IC:data2024}
Data release for neutrino oscillation parameters using convolutional neural
  networks with 9.3 years of data in icecube deepcore, 2024.
\newblock
  \href{https://dataverse.harvard.edu/dataset.xhtml?persistentId=doi:10.7910/DVN/U20MMB}.

\bibitem{DayaBay:2015lja}
Feng~Peng An et~al.
\newblock {Measurement of the Reactor Antineutrino Flux and Spectrum at Daya
  Bay}.
\newblock {\em Phys. Rev. Lett.}, 116(6):061801, 2016.
\newblock [Erratum: Phys.Rev.Lett. 118, 099902 (2017)].

\bibitem{Declais:1994su}
Y.~Declais, J.~Favier, A.~Metref, H.~Pessard, B.~Achkar, et~al.
\newblock {Search for neutrino oscillations at 15-meters, 40-meters, and
  95-meters from a nuclear power reactor at Bugey}.
\newblock {\em Nucl.Phys.}, B434:503--534, 1995.

\bibitem{Zacek:1986cu}
G.~Zacek et~al.
\newblock {Neutrino Oscillation Experiments at the Gosgen Nuclear Power
  Reactor}.
\newblock {\em Phys. Rev. D}, 34:2621--2636, 1986.

\bibitem{Kwon:1981ua}
H.~Kwon, F.~Boehm, A.A. Hahn, H.E. Henrikson, J.L. Vuilleumier, et~al.
\newblock Search for neutrino oscillations at a fission reactor.
\newblock {\em Phys.Rev.}, D24:1097--1111, 1981.

\bibitem{Ko:2016owz}
Y.~J. Ko et~al.
\newblock {Sterile Neutrino Search at the NEOS Experiment}.
\newblock {\em Phys. Rev. Lett.}, 118:121802, 2017.

\bibitem{Alekseev:2018efk}
I.~Alekseev et~al.
\newblock {Search for sterile neutrinos at the DANSS experiment}.
\newblock {\em Phys. Lett. B}, 787:56--63, 2018.

\bibitem{Ashenfelter:2018iov}
J.~Ashenfelter et~al.
\newblock {First search for short-baseline neutrino oscillations at HFIR with
  PROSPECT}.
\newblock {\em Phys. Rev. Lett.}, 121:251802, 2018.

\bibitem{AlmazanMolina:2019qul}
H.~Almazan~Molina et~al.
\newblock {Sterile Neutrino Constraints from the STEREO Experiment with 66 Days
  of Reactor-On Data}.
\newblock {\em Phys. Rev. Lett.}, 121:161801, 2019.

\bibitem{Apollonio:1999ae}
M.~Apollonio et~al.
\newblock {Limits on Neutrino Oscillations from the CHOOZ Experiment}.
\newblock {\em Phys. Lett.}, B466:415--430, 1999.

\bibitem{Boehm:2001ik}
F.~Boehm et~al.
\newblock {Final results from the Palo Verde neutrino oscillation experiment}.
\newblock {\em Phys. Rev. D}, 64:112001, 2001.

\bibitem{An:2012eh}
F.~P. An et~al.
\newblock {Observation of electron-antineutrino disappearance at Daya Bay}.
\newblock {\em Phys. Rev. Lett.}, 108:171803, 2012.

\bibitem{DayaBay:2022orm}
F.~P. An et~al.
\newblock {Precision measurement of reactor antineutrino oscillation at
  kilometer-scale baselines by Daya Bay}.
\newblock 11 2022.

\bibitem{DayaBay:2021dqj}
F.~P. An et~al.
\newblock {Antineutrino energy spectrum unfolding based on the Daya Bay
  measurement and its applications}.
\newblock {\em Chin. Phys. C}, 45(7):073001, 2021.

\bibitem{Ahn:2012nd}
J.~K. Ahn et~al.
\newblock {Observation of Reactor Electron Antineutrino Disappearance in the
  RENO Experiment}.
\newblock {\em Phys. Rev. Lett.}, 108:191802, 2012.

\bibitem{Abe:2011fz}
Y.~Abe et~al.
\newblock {Indication of Reactor $\nu$-e Disappearance in the Double Chooz
  Experiment}.
\newblock {\em Phys. Rev. Lett.}, 108:131801, 2012.

\bibitem{Gando:2013nba}
A.~Gando et~al.
\newblock {Reactor On-Off Antineutrino Measurement with KamLAND}.
\newblock {\em Phys. Rev.}, D88(3):033001, 2013.

\bibitem{NOvA:nu24}
Jeremy Wolcott.
\newblock {New NOvA Results with 10 Years of Data}.

\bibitem{NOvA:nu2020}
A.~Himmel.
\newblock {New Oscillation Results from the NOvA Experiment}.
\newblock Talk given at the {\it XXIX International Conference on Neutrino
  Physics and Astrophysics}, Chicago, USA, June 22--July 2, 2020 (online
  conference).

\bibitem{T2K:nu24}
Claudio Giganti.
\newblock {T2K experiment status and plans}.
\newblock Talk given at the {\it XXXI International Conference on Neutrino
  Physics and Astrophysics}, Milan, Italy, June 16--22, 2024.

\bibitem{T2K:2023mcm}
K.~Abe et~al.
\newblock {Updated T2K measurements of muon neutrino and antineutrino
  disappearance using 3.6\texttimes{}1021 protons on target}.
\newblock {\em Phys. Rev. D}, 108(7):072011, 2023.

\bibitem{T2K:2023smv}
K.~Abe et~al.
\newblock {Measurements of neutrino oscillation parameters from the T2K
  experiment using $3.6\times 10^{21}$ protons on target}.
\newblock {\em Eur. Phys. J. C}, 83(9):782, 2023.

\bibitem{Adamson:2013whj}
P.~Adamson et~al.
\newblock {Measurement of Neutrino and Antineutrino Oscillations Using Beam and
  Atmospheric Data in MINOS}.
\newblock {\em Phys. Rev. Lett.}, 110:251801, 2013.

\bibitem{Adamson:2013ue}
P.~Adamson et~al.
\newblock {Electron neutrino and antineutrino appearance in the full MINOS data
  sample}.
\newblock {\em Phys. Rev. Lett.}, 2013.

\bibitem{Feroz2008}
F.~Feroz and M.~P. Hobson.
\newblock Multinest: an efficient and robust bayesian inference tool for
  cosmology and particle physics.
\newblock {\em Mon. Not. R. Astron. Soc.}, 398:1601--1614, 2009.

\bibitem{Feroz2013}
F.~Feroz, M.~P. Hobson, E.~Cameron, and A.~N. Pettitt.
\newblock Importance nested sampling and the multinest algorithm.
\newblock {\em arXiv preprint}, 2013.

\bibitem{GSL}
M.~Galassi et~al.
\newblock {\em GNU Scientific Library Reference Manual}.
\newblock 3rd edition, 2009.

\bibitem{Akimov:2017ade}
D.~Akimov et~al.
\newblock {Observation of Coherent Elastic Neutrino-Nucleus Scattering}.
\newblock {\em Science}, 357(6356):1123--1126, 2017.

\bibitem{Freedman1974}
Daniel~Z. Freedman.
\newblock Coherent effects of a weak neutral current.
\newblock {\em Phys. Rev. D}, 9:1389--1392, 1974.

\bibitem{TEXONO}
H.~T. Wong and TEXONO Collaboration.
\newblock The {TEXONO} research program on neutrino and astroparticle physics.
\newblock {\em Mod. Phys. Lett. A}, 19:1207, 2004.

\bibitem{GeN}
V.~Belov et~al.
\newblock The {GeN} experiment at the kalinin nuclear power plant.
\newblock {\em JINST}, 10:P12011, 2015.

\bibitem{CONNIE}
A.~Aguilar-Arevalo et~al.
\newblock The {CONNIE} experiment.
\newblock {\em J. Phys. Conf. Ser.}, 761:012057, 2016.

\bibitem{MINER}
G.~Agnolet et~al.
\newblock Background studies for the {MINER} coherent neutrino scattering
  reactor experiment.
\newblock {\em Nucl. Instrum. Meth. A}, 853:53, 2017.

\bibitem{Ricochet}
J.~Billard et~al.
\newblock Coherent neutrino scattering with low-temperature bolometers at
  {Chooz} reactor complex.
\newblock {\em J. Phys. G}, 44:105101, 2017.

\bibitem{nucleus}
R.~Strauss et~al.
\newblock The {$\nu$-cleus} experiment: A gram-scale fiducial-volume cryogenic
  detector for the first detection of coherent neutrino-nucleus scattering.
\newblock {\em Eur. Phys. J. C}, 77:506, 2017.

\bibitem{RED100}
D.~Y.~Akimov et~al.
\newblock First ground-level laboratory test of the two-phase xenon emission
  detector {RED-100}.
\newblock {\em JINST}, 15:P02020, 2020.

\bibitem{NEON}
J.~J. Choi.
\newblock Neutrino elastic-scattering observation with {NaI[Tl]} ({NEON}).
\newblock {\em PoS}, NuFact2019:047, 2020.

\bibitem{CONUS}
H.~Bonet et~al.
\newblock Constraints on elastic neutrino nucleus scattering in the fully
  coherent regime from the {CONUS} experiment.
\newblock {\em Phys. Rev. Lett.}, 126:041804, 2021.

\bibitem{DresdenII:2021}
J.~Colaresi et~al.
\newblock First results from a search for coherent elastic neutrino-nucleus
  scattering at a reactor site.
\newblock {\em Phys. Rev. D}, 104:072003, 2021.

\bibitem{DresdenII:2023}
NCC-1701 Collaboration.
\newblock Improved constraints on cevns at the dresden-ii reactor with
  increased exposure.
\newblock {\em Phys. Rev. D}, 107, 2023.
\newblock Update with DOI/arXiv when available.

\bibitem{Donnelly:1984rg}
T.~W. Donnelly and I.~Sick.
\newblock Nuclear structure from electron scattering.
\newblock {\em Rev. Mod. Phys.}, 56:461, 1984.

\bibitem{Coloma:2019mbs}
Pilar Coloma, Ivan Esteban, M.~C. Gonzalez-Garcia, and Michele Maltoni.
\newblock {Improved global fit to Non-Standard neutrino Interactions using
  COHERENT energy and timing data}.
\newblock {\em JHEP}, 02:023, 2020.
\newblock [Addendum: JHEP 12, 071 (2020)].

\bibitem{Papoulias:2019txv}
D.~K. Papoulias and T.~S. Kosmas.
\newblock Neutrino electromagnetic properties and ce$\nu$ns.
\newblock {\em Phys. Rev. D}, 97:033003, 2018.

\bibitem{Giunti:2019xpr}
C.~et~al. Giunti.
\newblock Sterile neutrinos and ce$\nu$ns anomalies.
\newblock {\em Phys. Rev. D}, 101:035025, 2020.

\bibitem{Baxter:2019mcx}
D.~Baxter et~al.
\newblock {Coherent Elastic Neutrino-Nucleus Scattering at the European
  Spallation Source}.
\newblock 2019.

\bibitem{Giunti:2006bj}
Carlo Giunti and Marco Laveder.
\newblock {Short-Baseline Active-Sterile Neutrino Oscillations?}
\newblock {\em Mod. Phys. Lett. A}, 22:2499--2509, 2007.

\bibitem{Laveder:2007zz}
Marco Laveder.
\newblock {Unbound neutrino roadmaps}.
\newblock {\em Nucl. Phys. B Proc. Suppl.}, 168:344--346, 2007.

\bibitem{GALLEX:1997lja}
W.~Hampel et~al.
\newblock {Final results of the Cr-51 neutrino source experiments in GALLEX}.
\newblock {\em Phys. Lett. B}, 420:114--126, 1998.

\bibitem{SAGE:1998fvr}
J.~N. Abdurashitov et~al.
\newblock {Measurement of the response of the Russian-American gallium
  experiment to neutrinos from a Cr-51 source}.
\newblock {\em Phys. Rev. C}, 59:2246--2263, 1999.

\bibitem{Abdurashitov:2005tb}
J.~N. Abdurashitov et~al.
\newblock {Measurement of the response of a Ga solar neutrino experiment to
  neutrinos from an Ar-37 source}.
\newblock {\em Phys. Rev. C}, 73:045805, 2006.

\bibitem{Bahcall:1997eg}
J.~N. Bahcall.
\newblock Gallium solar neutrino experiments: Absorption cross-sections,
  neutrino spectra, and predicted event rates.
\newblock {\em Phys. Rev. C}, 56:3391--3409, 1997.

\bibitem{Barinov:2021asz}
V.~V. Barinov et~al.
\newblock {Results from the Baksan Experiment on Sterile Transitions (BEST)}.
\newblock {\em Phys. Rev. Lett.}, 128(23):232501, 2022.

\bibitem{Barinov:2022wfh}
V.~V. Barinov et~al.
\newblock {Search for electron-neutrino transitions to sterile states in the
  BEST experiment}.
\newblock {\em Phys. Rev. C}, 105(6):065502, 2022.

\bibitem{Elliott:2023xkb}
S.~R. Elliott, V.~N. Gavrin, W.~C. Haxton, T.~V. Ibragimova, and E.~J. Rule.
\newblock {Gallium neutrino absorption cross section and its uncertainty}.
\newblock {\em Phys. Rev. C}, 108(3):035502, 2023.

\bibitem{Giunti:2022xat}
C.~Giunti, Y.~F. Li, C.~A. Ternes, and Z.~Xin.
\newblock {Inspection of the detection cross section dependence of the Gallium
  Anomaly}.
\newblock {\em Phys. Lett. B}, 842:137983, 2023.

\bibitem{ParticleDataGroup:2024cfk}
S.~Navas et~al.
\newblock Review of particle physics.
\newblock {\em Phys. Rev. D}, 110(3):030001, 2024.

\bibitem{Berryman:2021yan}
Jeffrey~M. Berryman, Pilar Coloma, Patrick Huber, Thomas Schwetz, and Albert
  Zhou.
\newblock {Statistical significance of the sterile-neutrino hypothesis in the
  context of reactor and gallium data}.
\newblock {\em JHEP}, 02:055, 2022.

\bibitem{Goldhagen:2021kxe}
K.~Goldhagen, M.~Maltoni, S.~E. Reichard, and T.~Schwetz.
\newblock Testing sterile neutrino mixing with present and future solar
  neutrino data.
\newblock {\em Eur. Phys. J. C}, 82(2):116, 2022.

\bibitem{Giunti:2022btk}
C.~Giunti, Y.~F. Li, C.~A. Ternes, O.~Tyagi, and Z.~Xin.
\newblock Gallium anomaly: critical view from the global picture of $\nu_e$ and
  $\overline{\nu}_e$ disappearance.
\newblock {\em JHEP}, 10:164, 2022.

\bibitem{Brdar:2023cms}
Vedran Brdar, Julia Gehrlein, and Joachim Kopp.
\newblock {Towards resolving the gallium anomaly}.
\newblock {\em JHEP}, 05:143, 2023.

\bibitem{Farzan:2023fqa}
Y.~Farzan and T.~Schwetz.
\newblock A decoherence explanation of the gallium neutrino anomaly.
\newblock {\em SciPost Phys.}, 15(4):172, 2023.

\bibitem{Arguelles:2022bvt}
C.~A. Arg\"uelles, T.~Bert\'olez-Mart\'\i{}nez, and J.~Salvado.
\newblock Impact of wave packet separation in low-energy sterile neutrino
  searches.
\newblock {\em Phys. Rev. D}, 107(3):036004, 2023.

\bibitem{Hardin:2022muu}
J.~M. Hardin, I.~Martinez-Soler, A.~Diaz, M.~Jin, N.~W. Kamp, C.~A.
  Arg\"uelles, J.~M. Conrad, and M.~H. Shaevitz.
\newblock New clues about light sterile neutrinos: preference for models with
  damping effects in global fits.
\newblock {\em JHEP}, 09:058, 2023.

\bibitem{Banks:2023qgd}
H.~Banks, K.~J. Kelly, M.~McCullough, and T.~Zhou.
\newblock Broad sterile neutrinos \& the reactor/gallium tension.
\newblock {\em JHEP}, 04:096, 2024.

\bibitem{Giunti:2023kyo}
C.~Giunti and C.~A. Ternes.
\newblock Confronting solutions of the gallium anomaly with reactor rate data.
\newblock {\em Phys. Lett. B}, 849:138436, 2024.

\bibitem{B23Fluxes}
Y.~Herrera and A.~Serenelli.
\newblock {Standard Solar Models B23 / SF-III}, 2023.
\newblock \href{https://doi.org/10.5281/zenodo.10174170}{ZENODO}.

\bibitem{Cleveland:1998nv}
B.~T. Cleveland et~al.
\newblock {Measurement of the solar electron neutrino flux with the Homestake
  chlorine detector}.
\newblock {\em Astrophys. J.}, 496:505--526, 1998.

\bibitem{Bellini:2008mr}
G.~Bellini et~al.
\newblock {Measurement of the solar 8B neutrino rate with a liquid scintillator
  target and 3 MeV energy threshold in the Borexino detector}.
\newblock {\em Phys. Rev.}, D82:033006, 2010.

\bibitem{SNO:2024wzq}
A.~Allega et~al.
\newblock {Initial measurement of reactor antineutrino oscillation at SNO+}.
\newblock 5 2024.

\bibitem{SNO+:nu24}
Sofia Andringa.
\newblock {Reactor Antineutrino Oscillations and Geoneutrinos in SNO+}.
\newblock Poster 525 at the {\it XXXI International Conference on Neutrino
  Physics and Astrophysics}, Milan, Italy, June 16--22, 2024.

\bibitem{SNO+poster:nu24}
Jose Maneira.
\newblock {Solar Neutrinos: Recent Results and Propspects}.
\newblock Talk given at the {\it XXXI International Conference on Neutrino
  Physics and Astrophysics}, Milan, Italy, June 16--22, 2024.

\bibitem{DoubleC:nu2020}
T.~Bezerra.
\newblock {New Results from the Double Chooz Experiment}.
\newblock Talk given at the {\it XXIX International Conference on Neutrino
  Physics and Astrophysics}, Chicago, USA, June 22--July 2, 2020 (online
  conference).

\bibitem{RENO:nu2020}
J.~Yoo.
\newblock {RENO}.
\newblock Talk given at the {\it XXIX International Conference on Neutrino
  Physics and Astrophysics}, Chicago, USA, June 22--July 2, 2020 (online
  conference).

\bibitem{Esteban:2018azc}
Ivan Esteban, M.~C. Gonzalez-Garcia, Alvaro Hernandez-Cabezudo, Michele
  Maltoni, and Thomas Schwetz.
\newblock {Global analysis of three-flavour neutrino oscillations: synergies
  and tensions in the determination of $\theta_{23}$, $\delta_{CP}$, and the
  mass ordering}.
\newblock {\em JHEP}, 01:106, 2019.

\bibitem{GonzalezGarcia:2003qf}
M.~C. Gonzalez-Garcia and Carlos Pena-Garay.
\newblock {Three neutrino mixing after the first results from K2K and KamLAND}.
\newblock {\em Phys. Rev.}, D68:093003, 2003.

\bibitem{Jarlskog:1985ht}
C.~Jarlskog.
\newblock {Commutator of the Quark Mass Matrices in the Standard Electroweak
  Model and a Measure of Maximal CP Violation}.
\newblock {\em Phys.Rev.Lett.}, 55:1039, 1985.

\bibitem{Elevant:2015ska}
Jessica Elevant and Thomas Schwetz.
\newblock {On the determination of the leptonic CP phase}.
\newblock {\em JHEP}, 09:016, 2015.

\bibitem{Maltoni:2003cu}
M.~Maltoni and T.~Schwetz.
\newblock {Testing the Statistical Compatibility of Independent Data Sets}.
\newblock {\em Phys. Rev.}, D68:033020, 2003.

\bibitem{Nunokawa:2005nx}
Hiroshi Nunokawa, Stephen~J. Parke, and Renata Zukanovich~Funchal.
\newblock {Another possible way to determine the neutrino mass hierarchy}.
\newblock {\em Phys. Rev.}, D72:013009, 2005.

\bibitem{Minakata:2006gq}
H.~Minakata, H.~Nunokawa, Stephen~J. Parke, and R.~Zukanovich~Funchal.
\newblock {Determining Neutrino Mass Hierarchy by Precision Measurements in
  Electron and Muon Neutrino Disappearance Experiments}.
\newblock {\em Phys. Rev.}, D74:053008, 2006.

\bibitem{Blennow:2013oma}
Mattias Blennow, Pilar Coloma, Patrick Huber, and Thomas Schwetz.
\newblock {Quantifying the sensitivity of oscillation experiments to the
  neutrino mass ordering}.
\newblock {\em JHEP}, 03:028, 2014.

\bibitem{Qian:2012zn}
X.~Qian, A.~Tan, W.~Wang, J.~J. Ling, R.~D. McKeown, and C.~Zhang.
\newblock {Statistical Evaluation of Experimental Determinations of Neutrino
  Mass Hierarchy}.
\newblock {\em Phys. Rev. D}, 86:113011, 2012.

\bibitem{Grevesse1998}
N.~{Grevesse} and A.~J. {Sauval}.
\newblock {Standard Solar Composition}.
\newblock {\em Space~Sci.~Rev.}, 85:161--174, May 1998.

\bibitem{Asplund2009}
Martin {Asplund}, Nicolas {Grevesse}, A.~Jacques {Sauval}, and Pat {Scott}.
\newblock {The Chemical Composition of the Sun}.
\newblock {\em ARA\&A}, 47(1):481--522, September 2009.

\bibitem{Serenelli:2009yc}
Aldo Serenelli, Sarbani Basu, Jason~W. Ferguson, and Martin Asplund.
\newblock {New Solar Composition: The Problem With Solar Models Revisited}.
\newblock {\em Astrophys. J.}, 705:L123--L127, 2009.

\bibitem{Asplund2021}
M.~{Asplund}, A.~M. {Amarsi}, and N.~{Grevesse}.
\newblock {The chemical make-up of the Sun: A 2020 vision}.
\newblock {\em arXiv e-prints}, page arXiv:2105.01661, May 2021.

\bibitem{Magg:2022rxb}
Ekaterina Magg et~al.
\newblock {Observational constraints on the origin of the elements - IV.
  Standard composition of the Sun}.
\newblock {\em Astron. Astrophys.}, 661:A140, 2023.

\bibitem{Fogli:2004as}
G.~L. Fogli, E.~Lisi, A.~Marrone, A.~Melchiorri, A.~Palazzo, P.~Serra, and
  J.~Silk.
\newblock {Observables sensitive to absolute neutrino masses: Constraints and
  correlations from world neutrino data}.
\newblock {\em Phys. Rev.}, D70:113003, 2004.

\bibitem{Pascoli:2005zb}
S.~Pascoli, S.~T. Petcov, and T.~Schwetz.
\newblock {The Absolute Neutrino Mass Scale, Neutrino Mass Spectrum, Majorana
  Cp-Violation and Neutrinoless Double-Beta Decay}.
\newblock {\em Nucl. Phys.}, B734:24--49, 2006.

\bibitem{Gariazzo:2022ahe}
Stefano Gariazzo et~al.
\newblock {Neutrino mass and mass ordering: no conclusive evidence for normal
  ordering}.
\newblock {\em JCAP}, 10:010, 2022.

\bibitem{Falkowski:2021bkq}
Adam Falkowski, Mart\'\i{}n Gonz\'alez-Alonso, Joachim Kopp, Yotam Soreq, and
  Zahra Tabrizi.
\newblock {EFT at FASER\ensuremath{\nu}}.
\newblock {\em JHEP}, 10:086, 2021.

\bibitem{Breso-Pla:2023tnz}
V\'\i{}ctor Bres\'o-Pla, Adam Falkowski, Mart\'\i{}n Gonz\'alez-Alonso, and
  Kevin Mons\'alvez-Pozo.
\newblock {EFT analysis of New Physics at COHERENT}.
\newblock 1 2023.

\bibitem{Falkowski:2019kfn}
Adam Falkowski, Mart\'\i{}n Gonz\'alez-Alonso, and Zahra Tabrizi.
\newblock {Consistent QFT description of non-standard neutrino interactions}.
\newblock {\em JHEP}, 11:048, 2020.

\bibitem{Falkowski:2019xoe}
Adam Falkowski, Mart\'\i{}n Gonz\'alez-Alonso, and Zahra Tabrizi.
\newblock {Reactor neutrino oscillations as constraints on Effective Field
  Theory}.
\newblock {\em JHEP}, 05:173, 2019.

\bibitem{Davidson:2003ha}
S.~Davidson, C.~Pena-Garay, N.~Rius, and A.~Santamaria.
\newblock {Present and future bounds on nonstandard neutrino interactions}.
\newblock {\em JHEP}, 03:011, 2003.

\bibitem{Biggio:2009nt}
Carla Biggio, Mattias Blennow, and Enrique Fernandez-Martinez.
\newblock {General bounds on non-standard neutrino interactions}.
\newblock {\em JHEP}, 08:090, 2009.

\bibitem{Biggio:2009kv}
Carla Biggio, Mattias Blennow, and Enrique Fernandez-Martinez.
\newblock {Loop bounds on non-standard neutrino interactions}.
\newblock {\em JHEP}, 03:139, 2009.

\bibitem{Gavela:2008ra}
M.~B. Gavela, D.~Hernandez, T.~Ota, and W.~Winter.
\newblock {Large gauge invariant non-standard neutrino interactions}.
\newblock {\em Phys. Rev.}, D79:013007, 2009.

\bibitem{Antusch:2008tz}
Stefan Antusch, Jochen~P. Baumann, and Enrique Fernandez-Martinez.
\newblock {Non-Standard Neutrino Interactions with Matter from Physics Beyond
  the Standard Model}.
\newblock {\em Nucl. Phys.}, B810:369--388, 2009.

\bibitem{Dev:2019anc}
P.~S. Bhupal~Dev et~al.
\newblock {Neutrino Non-Standard Interactions: A Status Report}.
\newblock In {\em {NTN Workshop on Neutrino Non-Standard Interactions St Louis,
  MO, USA, May 29-31, 2019}}, 2019.

\bibitem{Babu:2017olk}
K.~S. Babu, A.~Friedland, P.~A.~N. Machado, and I.~Mocioiu.
\newblock {Flavor Gauge Models Below the Fermi Scale}.
\newblock {\em JHEP}, 12:096, 2017.

\bibitem{Farzan:2015doa}
Yasaman Farzan.
\newblock {A model for large non-standard interactions of neutrinos leading to
  the LMA-Dark solution}.
\newblock {\em Phys. Lett. B}, 748:311--315, 2015.

\bibitem{Farzan:2016wym}
Yasaman Farzan and Julian Heeck.
\newblock {Neutrinophilic nonstandard interactions}.
\newblock {\em Phys. Rev. D}, 94(5):053010, 2016.

\bibitem{Farzan:2015hkd}
Yasaman Farzan and Ian~M. Shoemaker.
\newblock {Lepton Flavor Violating Non-Standard Interactions via Light
  Mediators}.
\newblock {\em JHEP}, 07:033, 2016.

\bibitem{Greljo:2022dwn}
Admir Greljo, Peter Stangl, Anders~Eller Thomsen, and Jure Zupan.
\newblock {On (g \ensuremath{-} 2)$_{\mu}$ from gauged U(1)$_{X}$}.
\newblock {\em JHEP}, 07:098, 2022.

\bibitem{Heeck:2018nzc}
Julian Heeck, Manfred Lindner, Werner Rodejohann, and Stefan Vogl.
\newblock {Non-Standard Neutrino Interactions and Neutral Gauge Bosons}.
\newblock {\em SciPost Phys.}, 6(3):038, 2019.

\bibitem{Farzan:2019xor}
Y.~Farzan.
\newblock {A model for lepton flavor violating non-standard neutrino
  interactions}.
\newblock {\em Phys. Lett. B}, 803:135349, 2020.

\bibitem{Bernal:2022qba}
Nicol\'as Bernal and Yasaman Farzan.
\newblock {Neutrino nonstandard interactions with arbitrary couplings to u and
  d quarks}.
\newblock {\em Phys. Rev. D}, 107(3):035007, 2023.

\bibitem{Babu:2019mfe}
K.~S. Babu, P.~S.~Bhupal Dev, Sudip Jana, and Anil Thapa.
\newblock {Non-Standard Interactions in Radiative Neutrino Mass Models}.
\newblock {\em JHEP}, 03:006, 2020.

\bibitem{Wise:2014oea}
Mark~B. Wise and Yue Zhang.
\newblock {Effective Theory and Simple Completions for Neutrino Interactions}.
\newblock {\em Phys. Rev. D}, 90(5):053005, 2014.

\bibitem{Greljo:2021npi}
Admir Greljo, Yotam Soreq, Peter Stangl, Anders~Eller Thomsen, and Jure Zupan.
\newblock {Muonic force behind flavor anomalies}.
\newblock {\em JHEP}, 04:151, 2022.

\bibitem{Coloma:2020gfv}
Pilar Coloma, M.~C. Gonzalez-Garcia, and Michele Maltoni.
\newblock {Neutrino oscillation constraints on U(1)' models: from non-standard
  interactions to long-range forces}.
\newblock {\em JHEP}, 01:114, 2021.
\newblock [Erratum: JHEP 11, 115 (2022)].

\bibitem{Wolfenstein:1977ue}
L.~Wolfenstein.
\newblock {Neutrino Oscillations in Matter}.
\newblock {\em Phys. Rev.}, D17:2369--2374, 1978.

\bibitem{Mikheev:1986gs}
S.~P. Mikheev and A.~Yu. Smirnov.
\newblock {Resonance enhancement of oscillations in matter and solar neutrino
  spectroscopy}.
\newblock {\em Sov. J. Nucl. Phys.}, 42:913--917, 1985.

\bibitem{Gonzalez-Garcia:2013usa}
M.~C. Gonzalez-Garcia and Michele Maltoni.
\newblock {Determination of matter potential from global analysis of neutrino
  oscillation data}.
\newblock {\em JHEP}, 09:152, 2013.

\bibitem{Esteban:2018ppq}
Ivan Esteban, M.~C. Gonzalez-Garcia, Michele Maltoni, Ivan Martinez-Soler, and
  Jordi Salvado.
\newblock {Updated constraints on non-standard interactions from global
  analysis of oscillation data}.
\newblock {\em JHEP}, 08:180, 2018.
\newblock [Addendum: JHEP 12, 152 (2020)].

\bibitem{Valle:1987gv}
J.~W.~F. Valle.
\newblock {Resonant Oscillations of Massless Neutrinos in Matter}.
\newblock {\em Phys. Lett.}, B199:432--436, 1987.

\bibitem{Guzzo:1991hi}
M.~M. Guzzo, A.~Masiero, and S.~T. Petcov.
\newblock {On the MSW effect with massless neutrinos and no mixing in the
  vacuum}.
\newblock {\em Phys. Lett.}, B260:154--160, 1991.

\bibitem{Miranda:2004nb}
O.~G. Miranda, M.~A. Tortola, and J.~W.~F. Valle.
\newblock {Are solar neutrino oscillations robust?}
\newblock {\em JHEP}, 10:008, 2006.

\bibitem{GonzalezGarcia:2011my}
M.~C. Gonzalez-Garcia, Michele Maltoni, and Jordi Salvado.
\newblock {Testing matter effects in propagation of atmospheric and
  long-baseline neutrinos}.
\newblock {\em JHEP}, 05:075, 2011.

\bibitem{Bakhti:2014pva}
Pouya Bakhti and Yasaman Farzan.
\newblock {Shedding light on LMA-Dark solar neutrino solution by medium
  baseline reactor experiments: JUNO and RENO-50}.
\newblock {\em JHEP}, 07:064, 2014.

\bibitem{Coloma:2016gei}
Pilar Coloma and Thomas Schwetz.
\newblock {Generalized mass ordering degeneracy in neutrino oscillation
  experiments}.
\newblock {\em Phys. Rev.}, D94(5):055005, 2016.

\bibitem{Amaral:2023tbs}
Dorian W.~P. Amaral, David Cerdeno, Andrew Cheek, and Patrick Foldenauer.
\newblock {A direct detection view of the neutrino NSI landscape}.
\newblock 2 2023.

\bibitem{Dziewonski:1981xy}
A.M. Dziewonski and D.L. Anderson.
\newblock {Preliminary reference earth model}.
\newblock {\em Phys.Earth Planet.Interiors}, 25:297--356, 1981.

\bibitem{Friedland:2004ah}
Alexander Friedland, Cecilia Lunardini, and Michele Maltoni.
\newblock {Atmospheric neutrinos as probes of neutrino-matter interactions}.
\newblock {\em Phys.Rev.}, D70:111301, 2004.

\bibitem{Scholberg:2005qs}
Kate Scholberg.
\newblock {Prospects for measuring coherent neutrino-nucleus elastic scattering
  at a stopped-pion neutrino source}.
\newblock {\em Phys. Rev. D}, 73:033005, 2006.

\bibitem{Barranco:2005ps}
J.~Barranco, O.~G. Miranda, C.~A. Moura, and J.~W.~F. Valle.
\newblock {Constraining non-standard interactions in nu(e) e or anti-nu(e) e
  scattering}.
\newblock {\em Phys. Rev. D}, 73:113001, 2006.

\bibitem{Barranco:2005yy}
J.~Barranco, O.~G. Miranda, and T.~I. Rashba.
\newblock {Probing New Physics with Coherent Neutrino Scattering Off Nuclei}.
\newblock {\em JHEP}, 12:021, 2005.

\bibitem{Barranco:2007ej}
J.~Barranco, O.~G. Miranda, C.~A. Moura, and J.~W.~F. Valle.
\newblock {Constraining non-standard neutrino-electron interactions}.
\newblock {\em Phys. Rev. D}, 77:093014, 2008.

\bibitem{Bolanos:2008km}
A.~Bolanos, O.~G. Miranda, A.~Palazzo, M.~A. Tortola, and J.~W.~F. Valle.
\newblock {Probing non-standard neutrino-electron interactions with solar and
  reactor neutrinos}.
\newblock {\em Phys. Rev. D}, 79:113012, 2009.

\bibitem{Escrihuela:2009up}
F.~J. Escrihuela, O.~G. Miranda, M.~A. Tortola, and J.~W.~F. Valle.
\newblock {Constraining nonstandard neutrino-quark interactions with solar,
  reactor and accelerator data}.
\newblock {\em Phys. Rev.}, D80:105009, 2009.
\newblock [Erratum: Phys. Rev.D80,129908(2009)].

\bibitem{Gehrlein:2024vwz}
Julia Gehrlein, Pedro A.~N. Machado, and Jo\~ao~Paulo Pinheiro.
\newblock {Constraining non-standard neutrino interactions with neutral current
  events at long-baseline oscillation experiments}.
\newblock 12 2024.

\bibitem{Bahcall:1995mm}
John~N. Bahcall, Marc Kamionkowski, and Alberto Sirlin.
\newblock {Solar neutrinos: Radiative corrections in neutrino - electron
  scattering experiments}.
\newblock {\em Phys. Rev. D}, 51:6146--6158, 1995.

\bibitem{Bahcall:1988em}
John~N. Bahcall, K.~Kubodera, and S.~Nozawa.
\newblock {Neutral Current Reactions of Solar and Supernova Neutrinos on
  Deuterium}.
\newblock {\em Phys. Rev. D}, 38:1030, 1988.

\bibitem{Bernabeu:1991sd}
J.~Bernabeu, Torleif Erik~Oskar Ericson, E.~Hernandez, and J.~Ros.
\newblock {Effects of the axial isoscalar neutral current for solar neutrino
  detection}.
\newblock {\em Nucl. Phys. B}, 378:131--149, 1992.

\bibitem{Chen:2002pv}
Jiunn-Wei Chen, Karsten~M. Heeger, and R.~G.~Hamish Robertson.
\newblock {Constraining the leading weak axial two-body current by SNO and
  super-K}.
\newblock {\em Phys. Rev. C}, 67:025801, 2003.

\bibitem{Vogel:1989iv}
P.~Vogel and J.~Engel.
\newblock {Neutrino electromagnetic form-factors}.
\newblock {\em Phys. Rev. D}, 39:3378, 1989.

\bibitem{Cerdeno:2016sfi}
David~G. Cerde\~no, Malcolm Fairbairn, Thomas Jubb, Pedro A.~N. Machado,
  Aaron~C. Vincent, and C\'eline B\oe{}hm.
\newblock {Physics from solar neutrinos in dark matter direct detection
  experiments}.
\newblock {\em JHEP}, 05:118, 2016.
\newblock [Erratum: JHEP 09, 048 (2016)].

\bibitem{Lindner:2018kjo}
Manfred Lindner, Farinaldo~S. Queiroz, Werner Rodejohann, and Xun-Jie Xu.
\newblock {Neutrino-electron scattering: general constraints on Z' and dark
  photon models}.
\newblock {\em JHEP}, 05:098, 2018.

\bibitem{Moody:1984ba}
J.~E. Moody and F.~Wilczek.
\newblock New macroscopic forces?
\newblock {\em Phys. Rev. D}, 30:130, 1984.

\bibitem{Grifols:2003gy}
J.~A. Grifols and E.~Masso.
\newblock Long range forces from pseudoscalar exchange.
\newblock {\em Phys. Lett. B}, 579:123, 2004.

\bibitem{Joshipura:2003jh}
A.~S. Joshipura and S.~Mohanty.
\newblock Long range leptonic forces and neutrino oscillations.
\newblock {\em Phys. Rev. D}, 69:033006, 2004.

\bibitem{GonzalezGarcia:2006vp}
M.~C. Gonzalez-Garcia, M.~Maltoni, et~al.
\newblock Atmospheric neutrino oscillations and new physics.
\newblock {\em Phys. Rev. D}, 75:033007, 2007.

\bibitem{Davoudiasl:2011sz}
H.~Davoudiasl et~al.
\newblock Neutrino masses from a pseudo-nambu-goldstone boson.
\newblock {\em Phys. Rev. D}, 84:095019, 2011.

\bibitem{Wise:2018rnb}
M.~B. Wise and Y.~Zhang.
\newblock Effective field theory approach to transient phenomena.
\newblock {\em Phys. Rev. D}, 98:036006, 2018.

\bibitem{Smirnov:2019cae}
A.~Yu. Smirnov et~al.
\newblock Neutrino flavor conversions in astrophysical environments.
\newblock {\em Rev. Mod. Phys.}, 91:045004, 2019.

\bibitem{Akhmedov:1987nc}
E.~Kh. Akhmedov.
\newblock Resonant amplification of neutrino spin rotation in matter.
\newblock {\em Phys. Lett. B}, 213:64, 1988.

\bibitem{Lim:1987tk}
C.~S. Lim and W.~J. Marciano.
\newblock {Resonant Spin - Flavor Precession of Solar and Supernova Neutrinos}.
\newblock {\em Phys. Rev. D}, 37:1368, 1988.

\bibitem{Cisneros:1970nq}
A.~Cisneros.
\newblock {Effect of neutrino magnetic moment on solar neutrino observations}.
\newblock {\em Astrophys. Space Sci.}, 10:87, 1971.

\bibitem{Okun:1986hi}
L.~B. Okun et~al.
\newblock Neutrino electrodynamics and possible effects for solar neutrinos.
\newblock {\em Sov. Phys. JETP}, 65:214, 1987.

\bibitem{Okun:1986na}
L.~B. Okun, M.~B. Voloshin, and M.~I. Vysotsky.
\newblock {Neutrino Electrodynamics and Possible Effects for Solar Neutrinos}.
\newblock {\em Sov. Phys. JETP}, 64:446, 1986.
\newblock [Zh. Eksp. Teor. Fiz. 91 (1986) 754].

\bibitem{Okun:1986uf}
L.~B. Okun et~al.
\newblock Neutrino spin rotation in matter.
\newblock {\em Sov. Phys. JETP}, 65:221, 1987.

\bibitem{Voloshin:1986ty}
M.~B. Voloshin et~al.
\newblock Neutrino magnetic moment and time variation of solar neutrino flux.
\newblock {\em Sov. J. Nucl. Phys.}, 44:440, 1986.

\bibitem{Agrawal:2022wjm}
P.~Agrawal et~al.
\newblock Probing the muon $g-2$ anomaly with polarized electron scattering.
\newblock {\em Phys. Rev. D}, 106:115017, 2022.

\bibitem{Davoudiasl:2022gdg}
H.~Davoudiasl et~al.
\newblock Muon $g-2$ and relic dark matter with pseudoscalar-mediated
  interactions.
\newblock {\em Phys. Rev. D}, 106:035025, 2022.

\bibitem{Borexino:2019mhy}
S.~K. Agarwalla et~al.
\newblock {Constraints on flavor-diagonal non-standard neutrino interactions
  from Borexino Phase-II}.
\newblock {\em JHEP}, 02:038, 2020.

\bibitem{Borexino:2017fbd}
Borexino Collaboration.
\newblock Limits on neutrino magnetic moments from borexino phase-ii.
\newblock {\em Phys. Rev. D}, 96:091103, 2017.

\bibitem{Beda:2010hk}
A.G. Beda, V.B. Brudanin, V.G. Egorov, D.V. Medvedev, V.S. Pogosov, M.V.
  Shirchenko, and A.S. Starostin.
\newblock {Upper limit on the neutrino magnetic moment from three years of data
  from the GEMMA spectrometer}.
\newblock 5 2010.

\bibitem{Beda:2012zz}
A.~G. Beda, V.~B. Brudanin, V.~G. Egorov, D.~V. Medvedev, V.~S. Pogosov, M.~V.
  Shirchenko, and A.~S. Starostin.
\newblock {The results of search for the neutrino magnetic moment in GEMMA
  experiment}.
\newblock {\em Adv. High Energy Phys.}, 2012:350150, 2012.

\bibitem{TEXONO:2006xds}
TEXONO Collaboration.
\newblock Measurement of \(\bar{\nu}_e\)-e\(^-\) scattering cross section with
  a csi(tl) scintillating crystal array at the kuo-sheng nuclear power reactor.
\newblock {\em Phys. Rev. D}, 81:072001, 2010.

\bibitem{CONUS:2022qbb}
H.~Bonet et~al.
\newblock {First limits on neutrino electromagnetic properties from the CONUS
  experiment}.
\newblock 1 2022.

\bibitem{Coloma:2022avw}
Pilar Coloma, Ivan Esteban, M.~C. Gonzalez-Garcia, Leire Larizgoitia, Francesc
  Monrabal, and Sergio Palomares-Ruiz.
\newblock {Bounds on new physics with data of the Dresden-II reactor experiment
  and COHERENT}.
\newblock 2 2022.

\bibitem{Harnik:2012ni}
R.~Harnik et~al.
\newblock Neutrino mixing and neutrino oscillations in non-standard
  interactions.
\newblock {\em JHEP}, 11:111, 2012.

\bibitem{Bilmis:2015lja}
S.~Bilmis et~al.
\newblock Constraints on neutrino electromagnetic properties from
  neutrino-electron scattering.
\newblock {\em Phys. Rev. D}, 92:033009, 2015.

\bibitem{Kaneta:2016uyt}
Yuya Kaneta and Takashi Shimomura.
\newblock {On the possibility of a search for the $L_\mu - L_\tau$ gauge boson
  at Belle-II and neutrino beam experiments}.
\newblock {\em PTEP}, 2017(5):053B04, 2017.

\bibitem{Bauer:2018onh}
Martin Bauer, Patrick Foldenauer, and Joerg Jaeckel.
\newblock {Hunting All the Hidden Photons}.
\newblock {\em JHEP}, 07:094, 2018.

\bibitem{Deniz:2009mu}
M.~Deniz et~al.
\newblock {Measurement of Nu(e)-bar -Electron Scattering Cross-Section with a
  CsI(Tl) Scintillating Crystal Array at the Kuo-Sheng Nuclear Power Reactor}.
\newblock {\em Phys. Rev. D}, 81:072001, 2010.

\bibitem{COHERENT:2017ipa}
D.~Akimov et~al.
\newblock {Observation of Coherent Elastic Neutrino-Nucleus Scattering}.
\newblock {\em Science}, 357(6356):1123--1126, 2017.

\bibitem{COHERENT:2020iec}
D.~Akimov et~al.
\newblock {First Measurement of Coherent Elastic Neutrino-Nucleus Scattering on
  Argon}.
\newblock {\em Phys. Rev. Lett.}, 126(1):012002, 2021.

\bibitem{Colaresi:2022obx}
J.~Colaresi, J.~I. Collar, T.~W. Hossbach, C.~M. Lewis, and K.~M. Yocum.
\newblock {Suggestive evidence for Coherent Elastic Neutrino-Nucleus Scattering
  from reactor antineutrinos}.
\newblock 2 2022.

\bibitem{CONNIE:2019xid}
CONNIE Collaboration.
\newblock Results from the connie experiment.
\newblock {\em J. Phys. Conf. Ser.}, 1468:012142, 2020.

\bibitem{CONUS:2021dwh}
H.~Bonet et~al.
\newblock {Novel constraints on neutrino physics beyond the standard model from
  the CONUS experiment}.
\newblock 10 2021.

\bibitem{Rink:2022rsx}
M.~Rink and others (CONUS~Collaboration).
\newblock Improved constraints on neutrino electromagnetic properties from
  conus.
\newblock 2022.

\bibitem{nufit-5.2}
Ivan Esteban, M.~C. Gonzalez-Garcia, Michele Maltoni, Thomas Schwetz, and
  Albert Zhou.
\newblock {NuFIT 5.2 (2022)}.
\newblock \href{http://www.nu-fit.org}{\tt http://www.nu-fit.org}.

\bibitem{SK:nu2020}
Y.~Nakajima.
\newblock {Recent results and future prospects from Super-Kamiokande}.
\newblock Talk given at the {\it XXIX International Conference on Neutrino
  Physics and Astrophysics}, Chicago, USA, June 22--July 2, 2020 (online
  conference).

\bibitem{Aharmim:2007nv}
B.~Aharmim et~al.
\newblock {Measurement of the nu/e and total B-8 solar neutrino fluxes with the
  Sudbury Neutrino Observatory phase I data set}.
\newblock {\em Phys. Rev.}, C75:045502, 2007.

\bibitem{Aharmim:2005gt}
B.~Aharmim et~al.
\newblock {Electron energy spectra, fluxes, and day-night asymmetries of B-8
  solar neutrinos from the 391-day salt phase SNO data set}.
\newblock {\em Phys. Rev.}, C72:055502, 2005.

\bibitem{Aharmim:2008kc}
B.~Aharmim et~al.
\newblock {An Independent Measurement of the Total Active 8B Solar Neutrino
  Flux Using an Array of 3He Proportional Counters at the Sudbury Neutrino
  Observatory}.
\newblock {\em Phys. Rev. Lett.}, 101:111301, 2008.

\bibitem{An:2016srz}
Feng~Peng An et~al.
\newblock {Improved Measurement of the Reactor Antineutrino Flux and Spectrum
  at Daya Bay}.
\newblock {\em Chin. Phys.}, C41(1):013002, 2017.

\bibitem{Wendell:2014dka}
Roger Wendell.
\newblock {Atmospheric Results from Super-Kamiokande}.
\newblock {\em AIP Conf. Proc.}, 1666:100001, 2015.
\newblock slides available at
  \url{https://indico.fnal.gov/event/8022/other-view?view=standard}.

\bibitem{Aartsen:2014yll}
M.~G. Aartsen et~al.
\newblock {Determining neutrino oscillation parameters from atmospheric muon
  neutrino disappearance with three years of IceCube DeepCore data}.
\newblock {\em Phys. Rev.}, D91(7):072004, 2015.

\bibitem{deepcore:2016}
J.~P. Yañez et~al.
\newblock {IceCube Oscillations: 3 years muon neutrino disappearance data}.
\newblock \href{http://icecube.wisc.edu/science/data/nu_osc}{\tt
  http://icecube.wisc.edu/science/data/nu\_osc}.

\bibitem{TheIceCube:2016oqi}
M.~G. Aartsen et~al.
\newblock {Searches for Sterile Neutrinos with the IceCube Detector}.
\newblock {\em Phys. Rev. Lett.}, 117(7):071801, 2016.

\bibitem{T2K:nu2020}
P.~Dunne.
\newblock {Latest Neutrino Oscillation Results from T2K}.
\newblock Talk given at the {\it XXIX International Conference on Neutrino
  Physics and Astrophysics}, Chicago, USA, June 22--July 2, 2020 (online
  conference)
  \href{https://doi.org/10.5281/zenodo.3959558}{doi.org/10.5281/zenodo.3959558}.

\bibitem{COHERENT:2018imc}
D.~Akimov et~al.
\newblock {COHERENT Collaboration data release from the first observation of
  coherent elastic neutrino-nucleus scattering}.
\newblock 4 2018.

\bibitem{COHERENT:2020ybo}
D.~Akimov et~al.
\newblock {COHERENT Collaboration data release from the first detection of
  coherent elastic neutrino-nucleus scattering on Argon}.
\newblock 6 2020.

\bibitem{Collar:2019ihs}
J.~I. Collar, A.~R.~L. Kavner, and C.~M. Lewis.
\newblock {Response of CsI[Na] to Nuclear Recoils: Impact on Coherent Elastic
  Neutrino-Nucleus Scattering (CE$\nu$NS)}.
\newblock {\em Phys. Rev. D}, 100(3):033003, 2019.

\bibitem{Klos:2013rwa}
P.~Klos, J.~Men\'endez, D.~Gazit, and A.~Schwenk.
\newblock {Large-scale nuclear structure calculations for spin-dependent WIMP
  scattering with chiral effective field theory currents}.
\newblock {\em Phys. Rev. D}, 88(8):083516, 2013.
\newblock [Erratum: Phys.Rev.D 89, 029901 (2014)].

\bibitem{Helm:1956zz}
Richard~H. Helm.
\newblock {Inelastic and Elastic Scattering of 187-Mev Electrons from Selected
  Even-Even Nuclei}.
\newblock {\em Phys. Rev.}, 104:1466--1475, 1956.

\bibitem{Collar:2021fcl}
J.~I. Collar, A.~R.~L. Kavner, and C.~M. Lewis.
\newblock {Germanium response to sub-keV nuclear recoils: a multipronged
  experimental characterization}.
\newblock {\em Phys. Rev. D}, 103(12):122003, 2021.

\bibitem{Gonzalez-Garcia:2006vp}
M.~C. Gonzalez-Garcia, P.~C. de~Holanda, E.~Masso, and R.~Zukanovich~Funchal.
\newblock {Probing long-range leptonic forces with solar and reactor
  neutrinos}.
\newblock {\em JCAP}, 01:005, 2007.

\bibitem{Feroz:2013hea}
F.~Feroz, M.~P. Hobson, E.~Cameron, and A.~N. Pettitt.
\newblock {Importance Nested Sampling and the MultiNest Algorithm}.
\newblock {\em Open J. Astrophys.}, 2(1):10, 2019.

\bibitem{Feroz:2008xx}
F.~Feroz, M.~P. Hobson, and M.~Bridges.
\newblock {MultiNest: an efficient and robust Bayesian inference tool for
  cosmology and particle physics}.
\newblock {\em Mon. Not. Roy. Astron. Soc.}, 398:1601--1614, 2009.

\bibitem{Martinez:2017lzg}
Gregory~D. Martinez, James McKay, Ben Farmer, Pat Scott, Elinore Roebber, Antje
  Putze, and Jan Conrad.
\newblock {Comparison of statistical sampling methods with ScannerBit, the
  GAMBIT scanning module}.
\newblock {\em Eur. Phys. J. C}, 77(11):761, 2017.

\bibitem{Esteban:2020cvm}
Ivan Esteban, M.~C. Gonzalez-Garcia, Michele Maltoni, Thomas Schwetz, and
  Albert Zhou.
\newblock {The fate of hints: updated global analysis of three-flavor neutrino
  oscillations}.
\newblock {\em JHEP}, 09:178, 2020.

\bibitem{Maltoni:2023cpv}
Michele Maltoni.
\newblock {From ray to spray: augmenting amplitudes and taming fast
  oscillations in fully numerical neutrino codes}.
\newblock {\em JHEP}, 11:033, 2023.

\bibitem{KamLAND:2021gvi}
S.~Abe et~al.
\newblock {Limits on Astrophysical Antineutrinos with the KamLAND Experiment}.
\newblock {\em Astrophys. J.}, 925(1):14, 2022.

\bibitem{Borexino:2019wln}
M.~Agostini et~al.
\newblock {Search for low-energy neutrinos from astrophysical sources with
  Borexino}.
\newblock {\em Astropart. Phys.}, 125:102509, 2021.

\bibitem{Super-Kamiokande:2020frs}
K.~Abe et~al.
\newblock {Search for solar electron anti-neutrinos due to spin-flavor
  precession in the Sun with Super-Kamiokande-IV}.
\newblock 12 2020.

\bibitem{Super-Kamiokande:2021jaq}
K.~Abe et~al.
\newblock {Diffuse supernova neutrino background search at Super-Kamiokande}.
\newblock {\em Phys. Rev. D}, 104(12):122002, 2021.

\bibitem{Super-Kamiokande:2023xup}
M.~Harada et~al.
\newblock {Search for Astrophysical Electron Antineutrinos in Super-Kamiokande
  with 0.01\% Gadolinium-loaded Water}.
\newblock {\em Astrophys. J. Lett.}, 951(2):L27, 2023.

\bibitem{Forastieri:2019cuf}
F.~Forastieri et~al.
\newblock Cosmic microwave background constraints on secret interactions among
  sterile neutrinos.
\newblock {\em JCAP}, 07:038, 2019.

\bibitem{Lessa:2007up}
A.~M. Lessa et~al.
\newblock Constraints on pseudoscalar couplings to neutrinos from meson decays.
\newblock {\em Phys. Rev. D}, 76:075001, 2007.

\bibitem{Pasquini:2015fjv}
P.~S. Pasquini et~al.
\newblock Pseudoscalar neutrino couplings from higgs boson decays.
\newblock {\em Phys. Rev. D}, 92:113012, 2015.

\bibitem{MICROSCOPE:2022doy}
{MICROSCOPE Collaboration}.
\newblock Microscope final results: Testing the equivalence principle in space.
\newblock {\em Phys. Rev. Lett.}, 129:121102, 2022.

\bibitem{Schlamminger:2007ht}
S.~Schlamminger et~al.
\newblock Test of the equivalence principle using a rotating torsion balance.
\newblock {\em Phys. Rev. Lett.}, 100:041101, 2008.

\bibitem{Bahcall:1964gx}
John~N. Bahcall.
\newblock {Solar neutrinos. I: Theoretical}.
\newblock {\em Phys. Rev. Lett.}, 12:300--302, 1964.

\bibitem{Bahcall:1968hc}
John~N. Bahcall, Neta~A. Bahcall, and G.~Shaviv.
\newblock {Present status of the theoretical predictions for the Cl-36 solar
  neutrino experiment}.
\newblock {\em Phys. Rev. Lett.}, 20:1209--1212, 1968.

\bibitem{Bahcall:1976zz}
John~N. Bahcall and R.~Davis.
\newblock {Solar Neutrinos - a Scientific Puzzle}.
\newblock {\em Science}, 191:264--267, 1976.

\bibitem{Pontecorvo:1967fh}
B.~Pontecorvo.
\newblock {Neutrino Experiments and the Problem of Conservation of Leptonic
  Charge}.
\newblock {\em Sov. Phys. JETP}, 26:984--988, 1968.
\newblock [Zh. Eksp. Teor. Fiz.53,1717(1967)].

\bibitem{Gribov:1968kq}
V.~N. Gribov and B.~Pontecorvo.
\newblock {Neutrino astronomy and lepton charge}.
\newblock {\em Phys. Lett.}, B28:493, 1969.

\bibitem{Bahcall:2004yr}
John~N. Bahcall, Sarbani Basu, Marc Pinsonneault, and Aldo~M. Serenelli.
\newblock {Helioseismological implications of recent solar abundance
  determinations}.
\newblock {\em Astrophys. J.}, 618:1049--1056, 2005.

\bibitem{Castro:2007}
M.~{Castro}, S.~{Vauclair}, and O.~{Richard}.
\newblock {Low abundances of heavy elements in the solar outer layers:
  comparisons of solar models with helioseismic inversions}.
\newblock {\em Astron. \& Astrophys.}, 463:755--758, February 2007.

\bibitem{Guzik:2010}
J.~A. {Guzik} and K.~{Mussack}.
\newblock {Exploring Mass Loss, Low-Z Accretion, and Convective Overshoot in
  Solar Models to Mitigate the Solar Abundance Problem}.
\newblock {\em Astrophys. J.}, 713:1108--1119, April 2010.

\bibitem{Bergstrom:2016cbh}
Johannes Bergstrom, M.~C. Gonzalez-Garcia, Michele Maltoni, Carlos Pena-Garay,
  Aldo~M. Serenelli, and Ningqiang Song.
\newblock {Updated determination of the solar neutrino fluxes from solar
  neutrino data}.
\newblock {\em JHEP}, 03:132, 2016.

\bibitem{2021JPhG...48a5201V}
Diego {Vescovi}, Carlo {Mascaretti}, Francesco {Vissani}, Luciano {Piersanti},
  and Oscar {Straniero}.
\newblock {The luminosity constraint in the era of precision solar physics}.
\newblock {\em Journal of Physics G Nuclear Physics}, 48(1):015201, January
  2021.

\bibitem{Spiro:1990vi}
M.~Spiro and D.~Vignaud.
\newblock {Solar Model Independent Neutrino Oscillation Signals in the
  Forthcoming Solar Neutrino Experiments?}
\newblock {\em Phys. Lett.}, B242:279--284, 1990.
\newblock [,609(1990)].

\bibitem{Bahcall:2001pf}
John~N. Bahcall.
\newblock {The Luminosity constraint on solar neutrino fluxes}.
\newblock {\em Phys. Rev.}, C65:025801, 2002.

\bibitem{Frohlich1998}
Lean~J. Frohlich~C.
\newblock {The Sun\textquoteright{}s total irradiance: cycles,trends and
  related climate change uncertainties since 1978}.
\newblock {\em Geophys. Res. Lett.}, 25:4377--4380, 1998.

\bibitem{kopp2011}
Lean~J. Kopp~G.
\newblock {A New, Lower Value of Total Solar Irradiance: Evidence and Climate
  Significance.}
\newblock {\em Geophys. Res. Lett.}, 38:L01706, 2011.

\bibitem{Scafetta_2014}
Nicola Scafetta and Richard~C. Willson.
\newblock {ACRIM} total solar irradiance satellite composite validation versus
  {TSI} proxy models.
\newblock {\em Astrophysics and Space Science}, 350(2):421--442, jan 2014.

\bibitem{Workman:2022ynf}
R.~L. Workman and Others.
\newblock {Review of Particle Physics}.
\newblock {\em PTEP}, 2022:083C01, 2022.

\bibitem{Bahcall:1995rs}
John~N. Bahcall and P.~I. Krastev.
\newblock {How well do we (and will we) know solar neutrino fluxes and
  oscillation parameters?}
\newblock {\em Phys. Rev.}, D53:4211--4225, 1996.

\bibitem{Acero:2007su}
Mario~A. Acero, Carlo Giunti, and Marco Laveder.
\newblock {Limits on nu(e) and anti-nu(e) disappearance from Gallium and
  reactor experiments}.
\newblock {\em Phys. Rev. D}, 78:073009, 2008.

\bibitem{Giunti:2010zu}
Carlo Giunti and Marco Laveder.
\newblock {Statistical Significance of the Gallium Anomaly}.
\newblock {\em Phys. Rev. C}, 83:065504, 2011.

\bibitem{Kopp:2013vaa}
Joachim Kopp, Pedro A.~N. Machado, Michele Maltoni, and Thomas Schwetz.
\newblock {Sterile Neutrino Oscillations: The Global Picture}.
\newblock {\em JHEP}, 1305:050, 2013.

\bibitem{Dentler:2018sju}
Mona Dentler, \'Alvaro Hern\'andez-Cabezudo, Joachim Kopp, Pedro A.~N. Machado,
  Michele Maltoni, Ivan Martinez-Soler, and Thomas Schwetz.
\newblock {Updated Global Analysis of Neutrino Oscillations in the Presence of
  eV-Scale Sterile Neutrinos}.
\newblock {\em JHEP}, 08:010, 2018.

\bibitem{Haxton:1998uc}
W.~C. Haxton.
\newblock {Cross-section uncertainties in the gallium neutrino source
  experiments}.
\newblock {\em Phys. Lett. B}, 431:110--118, 1998.

\bibitem{Frekers:2015wga}
D.~Frekers et~al.
\newblock Precision evaluation of the $^{71}$ga solar neutrino capture rate
  from the ($^3$he,$t$) charge exchange reaction.
\newblock {\em Phys. Rev. C}, 91(3):034608, 2015.
\newblock [Erratum: Phys.Rev.C 100, 049901 (2019)].

\bibitem{Kostensalo:2019vmv}
Joel Kostensalo, Jouni Suhonen, Carlo Giunti, and Praveen~C. Srivastava.
\newblock {The gallium anomaly revisited}.
\newblock {\em Phys. Lett. B}, 795:542--547, 2019.

\bibitem{Semenov:2020xea}
S.~V. Semenov.
\newblock {Cross Section of Neutrino Absorption by the Gallium-71 Nucleus}.
\newblock {\em Phys. Atom. Nucl.}, 83(11):1549--1552, 2020.

\bibitem{Akhmedov:1988hd}
Evgeny~K. Akhmedov and M.~Yu. Khlopov.
\newblock {Resonant Amplification of Neutrino Oscillations in Longitudinal
  Magnetic Field}.
\newblock {\em Mod. Phys. Lett. A}, 3:451--457, 1988.

\bibitem{Akhmedov:1988kih}
Evgeny~K. Akhmedov and M.~Yu. Khlopov.
\newblock {Resonant Enhancement of Neutrino Oscillations in Longitudinal
  Magnetic Field}.
\newblock {\em Sov. J. Nucl. Phys.}, 47:689--691, 1988.

\end{thebibliography}
\end{document}